\documentclass{book}

\usepackage{fancyhdr}
\usepackage[table]{xcolor}
\usepackage{setspace}
\usepackage{rotating}
\usepackage{amsmath}
\usepackage{amssymb}

\usepackage{hyperref}
\usepackage[utf8]{inputenc}

\usepackage{graphicx}
\usepackage{physics}
\usepackage{tikz}
\usetikzlibrary{quantikz}

\usepackage{biblatex}
\usepackage[printonlyused,withpage,nolist]{acronym}

\usepackage{xcolor}

\usepackage[acronym,nomain,nonumberlist]{glossaries}
\makeglossaries

\definecolor{signaturegray}{gray}{0.98}
\DeclareUnicodeCharacter{2061}{}
\DeclareUnicodeCharacter{0301}{\'{e}}
\usepackage{parskip}

\begin{document}
\doublespacing

\baselineskip 12pt

\vspace*{25mm}

\begin{center}

\textbf{\Large Doctoral Program} \\
\vspace{10mm}
\textbf{In Information Technologies and the application to Management, Architecture and Geophysics. }\\
\vspace{20mm}
{\large La Salle - Ramon Llull University}\\
\vspace{20mm}

%According to Article 14 (g) of the\\
%{\scriptsize Regulation of Study Cycles Leading to the URL's Doctoral Degree\\DR, 2.ª série – N.º 133 – Regulamento n.º 656/2016 de 13 de julio de 2016}\\
\end{center}

\vspace{20mm}

{\scriptsize Candidate name:} %\tab[7cm]Number:}
\\
Parfait Atchadé Adélomou %\tab[7cm]se09062
\\
\vspace{3mm}

{\scriptsize Title:}\\
Quantum Algorithms for solving Hard Constrained Optimisation Problems\\
\vspace{3mm}

{\scriptsize Supervisor 1 name:}\\
Dra. Elisabet Golobardes Ribé\\
\vspace{3mm}

{\scriptsize Supervisor 2 name:}\\
Dr. Xavier Vilasís-Cardona\\
\vspace{3mm}

\pagebreak
\pagestyle{fancy}
\fancyhf{}
%\chead{\includegraphics[width=\textwidth]{media/UTADLOGO.jpg}}
\rfoot{Page \thepage}
%\fancyfoot{\leftmark}
%\fancyhead{\leftmark}
\lfoot{\leftmark}
%\rhead{Page \thepage}
%\lhead{\leftmark}

%\lfoot{PhD in Information Technologies and the application to Management, Architecture and Geophysics}

%%%%%%%%%%%%  Starting New Page here %%%%%%%%%%%%%%

\newpage
\section*{Acknowledgements}
    \textit{To Martina and Africa }

\vspace{\baselineskip}

This PhD thesis research has resulted from numerous encounters, experiences, consultations, conferences, networking, friendships and sleepless nights. There are many people to whom I owe a high debt of gratitude for their invaluable inspiration, comments, critiques, corrections, and references. All of which makes for an incredible and unforgettable academic culmination that now, notably and sadly, reaches the end after three years of hard, challenging and sometimes lonely research.

For this, one of my most sincere and profound thanks goes to the IBMQ team, especially to the \textit{aqua} division represented by \textit{Steve Wood}. For two years of correspondence, comments and technical exchange, we have come to understand many concepts and, above all feel, we are wrapped in a community that gave us all the facilities to achieve our goal.  To Simone Severini from AWS-Braket for his support to achieve our goal and to the quantum community and people I met during this quantum mechanics and computing journey. 

I make a special dedication to Martina to find a door to the quantum world through this work. A hope to understand and correctly model our environment and to have a tool to understand some of the problems or diseases that today have no cure. To Jenny, tireless, loving, respectful and helpful for all the tangible and intangible details to achieve this result. It reminds me of Bell's inequality; Two entangled particles, before the measurement, and even at an immeasurable distance, modifying one, the other remains faithful to the effects of the first and vice versa.

To Africa, since it will be, I believe, one of the indisputable pillars of quantum computing.

This thesis work is not the end but the beginning of a long journey in the quantum world.

I want to be open to any reader, researcher who wishes to share their point of view with me at pifparfait@gmail.com.

\vspace{\baselineskip}
Thank you

\vspace{\baselineskip}
Parfait Atchadé Adélomou

Barcelona 18 August 2021
\newpage
\section*{Abstract}
The thesis deals with Quantum Algorithms for solving Hard Constrained Optimization Problems. It shows how quantum computers can solve difficult everyday problems such as finding the best schedule for social workers or the path of a robot picking and batching in a warehouse. The path to the solution has led to the definition of a new artificial intelligence paradigm with quantum computing, quantum Case-Based Reasoning (qCBR) and to a proof of concept to integrate the capacity of quantum computing within mobile robotics using a Raspberry Pi 4 as a processor (qRobot), capable of operating with leading technology players such as IBMQ, Amazon Braket (D-Wave) and Pennylane.
To improve the execution time of variational algorithms in this NISQ era and the next, we have proposed EVA: a quantum Exponential Value Approximation algorithm that speeds up the VQE, and that is, to date, the flagship of the quantum computation.

\tableofcontents

\newpage
\section* {List of Publications}
\begin{itemize}
    \item The results of this thesis have been made public in seven papers.
    \subitem{*  Atchade-Adelomou P., Golobardes Ribé E., Vilasís Cardona X. (2020) Using the Variational-Quantum-Eigensolver (VQE) to Create an Intelligent Social Workers Schedule Problem Solver. In: de la Cal E.A., Villar Flecha J.R., Quintián H., Corchado E. (eds) Hybrid Artificial Intelligent Systems. HAIS 2020. Lecture Notes in Computer Science, vol 12344. Springer, Cham. https://doi.org/10.1007/978-3-030-61705-9\_21}
    \subitem{* Atchade-Adelomou P., Golobardes Ribé E., Vilasís Cardona X. Formulation of the social workers’ problem in quadratic unconstrained binary optimization form and solve it on a quantum computer. Journal of Computer and Communications. 2020 Nov 5;8(11):44-68. DOI: 10.4236/jcc.2020.811004}
    \subitem{* Atchade-Adelomou P., Alonso-Linaje G, Albo-Canals J, Casado-Fauli D. qRobot: A Quantum computing approach in mobile robot order picking and batching problem solver optimization. MDPI AG, vol 14, number = 7, pages = 194, 2021 May 11. Algorithms, url = https://doi.org/10.3390/a14070194}
     \subitem{* Atchade-Adelomou P, Casado-Fauli D, Golobardes Ribé E, Vilasís Cardona X. quantum Case-Based Reasoning (qCBR). arXiv preprint arXiv:2104.00409. 2021 Apr 1.}
    \subitem{*  Atchade-Adelomou P., Golobardes Ribé E., Vilasís Cardona X. Using the Parameterized Quantum Circuit combined with Variational-Quantum-Eigensolver (VQE) to create an Intelligent social workers' schedule problem solver. arXiv preprint arXiv:2010.05863. 2020 Oct 12.}
    \subitem{* Atchade-Adelomou P., Alonso-Linaje G. Quantum Enhanced Filter: QFilter. arXiv preprint arXiv:2104.03418. 2021 Apr 7.}
    \subitem{* Alonso-Linaje G, Atchade-Adelomou P. EVA: a quantum Exponential Value Approximation algorithm. arXiv preprint arXiv:2106.08731. 2021 Jun 16.}
\end{itemize}

\newpage
\newpage
\listoffigures
\newpage
\listoftables
\newpage

%\newpage

\newacronym{SWP}{SWP}{Social Workers' Problem}
\newacronym{TSP}{TSP}{Travelling Salesman Problem}
\newacronym{VRP}{VRP}{Vehicle Routing Problem}
\newacronym{JSP}{JSP}{Job Shop Problem}
\newacronym{JSSP}{JSSP}{Job Shop Scheduling Problem}

\newacronym{NISQ}{NISQ}{Noisy and Intermediate Scale Quantum}
\newacronym{VQE}{VQE}{Variational Quantum Eigensolver}
\newacronym{VQA}{VQA}{Variational Quantum Algorithm}
\newacronym{VQC}{VQC}{Variational Quantum Classifier}
\newacronym{QAOA}{QAOA}{Quantum Approximate Optimization Algorithm}
	
\newacronym{CBR}{CBR}{Case-Based Reasoning}
\newacronym{SVM}{SVM}{Super Vector Machine}
\newacronym{QSVM}{QSVM}{Quantum Super Vector Machine}
\newacronym{qCBR}{qCBR}{quantum Case-Based Reasoning}
\newacronym{qRobot}{qRobot}{quantum Robot}
\newacronym{QFilter}{QFilter}{quantum Filter}
\newacronym{EVA}{EVA}{quantum Exponential Value Approximation}
\newacronym{QML}{QML}{Quantum Machine Learning}
\newacronym{QEC}{QEC}{Quantum Error Correction}
\newacronym{QECC}{QECC}{Quantum Error Correction Concept}
\newacronym{QUBO}{QUBO}{Quadratic Unconstrained Binary Optimization}

%\end{acronym}

\glsaddall
\printglossary[type=\acronymtype,title=Acronyms]

\part{State of the Art}

%\chapter{Previous work, concepts and approaches to the state of the art}

 %%%%%%%%%%%%  Starting New Page here %%%%%%%%%%%%%%

\newpage

\chapter{Introduction}

One of the most frequent problems in our daily lives is optimisation problems, and for a long time, mathematics has always looked for a way to solve them. The differential calculus technique helps solve these problems for certain purposes that fulfil certain properties. However, many functions that appear when trying to solve or model these tasks do not satisfy the hypotheses of the differential calculus theorems, or have additional restrictions that nullify their usefulness.

One of the successes of the last century in the field of mathematics was the development of the simplex algorithm \cite{Dantzig1954} for solving optimisation problems. This success allowed us to find, and continues to allow us to find, the global optimum of a linear function with linear constraints. However, some limitations of this technique are observed when one wants to generalise the algorithm or to solve more complicated problems that are not exclusively summarised in finding the minimum of a linear function with linear restrictions \cite{dantzig1955generalized}. As a result, several numerical calculation tools were born to approximate optimisation tasks under certain regularity conditions. Still, they are not enough to try to solve all the problems.
Most of the techniques to solve these optimisation problems require continuity and differentiability conditions. However, many of the functions that appear in these tasks to be solved do not meet these conditions, and in some cases, it does not even make sense to talk about these properties because we are dealing with a discrete set. The field that deals with these last problems is known as combinatorial optimisation. \\

The \textit{Travelling Salesman Problem} (TSP) \cite {lenstra1975some} belongs to this kind of problem where the brute force algorithms, due to their computational impracticality cost, grows factorially as a function of $ N $, with $ N $ being the number of cities, making it difficult to have an efficient optimisation algorithm.

Due to this fact, there is then the need to develop specific algorithms for each problem, making an in-depth study of this field and each of the available techniques essential.

Some of these techniques are what are known as heuristic algorithms. The main objective of these heuristic algorithms is to approach the global minimum of a function, but without guaranteeing that it will be achieved. Instead, they find a method of finding good solutions for a certain process (task).
As we discussed earlier, the same argument that prevents the impossibility of finding global minimum search algorithms for any function makes it difficult to choose the appropriate heuristic algorithm for each problem as well. However, once again, the operation of this will depend on the nature of the function with which we are working, making it necessary to adapt it.

Combinatorial optimisation problems, in which social workers visit their patients in their respective homes and attend to them at a specific time, called \textit{Social Workers' Problem} (SWP), are similar to TSP due to its NP-Hard characteristic and require another focus if patient numbers grow considerably. The problem of optimising the schedules of social workers who visit their patients at home is a scheduling and routing problem of the NP-Hard class \cite{BOV94}. These difficulties will be observed with an increasing number of patients where the number of possible solutions grows exponentially with the size of the problem. The tasks subject to the labour of the social workers are \textit{combinatorial optimisation problems} and require an organisation based on time and its execution. Moreover, because of the importance of these problems, we usually find them in our daily life. These tasks may be subject to a finite set of resources and restrictions due, for instance, to the physical characteristics of the environment, temporary relationships, or the labour regulations. In addition to conditions, the goal is to optimise one or more criteria represented by an objective function that is usually related to cost, benefit, or execution time. 

The nature of the SWP invites us to analyse the main generalisations of the TSP in search of a suitable formulation for our problem. To do this, we will analyse some generalisations of TSP such as the \textit{Vehicle Routing Problem} (VRP) \cite {GCl64,MarPs2,PTo} and the \textit{Job Shop Scheduling Problem} (JSSP) \cite{Wbo}.

The generic VRP can be seen as the generalisation of the TSP, aiming to find a set of routes at a minimal cost. Usually, the path is defined from the beginning to the endpoint by passing through the depot to achieve the demand of all nodes.

VRP is an NP-hard combinatorial optimisation problem that can be exactly solved only for small cases of traditional computation problems. An approach that allows better results in practice is the heuristic approach. However, it does not guarantee optimisation. Thus, in recent decades, meta-heuristics have emerged as the most hopeful research focus for the VRP family of problems.

Even with the best restrictions, the VRP (and without the scheduling adaptation) continues to have exponential complexity depending on the growth in the number of input data. This concept leads us to explore new approaches to solve problems, such as large-scale social workers. \textit{Quantum computing} \cite{Ron} could be one of these approaches.

Since Richard Feynman suggested that computing could be done more efficiently by exploiting the power of \textit{quantum parallelism}, \cite{Ron,Har02}, today, there is evidence that makes quantum computing one of the most prominent candidates to replace conventional silicon-based systems partially. During the last three decades, one can find a series of quantum algorithms more efficient than the better-known classical counterparts. The most renowned algorithms are the \textit{Shor} \cite{Pet96} and \textit{Grover} \cite{Eva,Eva07} algorithms. However, there is extensive literature on quantum computing and techniques \cite{Adr18}.

Today’s classical computers operate on individual bits, which store information as binary 0 or 1 states. However, quantum computers use the physical laws of quantum mechanics to manipulate data. The unit of information is represented by a quantum bit or \textit{qubit} at this fundamental level. Tangibly, a qubit is any two-level quantum system. Mathematically, the state space of a single qubit can be associated with the complex projective line over the Hilbert space and satisfy that its euclidean norm is equal to one. A complex quantum computing system is constructed by combining one-qubit systems.

There are two dominant techniques for \textit{quantum computing}. \textit{Continuous-Time Quantum Computing} \cite{Kendon2020} used by D-Wave in which the problem to solve is mapped in quantum hamiltonians (Hamiltonian is the energy representing function, taking into account the energy of every possible configuration of the spins in the ferromagnetic material.) and the natural dynamics of physical systems. The \textit{Gate based Quantum Computing} \cite{McG14, Kir17} led by IBM,  shows the computation made through a series of discrete gate operations. While the following reference \cite{Kendon2020} argues how \textit{Quantum Walk} (QW), \textit{Quantum Annealing} (QA), and \textit{Adiabatic Quantum Computing} (AQC) are related, we highlight that the QW and AQC are pure quantum evolutions (unitary) as opposed to QA which involves external cooling.

The \textit{Adiabatic Quantum Computing} proposed by Farhi \cite{Edw,Edw19}, is based on the \textit{adiabatic theorem} \cite{McG14}. This theorem affirms that if a quantum system is driven by a progressively evolving Hamiltonian, which grows from one starting point of the Hamiltonian ($H_{init}$) to the final point $H_{final}$, and therefore, if the systems start in the ground state of $H_{init}$, the system will finish up in the ground state of $H_{final}$. Adiabatic computing was the first quantum computing technique \cite{McG14}.

Quantum Annealing, based on the adiabatic quantum computing paradigm, was initially introduced by Kadowaki and Nishimori \cite{Nis08}. Since its proposal, the QA technique has been a light for solving combinatorial optimisation problems. This technique tries to solve problems similar to how optimisation problems are solved using the classical simulated annealing \cite{McG14}; from a multivariate function formed from an energy landscape, the ground state corresponds to the optimal solution of the problem. The most significant advantage of quantum annealing is its high parallelism over classical code execution. It analyses all possible inputs in parallel to find the optimal solution using quantum tunnelling. This could be very useful when we want to reduce the complexity of the NP-complete problems.

QA has confirmed its ability to solve a broad range of combinatorial optimisation problems and also problems in other fields, such as quantum chemistry (One of the fields that are taking great advantage of capacity and the era in which quantum computing is right now) \cite{McG14}, bioinformatics\cite{McG14} and routing \cite{MarPs2}, to mention a few.

One of the widely used frameworks that helps to map our combinatorial optimisation problem is Quadratic Binary Optimisation Problems (QUBO) \cite{McG14, Nis08,KBe19}.

QUBO, as NP-hard, refers to a pattern matching technique that, among other applications, can be used in machine learning and optimisation, which involves minimising a quadratic polynomial on binary variables\cite{McG14}. QUBO has demonstrated its potential in solving some standard combinatorial optimisation problems such as the colouring of graphics, workshop planning, vehicle routing and programming, neural networks, the partition problem, 3-SAT, and machine learning where the parameters of the problem can be expressed as Boolean variables \cite{McG14,Nis08,KBe19}. 
As problems are constrained in real-life, constraints can be mapped using the penalty function \cite{lucas2014} taking advantage of the Lagrange multiplier \cite{boyd2004convex}. 
The QUBO formulation is suitable for annealing architecture and has a connection with finding the ground state of the generalised Ising Hamiltonian from statistical mechanics \cite{lucas2014}. That means QUBO can be mapped on the Ising model \cite{McG14}.

Advances in quantum computing offer a way forward for efficient solutions to many cases of substantial eigenvalue problems unsolvable in a traditional way \cite{Alb13}. Quantum approaches to finding eigenvalues previously relied on the Quantum Phase Estimation (QPE) algorithm. The QPE is one of the essential subroutines in quantum computation. It serves as a central building block for many quantum algorithms and offers exponential acceleration compared to classical methods, requiring several quantum operations  $O \left( \frac{1}{p} \right)$  to obtain an estimate with precision  $p$  \cite{Alb13,GGG19}. 

Variational Quantum Eigensolver (VQE), proposed by Peruzzo \cite{Alb13} based on the variational principle and form, estimates the ground state energy of the Hamiltonian \cite{Jer03}. The VQE is a hybrid quantum/classical algorithm originally proposed to approximate the ground state of a quantum system (the state attaining the minimum energy). The VQE can be used to solve approximately the optimisation problems.

Quantum Approximate Optimisation Algorithms (QAOA), based on the principles of adiabatic quantum computation \cite{McG14, GGG19, Qin18}, is used to solve QUBO problems. Farhi and Harrow showed the advantages of QAOA compared to classical approaches \cite{Edw, Edw19}, while Rebentrost \cite{Pat19} just debated the problems of constrained polynomial optimisation using adiabatic quantum computation methods. Other scientists such as Vyskocil and Djidjev \cite{Tom19} worked on how to apply restrictions in QUBO systems to avoid the use of large numbers of the coefficients, thus resulting in more qubits from the use of quadratic penalties, they proposed a new combinatorial design which involved solving problems of linear programming of mixed integers to adapt applications restitution. The following work \cite{Anu19} investigated and solved the Hamiltonian cycle problem in computational frameworks such as quantum circuits, quantum walks, and adiabatic quantum computing. \\

The methods mentioned earlier in quantum computing have been applied to routing and scheduling techniques. For example, the Ref. \cite{Seb19}  contributed an expansive vision and discussions on Ising formulations for various NP-complete and NP-hard optimisation problems, emphasising using as few as possible qubits. In the same way, there have been many works of literature on the VRP  and its variants.

The primary challenge is that existing quantum hardware does not yet seem capable of running algorithms on large enough problem instances. Furthermore, current quantum hardware is in the noisy intermediate-scale quantum (NISQ) era, meaning that the present quantum devices are under-powered and suffer from multiple issues.

Due to the quantum computing era, we are limited by the number of qubits and thus, noise; we have to look for strategies and formulations that help us solve real problems in this NISQ era.

We will take advantage of all these related works to define an appropriate strategy for our thesis project in this NISQ era, going from the formulation of the SWP, its implementation, experimentation, and the various comparisons and the quantum machine learning approach to solve the SWP to its generalisation with the qRobot.

\section{Motivation}
My primary motivation are social and technical. To design quantum algorithms that can be executed in this NISQ era (reducing the number of the qubits) to help African countries fight against extreme poverty and impact Western companies and society on combinatorial optimisation problems, such as last miles, location zone decision, routing, scheduling, picking and batching issues, etc.

\section{Work context}
To develop this PhD thesis consisting of a theoretical and a practical part, I have had to take some courses in the foundations of quantum physics. I have had to develop and deepen optimisation mathematics, visit a quantum laboratory, and work in the Boston ecosystem. I also needed to collaborate in varied workgroups such as IBMQ, Xanadu, google, Quantum World, AWS Braket, local groups such as Hispanic Computation, especially with the PhD Alberto Enciso from advanced mathematics (CSIC). I have had to take some courses on quantum computing work on the Qiskit textbook and  Pennylane tutorial. In addition, I have shared various results and research topics with the team led by Steve Wood and Jan Rainer Lahmann (the CTO of Lufthansa and IBMQ) and work with several different frameworks of payment, free, online and locally (Qiskit, Pennylane, Cirq, Qibo, AWS\_Braket, Raspberry Pi, etc.). I have participated in various conferences, forums, hackathons and classes and have advised on the writing of several quantum computing curricula. To finish, I have met and collaborated with several renowned scientists of the different computer techniques in marking and led collaborative projects and final degrees at Valladolid and La Salle - Ramon Llull university on algorithms and quantum approaches I designed within the framework of this thesis work.
\section{Goals}
This thesis aims to study art, and design a series of quantum algorithms on combinatorial optimisation problems and implement them in this NISQ era. To do this, we will define and create a combinatorial optimisation problem with restrictions and solve it in two different approaches. Conversely, we will formulate the problem mathematically in a top-down approach, both classically and quantum. On the other hand, we will pose the problem statistically where we will base its resolution on the machine learning technique of the base case reasoning algorithm. Later, we will generalise our formulation of the SWP to carry out efficient management of robots by substituting social workers for robots and patients for pick-up orders (We call this generalisation qRobot). Finally, we set another objective to design a collaborative didactic framework for quantum university students with our DS4DS group.

In the path to attain its primary objective, this PhD research will also have in mind the following goals:

\begin{itemize}
   
  \item Updated state of the art concerning Combinatorial Problem, Linear and Quadratic Constraints Solvers, Quantum Mechanic, Quantum Computers, Quantum Gates, Quantum Circuits, Quantum Computing, and Quantum Machine Learning. 
  \item State of the art of TSP, JSP and VRP.
  \item State of the art of heuristics for solving the combinatorial problems.
  \item State of the art of the concept of Hamiltonian of a system, Ising Hamiltonian Model, The Hamiltonian of a TSP, the Hamiltonian of a JSP, the Hamiltonian of a VRP.
  \item State of the art of the complexity class and Quantum Complexity class.
  \item Design and Model, mathematically the Social Workers' Problem (SWP) as CSP according to state of the art and the ones proposed in the scope of the PhD work
  \item Study of the players involved in SWP, namely quantum computing frameworks, quantum computers, and quantum research groups.
  \item Study and map the SWP in quantum computing.
  \item Developing and implementing the SWP in this NISQ era.
  \item Solve the SWP with QML. This methodology will be based on clustering and classification techniques, resulting in a rule base concerning the SWP. An extensive set of simulation results will be used as the basis for applying the proposed QML method.
  \item Designing and implementing a quantum Case-Based Reasoning (qCBR).
  \item Designing and implementing a Quantum Enhanced Filter: QFilter.
  \item Designing and implementing qRobot: A Quantum computing approach in mobile robot order picking and batching problem solver optimisation.
  \item Designing and implementing EVA: a quantum Exponential Value Approximation algorithm.
  \item Testing and validating the developed models and approaches.
  \item Passing on the knowledge acquired to the group of DS4DS collaborators.
\end{itemize}

\section{Hypothesis}
In this era of quantum computing with few numbers of qubits, we know that quantum optimisation algorithms based on the variational principle offer a good approximation to the optimal solution, taking advantage of the fundamentals of quantum mechanics such as superposition or quantum parallelism plus interference, reducing in some cases the computing's cost exponentially compared to classical computing. However, due to this era's technical problems, NISQ (few useful qubits, decoherence and noise), we have to ask,  to what extent could we use quantum computing to solve an optimisation problem with hard restrictions? Also, is it possible to solve quantum combinatorial optimisation problems in Top-down's philosophy and case-based reasoning?

\section{Development of the thesis}
This section will briefly present the technical results obtained in the thesis. After reviewing state of the art research, techniques to solve combinatorial optimisation problems, quantum fundamentals, current quantum technologies, and the leading companies in these fields, we have focused on defining combinatorial optimisation problems to respond to our thesis hypotheses. First, we mathematically design and formulate the \textit{Social Workers Problem} (SWP) \cite{Atc201,Atc20,Atc202,adelomou2020formulation,QSWPGithub,AllSWPGithub,Par20,AtchadeAdelomou2020,adelomou2020using,Adelomou2020} and define a heuristic to allow this problem, which includes inequality and time constraints, to be implemented in current quantum computers with very few qubits. To implement this problem, we have had to analyse and study the \textit{Ansatzes} and other various variational algorithms such as VQE and QAOA. 

Once the results were published at the \textit{16th International Conference on Hybrid Artificial Intelligence Systems (HAIS'21)}, we began to solve the same problem (SWP) with the Machine Learning approach; we called this solution \textit{quantum Case-Based Reasoning} (qCBR) \cite{atchadeadelomou2021quantum} inspired in classical \textit{Case-Based Reasoning}. The qCBR was plated as the sum of two large blocks (a variational classifier and a synthesiser). The qCBR uses a data representation model from a multidimensional vector in Hilbert space. It creates a vector subspace, where each vector has the information that defines the SWP on the Hilbert vector space and, when predicting whether an input vector (new case) corresponds to a previously analysed class, the qCBR calculates the probability that each type corresponds to the new vector from the proximity of each of the vector subspaces generated from each category. The qCBR, with its synthesiser block, refines the retrieved data in the case of not being the optimal result since the qCBR has the function of "generating" a new outcome based on the retrieved information.

To scale the algorithms developed and to be able to compare them in various technologies and quantum environments on the market (mainly \textit{Quantum Annealing} and \textit{Quantum Gate-Based computing}), a problem was defined where SWP replaced social workers with a mobile robot and had batches instead of patients. We call the result \textit{qRobot: A Quantum computing approach in mobile robot order picking and batching problem solver optimisation} \cite{a14070194}. We developed a new formulation, turned a Raspberry Pi 4 into a quantum "computer," and implemented the solution. Finally, we studied the comparisons in AWS-Braket (D-wave), Pennylane and Qiskit. The results were promising and the article was published and on the cover of Algorithms magazine.

In parallel, we have been working to improve the VQE algorithm, the flagship of quantum computing, and we published \textit{EVA: a quantum Exponential Value Approximation algorithm} \cite{alonsolinaje2021eva}. We proposed a different way to VQE to calculate the expected value given a quantum state. It should be remembered that the largest cloud services platforms for quantum computers charge by the number of shots and number of circuits. The VQE calculates the expected value by decomposing the Hamiltonian into Pauli operators and obtains this value for each of them, making simulations on cloud servers more expensive.
After completing our work, we designed an algorithm capable of calculating this value using a single circuit. Finally, we carried out a cost-time study and verified that it was possible to obtain a good performance in certain, more complex Hamiltonians compared to current methods.
Also, to consolidate our vision of hybrid computing, we contributed a publication on CNNs with the following article: \textit{Quantum Enhanced Filter, QFilter} \cite{qFilter2021}. We proposed a hybrid image classification model that took advantage of the potential of convolutional networks in classical computing and replaced the classical filters with variational quantum filters to reduce the computational cost of classical computing. Similarly, this work would be compared with other classification methods and system execution on different servers. It should be remembered that convolutional filters are based on the scalar product and represent a high cost for classical computation. In contrast, this operation (scalar product) is native to quantum computation (Hilbert vector space).

The feasibility of all the algorithms that we have proposed has been made in various environments such as Qiskit, AWS-Braket, D-Wave, Pennylane and Qibo.

\section{Thesis structure}
This thesis is organised as follows. In section \eqref{sec:2}, we review the Optimisation combinatorial problems related to our topic. Chapter \eqref{sec:3}, we introduce the systems and useful concepts needed for the methods to solve the optimisation problems. In section \eqref{sec:Solving_Combina_Pr}, we present the methods to solve combinatorial optimisation problems. The Quantum Mechanic is introduced in chapter \eqref{sec:5}, reviewing the quantum mechanics postulates. We introduce the Complexity Class in section \eqref{sec:6}. After analysing the quantum complexity in chapter \eqref{sec:7}, we revise the useful quantum gates. Quantum computers and the players are the subjecta of chapter \eqref{sec:8}. Chapter \eqref{sec:10}, we delve deeply into the most important technics in quantum computing. In section \eqref{sec:9}, we develop the research design and analyse our experiment tools, the quantum frameworks, and some experiments over quantum algorithms and technics.  Section \eqref{sec:11}, we present our approaches to solving the Social Workers' Problem. Section \eqref{sec:12} we present a quantum machine learning approach to solve the SWP. In section \eqref{sec:13} we generalise the SWP as qRobot. Section \eqref{sec:phd_results} and \eqref{sec:PhD_Discussions}, we show the results and emphases on a discussion over these results, and in our final chapter, we conclude and offer the potential for further work.

%%%%%%%%%%%%  Starting New Page here %%%%%%%%%%%%%%

\newpage

\chapter{Combinatorial Optimisation problems}\label{sec:2}
\section{Introduction}

Combinatorial optimisation means searching for an optimal solution in a finite or infinite set of potential solutions. Optimality is defined concerning some criterion function, which must be minimised or maximised is usually called the \textit{Cost function} or \textit{Objective function}.

The \textit{Social Workers' Problem} can be seen as the combination of a scheduling and routing problem. Because of this, we will explore some scheduling and routing algorithms. The first study will be the Travelling Salesman Problem (TSP) algorithm but with the mind of adding some time variables and redefining the variables to be optimised.

\section{Travelling Salesman Problem}

The \textit{Travelling Salesman Problem} known as TSP is defined as follows:
\textit{Given a list of cities and the distances between each pair of cities, what is the shortest possible route that visits each city exactly once and returns to the origin city?}

We label the cities by an integer $\{1,...,N\}$ according to their list order. We denote by $d_{ij}$ the distance between cities $i$ and $j$. We can associate each possible solution $p$ of the problem to a permutation $(x_1,...,x_N)$ that affects the order in which they are going to be tour the cities, and taking as function $f(p)=d_{x_1,x_2}+...+d_{x_{N-1},x_N}+ d_{{x_N},x_1}$, where $d_{x_i,x_{i+1} }$ corresponds to the distance between the cities $x_i$ and $x_{i+1}$, we have that our goal will be to find the minimum of $f$ in the set of permutations.

%For this problem we can associate each possible solution $p$ to a permutation $(x_1,...,x_N)$ of the numbers $\{1,...,N\}$ that affects the order in which they are going to be tour the cities, and taking as function $f(p)=d_{x_1,x_2}+...+d_{x_{N-1},x_N}+ d_{{x_N},x_1}$, where $d_{x_i,x_{i+1} }$ corresponds to the distance between the cities $x_i$ and $x_{i+1}$, we have that our goal will be to find the minimum of $f$ in the set of permutations of $\{1,... ,N\}$.

TSP is an NP-Hard problem within combinatorial optimisation. The problem was first formulated in 1930, and it is one of the most studied optimisation problems. Although the situation is computationally complex, many exact heuristics and methods are found to solve some instances from one hundred to thousands of cities.

In the TSP formulation \eqref{Tsp01} to \eqref{res3MTZ}, in this case, the distance matrix of the TSP will be determined by the elements $d_{ij}$ indicating the distance (cost) between node $i$ and node $j$. The decision variables are $x_{ij} = 1$, if the solution to the TSP goes from city $i$ to city $j$, $x_{ij} = 0$, otherwise. Then the solution of the problem is found by solving the formulation. This formulation is known as MTZ \cite{formulacion_MTZ}.

\begin{equation}
\label{Tsp01}
%\begin{aligned}
 \text{min}  \sum_{i=1}^{N} \sum_{j=1}^{N}d_{ij}x_{ij},
%\end{aligned},
\end{equation}

\begin{equation}
\label{Tsp02}
   \sum_{i=1}^{N}x_{ij}=1 \quad \forall j \in \{1, ..., N\},\\
\end{equation}

\begin{equation}
\label{Tsp03}
   \sum_{j=1}^{N} x_{ij}=1 \quad \forall i \in \{1, ..., N\},\\
\end{equation}

\begin{equation}
\label{Tsp04}
\begin{aligned}
    x_{i,j} \in \{0,1\} \quad \forall i \in \{0, \ldots,N\},\\
   \quad \forall j \in \{0, ..., N\},\\
\end{aligned}
\end{equation}

\begin{equation}
  u_i - u_j + Nx_{i,j} \leq N-1 \quad 1 \leq i\neq j \leq N.
  \label{res3MTZ}
\end{equation}

In this formulation, the objective function equation \eqref{Tsp01} minimises the cost function. The restrictions equations \eqref{Tsp02} and \eqref{Tsp03} declare that each salesman can only be in one node at any time. The restriction \eqref{Tsp04} describes that $x_{ij}$, are binary variables. The constraint \eqref{res3MTZ} are the route of continuity and the elimination of sub-courses, which ensure that the solution does not contain a sub-route disconnected from the exchange. 

There are  $N!$ possible routes (for the exact calculation). However, it can be simplified since the starting point does not matter; we can reduce the number of routes examined by a factor  $N$, leaving  $(N-1)!$ possible solutions. Also, if the direction in which the traveller is travelling does not matter, the number of routes to be examined is again reduced by a factor of $2$. Therefore, the number of the possible paths is $\frac{(N-1)!}{2}$.

\subsection{Constructive heuristics}

Following our analysis, we understand by a \textit{heuristic method}, an approach to problem-solving that uses a practical technic or various shortcuts to produce solutions that may not be optimal but are sufficient given a limited timeframe.

We find that the constructive heuristics are based on the \textit{Nearest Neighbour algorithm} (NN) \cite{YuD03} or also called Greedy algorithm \cite{YuD03}, which allows the traveller to choose the nearest unvisited city as the next move. This algorithm quickly returns a lower-cost route. For $N$  cities randomly distributed on a plane, the algorithm on average returns a path 25$\%$  longer than the smallest path possible. However, in many cases, they exist where the distribution of the given cities makes the NN  algorithm return the worst path \cite{Gutin2002}. This occurs for both symmetric and asymmetric TSPs \cite{gutin2007}. Rosenkrantz showed that the NN  algorithm has an approximation factor of order  $O(log N)$  for instances that satisfy the triangular inequality; with $N$ the number of the cities. A variation of the  NN  algorithm, called \textit{Nearest Fragment operator}, is the one that connects a group of closest unvisited cities and can find the shortest route with successive iterations. The NF  operator can also be applied to obtain initial solutions for the NN  algorithm and be improved in an elitist model where only the best solutions are accepted. There are other heuristic techniques in \textit{Local Search Algorithms} \cite{johnson1988easy}, but we will not address them classically in this thesis.

Another formulation approach is the Job Shop Scheduling Problem (JSSP) that we will analyse in the following section.

\section{Job Shop Scheduling Problems}
The \textit{Job Shop Scheduling problem} \cite{brucker1996improving, brucker1997improving, biegel1990genetic, carlier1990practical}, known as JSSP, consists of planning a set of  $N$  jobs $  \{ J_{1}\text{, ...,}J_{N} \}$  on a set of  $M$  resources or physical machines  $\{ R_{1}\text{, ...,}R_{M} \}$. Each $J_{i}$  job consists of a set of visits or operations  $\{ V_{i1}\text{, . . . , }V_{iM} \}$ that have to be executed sequentially. Each $V_{il}$  visit has a runtime of $p_{V_{il}}$ of time during which it requires the exclusive use of a single resource,$V_{il}$, starting from start time $st_{V_{il}}$ to be determined.

Each job has an earlier start time and is sometimes considered a later end time, forcing visit start times to take values in finite domains.

The  $JSSP$ has three restrictions: sequential or precedence restrictions, capacity restrictions and non-expulsion restrictions of the machines. Sequential constraints are defined by the sequence of visits to a job and can be described by linear inequalities of the type:  $st_{V_{il}}+p_{V_{il}} \leq st_{V_{i \left( l+1 \right)} }$ that is, the visit  $V_{il}$  must be carried out before the visit $V_{i(l+1)}$. The capacity constraints restrict the use of each resource to a single visit at each instant of time and can be described as a disjunctive constraint of the form:  $st_v + p_v \leq  st_w$ or  $st_w + p_w  \leq  st_v$ if $R_v = R_w$. This last constraint means that two tasks that use the same resource cannot overlap. Lastly, the non-expulsion restrictions express that once the start time has been determined,  $st_{V_{il}}$ for a visit $V_{il}$, this visit must be carried out during the time interval  $[ st_{V_{il}},st_{V_{il}}+p_{V_{il}}]$ without interruption.

The objective is to find optimal planning for a specific criterion, the most common being the following three:
\begin{itemize}
	\item Makespan is the end time of the last visit, and time is denoted as $C_{\max }$. This version of the problem is known, in the literature, as $J  \vert  \vert ~C_{\max }$ \cite{brucker1996improving, brucker1997improving}.

	\item Total Flow is the sum of the end times of all the works. This version is known as  $J  \vert  \vert  \sum _{}^{}C_{i}$, with  $C_{i}$  is the time of completion of the work $J_{i}$ \cite{brucker1996improving, brucker1997improving}.

	\item Tardiness is a delay time of the works. This version is known as $J \vert  \vert  \sum _{}^{}T_{i}$ with  $T_{i}$ as the tardiness, $T_{i} = \text{max} \{ 0, C_{i}-d_{i} \}$  with  $d_{i}$  is the due date of the visit $i$ \cite{brucker1996improving, brucker1997improving}.
\end{itemize}

Regarding the analogy, in our case, the $N$ jobs $\{ A_{1},...,A_{N}\}$  will be the assistants and the physical machines  $\{ P_{1},...,P_{M}\}$ will be patients. So from here, the same description and variables are respected.

We can consider a  $JSSP$  as a  $TSP$  executed  $M$ times or see it another way as the TSP is a particularisation of the JSSP taking into account $N$ jobs on a single machine. Being the travelling salesman, the machine and the works the cities.

Another approach to solving the problem is to use a generalisation of the TSP. In this case, the VRP.

\section{Vehicle Routing Problem}
The \textit{Vehicle Routing Problem} known as VRP which asks \textit{What is the optimal set of routes for a fleet of vehicles to traverse in order to deliver to a given group of customers?}

In VRP, the goal is to find optimal routes for multiple vehicles visiting a set of locations. (When there is only one vehicle, it is reduced to TSP). What do we understand by optimal routes for a  VRP? In the context of this thesis, the answer is the same as for a TSP; routes with the shortest total distance.

Since the study focuses on an optimisation problem, a good definition for optimal routes would be to minimise the most extended single path among all vehicles. This is the correct and best-known definition if the goal is to complete all deliveries as soon as possible.

It is essential to see that the basis of this formulation is TSP.  So, we can find a general formulation on the TSP that includes restrictions such as:
\begin{itemize}
	\item Maximum travelled distance restrictions.
	\item Visit time.
\end{itemize}

Let $G=  \left( V, E \right)$  be a complete directed graph with $V =  \{ 0, 1, 2, . . , N \},$  as the set of nodes and  $E=  \{  \left( i, j \right): i, j  \in  V, i \neq  j \}$ as the set of arcs, where node 0 represents the central, for the $K$ vehicles with the same maximum travelled distance $q$  and $N$ remaining nodes that represent geographically dispersed cities/locations. In this case, the distance matrix of the VRP will be determined by the elements $d_{ij}$ indicating the distance (cost) between node $i$ and node $j$. The VRP formulation is described as follow:

\begin{equation}
\label{VRPDev01}
%\begin{aligned}
 \text{min}  \sum_{i=1}^{N} \sum_{j=1}^{N}d_{ij}x_{ij},
%\end{aligned},
\end{equation}

\begin{equation}
\label{VRPDev02}
   \sum_{i=1}^{N}x_{ij}=1 \quad \forall j \in \{1, ..., N\},\\
\end{equation}

\begin{equation}
\label{VRPDev03}
   \sum_{j=1}^{N} x_{ij}=1 \quad \forall i \in \{1, ..., N\},\\
\end{equation}

\begin{equation}
\small \small
\label{VRPDev04}
 \sum _{j=1}^{N}x_{0j}=K ~~~\forall i  \in  1, \ldots ,N,
\end{equation}

\begin{equation}
\small \small
\label{VRPDev05}
    \sum _{j=1}^{N}x_{j0~}= K,~~~   \forall i  \in   1, \ldots ,N,
\end{equation}

\begin{equation}
\small \small
\label{VRPDev06}
    \sum_{i=1}^{N} \sum_{j=1}^{N}x_{ij}d_{ij} \leq q,
\end{equation}

\begin{equation}
  u_i - u_j + Nx_{i,j} \leq N-1 \quad 1 \leq i\neq j \leq N,
  \label{resVRPDevMTZ}
\end{equation}

\begin{equation}
\label{VRPDev07}
\begin{aligned}
    x_{i,j} \in \{0,1\} \quad \forall i \in \{0, \ldots,N\}\\
   \quad \forall j \in \{0, ..., N\}.\\
\end{aligned}
\end{equation}

In this formulation, the objective function equation \eqref{VRPDev01} minimises the cost function. The restrictions equations \eqref{VRPDev02} and \eqref{VRPDev03} declare that each vehicle can only be one node at any time. The constraint \eqref{VRPDev04} establishes that all the vehicles start from Depot and \eqref{VRPDev05} establishes that all the vehicles end at the Depot.  The restriction \eqref{VRPDev06} establishes that any vehicle can't travel more distance than allowed. In the case of wanting to measure the time, here, what we would do is change the matrix $d_{ij}$ for a matrix of the maximum contract time. The constraints \eqref{resVRPDevMTZ} are the route of continuity and the elimination of sub-courses, which ensure that the solution does not contain a sub-route disconnected from the exchange. Restrictions \eqref{VRPDev07} describes that $x_{ij}$, are binary variables. 

It is also observed that the Vehicle Routing Problem can serve to encode the difficulty of finding the optimal schedules of the  $N$  social assistants (vehicles) who visit the $m$  patients (locations).

Up to this point, the mathematical formulation of equations \eqref{VRPDev01} to \eqref{VRPDev07} represents a conventional CVRP. To solve a scheduling problem, we will need a time variable. The introduction of time (schedule) into the QUBO formulations of the CVRP is a significant obstacle to formulating several important VRP restrictions associated with the Vehicle Routing Problem with Time Windows (VRPTW)  \cite{papalitsas2019qubo}.

During the state of the art of these formulations carried out, we have found several articles \cite{papalitsas2019qubo, feld2019hybrid, irie2019quantum} that solve the TSP and VRP for annealing computers \cite{brooke1999quantum,boixo2014evidence,crosson2016simulated}. However, the number of variables is still intractable for the current size of quantum computers. 
The \textbf{number of qubits} of the TSPTW \cite{papalitsas2019qubo}, is proportional to $N^3+N^2\log_{2}N$, and for this VRPTW \cite{irie2019quantum}, is $N^4$.
For this reason, we will propose a new VRPTW formulation (\eqref{weight_SWP_eq} and \eqref{time_window_SWP_eq}) with a heuristic function executed by an classical algorithm that generates a description of a quantum circuit as advocate the following reference \cite{JayGambetta}. With this strategy, we aim to reduce the number of the qubits from $N^4$ to $N^2$ for our proposed VRPTW for solving our SWP.

The new formulation of the VRPTW will be developed in section \eqref{sec:11}.

\section{Summary}
In this section, we have introduced some classical combinatorial algorithms on which we will base the design of our SWP.
We have reviewed state of the art, both the definition and the formulations of the $TSP$ with its techniques of constructive heuristics. We also analysed the $JSSP$, which despite being an algorithm designed for factories, can be used to model the SWP and finally, the VRP.
We have also realised that since we want to implement this SWP with the VRPTW, we will need to define a strategy since the number of qubits of the VRPTW is proportional to $N^4$, making it intractable to implement today in a gate-based quantum computers. \\

To summarise, we observed that even with the best restrictions, heuristics or programming techniques, depending on the size of the data considerably high, the best $TSP$, $JSSP$  or  $VRP$  requires a very high computational cost for classical computing and would continue to approach an exponential cost. %When the number of the input increases, any polynomial algorithm is more efficient than any exponential \cite{impagliazzo2001problems}. Another positive feature of polynomial algorithms is that, in a sense, they take better advantage of technological advances.

In these cases of exponential computational cost, another approach would be needed, such as using the power of quantum computing to solve the problem of this magnitude.

Now we have to relate these classical algorithms with the usual concepts that can be of great help in the development of this thesis.

%%%%%%%%%%%%  Starting New Page here %%%%%%%%%%%%%%

\newpage

\chapter{Systems and useful concepts}\label{sec:3}
\section{Introduction }
This chapter will work on concepts and systems we need to solve combinatorial optimisation systems based on graph.
We will also analyse some necessary concepts that relate the resolution of the Ising model with graph theory and the basic Hamiltonian path of solving a routing problem such as SWP. %But, first, remember that the Hamiltonian trajectory problem is equivalent to that of a TSP since the task goes through all the nodes in a graph exactly once.

A \textit{Hamiltonian system} is a dynamic system, that is, a system whose state evolves with time, governed by Hamilton's equations. These systems are essential in dynamic optimisation techniques \cite{Ant10}. 

Next, we will analyse the Hamiltonian path, Hamiltonian cycles, and the Ising model.

\section{Hamiltonian path and cycles}

Let $G = (V, E)$, and  $N =  \vert  V  \vert$ be the graph and the Rank respectively.  
Having the following vertices  $\{1,. . . , N\}$, and the set of edges $(uv)$ to be directed (the order  $(uv)$  matters), it is trivial to extend to non-directed graphs, just considering a directed graph with $(vu)$ added to the set of arcs each time $(uv)$ is added to the arc set.

\begin{figure}[b!]
    \centering
    \includegraphics[width=0.5\textwidth]{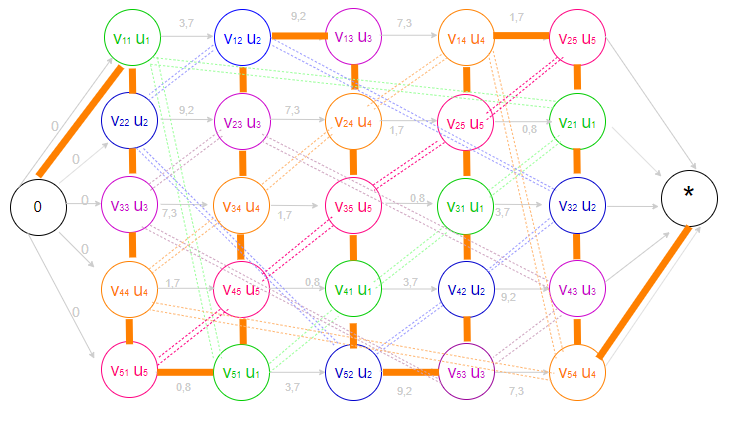}
    \caption{Shows a graph of 26 vertices in which a Hamiltonian cycle is highlighted}
    \label{fig:Hamilton_cycle}
\end{figure}

A \textit{Hamiltonian path}, in the mathematical field of graph theory, is a graph (Figure \eqref{fig:Hamilton_cycle}) path between two vertices of a graph that visits each vertex exactly once. When the last vertex visited is adjacent to the first, the path is a Hamiltonian cycle.

A \textit{Hamiltonian cycle} is defined as \textit{a closed-loop, graph cycle} on a graph where every node (vertex) is visited exactly once. 

A \textit{loop} is just an edge that connects a node to itself; a \textit{Hamiltonian cycle} is a path travelling from a point back to itself, visiting every node en route.

In the next paragraph, we will analyse the Ising model and its relationship with solving optimisation problems.

\section{Ising model and the Hamiltonian}
The \textit{Ising model} is defined as \textit{a mathematical model of ferromagnetism in statistical mechanics. The model consists of discrete variables representing magnetic dipole moments of atomic "spins" in one of two states $+1$ and $-1$.} 

Unfortunately, solving the spin system is costly because the energy landscape is equivalent to finding the global minimum of a function with several local minima. Thus, the Ising model describes the system's energy, and the operator or equation that describes the system is known as the Hamiltonian (see equation \eqref{Ising_H_eq}).

Finding the lowest energy of the spin system is equivalent to finding its best configuration (the global minimum of the associated function).

This problem is somewhat equivalent to a QUBO, which we will develop deeply in section \eqref{sec:10}. The only difference is that the QUBO works with binary variables and the Ising model with variables between $-1$ and $1$. Thus, we can summarise that the correspondence between a computer problem (QUBO) and a physical system is a change of a variable. \\

A \textit{Hamiltonian ($H$)} is the function that represents the energy of the every single possible configuration of the spins in the ferromagnetic material.

%Spin glasses are systems of  $N$  particles in which each particle can take two values: $-1$ and $+1$. The set of all possible values of the $N$  particles is called the configuration space and is designated as $\sum_{N}^{} \{ +1,-1 \}^{N}$. A configuration  $\sigma = \sigma _{1}, \sigma_{2}, \ldots , \sigma_{n}$ is an element of the summation in $N$, and each of the components $\sigma_{i}$  of this configuration is a spin. Each configuration has an associated energy $H_{N}(\sigma)$  from a function which is called the \textit{Hamiltonian}. The type of energy function that is chosen determines the model \eqref{Ising_H_eq}.

\begin{equation}
\label{Ising_H_eq}
     H = -\sum _{\langle i,j \rangle}^{} J_{ij}\sigma_i \sigma_j + \sum_{i}{}h_i\sigma_i.
\end{equation}

\begin{figure}[]
    \centering
    \includegraphics[width=0.35\textwidth]{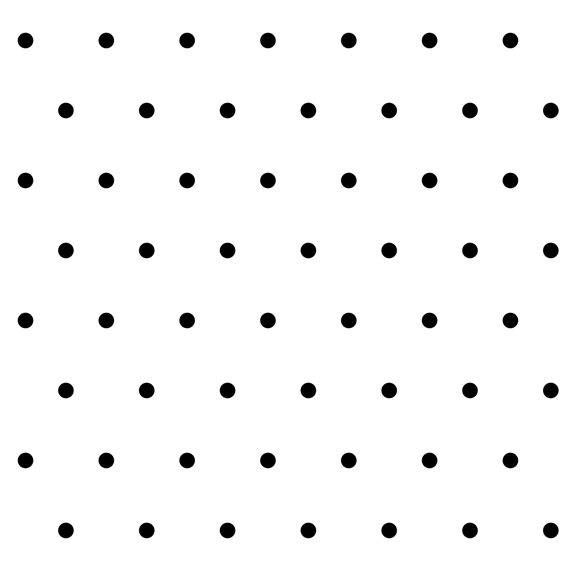}
    \caption{A Lattice in the Euclidean plane \cite{latticeGraph}}
    \label{fig:Lattice_eucludean}
\end{figure}

\begin{figure}[h!]
    \centering
    \includegraphics[width=0.8\textwidth]{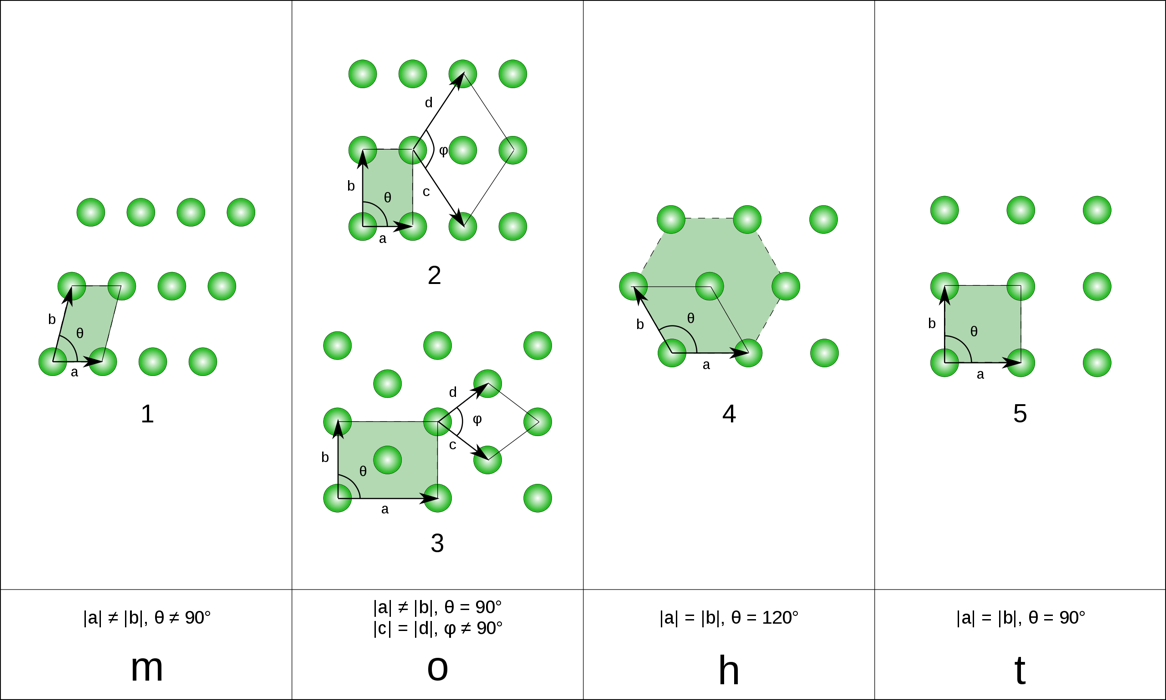}
    \caption{Five Lattices in the Euclidean plane \cite{latticeGraph}}
    \label{fig:Lattice_eucludean_plane}
\end{figure}

Where $J_{ij}$ represents the spin-spin interaction, $h_i$ represents the external field, and $\sigma_i$, $\sigma_j$ are the individual spins at each lattice site. The first sum is over all pairs of neighbouring lattice sites (also known as links); it represents the interactions between spins. The second sum is over all the lattice sites themselves; denotes the external field trying to align all spins in one direction. It is worth mentioning that, in the Ising model, each lattice site (figures \eqref{fig:Lattice_eucludean} and \eqref{fig:Lattice_eucludean_plane}) only interacts with the sites directly adjacent to it on the lattice; why the notation $\langle i,j \rangle$.

On one hand, the size of $h_i$ represents how strong the field is, so it tells us how much higher in energy one spin is than the other. Its sign tells us whether it's spin up or spin down that's preferred. On the other hand, the size of $J_{ij}$ tells us how strongly neighbouring spins are coupled to each other; That means how much they want to (anti-)align. Physically, the strength of spin-spin coupling could depend on the distance between them in the magnet's lattice, for instance. So its sign tells us whether neighbours prefer to align or anti-align. Physically, whether material is one or the other (or neither) depends on the exact quantum mechanical details of how the spins interact.

Spin glasses are systems of  $N$  particles in which each particle can take two values: $-1$ and $+1$. The set of all possible values of the $N$  particles is called the configuration space and is designated as $ \{ +1,-1 \}^{N}$.

The work of many researchers, mathematical, physical and electronic, has facilitated the conclusion that spin glasses can be used to model quantum states. The contribution of \textit{Manuel de la Rosa Fernández}\cite{rosa2018spin} reaches the same conclusion.
%Spin glasses are the basis of Hamiltonian systems that go from neural networks to artificial intelligence problems and go through numerous issues of a physical nature, such as; studies on states of matter and statistical mechanics.

%Mathematically, the fact that a problem is NP-complete means that a mapping to the decision form of the Ising model can be found with a polynomial number of steps. This assignment can be reinterpreted as a pseudo-boolean optimization problem \cite{Ham02}. 

There are constructions known as "p-spin lenses", which often lead to interactions of three bodies or more in  $H$, and which can later be used to reduce the problem to an Ising spin glass, introducing a polynomial number of turns that help reinforce the interaction of three bodies through multiple interactions of two bodies $s_{1}s_{2}$ \cite{babbush2013} \cite{biamonte2008}.

Based on Andrew Lucas's work\cite{lucas2014}, the author concludes that almost all famous $NP$ \cite{garey1979} \cite{karp1972} problems can be written as Ising models with a polynomial spin number that does not scale faster than  $N^{3}$. This work is essential for writing any quantum optimisation algorithm.

\section{Summary}
In this section,  we have reviewed the state of the art of useful concepts to develop this thesis. We have introduced  the Ising model and the Hamiltonian as the energy function.

All these concepts, theories and analyses are important to solve combinatorial optimisation systems, and they will help us fully when it comes to modelling our problem. In our case, we will further develop each constraint that we will map into the objective function. For all this, in the next section, we will analyse combinatorial optimisation problems and their resolution.

%%%%%%%%%%%%  Starting New Page here %%%%%%%%%%%%%%

\newpage

\chapter{Solving combinatorial optimisation problems}\label{sec:Solving_Combina_Pr}
\section{Introduction}
This chapter will review existing methods for solving constrained optimisation problems. For this, we need to analyse the existing programming techniques and improve them in the face of our mission.
\section{Linear Programming}
Almost all the approximation algorithms that we know are built based on Linear Programming (LP). \\

\textit{Linear Programming} is defined as \textit{the problem of optimising (minimising or maximising) a linear function subject to linear inequality constraints. The function being optimised is called the objective function.}

Linear Programming corresponds to an algorithm through which real situations are solved. It aims to identify and solve difficulties to increase productivity concerning resources (mainly limited and expensive), thus increasing benefits.

How to solve a problem using LP? \\

The first step in solving a linear programming problem is to identify the essential elements of a mathematical model; these are:
\begin{itemize}
	\item Objective Function\par
	\item Variables\par
	\item restrictions
\end{itemize}\par
The next step is to determine the following goals, for which we propose to follow the following methodology.
\begin{enumerate}
	\item Define the objective function criteria
	\item Identify and define variables
	\item Identify and define constraints
	\item Present the Objective Function
\end{enumerate}

\subsection{The Objective function, decision variables and the constraints}
The \textit{objective function} is directly related to the general question that one wishes to answer. If different questions arise in a model, the objective function would be associated with the higher-level question, the fundamental question. 

The \textit{decision variables} are identified, starting from a series of questions derived from the fundamental question. Decision variables are the controllable factors of the system that one is modelling. They can take various possible values, of which it is necessary to know their optimal weight, contributing to the achievement of the objective of the general function of the problem.

When we talk about \textit{constraints} in a linear programming problem, we are referring to everything that limits the freedom of the values that decision variables can take. The \textit{constraints} of the problem determine the conditions in which the use of resources is feasible, taking into account all the tasks of the problem.

The best way to find them is by thinking about a hypothetical case that was decided to give infinite value to the decision variables. For example, in the case of a TSP, one restriction will be to visit the city only once. 

In the case of the allocation algorithm, the decision variables $x_{ij}$ are incorporated, which are binary variables that take the value $x_{ij}=1$ as long as  $b_{i}$ is assigned to  $a_{j}$ and that $x_{ij}=0$. If these options were not given, the following restrictions \eqref{LP_const} and \eqref{Lp_const1} would be established:
\begin{equation}
\label{LP_const}
 \sum _{i=0 ~i \neq j}^{N}x_{ij} = 1,
\end{equation}
\begin{equation}
\label{Lp_const1}
 \sum _{j=0 ~i \neq j}^{N}a_{ij}x_{ij} \leq b_{i}.
\end{equation}

Where $a$ is a matrix and $b$ is a vector.
The first restriction, equation \eqref{LP_const}, means that visits will be made only by an agent. While the second restriction, equation \eqref{Lp_const1}, means that the set of visits made by an agent will not exceed the limit of resources that the agent has available. 
In the case of an exact TSP, the first set of equalities ensures that each exit city  $\{0, ..., N\}$  reaches exactly one city. The second set of equalities provides that from each city $\{1, ..., N\}$  exit exactly into a town (both restrictions also imply that there is precisely one exit from town 0). %Finally, the last limitation \eqref{subtour} forces a single road to cover all the cities, and it is not two or more disjoint roads that include all the towns together. 
%\begin{equation}
%\label{subtour}
%    u_{i}-u_{j}+Nx_{ij} \leq N-1.
%\end{equation}

For a positive scalar $P$ (\textbf{Lagrange Multiplier}) and $y_i$ as the \textbf{ancillary variables}. Equations \eqref{Penality_1} and \eqref{Penality_2} are the penalty functions corresponding to constraints \eqref{LP_const} and  \eqref{Lp_const1}.
\begin{equation}
    \label{Penality_1}
        P(\sum_{i=0 ~i \neq j}^{N}x_{ij} - 1)^2,
\end{equation}

\begin{equation}
    \label{Penality_2}
        P(\sum_{j=0 ~i \neq j}^{N}a_{ij}x_{ij} - b + \sum_{i=0}^{\lceil log_2b\rceil}2^iy_i)^2.
\end{equation}

%$$P(\sum _{i=0 ~i \neq j}^{N}x_{ij} - 1)^2 + P(\sum _{j=0 ~i \neq j}^{N}a_{ij}x_{ij} - b + \sum_{i=1}^{\lceil log_2b\rceil}2^iy_i)^2.$$

\subsection{Formulation of linear programming of binary integers}
Let us introduce a \textit{linear programming of binary integers} as follows: If  $z_{1}, z_{2}, z_{3}, \ldots, z_{N}$ are $N$ binary variables, which are ordered in a vector $z$, what is the most substantial value of $c \cdot z$, for some vector $c$, given a constraint?
\begin{equation}
\label{linearP}
 Sz=b.
\end{equation}

From \eqref{linearP} with $S$ an  $m \times N$  matrix and  $b$  a vector with  $m$  components, it is known that the resolution of this equation has an NP-hard complexity, with a corresponding NP-complete decision problem. As discussed above, almost most daily challenges are combinatorial optimisation problems, and in many cases, the framework use to model these problems is the ILP \cite{schrijver1998theory}. In our case, a social worker must maximise visits to a home patient, given the regulatory restrictions of her/his contract.

Let $H = H_{A} + H_{B}$ be the Hamiltonian (mathematical and physical model introduced above) defined by \eqref{Form_Ham_LP}.
\begin{equation}
\label{Form_Ham_LP}
 H_{A}=A \sum _{j=1}^{m} \left[ b_{j}- \sum _{i=1}^{N}S_{ji}z_{i} \right] ^{2}.
\end{equation}

With  $A \in  R^{+}$. The ground states of $H_{A}= 0$  enforce the constraint that  $Sz = b$. Then you get that
\begin{equation}
\label{LP_W_Fun}
    H_{B}=-B \sum_{i=1}^{N}c_{i}z_{i}.
\end{equation}
With $A \gg B$ and $B$ is a positive constant. \\

To find restrictions on the required  $A/B$  ratio, Let us proceed as follow. It is assumed that the constraint of the equation $Sz=b$  can be satisfied with a selection of  $z$. Taking this in account, the most meaningful possible value of  $-\Delta H_{B} = BC$  is shown as the constraint limits where:

\begin{equation}
\label{LP_cost}
    C= \sum _{i=1}^{N}\text{max}(c_{i},0). 
\end{equation}

One way of adjusting the Lagrange multiplier is given as described as the following. The lowest value of $\Delta H_{A}$  is leveraged on the properties of the matrix $S$ and would occur if only a single constraint were violated. That constraint was broken by the least amount possible, given by \eqref{matrix_s}.

\begin{equation}
\label{matrix_s}
S \equiv \mathop{{\mathop{\min }}}_{\mathop{ \sigma }_{i} \in  \{0,1\},j}⁡ \left( \max  \left[ 1,\frac{1}{2} \sum_{i}^{} \left( -1 \right)^{\sigma_{i}}S_{ji} \right]  \right). 
\end{equation}

This limit could be improved if more specific properties of  $S$  and  $b$  are known.

It can be concluded that if the coefficients  $c_{i}$  and  $S_{ji}$  are integers, we have that  $C \leq N$ max($c_{i}$) and $S \geq 1$, so it is concluded that  $A/B \geq N$.

\section{Quadratic Programming}
The \textit{Quadratic programming} can be stated as \textit{the procedure that minimises a quadratic function of  $n$  variables subject to $m$  linear constraints of equality or inequality.} \\

A quadratic program is the simplest nonlinear problem with inequality constraints. Quadratic programming is important because many issues appear naturally as quadratic (least-squares optimisation, with linear constraints). Besides, it is important because it appears as a sub-problem frequently to solve problems, not more complicated linear. The techniques proposed to solve quadratic issues have a lot in common with linear programming.

Specifically, each inequality must be satisfied as equality. The problem is then reduced to a search for vertices precisely as it was done in linear programming, where  $c$ is a vector of constant coefficients; $A$ is a matrix $( m \times n)$, and it is generally assumed that  $Q$ is a symmetric matrix.
Since the constraints are linear and presumably independent, the qualification of the restrictions is always satisfied, so the Karush-Kuhn Tucker conditions are also sufficient conditions to obtain an extreme that will also be a global minimum if  $Q$  is positively defined. On the other hand, if $Q$ is not defined as positive, the problem could be unbounded or lead to local minima.
Quadratic Programming plays a significant role in linear and nonlinear optimisation theory since it is closely related to LP and is an essential intermediate step to effectively solving general Nonlinear Programming problems.
In the case of solving the Hamiltonian of the Ising model, the two instances of solving the quadratic minimisation problems will be seen. % There are several, but they will not be listed in this thesis.

\subsection{Quadratic Formulation}
 Let ${f: }\mathbb{R}^{n}  \rightarrow  \mathbb{R}$,  $f$ be a quadratic function if: 
\begin{equation}
\label{quad_form}
 f \left( x_{1},~x_{2}\text{, . . . , }x_{n} \right) = \sum _{i,j=1}^{n}q_{ij}x_{i}x_{j}= \left( x_{1},~x_{2}\text{, . . . , }x_{n} \right)  \left( \begin{matrix}
q_{1,1}  &   \cdots   &  q_{1,n}\\
 \vdots   &  \ddots  &   \vdots \\
q_{n,1}  &   \cdots   &  q_{n,n}\\
\end{matrix}\right)  
\left( \begin{array}{c}
	{x_{1}} \\
	\vdots \\
	{x_{n}} \\
	\end{array} \right). 
\end{equation}

Let us focus on one of the properties of the quadratic function as follows:
\begin{equation}
\label{quad_funct}
 f \left(  \lambda x_{1},~ \lambda x_{2}\text{, \dots , } \lambda x_{n} \right) = \lambda ^{2}f \left( x_{1},~x_{2}\text{, \dots , }x_{n} \right). 
\end{equation}
On the one hand, \textit{Quadratic Minimisation Problems} subject to equality constraints require minimising the objective function $f \left( x \right)$ subject to linear equality constraints $Ax = b$.

But when these are subject to linear inequality constraints, they require minimising the objective function $f(x)$ subject to these constraints $Ax = b$, and may even contain equality constraints.

\begin{equation}
\label{qu_form}
    \text{min} f(x) =\frac{1}{2}x^{T}Qx+c^{T}x.
\end{equation}

Subject to:
\begin{equation}
\label{qu_restri}
    Ax\leq b,
\end{equation}
\begin{equation}
\label{qu_restri_}
    x\geq 0.
\end{equation}

Where  $c$  $\in R^{n}$  is a row vector,  $x$  $\in R^{n}$  and  $b$  $\in R^{n}$  are column vectors.  Let consider also that $Q= q_{ij}$  $\in M_{n \times n}$  and  $A= (a_{ij})  \in M_{n \times n}$ are matrices and the superscript $T$ denotes the transpose. The elements $q_{ij}$ of $Q$ are constants given such that $q_{ij} = q_{ji}$  (that is, $Q$ is a symmetric matrix,  $Q^{T}=Q$. Thus, the objective function is expressed in terms of these $q_{ij}$, the elements $c_{j}$ of $c$ and the variables $x_{j}$, as follows:
\begin{equation}
\label{qub_eq}
    f(x) =\frac{1}{2}x^{T}Qx+c^{T}x= \sum _{j=1}^{n}c_{i}x_{j}+\frac{1}{2} \sum _{i=1}^{n} \sum _{j=1}^{n}q_{ij}x_{i}x_{j}.
\end{equation}

The Quadratic unconstrained binary optimisation (QUBO) above, also known as an unconstrained binary quadratic programming (UBQP), is a combinatorial optimisation problem with a wide range of applications from finance and economics to machine learning is a specific case of Quadratic Programming with binaries variables. We will analyse it deeply in section \eqref{sec:quboChap}.

\subsection{Change of variables to simplify the quadratic formulation}

\begin{figure}[h!]
    \centering
    \includegraphics[width=0.7\textwidth]{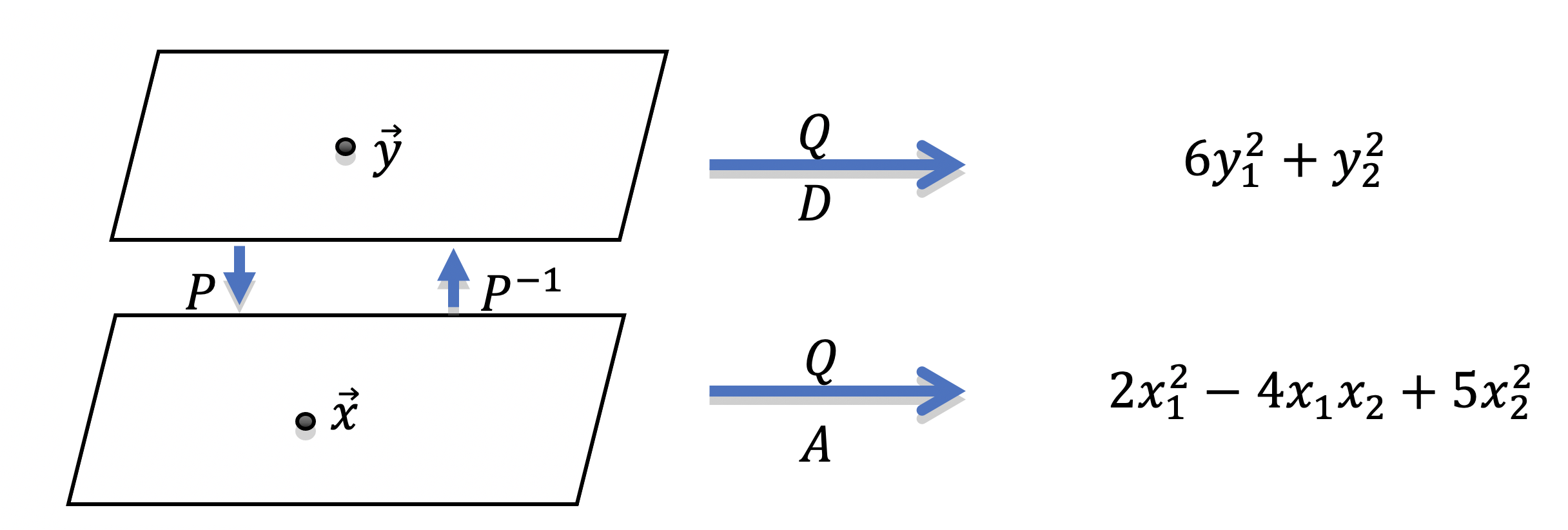}
    \caption{Variable change to simplify quadratic programming}
    \label{fig:Change_variable}
\end{figure}

In some cases, to simplify the objective function and therefore simplify the programming of the algorithm, it is required to make some variable changes (see Fig. \eqref{fig:Change_variable}). This contribution is analysed below.

Next, the change of variables is studied to simplify quadratic programming. These variable changes and linear algebra tricks will help model our problem into "quantum" programming. \\

The matrix that defines an arbitrary quadratic problem can be expressed as a symmetric matrix $A$. Because it is symmetric, it will be diagonalisable. That is, there will be an invertible matrix $P$ and another diagonal $D$ such that:

\begin{equation}
\label{DIAG_SVD}
  A = PDP^{-1}.
\end{equation}

The matrix $P$ acts as a base change, and if $A$ represents the quadratic problem in the initial base, $D$ will define the problem for the new base. Said matrix $D$ drastically simplifies the formulation and, therefore, the computation. \\

Given the quadratic function  $Q(x) =2x_{1}^{2}-4x_{1}x_{2}+5x_{2}^{2}$ we will perceive what happens with the variable changes. Let us imagine for a moment that the function $Q(x)$ is the Hamiltonian of the Ising model of the TSP.

Let us find the matrix $A$. 
\begin{equation}
 Q= \left[ \begin{matrix}
2  &  5\\
-2  &  5\\
\end{matrix}
 \right]
\end{equation}
 
 How is  $A$  found? Following this trick:

\begin{equation}
\label{matri_qad}
 A= \left[ \begin{matrix}
a  &  c\\
d  &  b\\
\end{matrix}
 \right]. 
\end{equation}

If we consider the function $Q(x)$,  $a$   is the constant that multiplies  $x_{1}^{2}$  and $b$  is the term that multiplies $x_{2}^{2}$;  $a = 2$  and  $a = 5$. So far, relatively easy. How are $c$  and $d$ determined?  $c$  and  $d$ are half the coefficient of  $x_{1}x_{2}$. So, $c= \frac{4}{2}=d$. From here are the values and eigenvectors to be able to change the variables.

Given the characteristic polynomial  $P (\lambda)=\text{det}( A- \lambda I_{n})$ if we want to calculate the eigenvalues, the characteristic polynomial must be zero. This translates to  $\det(A- \lambda I_{n}) =0$, where $I_{n}$  is the identity matrix of rank $n$ and $\lambda$  the eigenvalue.
Mathematically speaking, if we want the characteristic polynomial of matrix $A$ to be null, the following linear equation must be solved. The characteristic polynomial is of rank $n$, and its roots are the eigenvalues of matrix $A$. Since in our case matrix   $A$ is of rank 2, it will not be necessary to apply the Laplace expansion formula. Starting from a square matrix of degree $n$.
\begin{equation}
\label{det_eq}
    \det  \left( A- \lambda I_{n} \right) =0.
\end{equation}
According to Laplace's theorem, the value of its determinant is equal to the sum of the products of the elements of a row or column by their attachments, thus taking any row $f$, the determinant is:

\begin{equation}
\label{laplace_theorem}
    \det(P) = \sum _{j=1}^{n}a_{f,j}P_{f,j}.
\end{equation}
And taking a column $c$, it will be:
\begin{equation}
\label{gen_Laplace}
    \det  \left( P \right) = \sum _{i=1}^{n}a_{i,c}P_{i,c}.
\end{equation}

Returning to our case, solving the equation of the characteristic polynomial we get:
\begin{equation}
\label{cal_det}
 \det  \left( \begin{matrix}
2- \lambda   &  -2\\
-2  &  5- \lambda \\
\end{matrix}
 \right) =0.
\end{equation}
Since it is a rank two matrix, the calculation of the determinant is direct and is shown through the following quadratic expression:
\begin{equation}
\label{resolu_lamda}
  \lambda ^{2}-7 \lambda +6=0.
\end{equation}
Where the roots of the equation are  $\lambda _{1}=1$ and  $\lambda _{2}=6$. Now let us calculated the eingenvectors of the matrix $A$.
\begin{equation}
\label{Sol_Mat_change}
 \exists ~ v \left({\begin{matrix}
x_{1}\\
x_{2}\\
x_{3}\\
 \vdots \\
x_{n-1}\\
x_{n}\\
\end{matrix}
} \right)  \neq 0 \quad / \quad \exists  ~ \lambda  \in \mathbf{C}: \quad A \cdot v= \lambda  \cdot v.
\end{equation}

Where $\lambda$  is the eigenvalue, $v$ is the eigenvector, and, if this definition is applied, it arrives at \eqref{Res_bis}:

\begin{equation}
\label{Res_bis}
    \left[ \begin{matrix}
2  &  5\\
-2  &  5\\
\end{matrix}
 \right]  \left( \begin{matrix}
x_1\\
x_2\\
\end{matrix}
 \right) = \lambda  \left( \begin{matrix}
x_1\\
x_2\\
\end{matrix}
 \right). 
\end{equation}

With this expression $-2x_{1}=x_{2}$, we can find the two eigenvectors associated with matrix $A$.

\begin{equation}
\label{Sols_changes}
v_{1} = \left( \begin{matrix}
1\\
-2\\
\end{matrix}
 \right)  \text{and} ~v_{2} = \left(\begin{matrix}
2\\
1\\
\end{matrix} \right).
\end{equation}

Once this point has been reached, if the quadratic form is rewritten, the reader can understand how simple it is and, consequently, its resolution or programming will also be more straightforward. 

\begin{equation}
\label{prove_eq}
     P (y) = \lambda _{1}y_{1}^{2}+ \lambda _{2}y_{2}^{2}.
\end{equation}

Let us recall $Q(x) =2x_{1}^{2}-4x_{1}x_{2}+5x_{2}^{2}$, and find the matrix $P$.
\begin{equation}
\label{Matrix_P}
    P= \left[ \begin{matrix}
1  &  2\\
-2  &  1\\
\end{matrix}
 \right]. 
\end{equation}

If the matrix  \( P \)  is normalised, the following expression is reached,

\begin{equation}
\label{Matrix_P_bis}
    P=\frac{1}{\text{dist}} \left[ \begin{matrix}
1  &  2\\
-2  &  1\\
\end{matrix}
 \right]. 
\end{equation}

With  $\text{dist}=\sqrt[]{1^{2}+(-2)^{2}}$ as the distance. 

\begin{equation}
\label{Matrix_P_3}
    P=\frac{1}{\sqrt[]{5}} \left[ \begin{matrix}
1  &  2\\
-2  &  1\\
\end{matrix}
 \right]. 
\end{equation}
Retrieving the expression,  $A=PDP^{T}$  takes us to $D=P^{T}AP$, where D is the following matrix:
\begin{equation}
\label{Sol_final}
D= \left[ \begin{matrix}
6  &  0\\
0  &  1\\
\end{matrix}
 \right]. 
\end{equation}
These variable changes help simplify the objective function by undoing the crossed variables.

\section{Quadratic programming with quadratic constraints}

The \textit{quadratically constrained quadratic program} can be defined as \textit{an optimisation problem in which both the objective function and the constraints are quadratic functions.}\\

When the constraints are of order 1, the quadratic formulation just analysed works for us. But there are cases when the restrictions must be quadratic; of the type  $x_{1}  \left( x_{1} - 1 \right)   \leq  0$  and  $ x_{1}  \left( x_{1} - 1 \right)   \geq  0$  being equivalent to the constraint  $x_{1}  \left( x_{1} - 1 \right)  = 0$, and, in turn, equal to the restriction $~x_{1} \in   \{ 0, 1  \}$. In this case, a generalisation of quadratic programming must be used. Almost for all quantum algorithms where we want to improve resolution and computation time, \textit{Quadratically Constrained Quadratic Program} (QCQP) should be used. It is the same studied so far but with quadratic restrictions. For this, the new objective function  $P(x)$ must be minimised
\begin{equation}
\label{QPQC_form}
    \text{min}\quad P \left( x \right) =\frac{1}{2}x^{T}P_{0}x+q_{0}^{T}x+r_{0}.
\end{equation}
Subject to:
\begin{equation}
\label{rest_QPQC}
    \frac{1}{2}x^{T}P_{i}x+q_{i}^{T}x+r_{i} \leq 0 \quad \text{for} \quad i=1, \ldots ,m, 
\end{equation}
\begin{equation}
\label{LP_QPQC}
    Ax = b.
\end{equation}
$P_{i} \in S_{+}^{n}$, the objective and constraints are quadratic convex, where  $\{P_{0}{\dots,}P_{m}\}$  are matrices  $n \times n$ and $x \in R^{n}$ is the optimisation variable.
If  $\{P_{1}{\dots,} P_{m}\} \in S_{++}^{n}$, the feasible region is the intersection of $m$ ellipsoids and a cognate set.

When trying to model a problem into binary variables, a degree greater than two may appear in the function to minimise. As most of the quantum computing technique in this NISQ era is implemented to handle terms of degree less than or equal to 2, the degree of monomials that do not meet this condition must be reduced.\\
Let us analyse different ways to transform a cubic monomial of the cost function into a quadratic one. Once we have this technique, applying it repeatedly, we can decrease the degree of any monomial. \\

A first way is to transform the problem $ \min f(w)$ into the problem $\min g(k) $ subject to $ t = xy $ con $f(w)=xyz$ y $g(k)=tz.$ 

Later, we will realise that we cannot have a quadratic constraint, so let us find a way to substitute $t=xy$ for linear constraints.

Taking the constraints $ t \leq x $, $ t \leq y $ and $ t \geq x + y-1 $, we can transform the problem of $ \text {minimize} f(w) $ into the problem:

$$ \text {minimize}  \quad  tz $$, $$ \text {subject to} \left \lbrace \begin {array} {c} t \leq x \\ t \leq y \\ t \geq x + y-1. \end {array} \right.  $$

This remodelling involves introducing three inequality constraints. Inequality constraints, while often unavoidable, are somewhat inefficient. This is due to the need to include a set of auxiliary slack variables necessary to transform each inequality into equality. This procedure is similar to the one that occurs when modelling a linear programming problem to be solved using the simplex algorithm. There are more techniques, but we will not dive deeply into them in this thesis. Therefore, let us analyse for other better methods that allow us to reduce the degree of the monomials. \\

The first technique we will analyse is to exploit the relationship
$$ f(w) = \max_ {t \in \{0,1 \}} \{t (x + y + z-2) \}. $$
This relationship follows from the fact that, on the one hand, if any of the variables $ x, y, z $ is equal to $ 0 $, we have that $ (x + y + z-2) \leq 0 $ and therefore, the maximum on the right will be reached when $ w $ takes the value $ 0 $, resulting in $ 0 $. If, otherwise, all the variables are worth $ 1 $, we would have the product $ w \cdot 1 $, which takes the maximum value $ 1 $ when $ t = 1 $. \\

Now, if the monomial $ axyz $ from which we want to reduce the degree has coefficient $ a < 0 $, it is true that:
$$ \min_ {x, y, z} axyz = a \max_ {x, y, z} (xyz) = a \max_ {x, y, z} (\max_{t} (t (x + y + z-2))) = a \max_{x, y, z, t} (t (x + y + z-2)) = \min_{t, x, y, z} at (x + y + z -2). $$

Therefore, if $ a <0 $ we can simply substitute monomials of the form $ af(w) $ for $ a (t (x + y + z-2)) $.
If $ a> 0 $ we cannot use the same technique since, although
$$ f(w) = - \min_ {t \in \{0,1 \}} t (-x-y-z + 2), $$

The above equation with the minus sign does not allow this term to be entered in the cost function. \\

To solve this problem, let us return to the problem of expressing the equation $ z = xy $ in linear terms. The penalty functions are needed in these cases. We need to construct a \textit{penalty function} that takes large values when $ z \neq xy $ and the value $ 0 $ when $ z = xy $. Also, it is needed to work carefully on \textit{Lagrange multiplier}. This task is important for problems where the degree of the restriction can be greater than 2; Problems that go beyond those of this NISQ era.

Most problems of interest include additional constraints and many of
these problems can be re-formulated (re-casted) as a QUBO model by introducing quadratic penalties with a positive scalar $P$. The table \eqref{tab:penalties_table} summarises the widely used penalties function.

\begin{table}[h!]
\centering
\begin{tabular}{|c|c|}
\hline
 Classical Constraint & Equivalent Penalty \\ \hline
$x = y$ & $P(x+y-2xy)$ \\ \hline
$x + y = 1 $ & $P(1-x-y+2xy)$ \\ \hline
$x + y \leq 1$ & $P(xy)$\\ \hline
$x + y \geq 1 $ & $P(1-x-y+xy)$ \\ \hline
$x \leq y$ & $P(x-xy)$ \\ \hline
$x_1 + x_2 + x_3 \leq 1$ & $P(x_1x_2 + x_1x_3 + x_2x_3)$ \\ \hline
\end{tabular}
\caption{In this table, we see the most used and known penalties. It is worth mentioning that binary variables satisfy $x_{i}^2=x_{i}.$}
\label{tab:penalties_table}
\end{table}

\section{Summary}
In this chapter, we have analysed and studied the methods of solving combinatorial optimisation algorithms. Furthermore, we have studied \textit{Linear Programming} and its derivative of binary integers, \textit{Quadratic Programming} and \textit{Quadratic Programming with Quadratic Constraints}.
But here, we have not studied a lead combinatorial optimisation problem-solving algorithm, QUBO. However, the reader can find it in the section \eqref{sec:quboChap} on quantum computing because of its close connection with the Ising models. Moreover, QUBO is a problem central to adiabatic quantum computing, where it is solved by a physical process called quantum annealing. This model is very interesting for quantum computing for its easy mapping to the \textit{Ising model}.

We can now entirely focus and delve into quantum computing, but to do this, we will introduce some necessary concepts and theories of quantum mechanics.
%%%%%%%%%%%%  Starting New Page here %%%%%%%%%%%%%%

\newpage
\chapter{Introduction Quantum Mechanics}\label{sec:5}

\section{Introduction}\label{sec:5_}

\textit{Quantum mechanics} \cite{Bes12} is a theory that describes the physical properties of nature on an atomic scale. 

This section will give a complete description of the basic postulates of quantum mechanics. These postulates are important to connect the physical world and the mathematical formalism of quantum mechanics. First of all, we will review the limitation faced by Moore's law. Then we will analyze the state of quantum technologies. Finally, and before focusing on the postulates of quantum mechanics, we will briefly introduce the era in which quantum computing finds itself.

The first revolution \cite{Lar18} of quantum mechanics is the basis of several advances to our modern society. Almost all modern electronics in the last 50 years have been based on the properties of quantum mechanics. Semiconductors such as diodes, transistors, integrated circuits, etc., have led to the considerable growth of electronics today. 

\begin{figure}[h!]
    \centering
    \includegraphics[width=0.4\textwidth]{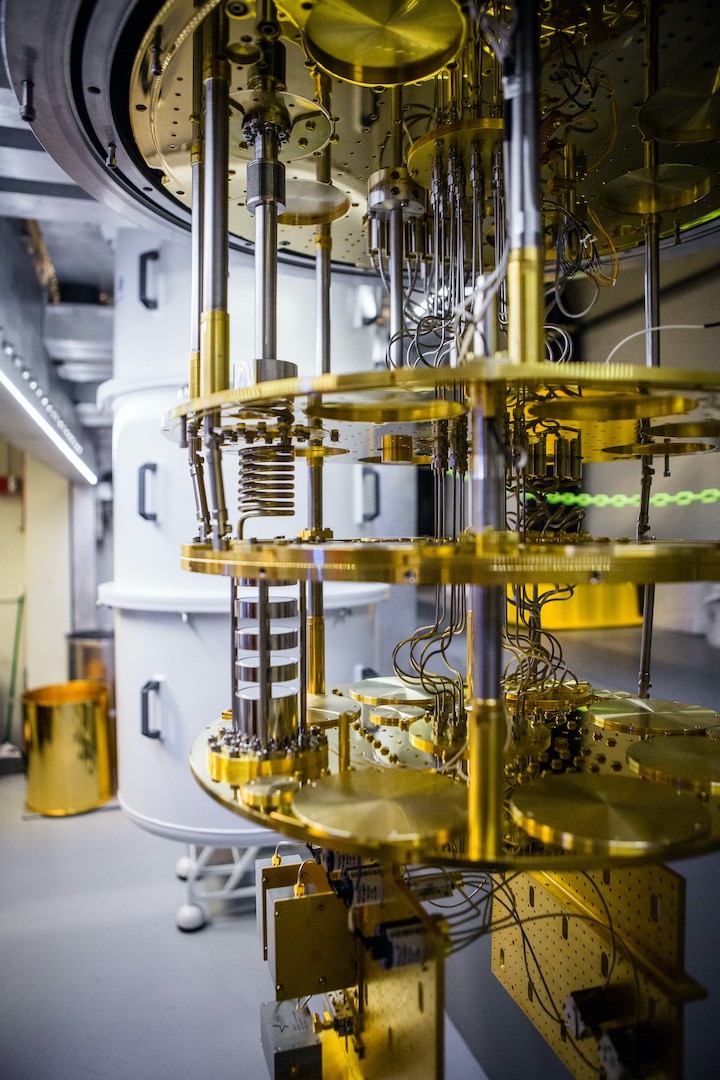}
    \caption{TBluefors dilution refrigerators used in the MIT Engineering Quantum Systems group to operate superconducting qubits at near-zero temperatures \cite{vasconcelos2020quantum}}
    \label{fig:True_QC_image}
\end{figure}

Quantum mechanics, in addition to studying the motion of particles, also allows us to understand the properties of materials and their features to manufacture transistors. These are the basis of all modern electronics and the control system of millions of devices that we use today. However, in this thesis, we focus on the second revolution of quantum mechanics, that is, in using the properties of quantum mechanics to empower computing \eqref{fig:True_QC_image}.

The \textit{Quantum information science} also known as QIS, is the area of information science that depends on the effects of quantum mechanics. During the last decades, the scientific community has dedicated a lot of time and has provided considerable resources to QIS \cite{TBe05}. As a result of this delivery, improvements have been achieved in the scientific advances that we see reflected in publications, conferences, and concrete solutions such as D-Wave, IBMQ, Xanadu, etc. Quantum computing, in particular, and quantum information science is a hot topic because of its novelty and the promising developments it predicts. \\

In this thesis work, all the postulates of quantum mechanics on which we rely are referenced by \cite{nielsen2002quantum}.

\section{Moore Law}
\begin{figure}[!h]
    \centering
    \includegraphics[width=0.5\textwidth]{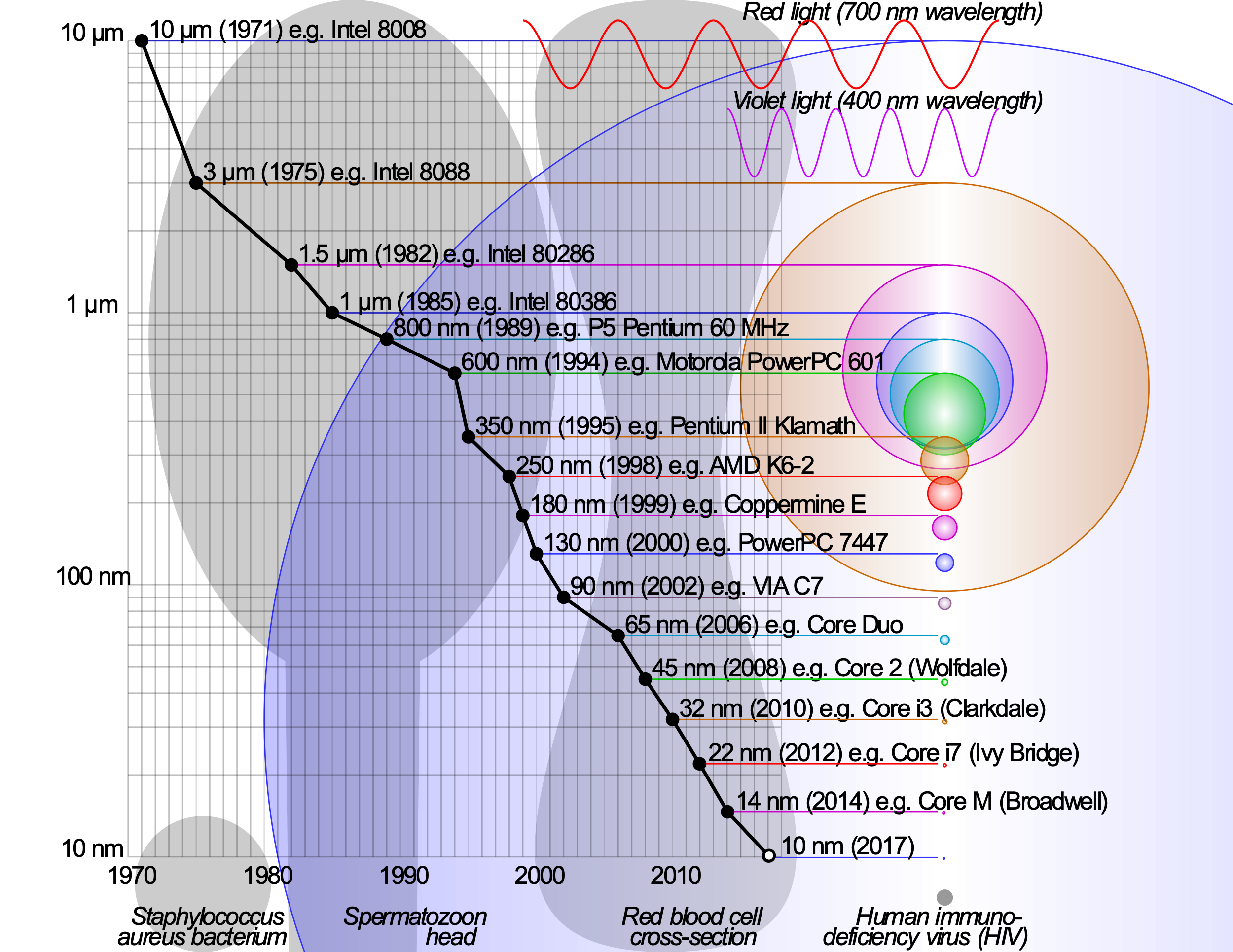}
    \caption{Moore law Roadmap: Quantum Computing Stack Exchange \cite{prati2017quantum}}
    \label{fig:MOORE_LAW}
\end{figure}

Another aspect to consider understanding the importance of the properties of quantum mechanics and thus empower computing is \textit{Moore's law}, which states that the number of transistors that can be integrated into a single chip doubles every 18 months. This process Figure \eqref{fig:MOORE_LAW} leads to a doubling of the memory and a doubling of the calculation speed. Therefore, it is expected that the size of the characteristic of the wafer (thin plate of semiconductor material) will be less than 10nm during this year (2020). At this point, the individual properties of atoms and electrons would begin to dominate, so Moore's law would no longer be valid. Therefore, the demands of the miniaturisation of electronics will eventually bring us to the point where quantum effects become important \cite{TBe05}. 

Because Moore's law will stop working today, there are even more reasons for the scientific community to resort to QIP much sooner than it seems \cite{TBe05,Iva12}.
On the other hand, multi-core architecture is becoming a practical approach; computational speed improvement can be achieved even without reducing the size of the feature by parallelisation. 
\begin{figure}[h!]
    \centering
    \includegraphics[width=0.7\textwidth]{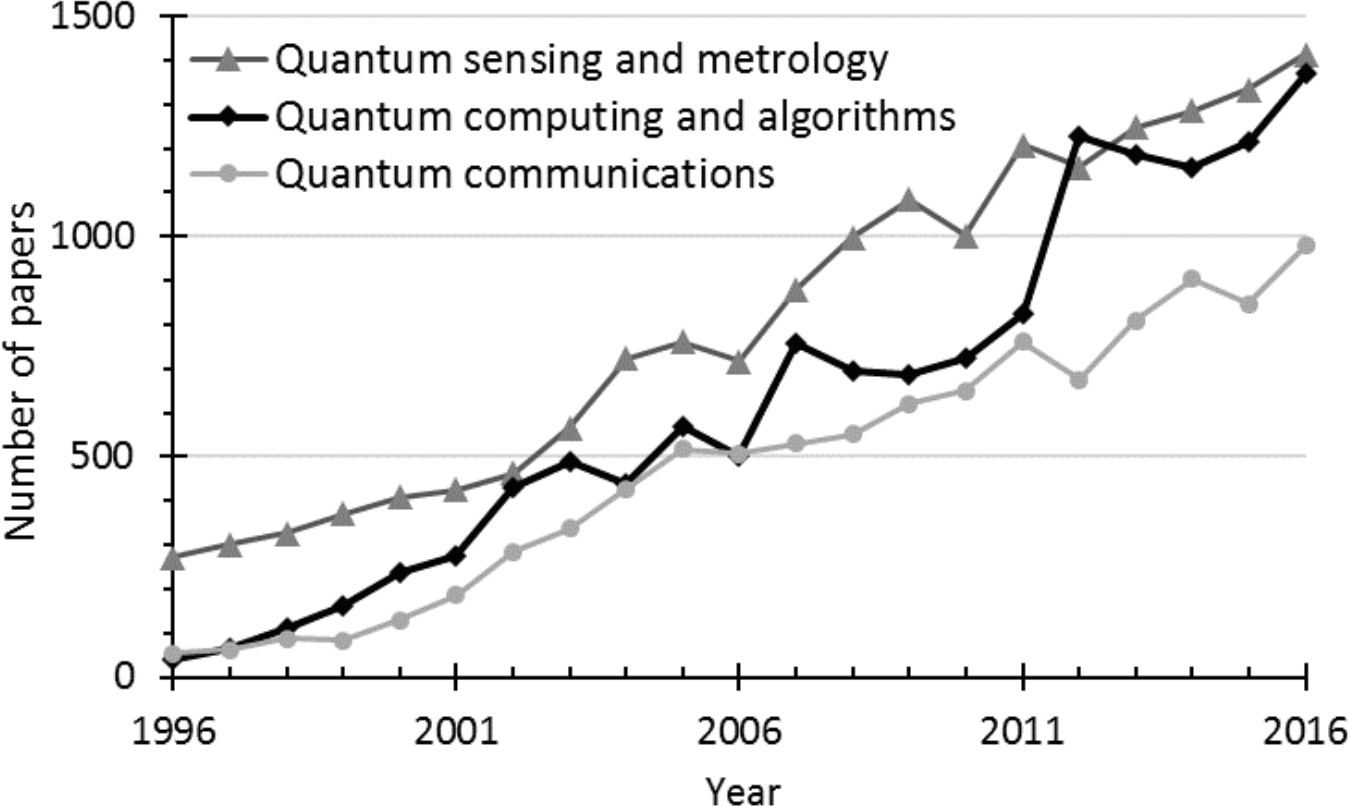}
    \caption{The number of research papers published per year in quantum computing and technologies \cite{engineering2019quantum}}
    \label{fig:Papers_evolution}
\end{figure}

\subsection{State of the Quantum Technologies}
Another point to highlight is the fact that, despite the intensive development of quantum algorithms, the number of available quantum algorithms is still small (Fig. \eqref{fig:Papers_evolution}) compared to that of classical algorithms basically because the current quantum gates are only several tens of quantum bits (qubits), which is relatively low for any significant quantum computing operation \cite{Joh18}. 

\begin{figure}[b!]
    \centering
    \includegraphics[width=0.7\textwidth]{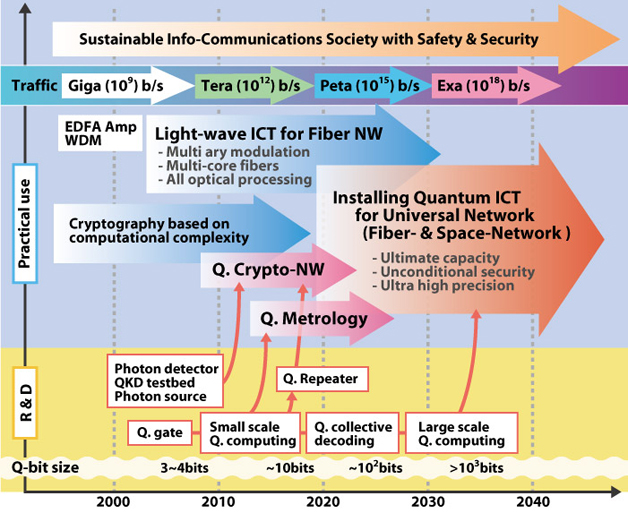}
    \caption{Quantum Computing is becoming the talk of the town \cite{aaronson2008limits}}
    \label{fig:QC_STATE}
\end{figure}

\begin{figure}[b!]
    \centering
    \includegraphics[width=0.7\textwidth]{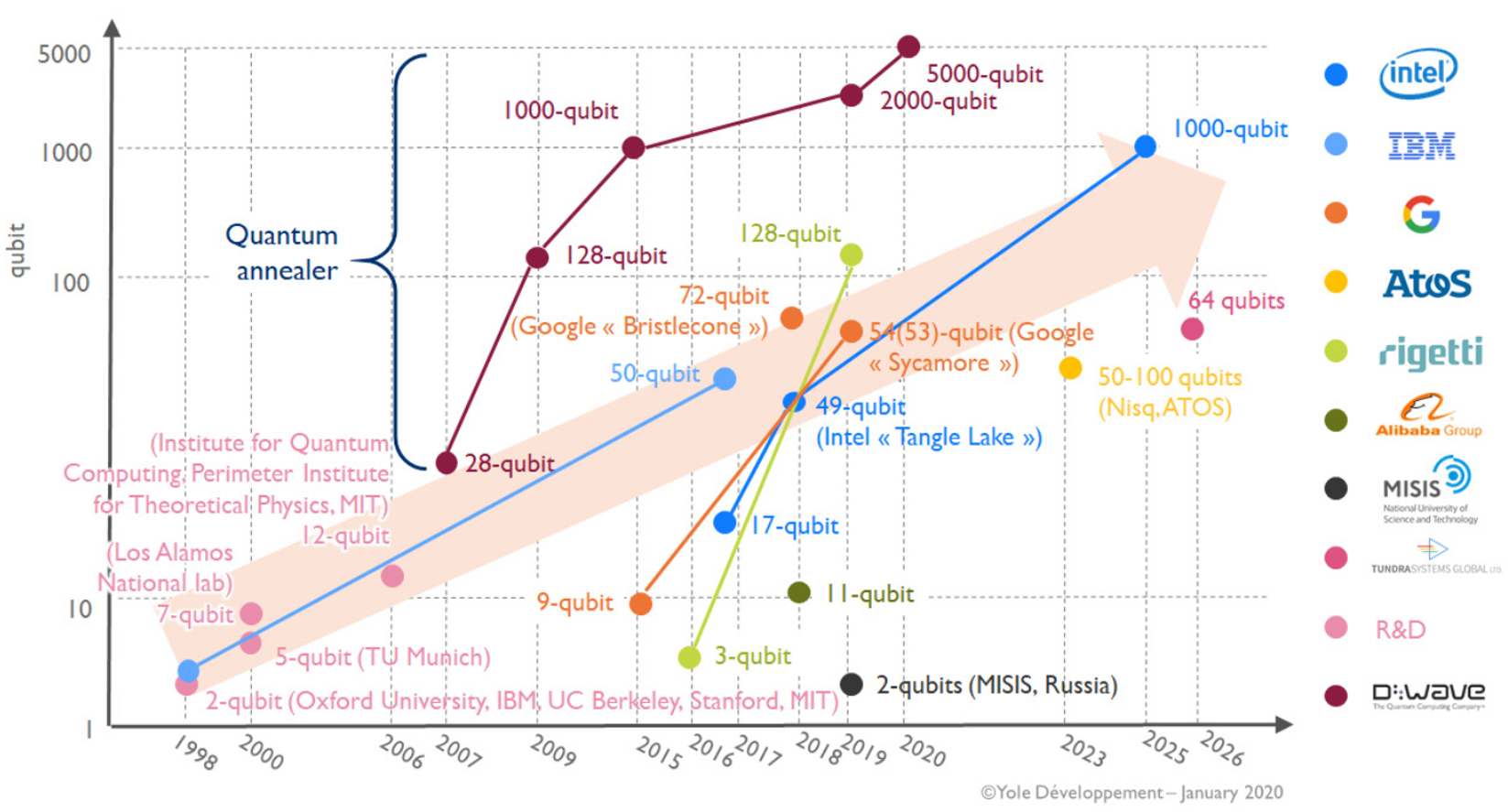}
    \caption{The number of qubits achieved by date and organisation. Roadmap by YOLE Development \cite{YOLED}}
    \label{fig:QUBITS_EVOLUTION}
\end{figure}

\begin{figure}[t!]
    \centering
    \includegraphics[width=0.5\textwidth]{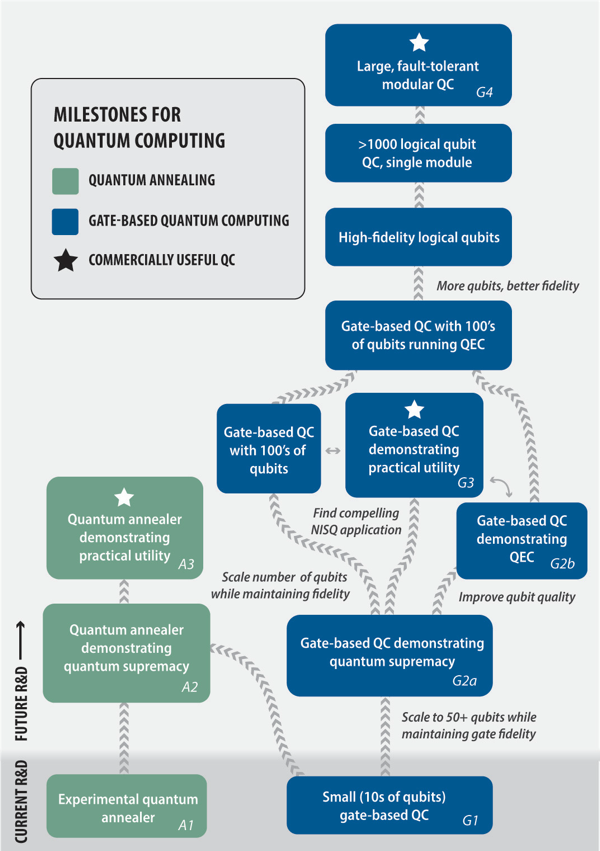}
    \caption{Metrics and milestones to help monitor the development of quantum computing. Roadmap by MIT \cite{engineering2019quantum}}
    \label{fig:QC_MILESTONES}
\end{figure}

\begin{figure}[!h]
    \centering
    \includegraphics[width=0.7\textwidth]{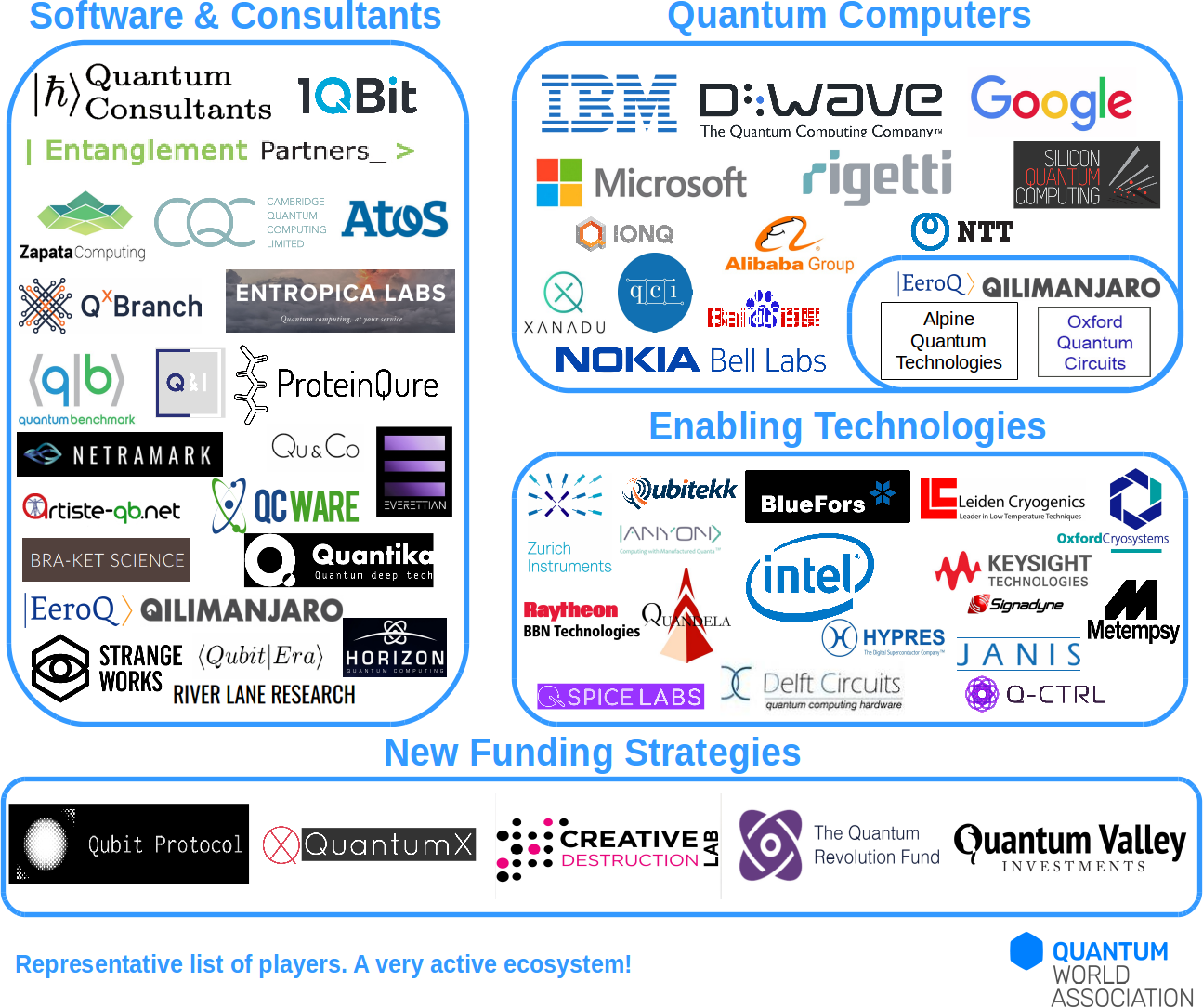}
    \caption{Companies and startups involved in the 2018 quantum computing ecosystem. Roadmap by Quantum World Association}
    \label{fig:QC_COMPANIES}
\end{figure}

This delay was partial until very recently, and it was believed that quantum computing would never or seldom come true. However, the latest advances in the different techniques (D-Wave, IBMQ, Xanadu, ...) of quantum gate implementations, as well as the proof of the precision threshold theorem \cite{Yan10}, give rise to optimism (Figures \eqref{fig:QC_STATE} and \eqref{fig:QUBITS_EVOLUTION}) that quantum computers \eqref{fig:True_QC_image} to large-scale could become a reality quite soon \cite{Joh18}. We can also appreciate (Figures \eqref{fig:QC_MILESTONES} and \eqref{fig:QC_COMPANIES} ) the metrics and milestones to help monitor the development of quantum computing and the companies and startups involved in the 2018 quantum computing ecosystem. 

One of the most significant problems that QIS faces is the physical deployment issues. There are many potential technologies, such as nuclear magnetic resonance (NMR), ion traps, quantum cavity electrodynamics, photonics, quantum dots, and superconducting technologies, to name just a few. However, it is not clear which technology will prevail. For example, Xanadu (photonics) technology seems to be most likely for quantum teleportation \cite{Sor20,8715261}.

Right now, there is a big battle going on between the significant manufacturers of quantum computers to achieve the highest number of qubits and 'impose' their scalability plan. In knowing who will be able to define the standard and, above all, get the quantum supremacy \cite{Aru19}.

\section{Noisy Intermediate-Scale Quantum}

The \textit{Noisy Intermediate-Scale Quantum} era known as (NISQ) era \cite{Joh18}, is defined \textit{as the era leading quantum processors to contain about 50 to a few hundred qubits. Still, it is not advanced enough to reach fault tolerance nor large enough to profit sustainably from quantum supremacy. It is used to describe the current state of the art in the fabrication of quantum processors.}

To summarise, the computers are challenging to achieve, and in the near term, there will NISQ computers with limited performance. To seize quantum computing during the NISQ era, algorithms with low resource demands and capable of returning approximate solutions are explored. In addition, quantum states cannot be indefinitely maintained over time, and the purely quantum properties are steadily lost during the execution of a quantum algorithm. Until the present time, the two greatest achievements are the double accomplishment of the so-called quantum supremacy, that is, using a quantum computer to solve a problem more efficiently and with better performance than any classical computer. This means that the devices in the NISQ era are not expected to be more powerful and change the world by themselves but rather to be an intermediate step towards a new generation of computers.

The real problem of this era is related to \textit{decoherence} \cite{Jan12}. \textit{Decoherence} is associated with the interaction of qubits with environments that blur the fragile states of overlap (entanglement). This results in the introduction of random errors due to the environment. However, there are quantum error correction techniques known as the \textit{Quantum Error Correction Concept} (QECC) \cite{Bes12,ARC96}. One of the powerful applications of quantum error correction is based on the protection of quantum information, as it is dynamically subjected to quantum computing. Imperfect quantum gates affect quantum computing by introducing errors into the computed data. Also, imperfect control gates add errors into the processed sequences as incorrect operations can be applied. However, this imperfection gives rise to exciting computing techniques and algorithms based on variational calculations. This opens up a world of possibility to the era of \textit{Quantum Machine Learning} (QML) \cite{Jac18,Pet14,Mar18}. The objective of QECC is to deal with errors introduced by quantum channels and those presented by (imperfect) quantum gates during the encoding and decoding process. Because of this, the reliability of the data processed by quantum computers depends on the probability of error per gate being below a certain threshold known as the precision threshold theorem. This is what NISQ defines as. \textit{'Noisy'} because we don't have enough qubits leftover for error correction, so we'll have to use the imperfect qubits directly on the physical layer and \textit{'Intermediate-Scale'} due to its reduced qubit (but not too small).
To understand quantum mechanics, we have to appreciate its postulates. These postulates provide a connection between the physical world and the mathematical formalism of quantum mechanics. It is important to emphasise that it is unnecessary to understand quantum mechanics in detail to know how to make a program on a quantum computer. But if we want to make intelligent and efficient algorithms, it is more than recommended to understand the fundamentals of quantum mechanics. The analogy to understand what we mean is the following: To program we can do it in \textit{PhP}, and \textit{python}, high-level languages, but we can also program in \textit{C} or in \textit{assembler} (machine language). If we want to be experts or make efficient algorithms, it would be useful to know the machine's architecture and to program in the machine language.

In the next section, we will present the postulates of quantum mechanics necessary to understand and, above all, delve into quantum computing. This thesis is based on the postulates listed in this reference \cite{nielsen2002quantum} \textit{Quantum Computation and Quantum Information}

\section{Postulate 1}\label{sec:postulate_1_S}

The \textit{postulate 1} is defined as \textit{associated with any isolated physical system is a complex vector space with an internal product (Hilbert space) known as the state space of the system. The system is fully described by its state vector, which is a unit vector in the state space of the system.}

This postulate explains the space of the quantum state and describes the area where quantum mechanics takes place. This area is nothing more than linear algebra in Hilbert vector space \cite{Nic88}. It is essential to know that, given a physical system, quantum mechanics does not tell us what the state space of that system is, nor can it tell us what the system's state vector is. Therefore, it is imperative to have this clear postulate. In other term, what we mean is that it is challenging to know the state of a quantum system at all times.

Let's consider that the simplest quantum mechanical system is the qubit. A qubit has a two-dimensional state space   $\vert 0 \rangle$  and  $\vert 1 \rangle$  that form an orthonormal basis for that state space. Then you can write an arbitrary state vector in the state space.

\begin{equation}
\label{quantum_state_1q}
    \vert  \psi  \rangle =a \vert 0 \rangle +b\vert 1 \rangle. 
\end{equation}

Where  $a$  and  $b$  are complex numbers. The condition that $ \vert  \psi  \rangle$  is a unit vector is given by  $\langle\psi \vert  \psi  \rangle =1$, is therefore equivalent to $\vert a \vert ^{2}+ \vert b \vert ^{2}=1$. The term  $ \langle \psi  \vert \psi \rangle =1$ is known as the normalisation condition for state vectors.

The way a qubit differs from a classic bit is that there are overlaps of these two states, in the form $a\vert 0 \rangle +b \vert 1 \rangle$ , where it is not possible to say that the qubit is definitely in the state  $\vert 0 \rangle$, or definitely in the state $\vert 1 \rangle$. In other words, what we mean is that the quantum state is a linear combination of the components  $a$  and $b$.

If we had a system of more than one qubit, the expression of the quantum state would be of the form  $\sum _{i}^{}a_{i} \vert  \psi _{i} \rangle$, and we would have a system with the superposition of the states $\vert  \psi _{i} \rangle$  and of amplitude  $a_{i}$ for the state $\vert  \psi _{i} \rangle$.

\section{Postulate 2}\label{sec:postulate_2_S}
The \textit{postulate 2} is defined as \textit{a unitary transformation describes the evolution of a closed quantum system. That is, the state  $\vert  \psi  \rangle$  of the system at the time  $t_{1}$  is related to the state  $\vert  \psi ^{'} \rangle$  of the system at the time  $ t_{2}$  by a unit operator  $U$  that depends only on times $t_{1}$  and  $t_{2}$, $\vert  \psi ^{'} \rangle = U \vert  \psi  \rangle$. This postulate gives a standard for describing quantum state changes over time.}\\

From Ref. \cite{Mic00} we know that quantum mechanics does not tell us the state space or quantum state of a particular quantum system; it also does not tell us which unit operators  $U$  describe real-world quantum dynamics. On the other hand, quantum mechanics assures us that the evolution of any closed quantum system can be described in this way.

\begin{equation}
\label{postulate_2}
    \vert  \psi ^{'} \rangle=U \vert  \psi  \rangle.
\end{equation}
In a way, we are saying that if we want to describe an $n$ qubit quantum system subject to time evolution, we must calculate its unit operators $U$. The challenge would be to find such unit operators. Quantum gates are somehow the operators that act on quantum states.
This postulate is very interesting and requires that the quantum system be closed \cite{Bes12}\cite{Pau18}. Moreover, various derivatives emerge from this postulate.

\subsection{Postulate 2’}\label{sec:postulate_22_S}
The \textit{postulate 2'} is defined as \textit{the time evolution of the state of a closed quantum system is described by the Schrödinger equation \eqref{shorodinger_eq}.}\\

\begin{equation}
\label{shorodinger_eq}
    i\hbar \frac{ \delta  \vert \psi  \rangle }{ \delta t}= H \vert \psi  \rangle. 
\end{equation}

Where $\hbar$ is Planck's constant and $H$ is a fixed hermitian operator known as the Hamiltonian of the closed system \cite{Bes12}.

In some way, we can say that if we know the Hamiltonian of a system, then we understand its dynamics completely. This concept is precious for modelling the system, and we will use it a lot throughout our thesis.

During the 20th century, much of the scientific (physical) community has been dedicated to discovering the Hamiltonian of any quantum system. And the conclusion they reached is that, in general, finding out that the Hamiltonian necessary to describe a particular physical system is challenging.

The Hamiltonian is static and a Hermitian operator that allows us to make its spectral decomposition with eigenvalues of energy E and the normalised eigenvectors corresponding to energy $E$ as eigenstates $\vert E \rangle$.

\begin{equation}
\label{Energie_P2}
    \vert H \rangle ~= \sum_{E}^{}E \vert E \rangle \langle E \vert, 
\end{equation}

\begin{equation}
\label{expon_energy} 
     \vert E \rangle =exp (-i\frac{Et}{\hbar})  \vert E \rangle .
\end{equation}

The equation \eqref{expon_energy} is valid only in the case of having a static Hamiltonian. The lowest energy is known as the ground state energy for the system, and the corresponding energy proper state (or adequate space or steady-state) is known as the ground state.

\begin{equation}
\label{Dem_Unitary}
    \vert \psi (t_{2}) \rangle = exp [\frac{-iH (t_{1}-t_{2})}{\hbar}] \vert  \psi(t_{1}) \rangle = U (t_{1}-t_{2}) \vert  \psi (t_{1})\rangle. 
\end{equation}

Where we define

\begin{equation}
\label{Generalization_U}
    U(t_{1}-t_{2}) \equiv exp [\frac{-iH (t_{1}-t_{2}) }{\hbar}]. 
\end{equation}

This interpretation is compelling and extremely useful since it is shown that any unitary operator $U$ can be written in the form \eqref{Simplify_U} tanking $\hbar=1$  and $t_{2}=0$ the equation comes out to be \eqref{Dem_Unitary}

\begin{equation}
\label{Simplify_U}
    U=exp(-iK). 
\end{equation}

With $K$, a Hermitian operator. 

\section{Postulate 3}\label{sec:postulate_3_S}
The \textit{postulate 3} is defined as \textit{quantum measurements are described using a  ${M_{m}}$  collection of measurement operators. These are the operators that operate in the state space of the system being measured. The $m$  index refers to the measurement results in the experiment. If the state of the quantum system is $\vert \psi  \rangle$  immediately before the measurement, the probability that the result $m$  will occur is given by the equation \eqref{Postulate3_eq}.}\\

\begin{equation}
\label{Postulate3_eq}
    p(m) = \langle \psi \vert  M_{m}^{T}M_{m} \vert  \psi  \rangle.
\end{equation}

The state of the system after the measurement is \eqref{Measurement_eq}.

\begin{equation}
\label{Measurement_eq}
    \frac{M_{m} \vert \psi  \rangle}{\sqrt[]{\langle \psi  \vert  M_{m}^{T}M_{m} \vert  \psi  \rangle}}.
\end{equation}

And, the measurement operator satisfies the completeness equation \cite{Bes12} that is given by the following equation \eqref{cpmpleteness_eq} for all the values of the quantum state $\vert \psi  \rangle$.

\begin{equation}
\label{cpmpleteness_eq}
    \sum_{m}^{}M_{m}^{T}M_{m}=Ip (m) =\langle \psi  \vert  M_{m}^{T}M_{m} \vert  \psi  \rangle =1.
\end{equation}

The value of the measure is the probability described by $\vert a \vert ^{2}+ \vert b \vert ^{2}=1$.

If we want to make the different $M_{m}$ observations, the operations we would be doing are the following:
\begin{equation}
\label{M0_eq}
    M_{0}= \vert 0 \rangle \langle 0 \vert = \begin{bmatrix}
        1 \\ 0
    \end{bmatrix}
    \begin{bmatrix}
        1 & 0 \\
    \end{bmatrix}
    = \begin{bmatrix}
        1  &  0\\
        0  &  0\\
    \end{bmatrix}.
\end{equation}

\begin{equation}
\label{M1_eq}
    M_{1}= \vert 1 \rangle \langle 1 \vert = \begin{bmatrix}
        0 \\ 1
    \end{bmatrix}
    \begin{bmatrix}
        0 & 1 \\
    \end{bmatrix}
    =\begin{bmatrix}
        0  &  0\\
        0  &  1\\
     \end{bmatrix}.
\end{equation}

With  $ \langle \psi \vert$  the conjugate of $\vert  \psi  \rangle $, represented by $ \vert  \psi  \rangle = \langle \psi   \vert ^{T} $ 

\begin{equation}
\label{measure_m0}
    M_{0}\vert 0 \rangle = \vert 0 \rangle \langle 0 \vert 0 \rangle =\begin{bmatrix}
        1  &  0\\
        0  &  0\\
    \end{bmatrix}
    \begin{bmatrix}
        1\\
        0\\
    \end{bmatrix}
    = \begin{bmatrix}
        1\\
        0\\
    \end{bmatrix}
    = \vert 0 \rangle.
\end{equation}
  
  \begin{equation}
\label{measure_m1}  
    M_{0} \vert 1 \rangle = \vert 0 \rangle \langle 0 \vert 1 \rangle = \begin{bmatrix}
        1  &  0\\
        0  &  0\\
    \end{bmatrix}
    \begin{bmatrix}
        0\\
        1\\
    \end{bmatrix}
    = \begin{bmatrix}
        0\\
        0\\
    \end{bmatrix}
    = 0.
\end{equation}

knowing that  $\vert 0 \rangle \bot \vert 1 \rangle$,  $\langle 0 \vert  1 \rangle$ disappears; It is cancelled. If we project on any state, we will have:

\begin{equation}
\label{projection}
    M_{0}\vert \psi  \rangle = \vert 0 \rangle \langle 0 \vert \psi \rangle = \vert 0 \rangle a \vert 0 \rangle +b \vert 1 \rangle = a \vert 0 \rangle. 
\end{equation}
We can generalise \eqref{measure_m0}: let  $M_{0}$, let $i$  be any qubit and let  $\vert  \psi  \rangle$ be a multi-state of qubits, the measurement or the projection on the state  $\vert 0 \rangle$ can be expressed as:

\begin{equation}
\label{Traza_Meas}
    Tr \begin{bmatrix}
    \vert \psi  \rangle  \langle \psi \vert M_{0}^{i}\\
    \end{bmatrix}.
\end{equation}

If the qubit $i$ is measured in the state $\vert 0 \rangle$, then the system will be in the state expressed by \eqref{Project_Tr}.

\begin{equation}
\label{Project_Tr}
    \frac{M_{0}^{i} \vert  \psi  \rangle}{\Vert M_{0}^{i}\vert  \psi  \rangle  \Vert_{2}}.
\end{equation}

Another way to get the result of the measurements is to remember that the measurement operator is Hermitian; this translates to  $M_{0}^{2}=M_{0}$ the same for $M_{1}^{2}=M_{1}$. Also remembering that the relationship of completeness obeys the equation $I=M_{0}^{T}M_{0}+M_{1}^{T}M_{1}=M_{0}+M_{1}$.

One of the crucial applications of \eqref{sec:postulate_3_S} is to distinguish quantum states. We recall that since the quantum system must be closed \eqref{sec:postulate_1_S}), getting to find out the quantum states represents a titanic and less intuitive task as in classical computing.

To clarify what we have just advanced, we will demonstrate the absurdity. We consider two non-orthogonal quantum states  $\vert  \psi_{1} \rangle$  and  $\vert  \psi_{2} \rangle$ and assume that the measurement is possible. If  $\vert  \psi_{1} \rangle$  and  $\vert  \psi_{2} \rangle$ are prepared, the measures (observations) will respond to the completeness \eqref{cpmpleteness_eq}. Defining $E_i \equiv \sum_{j:f(j)=i}^{}M_{j}^{\dagger}M_{j}$, where the probability of measuring $j$ such that $f(j)=1$ and $f(j)=2$ must be 1.  These observations $E_i$ may be written as:

\begin{equation}
\label{57}
    \langle \psi _{1} \vert  E_{1} \vert  \psi _{1} \rangle =1; \quad \langle \psi _{2} \vert  E_{2} \vert  \psi _{2} \rangle =1.
\end{equation}

Knowing that  $\sum _{i}^{}E_{i}=I$ therefore  $\langle \psi _{1} \vert  E_{1} \vert  \psi _{1} \rangle =1$  and must  $\langle \psi_{1} \vert  E_{2} \vert  \psi _{1} \rangle =0$,  and  $\sqrt[]{E_{2}} \vert  \psi  \rangle =0$. If we decompose  $\vert  \psi _{2} \rangle = \alpha  \vert  \psi _{1} \rangle + \beta  \vert  \varphi  \rangle $ with  $\vert  \psi _{1} \rangle$ and $\vert  \varphi  \rangle$ they are orthonormal. This leads us to the fact that $\vert  \alpha  \vert ^{2}+ \vert  \beta  \vert ^{2}=1$  and that  $\vert  \beta  \vert <1$  while $\vert  \psi_{1} \rangle$  and  $\vert  \psi _{2} \rangle$  are not orthonormal. With  $\sqrt[]{E_{2}} \vert  \psi  \rangle =  \beta \sqrt[]{E_{2}} \vert  \varphi  \rangle$. What contradicts completeness’s equation \eqref{Postulate3_eq}. 

\begin{equation}
\label{Measure_Post_Ob}
    \langle \psi_{2} \vert  E_{2} \vert  \psi_{2} \rangle = \vert  \beta  \vert ^{2} \langle \varphi  \vert  E_{2} \vert  \varphi  \rangle  \leq  \vert  \beta  \vert ^{2}<1.
\end{equation}

Continuing with \textit{postulate 3}, projective measurements form a particular case of it, being very useful and straightforward. Due to its simplicity, this measurement is the basis of many algorithms to know the observables.

A projective measurement is described by an observable, $M$, a Hermitian operator in the system's state space being observed. The observable has a spectral decomposition.

\begin{equation}
\label{59}
    M= \sum _{m}^{}mP_{m}.
\end{equation}

where $P_{m}$ is known as the projector in the proper space of $M$  with the right value  $m$. If we measure on the state $\vert  \psi  \rangle$, the probability of having the result $m$ is

\begin{equation}
\label{realObservable}
    p(m) =\langle \psi  \vert  P_{m} \vert  \psi  \rangle. 
\end{equation}
so, the quantum state just after the measurement will be:

\begin{equation}
\label{equation_Qstate}
    \frac{P_{m} \vert \psi \rangle }{\sqrt[]{\langle \psi  \vert  P_{m} \vert  \psi  \rangle }}.
\end{equation}

Otherwise, what we are saying is that  $\langle M \rangle =\langle \psi  \vert  M \vert  \psi  \rangle$. We can generalise this formula with the mean and variance:

\begin{equation}
\tag{62}
\label{measureProjection}
    [\Delta M] ^{2}= \langle ( M- \langle M \rangle) ^{2} \rangle = \langle M^{2} \rangle -  \langle M \rangle ^{2} \Delta M=\sqrt[]{ \langle M^{2} \rangle -  \langle M \rangle ^{2}}.
\end{equation}

One of the differences to highlight is differentiating \textit{measurement operations (M)} from \textit{projectors (P)}, since $P^2 = P$, but $M^2 != M$.

This formula \eqref{measureProjection} is the basis of Heisenberg's uncertainty principle \cite{Bes12}. The important thing about the Heisenberg principle is that we cannot simultaneously measure/know the position and velocity of the electron (a particle). Therefore, it is impossible to determine its trajectory. We are somehow telling ourselves that measurement is destructive (but this is not the basis of Heisenberg's principle) because the position and velocity of the particle are in overlap. Measuring one collapses the wave function and destroys the other component/s.

A mathematical explanation can be described as follows. Suppose  $A$ and  $B$  are two Hermitian operators, and $\vert  \psi  \rangle$ is a quantum state. Suppose that  $\langle \psi   \vert  AB  \vert  \psi  \rangle  = x + iy$, where  $x$  and $y$ are real. Applying the switches and anti-switches to the Hermitian operators (matrices)  $A$  and  $B$, we arrive at  $\langle \psi  [A,B] \psi  \rangle =2iy$ and  $\langle \psi  \vert  \{ A,B \}   \vert \psi  \rangle ~ = 2x$. This implies that:

\begin{equation}
\label{unvertainly_prin}
    \vert \langle \psi   \vert [A,B] \vert \psi  \rangle  \vert ^{2}+ \langle \psi   \vert ~ {A,B}  \vert  \psi  \rangle  \vert ^{2}=4 \vert \langle \psi   \vert AB \vert \psi  \rangle  \vert ^{2}~~.
\end{equation}

For the Cauchy-Schwarz inequality

\begin{equation}
\label{64}
    \vert \langle \psi  \vert AB \vert  \psi  \rangle  \vert ^{2} \leq \langle \psi   \vert A^{2} \vert \psi  \rangle \langle \psi  \vert B^{2} \vert  \psi  \rangle.
\end{equation}

underestimating the negative term, we arrive at:

\begin{equation}
\label{65}
     \vert \langle \psi   \vert [A,B]  \vert \psi  \rangle  \vert ^{2}  \leq 4 \langle \psi   \vert A^{2} \vert \psi  \rangle \langle \psi   \vert B^{2} \vert  \psi  \rangle. 
\end{equation}

if we now consider two observables  $M_{1}$  and  $M_{2}$, and substitute  $A = M_{1}-  \langle M_{1} \rangle$  and  $B =M_{2}-  \langle M_{2} \rangle$  in the last equation, we obtain the Heisenberg uncertainty principle from the equation \eqref{eq_unvertain_prin}.

\begin{equation}
\label{eq_unvertain_prin}
    \Delta M_{1} \Delta M_{2} \geq \frac{\langle \psi [M_{1},M_{2}]    \psi  \rangle }{2}.
\end{equation}

The correct interpretation of the uncertainty principle is that if we prepare a large number of quantum systems in identical states,  $\vert \psi  \rangle$, and then make measurements of $M_{1}$  in some of those systems, and of $M_{2}$  in others, then the standard deviation  $\Delta M_{1}$  of the results of  $M_{1}$  multiplied by the standard deviation  $\Delta M_{2}$  of the results of  $M_{2}$ will satisfy the inequality $\Delta M_{1} \Delta M_{2} \geq \frac{\langle \psi   [M_{1},M_{2}] \psi  \rangle }{2}$.

In the commutation relationship for the observable $X$ coordinates and moment $P$ is, $[X, P]  = j\hbar$. If we introduce it in the equation above, we get to \eqref{Heisenberg_dem_Eq}:

\begin{equation}
\label{Heisenberg_dem_Eq}
    \langle \Delta X \rangle^{2}  \langle \Delta P \rangle^{2} =\frac{\hbar^{2}}{4}.
\end{equation}

We make a parenthesis to explain the behaviour of the phase-in quantum mechanics. We already know that quantum mechanics is defined in the complex vector space (Hilbert space). We also know that a vector representation of a vector (wave) can be described with its argument and angle. Besides, the phase as an operator applied to a quantum state in quantum mechanics does not change the quantum state.

Now, let us consider that the state $e^{i \theta } \vert \psi  \rangle$, where $\vert \psi  \rangle$, is a state vector and $\theta$  is a real number. We say that the state $e^{i \theta } \vert \psi  \rangle$, is equal to  $\vert \psi  \rangle$ where the factor  $e^{i \theta }$ is known with the global phase of the system \cite{Bes12,Iva12}. This property is fascinating and useful when writing a quantum algorithm. Saving this operation entails a gain in time and computational cost.

Suppose we want to make the measurements of the observables  $M_{m}$ on the quantum state  $e^{i \theta } \vert \psi  \rangle$, with $\vert \psi  \rangle$  the state vector. Then, the respective measurements will be given by the following equations:

\begin{equation}
\label{general_eq_princ_unc}
    \langle \psi  \vert M_{m} \vert  \psi  \rangle  \langle \psi  \vert  e^{-i \theta }M_{m}^{T}M_{m}e^{i \theta } \vert  \psi  \rangle  = \langle \psi  \vert  M_{m}^{T}M_{m} \vert  \psi  \rangle.
\end{equation}

We see that the global phase does not affect measures, but another phase does. This phase is known as the relative phase \cite{Iva12}. We will use an example to explain it. Let's consider two quantum states:
\begin{equation}
\label{69}  
    \vert  \psi _{1} \rangle =a \vert 0 \rangle +b \vert 1 \rangle  \quad \text{and} \quad  \vert  \psi _{2} \rangle =a \vert 0 \rangle -b \vert 1 \rangle. 
\end{equation}
We see that the state amplitude  $\vert 1 \rangle$  of the first quantum state  $\vert  \psi _{1} \rangle$ is  $+b$  and $a$, the state amplitude $\vert 0 \rangle$ of the second quantum state $\vert  \psi _{2} \rangle$ is  $-b$. We see that the only difference is the sign, not the amplitude. If two quantum states have the same amplitudes and differ exclusively by the phase (sign), this phase is known by the relative phase of the system. And generically, we can write it as follows  $a=exp (i \theta)b$. This concept is fascinating and is the basis of quantum gates that will allow quantum computing.

\section{Postulate 4}\label{sec:Postulate_4_S}
The \textit{postulate 4} is defined as \textit{the state-space of a composite physical system is the tensor product of the state spaces of the components of physical systems. Furthermore, if we have systems numbered from 1 to  $n$, and the number of system $i$ is prepared in the state  $\vert  \psi _{i} \rangle$, then the joint state of the total system is $\vert  \psi _{1} \rangle \otimes \vert  \psi _{2} \rangle  \otimes  \cdot  \cdot   \cdot  \otimes \vert  \psi _{n} \rangle$.}

This postulate is the basis for creating the complex system of more than two qubits or quantum states. This postulate also allows us to define one of the most exciting and puzzling ideas associated with composite quantum systems: entanglement.
Consider the two-qubit states  $\vert \psi  \rangle =\frac{ \vert 00 \rangle + \vert 11 \rangle }{\sqrt[]{2}}$, this state has the remarkable property that there are no unique qubit states  $\vert a \rangle$ and $\vert b \rangle$ such that  $\vert  \psi  \rangle = \vert a \rangle  \vert b \rangle$. In other words, we say that $\vert \psi  \rangle  \neq  \vert a \rangle  \vert b \rangle$. This property is one of the reasons for the empowerment of quantum computing.

%\section{Postulate 5 and 6}\label{sec:Postulate_56_S}
%In this thesis work, we are based on the postulates listed in this reference \cite{nielsen2002quantum} \textit{Quantum Computation and Quantum Information}, where postulate 5 (The temporal evolution of a system) of quantum mechanics \cite{cohen1986quantum} is contemplated like our postulate 2' \eqref{sec:postulate_22_S} and the postulate 6 (commutation rules) of quantum mechanics that define the positional and momentum operators that satisfy the following commutation rules, they are treated as operators not like postulate.

\section{Summary}
This section's right conclusion highlights that these four postulates of quantum mechanics define how we can conceptualise and face a problem about quantum mechanics and its computation. The first postulate establishes the space for quantum mechanics by specifying how the state of an isolated quantum system should be described. The second postulate illuminates us on the dynamics of the closed quantum systems and its description through the Schrödinger equation and using the unitary evolution. In the third postulate, we are explained how to make the measurement describing the importance of the characteristics (restrictions) when extracting information (measuring) from our quantum systems. And finally, the fourth postulate reveals how we can create composite systems.
One of the most shocking and interesting ideas of quantum mechanics is that we cannot directly observe the state vector and that it is in charge of deciphering the behaviour of any quantum system.
\begin{figure}[h!]
    \centering
    \includegraphics[width=0.5\textwidth]{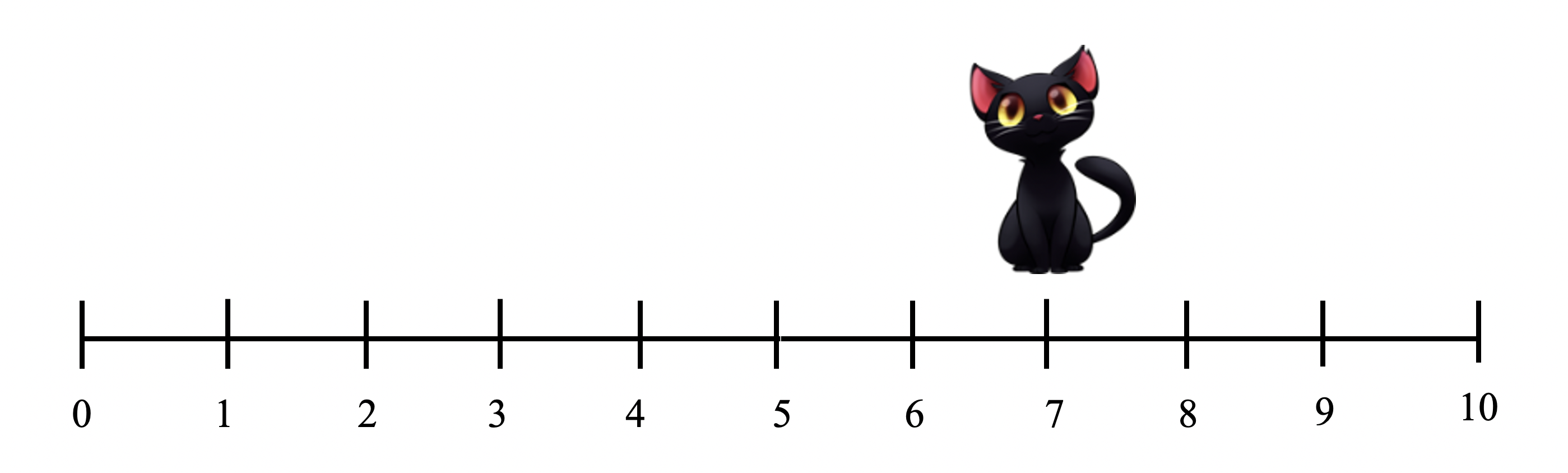}
    \caption{Observability of a classical system}
    \label{fig:OBS_CLAS}
\end{figure}
Let's imagine that we want to know what position the cat is in \eqref{fig:OBS_CLAS}. In a classic system, thinking about computing, the location (state vector) of the cat is given by the variable  $x= 7$

\begin{figure}[h!]
    \centering
    \includegraphics[width=0.5\textwidth]{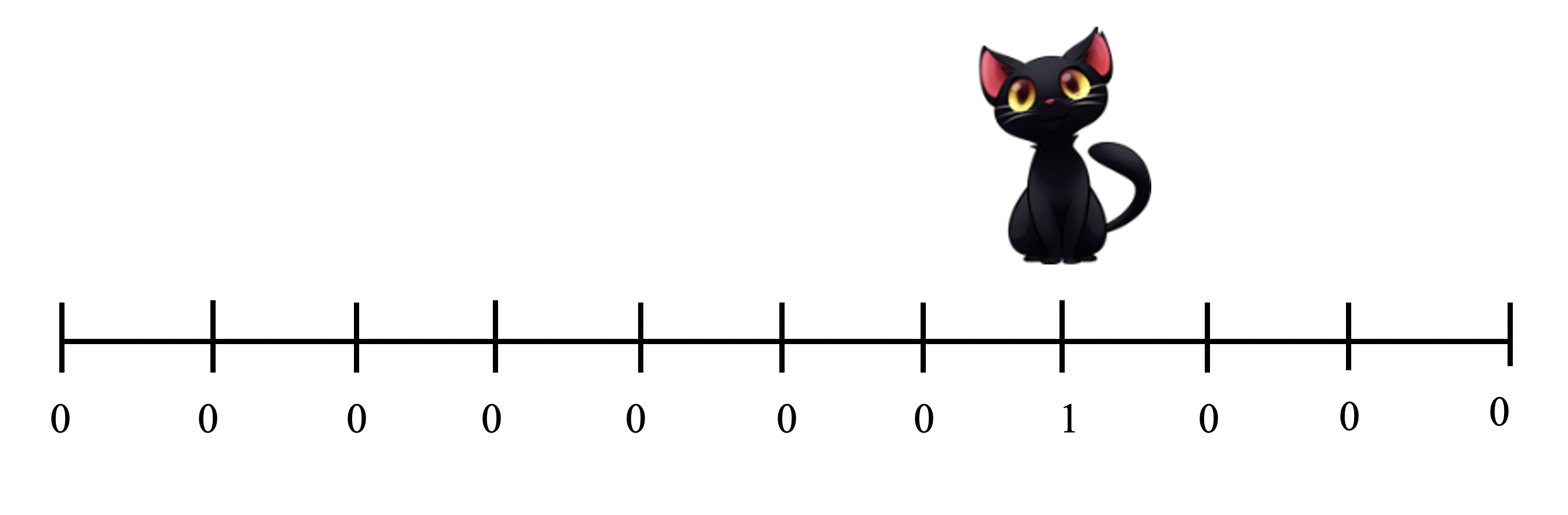}
    \caption{Observability of a quantum system}
    \label{fig:OBSV_QUANT}
\end{figure}
In a quantum system, the position (see figure \eqref{fig:OBSV_QUANT}) of the cat is not so quickly known. In quantum, each element in the state vector contains the probability of finding the cat in a specific place.
And it is represented by:
\begin{equation}
\label{M0_eq_}
     \vert x\rangle = \begin{bmatrix} 0\\ \vdots \\ 0 \\ \vdots \\1 \\ 0 \\ \vdots \\ 0 \end{bmatrix} 
            \begin{matrix} \\  \\  \\ \\ \leftarrow \\  \\  \\  \\ \end{matrix}
             \begin{matrix} \\  \\ \text{Probability of} \\ \text{cat being at} \\ \text{position 7} \\  \\  \\ \end{matrix}.
 \end{equation}
 
Classical physics, classical computation, and our intuition tell us that the fundamental properties of an object, such as energy, position, and velocity, are directly accessible to observation. However, these quantities no longer appear essential in quantum mechanics, depend on the state vector, and cannot be directly observed. Furthermore, merely observing a classical system does not necessarily change the system's state; instead, for a quantum system, observation is an invasive procedure that generally changes the state of the system (the state vector).

 %%%%%%%%%%%%  Starting New Page here %%%%%%%%%%%%%%

\newpage

\chapter{Complexity Class}\label{sec:6}
\section{Introduction}\label{sec:6_}

The \textit{complexity class} is defined as \textit{a set of computational problems of related resource-based complexity, and the two most commonly analysed resources are time and memory.}\\

Some problems appear to be "intractable" (complexity is exponential in the number of bits $n$) in classical computing, but they can be solved polynomially in $n$  by quantum computing. For example, the factorisation of prime numbers is an insoluble problem for classical computing, taking into account the size of the numbers to be factored. Still, solvable in quantum by Shor's algorithm \cite{Pet96}. Peter Shor's algorithm is based on the fact that a composite number can be represented in prime numbers in several steps being polynomial in $n$. The differential and key value of quantum algorithms are that they can explore all branches of a non-deterministic algorithm in parallel through the concept of quantum parallelism (superposition of basic states).

There are necessarily three properties that can vary in the definition of a complexity class. The resource of interest (time, space, etc.), the type of problem being considered (decision problem, optimisation problem, etc.), and the underlying computational model (deterministic Turing machine, probabilistic Turing machine, quantum computer, etc.).

We know that classical algorithms can be classified as useful when the number of steps is a polynomial function of size  $n$. Therefore, the computational complexity of these algorithms is typically denoted as  $P$. The class of problems that can be solved in a non-deterministic manner \cite{Iva12} in polynomial time are called $NP$ \cite{McG14}. The subclass of these problems, which are the most difficult, is complete $NP$ problems \cite{McG14}. For example, the VRP and its derivatives belong to this subclass, and if one of these problems can be solved efficiently, they can all be solved. The kind of problem that can be solved with the amount of the memory polynomial in the input size is called PSPACE \cite{McG14}. Also, the kinds of problems can be solved with high probability by using a random generator known as BPP, which originates from a time-limited error probability polynomial. In addition to all these kinds of complexity, there is a final category that can be solved in polynomial time if the exponential sums of many taxpayers are computable in polynomial time. This is denoted as $P^{\#P}$. In summary, we can establish this relationship between complexity classes.

\begin{equation}
\label{complexity_eq}
     P \subset BPP, P \subset NP \subset P^{\#P} \subset PSPACE.
\end{equation}
Being the quantum analogue of BPP, known as BQP \cite{McG14}, the most interesting in our study of quantum algorithms.

In computational complexity theory, Bounded error Quantum Polynomial-time BQP \cite{McG14} is the class of decision problems decidable by a quantum computer in polynomial time with an error probability of at least 1/3 for all instances. They are problems based on Large Linear Systems, Quantum material simulations and Quantum walk encompassing the SWP.

\section{Quantum complexity}

The \textit{quantum complexity theory} is defined as \textit{the sub-field of computational complexity theory that deals with complexity classes defined using quantum computers. This class studies the hardness of computational problems concerning these complexity classes and the relationship between quantum complexity classes and classical ones.} 

Alan Turing \cite{Bes12} defined a class of machines, known as Turing machines, that can be used to study the complexity of a computational algorithm. In particular, there are the so-called universal machines, which can be used to simulate any other Turing machine, and the latter can be used to simulate all the operations carried out on a modern computer.

This led to the formulation of the Church-Turing thesis \cite{Bes12}. One of the most powerful arguments in computing remains in force today. The study specifies that the class of functions that a Turing machine can calculate corresponds precisely to the class of functions that one would naturally consider computable by an algorithm. This thesis establishes the equivalence between the mathematical description defined by a Turing machine and an intuitive concept. Some problems are not foreseeable, and there is no known algorithm to solve them. The Church-Turing thesis \cite{Bes12} also applies to quantum algorithms.

We know from the complexity class \cite{Iva12} that an algorithm is competent if it can be solved in a polynomial number (P) of steps. We also know that Turing machines can describe these practical algorithms. With the fusion of these last two concepts, we can formulate a robust version of the Church-Turing thesis looking for a universality.  Any computational model can be simulated in a probabilistic Turing machine with a polynomial number of computational steps.

\section{The deterministic Turing machine}
The \textit{Turing deterministic machine} is described by an alphabet  $A$, a set of control states $Q$, and a transition function $\delta$.

\begin{equation}
\label{deterministic_TM}
    \delta :Q  \times  A \rightarrow Q  \times  A \times  D.  
\end{equation}

The alphabet elements are called letters ($A$), and by concatenating the letters, we get words. Set  $D$  is related to the read-write head, with elements  $D= \{-1,0,1\}$  $-1$,  $+1$ and $0$ that indicate the movement of the head to the left, to the right and the foot, respectively. The deterministic Turing machine can be defined as ($Q$, $A$, $\delta$, $q_{0}$, $q_{a}$, $q_{r}$), where the state of the machine is specified by $q \in Q$. In particular, $q_{0}$, $q_{a}$, $q_{r}$  $\in$ $Q$ denotes the initial state, the acceptance state and the rejection state, respectively.

The configuration of the Turing machine is given by $c= (q, x, y)$, where $x$, $y$  $\in A^{'}$, where $A^{'}$ is the set of all the words obtained by concatenating the letters of $A$.

The Turing machine has a tape (memory) specified by $xy$  with  $x$  being the scan (read). Somehow, we can define computation as a sequence of configurations that start with initial setup  $c_{0}$  (sometimes this initial configuration is called the initial state represented by  $\varepsilon _{0}$) until we reach the stop configuration. Computing stops after  $p$  calculation steps (interaction) when one of the configurations does not have a successor or if its state is $q_{a}$ o $q_{r}$.

\section{The probabilistic Turing machine.}
A \textit{probabilistic Turing machine} is defined as \textit{a non-deterministic Turing machine that chooses between the available transitions at each point according to some probability distribution.}\\

There are several more computationally dependent Turing machines. The most generic is the probabilistic Turing machine \cite{McG14}. Since the transition function assigns probabilities to possible operations, the deterministic machine is a specific case of probability. In other words, a probabilistic Turing machine is a non-deterministic Turing machine. That randomly selects possible transitions according to some probability distribution and is defined as follows:

\begin{equation}
\label{Probabilistic_TR}
    \delta :Q  \times  A \times Q  \times  A \times D.   
\end{equation}
In this case, a stochastic matrix can describe the machine state transitions $D= [0,1]$.
In the same way, as in the deterministic Turing machine, a given configuration is a successor configuration with probability  $\delta$. A terminal structure can be calculated from an initial setup with a probability given by the product of probabilities of intermediate configurations, which lead to it by a particular calculation defined by a sequence of states. For example, $A$  generalised case is the deterministic Turing machine of type m. Said $m$ tapes characterise machine, an alphabet  $A$, a finite state of control states $Q$ and the following transition function.

\begin{equation}
\label{full_Prob_TM}
    \delta  : (Q  \times  A^{m})  \rightarrow Q  \times (A \times  D) ^{m}.
\end{equation}

The configuration of a machine of type m is given by $(q, x_{1},~y_{1},.,x_{m},~y_{m})$, where $q$ is the current state of the machine,  $(x_{i},~y_{i})$  $\in$  $A^{'} \times A^{'}$, and  $x_{i}y_{i}$  denote the content of the \textit{i-th} type. Machines of type m are suitable for problems related to parallelism \cite{Pau18}.

If the computation time required for a machine of a type is $t$, then the computation time needed for a device of type $m$  is $O( t^{1/2})$, while in terms of computational complexity, they are comparable.

\section{Quantum Turing Machine}

The \textit{Quantum Turing Machine}, also known as the universal quantum computer, is \textit{an abstract machine used to model the effects of a quantum computer.}\\

The following transition function characterises the Turing quantum machine:

\begin{equation}
\label{Quantum_TM}
    \delta  :Q  \times  A \times Q  \times  A \times  D \rightarrow C.
\end{equation}

In quantum, the transition function moves a given configuration into a range of successor configurations. Each of which has a quantum probability amplitude, which corresponds to a unit transformation of a quantum state that is the set of complex numbers by $C$.
Scientist and physicist Charles Henry Bennett demonstrated that \textit{m-type Turing machines} could be simulated using reversible Turing machines with a specific reduction in efficiency \cite{Pau18}. Furthermore, Professor of Electrical Engineering and Automat Theory, Tommaso Toffoli, demonstrated that finite arbitrary mapping could be calculated reversibly by padding strings with zeros, permuting them, and projecting some of the bit strings to other bit strings. In other words, what Toffoli means is that elemental reversible gates can be used to implement bit string permutations. Finally, one of the most powerful contributions to quantum computing based on \textit{Postulate 2} \eqref{postulate_2} came from the physical scientist Paul Benioff, who demonstrated that the unitary evolution of the quantum state (which is reversible) is at least as powerful as a Turing machine \cite{Bes12,Iva12,Pau18}.

The classical probabilistic process can be represented by a tree, which grows exponentially with possible results. The key difference in quantum computing is that we assign the quantum probability amplitudes to the branches of a tree, which can interfere with each other.

\section{Study of complexity}

\begin{figure}[h!]
    \centering
    \includegraphics[width=0.6\textwidth]{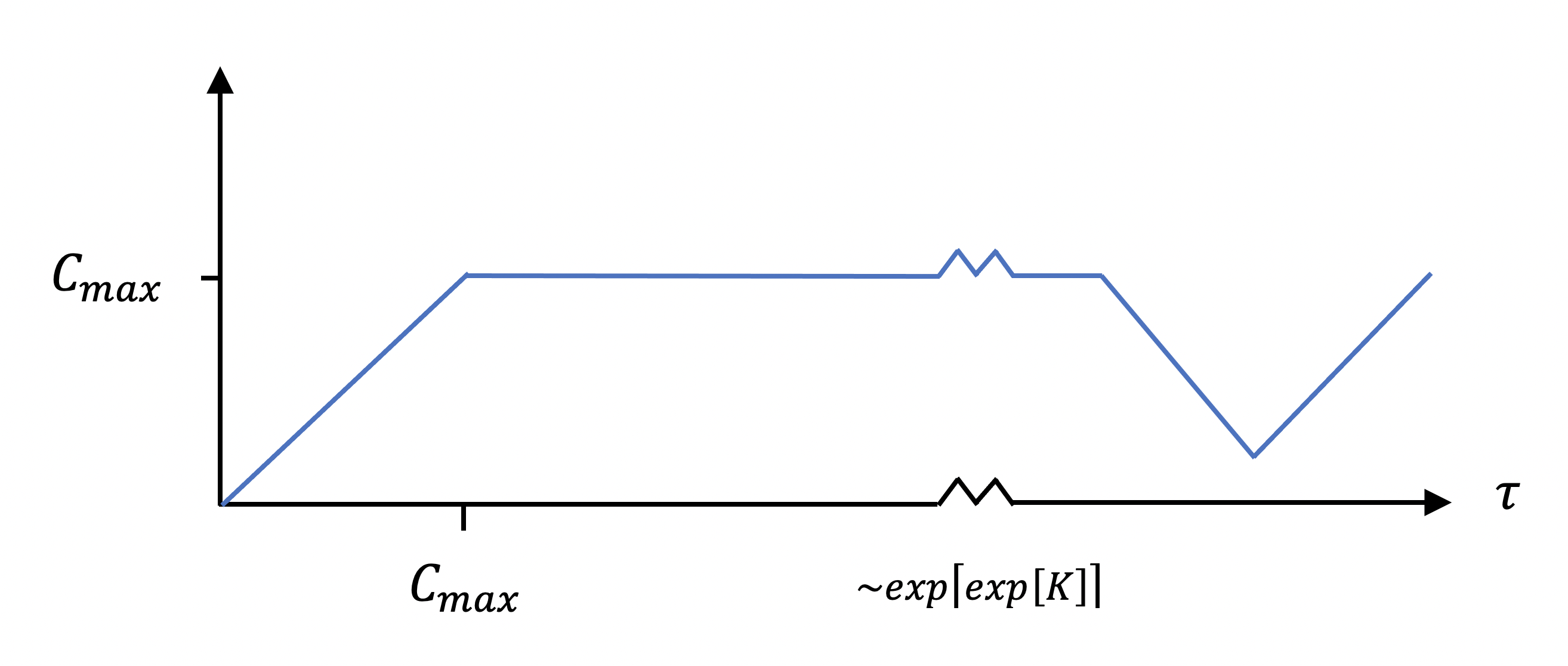}
    \caption{The conjectured evolution of the quantum complexity of the operator $e^{iHt}$ \cite{balasubramanian2020quantum}}
    \label{fig:study_Complexity}
\end{figure}

In our case, the concept of computational complexity that we use is essentially the same as the complexity of the quantum circuit, that is, the minimum number of quantum gates required to prepare a given unit operator \cite{Iva12}. The computational complexity of  $U(t)$ progresses over time. Both the black hole and the considerations of the quantum circuit suggest the following conjecture summarised in Figure \eqref{fig:study_Complexity} \cite{Ada18,Fel14}.

The complexity  $C(t)$ grows linearly as follow:

\begin{equation}
\label{Complexity_st_eq}
    C(t) =Kt.
\end{equation}

For an exponential time in  $K$ and $t^{K}$, the complexity reaches its maximum possible value $C_{\max }$ and flattens for a long time. This is the period of complexity. Equilibrium during which the complexity fluctuates over the maximum

\begin{equation}
\label{complexity_conjecture_Eq}
    C_{\text{max}}(t)\sim^{K}.
\end{equation}

On a much longer order time scale, quantum repeats $\exp [e^{K}]$ quasi-periodically return complexity to subexponential values. All this is a conjecture that, at the moment, cannot be proved, but that can be related to other complex assumptions.

The instance-averaging strategy sometimes allows conclusions about generic behaviour that would not be possible for specific cases. A particular example, which has generated recent interest, is the Sachdev-Ye-Kitaev (SYK) approach to coding. By averaging an appropriate set of time-independent Hamiltonians, it is possible to demonstrate that almost all Hamiltonians saturate the fast coding limit. Potentially, this averaging can also be applied to questions about the evolution of complexity \cite{Fel14}.

Another type of randomness is stochastic randomness, over which a statistically fluctuating (noisy) time-dependent Hamiltonian is averaged. The more averaging, the easier it is to conclude, and the stochastic average is easier than averaging over time independent Hamiltonians.
In our case, the unit matrix that describes our quantum system is written $U=e^{iH}$. 

\begin{equation}
\label{Hamiltonian_ising_eq}
    H= \sum _{i<j}^{}J_{ij}s_{i}s_{j}+ \sum _{i}^{}h_{i}s_{i}.
\end{equation}
With $H$, the Hamiltonian of our objective function. We express $H$  as a sum of Pauli's tensor products. This way is even very valid when it comes to finding the unit matrices of various quantum systems (section \eqref{sec:Postulate_4_S}).

There are many theories and strategies to determine the complexity of a quantum circuit. Going through approximation theory, based on entropy, or by the geometry of complexity. It can also be achieved based on the entropy of a classic system as we highlight in the following Lemma \cite{brown2018second}:

\textit{Quantum complexity for a $K$ qubit system behaves similarly to the entropy of a classical approach with  $2^{K}$ degrees of freedom}.

A practical criterion for calculating the complexity of a unit operator is the concept of the complexity of the quantum circuit. For example, in the case of running a quantum algorithm on IBMQ, the first thing the IBMQ framework does is to compile all the operations of a single qubit on universal gates $U_{1}$, $U_{2}$ and $U_{3}$.

Next, we will discuss the complexity of the circuit consists of all $K$  qubit circuits composed of k-local gates allow us to prepare $U$. For simplicity, we take the gates to act in series,

\begin{equation}
\label{complexity_U_eq}
    U= g_{N}g_{N-1} \ldots g_{1}.
\end{equation}

The complexity of the  $U$ circuit is denoted  $C(U)$. It is the minimum number of k-local gates needed to build  $U$. At last, it depends on the choice of allowable gates and some measures of the distance and concatenation on the gates. Hence, the concept of quantum relative complexity.

Relative complexity can also be defined for a pair of unit operators. This means that if we have a system composed of  $n = 2$  unit matrices $U_{n}$, the relative complexity would be the complexity $U_{1}U_{2}$ and, in a generic case, $U_{1}U_{2} \ldots U_{n}$. Inspired by Nielsen's ideas, we can build various auxiliary theories \cite{Mic00}. We consider an interesting approach is a theory based on a complexity metric. This theory is based on the idea of the complexity geometry that consists of defining a new parameter in $SU(2^{K})$ different from the standard metric, in which the distance between two elements of $SU(2^{K})$  reflects its complexity relative and where $SU$ is a space of the unit matrices of  $2^{K}$ dimensions.
To be precise, in our case, we are calculating the Kolmogorov complexity \cite{Dan00} of a time-independent Hamiltonian with a tolerance  $\delta$. The distance between two spins gives this tolerance $J_{ij}$ within the lattice model (see figures \eqref{fig:Lattice_eucludean} and \eqref{fig:Lattice_eucludean_plane}) \cite{Dor76,TDL52} when determining the Hamiltonian of Ising. 

In the Kolmogorov complexity calculation, time $t$  does not appear. In this case, the first part is a fixed overhead that does not scale with $~t$. The second part specifies the time required  $\log t~$ bits. And the third part establishes that the Hamiltonian also requires bits $\log t$, because to approximate $e^{-iHt}$ for a time $t$ requires precision in $H$ that is an inverse polynomial in $t$.

The simplest model is given when a single gate operates at every moment. So, the Kolmogorov complexity per gate would be ordering one insignificant. Although the choice also encompasses which group of  $k$ qubits performance between the gate at each stage. For example; in the case of  $k=2$, there are $ K(K- 1) /2$ possibilities to choose. That means, for each gate, we must add a Kolmogorov $\sim \log K^{2}$ complexity. We can quickly realise this by assigning a  $\log K^{2}$ of complexity to each gate. The total complexity of a unit operator in the stochastic model is  $\log K^{2}$  times the minimum number of gates required to prepare $U$. 

\section{Summary}
This chapter is significant because it touches on determining the complexity of a quantum algorithm. We have reviewed that the complexity of a unitary operator is from the concept of complexity of the quantum circuit, Nielsen's idea that marries the postulate 4 of quantum mechanics as the tremendous and pragmatic contribution Kolmogorov by the gate. Before all this, we have seen that the Turing machine (probabilistic) is useful for quantum computing. 

SWP is a problem that belongs to the BQP complexity class because it behaves like the Quantum Walk algorithm \cite{ambainis2007quantum}.

\newpage
\chapter{Quantum Gates}\label{sec:7}
\section{Introduction}

Quantum gates are the operators applied to a quantum system to modify its state (state vector).

As discussed above, Ref. \eqref{postulate_2} allows us to define quantum gates with these operators that act on quantum states $\vert  \psi ^{'} \rangle =U \vert  \psi \rangle$. 

\section{General U-gate}
\begin{figure}[h!]
    \centering
    \includegraphics[width=0.7\textwidth]{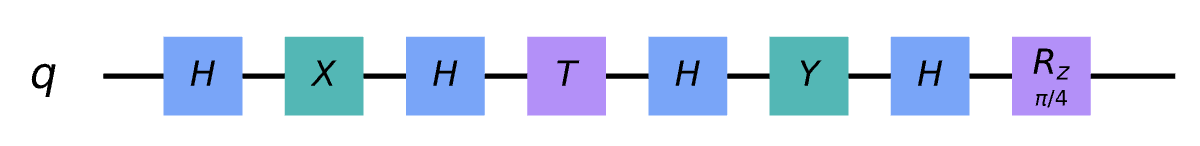}
    \caption{One-qubit circuit of quantum gate association to illustrate unit transformation matrix $U$}
    \label{fig:U-gate}
\end{figure}

One of the most important contributions to postulate two based quantum computing came from the physical scientist Paul Benioff \cite{Pet20} who has shown that the unitary evolution of the quantum state (which is reversible) is at least as powerful as a Turing\cite{Iva12} \cite{Mic00}. Both quantum and classical computing can be viewed as associations of universal quantum gates. For simplicity, we take the gates to act in series as is shown in figure \eqref{fig:U-gate},

\begin{equation}
\label{gates_serial_eq}
    U= g_{N}g_{N-1} \ldots g_{1}.
\end{equation}

Along these lines, there are two fundamental theorems related to the construction of quantum circuits.

\textit{\textbf{Theorem 1} \cite{Dmi05}\cite{Mic00}: All quantum circuits can be constructed using the only unit and NOT controlled gates}.

\textit{\textbf{Theorem 2} \cite{Dmi05}\cite{Mic00} : All quantum circuits can be constructed (in an approximate sense) using only Hadamard gates and Toffoli gates.}

In the same way, Emil Post announced that any circuit could be built with only three boolean AND, OR and NOT (fan-out) gates using the Toffoli gates. We won't go into detail in this demo, but the following reference validates it \cite{Mic00}.

Next, we will look at some new logic gates and then regain the notion of universality for quantum computing.

\textit{\textbf{Theorem 3} \cite{Dmi05}\cite{Mic00}: A map  $U:H_{n}  \rightarrow  H_{n}$  is unitary if, and only if, the matrix it represents is in some coordinate representation, $A$, satisfies $A^{\ast}A = AA^{\ast} = I_{n}$, where $\ast$  is the transposition of the complex conjugate vector. This property is independent of the chosen coordinate representation.}\\

Let $X$ be the operator associated with a square matrix, whose elements are $X_{mn}=\langle a^{(m)} \vert  X \vert a^{( n) } \rangle$, which can be written as follows:

\begin{equation}
\label{eq_Operator_sqM}
    \dot{X}= \left[ \begin{matrix}
\begin{matrix}
\begin{matrix}
\langle a^{ \left( 1 \right) } \vert  X \vert a^{ \left( 1 \right) } \rangle \\
\langle a^{ \left( 2 \right) } \vert  X \vert a^{ \left( 1 \right) } \rangle \\
\end{matrix}
  &  \begin{matrix}
\langle a^{ \left( 1 \right) } \vert  X \vert a^{ \left( 2 \right) } \rangle \\
\langle a^{ \left( 2 \right) } \vert  X \vert a^{ \left( 2 \right) } \rangle \\
\end{matrix}
\\
\end{matrix}
  &   \cdots   &  \begin{matrix}
\langle a^{ \left( 1 \right) } \vert  X \vert a^{ \left( n \right) } \rangle \\
\langle a^{ \left( 2 \right) } \vert  X \vert a^{ \left( n \right) } \rangle \\
\end{matrix}
\\
 \vdots   &  \ddots  &   \vdots \\
\begin{matrix}
\langle a^{ \left( m \right) } \vert  X \vert a^{ \left( 1 \right) } \rangle   & \langle a^{ \left( m \right) } \vert  X \vert a^{ \left( 2 \right) } \rangle \\
\end{matrix}
  &   \cdots   &  \langle a^{ \left( m \right) } \vert  X \vert a^{ \left( n \right) } \rangle \\
\end{matrix}
 \right]. 
\end{equation}

\section{Single qubit gates}

Let be a qubit with the matrix representation in the computational base   $\{ \vert 0 \rangle , \vert 1 \rangle \}$  we assign to these the representation of natural coordinates:

\begin{equation}
\label{single_qubit_eq}
    \vert 0 \rangle = \left[ \begin{matrix}
1\\
0\\
\end{matrix}
 \right]  ,  \vert 1 \rangle = \left[ \begin{matrix}
0\\
1\\
\end{matrix}
 \right]. 
\end{equation}
we can write based on the linear combination of the qubit by a matrix.
Let's define the fundamental quantum gates: the Hadamard  $H$  gate, the phase shift  $S$ gate, the  \( CNOT \)  gate, and the Pauli  \( \text{X, Y, Z} \)  operators.
Recall that the Hadamard H-gate, S-phase shift gate, and the NOT quantum gate have matrix representations in the computational base (CB)  $\{ \vert 0 \rangle , \vert 1 \rangle  \}$.

\section{Gate I}

\begin{equation}
\label{Identity_eq}
    I_{2\times 2}= \begin{bmatrix}
1  &  0\\
0  &  1\\
\end{bmatrix}
 = \vert 0 \rangle \langle 0 \vert + \vert 1 \rangle \langle 1 \vert. 
\end{equation}
\begin{figure}[b!]
    \centering
    \includegraphics[width=0.30\textwidth]{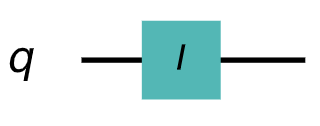}
    \caption{Circuit of the Quantum Identity gate $I$}
    \label{fig:I-gate}
\end{figure}

Gate $I$, figure \eqref{fig:I-gate}, is identity. Many kinds of literature consider it as the first operator of Pauli.

\section{Gate NOT}

\begin{equation}
\label{NOT_eq}
    X_{2\times 2}= \left[ \begin{matrix}
0  &  1\\
1  &  0\\
\end{matrix}
 \right] = \vert 0 \rangle \langle 1 \vert + \vert 1 \rangle \langle 0 \vert.
\end{equation}
\begin{figure}[h!]
    \centering
    \includegraphics[width=0.35\textwidth]{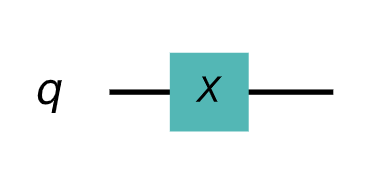}
    \caption{The Circuit of the quantum Pauli's first gate if we do not consider gate I}
    \label{fig:NOT-gate}
\end{figure}

We see that the NOT gate (figure \eqref{fig:NOT-gate}) maps  $\vert 0 \rangle$  to  $\vert 1 \rangle$  and $\vert 1 \rangle$ to  $\vert 0 \rangle$. We can also write the following operations.

\begin{equation}
\label{NOT_G_dem_eq}
    X \vert 0 \rangle =  \left[ \begin{matrix}
0  &  1\\
1  &  0\\
\end{matrix}
 \right]  \left[ \begin{matrix}
1\\
0\\
\end{matrix}
 \right] = \left[ \begin{matrix}
0\\
1\\
\end{matrix}
 \right] = \vert 1 \rangle.
\end{equation}

In the same way:
\begin{equation}
\label{Not_gate_Dem_1}
    X \vert 1 \rangle =  \left[ \begin{matrix}
0  &  1\\
1  &  0\\
\end{matrix}
 \right]  \left[ \begin{matrix}
0\\
1\\
\end{matrix}
 \right] = \left[ \begin{matrix}
1\\
0\\
\end{matrix}
 \right] =  \vert 0 \rangle. 
\end{equation}

The NOT gate is known as the $X$ gate.

\section{Gate Y}

\begin{equation}
\label{87}
    Y_{2\times 2}=  \left[ \begin{matrix}
0  &  -i\\
i  &  0\\
\end{matrix}
 \right] =  i \vert 0 \rangle \langle 1 \vert - i \vert 1 \rangle \langle 0 \vert. 
\end{equation}
\begin{figure}[h!]
    \centering
    \includegraphics[width=0.35\textwidth]{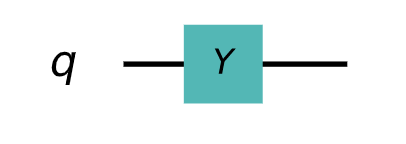}
    \caption{Circuit of the quantum gate Y. The second gate of Pauli}
    \label{fig:Y-gate}
\end{figure}
We see that the gate  $Y$ (figure \eqref{fig:Y-gate}) maps  $\vert 0 \rangle$ towards $\vert 1 \rangle$  and $\vert 1 \rangle$ towards $\vert 0 \rangle$  and at the same time, introduces a phase change. We can also write the following operations

\begin{equation}
\label{Y_gate_eq_Dem}
    Y \vert 0 \rangle = i \left[ \begin{matrix}
0  &  -1\\
1  &  0\\
\end{matrix}
 \right]  \left[ \begin{matrix}
1\\
0\\
\end{matrix}
 \right] = i \left[ \begin{matrix}
0\\
1\\
\end{matrix}
 \right] =i \vert 1 \rangle. 
\end{equation}

In the same way:

\begin{equation}\tag{89}
    Y  \vert 1 \rangle =  i \left[ \begin{matrix}
0  &  -1\\
1  &  0\\
\end{matrix}
 \right]  \left[ \begin{matrix}
0\\
1\\
\end{matrix}
 \right] = -i \left[ \begin{matrix}
1\\
0\\
\end{matrix}
 \right] =-i \vert 0 \rangle.
\end{equation}

\section{Gate Z}
\begin{equation}
\label{Gate_Z_eq}
    Z_{2\times 2}= \left[ \begin{matrix}
1  &  0\\
0  &  -1\\
\end{matrix}
 \right] = \vert 0 \rangle \langle 0 \vert -  \vert 1 \rangle \langle 1 \vert. 
\end{equation}
\begin{figure}[h!]
    \centering
    \includegraphics[width=0.30\textwidth]{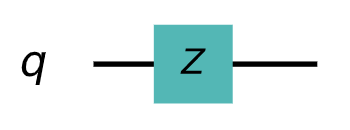}
    \caption{Circuit of the quantum gate Z. Third gate of Pauli}
    \label{fig:Z-gate}
\end{figure}

We see that the gate  $Z$ (figure \eqref{fig:Z-gate})  maps  $\vert 0 \rangle$ towards $\vert 0 \rangle$ and  $\vert 1 \rangle$ towards  $\vert -1 \rangle$  and at the same time, introduces a phase change. We can also write the following operations.

\begin{equation}
\label{Z_gate_dem}
    Z \vert 0 \rangle =  \left[ \begin{matrix}
1  &  0\\
0  &  -1\\
\end{matrix}
 \right]  \left[ \begin{matrix}
1\\
0\\
\end{matrix}
 \right] =  \left[ \begin{matrix}
1\\
0\\
\end{matrix}
 \right]  =  \vert 0 \rangle. 
\end{equation}

In the same way:

\begin{equation}
\label{92}
    Z \vert 1 \rangle =  \left[ \begin{matrix}
1  &  0\\
0  &  -1\\
\end{matrix}
 \right]  \left[ \begin{matrix}
0\\
1\\
\end{matrix}
 \right]  = - \left[ \begin{matrix}
0\\
1\\
\end{matrix}
 \right] =-  \vert 1 \rangle.
\end{equation}

If we look closely, we will realise that the operator  $Z$ applied to the quantum system does not alter the output. At most, add a 180-degree phase. This phase does not affect the system since it is an absolute phase. This is because states  $\vert 0 \rangle$  and  $\vert 1 \rangle$ are the two proper states of gate  $Z$. This is why the base $Z$ is used as the computational base (the base formed by states  $\vert 0 \rangle$  and $\vert 1 \rangle$. Many other bases are used as bases as follows.
\begin{equation}
    \label{ket_positive_eq}
    \vert + \rangle =\frac{1}{\sqrt[]{2}} (\vert 0 \rangle + \vert 1 \rangle) =\frac{1}{\sqrt[]{2}} \left[ \begin{matrix}
1\\
1\\
\end{matrix}
 \right]. 
\end{equation}
and 
\begin{equation}
    \label{ket_negative_eq}
   \vert - \rangle =\frac{1}{\sqrt[]{2}} (\vert 0 \rangle - \vert 1 \rangle)   = \frac{1}{\sqrt[]{2}} \left[ \begin{matrix}
1\\
-1\\
\end{matrix}
 \right].
\end{equation}

\section{The Pauli Operators}
The Pauli operators  ${X, Y \text{and}~ Z}$  (also known as  $\sigma _{x}, \sigma _{y}, \sigma _{z}$) correspond to the measurement of the turn along the $x^{-}$, $y^{-}$ and $z^{-}$  axes respectively. Its actions in the base states are given by where it is clear that the base states are elements of $Z$:
\begin{equation}
\label{pauli_feature_op_eq}
     X \vert 0 \rangle = \vert 1 \rangle, \quad X \vert 1 \rangle = \vert 0 \rangle, \quad Y \vert 0 \rangle = -i \vert 1 \rangle, \quad Y \vert 1 \rangle = i \vert 0 \rangle, \quad Z  \vert 0 \rangle = \vert 1 \rangle, \quad Z \vert 1 \rangle = - \vert 1 \rangle. 
\end{equation}
These operators fulfil the unit transformation property and the Hermitian property, that is, $A^{\ast}A = AA^{\ast} = I_{n}$.
\begin{equation}
\label{pauli_op_fufil_eq}
    XX^{\ast}=X^{2} = YY^{\ast}=Y^{2} = ZZ^{\ast}= Z^{2}=I.
\end{equation}
They also fulfil some interesting abelian group permutation properties \cite{Mic00}\cite{TBe05}\cite{Pie18}.
\begin{equation}
\label{pauli_permutation_eq}
    XY = iZYX = +iZ.
\end{equation}

\section{Hadamard gate}

It is one of the fundamental gates of quantum computing that is described as follows:
\begin{equation}
\label{Hadamard}
    H=\frac{1}{\sqrt[]{2}} \left[ \begin{matrix}
1  &  1\\
1  &  -1\\
\end{matrix}
 \right].
\end{equation}
 
\begin{equation}
\label{Hadamard_+}
    H \vert 0 \rangle = \vert + \rangle =\frac{1}{\sqrt[]{2}} (\vert 0 \rangle + \vert 1 \rangle). 
\end{equation}

\begin{equation}
\label{Hadamard_-}
    H \vert 1 \rangle = \vert - \rangle =\frac{1}{\sqrt[]{2}} (\vert 0 \rangle - \vert 1 \rangle).
\end{equation}
\begin{figure}[h!]
    \centering
    \includegraphics[width=0.30\textwidth]{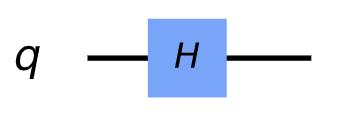}
    \caption{Hadamard Gate Circuit. A significant gate to get the superposition}
    \label{fig:H-gate}
\end{figure}
The Hadamard gate $H$ (figure \eqref{fig:H-gate}) is one of the universal gates of quantum computing. Therefore, it meets  $HH^{\ast}=H^{2} = I$.
What the Hadamard gate does when it is applied to $\vert 0 \rangle$, is convert a  $\vert 0 \rangle$ in $ \frac{1}{\sqrt[]{2}} (\vert 0 \rangle + \vert 1 \rangle)$, Halfway between $\vert 0 \rangle$, and $\vert 1 \rangle$  and converts $\vert 1 \rangle$ in  $\frac{1}{\sqrt[]{2}} (\vert 0 \rangle - \vert 1 \rangle)$ , which is also $``$halfway$"$  between $\vert 0 \rangle$, and  $\vert 1 \rangle$. This property is known as an overlay and will be very useful when performing the Bell EPR (entanglement). Geometrically, we can see it as a rotation on the Bolch sphere that responds to the transformation of the qubit state between the bases $X$  and $Z$.

\begin{figure}[h!]
    \centering
    \includegraphics[width=0.45\textwidth]{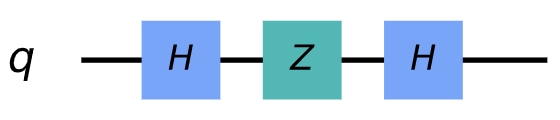}
    \caption{A circuit in Qiskit of the operation equivalent to $HZH=X$ }
    \label{fig:HZH-Circuit}
\end{figure}

This gate combined with the Pauli operators allows us to establish some exciting operations for quantum computing (See Fig. \eqref{fig:HZH-Circuit}).
\begin{equation}
\label{HZH_eq}
    HZH=\frac{1}{\sqrt[]{2}} \left[ \begin{matrix}
1  &  1\\
1  &  -1\\
\end{matrix}
 \right]  \left[ \begin{matrix}
1  &  0\\
0  &  -1\\
\end{matrix}
 \right] \frac{1}{\sqrt[]{2}} \left[ \begin{matrix}
1  &  1\\
1  &  -1\\
\end{matrix}
 \right] = \left[ \begin{matrix}
0  &  1\\
1  &  0\\
\end{matrix}
 \right] =X.
\end{equation}

\begin{equation}
\label{XHZH_eq}
    XHZH=\left[ \begin{matrix}
0  &  1\\
1  &  0\\
\end{matrix}
 \right]  \left[ \begin{matrix}
0  &  1\\
1  &  0\\
\end{matrix}
 \right] = \left[ \begin{matrix}
1  &  0\\
0  &  1\\
\end{matrix}
 \right] = I.
\end{equation}
where $HZH=X$.

%%%%%%%%%%%%%%%%%%%% Figure/Image No: 1 starts here %%%%%%%%%%%%%%%%%%%%

%%%%%%%%%%%%%%%%%%%% Figure/Image No: 1 Ends here %%%%%%%%%%%%%%%%%%%%

Generically, the Hadamard gate of $m$ components can be described by the following equation:

\begin{equation}
\label{hadamard_form_eq}
    H_{m}\vert x \rangle_{xy}=\frac{1}{2^{\frac{m}{2}}} \sum _{y \in F_{2}^{m}}^{} (-1)^{xy} \vert y \rangle. 
\end{equation}
If we observe a little, we can see that the Hadamard gate is a specific case of the Fourier quantum transform in the vector base  $F_{2}^{m}$ for $\vert x \rangle = \vert 0 \rangle$.

\begin{equation}
\label{hadamard_form_0_eq}
 H_{m} \vert 0 \rangle_{xy}=\frac{1}{2^{\frac{m}{2}}} \sum _{y \in F_{2}^{m}}^{} (-1) ^{xy} \vert y \rangle. 
\end{equation}
We realise that we have an identical superposition of all the states of the orthonormal base $F_{2}^{m}$. This is widely used to initialise quantum systems to have the same probability distribution.
\begin{equation}
\label{_Universality_eq_}
    U=e^{i\propto} \left[ \begin{matrix}
u_{0}  &  u_{1}\\
-\overline{u}_{1}  &  \overline{u}_{0}\\
\end{matrix}
 \right]. 
\end{equation}

With  $\propto \in R$  and with  $u_{i} \in C$  so it meets that,  $u_{0}\overline{u}_{0}+u_{1}\overline{u}_{1}=1$. This implies that there is only one $\theta  \in  \left[ 0, \pi  \right]$  such that  $\vert u_{0} \vert =cos\frac{ \theta }{2}$  and  $\vert u_{1} \vert =sin\frac{ \theta }{2}$. Therefore, we can write that  $u_{0}=e^{-i \lambda }\cos \frac{ \theta }{2}$  and $u_{1}=-e^{i \mu }\sin \frac{ \theta }{2}$. With $\lambda$ and $\mu$  $\in R$.
Rewriting our unit transformation matrix, we will have:
\begin{equation}
\label{U_gate_universal}
     U=e^{i\propto} \left[ \begin{matrix}
u_{0}  &  u_{1}\\
-\overline{u}_{1}  &  \overline{u}_{0}\\
\end{matrix}
 \right] = \left[ \begin{matrix}
e^{-i \lambda }cos (\frac{ \theta }{2})   &  -e^{i \mu }sin (\frac{ \theta }{2}) \\
-e^{i \mu }sin (\frac{ \theta }{2})   &  e^{-i \lambda }cos (\frac{ \theta }{2}) \\
\end{matrix}
 \right]. 
\end{equation}

If we note that   $\lambda = (\beta + \gamma) /2$  y  $\mu = (\beta - \gamma) /2 $,  $\beta , \gamma   \in R$, we can arrive at the generic expression of the unit matrix as an association of the rotation matrices.

\begin{equation}
\label{U_gate_eq_RxRyRz}
    \left[ \begin{matrix}
    e^{-i \lambda }cos⁡ ( \frac{ \theta }{2} )   &  -e^{i \mu }sin( \frac{\theta }{2}) \\
    -e^{i \mu } sin ⁡( \frac{ \theta }{2    })  & e^{-i \lambda }cos(\frac{\theta}{2} ) \\
\end{matrix}
 \right] =R_{z} \left(  \beta  \right) R_{y} \left(  \theta  \right) R_{z} \left(  \gamma  \right). 
\end{equation}
With

\begin{equation}
\label{Rz_gate_Eq}
    R_{z} \left(  \varphi  \right) = \left[ \begin{matrix}
e^{-i \varphi /2}  &  0\\
0  &  e^{i \varphi /2}\\
\end{matrix}
 \right] =\cos  \left( \frac{ \varphi }{2} \right) I_{2}-isin \left( \frac{ \varphi }{2} \right) Z=e^{-\frac{i \varphi }{2}Z}.
\end{equation}
\begin{equation}
\label{Ry_gate_eq}
     R_{y} \left(  \theta  \right) = \left[ \begin{matrix}
cos⁡ \left( \frac{ \theta }{2} \right) & -sin⁡ \left( \frac{ \theta }{2} \right) \\
sin⁡ \left( \frac{ \theta }{2} \right)   &  cos⁡ \left( \frac{ \theta }{2} \right) \\
\end{matrix}
 \right] =\cos  \left( \frac{ \theta }{2} \right) I_{2}-isin \left( \frac{ \theta }{2} \right) Y=e^{-\frac{i \theta }{2}Y}.
\end{equation}
\begin{equation}
\label{Rx_gate_eq}
    R_{x} \left(  \psi  \right) = \left[ \begin{matrix}
cos⁡ \left( \frac{ \psi }{2} \right)   & -isin⁡ \left( \frac{ \psi }{2} \right) \\
-isin⁡ \left( \frac{ \psi }{2} \right)   &  cos⁡ \left( \frac{ \psi }{2} \right) \\
\end{matrix}
 \right] =\cos  \left( \frac{ \psi }{2} \right) I_{2}-isin \left( \frac{ \psi }{2} \right) X =e^{-\frac{i \psi }{2}X}.
\end{equation}

The rotation operators are generated by exponentiation of the Pauli matrices according to:
 \begin{equation}
\label{Universal_eq_Mat}
    exp(iAx)=cos(x)I+isin(x)A,
\end{equation}
where $A$ is one of the three Pauli Matrices and only if  $A^2 = I$.

To\ finish\ writing that    \( U~ \left(  \theta ,  \beta ,~ \gamma  \right) =R_{z} \left(  \beta  \right) R_{y} \left(  \theta  \right) R_{z} \left(  \gamma  \right)  \) . We can also use the Euler decomposition\  of the unit transformation matrix  \( U \)  to arrive at the expression\cite{Mic00}:
\begin{equation}
\label{Universal_eq_Mat_}
    U =e^{i\propto}AXBXC.
\end{equation}

With $A$, $B$ and $C$ elements of the Clifford group\cite{Pie18}\cite{Mic00},\  $ABC= I_{2}$. We can now write $U=e^{i\propto}~R_{z}(\beta) R_{y}(\theta) R_{z}( \gamma)$ with:

\begin{equation}
\label{Cliffort_eq_}
    A=R_{z} \left(  \beta  \right) R_{y} \left( \frac{ \theta }{2} \right). 
\end{equation}
\begin{equation}
\label{Cliffort_eq_1}
    B=R_{y} \left( -\frac{ \theta }{2} \right) R_{z} \left( -\frac{ \beta + \gamma }{2} \right). 
\end{equation}

\begin{equation}
\label{Clifford_Eq_3}
    C=R_{z} (\frac{ \gamma - \beta }{2}).
\end{equation}

These operators fulfil the property of the unitary transformation and the Hermitian property, which means, $U^{\ast}U = UU^{\ast} = I_{n}$.

\section{Gate $R_{\phi}$}
\begin{figure}[h!]
    \centering
    \includegraphics[width=0.35\textwidth]{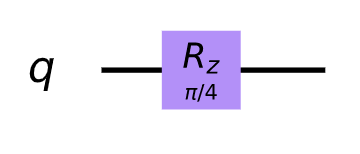}
    \caption{$R_{\phi}$ Gate of a qubit. This gate is significant for rotational operations}
    \label{fig:R_phi_qubit_}
\end{figure}

Let the universal gate $U_{3}$ be a simplification of the universal matrix described above, we define the rotate gate $R_{\phi}$ (Fig. \eqref{fig:R_phi_qubit_})  as a specific case of the universal gate $U$.
\begin{equation}
\label{def_U3_gate_eq}
    U_{3} (\theta , \phi , \lambda)  = \left[ \begin{matrix}
cos⁡ \left( \frac{ \theta }{2} \right)   &  -e^{i \lambda }sin⁡ \left( \frac{ \theta }{2} \right) \\
e^{i \phi }sin (\frac{ \theta }{2})   &  e^{i( \lambda +\phi) }cos (\frac{ \theta }{2}) \\
\end{matrix}
 \right]. 
\end{equation}

\begin{equation}
\label{R_Phi_qubit}
    R_{\phi }=U_{3}  \left(  \theta ,  \phi ,  \lambda  \right)  \vert _{ \theta = \phi =0}= U_{3} \left( 0, 0,  \lambda  \right) =U_{1}= \left[ \begin{matrix}
1  &  0\\
0  &  e^{i \lambda }\\
\end{matrix}
 \right]. 
\end{equation}

\section{Gate T}
\begin{figure}[h!]
    \centering
    \includegraphics[width=0.35\textwidth]{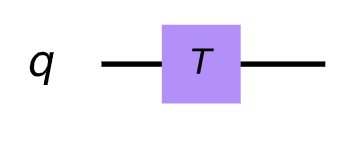}
    \caption{$T$ Gate of one qubit. Remember that the inverse of this gate is not itself }
    \label{fig:T_phi_qubit_}
\end{figure}

The gate $T$ (Fig. \eqref{fig:T_phi_qubit_}) it is equivalent to the gate $R_{\phi }$  with $\lambda = \pi /4$. So,
\begin{equation}
\label{T_gate_eq}
R_{ \phi  \vert  \lambda = \pi /4}=U_{1}= \left[ \begin{matrix}
1  &  0\\
0  &  e^{i \lambda }\\
\end{matrix}
 \right] = \left[ \begin{matrix}
1  &  0\\
0  &  e^{i\frac{ \pi }{4}}\\
\end{matrix}
 \right]. 
\end{equation}

\section{Gate S}
\begin{figure}[h!]
    \centering
    \includegraphics[width=0.35\textwidth]{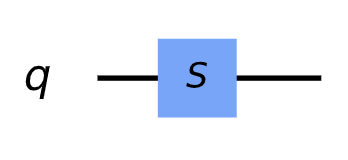}
    \caption{$S$ Gate of one qubit. Remember that the inverse of this gate is not itself }
    \label{fig:S_qubit_}
\end{figure}

The gate $S$ (Fig. \eqref{fig:S_qubit_}) is a particular cas of the $R_{ \phi }$ gate with $\phi = \pi /2$. So,
\begin{equation}
\label{gate_S_dem}
    R_{ \phi  \vert  \phi = \pi /2}= \left[ \begin{matrix}
1  &  0\\
0  &  e^{i\frac{ \pi }{2}}\\
\end{matrix}
 \right]. 
\end{equation}

We want to make two relevant comments about this gate to our path to quantum computing. First, gate $S$ makes a quarter turn around the Bloch sphere, and its inverse is not in itself like the other quantum gates described in this work.
We can see that gates $I$, $Z$, $S$ and $T$ were individual cases of gate $R_{\phi}$. Similarly, gate  $U_{3}$  is the most general of all single-qubit quantum gates. Each gate could be specified as $U_{3}(\theta ,  \phi ,  \lambda)$ and with this, we can conclude that the single-bit quantum universality resides at gate $U_{3} (  \theta ,  \phi ,  \lambda)$.
Qiskit provides gates  $U_{2}$  and $U_{1}$, which are specific cases of gate $ U_{3}$ where $\theta  = \frac{ \pi }{2}$ and $\theta = \phi =0$ respectively
\begin{equation}
\label{Universality_U3_gate_eq}
    U_{3}~ \left( \frac{ \pi }{2},  \phi ,  \lambda  \right) =U_{2}=\frac{1}{\sqrt[]{2}} \left[ \begin{matrix}
0  &  -e^{i \lambda }\\
e^{i \phi }  &  e^{i \lambda +i \phi }\\
\end{matrix}
 \right] U_{3} \left( 0, 0,  \lambda  \right) =U_{1}= \left[ \begin{matrix}
1  &  0\\
0  &  e^{i \lambda }\\
\end{matrix}
 \right]. 
\end{equation}
Before running on real IBMQ quantum hardware, all single-qubit operations are compiled on universal gates $U_{1}$,  $U_{2}$, and $U_{3}$. The universality of this quantum gate is seen from the fact that there is an infinite number of possible gates.

\section{More than one qubit system.}
\begin{figure}[h!]
    \centering
    \includegraphics[width=0.35\textwidth]{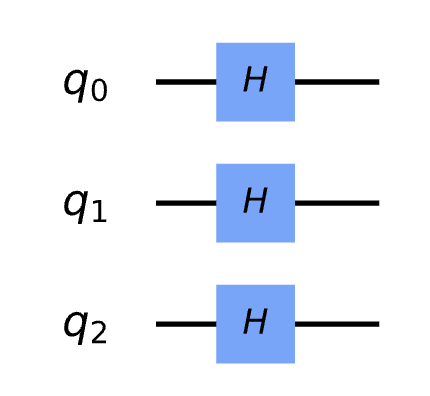}
    \caption{A 3-qubit system where we apply a Hadamard gate to each qubit. With this application, we assign each of the input qubits with the same probability}
    \label{fig:3_Hadarmard_}
\end{figure}

To express a qubit system, the tensor product is used. The tensor product \cite{Kir17} between more than two states is denoted as $\otimes$. If we write a three-qubit system (Fig. \eqref{fig:3_Hadarmard_}), it would be like the following case $\vert 011 \rangle = \vert 0 \rangle \otimes \vert 1 \rangle \otimes \vert 1 \rangle$ 
\begin{equation}
\label{composed_qcircuit}
 \vert 011 \rangle = \vert 0 \rangle \otimes \vert 1 \rangle \otimes \vert 1 \rangle = \left[ \begin{matrix}
1\\
0\\
\end{matrix}
 \right] \otimes \left[ \begin{matrix}
0\\
1\\
\end{matrix}
 \right] \otimes \left[ \begin{matrix}
0\\
1\\
\end{matrix}
 \right]. 
\end{equation}

\begin{equation}
\label{composed_qcircuit_1}
\vert 011 \rangle = \left[ \begin{matrix}
1 \left[ \begin{matrix}
0\\
1\\
\end{matrix}
 \right] \\
0 \left[ \begin{matrix}
0\\
1\\
\end{matrix}
 \right] \\
\end{matrix}
 \right] \otimes\left[ \begin{matrix}
0\\
1\\
\end{matrix}
 \right]  = \begin{bmatrix}
0\\
1\\
0\\
0\\
\end{bmatrix}
\otimes \left[ \begin{matrix}
0\\
1\\
\end{matrix}
 \right]  = \left[ \begin{matrix}
\begin{matrix}
0 \left[ \begin{matrix}
0\\
1\\
\end{matrix}
 \right] \\
1 \left[ \begin{matrix}
0\\
1\\
\end{matrix}
 \right] \\
0 \left[ \begin{matrix}
0\\
1\\
\end{matrix}
 \right] \\
\end{matrix}
\\
0 \left[\begin{matrix}
0\\
1\\
\end{matrix}
 \right]
\end{matrix}
 \right] = \begin{bmatrix}
0\\
0\\
0\\
1\\
0\\
0\\
0\\
0\\
\end{bmatrix}.
\end{equation}
For a two-qubit system, it would be  $\vert 01 \rangle = \vert 0 \rangle  \otimes \vert 1 \rangle =  \vert 0 \rangle   \vert 1 \rangle$. A binary quantum gate is a unitary operation in two qubits, that is, a unitary map  $F_{22} \rightarrow F_{2}\otimes F_{2}$  with the base $\{   \vert 00 \rangle, \vert 01 \rangle, \vert 10 \rangle , \vert 11 \rangle \}$ 
\begin{equation}
\label{2_qubits_eq}
    \vert 00 \rangle = \left[ \begin{matrix}
\begin{matrix}
1\\
0\\
0\\
\end{matrix}
\\
0\\
\end{matrix}
 \right], \quad \vert 01 \rangle = \left[ \begin{matrix}
\begin{matrix}
0\\
1\\
0\\
\end{matrix}
\\
0\\
\end{matrix}
 \right],\quad \vert 10 \rangle = \left[ \begin{matrix}
\begin{matrix}
0\\
0\\
1\\
\end{matrix}
\\
0\\
\end{matrix}
 \right], \quad \vert 11 \rangle = \left[ \begin{matrix}
\begin{matrix}
0\\
0\\
0\\
\end{matrix}
\\
1\\
\end{matrix}
 \right]. 
\end{equation}

The gate $C_{NOT}$ is the binary controlled gate defined by:
\begin{equation}
\label{CNOT_Eq}
    C_{NOT} = \left[ \begin{matrix}
1  &  0\\
0  &  0\\
\end{matrix}
 \right] \otimes I_{2}+ \left[ \begin{matrix}
0  &  0\\
0  &  1\\
\end{matrix}
 \right] \otimes X.
\end{equation}

\begin{equation}
\label{CNOT_demo_eq}
    C_{NOT}= \begin{bmatrix}
1  &  0\\
0  &  0\\
\end{bmatrix}
 \otimes \begin{bmatrix}
1  &  0\\
0  &  1\\
\end{bmatrix}
 + \begin{bmatrix}
0  &  0\\
0  &  1\\
\end{bmatrix}
\otimes X
\end {equation}

\begin{equation}
\label{CNOT_demo_eq_1}
 = \left[ \begin{matrix}
1 \left[ \begin{matrix}
1  &  0\\
0  &  1\\
\end{matrix}
 \right]   &  0 \left[ \begin{matrix}
1  &  0\\
0  &  1\\
\end{matrix}
 \right] \\
0 \left[ \begin{matrix}
1  &  0\\
0  &  1\\
\end{matrix}
 \right]   &  0 \left[ \begin{matrix}
1  &  0\\
0  &  1\\
\end{matrix}
 \right] \\
\end{matrix}
 \right] + \left[ \begin{matrix}
0 \left[ \begin{matrix}
0  &  1\\
1  &  0\\
\end{matrix}
 \right]   &  0 \left[ \begin{matrix}
0  &  1\\
1  &  0\\
\end{matrix}
 \right] \\
0 \left[ \begin{matrix}
0  &  1\\
1  &  0\\
\end{matrix}
 \right]   &  1 \left[ \begin{matrix}
0  &  1\\
1  &  0\\
\end{matrix}
 \right] \\
\end{matrix}
 \right]=  
              \begin{bmatrix} 1 & 0 & 0 & 0 \\
                              0 & 1 & 0 & 0 \\
                              0 & 0 & 0 & 0 \\
                              0 & 0 & 0 & 0 \\
              \end{bmatrix},
             +\begin{bmatrix} 0 & 0 & 0 & 0 \\
                              0 & 0 & 0 & 0 \\
                              0 & 0 & 0 & 1 \\
                              0 & 0 & 1 & 0 \\
              \end{bmatrix}.
\end{equation}
Where: 
\begin{equation}
\label{CNOT_F_eq_}
L_{C_{NOT}} = \begin{bmatrix} 1 & 0 & 0 & 0 \\
                              0 & 1 & 0 & 0 \\
                              0 & 0 & 0 & 1 \\
                              0 & 0 & 1 & 0 \\
              \end{bmatrix}.
\end{equation}

\begin{figure}[h!]
    \centering
    \includegraphics[width=0.35\textwidth]{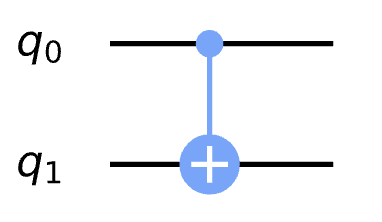}
    \caption{The two qubits $C_{NOT}$  gate}
    \label{fig:CNOT_Gate}
\end{figure}

What  $C_{NOT}$  gate (Fig. \eqref{fig:CNOT_Gate}) does is to map  $\vert 00 \rangle   \rightarrow \vert 00 \rangle,\vert 01 \rangle  \rightarrow \vert 01 \rangle,\vert 10 \rangle  \rightarrow  \vert 11 \rangle,\vert 11 \rangle  \rightarrow  \vert 10 \rangle$. Generically, $C_{NOT}$ is described as:

\begin{equation}
\label{CNOT_F_eq}
C_{NOT} = \begin{bmatrix} I_{2} & 0_{2} \\
                         0_{2} & X_{2} \\
                                \end{bmatrix}.
\end{equation}

The following list gives all the gates we need to create our algorithm $U_{1}$, $U_{2}$, $U_{3}$, $C_{NOT}$  and the identity gate $I$. Other types of quantum computers have different native gates, such as the two-qubit atomic gate \cite{SDe16}.

\section{Toffoli gate}
\begin{figure}[h!]
    \centering
    \includegraphics[width=0.7\textwidth]{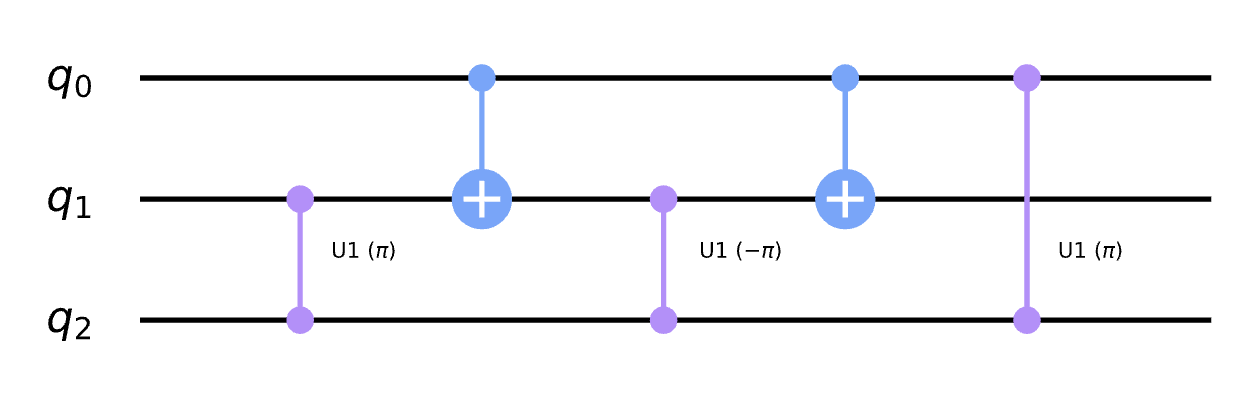}
    \caption{The Toffoli gate implemented with 2 $C_{NOT}$ gate, 2  $R_{ \phi  \vert  \phi = \pi }$  gates and another gate  $R_{\phi  \vert  \phi =- \pi}$.}
    \label{fig:Toffoli_Ga}
\end{figure}

\begin{figure}[h!]
    \centering
    \includegraphics[width=0.5\textwidth]{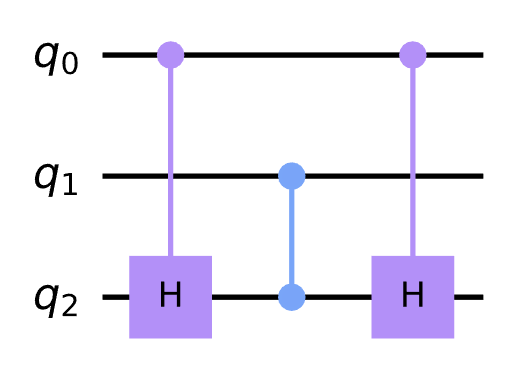}
    \caption{The Toffoli gate implemented with both the controlled-Hadamard and controlled-Z gates, which can be implemented with a single $C_{NOT}$.}
    \label{fig:Toffoli_Ga_bis}
\end{figure}

As we have discussed above, the Toffoli gate (Fig. \eqref{fig:Toffoli_Ga}) is a fundamental gate for the construction of quantum circuits. The operation performed by the Toffoli gate is described by the relationship $\vert q_{0} \rangle \vert q_{1} \rangle \vert q_{2} \rangle  \rightarrow  \vert q_{0} \rangle  \vert q_{1} \rangle   \vert q_{0}\oplus q_{12}  \rangle $. 
This gate can be built using the Hadamard gate, and phase-controlled rotation gate and CNOT gates. In the figure \eqref{fig:Toffoli_Ga}, we observe the implementation of this gate in Qiskit.
The Toffoli has no unique way of implementing an AND gate in quantum computing. Let us now propose another scheme for this fundamental gate given by Fig. \eqref{fig:Toffoli_Ga_bis}. Suppose we use both the controlled-Hadamard and controlled-Z gates, which can be implemented with a single $C_{NOT}$.

\section{Summary}
At this point, we are ready to create complex quantum circuits. We just have to remember that if we want to create a quantum system, we can combine two or more than two systems following \textit{postulate 2}. And, based on the works of Toffoli, Barenco and Emil Post, we can create any quantum circuit based on quantum gates fundamentals $C_{NOT}$, Hadamard $H$, the Paulis gates $X$, $Y$  and $Z$ and the $R_{ \phi }$ rotation gate). But before we get fully into quantum computing, we will review the state of the art of quantum computers in the next chapter.

\newpage

\chapter{Quantum Computers}\label{sec:8}
\section{Introduction}

From Feynman \cite{deM11} to today, quantum computers are computers based on quantum mechanics techniques to perform calculations. These computations are based on the probability of an object's state on a complete inner-product space known as the Hilbert space \cite{Kir17}. The states represent logical operations using an electron/photons spin. Spin-up is assigned to logic 1 and spin-down to logic 0. These discrete states allow for digital calculation \cite{Mic00,Mic06}. The quantum system with two states exploiting electron spin is a qubit. The calculations, or quantum mechanical representation, can process exponentially more data compared to classical computers. Because in quantum systems, the discrete states can exist in multiple states simultaneously.
\begin{figure}[h!]
    \centering
    \includegraphics[width=0.7\textwidth]{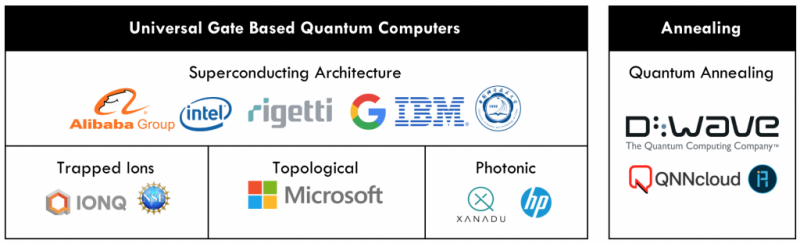}
    \includegraphics[width=0.7\textwidth]{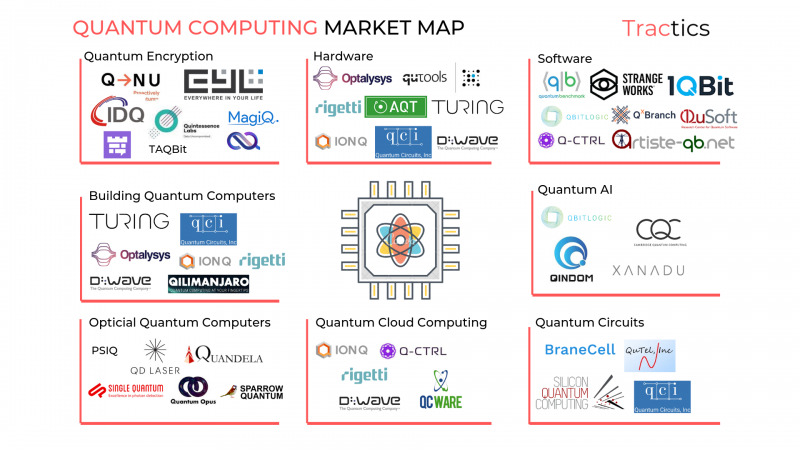}
    \caption{The quantum computing ecosystem \cite{cuomo2020towards}}
    \label{fig:Quantum_Computing_ecosystem}
\end{figure}
Quantum development kits – for coding in quantum assembly languages QCaaS subscription services – for use-case exploration, algorithm development, and simulation Partnerships with startups – to drive rapid innovation in QC initiatives Quantum hardware – to access quantum machines with enormous compute capacities.

One of the quantum mechanics principles used in quantum computing is superposition. The superposition propriety is when a qubit can be both $0$ and  $1$ simultaneously. As observed, the Hilbert space is the space of complex vectors. Therefore, the qubit's superposition can be comprehended like a linear combinatory of each vector of the basis.  Fundamentally, when a qubit is in a superposition of states, a logical operation applied to it will operate on both states simultaneously. Another principle of quantum computing that gives quantum computing one of the significant rewards is the entanglement \cite{RSh80}. This propriety is well-defined when the states of individual qubits are dependent on others.

What the companies (see Fig. \eqref{fig:Quantum_Computing_ecosystem}) do is mostly combine these principles (superposition and entanglement) with the foremost objective to create the core power of quantum computing \cite{Joh18}, hence quantum parallelism. With this combination, quantum computers perform computation on all possible inputs instantaneously. This enables quantum computers to explore and design algorithms that no classical computer will ever be able to create. This is a way to define the Quantum Supremacy\cite{McG14} \cite{Joh18}\cite{Aru19}.

\section{Simulated annealing}

The \textit{Simulated annealing} (SA) is described as \textit{a probabilistic technique for approximating the global optimum of a given function. Specifically, it is a metaheuristic to approximate global optimisation in a large search space for an optimisation problem. It is often used when the search space is discrete.} \\

\begin{figure}[h!]
    \centering
    \includegraphics[width=\textwidth]{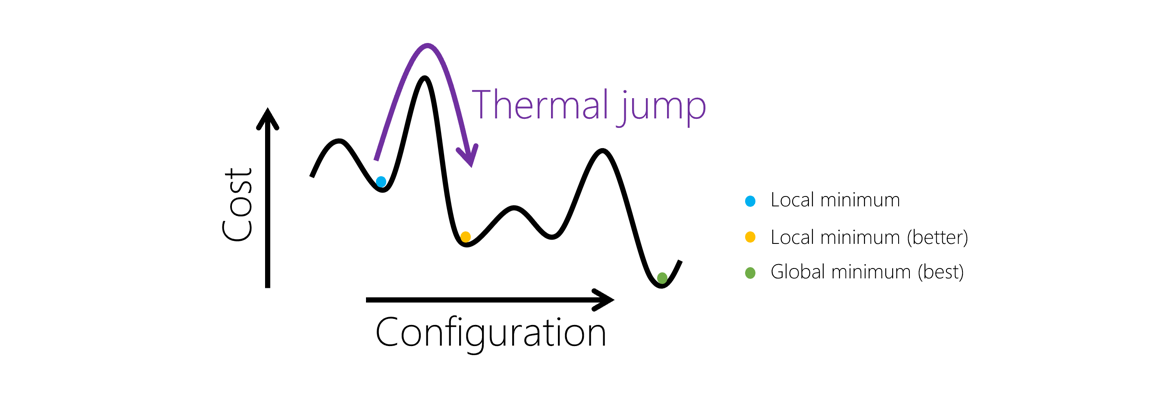}
    \caption{Simulated annealing concept.
The motion described by the particle but viewed uphill is described as a thermal jump. This is because Simulated annealing is a physics algorithm that mimics the behaviour of materials as they cool slowly. The "walker" is like an atom in a metal driven by temperature to reconfigure itself. These changes are random but are more likely to move to lower energy settings than higher energy configurations. That is why it is said that the "walker" follows a biased random movement.}
    \label{fig:Simulated_annealing}
\end{figure}

It is one of the most successful and used heuristic techniques. This method is advantageous and appropriate for optimisation problems in large (considerable) search spaces to solve the issues through an exhaustive search and with certain cost functions. Simulated annealing (Fig. \eqref{fig:Simulated_annealing}) can be seen as a random walk through the solution space, where each element (quantum particles) creates a path across the optimisation horizon. A heuristic is a technique to find an approximate solution. As we already know, they are useful in a scenario where the time needed to find an exact solution can be more than considerable. 
The motion described by the particle but viewed uphill is described as a thermal jump. This is because Simulated annealing is a physics algorithm that mimics the behaviour of materials as they cool slowly. The "walker" is like an atom in a metal driven by temperature to reconfigure itself. These changes are random but are more likely to move to lower energy settings than higher energy configurations. That is why it is said that the "walker" follows a biased random movement.

This technique is used to model, formulate, and solve combinatorial problems.

\section{Quantum annealing}

The \textit{Quantum annealing} (QA) is described as \textit{ a metaheuristic for finding the global minimum of a given objective function over a given set of candidate solutions that are stated by a process using quantum decoherence.} 

\begin{figure}[h!]
    \centering
    \includegraphics[width=\textwidth]{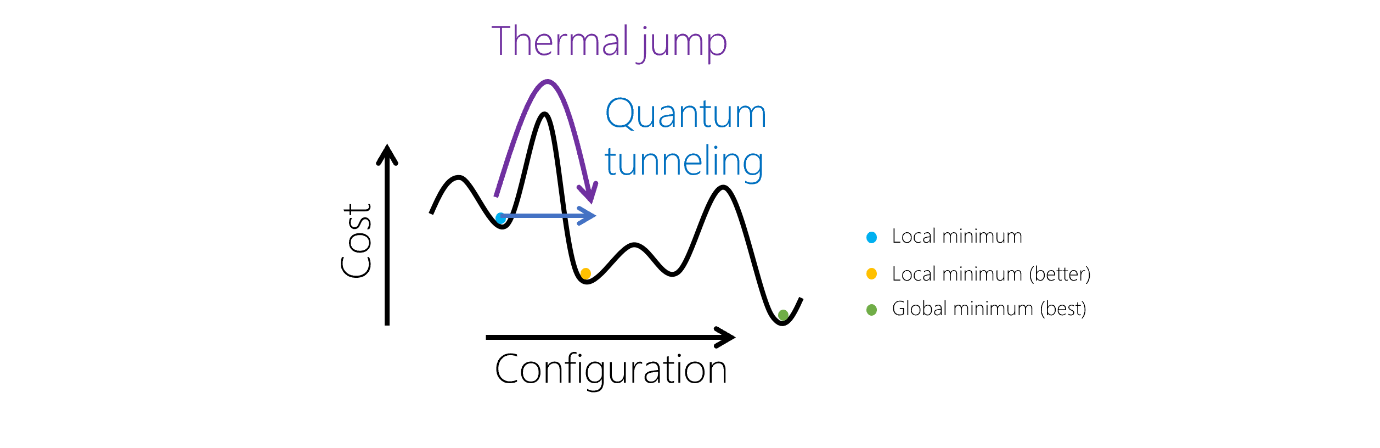}
    \caption{Advantage of quantum tunnelling over thermal jump taking advantage of the Simulated Annealing concept}
    \label{fig:Quantum_SA_Tun}
\end{figure}

Quantum annealing (Fig. \eqref{fig:Quantum_SA_Tun}) copies the simulated annealing philosophy by introducing a few small changes. In simulated annealing, the solution space is explored by doing thermal jumps from one solution to the next. But in quantum annealing, what you do is use the quantum effect called the quantum tunnel \cite{RSh80}, which allows us to travel through these energy barriers. It is the same effect that gives life to tunnel diodes.

\subsection{Quantum Tunneling}

The \textit{Quantum tunnelling} is described as \textit{ the quantum mechanical phenomenon where a wavefunction can propagate through a potential barrier.}\\

\begin{figure}[h!]
    \centering
    \includegraphics[width=0.6\textwidth]{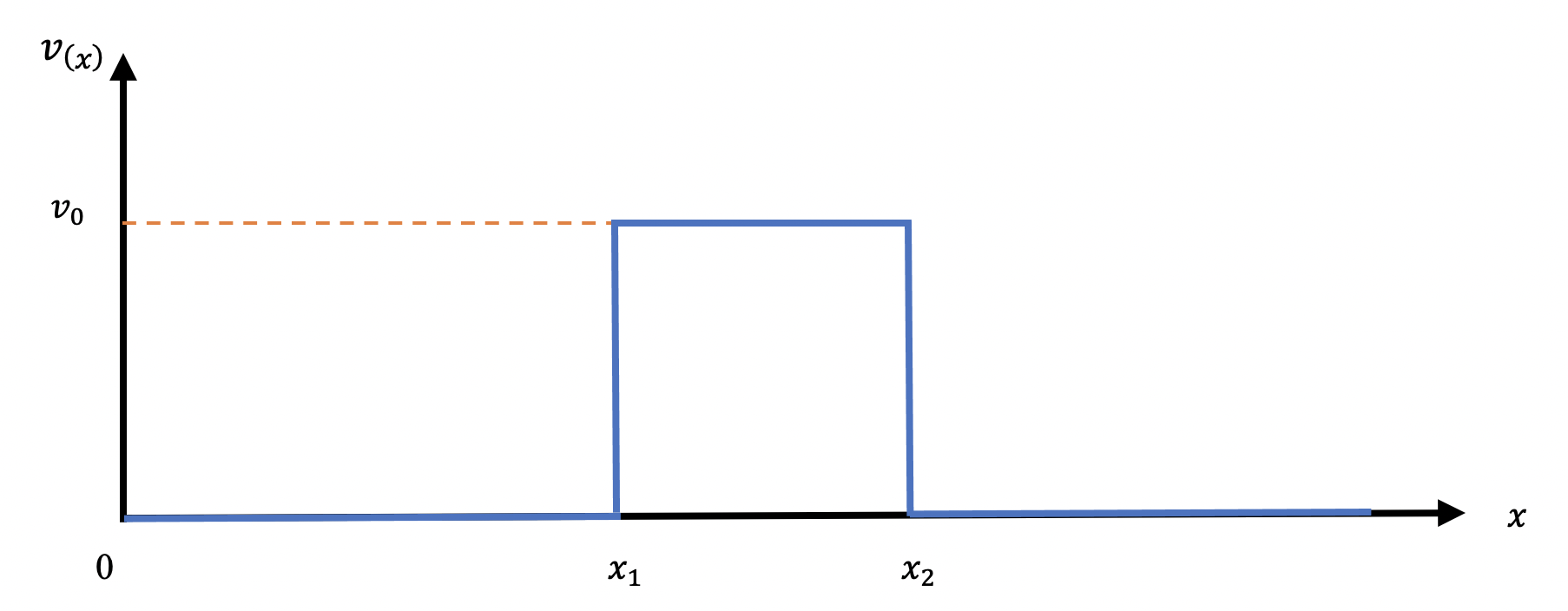}
    \caption{one-dimensional potential barrier}
    \label{fig:Quantum_Tunneling}
\end{figure}

To understand this characteristic of Quantum Tunneling, we can analyse the basic model to study the qualitative behaviour of a quantum system. In this model, we have the potential barrier seen in Figure \eqref{fig:Quantum_Tunneling} with height $V_{0}$ And width  $d = x_{2}- x_{1}$.
Let us consider that a particle of mass $m$ with energy  ${E <}V_{0}$  passes through this potential energy, and we want to know the probability of transition of the particle  $P \equiv \frac{ \vert  \psi _{b} \vert ^{2}}{ \vert  \psi _{a} \vert ^{2}}$  where  $\psi _{a}$  and  $\psi _{b}$  are the functions waveform on the left and right side of the barrier. Therefore, it is necessary to use the equation of motion of quantum systems, the Schrödinger equation, whose general form is:

\begin{equation}
\label{shrodinger_general}
    H \psi =i\hbar\frac{ \partial  \psi }{ \partial t}.
\end{equation}

In the case of the Figure \eqref{fig:Quantum_Tunneling}, the Hamiltonian is equal to 

\begin{equation}
\label{122}
    H=\frac{\hbar^{2}}{2m}\frac{d^{2} \psi }{dx^{2}}+ V_{0} \left[  \theta  \left( x_{1}- x \right) - \theta  \left( x_{2}- x \right)  \right]  \psi. 
\end{equation}

Now, applying the Schrödinger equation, we arrive at \eqref{shrodinger_general_1}:

\begin{equation}
\label{shrodinger_general_1}
    H=\frac{\hbar^{2}}{2m}\frac{d^{2} \psi }{dx^{2}}+ V_{0} \left[  \theta  \left( x_{1}- x \right) - \theta  \left( x_{2}- x \right)  \right]  \psi =i\hbar\frac{ \partial  \psi }{ \partial t}.
\end{equation}
The particle will behave like a flat wave on the left side of the barrier, remaining as  $\psi _{a}=Ae^{-ikx}+Be^{ikx}$ with $ k=\sqrt{\frac{2mE}{\hbar^{2}}}$. On the right side of the barrier, the particle is also a free particle  $\psi _{b}=Ce^{-ikx}+De^{ikx}$. Using normalisation and continuity conditions throughout the space, the probability of transition would be:

\begin{equation}
\label{Norm_Probability}
    P=\frac{1}{1+\frac{V_{0}^{2}\sin h^{2} \Delta }{4E \left( V_{0}-E \right) }}.
\end{equation}

With
 
\begin{equation}
\label{125}
    \Delta  \equiv d \sqrt {\frac{2m \left( V_{0}-E \right) }{\hbar^{2}}}.
\end{equation}

If we now apply the limit to  $P$, taking into account that $V_{0}$ constant and we increase $d$, the transition probability becomes smaller and, in the limit, $\mathop{\lim }_{d \rightarrow \infty}P=0$.
In Quantum annealing, the problem to be solved is modelled by mapping qubits, which are the variables. The energy of an allocation/mapping given to qubits is the value of the cost function that has a close relationship with the Hamiltonian.
How does it develop? Initially, the system begins with the quantum state in a full overlay over many possible qubit assignments. Then, the quantum field strength is varied instead of varying the temperature, as we did in simulated annealing.
The intensity of the quantum field is a parameter that defines the radius of the neighbouring states to which we can move. As time goes by and we get closer to a solution, this radius gets smaller and smaller. At the end of the annealing process, the system is established in a particularly low energy configuration (ground state) that can then be measured, giving us the optimisation problem solution. The cleanest mathematical formulation of quantum annealing is known for adiabatic quantum optimisation, and this is what almost all quantum optimisation methods attempt to emulate.

Adiabatic Quantum Computing (AQC) \cite{McG14} uses a concept of quantum physics known as the adiabatic theorem \cite{Wol15}. The process followed by the AQC can be summarised in two very recognisable steps:

\begin{enumerate}
	\item The first step is to prepare a system and initialise it to its lowest energy state, known as the ground state.

	\item The second step allows us to transform/map our problem in the system. 
\end{enumerate}

The adiabatic theorem states that as long as this transformation occurs slow enough, the system has time to adapt and will remain in the lowest energy configuration. When we are done with our transformations, we will find solutions.

To use the adiabatic theorem for computation, researchers H. Nishimori and T. Kadowaki demonstrated that it is necessary to use Ising's transversal model to code the problem to be optimised and activate it slowly. The Hamiltonian illustrating this is as follows,

\begin{equation}
\label{Ising_eq}
    H_{Ising}=-A \left( t \right)  \left(  \sum _{i<j}^{}J_{ij}Z_{i}Z_{j}+ \sum _{i}^{}h_{i}Z_{i} \right) +B \left( t \right)  \sum _{i}^{}h_{i}X_{i}.
\end{equation}

With $Z_{i}$ the Pauli matrix in $Z$. In many books, we can see instead of  $Z_{i}$, $\sigma _{z}$ and instead of  $X_{i}$, $\sigma _{x}$  or simply  $\sigma _{i}$ $\sigma _{x}$, $\sigma _{y}, \sigma _{z}$ generally representing the Pauli matrices or operators in $X$,  $Y$  and  $Z$. 

Taking into account that:

\begin{equation}
\label{Initial_condi_eq}
    A(t)  \vert _{t=0}=0 \quad \text{and} \quad B(0)\vert _{t=0}=1.
\end{equation}

With  $X_{i}$  the Pauli matrix in $X$  (generically $\sigma _{x}$).
As mentioned above, if we want the system to be in the ground state, the initial condition must be

\begin{equation}
\label{dem_system_eq}
    \vert  \psi(t)  \rangle  \vert _{t=0}= \vert +_{1} \rangle _{x}\otimes   \vert +_{2} \rangle _{x}\otimes \ldots \otimes \vert +_{n-1} \rangle _{x}\otimes \vert +_{n} \rangle _{x}.
\end{equation}

With  $ \vert +_{i} \rangle _{x}$ is the proper state of  $\sigma _{i}^{x}$. In some books or articles, we can see the following notification  $\vert  + \rangle ^{\otimes N}$  with $N$ is the number of qubits.

Now suppose that quantum annealing ends at ${T >> 1}$, where  $(t)  \vert _{t=T}=1$  and $B(t)  \vert _{t=T}=0$. The adiabatic theorem ensures that  $\vert  \psi(t) \rangle  \vert _{t=T}$  will be the ground state of $H(t)$ such that the energy of the system will be the energy destined for the ground state. This is the basis of adiabatic quantum computing. There are several derivatives of this technique, but it is not in line with this research.
\section{Quantum Computers}

The \textit{Quantum computers} can be defined as \textit{computers that use the quantum mechanic's properties to store data and perform computations that can be extremely worthwhile for certain tasks. As a result, they could vastly outperform even our best supercomputers and classical computer.} \\

There are several techniques for building a quantum computer, and the way it is currently done is by combining multiple various multicore processors \cite{KBe19}. All of this brings to life numerous models of quantum computing. Theoretical models, quantum circuit models, adiabatic quantum computing, measurement-based quantum computing, and topological quantum computing, are equivalent to each other within the reduction of polynomial-time. The most widespread and considerably developed model is the circuit model for gate-based quantum computation. As discussed in the chapter on quantum gates, the conceptual generalisation of Boolean logic gates (AND, OR, NOT, NAND, etc.) is used for classical computing works for quantum computing (Emil Post).  With the combination of these basic gates and the appropriate memory structures based on architecture, it gives life to the quantum computer.

Quantum annealing has a somewhat different software stack structure than gate-based model quantum computers. The annealing-based computer must be viewed as a specific case of a quantum accelerator based on quantum gate algorithms. So instead of a quantum circuit, as in the case of IBM and company, the level of abstraction is Ising's classic model. Let's take back everything we've already talked about solving problems with the Ising model. Like gate model superconducting quantum computers, superconducting quantum annealers also suffer from limited connectivity. This means we have to find a smaller embedded graph, combining several physical qubits into one logical qubit. Since this operation is already NP-Hard, it is necessary to use probabilistic heuristics.

\begin{figure}[h!]
    \centering
    \includegraphics[width=0.7\textwidth]{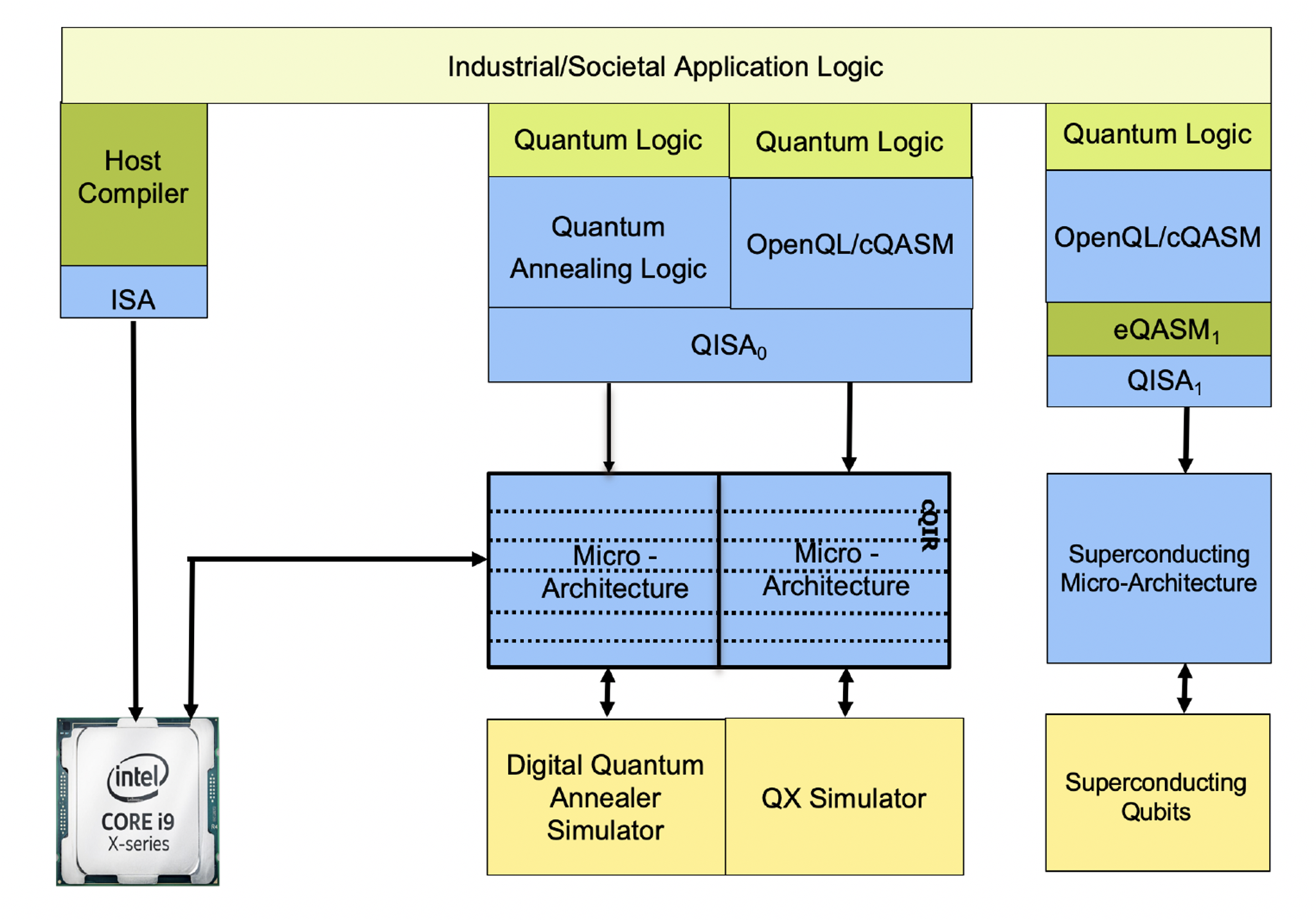}
    \caption{Full-Stack execution \cite{KBe19}}
    \label{fig:Full-Stack_execution}
\end{figure}

\begin{figure}[h!]
    \centering
    \includegraphics[width=0.7\textwidth]{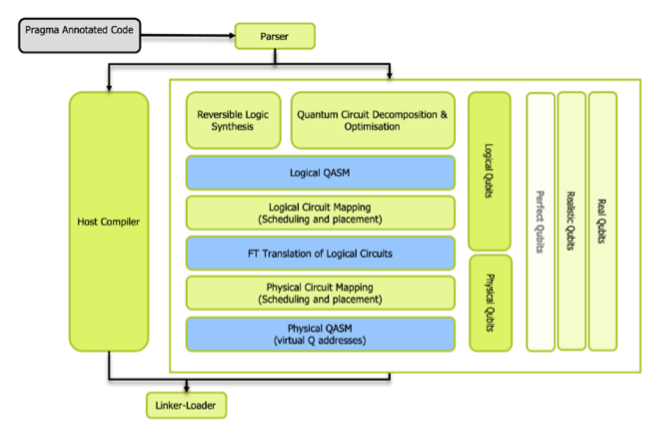}
    \caption{compiler infrastructure \cite{KBe19}}
    \label{fig:3_Hadarmard__}
\end{figure}

\begin{figure}[h!]
    \centering
    \includegraphics[width=\textwidth]{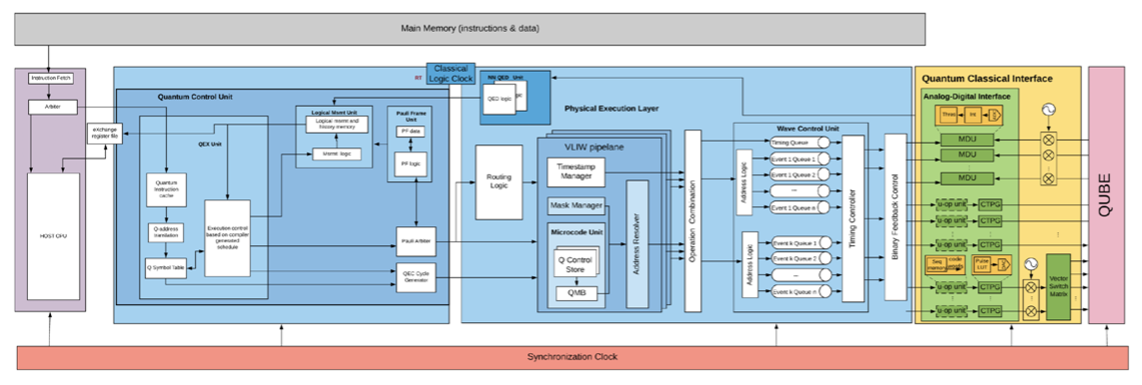}
    \caption{An example of general-purpose quantum microarchitecture \cite{KBe19}}
    \label{fig:general-purpose_QC}
\end{figure}

\begin{figure}[h!]
    \centering
    \includegraphics[width=\textwidth]{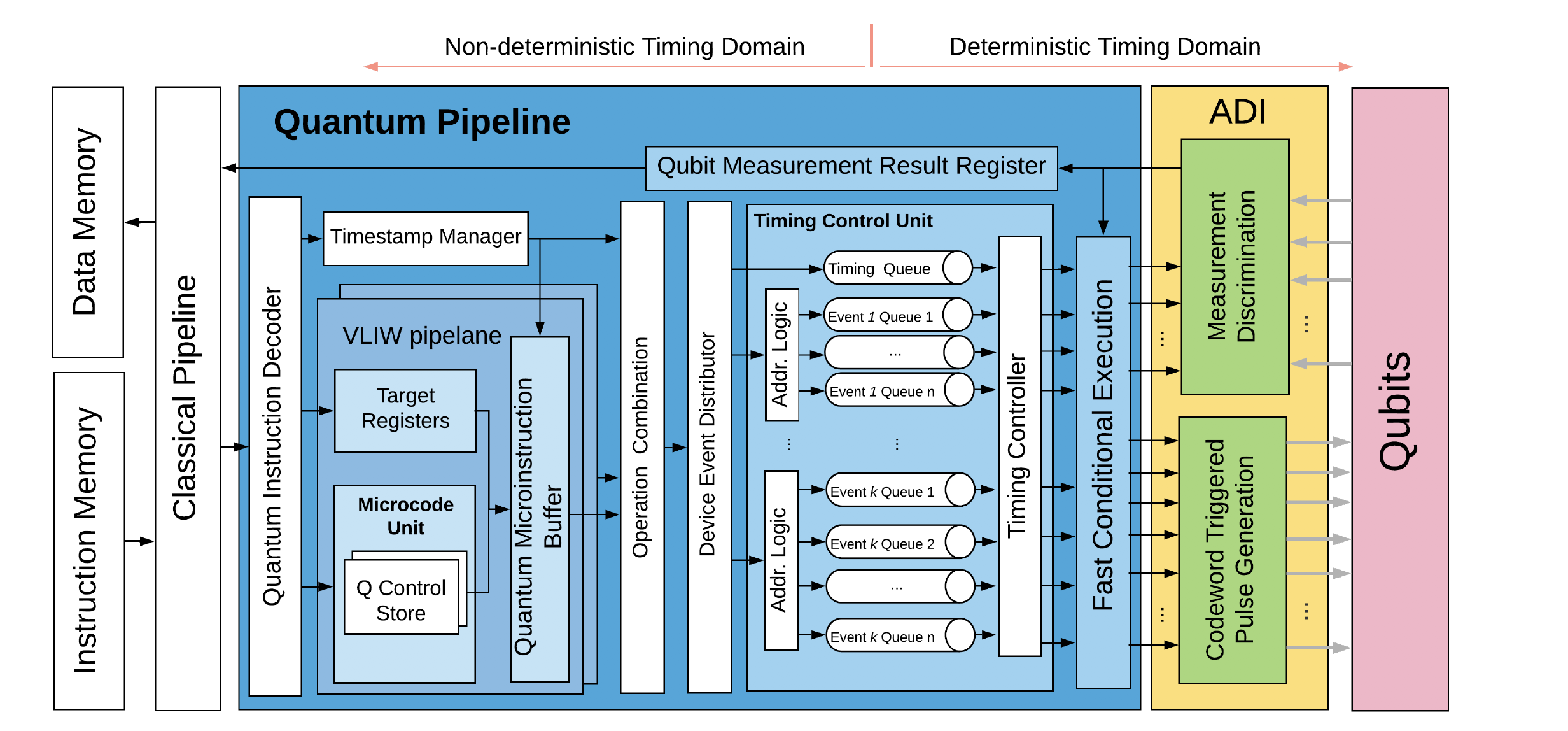}
    \caption{Example of experimental implementation of microarchitecture for (real) superconducting qubits \cite{KBe19}}
    \label{fig:EX_Implementation_QC}
\end{figure}

\begin{figure}[h!]
    \centering
    \includegraphics[width=\textwidth]{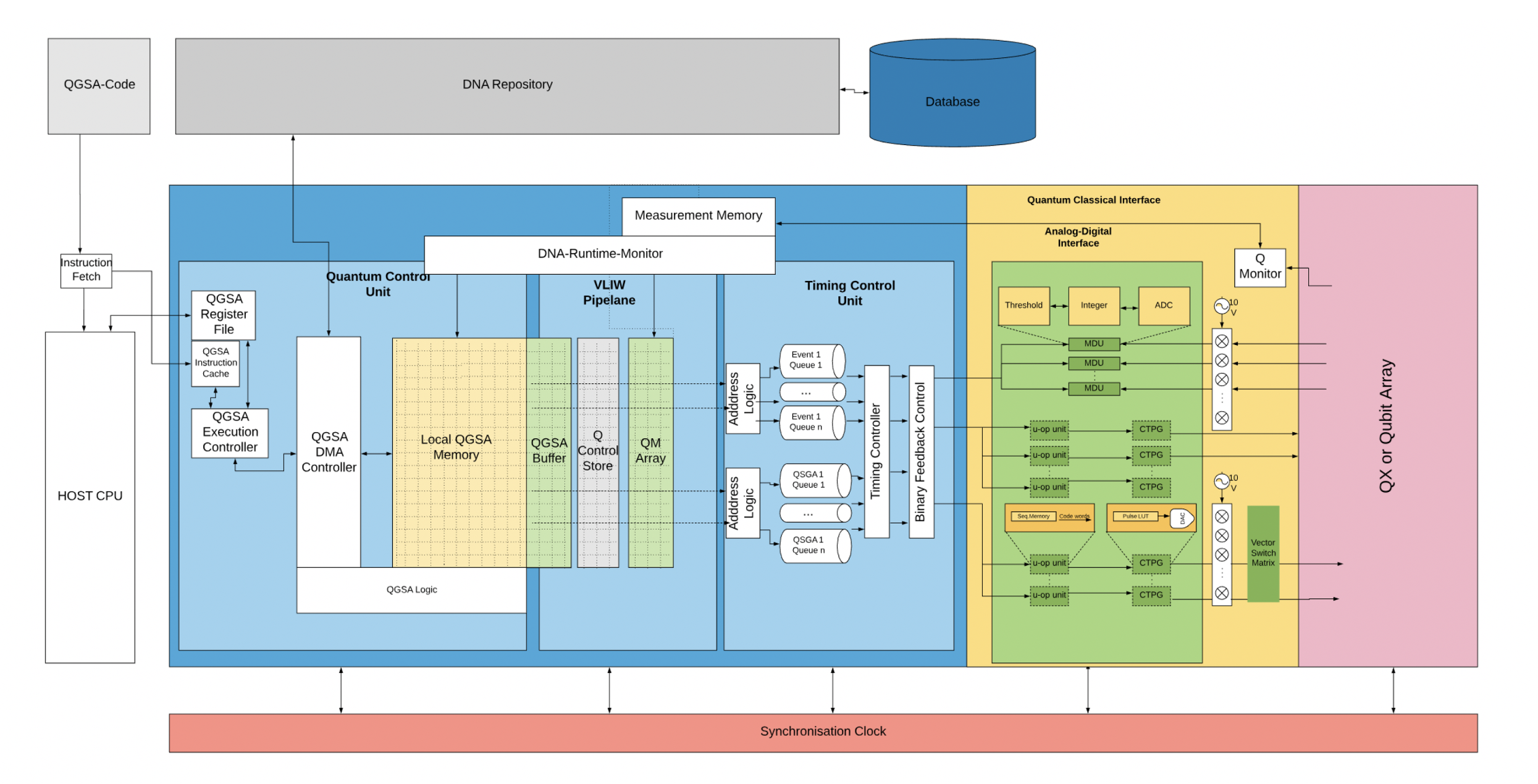}
    \caption{Example of the new microarchitecture for the quantum genome sequencing accelerator \cite{KBe19}}
    \label{fig:EX_New_Implementation_QC}
\end{figure}

Below, from Figure \eqref{fig:Full-Stack_execution} to \eqref{fig:EX_New_Implementation_QC}, we are presented with examples of Full-stack quantum computer architectures. To see the architecture being followed for the construction of the quantum independently of the Quantum Processing Unit (QPU). The QPU, also known as a quantum chip, is a physical chip (manufactured by players based on the techniques discussed above) that contains several interconnected qubits. It is the fundamental component of a full quantum computer, which includes the "motherboard\footnote{ In this case, to be a purist, it does not make sense to speak of the motherboard as in classic computers, but of housing thinking of a datacenter. Since today's quantum computers live in exceptional houses, another approach would be to think of a future with possible quantum laptops. }" environment for the QPU, the control electronics, and many other components.
Currently, there are two dominant configurations for quantum computing. Continuous-Time Quantum Calculus (D-Wave), in which problems are coded in quantum Hamiltonians and the natural dynamics of physical systems, and the Gate Model Quantum Calculus (IBM) \cite{McG14,Mic00,Kir17}, in which the calculation is made through a series of discrete gate operations.

In continuous-time quantum computation, optimisation is achieved by mapping the Hamiltonian optimisation problem of a controllable quantum system so that the low-energy states correspond to optimal solutions. The quantum superconducting circuit analysers produced by D-Wave Systems Inc are the most mature \cite{McG14,Joh18}.

A D-Wave machine is a physical representation of the Ising model and, as such, has a Hamiltonian "problem" of the form of the equation \eqref{Hamiltonian_ising_eq}.

Returning to our study, the main problem to solve is assigned to the previous Hamiltonian. The system begins with the Hamiltonian:  $H_{initial}= \sum _{i}^{}h_{i}Z_{i}$  and the annealing parameter, $s$  is used to assign the initial Hamiltonian  $H_{initial}$  to the Hamiltonian problem  $H_{ising}$  using  $H(s) =(1-s) H_{initial}+sH_{ising}$.

The process is done slow enough based on the annealing theorem to stay close to the system's ground state. At the same time, the Hamiltonian varies according to the problem, using tunnels to remain close to the ground state.

For quantum gate-based computers, one of the most promising algorithms for optimisation is the one known by Ansatz Alternative Quantum Operator, also known as the Approximate Quantum Optimisation Algorithm (QAOA) \cite{GGG19,Joh18,Alb13,GPa20,Pan19}. QAOA is exclusively designed to run in polynomial time on NISQ devices \cite{Joh18} and find optimal solutions for optimisation problems. This algorithm is used to solve critical optimisation problems that classically require exponential computational complexity to find the optimal solution exactly. Although, in principle, QAOA could be considered a gate model or a continuous-time configuration \cite{GGG19,Joh18}.

Based on the work of Andrew Lucas, Ising formulations of many NP problems \cite{And14}, any problem in NP can be assigned to an NP-hard in polynomial time, and integer factorisation is, in fact, an NP problem. Ising's Hamiltonian is a quadratic function that corresponds to a binary optimisation problem without quadratic constraints (QUBO), being in NP-hard. A universal Quantum Computing can solve the Hamiltonian of the same problem in polynomial time. The thing that a Quantum Annealing cannot do, since it cannot simulate a universal Quantum Computing in polynomial time \cite{Ale17}.

Gate-based quantum algorithms are designed, so that solution states interfere constructively. In contrast, non-solutions interfere destructively by skewing the final probability distribution to measure solutions. However, the error rates are still around  $10^{-2}$  and $10^{-3}$ and need to be substantially improved.

Adiabatic quantum computing (AQC) was the first quantum computation model to solve combinatorial optimisation problems. Unlike the gate, based on the quantum computation model, it was based on the adiabatic theorem of quantum mechanics. In this model, to perform any calculation, we need two Hamiltonians called  $H_{mixer}$  and $H_{cost}$. Among them, the fundamental state of $H_{mixer}$ must be easily prepared, like  $\vert + \rangle ^{\otimes N}$  and the ground state of $H_{cost}$ encode the solution to our problem.
\begin{figure}[h!]
    \centering
    \includegraphics[width=0.5\textwidth]{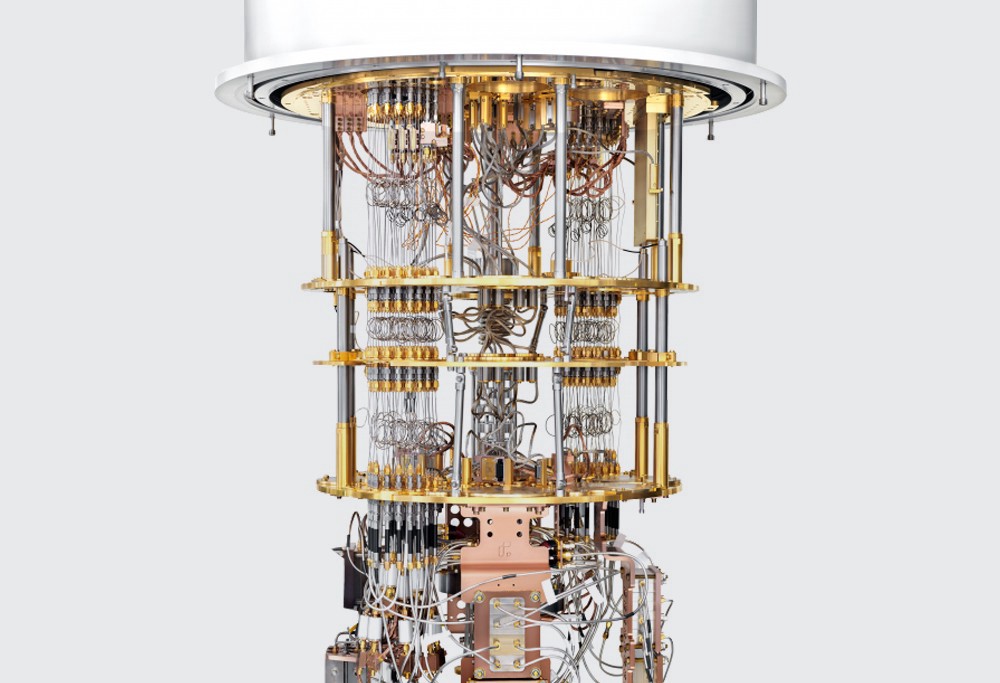}
    \caption{Microsoft Quantum Computer \cite{gibney2016inside}.}
    \label{fig:MS_QC}
\end{figure}
\begin{figure}[h!]
    \centering
    \includegraphics[width=0.5\textwidth]{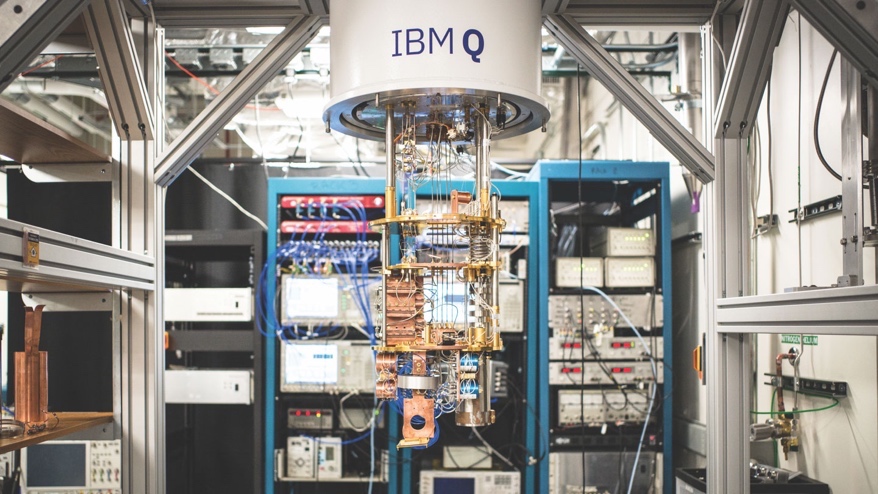}
    \caption{IBM Quantum Computer \cite{alvarez2018quantum}.}
    \label{fig:IBM_QC}
\end{figure}
\begin{figure}[h!]
    \centering
    \includegraphics[width=0.5\textwidth]{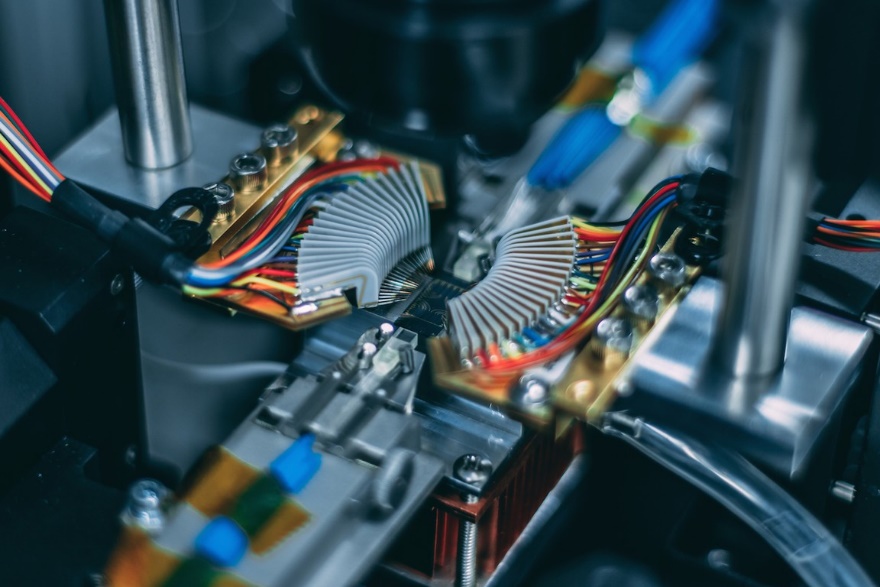}
    \caption{Xanadu Quantum Computer \cite{bourassa2021blueprint}.}
    \label{fig:Xanadu_QC}
\end{figure}
\begin{figure}[h!]
    \centering
    \includegraphics[width=0.5\textwidth]{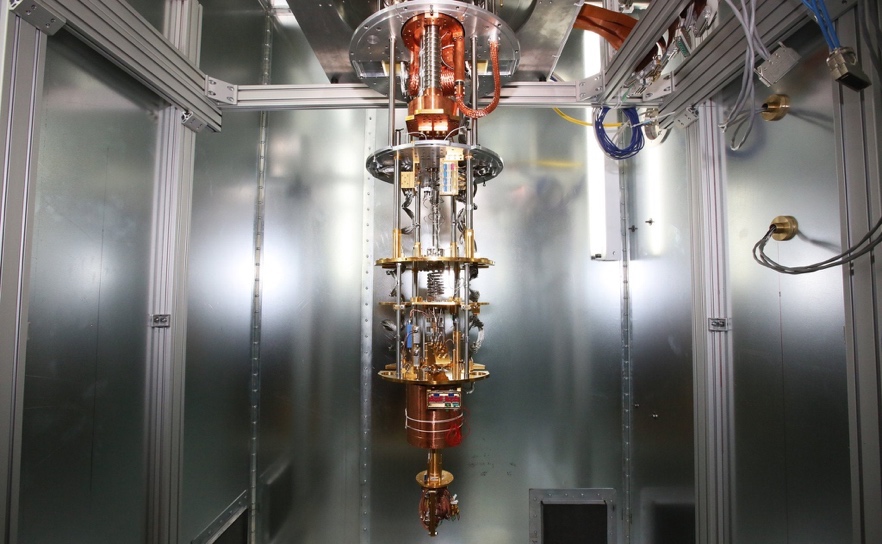}
    \caption{Inside D-Wave Quantum Computer \cite{gibney2017d}}
    \label{fig:D_Wave_QC}
\end{figure}
\subsection{Quantum gate-based computers}
In this section, we will present the leading companies that follow the career of the construction of the quantum computer based on gates.

IBM (Fig. \eqref{fig:IBM_QC}) is one of the most advanced technology companies in quantum computing with cloud computing; IBM Quantum Experience (IBMQ)\footnote{ https://quantum-computing.ibm.com/ }. To date, there are more than 250,000 users who have tried the IBMQ service \footnote{ Private data from slack and IBMQ profiles}. In addition, more than 100 companies are paying for its service, using IBM hardware, getting advice from the company's experts, and counting on all kinds of services.

Currently, IBM has 15 New York-based quantum cloud computers. The objective is to ensure that each country can have one, but while waiting for agreements with the different governments, IBM will begin its distribution in Germany and Japan.

Recently, IBM offered a quantum processor of up to 53 qubits. On the other hand:

Intel is analysing semiconductor and superconductor qubits, but fundamentally they are more interested in the qubit semiconductor processor. The essence is to focus on qubit production, partly backed by a solid microarchitecture.
Microsoft (Fig. \eqref{fig:MS_QC}) has some preference for the Majorana-based approach \cite{Chr18}, but they have yet to do the first qubit based on that quasi-particle. Nevertheless, they are very active in software development.
Alibaba is betting very heavily in the field of quantum computing. It has a quantum lab that focuses on a range of activities ranging from developing a quantum processor and classic quantum algorithms to simulating quantum physics \cite{Fan19}.
Google is also one of the leaders in superconducting qubits. The team led by John Martinis has developed the quantum computer Sycamore\footnote{ https://www.nytimes.com/2019/10/30/opinion/google-quantum-computer-sycamore.html } of 54 qubits and a specific algorithm to show quantum supremacy (the algorithm generates sequences of random numbers and has no relevant practical applications) \cite{Kal19}.
Rigetti is a startup company focused on the superconducting quantum processor. They are progressing well, but there is still no applicable processor on the market, even though a processor can be used for some testing purposes. Rigetti's Aspen-7-25Q-B Quantum Processing Unit (QPU) has 25 qubits with 24 programmable two-qubit gate interactions. The company released its last 31 qubits Aspen -8 QPU on May 5, 2020 \cite{Dan98,Dan981}.
Xanadu (Fig. \eqref{fig:Xanadu_QC}) focuses on continuous-variable quantum computing based on squeezed light photonics \cite{Vil20}. A slightly different model compared to qubits. The essential elements of its photonic system are qumodes, each of which can be represented by superimposing on different numbers of photons \cite{Set98,Sor20}.
Xanadu designs and incorporates photonic quantum silicon chips into existing hardware to create a complete quantum computing experience.

Following the race to get a quantum computer, Intel researchers (qHiPSTER\footnote{ Intel Quantum Simulator }), MIT and the University of Toronto are addressing the question of how the performance of next-generation quantum devices submodules can be designed and tested using existing quantum computers. Furthermore, the methods drawn with the researchers open a new path to the design of quantum processors, candidates when the demands to calculate the properties of the submodule exceed the capabilities of classical computing resources. All these steps lead us to the well-known Full-stack \cite{Mik16}.
\begin{figure}[h!]
    \centering
    \includegraphics[width=.7\textwidth]{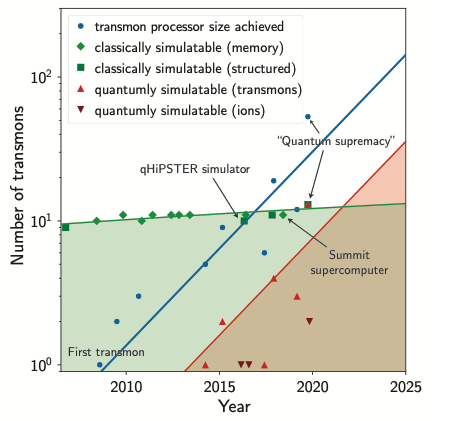}
    \caption{The challenge of simulating transmon quantum processors \cite{Thi}}
    \label{fig:ChallengeTransmons}
\end{figure}
With the increasing size of quantum processors, the submodules that make up the processor will be too large to simulate a classic computer accurately. Therefore, one will soon have to fabricate and test each new design primitive and parameter choice in time-consuming coordination between design, fabrication, and experimental validation.

\subsection{Quantum annealing-based computers}

This section will present the leading companies that follow the career of the construction of the quantum computer based on annealing.

D-Wave (Fig. \eqref{fig:D_Wave_QC}) is the pioneer company in the construction of quantum computers during this last decade. In 2018, its technology reached up to 2000 superconducting qubits. D-Wave Systems' commitment is in quantum hardware. However, it also offers a programming environment Ocean\footnote{ https://ocean.dwavesys.com/}, for basic operations compared with Qiskit or PennyLane. 
Fujitsu has invested in the development of digital annealing. Therefore, its bet is not entirely a quantum computer, but it is used to solve QUBO problems through simulation \cite{Ali00}.
Hitachi also specialises in making a quantum annealing-based quantum accelerator using semiconductor qubits. Its bet\footnote{https://www.hitachi.com/rd/sc/story/cmos\_annealing2/index.html} is similar to Fujitsu. 
\begin{figure}[h!]
    \centering
    \includegraphics[width=.5\textwidth]{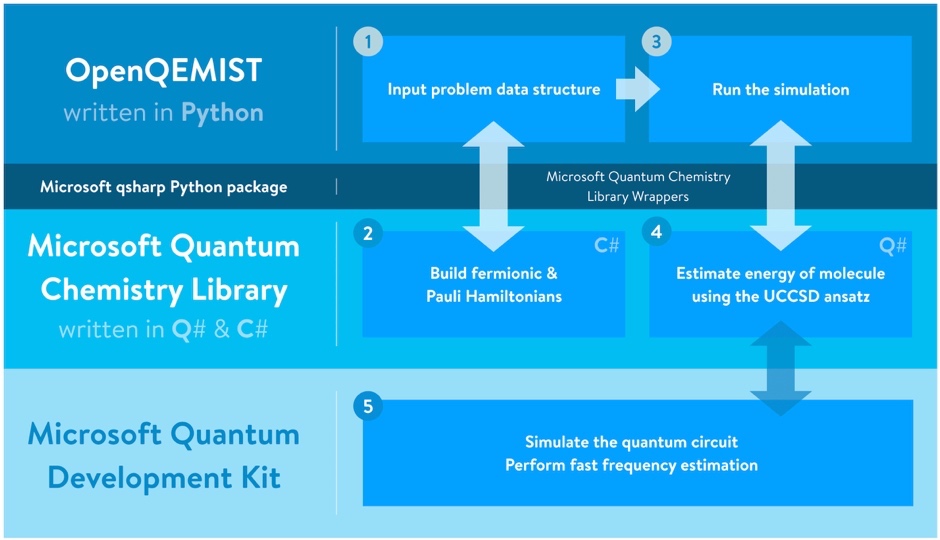}
    \caption{Microsoft Quantum Development Kit. Provides an end-to-end scalable quantum development environment and leverages the Q\# quantum programming language, which enables users to design, compile, and simulate quantum algorithms \cite{Azurre_MS}. }
    \label{fig:MS_SDK_QC}
\end{figure}
Seen from the outside, 1QBit is a software company that identifies insoluble industrial problems and creates the necessary software to take advantage of the best classical and quantum hardware technologies to solve them. 1QBit is instead focused on computational finance, materials science, quantum chemistry, and life sciences. Its algorithms serve both gate quantum computing and quantum annealing. Its framework, 1Qloud, focuses on optimisation issues that map to QUBO for quantum annealing computers and similar devices from D-Wave. Fujitsu, in contrast, its QEMIST\footnote{ https://1qbit.com/qemist/ } platform focuses on advanced materials and quantum chemistry research, with universal quantum computing processors. 

We can observe from the figure \eqref{fig:ChallengeTransmons} the challenge of simulating transmon quantum processors. We can also analyse Microsoft's framework that offers an end-to-end scalable environment from the figure \eqref{fig:MS_SDK_QC}.

\section{summary}

This chapter has analysed and defined the state of the art of quantum computers and the companies creating such computers. We have seen that to date, and there are two potential techniques; Quantum Annealing and Quantum Gate-Based. Furthermore, we have revised several strategies for building a quantum computer and are currently done by combining multiple multicore processors. Finally, we also focus on the challenge of simulating transmon quantum processors. \\

We can focus on research since we have the necessary ingredients to experiment and implement a quantum algorithm for the combinatorial optimisation problem.

%%%%%%%%%%%%  Starting New Page here %%%%%%%%%%%%%%

\newpage

\chapter{Quantum Computing}\label{sec:10}

\section{Introduction }
In this chapter, we will experiment with quantum computing, mainly on Qiskit (Fig. \eqref{fig:Qiskit_framework}). We will explore variational techniques and some of their libraries to do this.
\subsection{Quadratic Unconstrained Binary Optimisation Problems}\label{sec:quboChap}

The \textit{Quadratic Unconstrained Binary Optimisation}, commonly called QUBO, is known as \textit {a combinatorial optimisation problem with many applications, from finance and economics to machine learning.}\\

QUBO problems are traditionally used in computer science. True and False variables, states that correspond to $1$ and $0$ values.
A QUBO problem is defined using an upper-diagonal matrix  $Q$, which is an  ${N \times N}$ upper-triangular matrix of real weights, and $x$, a vector of binary variables, as minimising the function.

\begin{equation}
\label{Qubo_form_eq}
    f(x)_{\text{QUBO}}= \sum _{i}^{N} \sum_{j<i}^{N}Q_{ij}x_{i}x_{j}+ \sum _{i}^{N}Q_{i}x_{i}.
\end{equation}

Where the diagonal terms  $Q_{i}$ are the linear coefficients and the nonzero off-diagonal terms are the quadratic coefficients $Q_{ij}$. The QUBO is unconstrained in that there are no constraints on the variables other than those expressed in $Q$. 
The diagonal entries of $Q$ are the linear coefficients that bias the qubits. The nonzero off-diagonal terms are the quadratic coefficients that define the strength of the coupling between variables.

The input may be full or sparse. Both upper- and lower-triangular values can be used;  $(i, j)$ and $(j,i)$ entries are added together. An exception is raised if a $Q$ value is given to a coupler not present. Only entries indexed by working couplers may be nonzero.
This concept is so powerful that we will use it in our formulation, taking into account that with the docplex, we can't formulate inequation constraints. 

The QUBO is usually represented more concisely as equation \eqref{Qubo_obj_funct_eq}

\begin{equation}
\label{Qubo_obj_funct_eq}
    \mathop{\min }_{x \in  \{ 0,1\mathop{ \} }^{N}}x^{T}Qx.
\end{equation}

We can also find the Hamiltonian of the objective function expressed as an in the scalar notation of the QUBO form as follows:

\begin{equation}
\label{Ham_Qubo_Form_eq}
    H(a_{i},b_{ij};q_{i})_{\text{QUBO}}= \sum _{i}^{N} \sum _{j<i}^{N}b_{ij}q_{i}q_{j}+ \sum _{i}^{N}a_{i}q_{i}.
\end{equation}

This form can be easily found in the D-Wave formulation.
The transformation between Ising and QUBO is $s=2x-1$. Let demonstrate that QUBO and the Ising of the Hamiltonian are similar. This means that we will have the algorithm in Ising form by writing an algorithm for QUBO with a single variable change. That is very useful to have the algorithm for various platforms based on quantum gates or quantum annealing.

\begin{equation}
\label{Ising_form_qubo_eq}
    H(s) = \sum _{i}^{N} \sum _{j<i}^{N}J_{ij}s_{i}s_{j}+ \sum _{i}^{N}h_{i}s_{i}.
\end{equation}

With  $s_{i}$  and $s_{j} \in \left \{-1,1 \right\}$. For the translation in QUBO formulation, let’s consider  $x_{i}$ and  $x_{j}  \in  \left\{ 0,1 \right\}$  and using the spin relation  $s=2x-1$, we can remap the Hamiltonian as follow:

\begin{equation}
\label{descomp_qubo_Ising_eq}
\begin{aligned}
  &H(s)= \sum _{i}^{N} \sum _{j<i}^{N}J_{ij} (2x_{i}-1)( 2x_{j}-1) + \sum _{i}^{N}h_{i} (2x_{i}-1) \\
     &= \sum _{i}^{N} \sum _{j<i}^{N}J_{ij} (4x_{i}x_{j}-2x_{i}-2x_{j}+1) + \sum _{i}^{N}(2h_{i}x_{i}-h_{i}). 
\end{aligned}
\end{equation}

Regrouping all the constants into  $C_{0}$ we have the following expression:

\begin{equation}
\label{desc_qubo_Ising_1_eq}
    H \left( s \right) = \sum _{i}^{N} \sum _{j<i}^{N}4J_{ij}x_{i}x_{j}- \sum _{i}^{N}x_{i} \sum _{j<i}^{N}2J_{ij}- \sum _{i}^{N} \sum _{j<i}^{N}2J_{ij}x_{j}+ \sum _{i}^{N}2h_{i}x_{i}-C_{0}.
\end{equation}

By grouping $x_{i}$ terms together, we can write, $a_{i}= \sum _{j<i}^{N}2J_{ij}+2h_{i}$, so:

\begin{equation}
\label{desc_qubo_Ising_2_eq}
    H \left( s \right) = \sum _{i}^{N} \sum _{j<i}^{N}4J_{ij}x_{i}x_{j}- \sum _{i}^{N}a_{i}x_{i}-2 \sum _{i}^{N} \sum _{j<i}^{N}J_{ij}x_{j}+C_{0}.
\end{equation}

Let us develop the term  $\sum _{i}^{N} \sum _{j<i}^{N}J_{ij}x_{j}$  in term of  $x_{i}$.

\begin{equation}
\label{desc_qubo_Ising_3_eq}
\begin{aligned}
    &\sum _{i}^{N} \sum _{j<i}^{N}J_{ij}x_{j}=J_{10}x_{0}+ J_{20}x_{0}+J_{21}x_{1}+J_{30}x_{0}+J_{31}x_{1}+J_{32}x_{2}+ \ldots \\
    &+J_{N0}x_{0}+J_{N1}x_{1}+J_{N2}x_{2}+J_{N2}x_{3}+J_{N,N-1}x_{N-1}.
\end{aligned}
\end{equation}

We can observe that each column has a term:

\begin{equation}
\label{desc_qubo_Ising_4_eq}
      \left( J_{10}+J_{20}+J_{30}+ \ldots +J_{N0} \right) x_{0}+ \left( J_{21}+J_{31}+ \ldots +J_{N1} \right) x_{1}+ \ldots + \left( J_{N,N-1} \right) x_{N-1}.
\end{equation}

Let’s cast the  $J_{ij}$  as some constants $b_{i}$, so:

\begin{equation}
\label{desc_qubo_Ising_5_eq}
     b_{0}x_{0}+b_{1}x_{1}+b_{2}x_{2}+ \ldots +b_{N-1}x_{N-1}.
\end{equation}

Since there is no $x_{N}$ in the original sum, so $b_{N}=0$. So we can write that:

\begin{equation}
\label{desc_qubo_Ising_6_eq}
    \sum _{i}^{N} \sum _{j<i}^{N}J_{ij}x_{j}= \sum _{i}^{N}b_{i}x_{i}.
\end{equation}

So, our Hamiltonian can be written as:

\begin{equation}
\label{desc_qubo_Ising_7_eq}
    H \left( s \right) = \sum _{i}^{N} \sum _{j<i}^{N}4J_{ij}x_{i}x_{j}- \sum _{i}^{N}a_{i}x_{i}-2 \sum _{i}^{N}b_{i}x_{i}+C_{0}.
\end{equation}

Let $J_{ij}^{'}=4J_{ij}$  and  $h_{i}^{'}=a_{i}-2b_{i}$, therefore,

\begin{equation}
\label{desc_qubo_Ising_8_eq}
    H \left( s \right) = \sum _{i}^{N} \sum _{j<i}^{N}J_{ij}^{'}x_{i}x_{j}- \sum _{i}^{N}h_{i}^{'}x_{i}+C_{0}H \left( s \right) =H \left( x \right) +C_{0}.
\end{equation}

Where we can experiment that Ising  $H(s)$ and QUBO $H(x)$ are similar in form and relations; they are isomorphic.
%%%%%%%%%%%%%%%%%%%% Table No: 6 starts here %%%%%%%%%%%%%%%%%%%%
\begin{table}[t!]
\centering
\begin{tabular}{ |c|c|c|c|c|  }
 \hline
   \multicolumn{5}{|c|}{Terms} \\
 \hline
 Problem Expression & Linear Coefficient & Quadratic Coefficient & Variable & States\\
 \hline
 QUBO (scalar)   & $a_{i}$   &  $a_{i,j}$  & $q_{i}$ & $\{0,1\}$\\
 QUBO (matrix)   & $Q_{i,j}$ &  $Q_{i,j}$  & $x_{i}$ & $\{0,1\}$\\
 Ising           & $h_{i}$   &  $J_{i,j}$  & $s_{i}$ & $\{-1,1\}$ \\
 \hline
\end{tabular}
\caption{Comparation between Ising and QUBO representation and related terminology.}
\label{tab:Translator_Qubo_Ising}
\end{table}
%%%%%%%%%%%%%%%%%%%% Table No: 6 ends here %%%%%%%%%%%%%%%%%%%%

To convert the coefficients from QUBO to Ising:
\begin{equation}
\label{desc_qubo_Ising_9_eq}
\begin{aligned}
 &J_{ij}=\frac{1}{4}Q_{ij}\\
 &h_{i}=\frac{1}{2}Q_{ii}+\frac{1}{4} \sum _{i<j}^{N}Q_{ij}.
\end{aligned}
\end{equation}

Or from Ising to QUBO:
\begin{equation}
\label{desc_qubo_Ising_10_eq}
\begin{aligned}
    &Q_{ij}=4J_{ij}\\
    &Q_{ii}=2h_{i}-\frac{1}{2} \sum _{i<j}^{N}Q_{ij}.
\end{aligned}
\end{equation}

\section{Variational calculation} \label{sec:variational_Cal}
The \textit {variational method} is defined as \textit{the way to find approximations to the lowest energy eigenstate or ground state and some excited states. It allows calculating approximate wavefunctions such as molecular orbitals. The basis for this method is the variational principle.}\\

The \textit{variational calculation} \cite{TRo11} is the basis of the variational principle \cite{IEk74}. We can say that the variational calculus consists of looking for maximums and minimums or extensively looking for relative ends of a function of functions (functional) over a space of functions. This calculation can be seen as a generalisation of the elementary calculus of a variable's maximum and minimum real functions.

Mathematically speaking, when we talk about optimisation, we are talking in some way to find the maximum or minimum of the function that models our scenario; Our objective function. That is, calculate the minimum or maximum of our objective function. Although it seems easy, in many cases, the calculation of the minimum or ceiling is not entirely trivial because of the structure of the data, the size of the data or basically or for the computational cost required to make this calculation makes it non-trivial. The computational cost \cite{RKi12} is one of the limits of all scientific advances. For that same reason, the scientific community is working to equip itself with machines that can give it the most significant computational capacity \cite{Ron}.

Several branches defined and designed alternatives in solving optimisation problems by calculating variations. One of the most contemplated approaches is from Richard Bellman \cite{SDr02}, who developed dynamic programming \cite{SDr02,BBh10} with apparent alternatives to the calculation of variations. 

The work of Sturm-Liouville \cite{RPA99} and Rayleigh-Ritz method \cite{RBB85}, are the basis of the Variational Quantum Eignesolver; the VQE. A dynamic control system that allows us to make the variational calculation of a quantum state{$ \vert  \psi \left( \theta  \right)  \rangle$} associated with its expectation value $H$.

Let $\vert  \psi _{i} \rangle$ be an eigenvector, of a matrix $A$ which is invariant under transformation by $A$  up to a scalar multiplicative constant (the eigenvalue $\lambda _{i}$. That is  $A \vert  \psi _{i} \rangle = \lambda _{i} \vert  \psi _{i} \rangle$. If we define the Hamiltonian $H$ a matrix that is Hermitian,

\begin{equation}
\label{Hermitian_eq}
    H= \sum _{i=1}^{N} \lambda _{i} \vert  \psi _{i} \rangle \langle \psi _{i} \vert. 
\end{equation}

where each  $\lambda_{i}$ is the eigenvalue corresponding to the eigenvector $\vert  \psi_{i} \rangle$ . Furthermore, the expectation value of the observable on $H$  on an arbitrary quantum state $\vert  \psi  \rangle$  is given by:

\begin{equation}
\label{H_Expected_Value_eq}
    \langle H \rangle = \langle \psi\left(  \theta  \right)\vert  H \vert \psi \left(\theta  \right)  \rangle. 
\end{equation}

Substituting the value of  $H$ and using the Hermitian property in the equation,

\begin{equation}
\label{demo_Hermitian_Expectation_Value_eq}
\begin{aligned}
 &\langle H \rangle = \langle \psi \left(\theta  \right) \vert   \left(\sum_{i=1}^{N} \lambda_{i} \vert  \psi_{i} \rangle \langle \psi _{i} \vert  \right)   \vert \psi   \left(  \theta  \right)  \rangle =  \left(  \sum _{i=1}^{N} \lambda_{i}\langle \psi   \left(\theta  \right)   \vert  \psi_{i} \rangle \langle \psi_{i} \vert   \psi   \left(  \theta  \right)  \rangle  \right) \\
 &= \sum_{i=1}^{N} \lambda _{i} \vert \langle \psi_{i} \vert   \psi \left(  \theta  \right)  \rangle  \vert ^{2} \langle H \rangle =  \sum_{i=1}^{N} \lambda_{i} \vert \langle \psi_{i} \vert   \psi  \left(\theta  \right)  \rangle \vert ^{2}.
 \end{aligned}
\end{equation}

This $\langle H \rangle = \sum _{i=1}^{N} \lambda _{i} \vert \langle \psi _{i} \vert   \psi   \left(  \theta \right)  \rangle  \vert ^{2}$ demonstrates that the expectation value of an observable on any state can be expressed as a linear combination using the eigenvalues associated with $H$ as the weights. Moreover, each of the weights in the linear combination is greater than or equal to 0, as $\vert \langle \psi _{i} \vert   \psi   \left(\theta  \right)  \rangle  \vert ^{2} \geq 0$ and so it is clear that $\langle H \rangle _{ \psi\left( \overrightarrow{ \theta } \right) } \geq  \lambda _{\min }$.

This is very powerful; it implies that the expectation value of any wave function will always be at least the minimum eigenvalue associated with a given $H$. Moreover, the expectation value of the state $\vert  \psi _{\min } \rangle$ is provided by  $ \langle H \rangle _{ \psi _{\min }  \left( \overrightarrow{ \theta } \right) }= \lambda _{\min }=E_{gs}$. This is the base of the VQE, where $E_{gs}$ is the ground state energy of that system related to the Hamiltonian $H$. 

A fundamental and relevant concept is knowing that a fixed variational form with a polynomial number of parameters can only generate transformations to a polynomially dimensioned subspace of all states in a Hilbert space of exponential size \cite{Mic00}. This is very important for all algorithms based on variational calculation. It is also imperative to mention that the ability to generate an arbitrary state ensures that during the optimisation process, the variational form does not limit the set of achievable states on which the expected value can be taken. Usually, this limitation is given by the classical algorithm. This leads us to see that there are several variational forms. Some, like  $Ry$ and $RyRz$, are heuristically designed, regardless of the target domain.

While building a variational form, we must balance two different objectives.

Normally, a variational form $n$ qubits could generate any possible state $\vert \psi  \rangle$ where  $\vert \psi  \rangle  \in \mathbb{C}^{\bigotimes N}$  and $N = 2^{n}$. Nevertheless, we would like the variational form to use as few variables/parameters as possible. This is the contradiction. We must have expertise when creating the ideal variational form of our circuit.

In the experimentation chapter, we will work intuitively until we reach our algorithm, fulfilling the two mentioned objectives.

\subsection{Variational Quantum Eigensolver} \label{sec:VQE_Sec}

The \textit{Variational Quantum Eigensolver}, most known as VQE, is a \textit{flagship algorithm for quantum chemistry using near-term quantum computers. It is an application of the Ritz variational principle, where a quantum computer is trained to prepare the ground state of a given molecule.}\\

\begin{figure}[h!]
    \centering
    \includegraphics[width=0.8\textwidth]{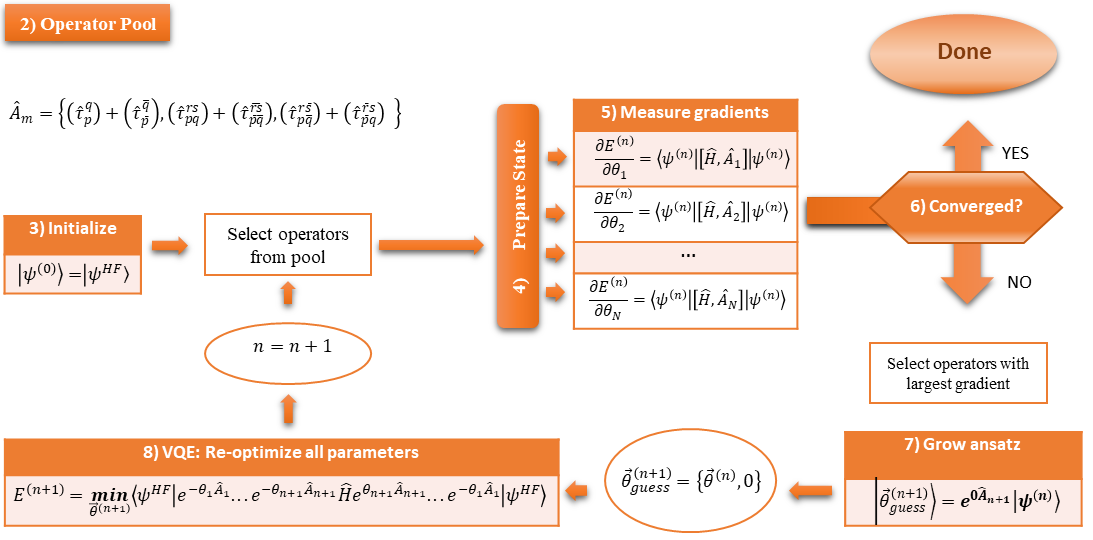}
    \caption{Adaptative Variational Quantum Eigensolver \cite{grimsley2019adaptive}.}
    \label{fig:VQE_ADPT_}
\end{figure}

\begin{figure}[h!]
    \centering
    \includegraphics[width=0.8\textwidth]{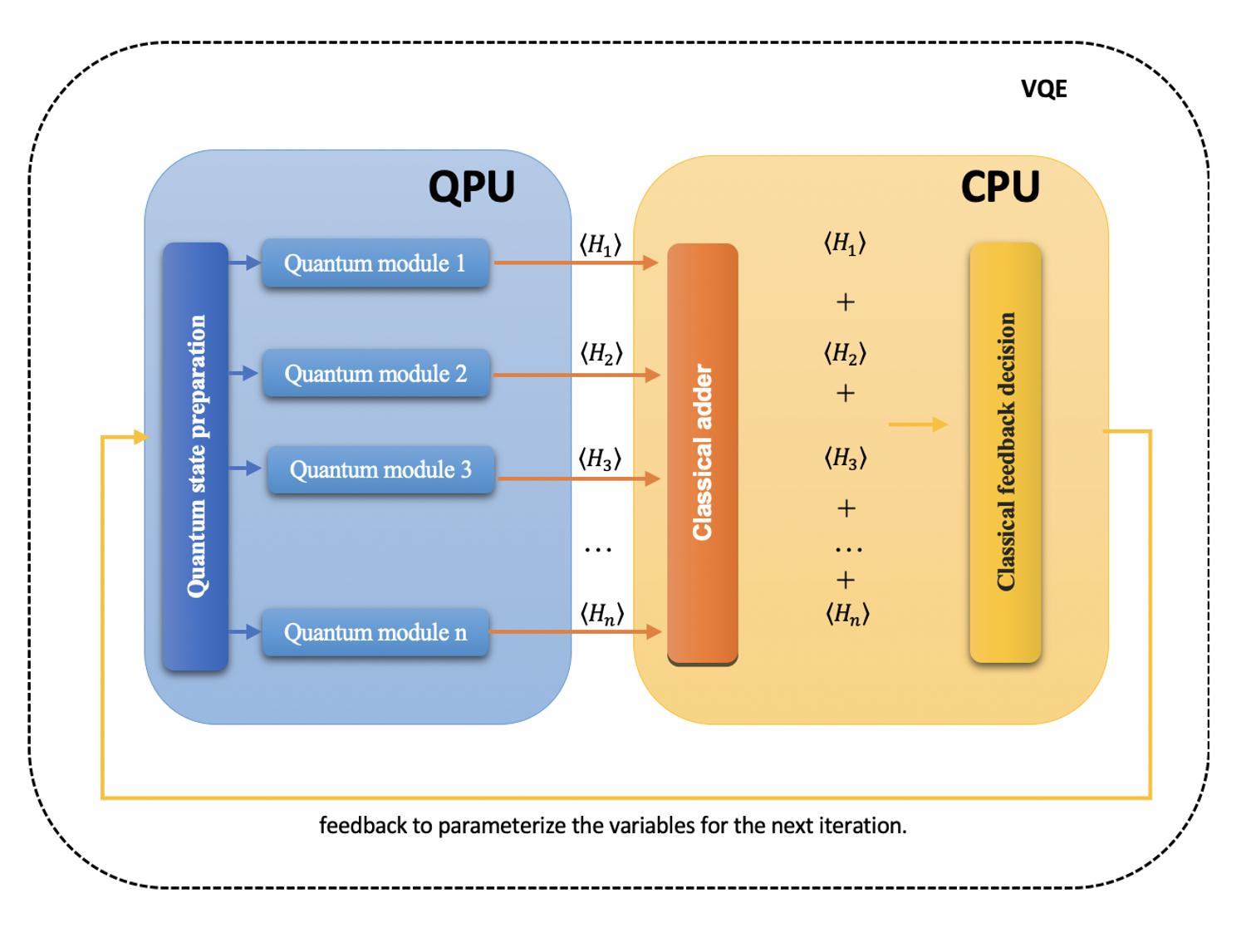}
    \caption{Architecture of the Variational Quantum Eigensolver. The architecture of the quantum-variational eigensolver. Algorithm 1: Quantum states that have been previously prepared are fed into the quantum modules, which compute $\langle H_i\rangle$, where $ H_{i} $ is any given term in the sum defining H. The results are passed to the CPU, which computes $\langle H \rangle$. Algorithm 2: The classical minimisation algorithm, run on the CPU, takes $\langle H\rangle$ and determines the new state parameters, which are then fed back to the QPU.}
    \label{fig:VQE_STANDAR}
\end{figure}

Unfortunately, we're still in the Noisy Intermediate-Scale Quantum (NISQ) \cite{Joh18} era because we don't have a perfect quantum computer yet. To compensate for the fact that quantum isn't perfect yet, researchers started developing algorithms that work both quantum and classical to solve problems. The VQE \cite{Alb13,Dao19,Har19} is based on the variational principle, which is dynamic programming. Using the VQE, we can make smart and adaptive algorithms (Figures \eqref{fig:VQE_ADPT_} and \eqref{fig:VQE_STANDAR}). This area is known as Quantum Machine Learning (QML) \cite{JBi17} and one of the warmest QML algorithms nowadays is the Variational Quantum Eigensolver. This is because its applications range from finance, biology, scheduling and chemistry. One of the essential characteristics of molecules is their ground state energy. The ground state energy is just the lowest possible energy state that a molecule can be in. The ground state energy of a molecule is vital because it gives us more information about the electron configuration of that molecule.

The VQE is a hybrid quantum/classical algorithm originally proposed to approximate the ground state of a quantum chemical system (the state attaining the minimum energy).

By varying the experimental parameters in the preparation of the state and calculating the Rayleigh-Ritz ratio \cite{Cha05} using the subroutine in a classical minimisation, unknown eigenvectors can be prepared. At the end of the algorithm, the reconstruction of the eigenvector stored in the final set of experimental parameters that define the state will be done.
The variational method in quantum mechanics is used, which is a way of finding approximations to the energetic state of lower energy or fundamental state, and some excited states. This allows to calculate approximate wave functions, such as molecular orbitals and is the basis of this method. It is the variational principle that will enable us to write the following equation  $\langle H \rangle _{ \psi   \left( \overrightarrow{ \theta } \right) } \geq  \lambda _{i}$. With  $\lambda _{i}$ as eigenvector and $\langle H \rangle _{ \psi   \left( \overrightarrow{ \theta } \right) }$  as the expected value. The problem that the VQE solves is reduced to finding such an optimal choice of parameters $\overrightarrow{ \theta }$, that the expected value is minimised and that a lower eigenvalue is located. $\langle H \rangle =\langle \psi   \left(  \theta  \right)   \vert  H  \vert   \psi   \left(  \theta  \right)  \rangle$.

Architecture of the quantum-variational eigensolver (See Fig. \eqref{fig:VQE_STANDAR}). \textbf{Algorithm 1:} Quantum states that have been previously prepared are fed into the quantum modules, which compute $ \langle H_i \rangle$, where  $H_{i}$  is any given term in the sum defining $H$. The results are passed to the CPU, which computes $\langle H \rangle$. \textbf{Algorithm 2:} The classical minimisation algorithm, run on the CPU, takes  $\langle H \rangle$ and determines the new state parameters, which are then fed back to the QPU.
First, we prepare the trial wavefunction on a quantum processor. Then, we measure the qubits, resulting in an  $n$ -bit string $x_{0}...x_{n-1}$. Each observed string easily translates into a sample from $\langle \psi   \left(  \theta  \right)   \vert  H  \vert   \psi   \left(  \theta  \right)  \rangle$, because $H$ is a weighted summation of tensor products of Pauli Z-matrices, and each such term can be computed with a simple parity check. We denote these samples by $H_{k} \left(  \theta  \right)$, $k = 1, \ldots ,K$, where $K$ is a natural number and is the number of samples.

The sample mean \eqref{mean_eq} is an estimator for $\langle H \rangle =\langle \psi   \left(  \theta  \right)   \vert  H  \vert   \psi   \left(  \theta  \right)  \rangle$ and is used as the objective function for the classical optimisation algorithm.

\begin{equation}
\label{mean_eq}
    \text{mean}=\frac{1}{K} \sum _{k=1}^{K}H_{k} \left(  \theta  \right). 
\end{equation}

The need for the VQE algorithm is seen when solving an optimisation problem related to a Hamiltonian. Once the Hamiltonian is built, we use/apply VQE to determine the ground state, from which an optimal solution to our objective function can be sampled with a probability of $1$.

Assuming $n$ qubits, we start by applying single qubit  $Y$ rotations to every qubit, parametrised by an angle  $\theta_{0,i}$  for qubit  $i$. We then repeat the following $p times$.  We apply controlled $Z$ gates to all qubit pairs $(i,j)$ satisfying  ${i < j}$, where $i$ denotes the control qubit and $j$  the target qubit; and we add another layer of single-qubit $Y$ rotations to every qubit, parametrised by $\theta _{k,i}$ for qubit $i$ and repetition  $k \in  \{ 1,  \ldots ,p \}$.

\subsection{Noisy VQE for the Optimisation problem}
In Quantum computing, we can classify the algorithms in 3 groups. Gate circuits \cite{Nik15}, Annealing \cite{Wbo} and Variational \cite{Dao19,Aru19,TRo11}.
\begin{figure}[h!]
    \centering
    \includegraphics[width=0.25\textwidth]{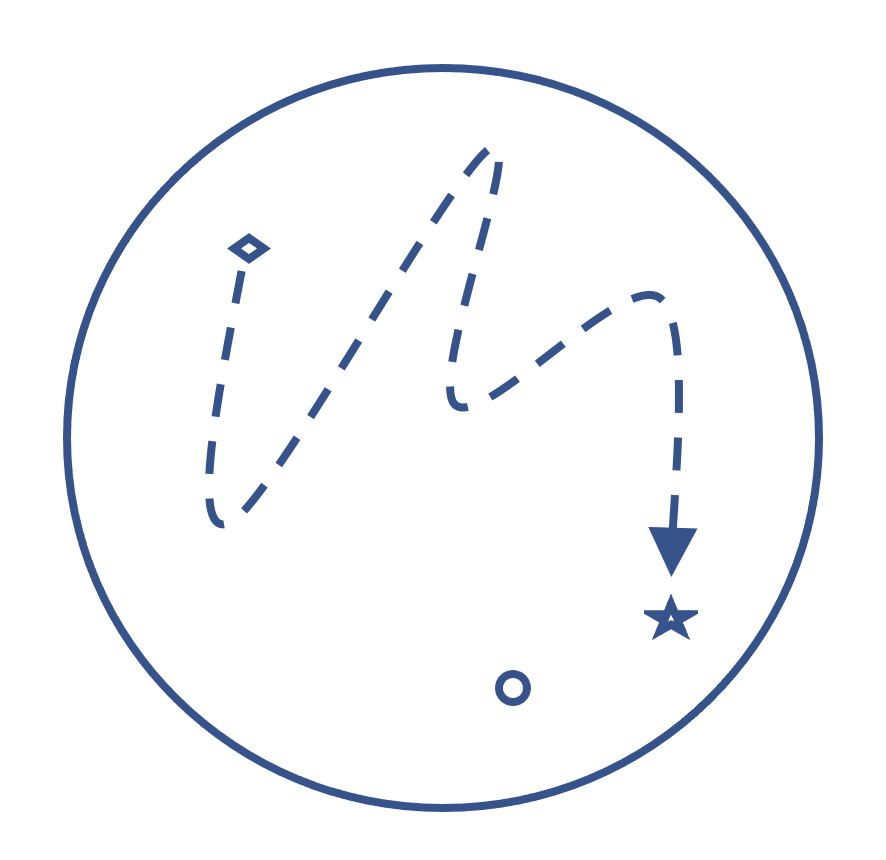}
    \caption{Minimisation principle used in Variational Quantum Eigensolvers to empower Quantum Machine Learning.}
    \label{fig:NOISY_VQE}
\end{figure}
Although the logic gates are not perfect and have noise by having the variational principle, we can see in NISQ devices a golden opportunity to have a machine that analyses Hilbert's vector space. 

The term "Noisy Intermediate Scale Quantum" describes the era in which we find ourselves today. Noisy, because the computer still does not offer enough qubits to save for error correction. So, we have imperfect qubits at the physical layer, last but not at least "intermediate scale" because of their small number of qubits.

This leads us to forget about the imperfections of the gates and only think of the variational ansatz \cite{Dao19} and that this ansatz can be analysed efficiently on a quantum computer. Something we can't do with a classic computer. Fig. \eqref{fig:NOISY_VQE}. summarises the idea. In short, we can say that we have a quantum computer that generates variational states, known as a Variational Quantum Computer \cite{Dao19}. Another way to see it could be, in each of the iterations, we have a quantum circuit close to the solution we would be looking for. This is the basis of Quantum learning \cite{JBi17}. We are doing machine learning (ML) on circuit design.

With this vision, we are developing a perfect machine to solve optimisation and classification problems by exploring the entire configuration space of quantum molecules.  

As commented, the VQE are useful because they find the lowest possible eigenvalue of a given Hermitian matrix  $H$  (it doesn't matter the size of  $H$) using the variational method or the variational principle. It's also known that the expected value must always be equal or greater than the lowest possible eigenvalue. This means that if we just keep minimising that expectation value, we only get closer and closer to the minimum eigenvalue of the given  $H$  matrix and never below it \cite{Dao19}. With this powerful concept, the great clue is how to map our objective function into a Hamiltonian model of a given molecular system. To do that, first, we map the molecular Hamiltonian into a qubit. This essentially means that we are mapping the electron orbital interactions inside the molecules onto our qubits. Next, we prepare the set. Our set to be shallow has to cover a good enough range for our trial wave functions, so since we don't face our ground state energy. With the information given by a specific Hamiltonian, now, we calculate the energy of that electron configuration.

At this point, the algorithm measures those values and send them through to the classical optimiser. The classical optimiser minimises our parameters, getting a lower expectation value $H$. After that, we feed all these values back into the quantum part and reiterate it with this loop many times until it converges onto the lowest possible energy state for that interatomic distance to follow all the described steps. All this is achieved regardless of the noise or imperfection of the logic gates.

\subsection{Quantum approximate optimisation algorithm}

The \textit{Quantum Approximate Optimisation Algorithm} most known as QAOA, \textit{is now as far as the best quantum approximate optimisation algorithm developed and introduced by Farhi, Goldstone, and Gutmann, that finds a right solution (not always find the exact solution) to an optimisation problem in polynomial time \cite{Edw,Edw19}. }\\

\begin{figure}[h!]
    \centering
    \includegraphics[width=\textwidth]{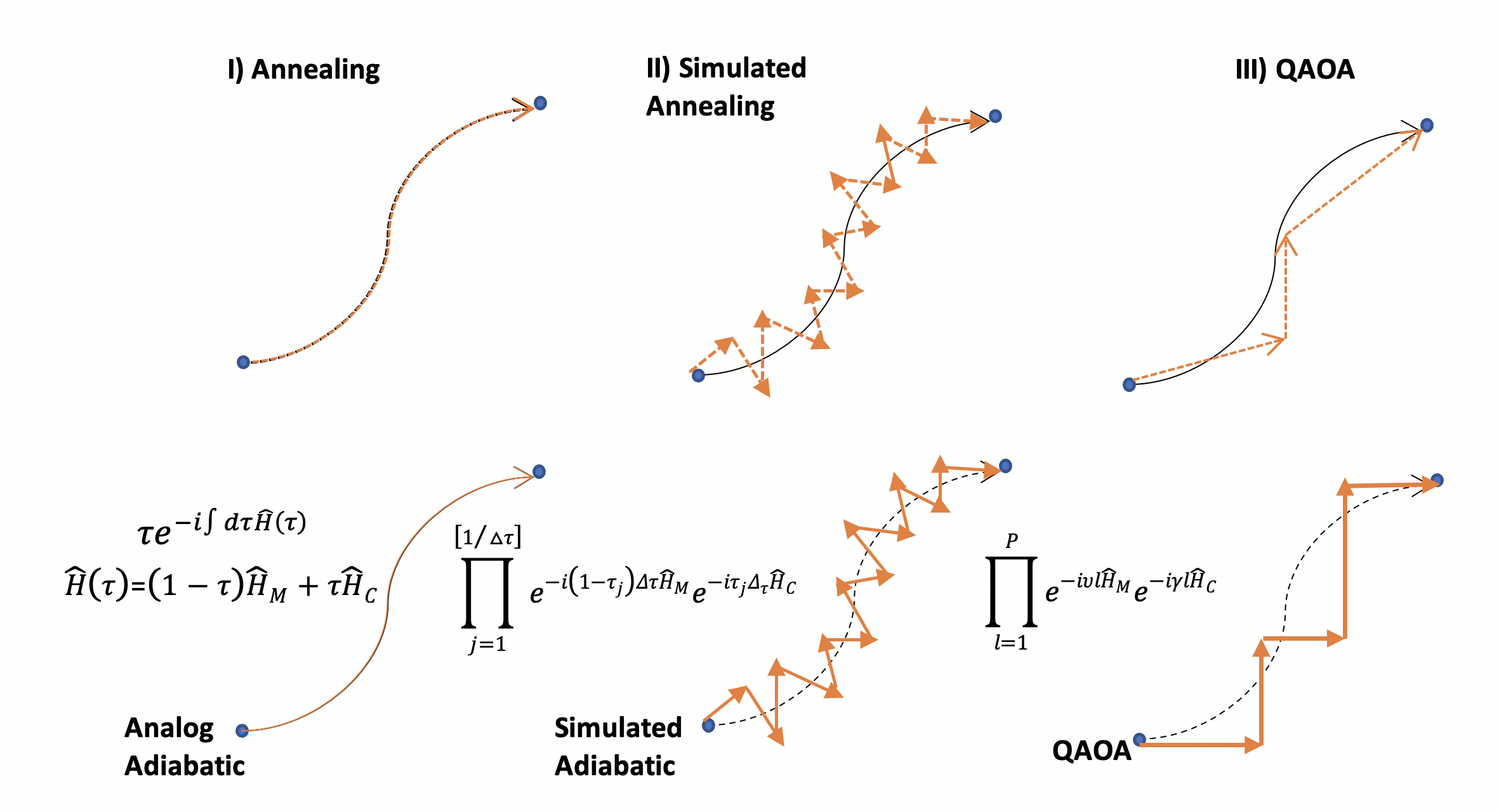}
    \caption{From Annealing to QAOA. These graphs conceptually show us the differences/evolution between quantum annealing (left) that follows this adiabatic time evolution path. Simulated annealing (middle) follows the same way in discrete steps and QAOA that follows adiabatic time evolution path in p steps. These define the accuracy of the solution.
}
    \label{fig:QAOAIlustr}
\end{figure}

\begin{figure}[!h]
    \centering
    \includegraphics[width=0.6\textwidth]{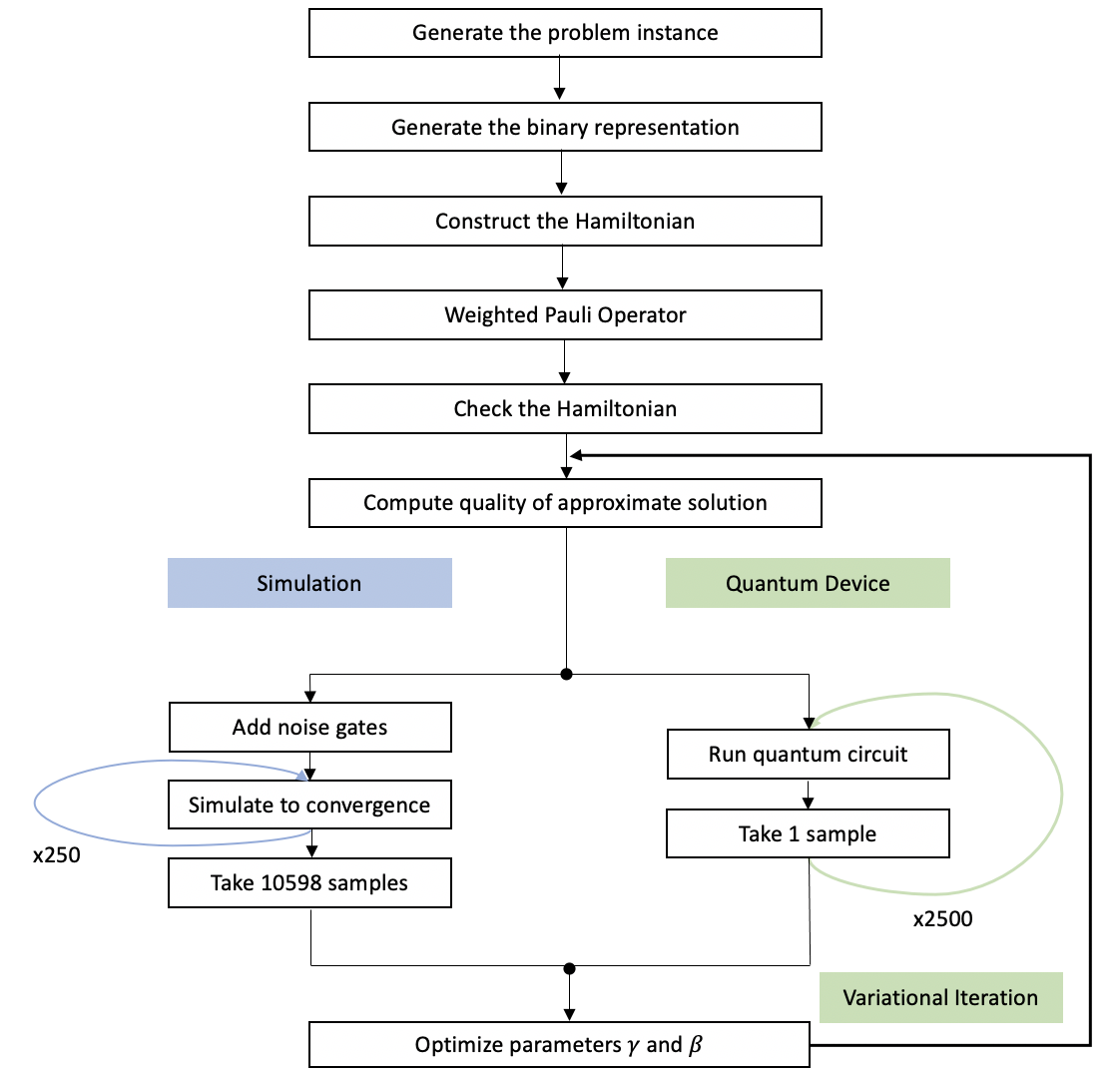}
    \caption{QAOA’s diagram process to perform it experimentally with quantum devices.
In our study, 1000 samples are used to estimate the value of the cost function $E ( \protect \overrightarrow{\gamma},\protect \overrightarrow{\beta}) =\langle \protect \overrightarrow{\gamma},\protect \overrightarrow{\beta} \vert  H \vert  \protect \overrightarrow{\gamma},\protect \overrightarrow{\beta} \rangle$ at each variational iteration.}
    \label{fig:QAOA_Steps}
\end{figure}

One reason why the work done by researchers makes QAOA (see Figures \eqref{fig:QAOAIlustr}, \eqref{fig:QAOA_Block} and \eqref{fig:QAOA_Steps}) very interesting and highly useful for the NISQ era is its potential to display quantum supremacy \cite{Kal19}.

This algorithm belongs to the NISQ era algorithm class already discussed, which are from the hybrid algorithm class. More precisely, classical-quantum hybrid variational algorithms. We can safely consider it an evolution of adiabatic time trotted in $p$  steps towards the ground state of a Hamiltonian (characteristics of a quantum circuit/system) that encodes the problem. The algorithm does "discretise" by introducing steps in the time variable (time evolution). Where the steps variable $p$ defines the accuracy of the solution. 

These graphs conceptually show us the differences/evolution between quantum annealing (left) that follows this adiabatic time evolution path. Simulated annealing (middle) follows the same way in discrete steps and QAOA that follows adiabatic time evolution path in $p$  steps. These define the accuracy of the solution.

This algorithm builds on and improves run time on Simulated Annealing. Therefore, intuitively, we can believe before any demonstration that QAOA is the best candidate to solve combinatorial optimisation problems assigned to the minimisation of a Hamiltonian Ising in NISQ devices.

The Adiabatic Quantum Computing (AQC) has one serious problem when the computation time to solve any problem rises exponentially as  $\Delta E$ becomes infinitesimally small. This bound ACQ's capability to solve a specific instance of hard optimisation problems \cite{Qin18}.

This means that the difference between the ground state and first excited state energy of the Hamiltonian  $H \left( t \right)$  of the problem we will resolve bounds the step size one can take to follow the adiabatic pathway. QAOA resides in this limitation of AQC. So, and according to the Fig. \eqref{fig:QAOAIlustr}, we trotterise in  $p$  step relating to some configuration parameters. The Trotterisation \cite{Qin18}  formula is known by the equation \eqref{trotterization_eq}.

\begin{equation}
\label{trotterization_eq}
    e^{-iHt}= \left( e^{-iH_{0}\frac{t}{p}}\ast^{-iH_{1}\frac{t}{p}}\ast \ldots \ast^{-iH_{k-1}\frac{t}{p}} \right) ^{p}+ f \left(  \gamma  \right).
\end{equation}

With $f \left(  \gamma  \right)$ as some polynomial factors equal to zero. 

As Hamiltonians are Hermitian operators that are usually a sum of a large number of individual Hamiltonians $\sum _{j}^{}H_{j}$, we can use Lie product formula \cite{Tos74} as shown by the equation \eqref{trotterization_1_eq}. 

\begin{equation}
\label{trotterization_1_eq}
    e^{-i \left(  \sum _{j}^{}H_{j} \right) t}=\mathop{\lim }_{N \rightarrow \infty} \left( e^{-iH_{0}\frac{t}{N}}\ast^{-iH_{1}\frac{t}{N}}\ast \ldots \ast^{-iH_{m-1}\frac{t}{N}} \right) ^{N}.
\end{equation}

Since the limit of this series is infinite with $N \in Z^+$, when we implement this in quantum computing, we must truncate the function by introducing a quantifiable bounded error  $\varepsilon$ refers to  $\Vert e^{-iHt}-U \Vert  \leq  \varepsilon$. This truncation is known as Trotterisation, and it's widely used to simulate non-commuting Hamiltonians on quantum computers. We will take advantage of this technique in the development of EVA \cite{alonsolinaje2021eva}.

Suppose we want to simulate one circuit given by the following Hamiltonian: 

\begin{equation}
\label{Hamiltonian_proof_eq}
    H= X_{0}+Y_{1}+Z_{2}.
\end{equation}

Where $X$,  $Y$ and  $Z$ are, Pauli matrices (In some books, we can find this notation relating to the Pauli matrices $\sigma_i$ ($\sigma_x,\sigma_y,\sigma_z$)) and the subscripts label the qubits that the Hamiltonians apply to. We can't simulate each Hamiltonian separately because they don't commute(the anti-commutation can be achieved by noting that $XZ=ZX$ and $-YZ=ZY$) \cite{Mic00}. 
This is the main reason why we use Trotterisation, where we evolve the whole Hamiltonian by repeatedly switching between evolving  $X$, $Y$  and $Z$ each for a small period. The first step from the given system is finding the quantum gates (we are developing on Quantum gate-based computers) that implement each of its terms.

This case is simple since the quantum gates will implement the individual terms correctly.

\begin{equation}
\label{Rotation_Rx_eq}
     R_{x} \left(  \theta  \right) =e^{-i\frac{ \theta }{2}X}, R_{y} \left(  \theta  \right) =e^{-i\frac{ \theta }{2}Y},  R_{z} \left(  \theta  \right) =e^{-i\frac{ \theta }{2}Z}.
\end{equation}

Where $\theta$ specifies the angle by which to rotate the state in a specified axis, this notion is pretty powerful because it determines how long (as a time ($t$)) to apply the Hamiltonian to the qubit.

If we consider, in this case, that $p=2$ and $t=1$ we can simulate/compute our circuit (the Hamiltonian) by using the Lie formula as shown:
\begin{figure}[!h]
    \centering
    \includegraphics[width=0.8\textwidth]{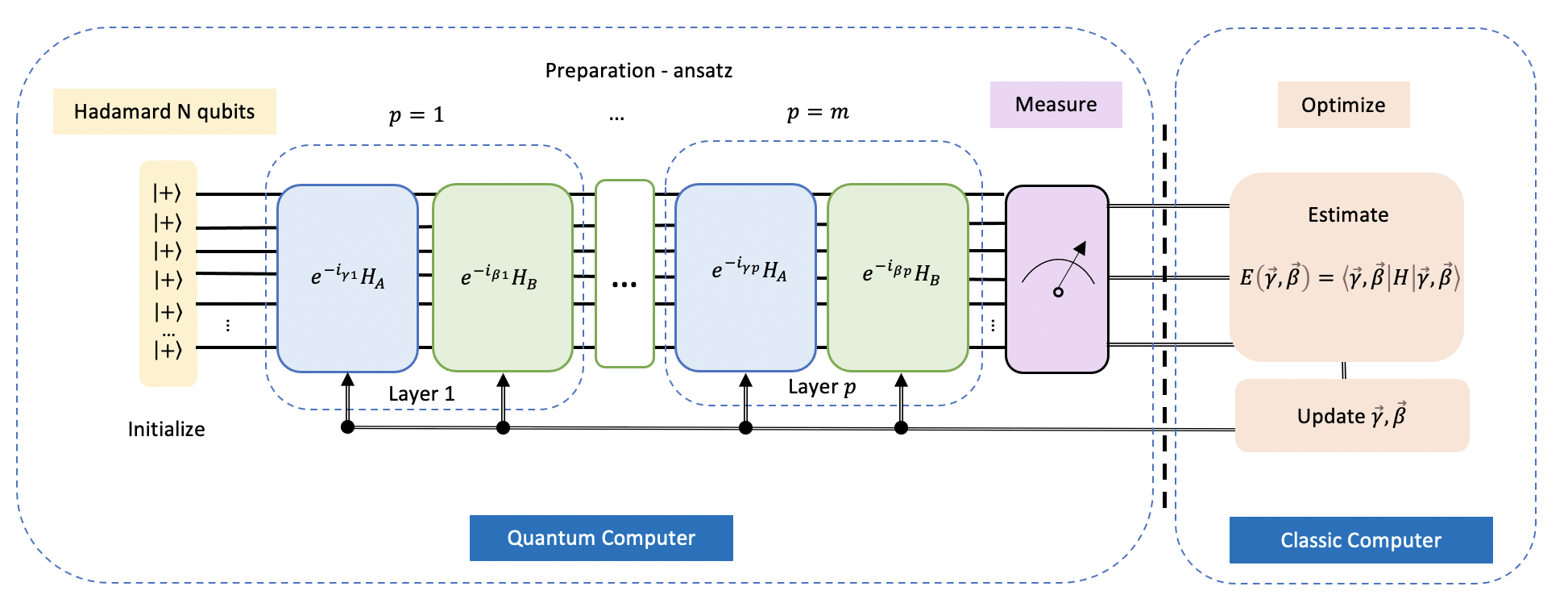}
    \caption{QAOA algorithm operation.The system is initialised along the y-direction in the Bloch sphere in the $\vert +\rangle^{\otimes N}$ state. The unitary evolution under $H = X_{0} + Y_{1} + Z_{2}$  is implemented for $\gamma_{i} \beta_{i}$ angles for p times. At the end of the algorithm, global measurements in the x and the y basis are performed to compute the average energy $\langle H\rangle=E(\protect \overrightarrow{\gamma}, \protect \overrightarrow{\beta})$, which is compared to the theoretical ground state energy $E_{\text{groundstate}}$.}
    \label{fig:QAOA_Block}
\end{figure}
\begin{equation}
\label{demo_trottetization_eq}
    e^{X_{0}+Y_{1}+Z_{2}}=  \left( e^{-iX_{0}}\ast^{-iY_{1}}\ast^{-iZ_{2}} \right)  \left( e^{-i\frac{X_{0}}{2}}\ast^{-i\frac{Y_{1}}{2}}\ast^{-i\frac{Z_{2}}{2}} \right). 
\end{equation}
In a generic case, the Hamiltonian considered is from the cross-field. The algorithm divides it into two components  $H = H_{A} + H_{B}=H_{Ising}$.
With \eqref{hamilton_trotte_eq} and \eqref{Transversal_field_eq}.
\begin{equation}
\label{hamilton_trotte_eq}
    H_{A}=-A \left( t \right)  \left(  \sum _{i<j}^{}J_{ij}Z_{i}Z_{j}+ \sum _{i}^{}h_{i}Z_{i} \right). 
\end{equation}
\begin{equation}
\label{Transversal_field_eq}
    H_{B}=B \left( t \right)  \sum _{i}^{}h_{i}X_{i}.
\end{equation}
And execute the steps mentioned above. The graphical representation of these steps is that of the figure \eqref{fig:QAOA_Block} and following the processes defined by the figure \eqref{fig:QAOA_Steps}. 
The state obtained after players of the QAOA can be written as: 
\begin{equation}
\label{QAOA_eq}
    \vert \overrightarrow{ \gamma },\overrightarrow{ \beta } \rangle = \prod_{k=1}^{p}e^{-i \beta _{k} \left( H_{B}/J_{0} \right) }e^{-i \gamma _{k} \left( H_{A}/J_{0} \right) } \vert \psi _{0} \rangle. 
\end{equation}
Where  $\vert \psi _{0} \rangle$ is the initial state, where the evolution times $\beta _{k}$  and  $\gamma _{k}$  are variational parameters used in the k-th QAOA layer to minimise the final energy $E \left( \overrightarrow{ \gamma },\overrightarrow{ \beta } \right) =\langle \overrightarrow{ \gamma },\overrightarrow{ \beta } \vert  H \vert  \overrightarrow{ \gamma },\overrightarrow{ \beta } \rangle$ and where  $J_{0}$ is the average nearest neighbour coupling, and in the case of $B = 0$. 
For each  $p$ (interaction), represent the unusual optimal angles by  $\{  ( \beta ^{\ast \left( p \right) }, \gamma ^{\ast \left( p \right) } )  \}$ which we can also think of as a pair of angle curves, as a function of step-index $i$. As $p$ is varied, we may think of these minima as a set. And one can do several studies on this set, for example, the convergence study of variables  $p$  and $N$. This interesting article \cite{Pan19} reviews how to improve even the QAOA algorithm.
At this point, we just need to experiment and focus on our problem.
The next chapter will experiment and solve the Quantum Social Assistant Workers Scheduling Problem.

\section{Summary}
In this chapter, we have experimented with quantum computing in the Qiskit environment. We have experimented with variational techniques, variational algorithms like VQE, QAOA, and QUBO techniques. 
We tested the VQE as one of the best algorithms of this quantum era, and we have also verified the detailed operation of the QAOA, which is the bet of the gate-based computers that work as an annealing algorithm. As stated in the QAOA, one of the advantages is that we can increase the precision arbitrarily. In contrast, QA will only find the solution with probability one when the time goes to infinity, which is impractical. In addition, if $T$ is too long, it is possible not to find the outcome as the probability is not monotonic.

Now we have the basis to contribute to the scientific community.

\part{Contributions}

%\chapter{My main contributions}

\newpage

\chapter{Research Design}\label{sec:9}
\section{Introduction}\label{sec:9_}

\begin{figure}[h!]
    \centering
    \includegraphics[width=0.85\textwidth]{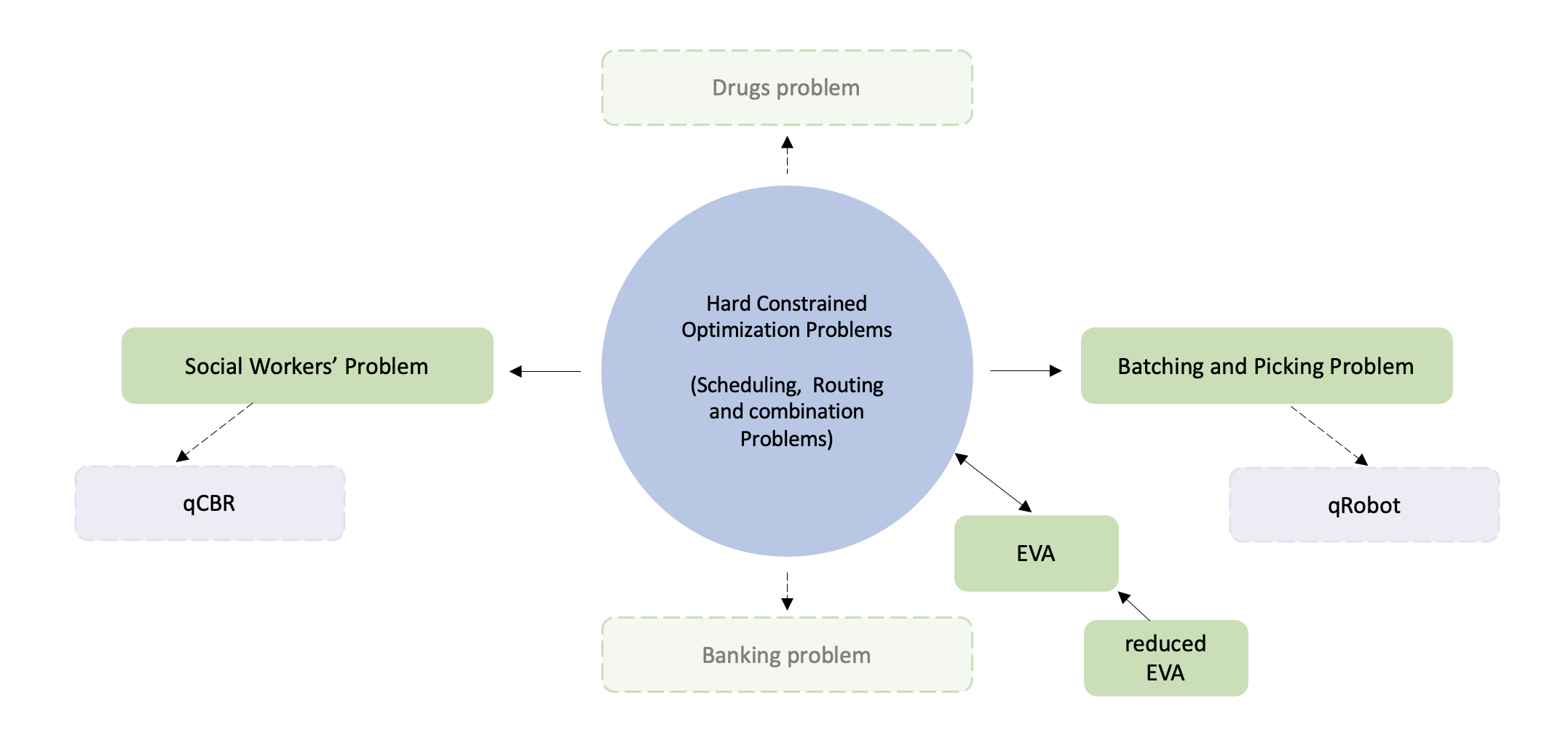}
    \caption{My research's path over these 3 years}
    \label{fig:Phd_overview}
\end{figure}

In this section, we define our research scenario, which is illustrated by figure \eqref{fig:Phd_overview}, but, first, we will describe the problem and the necessary tools.

The figure \eqref{fig:Phd_overview} summarises the work scenario on the tasks proposed for the thesis. First, we define a problem that helps us answer our main hypothesis. Then, we ask whether it is possible to design a quantum algorithm to solve combinatorial optimisation problems with hard constraints.

The designed task is a Social Worker Problem (SWP) project, which combines routing, planning, and combinatorial tasks. From this point, we will solve it using two different techniques: on the one hand, the top-down approach, which means that we know how to write down our objective function and then solve it, and on the other hand,  the machine learning technique, by finding a model that generalises our problem; this model is called, quantum Case-Based Reasoning (qCBR). The qCBR is an artificial intelligence approach to problem-solving with a good record of success. The main idea of qCBR is to interpret the \textbf{statement of the problem} as an \textbf{input object}, and the \textbf{solution to the problem} as an output (\textbf{label}).

After solving the SWP, we propose and solve the Batching and Picking Problem as the generalisation of the SWP called qRobot.

Normally, all cloud services have two pricing components when using a quantum computer or quantum processing unit (QPU); on Amazon Braket: \textbf{a per-shot fee} and \textbf{a per-circuit fee}. To improve the computation time, reduce the economic cost and leap in quality in chemistry, we have proposed a quantum Exponential Value Approximation algorithm (EVA) and reduced-EVA \cite{alonsolinaje2021eva}; the latter is a new proposal for calculating the expected value to improve the flagship quantum algorithm VQE in quantum cloud computing. 

Therefore, we will continue working to improve the routing and optimisation algorithms for another sector such as banking in this era of very few useful qubits, pending the lecture of this doctoral thesis.

\section{Proposed problem}
\begin{figure}[h!]
    \centering
    \includegraphics[width=.7\textwidth]{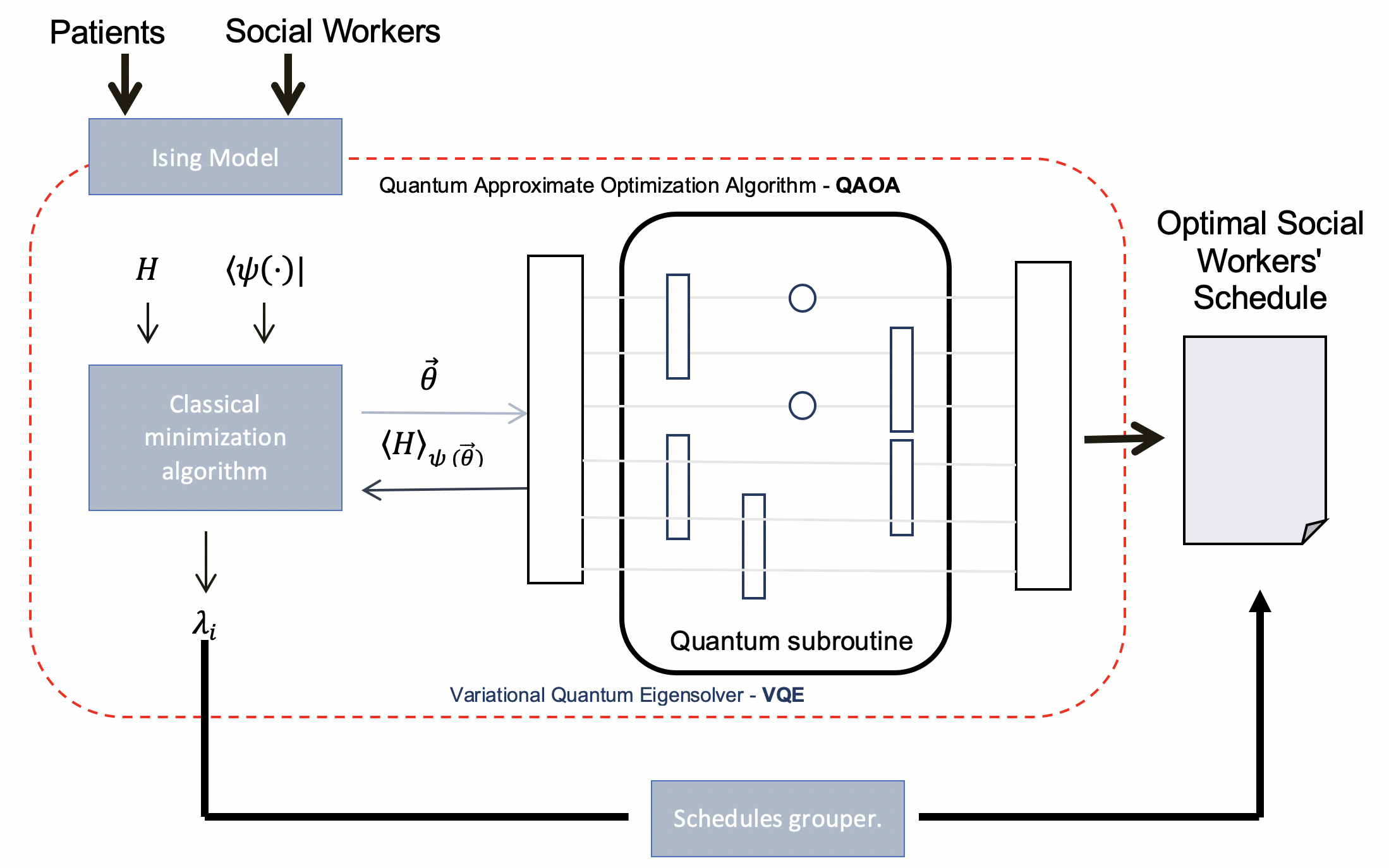}
    \caption{Using the Variational-Quantum-Eigensolver (VQE) as Quantum Machine Learning (QML) to creates an Intelligent social workers schedule problem solver.}
    \label{fig:VQE_Bloc}
\end{figure}

Let  $n$ be the number of patients (users) and consider a weekly calendar of visits for each of them. Our objective is to find an optimal schedule that minimises the travelled distance according to the time windows. Figure \eqref{fig:VQE_Bloc} illustrates the resolution of SWP with VQE. In the end, we assign the social workers to the group with the resultant optimal hours.

In our case study, the daily schedule table \eqref{tab:Sample_SWP_Schedu}, is set at 8 hours, and the distance between patients is at least 15 minutes. Let us remind that our main motivation is to find a feasible formulation of the SWP that use the minimum numbers of qubits and solve this proposed problem in this NISQ era.

%%%%%%%%%%%%%%%%%%%% Table No: 3 starts here %%%%%%%%%%%%%%%%%%%%
\begin{table}[h!]
\centering
\begin{tabular}{|c|c|c|c|c|c|}
 \hline
 \multicolumn{6}{|c|}{A sample of the social workers' schedule}     \\
 \hline
      Time     & Monday   & Tuesday& Wednesday& Thursday & Friday   \\
 \hline
9:00 – 10:00   &  $U_1$   & $U_1$ &  $U_1$    & $U_1$    & $U_1$    \\
9:30 – 10:30   &  $U_4$   &       &           &          &          \\
10:15 – 11:15  &          & $U_4$ &           &          &          \\
11:30 – 12:30  &          &       &           &          & $U_4U_5$ \\
11:45 – 12:45  &  $U_5$   &       &           &          &          \\
12:00 – 13:00  &  $U_2$   &       &           &          &           \\
14:45 – 15:45  &          & $U_2$ &           &          &           \\
15:00 – 16:00  &          &       &  $U_3$    &          &           \\
15:15 – 16:15  & $U_3$    &       &           &          &           \\
15:45 – 16:45  &          &       &  $U_3$    &          &           \\
16:00 – 17:00  &          &       &  $U_2$    &          &           \\
16:30 – 17:30  &          & $U_3$ &           &          &           \\
17:00 – 18:00  &          &       &           &          & $U_3$    \\
 \hline
\end{tabular}
\caption{Weekly patient care schedules. It is not necessary to visit all registered patients every day of the week, as there can also be more than one patient assigned simultaneously on the same day. Where $U_{1}$ to $U_{2}$ are the patients (users) and equal to the variable $i$ or $j$ of the mathematical formulation}
\label{tab:Sample_SWP_Schedu}
\end{table}
%%%%%%%%%%%%%%%%%%%% Table No: 3 ends here %%%%%%%%%%%%%%%%%%%%

\vspace{\baselineskip}
\begin{itemize}
	\item A set of social workers ($N_{1},~N_{2},~N_{3},  \ldots, N_{N}$ ).

	\item A set of patients ($P_{1},~P_{2},~P_{3},  \ldots, P_{N}$).

	\item A set of visits ($ U_{1},~U_{2},~U_{3},  \ldots, U_{N}$).

	\subitem each visit is linked to a patient: a patient can have multiple appointments on a day.

    \subitem for each visit, we know the start time and duration.

	\item The social workers can work at most 8 hours per day.

	\item We know the cost of travelling between each pair of patients. The cost can be seen as a function of travel time and distance.
\end{itemize}

The objective is the following:
\begin{itemize}
	\item Find a schedule where each visit is assigned to a social worker

	\item We minimise the travel cost while also respecting that a social worker does not work more than 8 hours per day.
\end{itemize}

%Taking into account, the advances on the speedup of the \textit {Quantum Approximate Optimization Algorithm} (QAOA)\cite{GGG19} \cite{Joh18}\cite{Alb13}\cite{GPa20}\cite{Pan19} in the NISQ era, we can affirm that the experiment's scenario and its design are representatives. Because the only difficulty that can be added here, for this combinatorial optimisation problem, is the number of patients and some restrictions. These difficulties have to do directly with the computational cost (not with the formulation/algorithm) where quantum computing is called to be more efficient \cite{Joh18}. Hoping to access a more powerful computer, we limited our test to a 20 qubits computer, the most potent public quantum computer.

In this work, our upper bound is given by the number of the qubits from the gate-based quantum computer; thus, 20 qubits. We need a strategy to tackle this problem. But before this, in the next section, we will define the experimentation tools we need.

We need to analyse and choose a programming environment to solve our problem. Next, we will focus on the tools to carry out our experiments.\par

\section{Experimentation tools}
\subsection{Programming environments }
The race to leadership in quantum computing involves determining standards. However, everything currently points to the NISQ era being quite stable than public opinion imagine. Based on the work done by Mark Fingerhuth, Toma'sˇ Babej and Peter Wittek, \textit{Open source software in quantum computing} \cite{Pet14,Mar18,Mar181}, we can observe that several players are working hard in high-level programming environments. Furthermore, a change in philosophy was also seen in mid-2019 with companies specialising in compiler writing and different collaborations between technology giants such as IBM, Google, Microsoft and AWS-Braket with high-tech startups such as Xanadu. All this leads us to think that we will see many more applications in quantum computing and, above all, quantum datacenters in a few months.

\begin{figure}[h!]
    \centering
    \includegraphics[width=\textwidth]{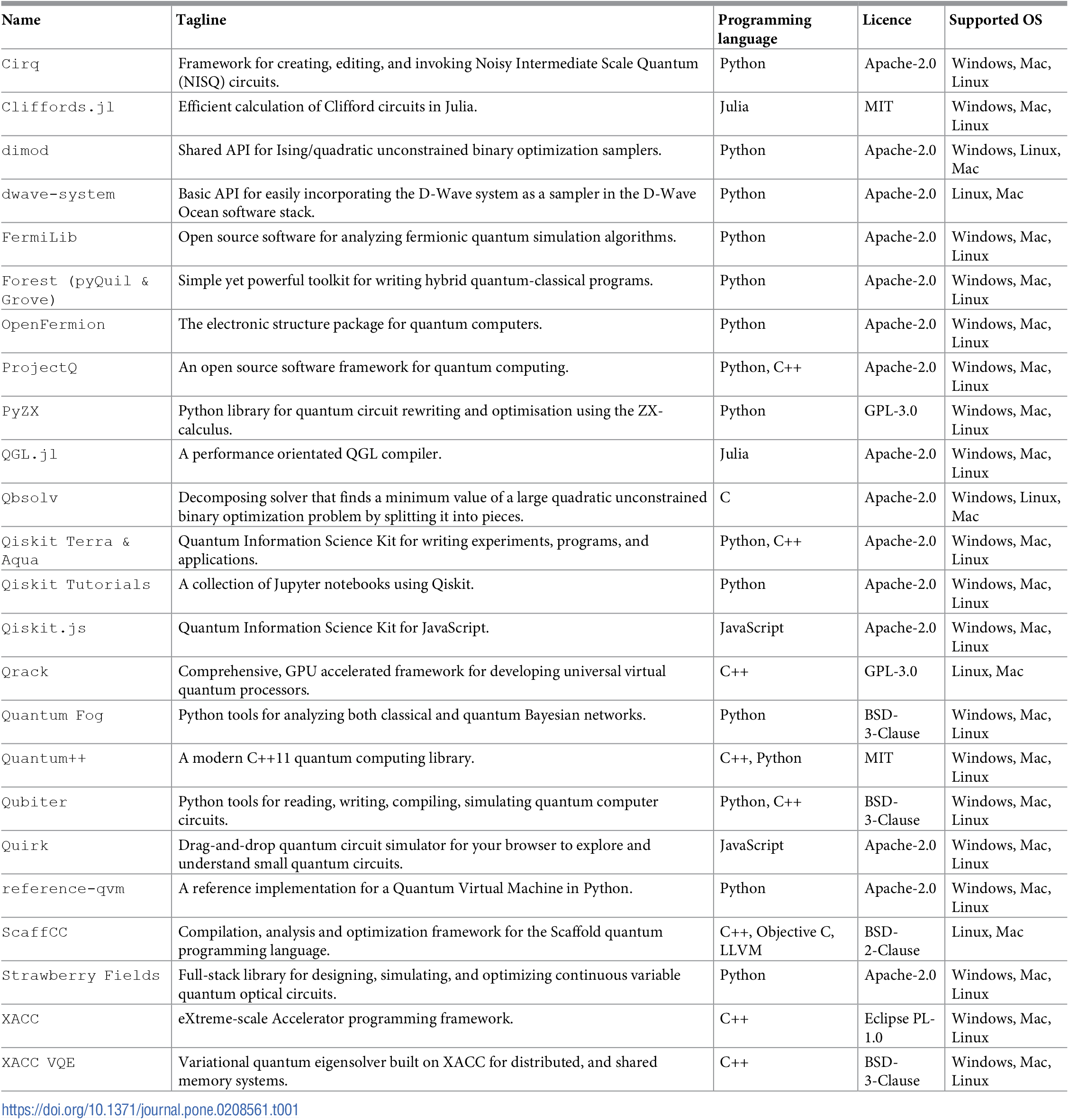}
    \caption{Overview of all published quantum computing programming environments \cite{Mar181}.}
    \label{fig:overview_QC_environments}
\end{figure}

\begin{figure}[h!]
    \centering
    \includegraphics[width=\textwidth]{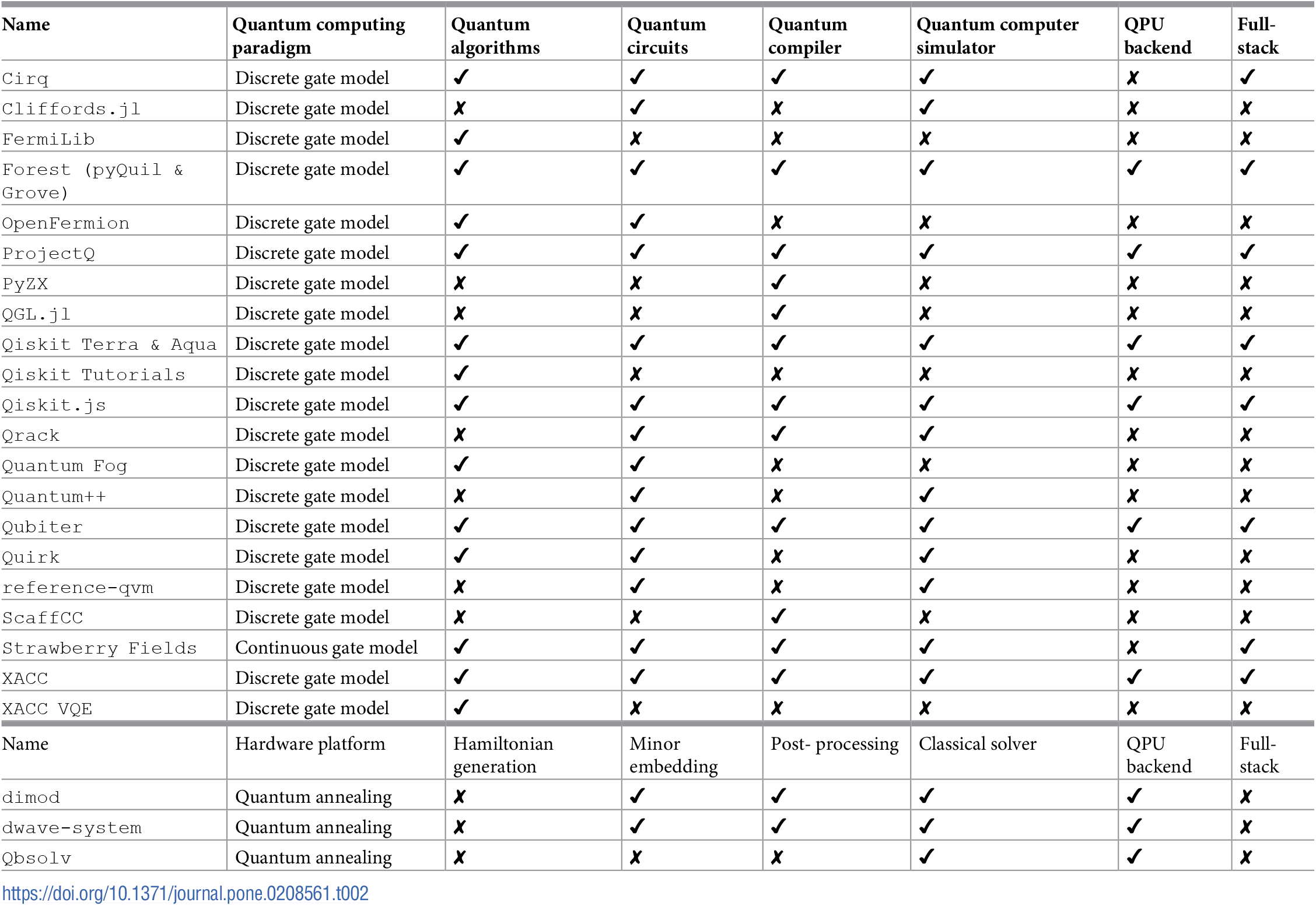}
    \caption{Detailed characteristics of all published quantum computing programming environments. Data, August 2018 \cite{Mar181}.}
    \label{fig:Feature_environment_QC}
\end{figure}

\begin{figure}[h!]
    \centering
    \includegraphics[width=\textwidth]{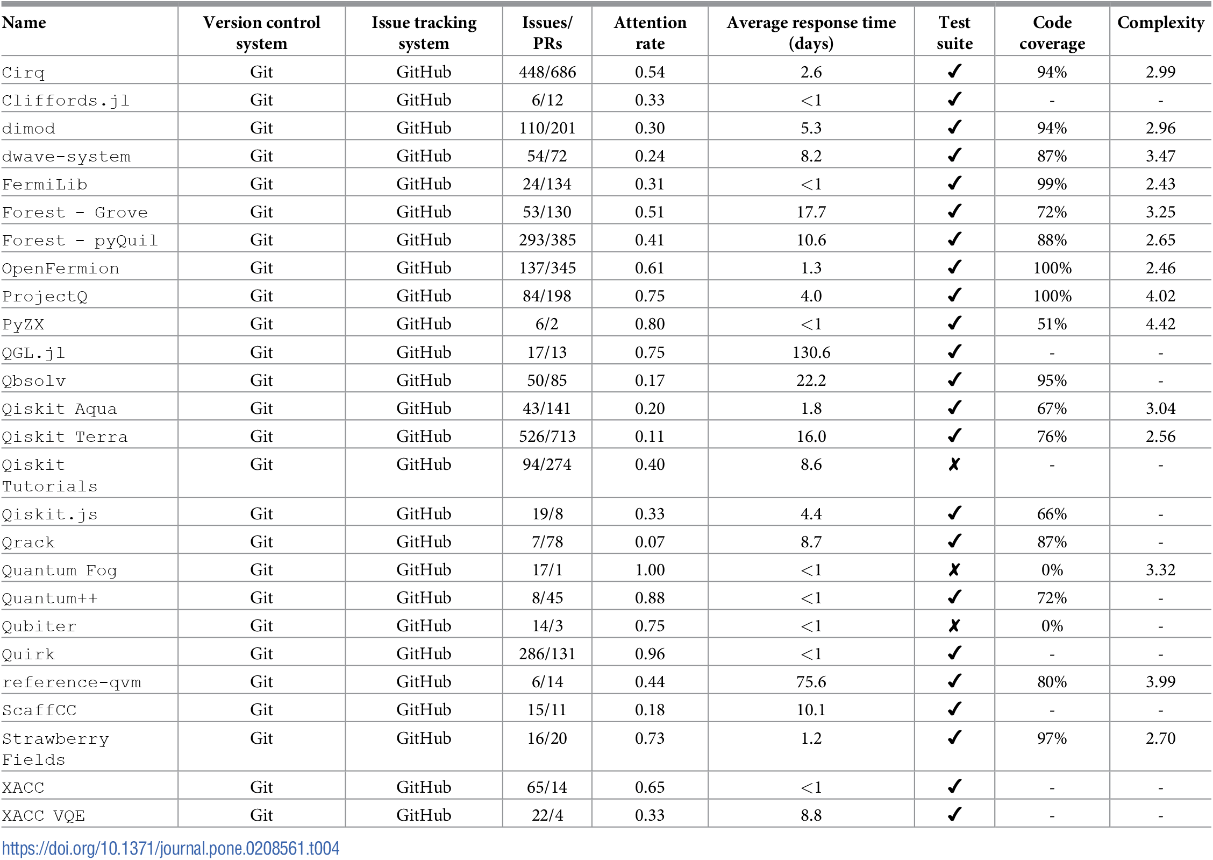}
    \caption{Evaluation results for the static analysis of each project and its source code \cite{Mar181}.}
    \label{fig:Evualuation_source_code}
\end{figure}

Next, we will analyse the frameworks (Figure \eqref{fig:overview_QC_environments}, \eqref{fig:Feature_environment_QC} and \eqref{fig:Evualuation_source_code} ) and decide on an environment for our experimentation.

\subsubsection{PennyLane:}
\begin{figure}[h!]
    \centering
    \includegraphics[width=0.7\textwidth]{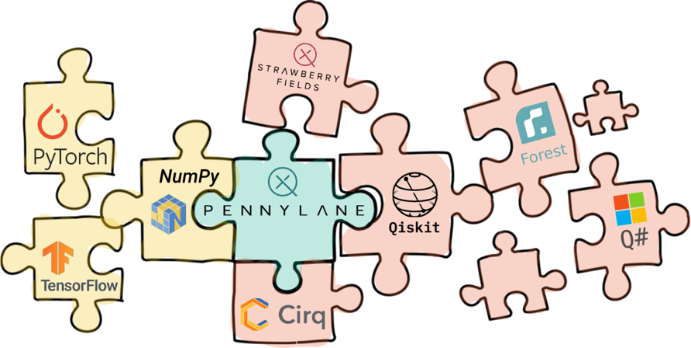}
    \caption{Xanadu bets with its PennyLane platform \cite{bergholm2018pennylane}.}
    \label{fig:Pennylane_Platform}
\end{figure}

\textit{PennyLane}\footnote{ https://github.com/XanaduAI/pennylane/} \cite{Vil20} (See the Fig. \eqref{fig:Pennylane_Platform}) is a \textit{Python 3-based programming environment for optimisation and machine learning in quantum and hybrid (quantum-classical) computing. The library provides a unified architecture for quantum computing devices of the era we are in and supports qubit and continuous variable paradigms. The main feature of PennyLane is the capability to calculate gradients of variational quantum circuits that are compatible with classical techniques such as backpropagation.}

PennyLane extends the standard machine differentiation algorithms in optimisation and machine learning to include quantum and hybrid calculations. A plugin system makes the framework compatible with any simulator or gate-based quantum hardware. For example, Xanadu, through its framework, provides plugins for Strawberry Fields\cite{Vil20}, Rigetti Forest\footnote{ http://docs.rigetti.com/en/stable/ }, Qiskit, Cirq\footnote{ https://cirq.readthedocs.io/en/stable/ } and ProjectQ, allowing PennyLane optimisations to run on publicly accessible quantum devices provided by Rigetti and IBMQ. Furthermore, PennyLane can be used as a classic environment thanks to its interaction with accelerated machine learning libraries such as TensorFlow, PyTorch and autograd. Also, PennyLane can be used for optimisation problem solving using VQE, QAOA, QML or any quantum formulation.
The following is a list of the plugins already developed to make PennyLane universal.
\textit{PennyLane-SF}: Supports integration with Strawberry Fields, a full-stack Python library to simulate photon quantum computing.
\textit{PennyLane-qiskit }: Supports integration with Qiskit, an open-source quantum computing environment from IBM that provides support for Qiskit Aer quantum simulator devices and IBMQ hardware devices.
\textit{PennyLane-cirq}: Supports integration with Cirq, an open-source Google quantum computing environment.
\textit{PennyLane-Forest}: Supports integration with PyQuil, Rigetti Forest SDK and Rigetti QCS, an open-source quantum computing environment from Rigetti that provides device support with \textit{Quantum Virtual Machine} (QVM) and \textit{Quantum Processing Units} (QPU) hardware.
\textit{PennyLane-Qsharp}: Supports integration with Microsoft Quantum Development Kit, a quantum computing framework that uses the Q$\#$  quantum programming language.

\subsubsection{Cirq}
\begin{figure}[h!]
    \centering
    \includegraphics[width=0.20\textwidth]{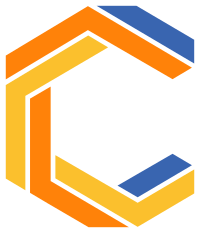}
    \caption{Google bets with its Cirq platform.}
    \label{fig:Cirq_Platform}
\end{figure}

\textit{Cirq} is Google's commitment (Fig. \eqref{fig:Cirq_Platform}) as \textit{a framework for the programming and simulation of quantum computers. Cirq is a software library for writing, manipulating and optimising quantum circuits and then running them on and simulating quantum computers.}\\ 

Cirq takes into account the NISQ era to get the most out of the hardware, analysing all the details of it before executing the quantum circuit or not.
The current version of Cirq is alpha, ready to test\footnote{ https://cirq.readthedocs.io/en/stable/ }. 

\subsubsection{Qsharp}
\begin{figure}[h!]
    \centering
    \includegraphics[width=0.20\textwidth]{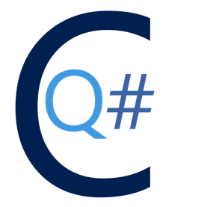}
    \caption{Bet on Microsoft with its Qsharp platform \cite{qsharp_}.}
    \label{fig:Micr_Platform1}
\end{figure}

\begin{figure}[h!]
    \centering
    \includegraphics[width=0.7\textwidth]{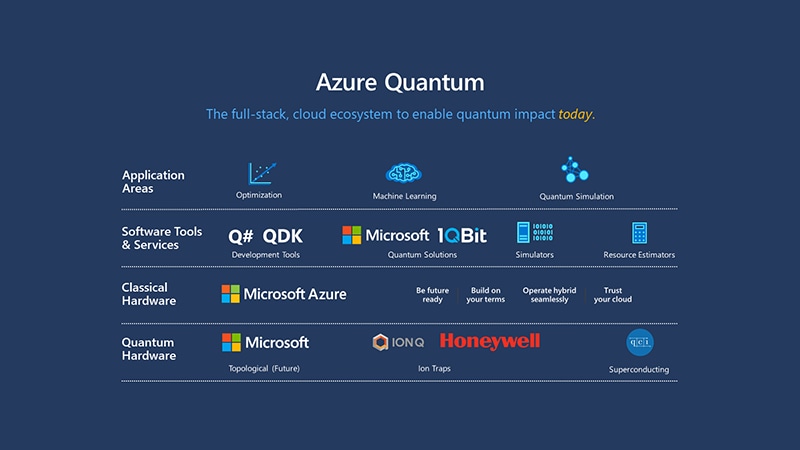}
    \caption{Azure Quantum Computing \cite{Azurre_MS}.}
    \label{fig:Micr_Platform_Az}
\end{figure}

\textit{Qsharp}\footnote{ https://docs.microsoft.com/es-es/quantum/overview/what-is-qsharp-and-qdk } is \textit{Microsoft's push for the bet on its quantum hardware (Fig. \eqref{fig:Micr_Platform1} and \eqref{fig:Micr_Platform_Az}). Q$\#$  is a new high-level programming language focused on the quantum.}

Its features rich integration with \textit{Visual Studio}, \textit{Visual Studio Code} and interoperability with the Python programming language. Enterprise-grade development tools provide the fastest path to quantum programming on Windows, macOS, or Linux.

Microsoft as IBM, bets for the cloud solution in the short term. In the case of Microsoft, its bet is \textit{Azure Quantum}. \\
\textit{Azure Quantum} is a \textit{cloud platform for quantum computing where developers use the QDK to write Q$\#$  programs and run on quantum hardware or formulate problems to run on quantum-inspired solvers.} \\
%%%%%%%%%%%%%%%%%%%% Table No: 4 starts here %%%%%%%%%%%%%%%%%%%%
\begin{figure}[h!]
    \centering
    \includegraphics[width=\textwidth]{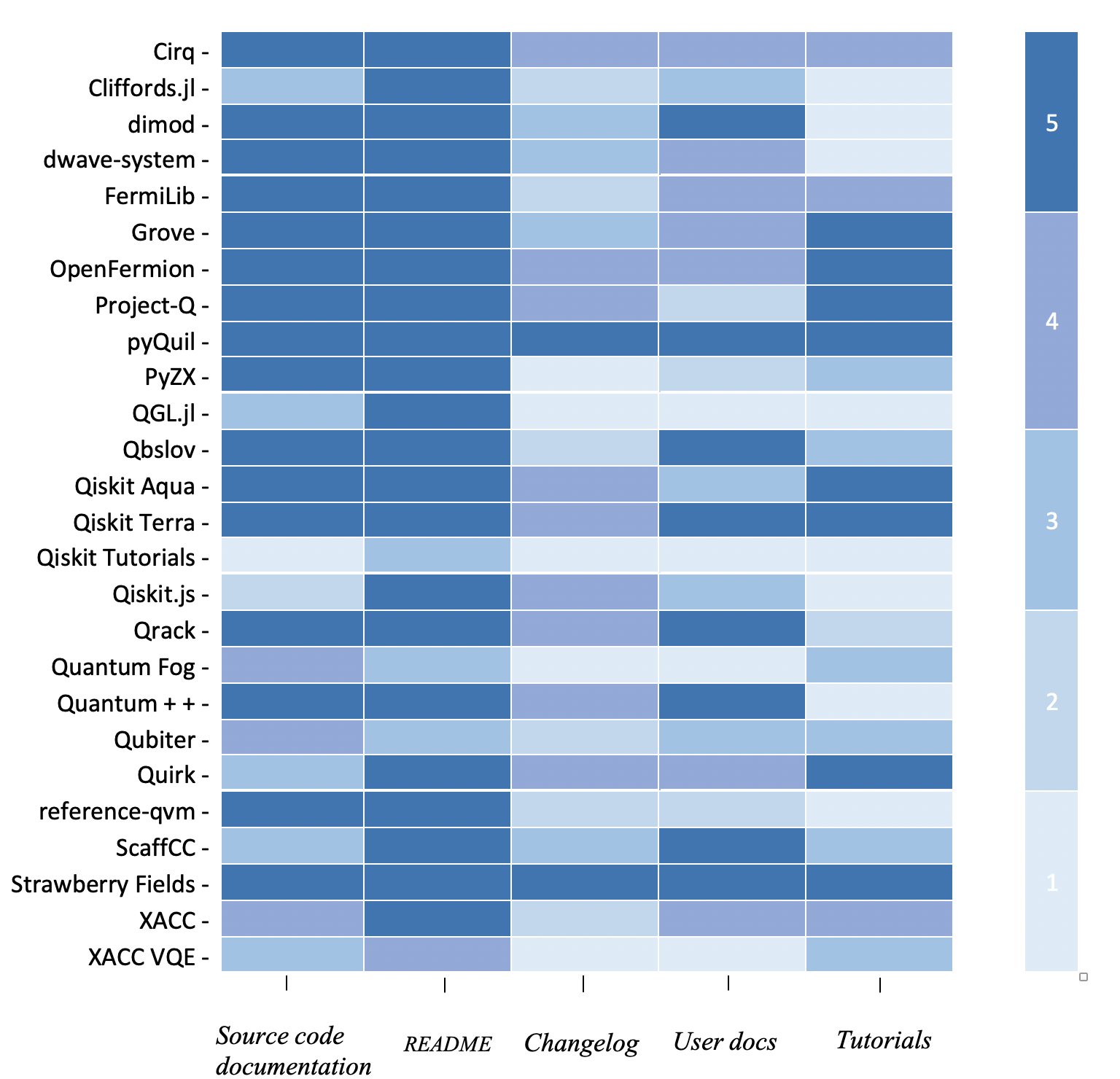}
    \caption{The heat map on the results of analysis and documentation.}
    \label{fig:The_Heat_Documentation}
\end{figure}

%%%%%%%%%%%%%%%%%%%% Table No: 4 ends here %%%%%%%%%%%%%%%%%%%%

The \textit{heatmap}, figure \eqref{fig:The_Heat_Documentation}, displays the evaluation results from source code documentation, README files, changelogs, user documentation, and tutorials on a scale of 1 (bad) to 5 (good). The data was obtained in August 2018 \cite{Mar181}.
\begin{figure}[h!]
    \centering
    \includegraphics[width=\textwidth]{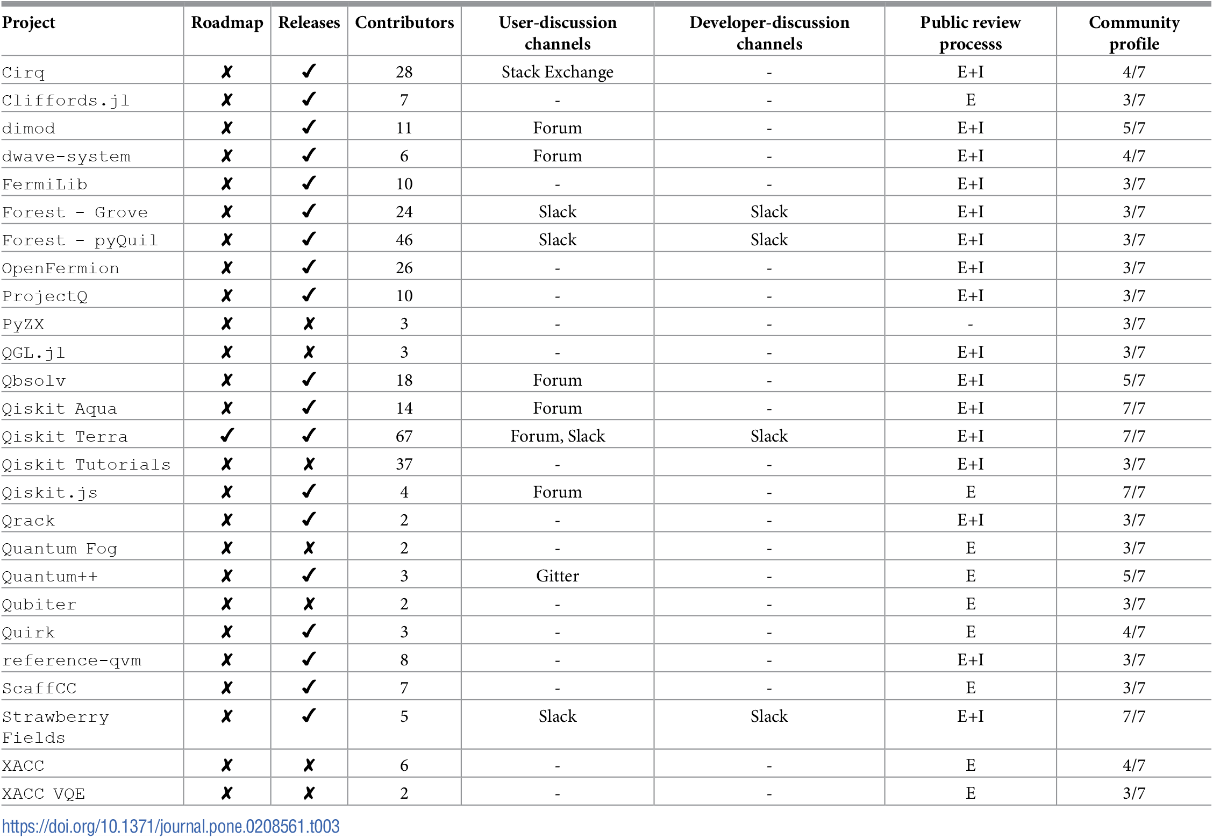}
    \caption{Results of the evaluation for the community analysis. For each project, we indicate whether there is a public development roadmap and whether the software is published in the form of releases. Besides, we report the GitHub community profile score, the total number of contributors, the type of discussion channel focused on the user and the developer, and the type of public code review process, specifically if it applies to internal (I) external (E) taxpayers. Data were obtained in August 2018 \cite{Mar181}.}
    \label{fig:Development_Roadmap}
\end{figure}

\textit{Qiskit} is  \textit{the IBM programming environment consists of Terra\footnote{ https://qiskit.org/terra } (central compiler and libraries for quantum programming), Aer\footnote{ https://qiskit.org/aer } (noise modelling and simulators without noise), Ignis (error characterisation and QEC) and Aqua\footnote{ https://qiskit.org/aqua } (Applications).} \\

\begin{figure}[h!]
    \centering
    \includegraphics[width=0.5\textwidth]{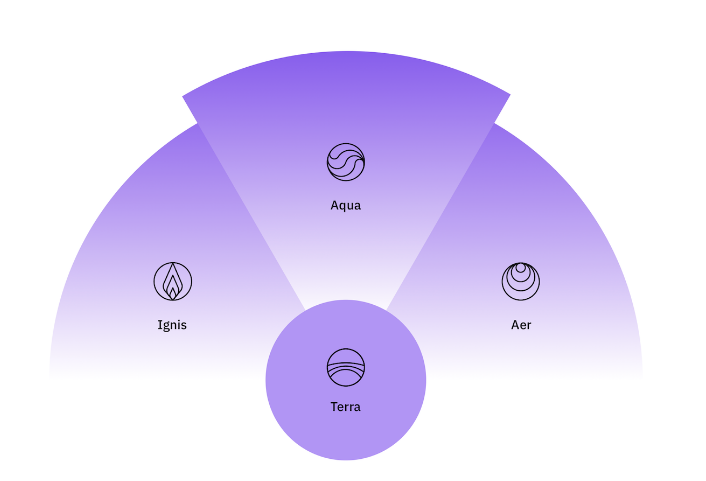}
    \caption{Components of Qiskit’s framework \cite{Qis21}.}
    \label{fig:Qiskit_framework}
\end{figure}

Next, I will go into more detail with each of the components to understand it better, as is shown in figure \eqref{fig:Qiskit_framework}.
\textit{Terra} provides the fundamental roots for our software stack. Within Terra, there is a set of tools for composing quantum programs at the circuit and pulse levels, to optimise the constraints of a particular quantum physical processor, and to manage batch execution of experiments on remote access backends.

\textit{Aqua}, the "water" element, is the element of life. This library will solve everything related to Chemistry, Optimisation or AI. Aqua is accessible to experts in chemistry, optimisation, or artificial intelligence domains who want to explore the benefits of using quantum computers as accelerators for specific computational tasks without worrying about how to translate the problem into the language of quantum machines.

\textit{Ignis}, the "fire" element, is dedicated to fighting noise and mistakes to forge a new path. While Aer, the item "air", permeates all the aspects of Qiskit. Simulators, Emulators, and Debuggers.

\subsubsection{IBM Quantum Experience}
\begin{figure}[h!]
    \centering
    \includegraphics[width=0.8\textwidth]{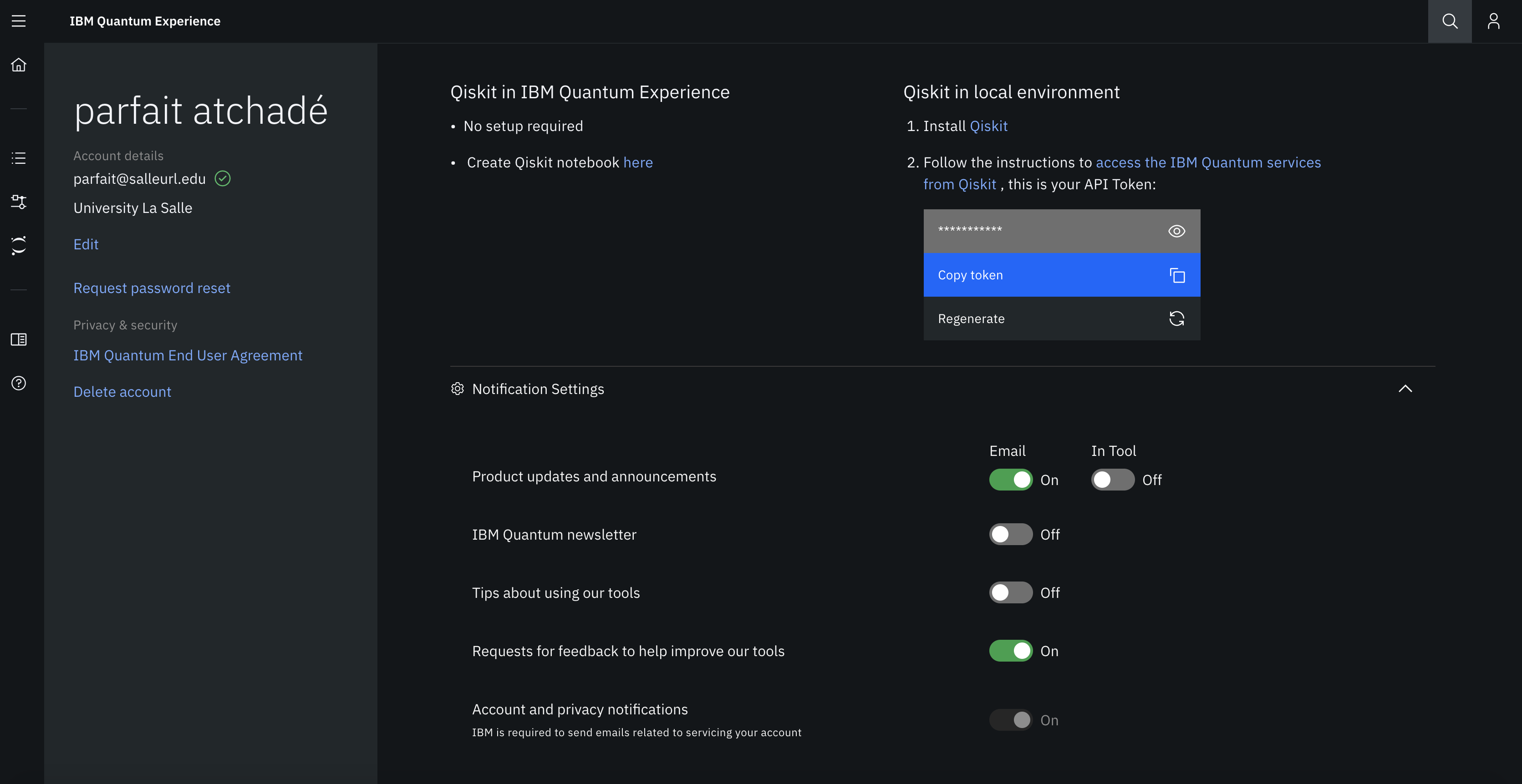}
    \caption{IBMQ User Profile.}
    \label{fig:IBMQ_1}
\end{figure}

\begin{figure}[h!]
    \centering
    \includegraphics[width=0.8\textwidth]{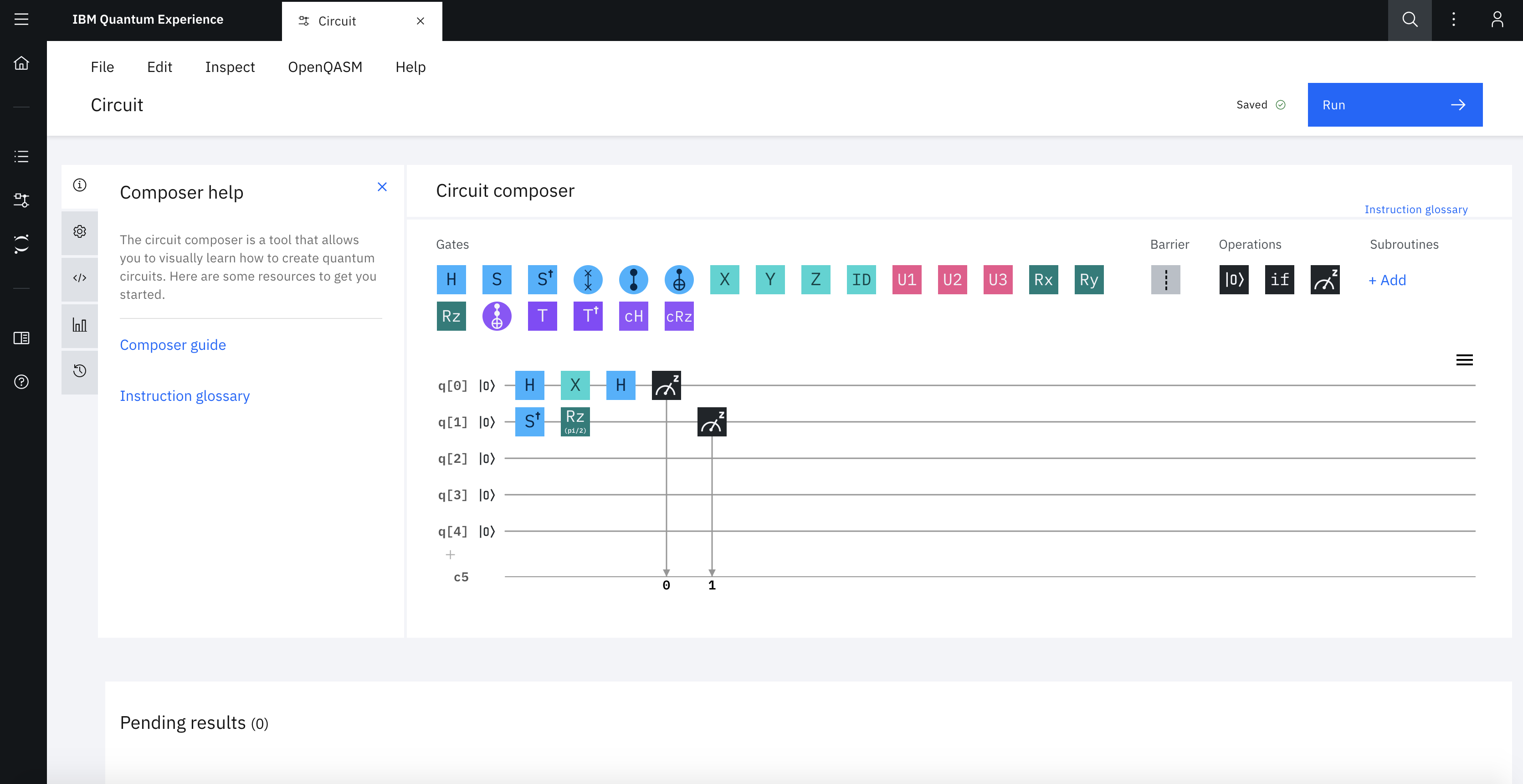}
    \caption{IBMQ Circuit Composer.}
    \label{fig:IBMQ_2}
\end{figure}

\begin{figure}[h!]
    \centering
    \includegraphics[width=0.8\textwidth]{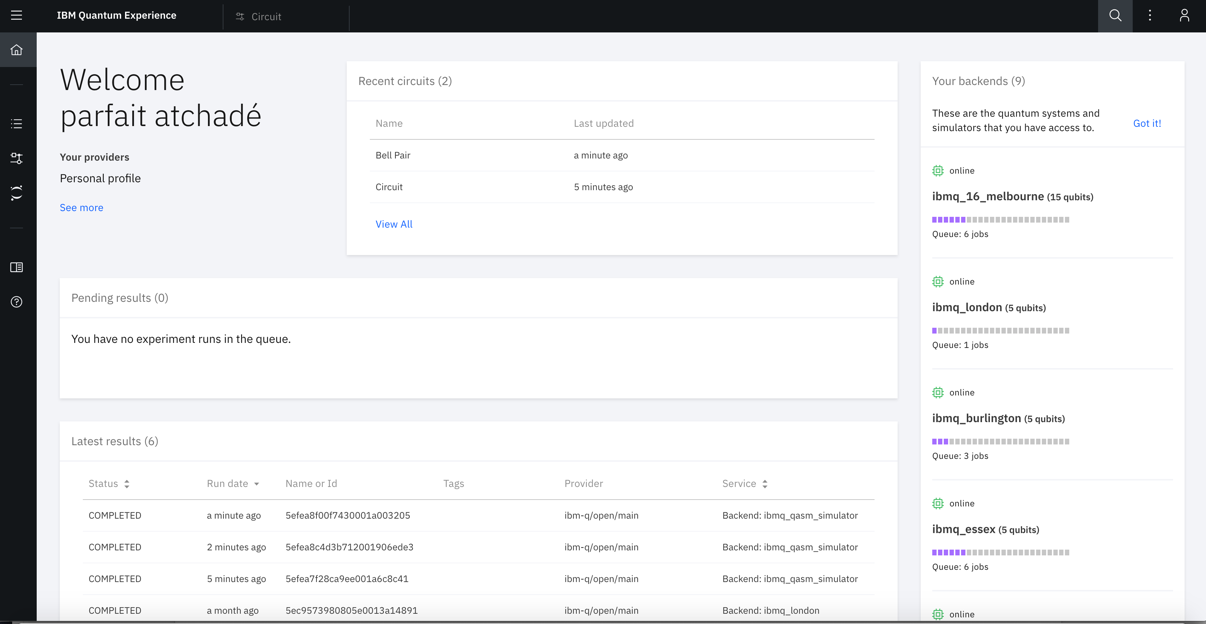}
    \caption{IBMQ Laboratory Profile.}
    \label{fig:IBMQ_3}
\end{figure}

\begin{figure}[h!]
    \centering
    \includegraphics[width=0.8\textwidth]{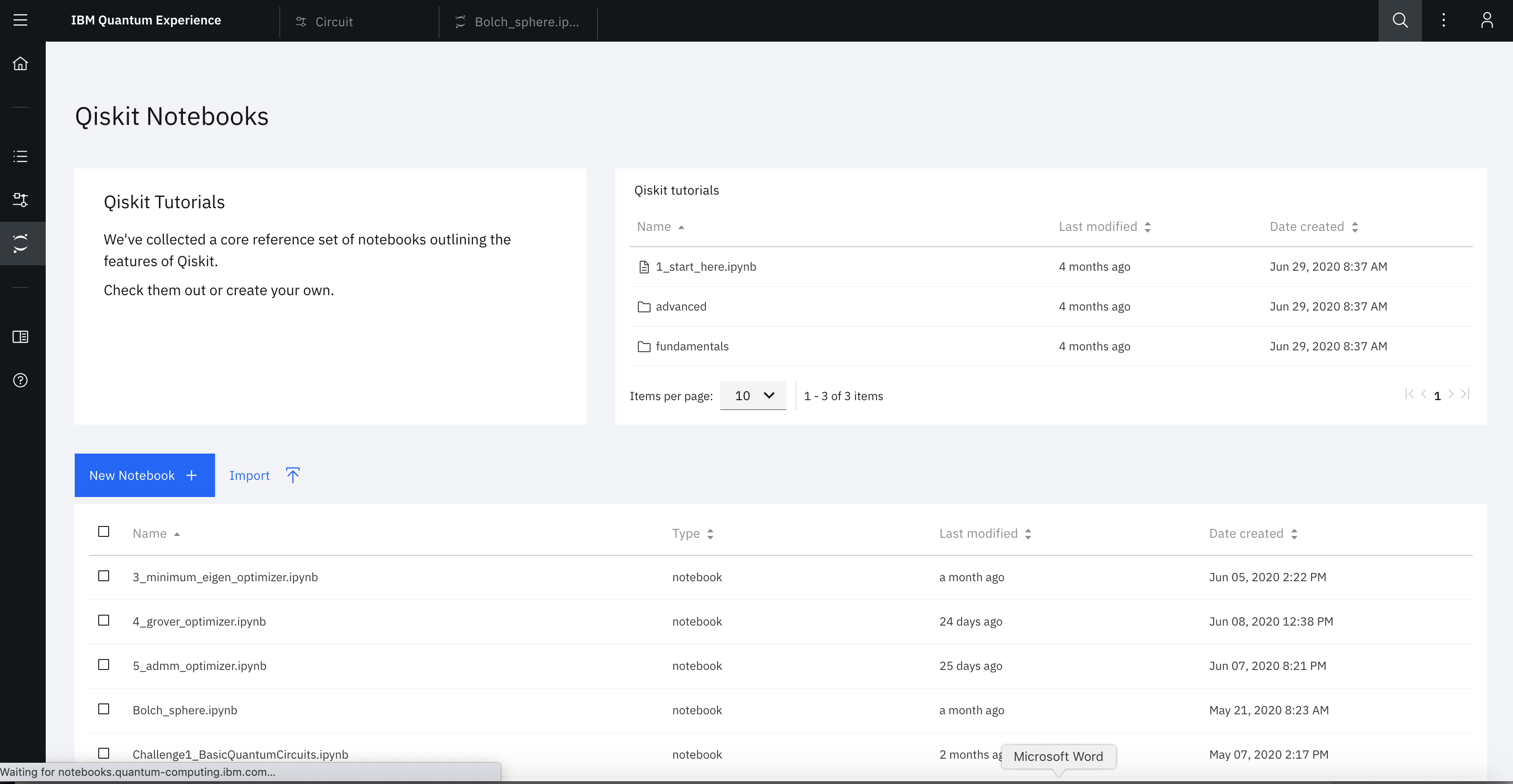}
    \caption{IBMQ Jupyter Notebook Files view.}
    \label{fig:IBMQ_4}
\end{figure}

\begin{figure}[h!]
    \centering
    \includegraphics[width=0.8\textwidth]{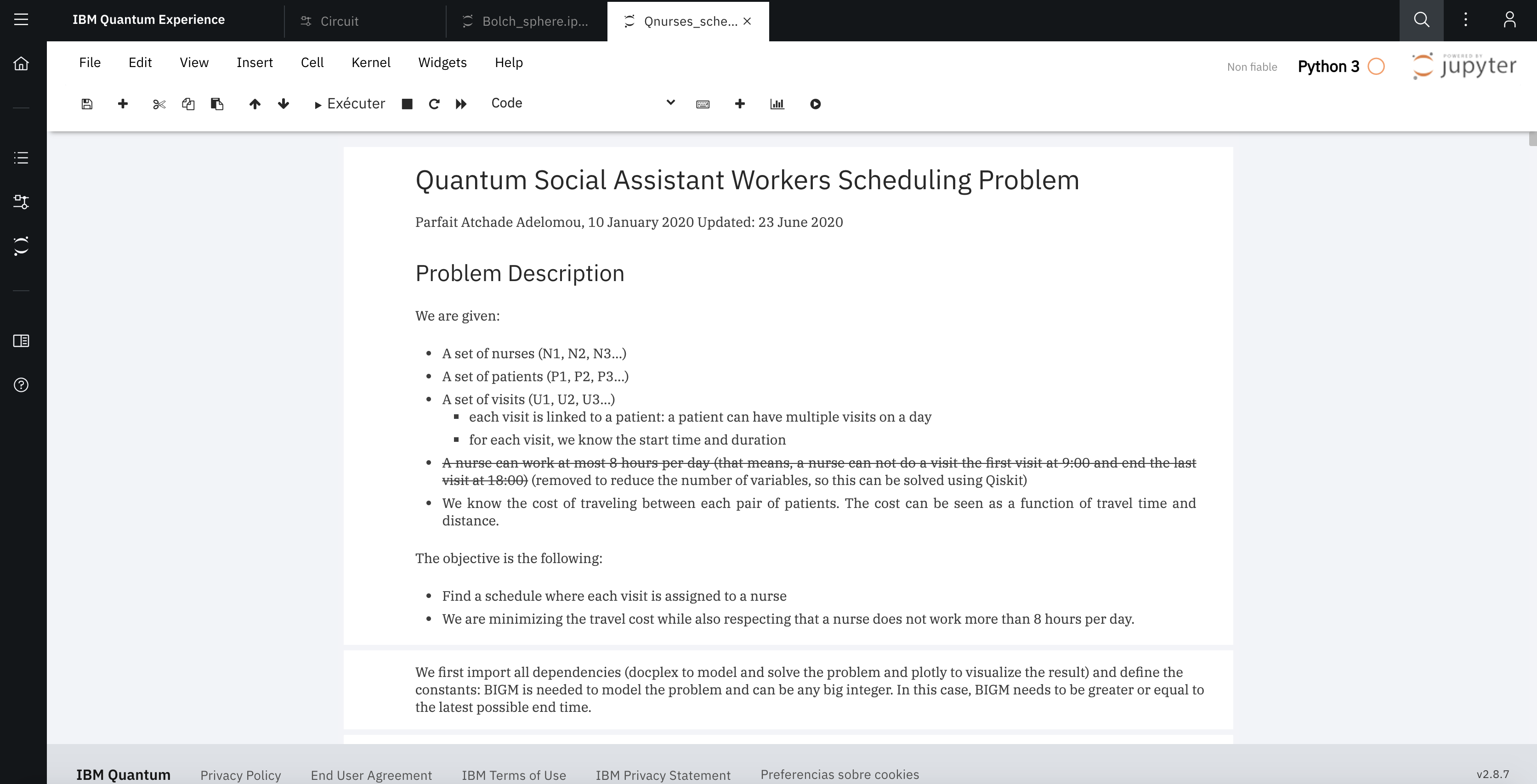}
    \caption{IBMQ Jupyter Notebook view.}
    \label{fig:IBMQ_5}
\end{figure}

\begin{figure}[h!]
    \centering
    \includegraphics[width=0.8\textwidth]{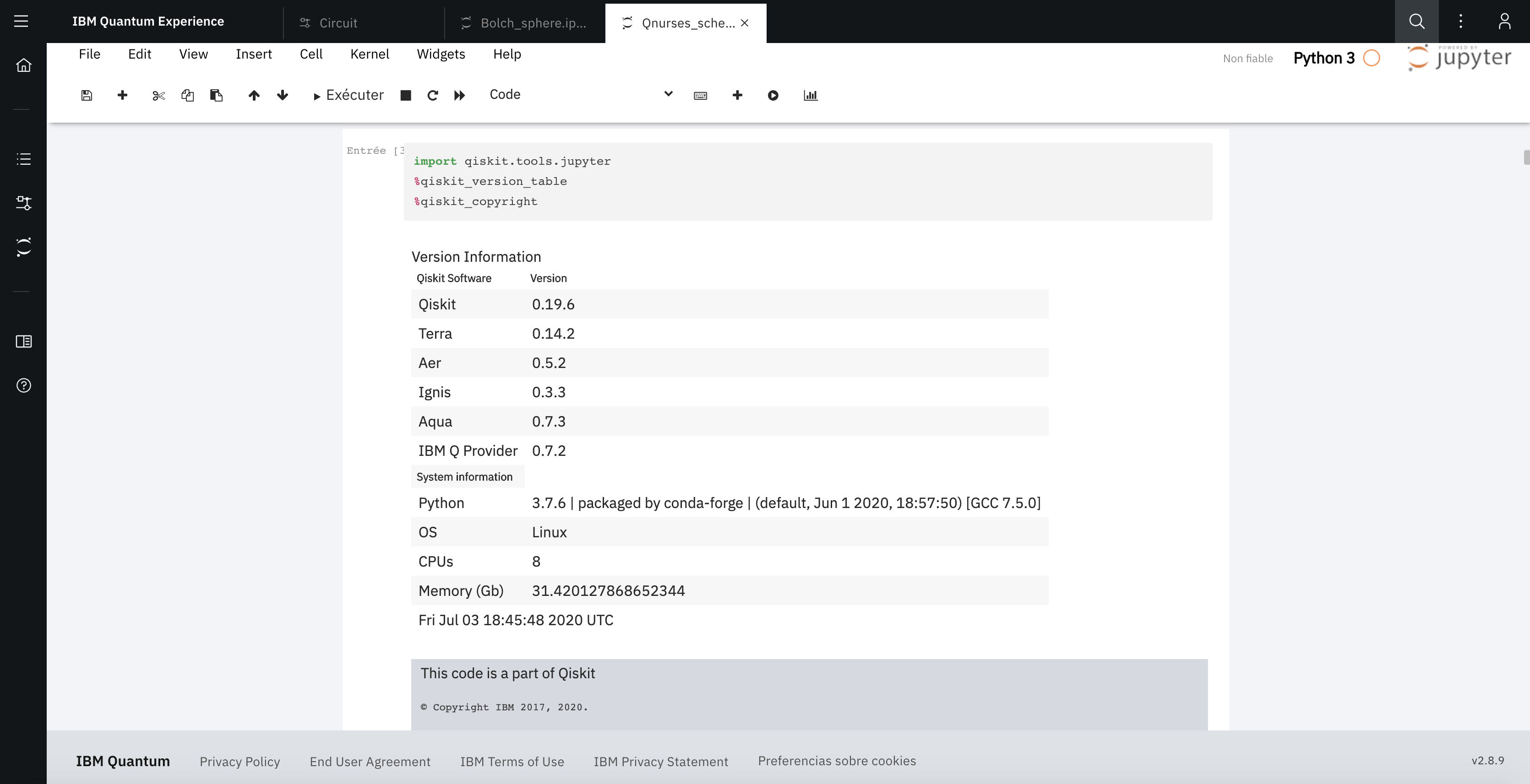}
    \caption{From IBMQ's Jupyter Notebook, the Qiskit version.}
    \label{fig:IBMQ_6}
\end{figure}

\begin{figure}[h!]
    \centering
    \includegraphics[width=0.8\textwidth]{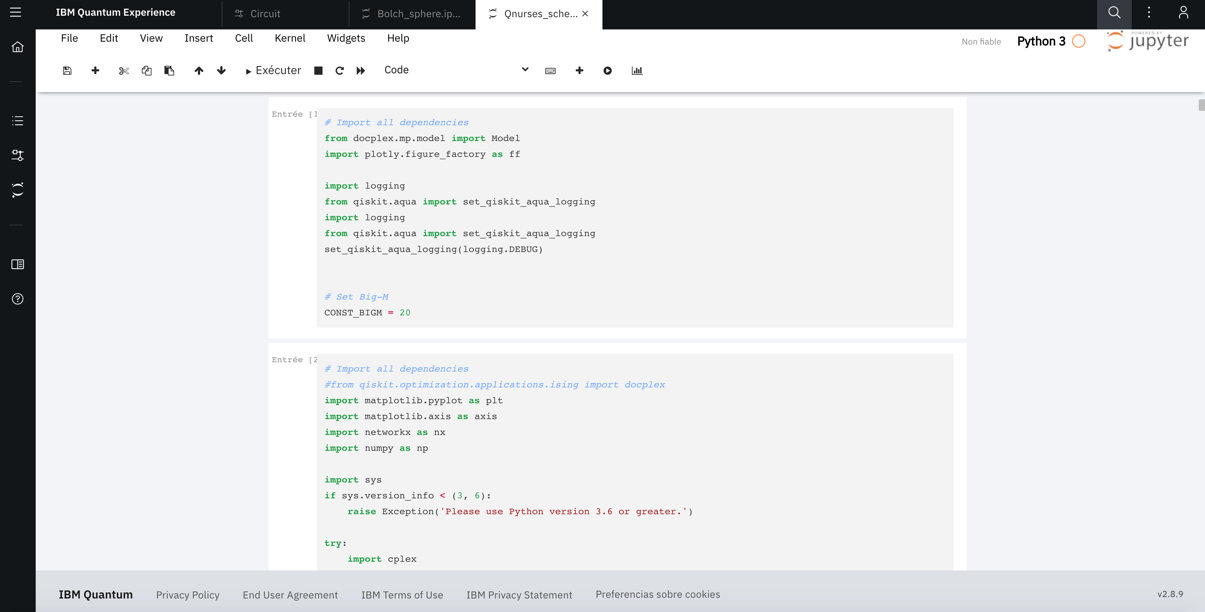}
    \caption{From IBMQ's a code view from Jupyter Notebook.}
    \label{fig:IBMQ_7}
\end{figure}

\textit{IBM Quantum Experience}\footnote{ https://quantum-computing.ibm.com/ } (IBMQ) (Figures \eqref{fig:IBMQ_1} to \eqref{fig:IBMQ_7}) is \textit{an online platform that provides users of the general public with access to a prototype suite of IBM quantum processors via the cloud.}\\

An online Internet forum to discuss relevant quantum computing issues, a set of tutorials on how to program IBMQ devices and other educational materials on quantum computing.


It is of utmost importance to perceive and understand the strategy of IBM when it describes the Qiskit roadmap as a platform and business bet. This strategy determines how to propose and design the algorithm for the end-user. In just two years of research that I have been following the firm, IBM is already beginning to encapsulate many libraries as a black box to allow users to work more at a high level. This has enormous benefits for the scientific community since quantum computing developers will increase exponentially. The complexity of understanding quantum computing fundamentals will be reduced notably; the part of the quantum mechanics and, thus, would not be necessary. But this will also limit the good understanding of some quantum phenomena because it will be increasingly difficult to reach the low level. However, the low level will be reserved at the level of the circuit composer.

%%%%%%%%%%%%%%%%%%%% Table No: 5 starts here %%%%%%%%%%%%%%%%%%%%
\begin{table}[t!]
\centering
\begin{tabular}{|c|c|c|}
 %\hline
 %\multicolumn{3}{|c|} {DadaS}    \\
 \hline
      Name              & Status        & Status   \\
 \hline
Assembler               &	Stable      &	Completed in version 0.9 \\
Circuit	                &   Unstable    &	          -              \\
Copiler	                &      Stable	&  Completed in version 0.9 \\
Converters              &	Unstable	&             -                \\
Dagcircuit	            & Remove        &	Will be part of circuits \\
Extensions	            & Remove	    &  Will be part of circuits   \\
Ignis.characterization  &     -         &		       -                 \\
Ignis.mitigation	    &    -          &	          -                  \\
Ignis.characterization	&     -         &	         -                   \\
Providers               &  Stable       &	Completed in version 0.7    \\
Pulse                   &	Unstable    &	           -                 \\
Gasm	                & Remove	    &  Passer location to be determined \\
Gobj	                & Remove	    &  Moved into the provider             \\
Quantum\_info	        &   Unstable    &	          -                  \\
Result	                &   Remove	    &   Moved into the provider     \\
Schemas	                &   Stable	    &   Completed in version 0.7    \\
Tests	                &   Unstable	&            -                   \\
Tools	                &   Unstable	&   Various elements to be removed\\
Transpiler	            &   Stable	    &   Completed in version 0.9        \\
Validation          	&   Stable	    &   Completed in version 0.7    \\
Visualization	        &   Stable  	&   Completed in version 0.9 \\
 \hline
\end{tabular}
\caption{Status and release of the modules towards Qiskit 1.0}
\label{tab:Status_Qiskit_v1.0}
\end{table}
%%%%%%%%%%%%%%%%%%%% Table No: 5 ends here %%%%%%%%%%%%%%%%%%%%

\subsection{Experimentation algorithms}
For its simplicity, the optimisation based on the decent gradient, where each parameter is updated in the direction that produces the most significant local change in energy (in our case), is used intuitively. We also know that the number of evaluations carried out is proportional to selected optimisation parameters. This will quickly help the algorithm find a local optimum in the proposed search space. However, this optimisation method is often stuck at a local optimum in many cases. This and the relative cost in the number of circuit evaluations performed make the descent gradient method not recommended for use when solving the problem in a scenario of many qubits and a noisy environment. The next sections will analyse the widely used classical optimisers usually combined with the variational method.

\subsubsection{SPSA}
Introducing the \textit{Simultaneous Perturbation Stochastic Approximation Optimiser} known as \textit{SPSA} \cite{Hil97} which can approximate the gradient of the objective function with only two measurements. It does this by simultaneously disturbing all parameters randomly, in contrast to the decent slope where each parameter is independently concerned.
When we need to add a circuit in real condition, that is, with noise, it is strongly recommended to use SPSA as the classic optimiser.
The above description makes the SPSA the appropriate optimiser for optimising a noisy objective function.

\subsubsection{SLSQP}
The \textit{Sequential Least-Squares Programming} \cite{bonnans2006numerical} also known as \textit{SLSQP} is a sequential least squares programming algorithm that uses the \textit{Han–Powell quasi-Newton method} \cite{gabay1982reduced} with a \textit{BFGS} \cite{BFGS_Limted} update of the B – matrix and an L1 – test function in the step–length algorithm. 

\subsubsection{COBYLA}
The classical optimiser \textit{Constrained Optimisation BY Linear Approximation} \cite{The21} commonly called \textit{COBYLA} only evaluates the objective function per optimisation iteration. The number of evaluations is independent of the cardinality of the set of parameters. Therefore, it is recommended to use COBYLA if the objective function does not produce noise, and it is convenient to minimise the number of evaluations carried out.

Qiskit, and more specifically in the Aqua library, and PennyLane provide us with several classical optimisers.

\subsection{Variational Quantum Algorithms Experimentation}
\begin{figure}[h!]
    \centering
    \includegraphics[width=\textwidth]{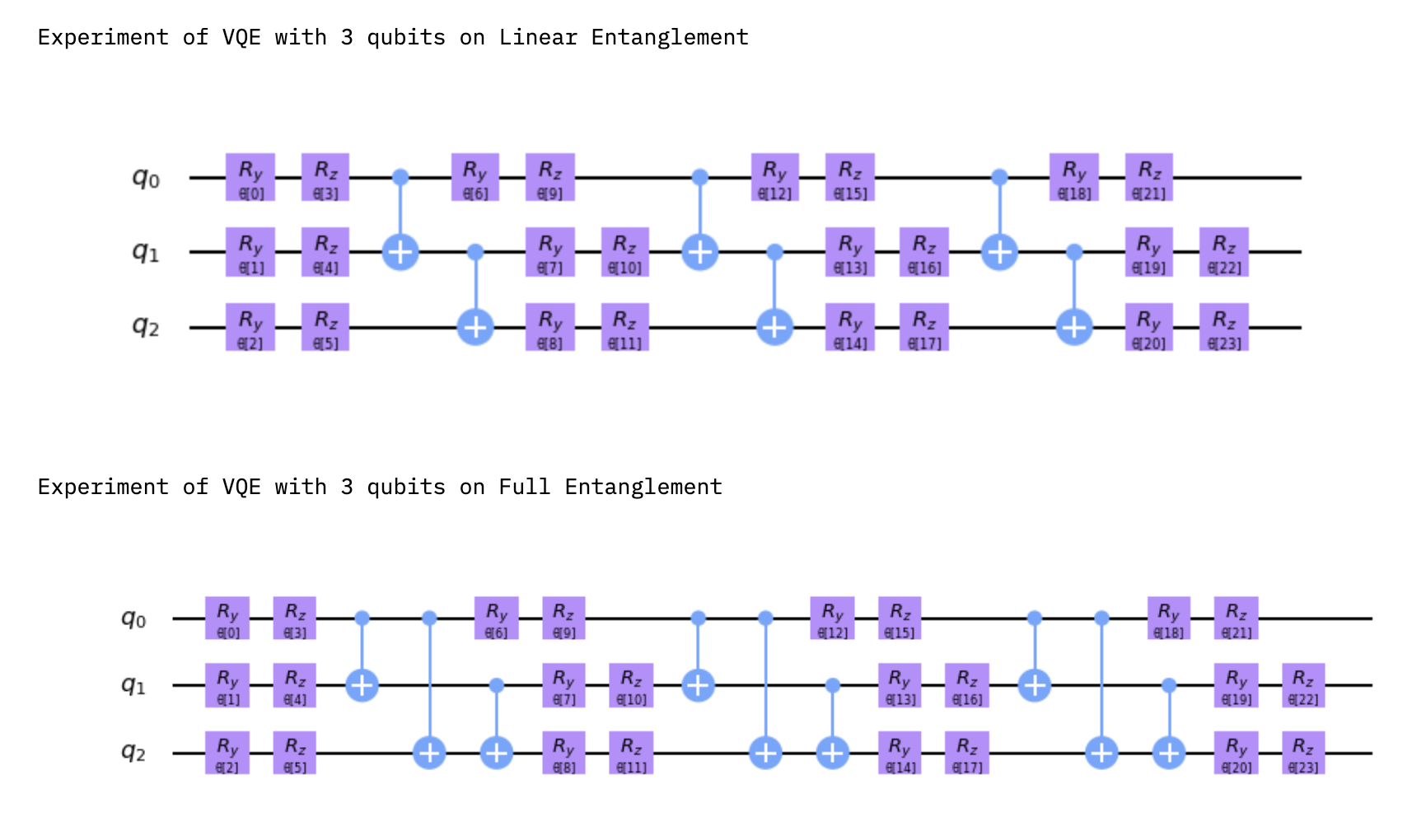}
    \caption{VQE experiment with three qubits on linear and full entanglement}
    \label{fig:VQE_Experiment}
\end{figure}
It is known that solving the Schrödinger equation analytically is very hard, where the variational principle is a way to approximate the solution. However, as discussed above, it is not possible for a polynomial parameterised variational form to generate a transformation to any state. Variational forms can be grouped into two categories, depending on how they address this limitation. The first category of variational forms uses a domain or application-specific knowledge to limit the set of possible output states. The second approach uses a heuristic circuit without prior mastery or application-specific expertise.

Therefore, variational shapes have been built for specific quantum computing architectures. Furthermore, the circuits are tuned to take full advantage of the connectivity and gates natively available from a given quantum device. This leads us to think about the importance of compilers if one wants to make multiplatform algorithms (thinking of using them on several different quantum computers.).

In the second approach, the gates are layered to obtain a good approximation in a wide range of states. Qiskit Aqua supports three such variational forms (see Fig. \eqref{fig:VQE_Experiment}): $RYRZ$, $RY$ and $SwapRZ$ (we will only discuss the first two). All of these variational forms accept multiple user-specified configurations. Three necessary arrangements are the number of qubits in the system, the depth setting, and the entanglement setting. A single layer of a variational form specifies a specific pattern of single-qubit rotations and $CX$ gates. The depth setting says how frequently the variational form should repeat this pattern. By increasing the depth setting, at the cost of increasing the number of parameters that must be optimised, the set of states in the variational form can generate increases. Finally, the entanglement setting selects the configuration, and implicitly the number, of $CX$ gates. For example, when the entanglement setting is linear, $CX$ gates are applied to adjacent qubit pairs in order (and thus $(n-1)CX$ gates are added per layer). When the entanglement setting is full, a $CX$ gate is applied to each qubit pair in each layer. The circuits for $RYRZ$ corresponding to entanglement = "\textit{full}" and entanglement = "\textit{linear}" can be seen by executing the following code snippet:

In the second approach, the quantum gates are layered to obtain good approximation in a wide range of states. Qiskit Aqua supports three variational forms:  $RYRZ$, $RY$ and $SwapRZ$. All of these variational forms accept multiple user-specified settings. Three essential settings/configurations are:
\begin{itemize}
	\item the number of qubits in the system, 
	\item the depth setting, 
	\item and the interlace setting. 
\end{itemize}
A single layer of a variational shape specifies a specified pattern of unique qubit rotations and $CX$ gates. The depth setting indicates how frequently the variational form should repeat this pattern by increasing the depth setting. The set of states that the variational form can generate increases at the cost of increasing the number of parameters to be optimised. Finally, the entanglement pattern selects the configuration, and implicitly the number, of $CX$  gates. For example, when the entanglement configuration is linear, $CX$  gates are applied to adjacent qubit pairs in order and thus  $ (n-1) CX$ gates per layer are added. When the entanglement configuration is full, a  $CX$ gate is applied to each pair of qubits in each layer. The circuits for $RYRZ$  corresponding to entanglement = "full" and entanglement = "linear" can be seen in Fig. \eqref{fig:VQE_Experiment}.
Let $d$ be the depth setting and $n$ the number of qubits, the $RYRZ$  parameters have  $n  \times (d + 1) \times  2$, $RY$  with linear entanglement has  $2n  \times   d +\frac{1}{2}$  parameters, and $RY$ with full entanglement has  $d   \times  n  \times  \frac{(n+1) }{2}+ n$  parameters.

\subsection{CPLEX}
\begin{figure}[h!]
    \centering
    \includegraphics[width=0.8\textwidth]{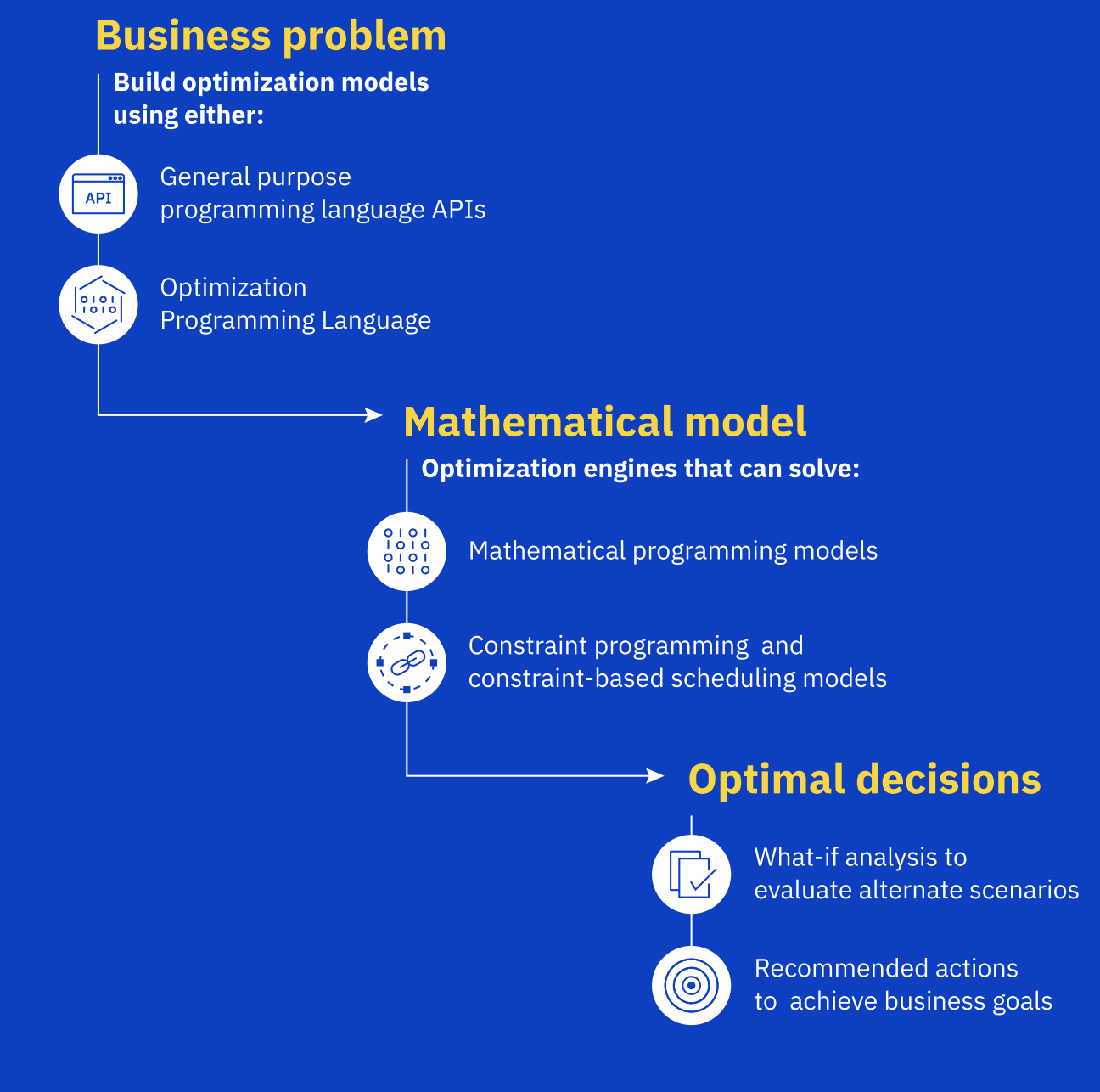}
    \caption{IBM ILOG CPLEX Optimisation Studio uses decision optimisation technology to create optimal plans, and schedules \cite{manual1987ibm}.}
    \label{fig:ILOG_CPLEX}
\end{figure}

From IBM literature, Linear Programming (LP) was revolutionised when the \textit{CPLEX®} software was developed. CPLEX (Fig. \eqref{fig:ILOG_CPLEX}) was the first commercial linear optimiser on the market to be written in the $C$ programming language. CPLEX gave operations researchers unprecedented flexibility, reliability, and performance, enabling them to create new optimisation algorithms, models, and applications.
The term CPLEX itself is based on the concept of a Simplex algorithm\cite{Rob14}, invented by George Dantzig in 1947, that is written in C: C-Simplex resulted in CPLEX.
The Simplex algorithm became the groundwork of the entire field of mathematical optimisation and provided the first practical method of solving a Linear Programming problem.
CPLEX developed over time to adapt and became a leader in the secondary categories of Linear Programming, such as \textit{Integer Programming} (IP), \textit{Mixed Integer Programming} (MIP), and the latest \textit{Quadratic Programming} (QP).
The integer programming is much more challenging to solve than linear programming, but they have a lot of essential business applications. This was IBM's bet 20 years ago, and they provided CPLEX with sophisticated mathematical techniques to solve these very hard integer programs. These techniques involve the systematic search for possible combinations of the discrete decision variables, using linear or quadratic programming relaxations to calculate the limits of the value of the optimal solution.
MIP includes formulation of problems such as vehicle routing, facility location, personnel scheduling, etc. These techniques are interesting and necessary when one does not want to program from scratch. 
We will propose five different ways of solving our problem in our case.
IBM ILOG CPLEX offers C, C++, Java, .NET, and Python libraries that solve linear programming (LP) and related problems. Correctly, it solves linearly or quadratically constrained optimisation problems where the objective to be optimised can be expressed as a linear function or a convex quadratic function. In addition, the variables in the model may be declared as continuous or further constrained to take only integer values.
CPLEX includes the following technics for high mathematical programming solvers:
\begin{itemize}
	\item \textit{Problem modelling}: IBM® ILOG® CPLEX® Optimiser provides a framework to model business issues mathematically.\par

	\item \textit{Improved profits}: IBM ILOG CPLEX Optimiser's mathematical programming provides technology to help improve efficiency, reduce costs, and increase profitability.\par

	\item \textit{Fundamental algorithms}: IBM ILOG CPLEX Optimiser provides flexible, high-performance mathematical programming solvers for linear programming, mixed-integer programming, quadratic programming, and quadratically constrained programming problems. These include a distributed parallel algorithm for mixed integer programming to leverage multiple computers to solve difficult problems.\par

	\item \textit{Robust algorithms for large problems}: IBM ILOG CPLEX Optimiser has solved issues with millions of constraints and variables.\par

	\item \textit{Industry-leading support}: IBM has an impressive rate of product improvement and ample support resources to serve you.\par

	\item \textit{High performance}: IBM ILOG CPLEX Optimiser delivers the power needed to solve large, real-world optimisation problems, as well as the speed required for today's interactive decision optimisation applications.\par

	\item \textit{Robust and reliable}: A large installed base helps us make IBM ILOG CPLEX Optimiser better with each release. Every new feature is tested on the world's biggest, most diverse model library.
\end{itemize}\par
\subsection{DOCPLEX}
The \textit{Decision Optimisation CPLEX Modelling Python}, commonly known as docplex from IBM, is a tool with which data scientists can use their preferred scientific tools (Python, with NumPy/pandas/scipy stack) to develop optimisation-based projects.
The Docplex library is composed of 2 modules:
\begin{itemize}
	\item IBM® Decision Optimisation CPLEX Optimiser Modelling for Python - with namespace docplex.mp. This library is for Mathematical Programming Modeling for Python.
	\item IBM® Decision Optimisation CP Optimiser Modelling for Python - with namespace docplex.cp. This library is for Constraint Programming Modelling for Python.
\end{itemize}

On May 1, 2020, IBM managed to integrate into its docplex the quadratic programming for MIP, which helps very much and makes mapping easier into the Ising model.
Last July 9, IBM unveiled a new module that will boost research, development, and benchmarking of quantum optimisation algorithms for this NISQ era\cite{Joh18}.
To map a classical model to quantum computing, we need to find the Hamiltonian of the Ising model. Nevertheless, the Hamiltonians of the Ising model are highly complicated and have no intuitive\cite{McG14}\cite{Jer03}\cite{Mic00}. So, mapping a combinatorial optimisation problem to the Hamiltonian of the Ising model can be very challenging, complicated, and may require specialised knowledge as the vectorisation of the matrices with the Kronecker product \cite{Hua13} to express matrix multiplication as a linear transformation in matrices\cite{Jua13}.
With the translator, all kinds of users can write a quantum optimisation algorithm model using docplex. With this tool, many things become much easier than writing the Hamiltonian of the Ising model manually basically because the model is short and intuitive.

\section{Summary}
In this chapter, we have experimented on the different quantum programming environments, and we are ready to choose the background on which we will focus our efforts based on figure \eqref{fig:Development_Roadmap} and \eqref{tab:Status_Qiskit_v1.0}. Of course, this does not mean that our algorithms are only for that environment. Still, we need to make a strategic decision to move forward and, above all, make the most of the community and the libraries that offer us to do some or other tests.

Let us take a look at Qiskit for its community, the facilities that IBM provides, the scientific community and the network of talent it has. As mentioned above, Qiskit already has a community of more than 25,000 developers and an available and affordable team for any questions. It has at any time for all the libraries, especially for its scalability plan. IBMQ is our bet.
However, we experiment and will continue to research and monitor the other frameworks such as Dwave, AWS-Braket, Cirq, PennyLane, etc.

%%%%%%%%%%%%  Starting New Page here %%%%%%%%%%%%%%

\newpage
\chapter{Approaches to solve the SWP}\label{sec:11}
\section{Introduction}
We design this formulation keeping in mind the limitation of our device in this NISQ era. Therefore, we will encode the time information to represent the limits of the constraints in the said formula.

This section will work on the various formulation proposals to solve the proposed problem adequately on the chosen framework (Qiskit). But one of the algorithm proposals will assess the generality of the platforms. This algorithm could be executed (with very few splices) either in a Dwave Ocean, PennyLanne or a Qibo.

\section{Social Workers Problem based on a new VRPTW}

In this proposal, we take advantage of the formulation of the CVRP to establish our proof of the concept. It is worth saying that we pursue to avoid using the inequality constraints and use the least number of qubits according to this NISQ era. Nevertheless, to do some comparative studies, we would rather use simulators with more qubits than quantum computers to test our algorithm. Therefore, we will base our algorithm on techniques (TSP, VRP) already consolidated to achieve efficiency in many qubits. We use VRP's universal formulation to model the routing part of the proposed VRPTW. \\

Let $G=  \left( V, E \right)$  be a complete graph directed with $V =  \{ 0, 1, 2, . . , n \},$  as the set of nodes and  $E=  \{  \left( i, j \right): i, j  \in  V, i \neq  j \}$ as the set of edges, where node 0 represents the central, for a team of $K$ social workers with the same maximum number of kilometres to travel  $q$  and $n$ remaining nodes represent geographically dispersed patients.

The non-negative travel cost $W_{ij}$ is associated with each arc $\left( i,j \right)   \in  E$. Let $d_{ij}$  be our distance matrix and to simplify, we consider that the distances are symmetrical. Where $x_{ij}$ are the decision variables of the paths between the patient $i$ and $j$.

Minimise:
\begin{equation}
\label{SWP_VRP_form_eq}
    \sum _{i=1}^{n} \sum _{j=1,i \neq j}^{n}W_{ij}x_{ij,}.
\end{equation}

Subject to:
\begin{equation}
\label{SWP_const_1_eq}
    \sum _{j=1}^{n}x_{ij}=1 \quad \forall i  \in  1, \ldots ,n,
\end{equation}

\begin{equation}
\label{SWP_const_2_eq}
    \sum _{i=1}^{n}x_{ji}=1 \quad \forall j \in  1, \ldots ,n,
\end{equation}

\begin{equation}
\label{SWP_const_3_eq}
      \sum_{i=1}^{n} \sum_{j=1}^{n}x_{ij}d_{ij} \leq q, 
\end{equation}

\begin{equation}
\label{SWP_const_4_eq}
 \sum _{j=1}^{n}x_{0j}=K ~~~\forall i  \in  1, \ldots ,n,
\end{equation}

\begin{equation}
\label{SWP_const_5_eq}
    \sum _{j=1}^{n}x_{j0~}= K,~~~   \forall i  \in   1, \ldots ,n,
\end{equation}

\begin{equation}
\label{SWP_const_6_eq}
     u_i - u_j + nx_{i,j} \leq n-1 \quad 1 \leq i\neq j \leq n.
\end{equation}
 
\begin{equation}
\label{SWP_const_7_eq}
    x_{ij} \in  \{0,1\} , ~~~ \forall i,j \in 0, \ldots ,n, i  \neq j. 
\end{equation}

In this formulation, the objective function equation \eqref{SWP_VRP_form_eq} minimises the new cost function with the time window \eqref{weight_SWP_eq}. The restrictions equations \eqref{SWP_const_1_eq} and \eqref{SWP_const_2_eq} declare that each social worker can only be one node at any time (that means we will only visit once the patient). The restriction \eqref{SWP_const_3_eq} establishes that any social worker can travel more distance than allowed. In the case of wanting to measure the time, here, what we would do is change the matrix $d_{ij}$ for a matrix of the maximum contract time. The constraint \eqref{SWP_const_4_eq} establishes that all the social workers start from Depot and \eqref{SWP_const_5_eq} establishes that all the social workers end at the Depot. The constraints \eqref{SWP_const_6_eq} are the route of continuity and the elimination of sub-courses, which ensure that the solution does not contain a sub-route disconnected from the exchange. The equation \eqref{SWP_const_7_eq} describes that $x_{ij}$, are binary variables. 

Up to this point, the mathematical formulation from equations \eqref{SWP_VRP_form_eq} to \eqref{SWP_const_7_eq} represents a conventional CVRP. However, to solve a scheduling problem, we need a time variable. The introduction of time (schedule) into the QUBO formulations of the CVRP is a significant obstacle to formulating several important VRP restrictions associated with the VRPTW time window \cite{papalitsas2019qubo}.

Our CVRP formulation proposal must incorporate the schedule (calendar) of table \eqref{tab:Sample_SWP_Schedu}.

During the state of the art of these formulations carried out, we have found several articles \cite{papalitsas2019qubo, feld2019hybrid, irie2019quantum} that solve the TSP and VRP for annealing computers \cite{brooke1999quantum,boixo2014evidence,crosson2016simulated}. However, the number of variables is still intractable for the current size of quantum computers. 
The \textbf{number of qubits} of the TSPTW \cite{papalitsas2019qubo}, is proportional to $N^3+N^2\log_{2}N$, and for this VRPTW \cite{irie2019quantum}, is $N^4$.
For this reason, we propose a new VRPTW formulation (from \eqref{SWP_VRP_form_eq} to \eqref{time_window_SWP_eq}) with a heuristic function executed by an classical algorithm that generates a description of a quantum circuit as advocate the following reference \cite{JayGambetta}. With this strategy, we aim to reduce the number of the qubits from $N^4$ to $N^2$ for our proposed VRPTW for solving our SWP.

For example, a possible formulation of the VRP uses $ N ^ 4 $, the number of the qubits would be $625$, which is more than the most powerful quantum computer based on the gate model has to date. The gate-based computers have around 100 qubits making this task intractable today. The number of qubits is higher for computers based on quantum annealing, reaching 2000 qubits like the D-Wave computer. However, the correspondence between variables and qubits will not be one-to-one due to the architecture of these computers, so that we will have a smaller number of useful qubits. The following reference \cite{headquarters2020technical} deals with the topology and graph problem mapping on the D-Wave 2000Q QPU computer in detail.

The new time window formulation of our VRPTW is expressed by the equations \eqref{weight_SWP_eq} and \eqref{time_window_SWP_eq}.

\begin{equation}
\label{weight_SWP_eq}
    W_{ij}=d_{ij}+f \left( t_{ij} \right). 
\end{equation}

\begin{equation}
\label{time_window_SWP_eq}
    f \left( t_{ij} \right) = \gamma \frac{ \left(  \tau_{i-} \tau_{j} \right) ^{2}}{d_{\max }-d_{\min }}.
\end{equation}

Where $W_{ij}$ is our cost/weight and time window function, $d_{ij}$ is the distance between the patient $i$ and $j$ and $f(t_{ij})$ is our time window’s function. $f(t_{ij})$ is a growing function, and we model it by a quadratic function to weigh short distances concerning large ones. We are taking into account that the first weight function $W_{ij}=d_{ij}$  is a distance function, we want to make $f(t_{ij})$  behave like $d_{ij}$, and thus, be able to take full advantage of the behaviour of the primary objective function. 
$\gamma >0$ is a weighted degree parameter of our time window function;  $\tau_{i}$  is the start time of a time slot for patient  $i$  and  $\tau_{j}$  for the patient  $j$. where $d_{\max }$ represents the maximum distance between all patients and,  $d_{\min }$  is the minimum distance between the gaps of all patients. The term $T_{ij}= (\tau_{i-} \tau_{j}) >0$ is the time window.

The simplified Hamiltonian resulting from the schedule optimisation problem is as follows:
\begin{equation}
\label{SWP_FORM_eq}
\begin{aligned}
 &H= \sum _{i=1}^{n} \sum _{j=1,i \neq j}^{n} \left( d_{ij}+ \gamma \frac{ \left(  \tau_{i-} \tau_{j} \right) ^{2}}{d_{\max }-d_{\min }} \right) x_{i,j}+ A \sum _{i=1}^{n} (1- \sum _{j=1 }^{n}x_{i,j}) ^{2}+A \sum _{i=1}^{n} (1- \sum _{j=1}^{n}x_{ji}) ^{2} \\
 &+ A (k- \sum _{i=1}^{n}x_{0,i}) ^{2}+ A(k- \sum _{j=1}^{n}x_{j,0}) ^{2}.
\end{aligned}
\end{equation}

Where  $A$ is the Lagrange multiplier which is a free parameter such that $A > \text{max}\left( d_{ij}+ \gamma \frac{ \left(  \tau_{i-} \tau_{j} \right) ^{2}}{d_{\max }-d_{\min }} \right)$.

The number of the qubits after applying our strategy will be:

\begin{equation}
\small 
\label{numQub}
   \textbf{Num qubits}_{\text{SWP}} = \binom{n}{2} = \frac{n!}{2!(n-2)!} = \frac {n(n-1)}{2}+ \epsilon.
\end{equation}

Where the $\epsilon$ is the ancillary variables given by $\sum_{i=0}^{\lceil log_2b\rceil}2^iy_i$ according to the capacity restriction\eqref{SWP_const_3_eq}. In the case of removing the symmetry; the number of qubits will be $n(n-1)+\epsilon$.

\subsection{Social Workers' Problem based on a QUBO approach }\label{sec:SWPPaperQubo}
Within the pedagogical nature of this work, we will detail the resolution of the problem step by step to meet one of the objectives of the thesis.\\

Let us solve our formulation in QUBO form by considering $n=4$ patients. Where $Q$ is a  $2^{N}\times 2^{N}$ matrix with $N=n(n-1)$ as the number of qubits. So, in this case, $N=12$ qubits. Let's remember that we are using binary variables so, $( x_{u, \upsilon }^{i}) ^{2}=x_{u, \upsilon }^{i}$.
\begin{equation}
%\begin{justify}
\label{SWP_Q_F_SOL_eq}
    \sum _{ij \in E}^{3}W_{ij}x_{i,j}= \sum _{i}^{3} \sum_{\substack{j \\ i\neq j}}^{3}W_{ij}x_{i,j} = \left( \begin{array}{c}
    W_{0,1}x_{0,1} + W_{0,2}x_{0,2} + W_{0,3}x_{0,3} + \\
    W_{1,0}x_{1,0} + W_{1,2}x_{1,2} + W_{1,3}x_{1,3} + W_{2,0}x_{2,0} + W_{2,1}x_{2,1} +\\ W_{2,3}x_{2,3} + W_{3,0}x_{3,0} + W_{3,1}x_{3,1} + W_{3,2}x_{3,2} \end{array}\right).
%\end{justify}
\end{equation}
\begin{equation}
\label{2nd_term}
\begin{aligned}
  & A \sum_{i=1}^{3} \left( 1- \sum_{j \in  \delta  \left( i \right) ^{-}}^{3}x_{ji} \right) ^{2}\\ 
  &=A \left( 1-x_{0,1}-x_{2,1}-x_{3,1}+1-x_{0,2}-x_{1,2}-x_{3,3}+1-x_{0,3}-x_{1,3}-x_{2,3} \right) ^{2}\\
 &=A \left( 3+-x_{0,1}-x_{2,1}-x_{3,1}-x_{0,2}-x_{1,2}-x_{3,2}-x_{0,3}-x_{1,3}-x_{2,3} \right) ^{2}\\
 &=A \left( \begin{array}{c}
	3^{2}+ \left( x_{0,1} \right) ^{2}+ \left( x_{2,1} \right) ^{2}+ \left( x_{3,1} \right) ^{2}+ \left( x_{0,2} \right) ^{2}+ \left( x_{1,2} \right) ^{2}+ \left( x_{3,2} \right) ^{2}+ \left( x_{0,3} \right) ^{2}+ \left( x_{1,3} \right) ^{2}+ \left( x_{2,3} \right) ^{2}\\
	-6x_{0,1}-6x_{2,1}-6x_{3,1}-6x_{0,2}-6x_{1,2}-6x_{3,2}-6x_{0,3}-6x_{1,3}-6x_{2,3}\\
	
	+2x_{0,1}x_{2,1}+2x_{0,1}x_{3,1}+2x_{0,1}x_{0,2}+2x_{0,1}x_{1,2}+2x_{0,1}x_{3,2}+2x_{0,1}x_{0,3}+2x_{0,1}x_{1,3}+2x_{0,1}x_{2,3}\\
	
	+2x_{2,1}x_{3,1}+2x_{2,1}x_{0,2}+2x_{2,1}x_{1,2}+2x_{2,1}x_{3,2}+2x_{2,1}x_{0,3}+2x_{2,1}x_{1,3}+2x_{2,1}x_{2,3}\\
	
	+2x_{3,1}x_{0,2}+2x_{3,1}x_{1,2}+2x_{3,1}x_{3,2}+2x_{3,1}x_{0,3}+2x_{3,1}x_{1,3}+2x_{3,1}x_{2,3}\\
	
	+2x_{0,2}x_{1,2}+2x_{0,2}x_{3,2}+2x_{0,2}x_{0,3}+2x_{0,2}x_{1,3}+2x_{0,2}x_{2,3}\\
	
	+2x_{1,2}x_{3,2}+2x_{1,2}x_{0,3}+2x_{1,2}x_{1,3}+2x_{1,2}x_{2,3}\\
	
	+2x_{3,2}x_{0,3}+2x_{3,2}x_{1,3}+2x_{3,2}x_{2,3}\\
	
	+2x_{0,3}x_{1,3}+2x_{0,3}x_{23}\\
	
	+2x_{1,3}x_{2,3}\\
	\end{array} \right),
\end{aligned}
\end{equation}

\begin{equation}
\label{3r_term}
\begin{aligned}
  &A \sum_{i=1}^{3} \left( 1- \sum_{j \in  \delta  \left( i \right) ^{+}}^{3}x_{i,j} \right) ^{2} \\
 & =A \left( 1-x_{1,0}-x_{1,2}-x_{1,3}+1-x_{2,0}-x_{2,1}-x_{2,3}+1-x_{3,0}-x_{3,1}-x_{3,2} \right) ^{2}\\ 
 & =A \left( 3+-x_{1,0}-x_{1,2}-x_{1,3}-x_{2,0}-x_{2,1}-x_{2,3}-x_{3,0}-x_{3,1}-x_{3,2} \right) ^{2} \\ 
 & =A \left( \begin{array}{c}
	3^{2}+ \left( x_{1,0} \right) ^{2}+ \left( x_{1,2} \right) ^{2}+ \left( x_{1,3} \right) ^{2}+ \left( x_{2,0} \right) ^{2}+ \left( x_{2,1} \right) ^{2}+ \left( x_{2,3} \right) ^{2}+ \left( x_{3,0} \right) ^{2}+ \left( x_{3,1} \right) ^{2}+ \left( x_{3,2} \right) ^{2}\\
	-6x_{1,0}-6x_{1,2}-6x_{1,3}-6x_{2,0}-6x_{2,1}-6x_{2,3}-6x_{3,0}-6x_{3,1}-6x_{3,2}\\
	+2x_{1,0}x_{1,2}+2x_{1,0}x_{1,3}+2x_{1,0}x_{2,0}+2x_{1,0}x_{2,1}+2x_{1,0}x_{2,3}+2x_{1,0}x_{3,0}+2x_{1,0}x_{3,1}+2x_{1,0}x_{3,2}\\
	+2x_{1,2}x_{1,3}+2x_{1,2}x_{2,0}+2x_{1,2}x_{2,1}+2x_{1,2}x_{2,3}+2x_{1,2}x_{3,0}+2x_{1,2}x_{3,1}+2x_{1,2}x_{3,2}\\
	+2x_{1,3}x_{2,0}+2x_{1,3}x_{2,1}+2x_{1,3}x_{2,3}+2x_{1,3}x_{3,0}+2x_{1,3}x_{3,1}+2x_{1,3}x_{3,2}\\
	+2x_{2,0}x_{2,1}+2x_{2,0}x_{2,3}+2x_{2,0}x_{3,0}+2x_{2,0}x_{3,1}+2x_{2,0}x_{3,2}\\
	+2x_{2,1}x_{2,3}+2x_{2,1}x_{3,0}+2x_{2,1}x_{3,1}+2x_{2,1}x_{3,2}\\
	+2x_{2,3}x_{3,0}+2x_{2,3}x_{3,1}+2x_{2,3}x_{3,2}\\
	+2x_{3,0}x_{3,1}+2x_{3,0}x_{3,2}\\
	+2x_{3,1}x_{3,2}\\
	\end{array} \right),
\end{aligned}	
\end{equation}
\begin{equation}
\label{4th_term}
\begin{aligned}
 & A \left( k- \sum_{i \in  \delta  \left( 0 \right) ^{+}}^{3}x_{0,i} \right) ^{2}=A \left( k-x_{0,1}-x_{0,2}-x_{0,3} \right) ^{2} \\
 & =A \left( \begin{array}{c}
	k^{2}+ \left( x_{0,1} \right) ^{2}+ \left( x_{0,2} \right) ^{2}+ \left( x_{0,3} \right) ^{2}-2kx_{0,1}-2kx_{0,2}-2kx_{0,3}\\
	+2x_{0,1}x_{0,2}+2x_{0,1}x_{0,3}+2x_{0,2}x_{0,3}
	\end{array} \right) \\
  &=A \left( k^{2}+ \left( x_{0,1} \right) ^{2}+ \left( x_{0,2} \right) ^{2}+ \left( x_{0,3} \right) ^{2}-2kx_{0,1}-2kx_{0,2}-2kx_{0,3}+2x_{0,1}x_{0,2}+2x_{0,1}x_{0,3}+2x_{0,2}x_{0,3} \right), 
\end{aligned}
\end{equation}
\begin{equation}
\label{5th_term}
\begin{aligned}
 & A \left( k- \sum_{j \in  \delta  \left( 0 \right) ^{+}}^{3}x_{j,0} \right) ^{2}=A \left( k-x_{1,0}-x_{2,0}-x_{3,0} \right) ^{2} \\
 & =A \left( \begin{array}{c}
	k^{2}+ \left( x_{1,0} \right) ^{2}+ \left( x_{2,0} \right) ^{2}+ \left( x_{3,0} \right) ^{2}-2kx_{1,0}-2kx_{2,0}-2kx_{3,0}\\
	+2x_{1,0}x_{2,0}+2x_{1,0}x_{3,0}+2x_{2,0}x_{3,0}\\
	\end{array} \right) \\
 & =A \left( k^{2}+ \left( x_{1,0} \right) ^{2}+ \left( x_{2,0} \right) ^{2}+ \left( x_{3,0} \right) ^{2}-2kx_{1,0}-2kx_{2,0}-2kx_{3,0}+2x_{1,0}x_{2,0}+2x_{1,0}x_{3,0}+2x_{2,0}x_{3,0} \right),
\end{aligned}
\end{equation}

Grouping the terms \eqref{SWP_Q_F_SOL_eq} to \eqref{5th_term}) we reach out to the following expression:

\begin{equation}
\label{1_resultant_Q_SWF_eq}
\begin{aligned}
    & W_{0,1}x_{0,1} + W_{0,2}x_{0,2} + W_{0,3}x_{0,3} + W_{1,2}x_{1,2} + W_{1,3}x_{1,3} + W_{2,1}x_{2,1} + W_{2,3}x_{2,3} + \\
    & A \left( \begin{array}{c}
    + 3^{2} + x_{0,1}^{2} + x_{2,1}^{2} + x_{3,1}^{2} + x_{0,2}^{2} + x_{1,2}^{2} + x_{3,2}^{2} + x_{0,3}^{2} + x_{1,3}^{2} + x_{2,3}^{2}\\
	- 6x_{0,1} - 6x_{2,1} - 6x_{3,1} - 6x_{0,2} - 6x_{1,2} - 6x_{3,2} - 6x_{0,3} - 6x_{1,3} - 6x_{2,3}\\
	+ 2x_{0,1}x_{2,1} + 2x_{0,1}x_{3,1} + 2x_{0,1}x_{0,2} + 2x_{0,1}x_{1,2} + 2x_{0,1}x_{3,2} + 2x_{0,1}x_{0,3} + 2x_{0,1}x_{1,3} + 2x_{0,1}x_{2,3}\\
	+ 2x_{2,1}x_{3,1} + 2x_{2,1}x_{0,2} + 2x_{2,1}x_{1,2} + 2x_{2,1}x_{3,2} + 2x_{2,1}x_{0,3} + 2x_{2,1}x_{1,3} + 2x_{2,1}x_{2,3}\\
	+ 2x_{3,1}x_{0,2} + 2x_{3,1}x_{1,2} + 2x_{3,1}x_{3,2} + 2x_{3,1}x_{0,3} + 2x_{3,1}x_{1,3} + 2x_{3,1}x_{2,3}\\
	+ 2x_{0,2}x_{1,2} + 2x_{0,2}x_{3,2} + 2x_{0,2}x_{0,3} + 2x_{0,2}x_{1,3} + 2x_{0,2}x_{2,3}\\
	+ 2x_{1,2}x_{3,2} + 2x_{1,2}x_{0,3} + 2x_{1,2}x_{1,3} + 2x_{1,2}x_{2,3}\\
	+ 2x_{3,2}x_{0,3} + 2x_{3,2}x_{1,3} + 2x_{3,2}x_{2,3}\\
	+ 2x_{0,3}x_{1,3} + 2x_{0,3}x_{23}\\
	+ 2x_{1,3}x_{2,3} \end{array} \right),\\
	 & +A \left( \begin{array}{c}
	3^{2}+ \left( x_{1,0} \right) ^{2}+ \left( x_{1,2} \right) ^{2}+ \left( x_{1,3} \right) ^{2}+ \left( x_{2,0} \right) ^{2}+ \left( x_{2,1} \right) ^{2}+ \left( x_{2,3} \right) ^{2}+ \left( x_{3,0} \right) ^{2}+ \left( x_{3,1} \right) ^{2}+ \left( x_{3,2} \right) ^{2}\\
	-6x_{1,0}-6x_{1,2}-6x_{1,3}-6x_{2,0}-6x_{2,1}-6x_{2,3}-6x_{3,0}-6x_{3,1}-6x_{3,2}\\
	+2x_{1,0}x_{1,2}+2x_{1,0}x_{1,3}+2x_{1,0}x_{2,0}+2x_{1,0}x_{2,1}+2x_{1,0}x_{2,3}+2x_{1,0}x_{3,0}+2x_{1,0}x_{3,1}+2x_{1,0}x_{3,2}\\
	+2x_{1,2}x_{1,3}+2x_{1,2}x_{2,0}+2x_{1,2}x_{2,1}+2x_{1,2}x_{2,3}+2x_{1,2}x_{3,0}+2x_{1,2}x_{3,1}+2x_{1,2}x_{3,2}\\
	+2x_{1,3}x_{2,0}+2x_{1,3}x_{2,1}+2x_{1,3}x_{2,3}+2x_{1,3}x_{3,0}+2x_{1,3}x_{3,1}+2x_{1,3}x_{3,2}\\
	+2x_{2,0}x_{2,1}+2x_{2,0}x_{2,3}+2x_{2,0}x_{3,0}+2x_{2,0}x_{3,1}+2x_{2,0}x_{3,2}\\
	+2x_{2,1}x_{2,3}+2x_{2,1}x_{3,0}+2x_{2,1}x_{3,1}+2x_{2,1}x_{3,2}\\
	+2x_{2,3}x_{3,0}+2x_{2,3}x_{3,1}+2x_{2,3}x_{3,2}\\
	+2x_{3,0}x_{3,1}+2x_{3,0}x_{3,2}\\
	+2x_{3,1}x_{3,2}\\
	\end{array} \right) \\
	& n+ A \left( k^{2}+ \left( x_{0,1} \right) ^{2}+ \left( x_{0,2} \right) ^{2}+ \left( x_{0,3} \right) ^{2}-2kx_{0,1}-2kx_{0,2}-2kx_{0,3}+2x_{0,1}x_{0,2}+2x_{0,1}x_{0,3}+2x_{0,2}x_{0,3} \right) \\
	& + A \left( k^{2}+ \left( x_{1,0} \right) ^{2}+ \left( x_{2,0} \right) ^{2}+ \left( x_{3,0} \right) ^{2}-2kx_{1,0}-2kx_{2,0}-2kx_{3,0}+2x_{1,0}x_{2,0}+2x_{1,0}x_{3,0}+2x_{2,0}x_{3,0} \right), 
	\end{aligned}
\end{equation}

Now let’s apply the binary variable property  $x_{i,j}^{2} = x_{i,j}$ so,

\begin{equation}
\label{1_resultant_Q_SWF_1_eq}
\begin{aligned}
    & W_{0,1}x_{0,1} + W_{0,2}x_{0,2} + W_{0,3}x_{0,3} + W_{1,2}x_{1,2} + W_{1,3}x_{1,3} + W_{2,1}x_{2,1} + W_{2,3}x_{2,3} 
    \\
    &+ A \left( \begin{array}{c}
	3^{2} + x_{0,1} + x_{2,1} + x_{3,1} + x_{0,2} +x_{1,2} + x_{3,2} + x_{0,3} +x_{1,3} + x_{2,3}\\
	-6x_{0,1}-6x_{2,1}-6x_{3,1}-6x_{0,2}-6x_{1,2}-6x_{3,2}-6x_{0,3}-6x_{1,3}-6x_{2,3}\\
	+2x_{0,1}x_{2,1}+2x_{0,1}x_{3,1}+2x_{0,1}x_{0,2}+2x_{0,1}x_{1,2}+2x_{0,1}x_{3,2}+2x_{0,1}x_{0,3}+2x_{0,1}x_{1,3}+2x_{0,1}x_{2,3}\\
	+2x_{2,1}x_{3,1}+2x_{2,1}x_{0,2}+2x_{2,1}x_{1,2}+2x_{2,1}x_{3,2}+2x_{2,1}x_{0,3}+2x_{2,1}x_{1,3}+2x_{2,1}x_{2,3}\\
	+2x_{3,1}x_{0,2}+2x_{3,1}x_{1,2}+2x_{3,1}x_{3,2}+2x_{3,1}x_{0,3}+2x_{3,1}x_{1,3}+2x_{3,1}x_{2,3}\\
	+2x_{0,2}x_{1,2}+2x_{0,2}x_{3,2}+2x_{0,2}x_{0,3}+2x_{0,2}x_{1,3}+2x_{0,2}x_{2,3}\\
	+2x_{1,2}x_{3,2}+2x_{1,2}x_{0,3}+2x_{1,2}x_{1,3}+2x_{1,2}x_{2,3}\\
	+2x_{3,2}x_{0,3}+2x_{3,2}x_{1,3}+2x_{3,2}x_{2,3}\\
	+2x_{0,3}x_{1,3}+2x_{0,3}x_{23}\\
	+2x_{1,3}x_{2,3}\\
	\end{array} \right) \\
	&+ A \left( \begin{array}{c}
	3^{2}+x_{1,0}+x_{1,2}+x_{1,3}+x_{2,0}+x_{2,1}+x_{2,3}+x_{3,0} +x_{3,1}+x_{3,2}\\
	-6x_{1,0}-6x_{1,2}-6x_{1,3}-6x_{2,0}-6x_{2,1}-6x_{2,3}-6x_{3,0}-6x_{3,1}-6x_{3,2}\\
	+2x_{1,0}x_{1,2}+2x_{1,0}x_{1,3}+2x_{1,0}x_{2,0}+2x_{1,0}x_{2,1}+2x_{1,0}x_{2,3}+2x_{1,0}x_{3,0}+2x_{1,0}x_{3,1}+2x_{1,0}x_{3,2}\\
	+2x_{1,2}x_{1,3}+2x_{1,2}x_{2,0}+2x_{1,2}x_{2,1}+2x_{1,2}x_{2,3}+2x_{1,2}x_{3,0}+2x_{1,2}x_{3,1}+2x_{1,2}x_{3,2}\\
	+2x_{1,3}x_{2,0}+2x_{1,3}x_{2,1}+2x_{1,3}x_{2,3}+2x_{1,3}x_{3,0}+2x_{1,3}x_{3,1}+2x_{1,3}x_{3,2}\\
	+2x_{2,0}x_{2,1}+2x_{2,0}x_{2,3}+2x_{2,0}x_{3,0}+2x_{2,0}x_{3,1}+2x_{2,0}x_{3,2}\\
	+2x_{2,1}x_{2,3}+2x_{2,1}x_{3,0}+2x_{2,1}x_{3,1}+2x_{2,1}x_{3,2}\\
	+2x_{2,3}x_{3,0}+2x_{2,3}x_{3,1}+2x_{2,3}x_{3,2}\\
	+2x_{3,0}x_{3,1}+2x_{3,0}x_{3,2}\\
	+2x_{3,1}x_{3,2}\\
	\end{array} \right) \\
	& + A \left( k^{2}+x_{0,1}+x_{0,2}+x_{0,3}-2kx_{0,1}-2kx_{0,2}-2kx_{0,3}+2x_{0,1}x_{0,2}+2x_{0,1}x_{0,3}+2x_{0,2}x_{0,3} \right) \\
	& + A \left( k^{2}+x_{1,0}+x_{2,0}+x_{3,0}-2kx_{1,0}-2kx_{2,0}-2kx_{3,0}+2x_{1,0}x_{2,0}+2x_{1,0}x_{3,0}+2x_{2,0}x_{3,0} \right) 
\end{aligned}
\end{equation}
\newpage
Grouping similar terms:
\begin{equation}
\label{1_resultant_Q_SWF_2_eq}
\begin{aligned}
& W_{0,1}x_{0,1}+W_{0,2}x_{0,2}+W_{0,3}x_{0,3}+W_{1,2}x_{1,2}+W_{1,3}x_{1,3}+W_{2,1}x_{2,1}+W_{2,3}x_{2,3} \\
& +A \left( \begin{array}{c}
	3^{2}~\\
	-5x_{0,1}-5x_{2,1}-5x_{3,1}-5x_{0,2}-5x_{1,2}-5x_{3,2}-5x_{0,3}-5x_{1,3}-5x_{2,3}\\
	+2x_{0,1}x_{2,1}+2x_{0,1}x_{3,1}+2x_{0,1}x_{0,2}+2x_{0,1}x_{1,2}+2x_{0,1}x_{3,2}+2x_{0,1}x_{0,3}+2x_{0,1}x_{1,3}+2x_{0,1}x_{2,3}\\
	+2x_{2,1}x_{3,1}+2x_{2,1}x_{0,2}+2x_{2,1}x_{1,2}+2x_{2,1}x_{3,2}+2x_{2,1}x_{0,3}+2x_{2,1}x_{1,3}+2x_{2,1}x_{2,3}\\
	+2x_{3,1}x_{0,2}+2x_{3,1}x_{1,2}+2x_{3,1}x_{3,2}+2x_{3,1}x_{0,3}+2x_{3,1}x_{1,3}+2x_{3,1}x_{2,3}\\
	+2x_{0,2}x_{1,2}+2x_{0,2}x_{3,2}+2x_{0,2}x_{0,3}+2x_{0,2}x_{1,3}+2x_{0,2}x_{2,3}\\
	+2x_{1,2}x_{3,2}+2x_{1,2}x_{0,3}+2x_{1,2}x_{1,3}+2x_{1,2}x_{2,3}\\
	+2x_{3,2}x_{0,3}+2x_{3,2}x_{1,3}+2x_{3,2}x_{2,3}\\
	+2x_{0,3}x_{1,3}+2x_{0,3}x_{23}\\
	+2x_{1,3}x_{2,3}\\
	\end{array} \right)\\
	&+A \left( \begin{array}{c}
	3^{2}\\
	-5x_{1,0}-5x_{1,2}-5x_{1,3}-5x_{2,0}-5x_{2,1}-5x_{2,3}-5x_{3,0}-5x_{3,1}-5x_{3,2}\\
	+2x_{1,0}x_{1,2}+2x_{1,0}x_{1,3}+2x_{1,0}x_{2,0}+2x_{1,0}x_{2,1}+2x_{1,0}x_{2,3}+2x_{1,0}x_{3,0}+2x_{1,0}x_{3,1}+2x_{1,0}x_{3,2}\\
	+2x_{1,2}x_{1,3}+2x_{1,2}x_{2,0}+2x_{1,2}x_{2,1}+2x_{1,2}x_{2,3}+2x_{1,2}x_{3,0}+2x_{1,2}x_{3,1}+2x_{1,2}x_{3,2}\\
	+2x_{1,3}x_{2,0}+2x_{1,3}x_{2,1}+2x_{1,3}x_{2,3}+2x_{1,3}x_{3,0}+2x_{1,3}x_{3,1}+2x_{1,3}x_{3,2}\\
	+2x_{2,0}x_{2,1}+2x_{2,0}x_{2,3}+2x_{2,0}x_{3,0}+2x_{2,0}x_{3,1}+2x_{2,0}x_{3,2}\\
	+2x_{2,1}x_{2,3}+2x_{2,1}x_{3,0}+2x_{2,1}x_{3,1}+2x_{2,1}x_{3,2}\\
	+2x_{2,3}x_{3,0}+2x_{2,3}x_{3,1}+2x_{2,3}x_{3,2}\\
	+2x_{3,0}x_{3,1}+2x_{3,0}x_{3,2}\\
	+2x_{3,1}x_{3,2}\\
	\end{array} \right) \\
	&+ A \left( k^{2}+x_{0,1}+x_{0,2}+x_{0,3}-2kx_{0,1}-2kx_{0,2}-2kx_{0,3}+2x_{0,1}x_{0,2}+2x_{0,1}x_{0,3}+2x_{0,2}x_{0,3} \right) \\
	&+ A ( k^{2}+x_{1,0}+x_{2,0}+x_{3,0}-2kx_{1,0}-2kx_{2,0}-2kx_{3,0}+2x_{1,0}x_{2,0}+2x_{1,0}x_{3,0}+2x_{2,0}x_{3,0}) 
	\end{aligned}.
\end{equation}
Now let's group the linear terms as following:
\begin{equation}
\label{1_resultant_Q_SWF_EN_eq}
\begin{aligned}
&2A3^{2}+2Ak^{2} +  \left( W_{0,1}+A \left( 1-5-2k \right)  \right) x_{0,1}+ \left( W_{0,2}+A \left( 1-5-2k \right)  \right) x_{0,2}+ \\ 
&\left( W_{0,3}+A \left( 1-5-2k \right)  \right)
x_{0,3} +A \left( W_{1,0}-A \left( 1-5-2k \right)  \right) x_{1,0}+A \left( W_{2,0}-A \left( 1-5-2k \right)  \right) x_{2,0} \\
&+A \left( W_{3,0}-A \left( 1-5-2k \right)  \right) x_{3,0}+ \left( W_{1,2}-10A \right) x_{1,2}+ \left( W_{1,3}-10A \right) x_{1,3}+ \left( W_{2,1}-10A \right) x_{2,1} \\
&+ \left( W_{2,3}-10A \right) x_{2,3}- \left( W_{3,1}-10A \right) x_{3,1}- \left( W_{3,2}-10A \right) x_{3,2}
\end{aligned}
\end{equation}
Now let's group Quadratic terms.
\begin{equation}
\label{Quad_terms_SWP_eq}
 \begin{aligned}
 & +4Ax_{0,1}x_{0,2}+4Ax_{0,1}x_{0,3}+4Ax_{0,2}x_{0,3}\\
 &+2Ax_{0,1}x_{1,2}+2Ax_{0,1}x_{1,3} +2Ax_{0,1}x_{2,1}+2Ax_{0,1}x_{2,3}+2Ax_{0,1}x_{3,1}+2Ax_{0,1}x_{3,2}+4Ax_{1,0}x_{2,0}+ 4x_{1,0}x_{3,0}\\
 &+4Ax_{2,0}x_{3,0}~+2Ax_{1,0}x_{1,2}+2Ax_{1,0}x_{1,3}~+2Ax_{1,0}x_{2,1}+2Ax_{1,0}x_{2,3}+2Ax_{1,0}x_{3,1}+2Ax_{1,0}x_{3,2}\\
 & + 2Ax_{2,1}x_{1,2}~+2Ax_{2,1}x_{1,3}+2Ax_{2,1}x_{3,0}+4Ax_{2,1}x_{3,1}+4Ax_{2,1}x_{3,2}+2Ax_{2,1}x_{0,2}+ 2Ax_{2,1}x_{0,3}+4Ax_{2,1}x_{2,3}\\
 &+2Ax_{1,2}x_{0,3}+2Ax_{1,2}x_{2,0}+2Ax_{1,2}x_{2,1}+2Ax_{1,2}x_{3,0}+2Ax_{1,2}x_{3,1}+4Ax_{1,2}x_{1,3}+4Ax_{1,2}x_{3,2}+4Ax_{1,2}x_{2,3}\\
 &+2Ax_{3,1}x_{0,2}+2Ax_{3,1}x_{1,2}+4Ax_{3,1}x_{3,2}+2Ax_{3,1}x_{0,3}+2Ax_{3,1}x_{1,3}+2Ax_{3,1}x_{2,3}\\
 &+2Ax_{1,3}x_{2,0}+ 2Ax_{1,3}x_{2,1}+4Ax_{1,3}x_{2,3} + 2Ax_{1,3}x_{3,0}+2Ax_{1,3}x_{3,1}+2Ax_{1,3}x_{3,2}\\
 &+2Ax_{0,2}x_{1,2}+ 2Ax_{0,2}x_{3,2}+ 2Ax_{0,2}x_{1,3}+2Ax_{0,2}x_{2,3}\\
 &+2Ax_{2,0}x_{2,1}+2Ax_{2,0}x_{2,3}+2Ax_{2,0}x_{3,1}+2Ax_{2,0}x_{3,2}\\ 
 &+2Ax_{3,2}x_{0,3}+ 2Ax_{3,2}x_{1,3}+ 2Ax_{3,2}x_{2,3~}\\
 &+2Ax_{2,3}x_{3,0}+2Ax_{2,3}x_{3,1}+2Ax_{2,3}x_{3,2}\\
 &+2x_{0,3}x_{1,3}+2x_{0,3}x_{23}\\ 
 & +2x_{3,0}x_{3,1}+2x_{3,0}x_{3,2} \\ 
 \end{aligned}
\end{equation}

Recalling the quadratic form  $x^{T}Qx+g^{T}x+C$ with the following terms.

\begin{equation}
\label{Sol_q_SWP_eq}
X= \left[ \begin{matrix}
x_{0,1}\\
x_{0,2}\\
x_{0,3}\\
x_{1,0}\\
x_{1,2}\\
x_{1,3}\\
x_{2,0}\\
x_{2,1}\\
x_{2,3}\\
x_{3,0}\\
x_{3,1}\\
x_{3,2}\\
\end{matrix}
 \right] 
Q = \left[ \begin{array}{cccccccccccc}
0 & 4 & 4 & 0 & 2 & 2 & 0 & 2 & 2 & 0 & 2 & 2\\
0 & 0 & 4 & 0 & 2 & 2 & 0 & 2 & 2 & 0 & 2 & 2\\
0 & 0 & 0 & 0 & 2 & 2 & 0 & 2 & 2 & 0 & 2 & 2\\
0 & 0 & 0 & 0 & 2 & 2 & 4 & 2 & 2 & 4 & 2 & 2\\
0 & 0 & 0 & 0 & 0 & 4 & 2 & 2 & 4 & 2 & 2 & 4\\
0 & 0 & 0 & 0 & 0 & 0 & 2 & 2 & 4 & 2 & 2 & 2\\
0 & 0 & 0 & 0 & 0 & 0 & 0 & 2 & 2 & 4 & 2 & 2\\
0 & 0 & 0 & 0 & 0 & 0 & 0 & 0 & 4 & 2 & 4 & 4\\
0 & 0 & 0 & 0 & 0 & 0 & 0 & 0 & 0 & 2 & 2 & 2\\
0 & 0 & 0 & 0 & 0 & 0 & 0 & 0 & 0 & 0 & 2 & 2\\
0 & 0 & 0 & 0 & 0 & 0 & 0 & 0 & 0 & 0 & 0 & 4\\
0 & 0 & 0 & 0 & 0 & 0 & 0 & 0 & 0 & 0 & 0 & 0\\
\end{array}\right]
g= \left[ \begin{matrix}
 W_{0,1}-A(4+2k)  \\
 W_{0,2}-A(4+2k)\\
 W_{0,3}-A(4+2k)  \\
 W_{1,0}-A(4+2k)  \\
 W_{1,2}-10A  \\
 W_{1,3}-10A  \\
 W_{2,0}-A(4+2k) \\
 W_{2,1}-10A  \\
 W_{2,3}-10A  \\
 W_{3,0}-A(4+2k)  \\
 W_{3,1}-10A \\
 W_{3,2}-10A  \\
\end{matrix}
\\
\right], 
\end{equation}
and
\begin{equation}
\label{costant_C_eq}
    C=2Ak^{2}+2A(n-1)^{2}, 
\end{equation}

%%%%%%%%%%%%%%%%%%%% Table No: 8 starts here %%%%%%%%%%%%%%%%%%%%

\begin{table}[]
    \centering
    \begin{tabular}{c|c|c|c|c|c|c|c|c|c|c|c|c|}

& $x_{0,1}$ & $x_{0,2}$ & $x_{0,3}$ & $x_{1,0}$ & $x_{1,2}$ & $x_{1,3}$ & $x_{2,0}$ & $x_{2,1}$ & $x_{2,3}$ & $x_{3,0}$ & $x_{3,1}$ & $x_{3,2}$ \\ \hline
$x_{0,1}$ & 0 & 4 & 4 & 0 & 2 & 2 & 0 & 2 & 2 & 0 & 2 & 2\\
$x_{0,2}$ & 0 & 0 & 4 & 0 & 2 & 2 & 0 & 2 & 2 & 0 & 2 & 2\\
$x_{0,3}$ & 0 & 0 & 0 & 0 & 2 & 2 & 0 & 2 & 2 & 0 & 2 & 2\\
$x_{1,0}$ & 0 & 0 & 0 & 0 & 2 & 2 & 4 & 2 & 2 & 4 & 2 & 2\\
$x_{1,2}$ & 0 & 0 & 0 & 0 & 0 & 4 & 2 & 2 & 4 & 2 & 2 & 4\\
$x_{1,3}$ & 0 & 0 & 0 & 0 & 0 & 0 & 2 & 2 & 4 & 2 & 2 & 2\\
$x_{2,0}$ & 0 & 0 & 0 & 0 & 0 & 0 & 0 & 2 & 2 & 4 & 2 & 2\\
$x_{2,1}$ & 0 & 0 & 0 & 0 & 0 & 0 & 0 & 0 & 4 & 2 & 4 & 4\\
$x_{2,3}$ & 0 & 0 & 0 & 0 & 0 & 0 & 0 & 0 & 0 & 2 & 2 & 2\\
$x_{3,0}$ & 0 & 0 & 0 & 0 & 0 & 0 & 0 & 0 & 0 & 0 & 2 & 2\\
$x_{3,1}$ & 0 & 0 & 0 & 0 & 0 & 0 & 0 & 0 & 0 & 0 & 0 & 4\\
$x_{3,2}$ & 0 & 0 & 0 & 0 & 0 & 0 & 0 & 0 & 0 & 0 & 0 & 0\\
    \end{tabular}
    \caption{QUBO $Q$ Matrix for the Social Workers' Problem}
    \label{tab:QUBO_Q_SWP}
\end{table}

We put equations\eqref{weight_SWP_eq} and \eqref{time_window_SWP_eq} together, and we arrive at the following expression:

\begin{equation}
\label{time_wind_eq}
    W_{ij}=d_{ij}+ \gamma \frac{ \left(  \tau_{i-} \tau_{j} \right) ^{2}}{d_{\max }-d_{\min }},
\end{equation}
There are several strategies to implement the objective function with inequality constraints, but these strategies require additional variables that work as a stack. The approach most frequently used in the scientific community is to use auxiliary binary variables to convert inequality to equality and then proceed, as usual, by squaring the equality constraint following penalty theory.
These additional variables are translated into qubits; today's extra qubits are scarce for some experiments.
To do this, we define a strategy that allows us to solve combinatorial optimisation problems with inequality constraints without increasing the number of qubits required.
Our strategy is based on coding the variables of time inequality, the time window following the formulation \eqref{time_wind_eq}.

However, IBM's significant contribution\footnote{ https://medium.com/qiskit/towards-quantum-advantage-for-optimization-with-qiskit-9a564339ef26} opens up promising horizons in quadratic programming with inequality constraints.

Now we only have to code the information in table \eqref{tab:Sample_SWP_Schedu}, the data (distance, costs and correction) of each patient and generate our weight matrix, which in turn will serve to calculate our variables linear $g$.
With all this, we already have all the components for one, write our objective function in the form QUBO and second solve it.

As we already have our objective function as a QUBO in the form  $\langle x^{T} \vert  Q \vert  x \rangle$  and as we have shown in chapter \eqref{sec:10}, based on equations \eqref{desc_qubo_Ising_9_eq} and \eqref{desc_qubo_Ising_10_eq}, and summarised in table \eqref{tab:Translator_Qubo_Ising}, going from the QUBO to Ising formulation leads to calculating the values of $J_{ij}$ and $h_{i}$.

Next, we calculate these variables. We start with $J_{ij}$  as is summarised in table \eqref{tab:J_ij for SWP}.
\begin{equation}
\label{J_i_j_cond_eq}
    J_{ij} = \left \{ \begin{matrix} \frac{q_{i,j}+q_{i,j}}{4} & i < j
\\ 0 & \text{otherwise.} \end{matrix}\right. 
\end{equation}

\begin{table}[]
    \centering
    \begin{tabular}{c|c|c|c|c|c|c|c|c|c|c|c|c|}

& $J_{1}$ & $J_{2}$ & $J_{3}$ & $J_{4}$ & $J_{5}$ & $J_{6}$ & $J_{7}$ & $J_{8}$ & $J_{9}$ & $J_{10}$ & $J_{11}$ & $J_{12}$ \\ \hline
$J_{1}$ & 0 & 4 & 4 & 0 & 2 & 2 & 0 & 2 & 2 & 0 & 2 & 2\\
$J_{2}$ & 0 & 0 & 4 & 0 & 2 & 2 & 0 & 2 & 2 & 0 & 2 & 2\\
$J_{3}$ & 0 & 0 & 0 & 0 & 2 & 2 & 0 & 2 & 2 & 0 & 2 & 2\\
$J_{4}$ & 0 & 0 & 0 & 0 & 2 & 2 & 4 & 2 & 2 & 4 & 2 & 2\\
$J_{5}$ & 0 & 0 & 0 & 0 & 0 & 4 & 2 & 2 & 4 & 2 & 2 & 4\\
$J_{6}$ & 0 & 0 & 0 & 0 & 0 & 0 & 2 & 2 & 4 & 2 & 2 & 2\\
$J_{7}$ & 0 & 0 & 0 & 0 & 0 & 0 & 0 & 2 & 2 & 4 & 2 & 2\\
$J_{8}$ & 0 & 0 & 0 & 0 & 0 & 0 & 0 & 0 & 4 & 2 & 4 & 4\\
$J_{9}$ & 0 & 0 & 0 & 0 & 0 & 0 & 0 & 0 & 0 & 2 & 2 & 2\\
$J_{10}$ & 0 & 0 & 0 & 0 & 0 & 0 & 0 & 0 & 0 & 0 & 2 & 2\\
$J_{11}$ & 0 & 0 & 0 & 0 & 0 & 0 & 0 & 0 & 0 & 0 & 0 & 4\\
$J_{12}$ & 0 & 0 & 0 & 0 & 0 & 0 & 0 & 0 & 0 & 0 & 0 & 0\\
    \end{tabular}
    \caption{$J_{ij}$  Interaction forces between grid neighbours. We assume that $J_{ij}=0$ for $i$ and $j$ are not adjacent.}
    \label{tab:J_ij for SWP}
\end{table}

Let us calculate the external forces  $h_{i}$:

\begin{equation}
\label{h_i_eq}
    h_{i}=\frac{1}{4} \sum _{k=1}^{N} \left( q_{i,k}+q_{k,i} \right) h_{i}=\frac{1}{4} \left[ \begin{array}{c}
	 \left( q_{i,1}+q_{1,i} \right) + \left( q_{i,2}+q_{3,i} \right) + \left( q_{i,3}+q_{3,i} \right) \\
	+ \left( q_{i,4}+q_{4,i} \right) + \left( q_{i,5}+q_{5,i} \right) + \left( q_{i,6}+q_{6,i} \right) \\
	+ \left( q_{i,7}+q_{7,i} \right) + \left( q_{i,8}+q_{8,i} \right) + \left( q_{i,9}+q_{9,i} \right) \\
	+ \left( q_{i,10}+q_{10,i} \right) + \left( q_{i,11}+q_{11,i} \right) + \left( q_{i,12}+q_{12,i} \right) \\
	\end{array} \right]. 
\end{equation}

Now let calculate $i=1$.

\begin{equation}
\label{h_i_1_eq}
\begin{aligned}
&h_{1}=\frac{1}{4} \left[ \begin{array}{c}
	 \left( q_{1,1}+q_{1,1} \right) + \left( q_{1,2}+q_{2,1} \right) + \left( q_{1,3}+q_{3,1} \right) \\
	+ \left( q_{1,4}+q_{4,1} \right) + \left( q_{1,5}+q_{5,1} \right) + \left( q_{1,6}+q_{6,1} \right) \\
	+ \left( q_{1,7}+q_{7,1} \right) + \left( q_{1,8}+q_{8,1} \right) + \left( q_{1,9}+q_{9,1} \right) \\
	+ \left( q_{1,10}+q_{10,1} \right) + \left( q_{1,11}+q_{11,1} \right) + \left( q_{1,12}+q_{12,1} \right) \\
	\end{array} \right] \\
	&=\frac{1}{4} \left[ \begin{array}{c}
	 \left( 0+0 \right) + \left( 4+0 \right) + \left( 4+0 \right) \\
	+ \left( 0+0 \right) + \left( 2+0 \right) + \left( 2+0 \right) \\
	+ \left( 0+0 \right) + \left( 2+0 \right) + \left( 2+0 \right) \\
	+ \left( 0+0 \right) + \left( 2+0 \right) + \left( 2+0 \right) \\
	\end{array} \right] =5.
	\end{aligned}
\end{equation}
Where  $h_{1}=5$. 
Now let calculate  $i=2$.
\begin{equation}
\label{h_1_eq}
\begin{aligned}
&h_{2}=\frac{1}{4} \left[ \begin{array}{c}
	 \left( q_{2,1}+q_{1,2} \right) + \left( q_{2,2}+q_{2,2} \right) + \left( q_{2,3}+q_{3,2} \right) \\
	+ \left( q_{2,4}+q_{4,2} \right) + \left( q_{2,5}+q_{5,2} \right) + \left( q_{2,6}+q_{6,2} \right) \\
	+ \left( q_{2,7}+q_{7,2} \right) + \left( q_{2,8}+q_{8,2} \right) + \left( q_{2,9}+q_{9,2} \right) \\
	+ \left( q_{2,10}+q_{10,2} \right) + \left( q_{2,11}+q_{11,2} \right) + \left( q_{2,12}+q_{12,2} \right) \\
	\end{array} \right] \\
	&=\frac{1}{4} \left[ \begin{array}{c}
	 \left( 0+4 \right) + \left( 0+0 \right) + \left( 4+0 \right) \\
	+ \left( 0+0 \right) + \left( 2+0 \right) + \left( 2+0 \right) \\
	+ \left( 0+0 \right) + \left( 2+0 \right) + \left( 2+0 \right) \\
	+ \left( 0+0 \right) + \left( 2+0 \right) + \left( 2+0 \right) \\
	\end{array} \right] =5.
	\end{aligned}
\end{equation}

Where $h_{2}=5$

Now let calculate $i=3$.

\begin{equation}
\label{h_3_eq}
\begin{aligned}
&h_{3}=\frac{1}{4} \left[ \begin{array}{c}
	 \left( q_{3,1}+q_{1,3} \right) + \left( q_{3,2}+q_{2,3} \right) + \left( q_{3,3}+q_{3,3} \right) \\
	+ \left( q_{3,4}+q_{4,3} \right) + \left( q_{3,5}+q_{5,3} \right) + \left( q_{3,6}+q_{6,3} \right) \\
	+ \left( q_{3,7}+q_{7,3} \right) + \left( q_{3,8}+q_{8,3} \right) + \left( q_{3,9}+q_{9,3} \right) \\
	+ \left( q_{3,10}+q_{10,3} \right) + \left( q_{3,11}+q_{11,3} \right) + \left( q_{3,12}+q_{12,3} \right) \\
	\end{array} \right] \\
	&=\frac{1}{4} \left[ \begin{array}{c}
	 \left( 0+4 \right) + \left( 0+4 \right) + \left( 0+0 \right) \\
	+ \left( 0+0 \right) + \left( 2+0 \right) + \left( 2+0 \right) \\
	+ \left( 0+0 \right) + \left( 2+0 \right) + \left( 2+0 \right) \\
	+ \left( 0+0 \right) + \left( 2+0 \right) + \left( 2+0 \right) \\
	\end{array} \right] =5.
	\end{aligned}
\end{equation}

Where $h_{3}=5$ 
Now let calculate $i=4$.

\begin{equation}
\label{h_4_eq}
\begin{aligned}
&h_{4}=\frac{1}{4} \left[ \begin{array}{c}
	 \left( q_{4,1}+q_{1,4} \right) + \left( q_{4,2}+q_{2,4} \right) + \left( q_{4,3}+q_{3,4} \right) \\
	+ \left( q_{4,4}+q_{4,4} \right) + \left( q_{4,5}+q_{5,4} \right) + \left( q_{4,6}+q_{6,4} \right) \\
	+ \left( q_{4,7}+q_{7,4} \right) + \left( q_{4,8}+q_{8,4} \right) + \left( q_{4,9}+q_{9,4} \right) \\
	+ \left( q_{4,10}+q_{10,4} \right) + \left( q_{4,11}+q_{11,4} \right) + \left( q_{4,12}+q_{12,4} \right) \\
	\end{array} \right] \\
	&=\frac{1}{4} \left[ \begin{array}{c}
	 \left( 0+0 \right) + \left( 0+0 \right) + \left( 0+0 \right) \\
	+ \left( 0+0 \right) + \left( 2+0 \right) + \left( 2+0 \right) \\
	+ \left( 4+0 \right) + \left( 2+0 \right) + \left( 2+0 \right) \\
	+ \left( 4+0 \right) + \left( 2+0 \right) + \left( 2+0 \right) \\
	\end{array} \right] =5.
	\end{aligned}
\end{equation}

Where $h_{4}=5$ 
Now let calculate $i=5$.
\begin{equation}
\label{h_5_eq}
\begin{aligned}
&h_{5}=\frac{1}{4} \left[ \begin{array}{c}
	 \left( q_{5,1}+q_{1,5} \right) + \left( q_{5,2}+q_{2,5} \right) + \left( q_{5,3}+q_{3,5} \right) \\
	+ \left( q_{5,4}+q_{4,5} \right) + \left( q_{5,5}+q_{5,5} \right) + \left( q_{5,6}+q_{6,5} \right) \\
	+ \left( q_{5,7}+q_{7,5} \right) + \left( q_{5,8}+q_{8,5} \right) + \left( q_{5,9}+q_{9,5} \right) \\
	+ \left( q_{5,10}+q_{10,5} \right) + \left( q_{5,11}+q_{11,5} \right) + \left( q_{5,12}+q_{12,5} \right) \\
	\end{array} \right] \\
	&=\frac{1}{4} \left[ \begin{array}{c}
	 \left( 0+2 \right) + \left( 0+2 \right) + \left( 0+2 \right) \\
	+ \left( 0+2 \right) + \left( 0+0 \right) + \left( 4+0 \right) \\
	+ \left( 0+2 \right) + \left( 2+0 \right) + \left( 4+0 \right) \\
	+ \left( 2+0 \right) + \left( 2+0 \right) + \left( 4+0 \right) \\
	\end{array} \right] =7.
	\end{aligned}
\end{equation}

Where  $h_{5}=7$ 
Now let calculate $i=6$.

\begin{equation}
\label{h_6_eq}
\begin{aligned}
&h_{6}=\frac{1}{4} \left[ \begin{array}{c}
	 \left( q_{6,1}+q_{1,6} \right) + \left( q_{6,2}+q_{2,6} \right) + \left( q_{6,3}+q_{3,6} \right) \\
	+ \left( q_{6,4}+q_{4,6} \right) + \left( q_{6,5}+q_{5,6} \right) + \left( q_{6,6}+q_{6,6} \right) \\
	+ \left( q_{6,7}+q_{7,6} \right) + \left( q_{6,8}+q_{8,6} \right) + \left( q_{6,9}+q_{9,6} \right) \\
	+ \left( q_{6,10}+q_{10,6} \right) + \left( q_{6,11}+q_{11,6} \right) + \left( q_{6,12}+q_{12,6} \right) \\
	\end{array} \right] \\
	&=\frac{1}{4} \left[ \begin{array}{c}
	 \left( 0+2 \right) + \left( 0+2 \right) + \left( 0+2 \right) \\
	+ \left( 0+2 \right) + \left( 0+4 \right) + \left( 0+0 \right) \\
	+ \left( 2+0 \right) + \left( 2+0 \right) + \left( 4+0 \right) \\
	+ \left( 2+0 \right) + \left( 2+0 \right) + \left( 2+0 \right) \\
	\end{array} \right] =\frac{13}{2}.
	\end{aligned}
\end{equation}

Where $h_{6}=6,5$ 

Now let calculate $i=7$.

\begin{equation}
\label{h_7_eq}
\begin{aligned}
&h_{7}=\frac{1}{4} \left[ \begin{array}{c}
	 \left( q_{7,1}+q_{1,7} \right) + \left( q_{7,2}+q_{2,7} \right) + \left( q_{7,3}+q_{3,7} \right) \\
	+ \left( q_{7,4}+q_{4,7} \right) + \left( q_{7,5}+q_{5,7} \right) + \left( q_{7,6}+q_{6,7} \right) \\
	+ \left( q_{7,7}+q_{7,7} \right) + \left( q_{7,8}+q_{8,7} \right) + \left( q_{7,9}+q_{9,7} \right) \\
	+ \left( q_{7,10}+q_{10,7} \right) + \left( q_{7,11}+q_{11,7} \right) + \left( q_{7,12}+q_{12,7} \right) \\
	\end{array} \right] = \\
	&\frac{1}{4} \left[ \begin{array}{c}
	 \left( 0+0 \right) + \left( 0+0 \right) + \left( 0+0 \right) \\
	+ \left( 0+4 \right) + \left( 0+2 \right) + \left( 0+2 \right) \\
	+ \left( 0+0 \right) + \left( 2+0 \right) + \left( 2+0 \right) \\
	+ \left( 4+0 \right) + \left( 2+0 \right) + \left( 2+0 \right) \\
	\end{array} \right] =5.
	\end{aligned}
\end{equation}
Where  $h_{7}=5$ 

Now let calculate $i=8$.

\begin{equation}
\label{h_8_eq}
\begin{aligned}
& h_{8}=\frac{1}{4} \left[ \begin{array}{c}
	 \left( q_{8,1}+q_{1,8} \right) + \left( q_{8,2}+q_{2,8} \right) + \left( q_{8,3}+q_{3,8} \right) \\
	+ \left( q_{8,4}+q_{4,8} \right) + \left( q_{8,5}+q_{5,8} \right) + \left( q_{8,6}+q_{6,8} \right) \\
	+ \left( q_{8,7}+q_{7,8} \right) + \left( q_{8,8}+q_{8,8} \right) + \left( q_{8,9}+q_{9,8} \right) \\
	+ \left( q_{8,10}+q_{10,8} \right) + \left( q_{8,11}+q_{11,8} \right) + \left( q_{8,12}+q_{12,8} \right) \\
	\end{array} \right] \\
	&=\frac{1}{4} \left[ \begin{array}{c}
	 \left( 0+2 \right) + \left( 0+2 \right) + \left( 0+2 \right) \\
	+ \left( 0+2 \right) + \left( 0+2 \right) + \left( 0+2 \right) \\
	+ \left( 0+2 \right) + \left( 0+0 \right) + \left( 4+0 \right) \\
	+ \left( 2+0 \right) + \left( 4+0 \right) + \left( 4+0 \right) \\
	\end{array} \right] =7.
	\end{aligned}
\end{equation}
Where  $h_{8}=7$

Now let calculate  $i=9$.
\begin{equation}
\label{h_9_eq}
\begin{aligned}
&h_{9}=\frac{1}{4} \left[ \begin{array}{c}
	 \left( q_{9,1}+q_{1,9} \right) + \left( q_{9,2}+q_{2,9} \right) + \left( q_{9,3}+q_{3,9} \right) \\
	+ \left( q_{9,4}+q_{4,9} \right) + \left( q_{9,5}+q_{5,9} \right) + \left( q_{9,6}+q_{6,9} \right) \\
	+ \left( q_{9,7}+q_{7,9} \right) + \left( q_{9,8}+q_{8,9} \right) + \left( q_{9,9}+q_{9,9} \right) \\
	+ \left( q_{9,10}+q_{10,9} \right) + \left( q_{9,11}+q_{11,9} \right) + \left( q_{9,12}+q_{12,9} \right) \\
	\end{array} \right] \\
	&=\frac{1}{4} \left[ \begin{array}{c}
	 \left( 0+2 \right) + \left( 0+2 \right) + \left( 0+2 \right) \\
	+ \left( 0+2 \right) + \left( 0+4 \right) + \left( 0+4 \right) \\
	+ \left( 0+2 \right) + \left( 0+4 \right) + \left( 0+0 \right) \\
	+ \left( 2+0 \right) + \left( 2+0 \right) + \left( 2+0 \right) \\
	\end{array} \right] =7.
	\end{aligned}
\end{equation}
Where $h_{9}=7$ 

Now let calculate $i=10$.
\begin{equation}
\label{h_10_eq}
\begin{aligned}
&h_{10}=\frac{1}{4} \left[ \begin{array}{c}
	 \left( q_{10,1}+q_{1,10} \right) + \left( q_{10,2}+q_{2,10} \right) + \left( q_{10,3}+q_{3,10} \right) \\
	+ \left( q_{10,4}+q_{4,10} \right) + \left( q_{10,5}+q_{5,10} \right) + \left( q_{10,6}+q_{6,10} \right) \\
	+ \left( q_{10,7}+q_{7,10} \right) + \left( q_{10,8}+q_{8,10} \right) + \left( q_{10,9}+q_{9,10} \right) \\
	+ \left( q_{10,10}+q_{10,10} \right) + \left( q_{10,11}+q_{11,10} \right) + \left( q_{10,12}+q_{12,10} \right) \\
	\end{array} \right] \\
	&=\frac{1}{4} \left[ \begin{array}{c}
	 \left( 0+0 \right) + \left( 0+0 \right) + \left( 0+0 \right) \\
	+ \left( 0+4 \right) + \left( 0+2 \right) + \left( 0+2 \right) \\
	+ \left( 0+4 \right) + \left( 0+2 \right) + \left( 0+2 \right) \\
	+ \left( 0+0 \right) + \left( 2+0 \right) + \left( 2+0 \right) \\
	\end{array} \right] =\frac{9}{2}.
\end{aligned}
\end{equation}
Where $h_{10}=4,5$ 

Now let calculate $i=11$.
\begin{equation}
\label{h_11_eq}
\begin{aligned}
& h_{11}=\frac{1}{4} \left[ \begin{array}{c}
	 \left( q_{11,1}+q_{1,11} \right) + \left( q_{11,2}+q_{2,11} \right) + \left( q_{11,3}+q_{3,11} \right) \\
	+ \left( q_{11,4}+q_{4,11} \right) + \left( q_{11,5}+q_{5,11} \right) + \left( q_{11,6}+q_{6,11} \right) \\
	+ \left( q_{11,7}+q_{7,11} \right) + \left( q_{11,8}+q_{8,11} \right) + \left( q_{11,9}+q_{9,11} \right) \\
	+ \left( q_{11,10}+q_{10,11} \right) + \left( q_{11,11}+q_{11,11} \right) + \left( q_{11,12}+q_{12,11} \right) \\
	\end{array} \right]\\
&	=\frac{1}{4} \left[ \begin{array}{c}
	 \left( 0+2 \right) + \left( 0+2 \right) + \left( 0+2 \right) \\
	+ \left( 0+2 \right) + \left( 0+2 \right) + \left( 0+2 \right) \\
	+ \left( 0+2 \right) + \left( 0+4 \right) + \left( 0+2 \right) \\
	+ \left( 0+2 \right) + \left( 0+0 \right) + \left( 4+0 \right) \\
	\end{array} \right] =\frac{13}{2}.
	\end{aligned}
\end{equation}
Where  $h_{11}=6,5$ 

Now let calculate $i=12$.
\begin{equation}
\label{h_12_eq}
\begin{aligned}
& h_{12}=\frac{1}{4} \left[ \begin{array}{c}
	 \left( q_{12,1}+q_{1,12} \right) + \left( q_{12,2}+q_{2,12} \right) + \left( q_{12,3}+q_{3,12} \right) \\
	+ \left( q_{12,4}+q_{4,12} \right) + \left( q_{12,5}+q_{5,12} \right) + \left( q_{12,6}+q_{6,12} \right) \\
	+ \left( q_{12,7}+q_{7,12} \right) + \left( q_{12,8}+q_{8,12} \right) + \left( q_{12,9}+q_{9,12} \right) \\
	+ \left( q_{12,10}+q_{10,12} \right) + \left( q_{12,11}+q_{11,12} \right) + \left( q_{12,12}+q_{12,12} \right) \\
	\end{array} \right] \\
	&=\frac{1}{4} \left[ \begin{array}{c}
	 \left( 0+2 \right) + \left( 0+2 \right) + \left( 0+2 \right) \\
	+ \left( 0+2 \right) + \left( 0+4 \right) + \left( 0+2 \right) \\
	+ \left( 0+2 \right) + \left( 0+4 \right) + \left( 0+2 \right) \\
	+ \left( 0+2 \right) + \left( 0+4 \right) + \left( 0+0 \right) \\
	\end{array} \right] =7.
	\end{aligned}
\end{equation}

Where $h_{12}=7$ 
\begin{table}[]
    \centering
    \begin{tabular}{|c|c|} \hline
        $h_{1}$ & 5 \\ \hline
        $h_{2}$ & 5 \\ \hline
        $h_{3}$ & 5 \\ \hline
        $h_{4}$ & 5 \\ \hline
        $h_{5}$ & 7 \\ \hline
        $h_{6}$ & 6.5 \\ \hline
        $h_{7}$ & 5 \\ \hline
        $h_{8}$ & 7 \\ \hline
        $h_{9}$ & 7 \\ \hline
        $h_{10}$ & 4.5 \\ \hline
        $h_{11}$ & 6.5 \\ \hline
        $h_{12}$ & 7 \\ \hline
    \end{tabular}
    \caption{Calculated values of the external force $h_{i}$}
    \label{tab:h_i_coeffs}
\end{table}

With the calculated $J_{ij}$ (table \eqref{tab:J_ij for SWP}) and  $h_{i}$ (table \eqref{tab:h_i_coeffs}), we can now solve our Social Workers' Problem with VQE $\langle \psi \left(\theta  \right)   \vert  H  \vert   \psi   \left( \theta  \right) \rangle$  or QAOA $\langle \overrightarrow{ \gamma },\overrightarrow{ \beta } \vert  H \vert  \overrightarrow{ \gamma },\overrightarrow{ \beta } \rangle$, or with any mentioned variational method. We are considering that the superposition state  $N$  qubits of   $Q= \{q_{1} \cdots q_{N}\}$  is described by  $\vert   \psi  \rangle = \vert  \psi_{1} \cdots   \psi_{N} \rangle$.

\begin{equation}
\label{Ising_HP_eq}
    H_{P} \vert  \psi  \rangle = \sum _{i}^{N}h_{i} \sigma _{i}^{z}+ \sum _{i<j}^{N}J_{ij} \sigma _{i}^{z} \sigma _{j}^{z}  \vert \psi  \rangle. 
\end{equation}
The energy function is given by the equation \eqref{Ising_HP_eq} and where the notation $\sigma _{i}^{z}$ means that the Pauli-$Z$ operator is applied to the single-qubit following this approach $\vert  \cdots   \psi_{i} \cdots  \rangle$ according to \eqref{sigma_Z_i} and \eqref{sigma_Z_ij}.

%\begin{equation}
%\label{Pauli_Z_Operator_eq}
% \in_{I}= \sum _{i \in V}^{}h_{i} \left( -1 \right) ^{ \psi i}+ \sum _{ \left( i,j %\right)  \in V}^{}J_{ij} \left( -1 \right) ^{ \psi i} \left( -1 \right) ^{ \psi j}.
%\end{equation}

%Another view of the equation \eqref{Pauli_Z_Operator_eq} is given by \eqref{sigma_Z_i} and \eqref{sigma_Z_ij}.

\begin{equation}
\label{sigma_Z_i}
 \sigma _{i}^{z} \longrightarrow \mathop{\underbrace{\mathop{I}}}_{\mathop{1}^{st}~position}\otimes \cdots \otimes\mathop{\underbrace{\mathop{\mathop{ \sigma }^{z}}}}_{\mathop{i}^{th}~position}\otimes \cdots \otimes\mathop{\underbrace{\mathop{I}}}_{\mathop{N}^{th}~position}.
\end{equation}

\begin{equation}
\label{sigma_Z_ij}
 \sigma _{i}^{z} \sigma _{j}^{z}\xrightarrow{i<j}\mathop{\underbrace{\mathop{I}}}_{\mathop{1}^{st}~position}\otimes \cdots \otimes\mathop{\underbrace{\mathop{\mathop{ \sigma }^{z}}}}_{\mathop{i}^{th}~position}\otimes \cdots \otimes\mathop{\underbrace{\mathop{\mathop{ \sigma }^{z}}}}_{\mathop{j}^{th}~position}\otimes \cdots \otimes\mathop{\underbrace{\mathop{I}}}_{\mathop{N}^{th}~position}.
\end{equation}

\subsection{The algebraic approach of the SWP formulation}

We are looking for a more compact formulation to make it easier to codify in quantum, and to achieve this, we vectorise the decision variables $x_{ij}$.
We know that vectorisation \cite{Hug12,Wil19} of a matrix is a linear transformation that converts the matrix into a column vector. So, let  $\text{vec} \left( X \right)$, the column vector  $mn \times 1$  obtained by stacking the columns of the matrix $X$ one on top of the other:
\begin{equation}
\label{vect_form_eq}
    \text{vec}(X) = \left[ x_{1,1}, \ldots , x_{1,m}, \ldots , x_{1,n}, \ldots ,x_{m,n} \right]^{T}.
\end{equation}
We also know that vectorisation is frequently used with the Kronecker product to express matrix multiplication as a linear transformation in matrices. In particular:

\begin{equation}
\label{Kronecker_eq1}
\begin{aligned}
    & \text{vec}(ABC) = (C^{T}\otimes A) \text{vec}(B) = (I^{T}\otimes AB) \text{vec} (C). 
\end{aligned}
\end{equation}

\begin{equation}
\label{Kronecker_eq2}
\begin{aligned}
    \text{vec}(BC) = (I\otimes B) \text{vec}(C). 
\end{aligned}
\end{equation}

\begin{equation}
\label{Kronecker_eq3}
\begin{aligned}
    \text{vec}(CB) = (B^T \otimes I) \text{vec}(C). 
\end{aligned}
\end{equation}

If we apply vectorisation as a linear sum, the matrix vectorisation operation can be written in terms of a linear sum. Let $e_{i}$ be the $n^{th}$ canonical base vector for $n$ dimensional space, that is:

\begin{equation}
\label{canonical_base_eq}
    e_{i}= \left[ 0 , \ldots , 0 , \ldots ,0 , \ldots ,1 , \ldots ,0 , \ldots ,0 \right] ^{T}.
\end{equation}
Let  $B_{i}$ a block matrix  $nm \times m$  defined as follows:

\begin{equation}
\label{197}
B_{i}= \begin{bmatrix}
0\\
 \vdots \\
0\\
I_{m}\\
0\\
 \vdots \\
0\\
\end{bmatrix}
 =e_{i}\otimes I_{m}.
\end{equation}
$B_{i}$  consists of $n$ block matrices of size $m  \times  m$, stacked in columns, and all these matrices are all zero except the $i^{th}$, which is an identity matrix $m \times mI_{m}$.

Then the vectorised version of  $X$  can be expressed as follows:
\begin{equation}
\label{vectorization_form_eq}
    \text{vec}(X) = \sum _{i=1}^{n}B_{i}Xe_{i}.
\end{equation}
The multiplication of  $X$ by  $e_{i}$  extracts the $i^{th}$ column, while the multiplication by  $B_{i}$  places it in the desired position in the final vector. Alternatively, the linear sum can be expressed using the Kronecker product:

\begin{equation}
\label{Base_Kronecker_pd_eq}
\text{vec} (X) = \sum _{i=1}^{n} (e_{i}\otimes I_{m}) Xe_{i}= \sum _{i=1}^{n} (e_{i}\otimes I_{m})X^{T}.
\end{equation}

With  $Xe_{i}= X^{T}$. Where \textsuperscript{T} is the transpose.

Now, let us define $Z$ with $n^2$ dimensions as follows:

\begin{equation}
\label{Z_definition}
   Z = (X_{11}X_{12}X_{13} \dots X_{1n}X_{21}X_{22}\dots X_{2n} \dots X_{nn} )^T.
\end{equation}

We can simplify our objective function to be implemented in quantum algebraically. Let us demonstrate that the following expression holds:
\begin{equation}
\label{Simplificatio_ZTOX}
    \sum_{i=1}^{n}(\sum_{j=1}^{n}X_{ij}-1)^2 = \sum_{i=1}^{n}(e_i\otimes 1_{n}^TZ-1)^2.
\end{equation}
Thus, we can say:
\begin{equation}
\label{ZequaltoX}
    (e_i \otimes 1_{n}^T)Z = \sum_{j=1}^{n}X_{ij}.
\end{equation}

%Knowing that:
%$e_i = (0 \dots 0  \ldots 1 \dots 0 \dots 0)$ , $1_{n}^T = (1 \dots 1)$, both have dimension $n$ and $Z = (X_{11}X_{12}X_{13} \dots X_{1n}X_{21}X_{22}\dots X_{2n} \dots X_{nn} )$ with $n^2$ dimensions.

With 
\begin{equation}
\label{dem_11}
    e_i = (0 \dots 0  \ldots 1 \dots 0 \dots 0),
\end{equation}

and 
\begin{equation}
\label{dem_12}
    1_{n}^T = (1 \dots 1),
\end{equation}
both have dimension $n$.

Let us write down $e_i \otimes 1_{n}^T$ taking in account all the definitions. 

%$e_i = (0 \dots 0  \ldots 1 \dots 0 \dots 0)$, $1_{n}^T = (1 \dots 1)$, both have dimension $n$ and $Z = (X_{11}X_{12}X_{13} \dots X_{1n}X_{21}X_{22}\dots X_{2n} \dots X_{nn} )^T$ with $n^2$ dimensions.

\begin{equation}
\label{Z_algebraic}
\begin{aligned}
  &e_i \otimes 1_{n}^T = (0\cdot (1 \ldots 1) \dots 1\cdot (1 \ldots 1)\ldots 0\cdot (1 \ldots 1) \dots ) \\
  &= (0000 \ldots 0000 \ldots 1111 \ldots 0000 \ldots 0000),
\end{aligned}
\end{equation}

%$e_i \otimes 1_n = (0\cdot (1 \ldots 1) \dots 1\cdot (1 \ldots 1)\ldots 0\cdot (1 \ldots 1) \dots ) = (0000 \ldots 0000 \ldots 1111 \ldots 0000 \ldots 0000)$.

According to the equation \eqref{Z_definition}, we can calculate the following expression as:

\begin{equation}
\label{SWP_algebraically}
\begin{aligned}
  &(e_{i} \otimes 1_{n}^{T})\cdot Z = 0\cdot X_{11} + 0\cdot X_{12} + \ldots 0\cdot X_{1n} \ldots + \ldots 1\cdot X_{i1} + 0\cdot X_{i2} + \ldots + 1\cdot X_{in} \\
  &+ 0\cdot X_{in+1}+ \ldots = X_{i1}+ \ldots + X_{in} = \sum_{j=1}^{n}X_{ij}.
\end{aligned}
\end{equation}
%$(e_i \otimes 1_n)\cdot Z = 0\cdot X_{11} + 0\cdot X_{12} + \ldots 0\cdot X_{1n} \ldots + \ldots 1\cdot X_{i1} + 0\cdot X_{i2} + \ldots + 1\cdot X_{in} + 0\cdot X_{in+1}+ \ldots = X_{i1}+ \ldtos + X_{in} = \sum_{j=1}^{n}X_{ij}$. 

In the end, we see that we arrive at the expression \eqref{ZequaltoX} we want. So, now we only have to substitute it into our simplified SWP formulation \eqref{SWP_FORM_eq}.

\subsection{Vectorisation form of our formulation}\label{sec:SWP-Vector-QUBO}
This section will develop the vectorisation form of the SWP.  Let $Z$  the vector of the decision variables $X_{ij}$  with  $Z \in  \{0,1\}^{N}$  and  $N=n(n-1)$:
\begin{equation}
\label{Decision_var_eq}
Z= [X_{01},X_{02},X_{03}, \ldots ,X_{10},X_{12},X_{13}, \ldots ,X_{n \left( n-1 \right) }]^{T},
\end{equation}

In addition, let us denote $v_i = (Z_{i^{'}j}^{'})_{i^{'}j}$ with:
\begin{equation}
\label{V_def}
   Z_{i^{'}j}^{'} = \left \{ \begin{matrix}1 \quad & \text{if} \quad j=i \quad \text{for any} \quad i^{'}\\
 0 & \text{otherwise,} \end{matrix}\right. 
\end{equation}
 and let us denote $v_0 = (Z_{ij}^{''})$ with:
 \begin{equation}
\label{V_0_def}
   Z_{ij}^{''} = \left \{ \begin{matrix}1 \quad & \text{if} \quad j=0  
\\ 0 & \text{otherwise.} \end{matrix}\right. 
\end{equation}
 
Now applying equations \eqref{Decision_var_eq}, \eqref{Base_Kronecker_pd_eq}, \eqref{V_def} and \eqref{V_def}  into the simplified SWP formulation \eqref{SWP_FORM_eq}, we arrive at the next equation.
 
\begin{equation}
\label{descomp_Funct_Objective_eq}
H=W^{T}Z+A \sum _{i=1}^{n} \left( 1- \left( e_{i}\otimes I_{n}^{T} \right) Z \right) ^{2}+A \sum _{i=1}^{n} \left( 1-v_{i}^{T}Z \right) ^{2}+A \left( k- \left( e_{0}\otimes I_{n} \right) ^{T}Z \right) ^{2}+A \left( k-v_{0}^{T}Z \right) ^{2}.
\end{equation}
With:
\begin{equation}
\label{T_W_SWP_eq}
W_{ij}=d_{ij}+ \gamma \frac{ \left(  \tau_{i-} \tau_{j} \right) ^{2}}{d_{\max }-d_{\min }}.
\end{equation}
Now, let’s develop the resultant equation.
\begin{equation}
\label{descomp_swp_Kro_eq}
\begin{aligned}
&H = W^{T}Z+ A \sum _{i=1}^{n} \left(  \left(  \left( e_{i}\otimes I_{n}^{T} \right) Z \right) ^{2}-2 \left( e_{i}\otimes I_{n}^{T} \right) Z +1 \right)  + A \sum _{i=1}^{n} \left(  \left( v_{i}^{T}Z \right) ^{2} -2v_{i}^{T}Z+1 \right) \\
&+ A \left[  \left(  \left( e_{0}\otimes I_{n} \right) ^{T}Z \right) ^{2}-2k \left( e_{0}\otimes I_{n} \right) ^{T}Z+k^{2} \right] + A \left[  \left( v_{0}^{T}Z \right) ^{2} -2k  \left( v_{0}^{T}Z \right) +k^{2} \right], \\
\\
&H= W^{T}Z+A \sum _{i=1}^{n} \left[  \left( e_{i}\otimes I_{n}^{T} \right) Z \right] ^{2}+ \left[ v_{i}^{T}Z \right] ^{2}-2A \left[  \left( e_{i}\times I_{n}^{T} \right) +v_{i}^{T} \right] Z+2A+A [  \left[  \left( e_{0}\otimes I_{n}^{T} \right) +v_{0}^{T} \right] ^{2}Z \\
&- 2AK  (e_{0}\otimes I_{n}^{T})+v_{0}^{T}Z  + 2Ak^{2}] .
\end{aligned}
\end{equation}

Regrouping the terms in the quadratic formulation $Z^{T}QZ +g^{T}Z~ +C$:

\begin{equation}
\label{decompo_suite_Eq}
\begin{aligned}
&H= A \sum _{i=1}^{n} \left[  \left( e_{i}\otimes I_{n} \right) ^{2}Z^{2}+ \left[ v_{i}^{T} \right] ^{2}Z^{2} \right] +w-2A \sum _{i=1}^{n} \left[  \left( e_{i}\otimes I_{n}^{T} \right) +v_{i}^{T} \right]\\
&-2Ak  \left[  \left( e_{0}\otimes 1_{n} \right) ^{T}+v_{0}^{T} \right] +2An+2Ak^{2}.
\end{aligned}
\end{equation}

With the variables  $Q$,  $g$ and  $C$:

\begin{equation}
\label{Q_Matrix_eq}
Q= \sum_{i=1}^{n} \left[  \left( e_{i}\otimes I_{n} \right) ^{2}+ \left[ v_{i}^{T} \right]^{2} \right] = \sum_{i=1}^{n} [ (e_{i}\otimes I_{n})  (e_{i}\otimes I_{n} ^{T})+ [ v_{i}v_{i}^{T}]]. 
\end{equation}

\begin{equation}
\label{g_matrix_eq}
g=w-2A \sum_{i=1}^{n} \left[  \left( e_{i}\otimes I_{n}^{T} \right) +v_{i}^{T} \right] -2Ak  \left[ ( e_{0}\otimes 1_{n})^{T} + v_{0}^{T} \right]. 
\end{equation}

\begin{equation}
\label{C_matrix_eq}
C=+2An+2Ak^{2}.
\end{equation}

From this point, we can use any solver based on annealing to solve we formulation. Then, if we want to translate it into the gate-based computer, we will only need to map it to the Ising model and select which solver could be adequate. 

\subsection{Mapping the SWP's formulation into a list of Pauli operators}

There is a very close relationship between \textit{Clifford's algebras} \cite{Per97} with Quantum computing. Clifford's algebras are abstract structures that have been widely used in various theories of the physical and mathematical types. In particular, they favoured a comprehensive treatment of operators involved in Quantum Computing. For example, Clifford's group defines the computational subspace operations on the Hilbert vector space.
\begin{equation}
\label{Paulis_Matrix_eq}
\begin{aligned}
 \sigma _{0}=I= \left( \begin{matrix}
1  &  0\\
0  &  1\\
\end{matrix}
 \right) , \sigma _{1}= X= \left( \begin{matrix}
0  &  1\\
1  &  0\\
\end{matrix}
 \right) ,~ \sigma _{2}=Y= \left( \begin{matrix}
0  &  -i\\
i  &  0\\
\end{matrix}
 \right) ~\text{and}~  \sigma _{3}=Z= \left( \begin{matrix}
1  &  0\\
0  &  -1\\
\end{matrix}
 \right). 
\end{aligned}
\end{equation}
The \textit{Clifford group} \cite{Mic00} is the group of unitaries that normalise the \textit{Pauli group} \cite{Mic00} and use the Pauli group for quantum computing. The Pauli group on one qubit is the 16-element matrix group consisting of the  $2 \times 2$ identity matrix $~I$  and all of the Pauli matrices together with the products of these matrices with the factors  $\pm 1$  and  $\pm i$: 

\begin{equation}
\label{Grup_Pauli_eq}
G_{1}= \{  \pm I,  \pm iI, \pm X, \pm iX, \pm Y, \pm iY, \pm Z, \pm iZ, \}  \equiv  \langle X,Y,Z \rangle. 
\end{equation}

In the case of  $n$  qubits, the Pauli group is defined as $G_{n}$  in the tensor product of the Hilbert space  $\left( C^{2} \right) ^{\bigotimes n}$.

\begin{equation}
\label{Grup_Pauli_n_eq}
G_{n}= \{  \pm I,  \pm iI, \pm X, \pm iX, \pm Y, \pm iY, \pm Z, \pm iZ, \} ^{\bigotimes n}.
\end{equation}

With  $2^{n} \times 2^{n}$  matrix acting on an $n$ qubits Hilbert space. Using the Pauli group properties, we can avoid doing costly matrix operations.

%The Pauli group is also used for the correction of the quantum error.

\subsubsection{Mapping of quadratic variables of the Hamiltonian in the Z-basis.}
With the QUBO form of the SWP (equation \eqref{decompo_suite_Eq}) from the Kronecker product and vectorisation, we can now map this $Q$ \eqref{Q_Matrix_eq}, $g$ \eqref{g_matrix_eq} and $C$ \eqref{C_matrix_eq} into the computational Z-basis by using the Pauli’s operators.

\begin{equation}
\label{Qz_Mapping_eq}
Q_{z} = \left( \frac{Q}{4} \right), 
\end{equation}
\begin{equation}
\label{gz_Mapping_eq}
    g_{z}= \left( -\frac{g~}{2}- \left( I_{v}~\otimes~ \frac{Q}{4} \right)  -  \left( \frac{Q}{4}~\otimes~ I_{v} \right)  \right), 
\end{equation}
\begin{equation}\tag{213}
c_{z} =  \left( c+ \left( \frac{g}{2}~\otimes~ I_{v} \right) + \left( I_{v}~\otimes  \left( \frac{Q}{4}~\otimes~ I_{v} \right)  \right)  \right), 
\end{equation}
\begin{equation}
\label{cz_Mapping_eq}
    c_{z}= c_{z} + tr \left( Q_{z} \right),
\end{equation}

\begin{equation}
\label{Qz_eq}
Q_{z}=Q_{z} - \left( \text{diag} \left( Q_{z} \right)  \right). 
\end{equation}
\subsubsection{Coding the Hamiltonian into the Pauli Operators}
We can now implement our algorithm with the Aqua library of the Qiskit framework. Using the group theory that forms the Pauli operators to reduce computational costs by defining operation over the Pauli components. To do this, we can decompose the Hamiltonian of Ising into a list of the Pauli operators that make up the Clifford group (a set of mathematical transformations which affect permutations of the Pauli operators.) $P_{n}= { I ,( X,Y,Z)}$.

\subsubsection{Quantum solution from the ground up and closer to the Quantum Native form.}
For quantum solver, we used Qiskit. Here, we embedded some, IBMQ functions as \textit{WeightedPauliOperator}, \textit{NumPyMinimumEigensolver} and \textit{TwoLocal}. Nevertheless, for the \textit{generic solver}, all these functions can be rewritten quickly.

First, we derived the solution from the ground up, using a class \textit{QuantumOptimiser} that encodes the quantum approach to solve the problem. Then we instantiated it and solved it. We defined the following methods inside the class:

\begin{enumerate}
	\item \textit{binary\_representation}: encodes the proposed problem into the Hamiltonian of the Ising model according to the quadratic programming form.

	\item construct\_hamiltonian: constructs the Hamiltonian of the Ising model in terms of the $Z$ basis.

	\item check\_hamiltonian: makes sure that the Hamiltonian of the Ising model is correctly encoded in the $Z$ basis: to do this, the function solved an eigenvalue-eigenvector problem for a symmetric matrix of dimension  $2^{N} \times 2^{N}$. For the complexity limit $N=n \left( n-1 \right)$.

	\item vqe\_solution: solves the proposed problem via VQE by using the SPSA solver with default parameters.

	\item \_q\_solution: internal routine to represent the solution in a usable format.

	\item qaoa\_solution: solves the proposed problem via QAOA by using the SPSA solver with default parameters.
\end{enumerate}

%All these six steps, according to our objective function defined by the equations \eqref{decompo_suite_Eq}. These steps are necessary for quantum programming.

With these steps, we solve our problem through the exact result, VQE and QAOA. Next, we will discuss the other techniques used to check the good functioning of the code in quantum computing. But for this, we need the docplex and the library of quadratic programming.

\subsection{Resolution of our formulation with docplex}
We apply the docplex model to the proposed question in equations \eqref{SWP_FORM_eq}, \eqref{SWP_const_1_eq}, \eqref{SWP_const_2_eq}, \eqref{SWP_const_4_eq}, \eqref{SWP_const_5_eq} and \eqref{SWP_const_7_eq}. All the code can be found in  \cite{QSWPGithub,Ade21,qRobotP,EVA_} .

Now, we can solve the model (the problem) with all classical and quantum techniques.

\subsection{CPLEX solver}

We used the \textit{CPLEX solver} to solve our algorithm. From May 2020, Qiskit supports Quadratic Constrained Programming under docplex with binary, integer, and continuous variables, as well as equality and inequality constraints.

\begin{figure}[!h]
    \centering
    \includegraphics[width=0.8\textwidth]{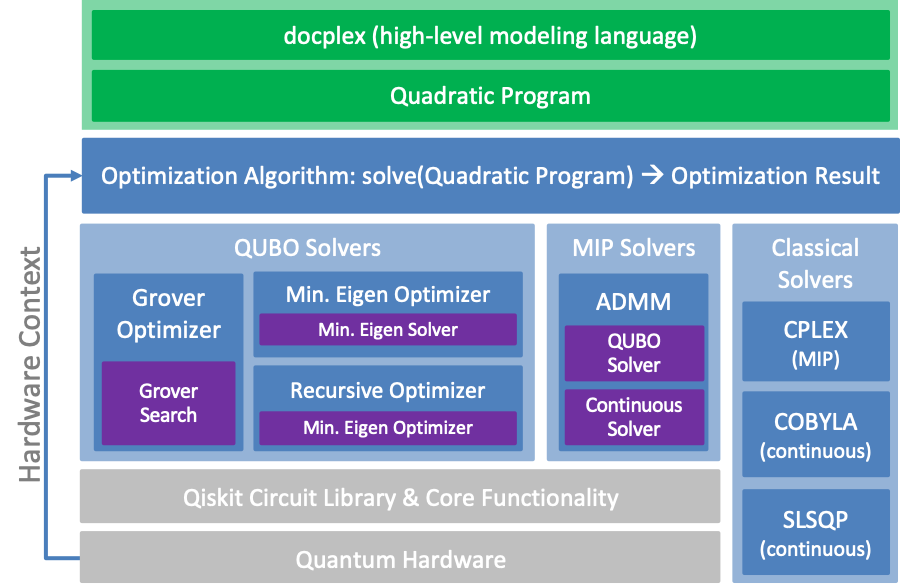}
    \caption{Quantum Optimisation in Qiskit. A high-level modelling language, a modular set of state-of-the-art quantum optimisation algorithms for different problem classes, leveraging Qiskit's fundamental quantum algorithms as well as core circuit functionality; combined with a uniform and flexible interface for easy testing, benchmarking, and validation of models and algorithms \cite{MinimumEigenOptimizer}.}
    \label{fig:Quantum_Opt_qiskit}
\end{figure}

In Chapter \eqref{sec:Solving_Combina_Pr}, we have seen how to solve a quadratic function. During these last months, as discussed above, Qiskit has developed a library \eqref{fig:Quantum_Opt_qiskit} to solve quadratic functions, which we will analyse below.

\subsection{Solving our Quantum SWP with MinimumEigenOptimiser}
To solve the SWP, we needed to use the Quadratic Program from IBM, so let us introduce the MinimumEigenSolver and MinimumEigenOptimiser \cite{MinimumEigenOptimizer}. 
Qiskit provides automatic conversion from a suitable Quadratic Program to an Ising Hamiltonian, permitting the leverage of all the QUBO solvers and the Minimum Eigen Solver VQE, QAOA, or any other classical exact method solver (Fig.\eqref{fig:Quantum_Opt_qiskit}).

Qiskit wraps the translation to an Ising Hamiltonian (in Qiskit Aqua, also called Operator), the call to a \textit{Minimum Eigensolver} as well as the conversion of the results back to Optimisation Result in the \textit{Minimum Eigen Optimiser}.

In the following solver, we first illustrate the conversion from a \textit{Quadratic Program} to an \textit{Operator} and then show how to use the Minimum Eigen Optimiser with different \textit{Minimum Eigen solver} to solve a given Quadratic Program; let us focus specifically on QUBO. The algorithms in Qiskit automatically try to convert a given problem to the supported problem class if possible; for instance, the Minimum Eigen Optimiser will automatically translate integer variables to binary variables or add a linear equality constraint as a quadratic penalty term to the objective. It also calculates the best Lagrange multiplier for that formulation. It is worth recalling that the user needs to install CPLEX packages for the classical exact solver.

The circuit depth of QAOA potentially has to be increased with the problem size, which might be prohibitive for NISQ devices. A possible alternative is Recursive QAOA or also the recursive Grover. Qiskit generalises this concept to the Recursive Minimum Eigen Optimiser.

\subsection{Solving our quantum SWP by using the ADMM Optimiser}

Another technique we used to solve our proposed problem is the ADMM Optimiser \cite{Cla20}. The ADMM optimiser can solve classes of mixed-binary constrained optimisation problems (MBCO) as a QUBO subproblem on the quantum device via variational algorithms (VQE, QAOA) and continuous convex constrained subproblem, which can be efficiently solved with classical optimisation solvers.

To solve our formulation with the ADMM module, we follow these steps:

We\ first initialise all the algorithms we plan to use.  We initialise this case by setting the phase's parameters and the QUBO and convex solvers. The default parameters for ADMM modules  $  \rho =1001,~~ \beta =1000$ can be used as starting points with the penalisation factor\textit{\_c= 900} of equality constraints  $Gx=b$. The tolerance for primal residual convergence is set to $e^{-6}$. The 3-block implementation of the ADMM is guaranteed to converge for Theorem number 4 of the ref: \cite{Cla20}, because we active the inequality constraint of the continuous variable as recommended by \cite{Cla20}. More details can be found at \cite{MinimumEigenOptimizer}.

%As we introduced above, to solve the Quadratic Programming from qiskit as QUBO problems can be chosen between all the following packages. We are talking about Minimum Eigen Optimizer using different Minimum Eigensolver, such as VQE, QAOA or Numpy Minimum Eigensolver (classical), GroverOptimizer, CplexOptimizer (the user needs to install CPLEX is installed. Remember that it is for classical). 

For the classical optimiser (COBYLA, SLSQP, etc.)  as shown in the Fig. \eqref{fig:Quantum_Opt_qiskit} and analysed in chapter 9, we used the \textit{COBYLA Optimiser} and the \textit{Minimum Eigen Optimiser} using the Numpy Minimum Eigensolver for QUBOs as classical alternatives to CPLEX for testing, validation, and benchmarking.
In the next step, we followed each ADMM iteration ($ k=1,2, \ldots,$) until termination; we solved our proposed QUBO problem with the quantum solver and updated the dual variables. The last step is to return optimisers and cost variables.
All the experiments code can be found at \cite{QSWPGithub,Ade21,qRobotP,EVA_, QSWPGithub, AllSWPGithub}, where the reader can analyse the files of the resolution of our algorithms.

\subsection{Backtracking}
One of the classical techniques that we experimented with to solve our proposed problem is Backtracking \cite{Schmidt1976,Civicioglu2013}. We also did this study to make comparisons of the methods that we will review in the chapter reserved for discussions. However, the Social Workers Problems that are discussed can be solved classically using other known algorithms like the brute force; Dynamic Programming \cite{Bellman1966,SDr02} or Greedy Algorithms. %In logarithmic, linear-logarithmic time complexity in input data size, and therefore, in some cases (without regard to the brute force), outshine the backtracking algorithm in every respect.

Backtracking is not the most efficient algorithm to solve this problem, but it is suitable for the type of CSP problem (Constraint Satisfaction Problems). It also allows us to analyse the heuristics that we have designed to reduce the number of qubits. Backtracking depends on a user-given scenario that defines the problem to be solved, the nature of the partial solutions, and how they are scaled into complete solutions. It is, therefore, a metaheuristic rather than a specific algorithm; it is guaranteed to find all outcomes to a finite (limited) problem in a bounded amount of time since backtracking algorithms are generally exponential in both time and space. This algorithm is far from what could be considered an optimal resolution in terms of computational cost since it continues to offer an exponential cost of $O(M^N)$, where $N$ is the maximum depth of the search tree and $M$ the number of the social workers.

In our case, we use the Backtracking algorithm to consider all possible issues within the constraints. 

\subsection{First approach of the SWP's generalisation}\label{sec:SWP-first-generalization}
This section will propose a more generic formulation that does not consider the limitations of the era in which we are. This formulation can be programmed on any Quantum Annealing computer or Based-Gates quantum computer.

Let  $H$ be our new Hamiltonian as follow,
\begin{equation}
\label{Hamiltonian_eq}
H=H_{m}+H_{c}+H_{t},
\end{equation}
with
\begin{equation}
\label{217}
H_{m}=A \sum _{i=1}^{n+1} \sum _{u=0}^{n} \sum _{\begin{matrix}
 \upsilon =0\\
 \upsilon  \neq u\\
\end{matrix}
}^{n}x_{u, \upsilon }^{i}W_{u, \upsilon }.
\end{equation}
\begin{equation}
\label{Full_hamiltonian_eq}
\begin{aligned}
&H_{c}=B(1- \sum _{ \upsilon =1}^{n}x_{0, \upsilon }^{1}) ^{2}+B \sum _{i=2}^{n} \left( 1- \sum _{u=1}^{n} \sum _{\begin{matrix}
 \upsilon =1\\
 \upsilon  \neq u\\
\end{matrix}
}^{n}x_{u, \upsilon }^{i} \right) ^{2}+B \left( 1- \sum _{u=1}^{n}x_{ \upsilon ,0}^{n+1} \right) ^{2}\\
&+B \sum _{u=1}^{n} \left( 1- \left( x_{u,0}^{n+1}+ \sum _{i=2}^{n} \sum _{\begin{matrix}
 \upsilon =1\\
 \upsilon  \neq u\\
\end{matrix}
}^{n}x_{u, \upsilon }^{i} \right)  \right) ^{2}+B \sum _{ \upsilon =1}^{n} \left( 1- \left( x_{0, \upsilon }^{1}+ \sum _{i=2}^{n} \sum _{\begin{matrix}
u=1\\
u \neq  \upsilon \\
\end{matrix}
}^{n}x_{u, \upsilon }^{i} \right)  \right) ^{2}{\fontsize{10pt}{12.0pt}\selectfont }.
\end{aligned}
\end{equation}

\begin{equation}
\label{Full_H_SWP_eq}
\begin{aligned}
H_{t}= \tau_{1}+ \tau \sum _{i=2}^{n} \tau_{i}W_{i}+ \tau \sum _{i=1}^{n} \Lambda_{i},
\end{aligned}
\end{equation}

where
\begin{equation}
\label{Time_Windows_Full_eq}
 \Lambda _{i}= \left( 1- \sum _{k=1}^{k}t_{k,i} \right) ^{2}~~~  \left( 1 \leq i \leq n \right).
\end{equation}

Where   $\upsilon$ \textit{ }is with a time window\textit{   $\left[ e_{ \upsilon },l_{ \upsilon } \right]$}

\begin{equation}
\label{1_Time_Windows_eq}
T_{1}= \left(  \sum _{ \upsilon =1}^{n}x_{0, \upsilon }^{1}w_{0, \upsilon }+ \sum _{k=1}^{k}kt_{k,i}- \sum _{ \upsilon =1}^{n}x_{0, \upsilon }^{1}l_{ \upsilon } \right) ^{2}.
\end{equation}

\begin{equation}
\label{Full_Time_windows_eq}
T_{i}= \left(  \sum _{d=1}^{i} \sum _{u=1}^{n} \sum _{\begin{matrix}
 \upsilon =1\\
 \upsilon  \neq 1\\
\end{matrix}
}^{n}x_{u, \upsilon }^{d}w_{u, \upsilon }+ \sum _{k=1}^{k}kt_{k,i}- \sum _{u=1}^{n} \sum _{\begin{matrix}
 \upsilon =1\\
 \upsilon  \neq u\\
\end{matrix}
}^{n}x_{u, \upsilon }^{i}l_{ \upsilon } \right) ^{2}  \left( 2 \leq i \leq n \right). 
\end{equation}

\begin{equation}
\label{Binary_variable_t_ij}
t_{k,i}= \left \{ \begin{matrix}
1,~~\\
0,\\
\end{matrix}
 \begin{matrix}
\text{the time margin in the tour k}\\
\text{otherwise}. \\
\end{matrix}\right.
\end{equation}

The binary variable $i$ represent the position of the tour. The constants  $B$, $A$ and  $\tau$ are positive constants, which must be chosen according to our requirements. Therefore, $A<B$, to ensure that the constraints of $H_{c}$ are fulfilled. At last, $ \tau<B$ ranks the time windows constraints over the objective to minimise the path (tour).

%%%%%%%%%%%%  Starting New Page here %%%%%%%%%%%%%%

\newpage
\section{Research Results }\label{sec:SWP_results}

\subsection{Experimentation} \label{sec:SWP_experimentation}
We tested our algorithm in QUBO form (and mapped it into the Ising model) on VQE, QAOA, Numpy Minimum Eigensolver (classical), GroverOptimiser, CplexOptimiser, Numpy Minimum Eigensolver, Backtracking and CP-SAT Solver from google on the \textit{ibmq-16-melbourne v1.0.0}, \textit{ibmq-qasm-simulator (up to 32 qubits)} with COBYLA and SLSQP as the classical optimiser. 

%The Figures \eqref{fig:10_SW_schedules} to \eqref{fig:SWP_Quantum_MinimumEigenOptimizer} show the results of our algorithm once we executed our algorithm under the IBMQ. The optimal visit considers the hours of visits to form the optimal schedule. We have done several experiments with the QML defining different scenes using shot configuration. With our quantum machine, we can configure the number of repetitions of each circuit for sampling we need. With that, we will be doing machine learning on circuit design for each shot. And when the loop ends, we will get to the ground state energy. Consequently, we solved our problem by creating one quantum circuit for each shot, and the best circuit will be the one that optimizes our Social Workers' Problem. 

\begin{figure}[!h]
    \centering
    \includegraphics[width=0.45\textwidth]{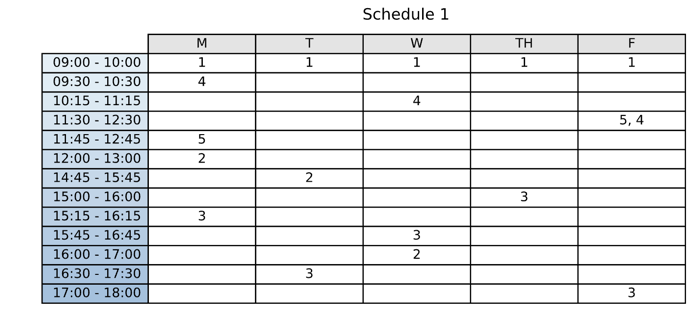}
    \includegraphics[width=0.45\textwidth]{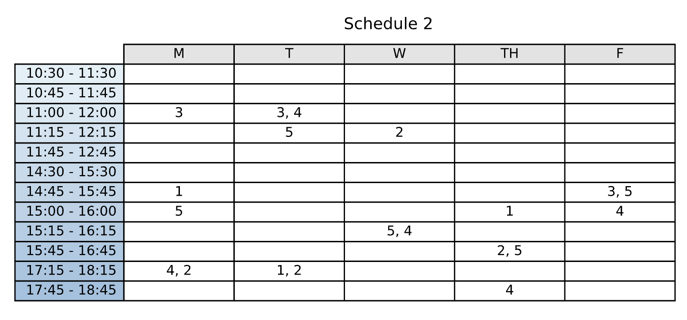}
    \includegraphics[width=0.45\textwidth]{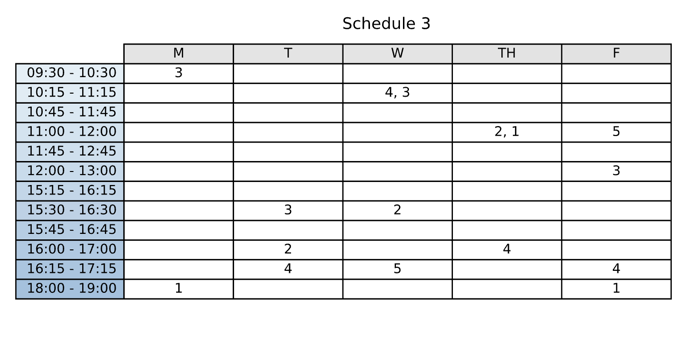}
    \includegraphics[width=0.45\textwidth]{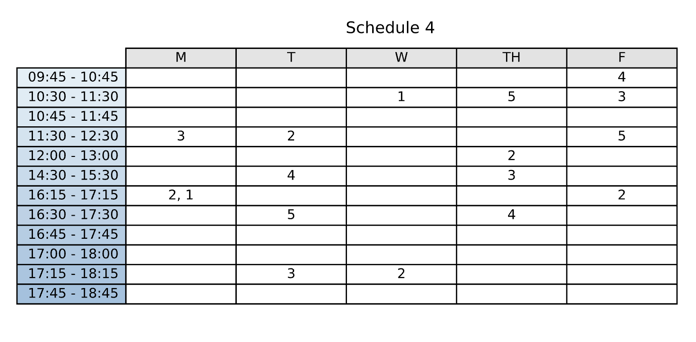}
    \includegraphics[width=0.45\textwidth]{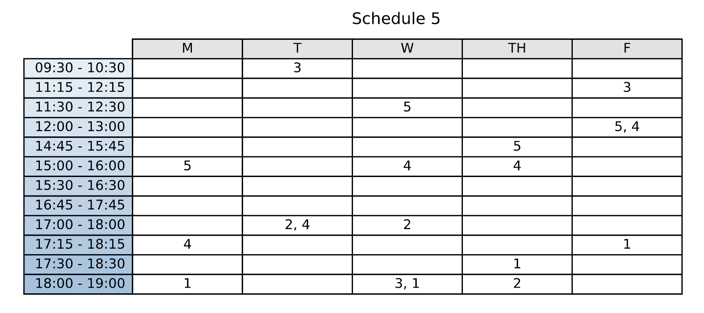}
    \includegraphics[width=0.45\textwidth]{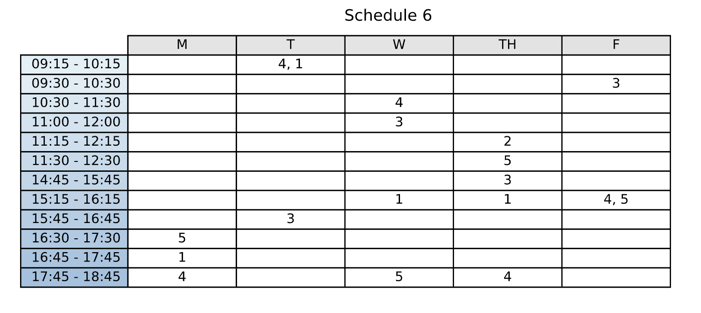}
    \includegraphics[width=0.45\textwidth]{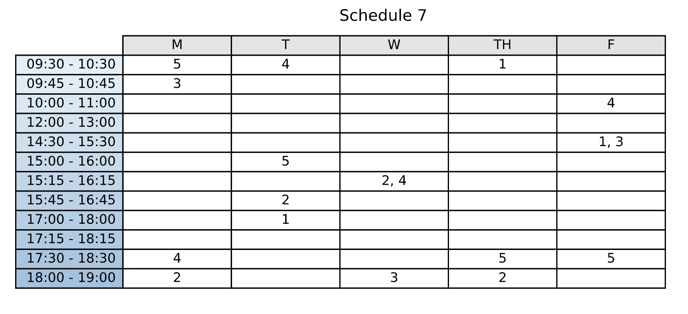}
    \includegraphics[width=0.45\textwidth]{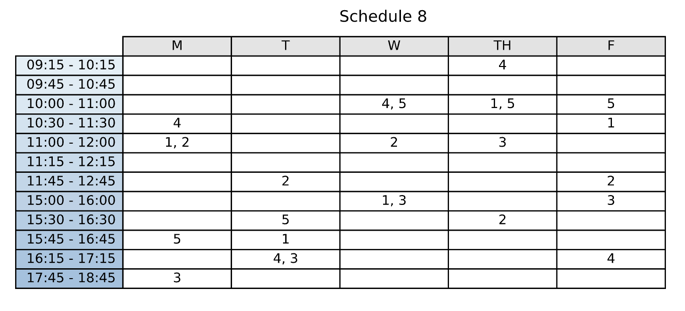}
    \includegraphics[width=0.45\textwidth]{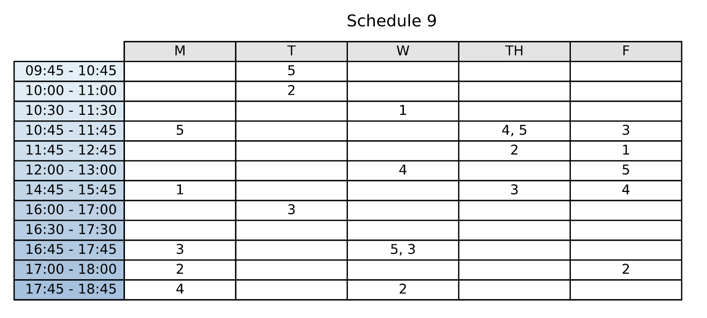}
    \includegraphics[width=0.45\textwidth]{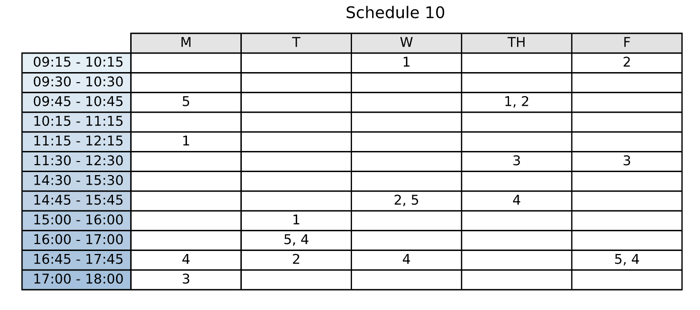}
    \caption{Generated 10 Social Workers' schedules}
    \label{fig:10_SW_schedules}
\end{figure}

\begin{figure}[!h]
    \centering
    \includegraphics[width=0.45\textwidth]{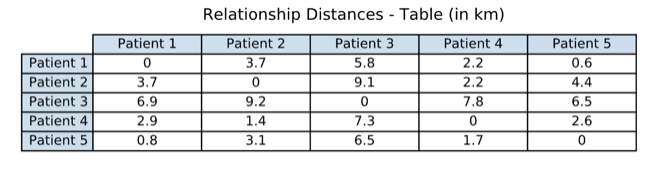}
    \includegraphics[width=0.45\textwidth]{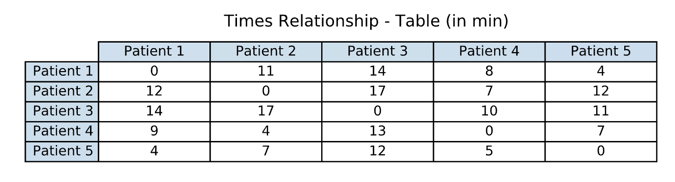}
    \caption{Generated 10 Social Workers' Distance Matrix}
    \label{fig:10_SW_Dist_Matrix}
\end{figure}

\begin{figure}[!h]
    \centering
    \includegraphics[width=0.45\textwidth]{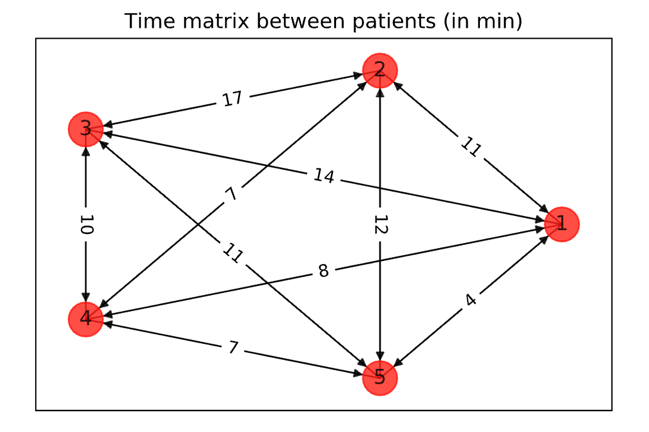}
    \includegraphics[width=0.45\textwidth]{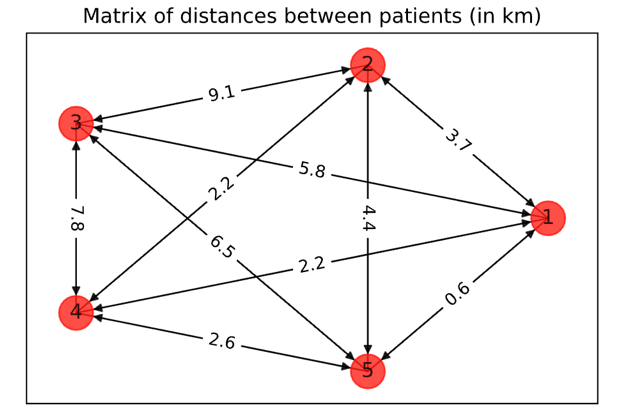}
    \caption{Generated 10 Social Workers' Distance Nodes}
    \label{fig:10_SW_Dist_Nodes}
\end{figure}

\begin{figure}[!h]
    \centering
    \includegraphics[width=0.4\textwidth]{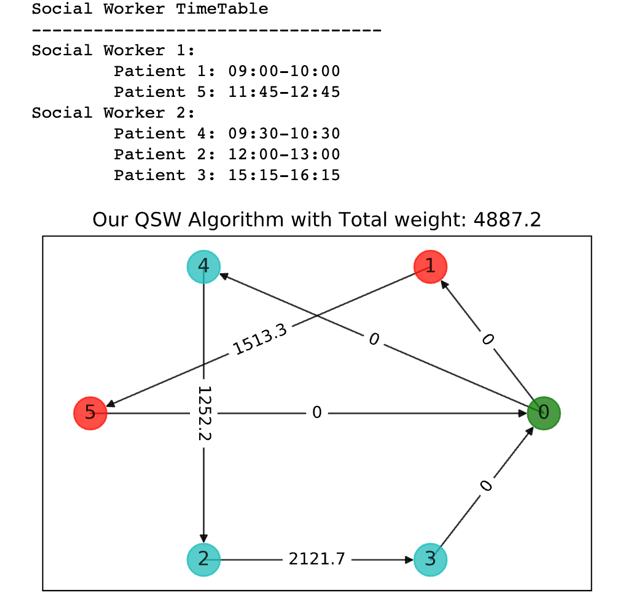}
    \includegraphics[width=0.4\textwidth]{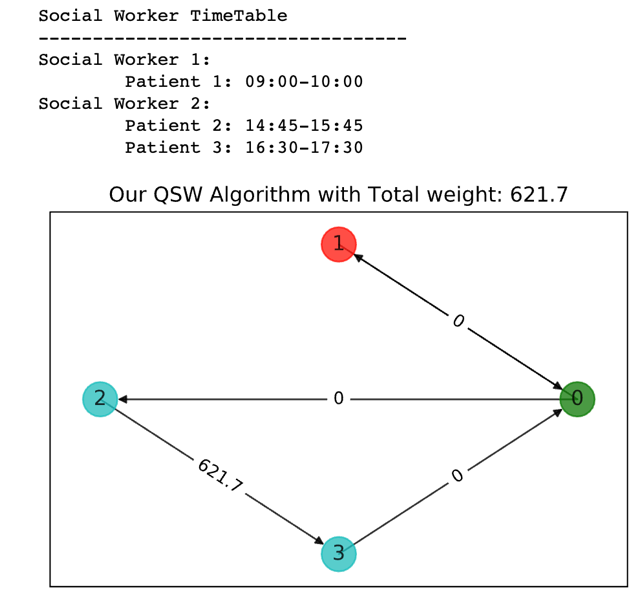}
    \includegraphics[width=0.4\textwidth]{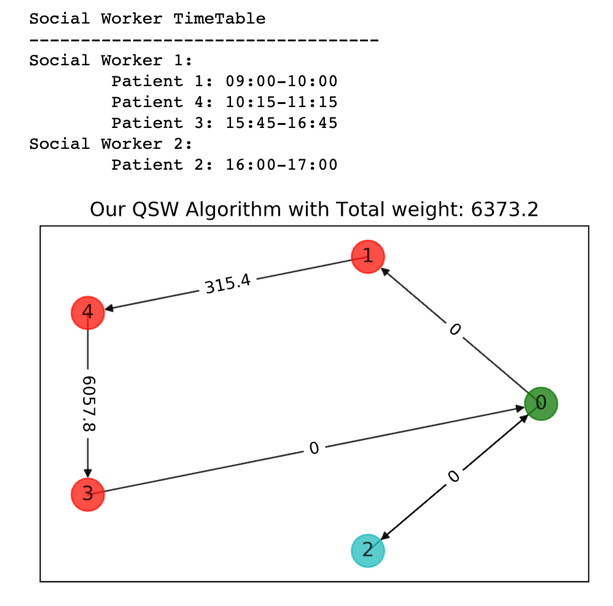}
    \includegraphics[width=0.4\textwidth]{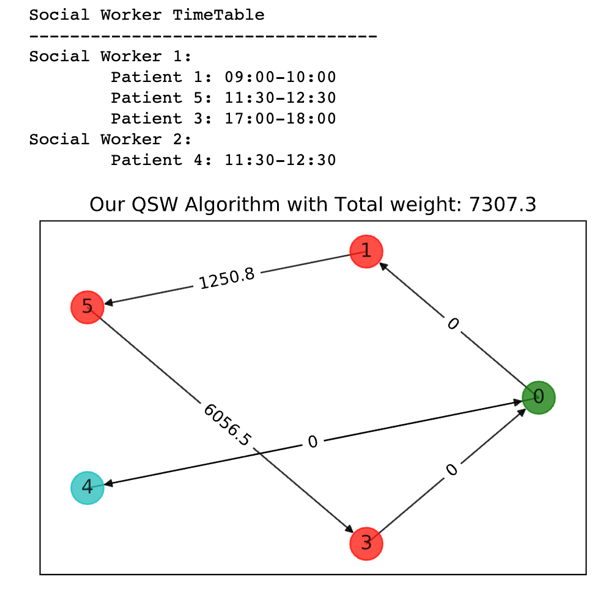}
    \includegraphics[width=0.4\textwidth]{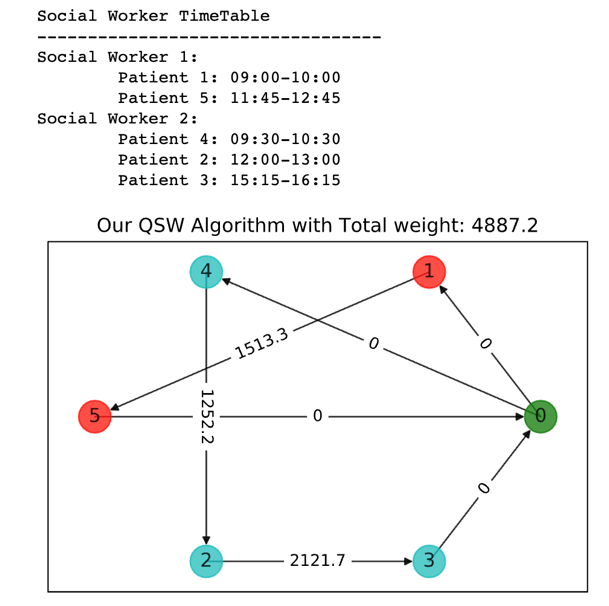}
    \includegraphics[width=0.4\textwidth]{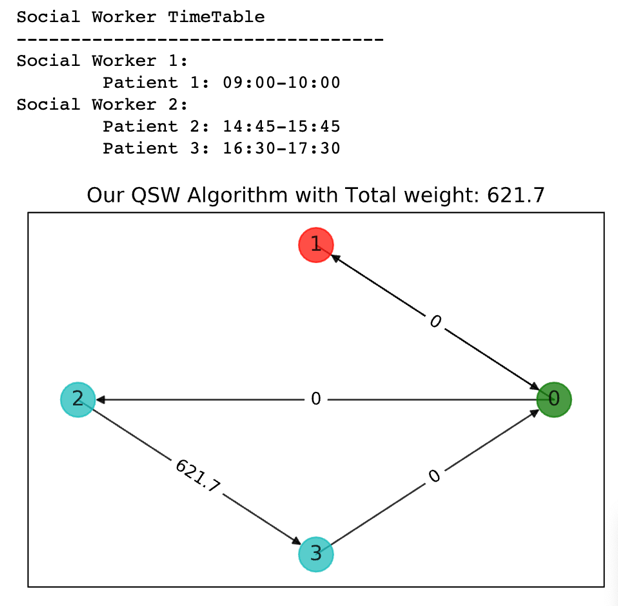}
    \caption{Outcomes from the SWP by the Backtracking algorithm. From Monday to Wednesday}
    \label{fig:SWP_Classical_BackT_Mon-Wed}
\end{figure}

\begin{figure}[!h]
    \centering
    \includegraphics[width=0.4\textwidth]{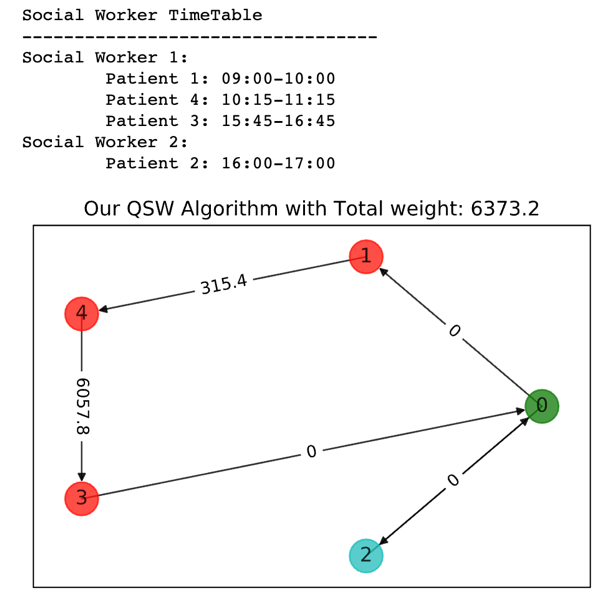}
    \includegraphics[width=0.4\textwidth]{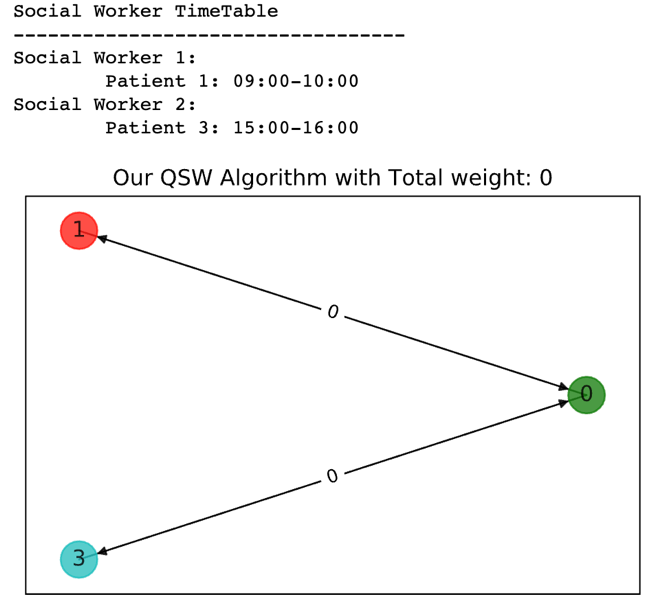}
    \includegraphics[width=0.4\textwidth]{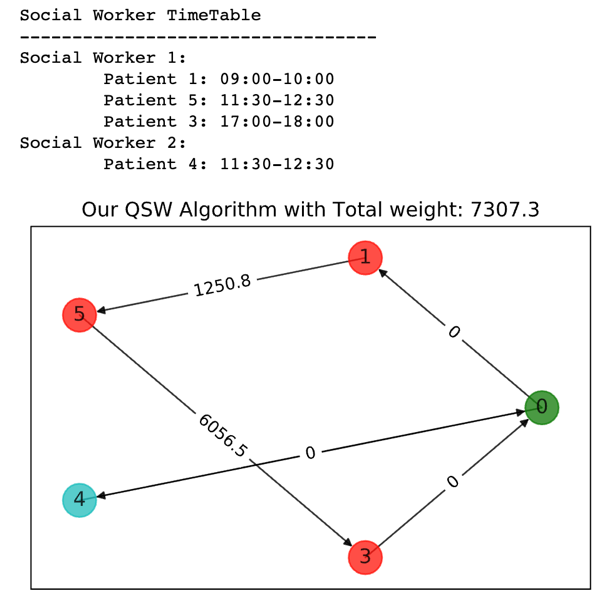}
    \caption{Outcomes from the SWP by the Backtracking algorithm. From Thursday to Friday}
    \label{fig:SWP_Classical_BackT_Th-Fri}
\end{figure}

\begin{figure}[!h]
    \centering
    \includegraphics[width=0.4\textwidth]{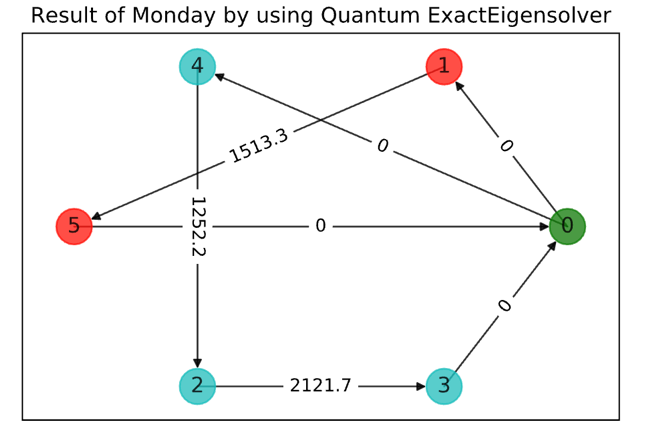}
    \includegraphics[width=0.4\textwidth]{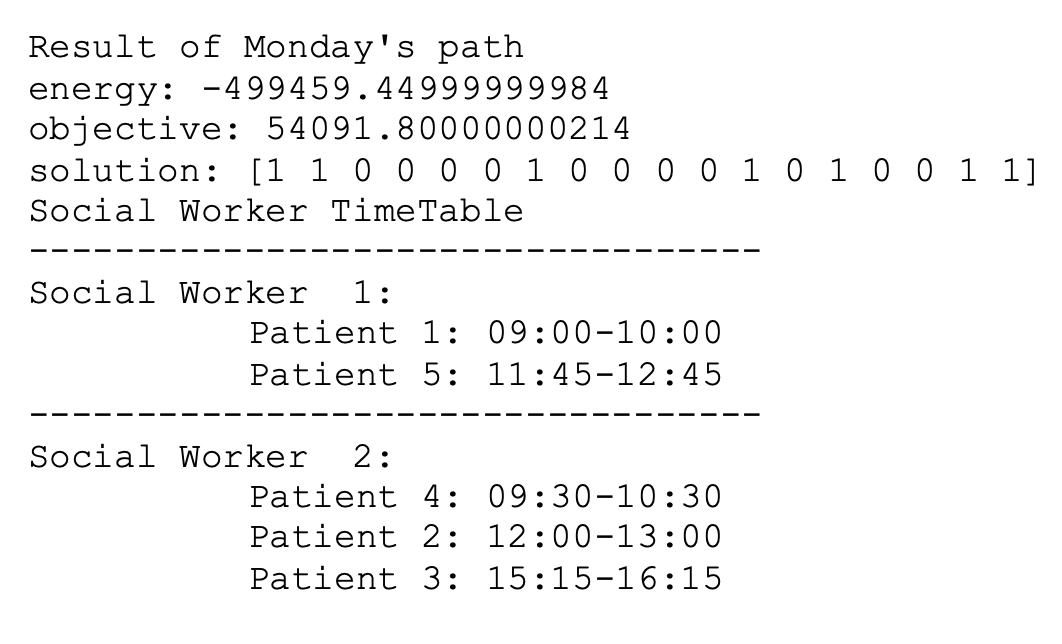}
    \includegraphics[width=0.4\textwidth]{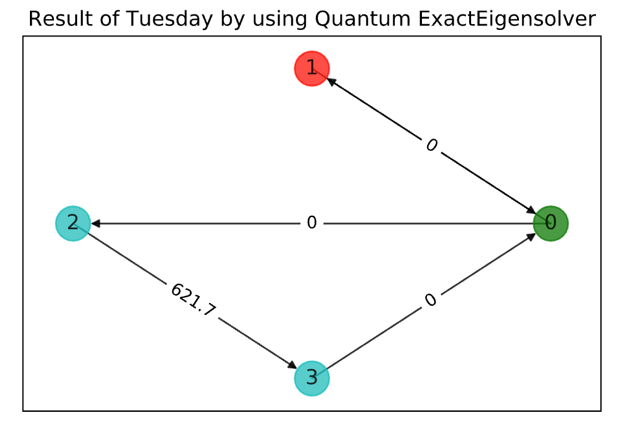}
    \includegraphics[width=0.4\textwidth]{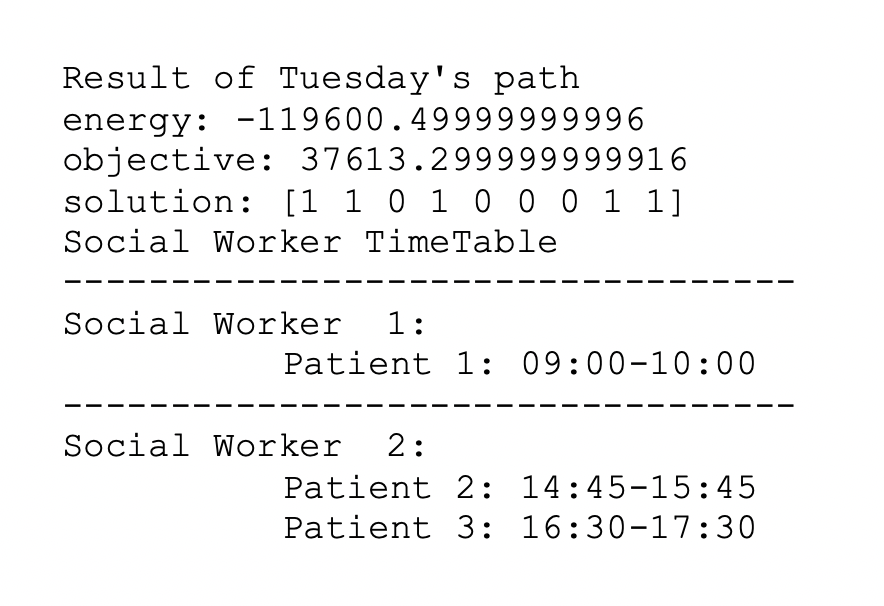}
    \includegraphics[width=0.4\textwidth]{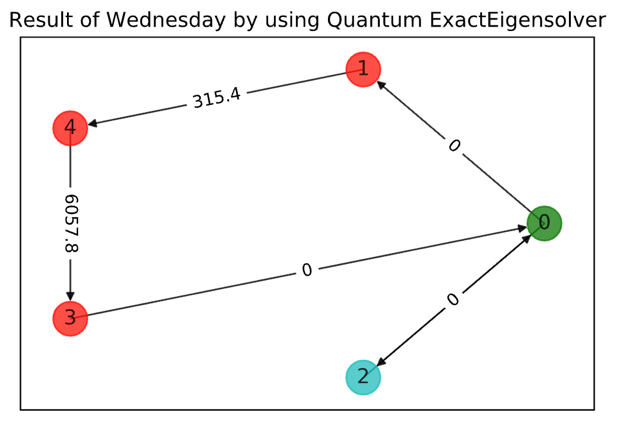}
    \includegraphics[width=0.4\textwidth]{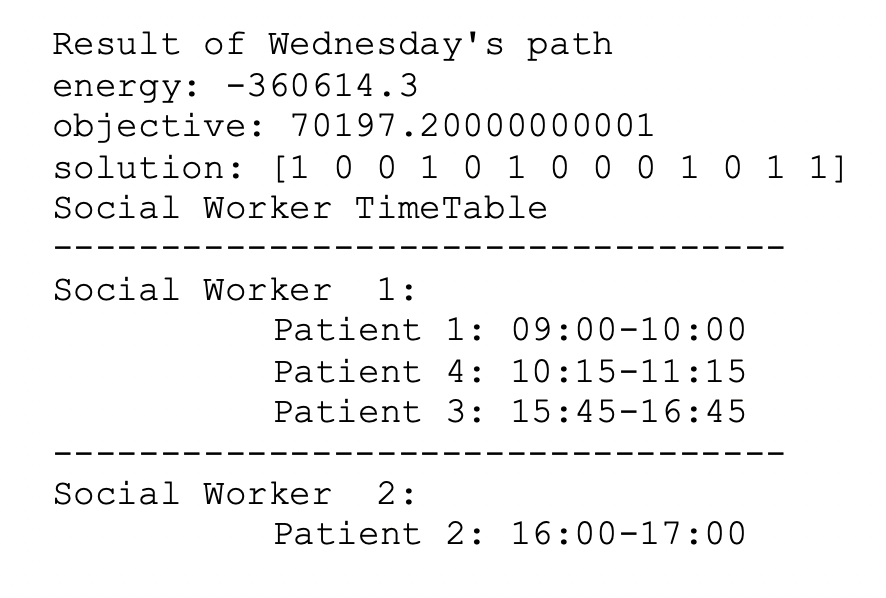}
    \caption{Outcomes from the SWP by the Quantum Exact solver algorithm. From Monday to Wednesday}
    \label{fig:SWP_Quantum_Exact_Solver_Mon-Wed_}
\end{figure}

\begin{figure}[!h]
    \centering
    \includegraphics[width=0.4\textwidth]{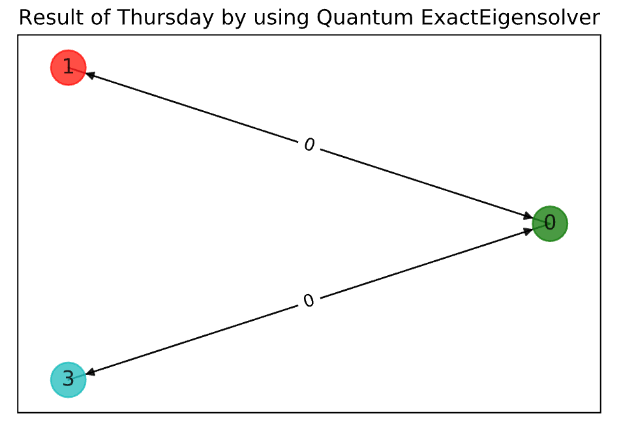}
    \includegraphics[width=0.4\textwidth]{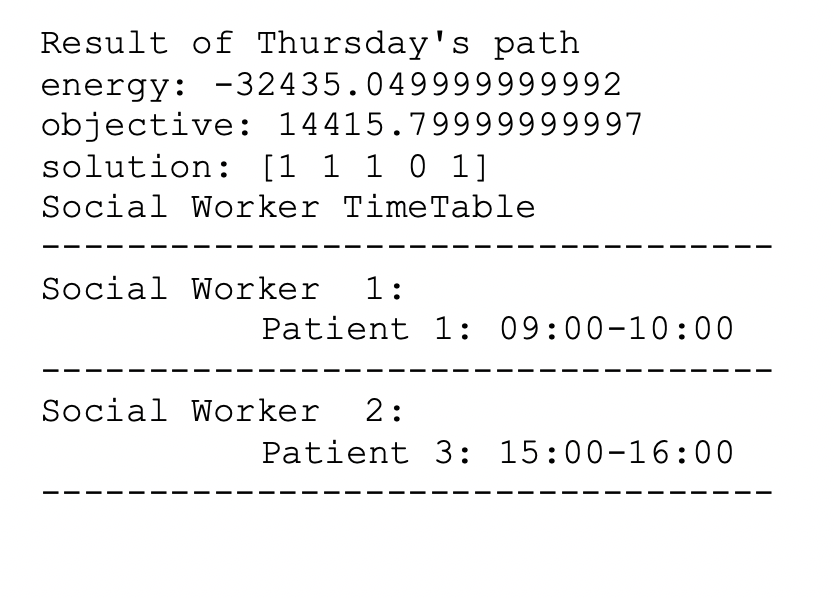}
    \includegraphics[width=0.4\textwidth]{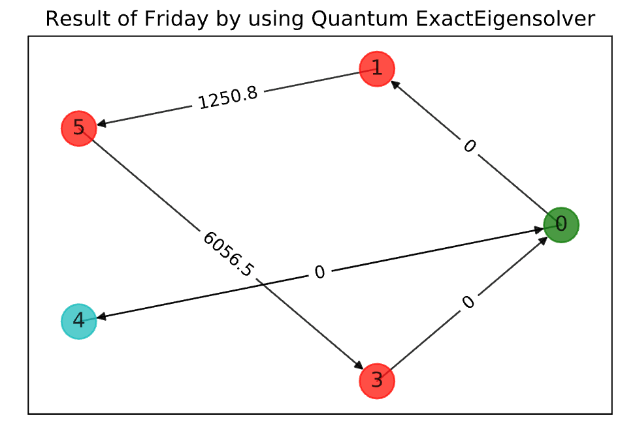}
    \includegraphics[width=0.4\textwidth]{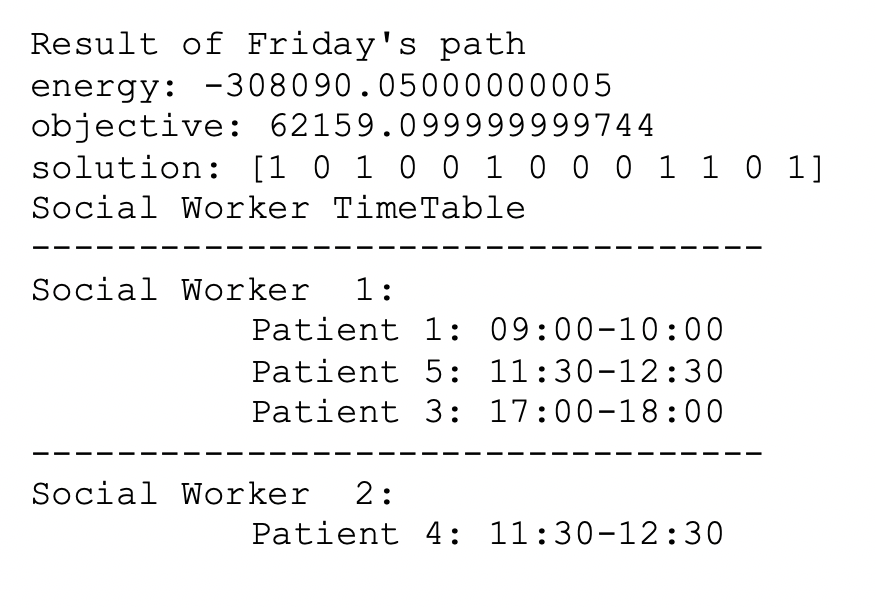}
 \caption{Outcomes from the SWP by the Quantum Exact solver algorithm. From Thursday to Friday}
    \label{fig:SWP_Quantum_Exact_Solver_Thur-Fri}
\end{figure}

\begin{figure}[!h]
    \centering
    \includegraphics[width=0.4\textwidth]{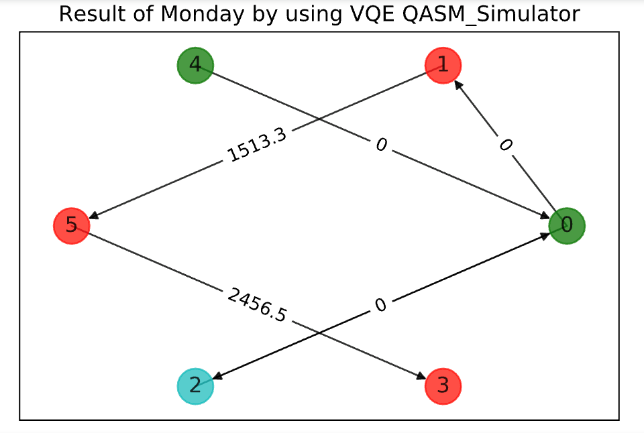}
    \includegraphics[width=0.4\textwidth]{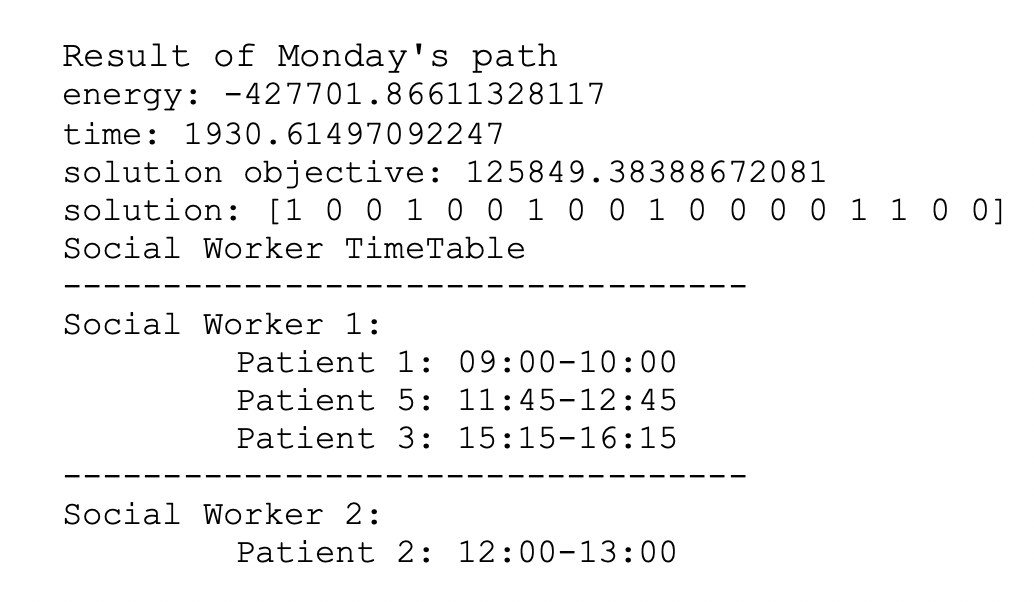}
    \includegraphics[width=0.4\textwidth]{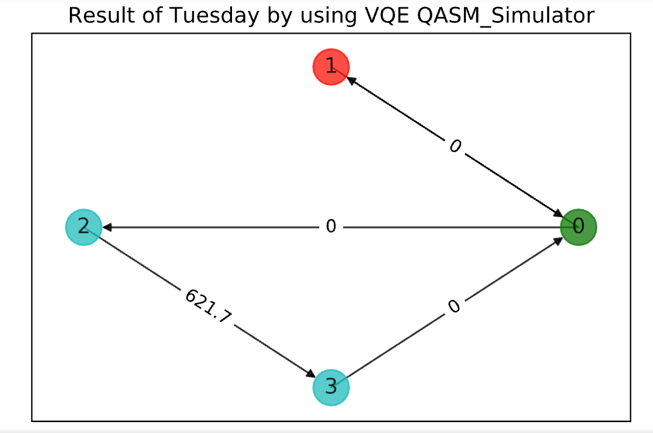}
    \includegraphics[width=0.4\textwidth]{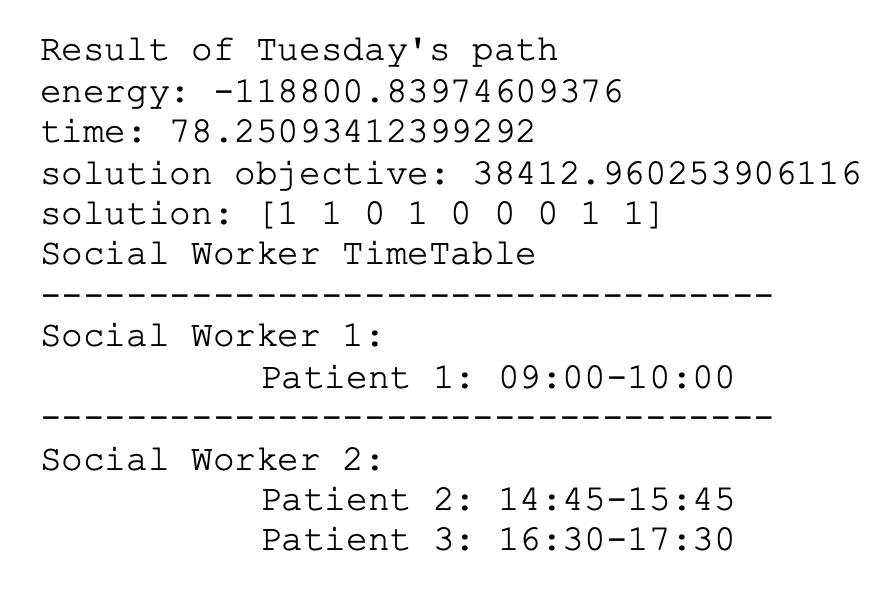}
    \includegraphics[width=0.4\textwidth]{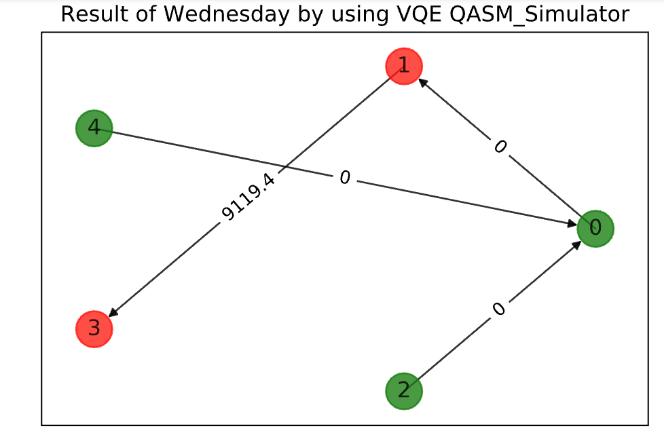}
    \includegraphics[width=0.4\textwidth]{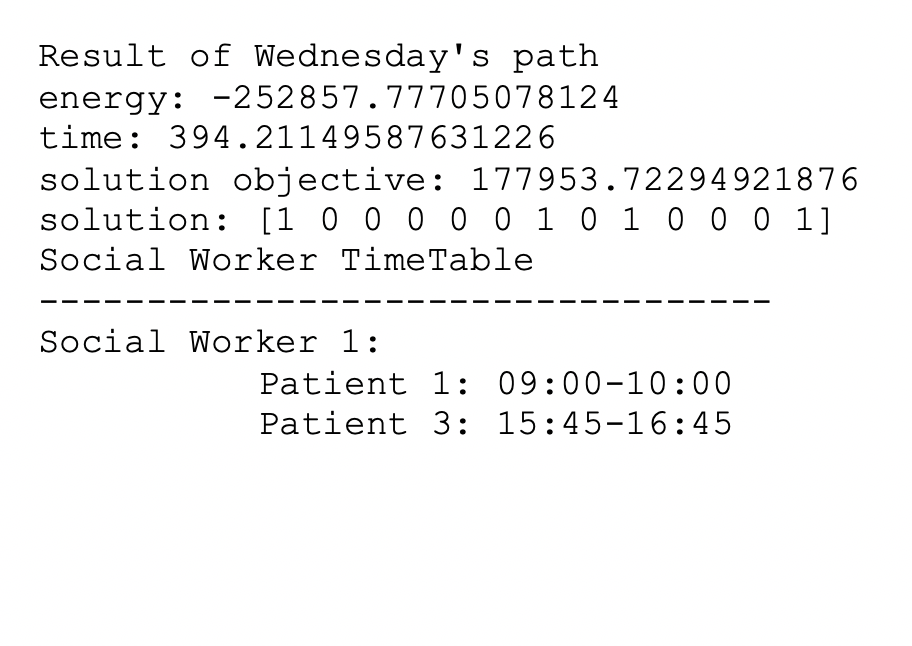}
    \caption{Outcomes from the SWP by the Quantum ASM Simulator from IBMQ. From Monday to Wednesday}
    \label{fig:SWP_Quantum_Exact_Solver_Mon-Wed}
\end{figure}

\begin{figure}[!h]
    \centering
    \includegraphics[width=0.4\textwidth]{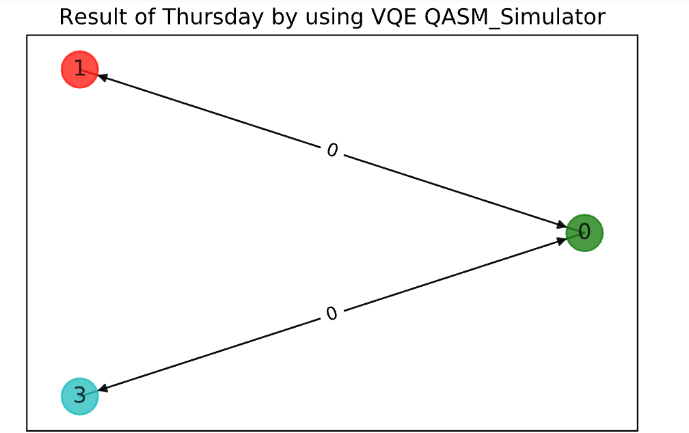}
    \includegraphics[width=0.4\textwidth]{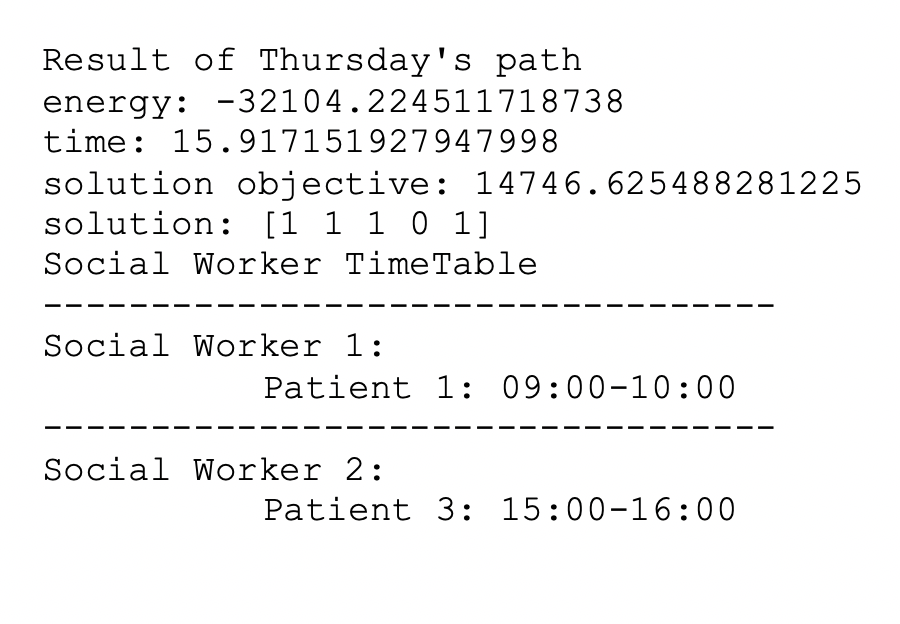}
    \includegraphics[width=0.4\textwidth]{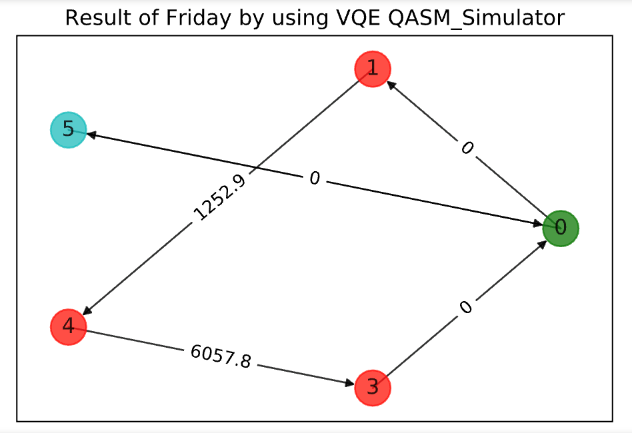}
    \includegraphics[width=0.4\textwidth]{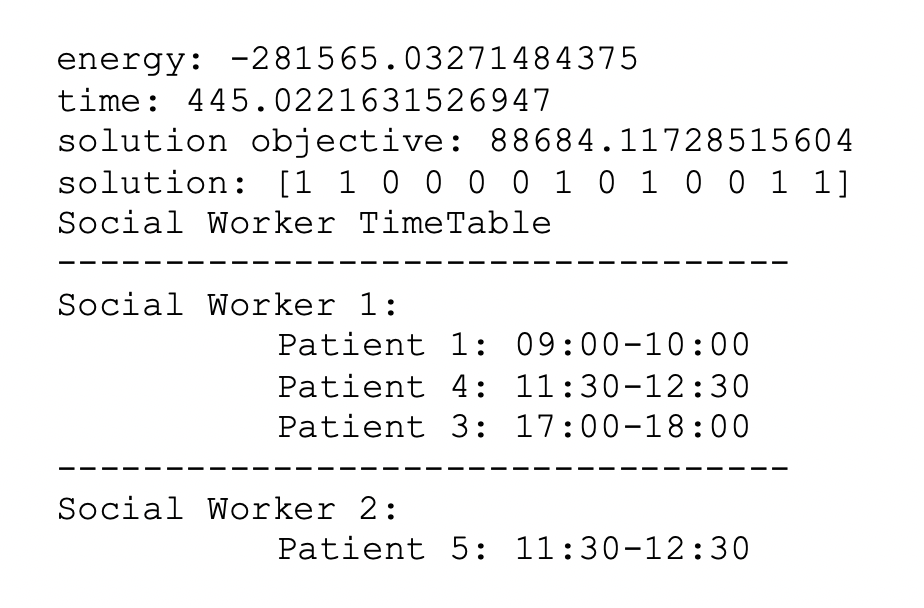}
 \caption{Outcomes from the SWP by the Quantum ASM Simulator from IBMQ. From Thursday to Friday}
    \label{fig:SWP_Quantum_Exact_Solver_Thur-Fri_}
\end{figure}

\begin{figure}[!h]
    \centering
    \includegraphics[width=0.6\textwidth]{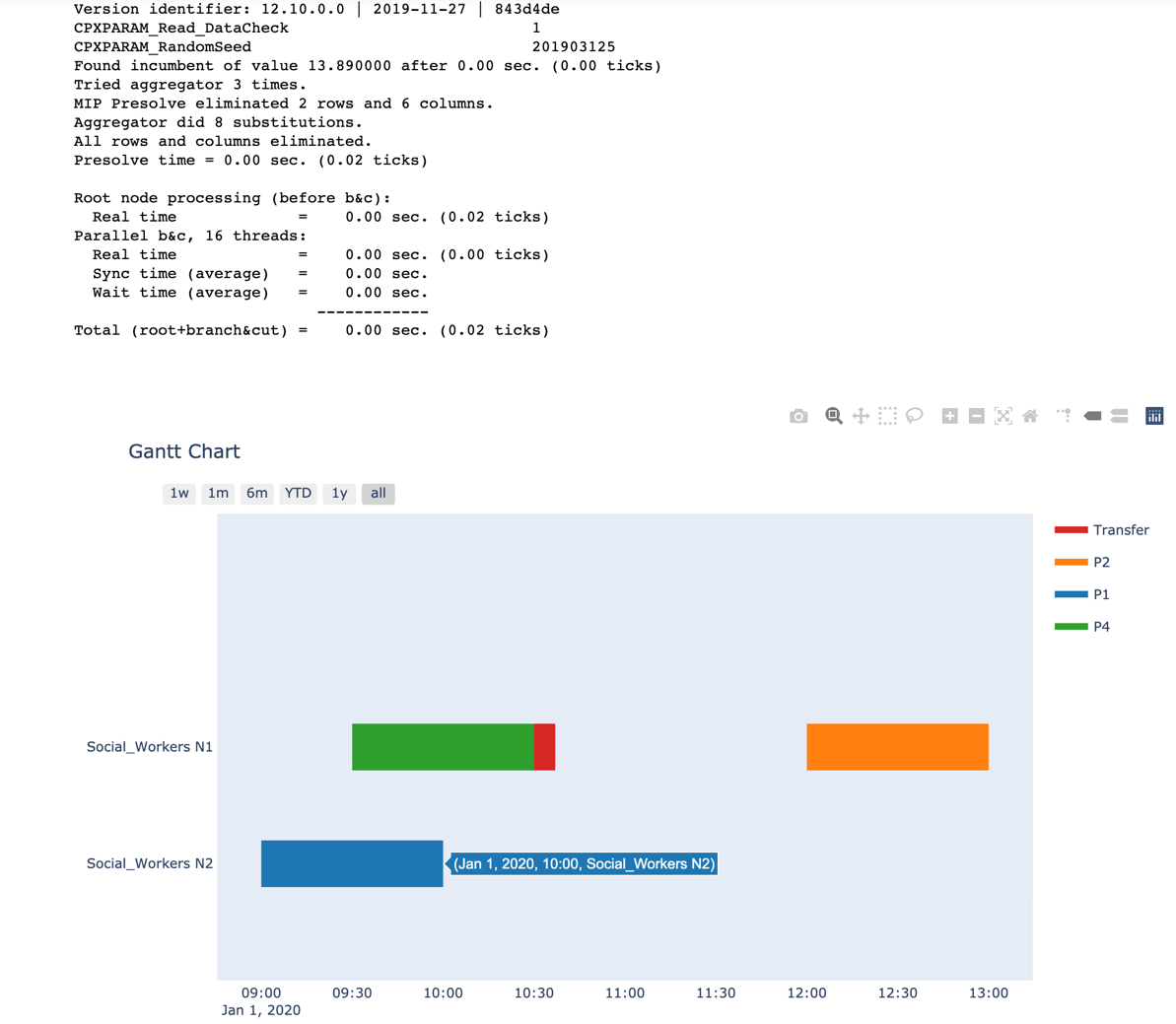}
 \caption{Outcomes from the SWP by ibmq\_16\_melbourne v2.1.0 from IBMQ.}
    \label{fig:SWP_Quantum_ibmq_melbourne}
\end{figure}

\begin{figure}[!h]
    \centering
    \includegraphics[width=0.6\textwidth]{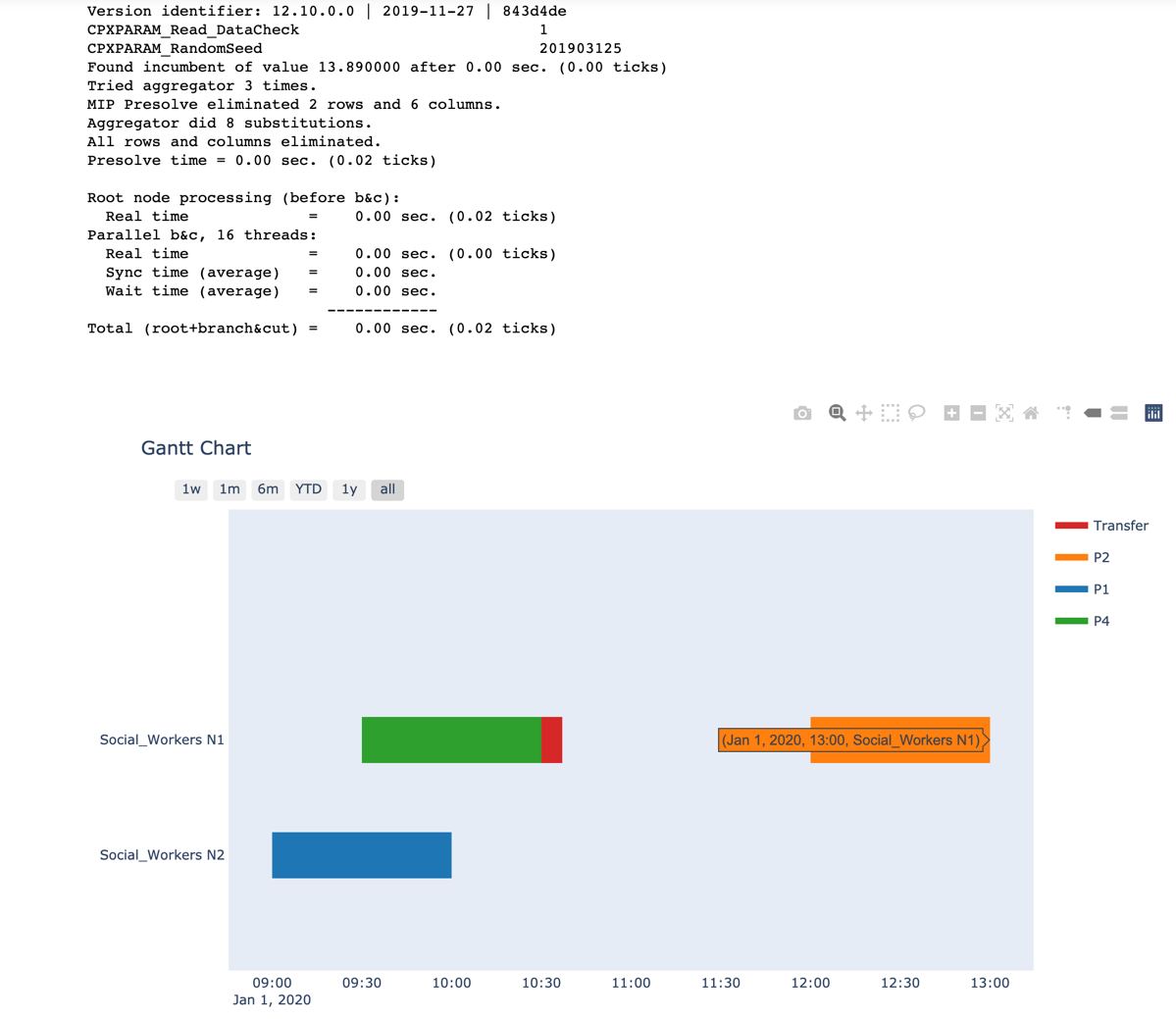}
 \caption{Outcomes from the SWP by ADMM optimiser from IBMQ.}
    \label{fig:SWP_Quantum_ADMM}
\end{figure}

\begin{figure}[!h]
    \centering
    \includegraphics[width=0.6\textwidth]{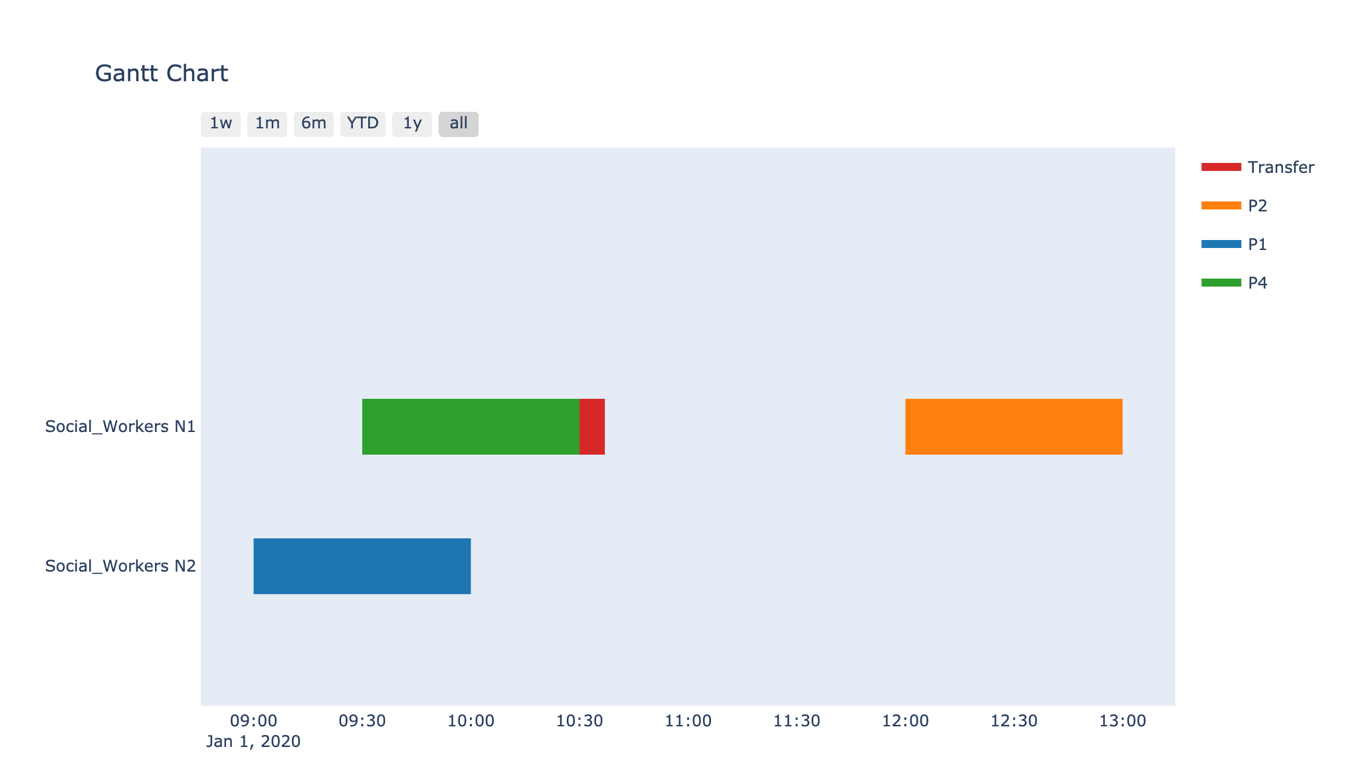}
 \caption{Outcomes from the SWP by Minimum Eigen Optimiser from IBMQ.}
    \label{fig:SWP_Quantum_MinimumEigenOptimizer}
\end{figure}

Figures \eqref{fig:10_SW_schedules} to \eqref{fig:SWP_Quantum_MinimumEigenOptimizer} show the results of our algorithm executed on the IBMQ environment. We have done several experiments with the QML defining different scenes using shot configuration and each circuit's number of repetitions for sampling. Consequently, we solved our problem by creating one quantum circuit for each shot, and the best circuit was the one that optimised our Social Workers Problem. 

Next, we will analyse the results (figures and tables) in each case studied and presented above.

%\subsubsection{Results by Classical Backtracking }
%\subsubsection{Results by Quantum Exact EigenSolver}
%\subsubsection{Results for Quantum ASM Simulator}
%\subsubsection{Results by Real Quantum Computer (ibmq\_16\_melbourne v2.1.0 (15 qubits))}
%\subsubsection{Results by ADMM Optimizer}
%\subsubsection{Results by MinimumEigenOptimizer}

%%%%%%%%%%%%  Starting New Page here %%%%%%%%%%%%%%

\newpage
\section{Discussions}\label{sec:SWP_Discussions}

Before analysing the practical results, let us focus on discussing the impact of the formulation. First, let us recall that one of our aims was to implement the proposed problem in this era of NISQ. So, we needed to find an optimal formulation (\eqref{weight_SWP_eq} and \eqref{time_window_SWP_eq}) that reduced the qubit number. On the one hand, the found formulation leveraged the heuristic function we designed and, on the other hand, used a classical algorithm that generated a description of a quantum circuit from the found heuristic function.

The VQE worked very well and empowers the QML era. From what we expected, the VQE has solved it by far. With the help of the IBMQ Aqua\footnote{ www.qiskit.org/aqua } program environment (qiskit\footnote{ www.qiskit.org }), we were able to test our algorithm. Too bad that we do not have access to a quantum computer with more qubits. Since in our case for $ n = 6$, the number of qubits necessary is $n(n-1)=30$ qubits  for the QASM\_Simulator and  $n=4$  for the real quantum computer (15 qubits). We have not been able to do many more tests for values of $n$  greater than six since the computer on which we test our algorithm takes too long due to the simulation of the Hamiltonian in a classical computer.

The VQE as a variational algorithm can be useful to empower intelligent solutions as the objective of this article: Using the Variational-Quantum-Eigensolver (VQE) to create an Intelligent social workers schedule problem solver.

The evaluation of the algorithm on an \textit{ibmq\_16\_melbourne v1.0.0} from IBM was fulfilled. With any change in the input, variables are mapped proportionally to our cost variable with a time window. 

With this work, we are trying to offer to cities an instrument \cite{XWa18} which could optimise the costs allied to the management of social workers and improve social gains. Moreover, this work could also be a starting point for many countries in Africa that are seizing the opportunity of the mobile technology revolution \cite{Ali18} to develop and increase their industrious and e-health system \cite{Sal17}.

We would like to add that the suggested formulation \eqref{descomp_Funct_Objective_eq} and \eqref{T_W_SWP_eq} is not only specific to the proposed problem. It can be used to solve any family planning, scheduling and routing problem related to a list of tasks, restrictions, allocation of resources on location and time. The test performed and showed in Fig. \eqref{fig:Analysis_epsilon} allows us to see the behaviour of our formulation with the variation of the correction factor  $\varepsilon$. We understood how our time window $T_{ij}= \left(  \tau_{i-} \tau_{j} \right)$ adapted perfectly at the extremes to the cost variable in the distance. This achievement is due to the chosen quadratic function \eqref{T_W_SWP_eq}. We wanted it to be adapted in this way so that our resultant function weighted together the short distances and time and the long distances and late times.
Other functions can be studied to have a test bench to compare the final results.
QAOA, like VQE, takes a qubit operator from the Hamiltonian of the Ising model. The only mapping that gets done is when QAOA builds a variational form based on the qubit operator and what we figured was that does not change the original qubit operator input. Figure \eqref{fig:Comparing_5_SW_VQE_QAOAS} and \eqref{fig:Comparing_10_SW_VQE_QAOA} reveal the comparison work between VQE and QAOA algorithms for the same configuration parameters. After several tests, we confirmed that our algorithm took less execution time with the QAOA than the VQE and required fewer samples for the optimal solution. But in many cases, we have had to increase the shot value to get a reasonably stable result.

\section{Benchmarks of SWP}
In this section, we are going to make a comparison of all the techniques we used to solve the SWP.

\subsection{Comparing VQE solution with classical solver}\label{sec:SWP_Bench_VQE_Classical}
\begin{figure}[!h]
    \centering
    \includegraphics[width=0.6\textwidth]{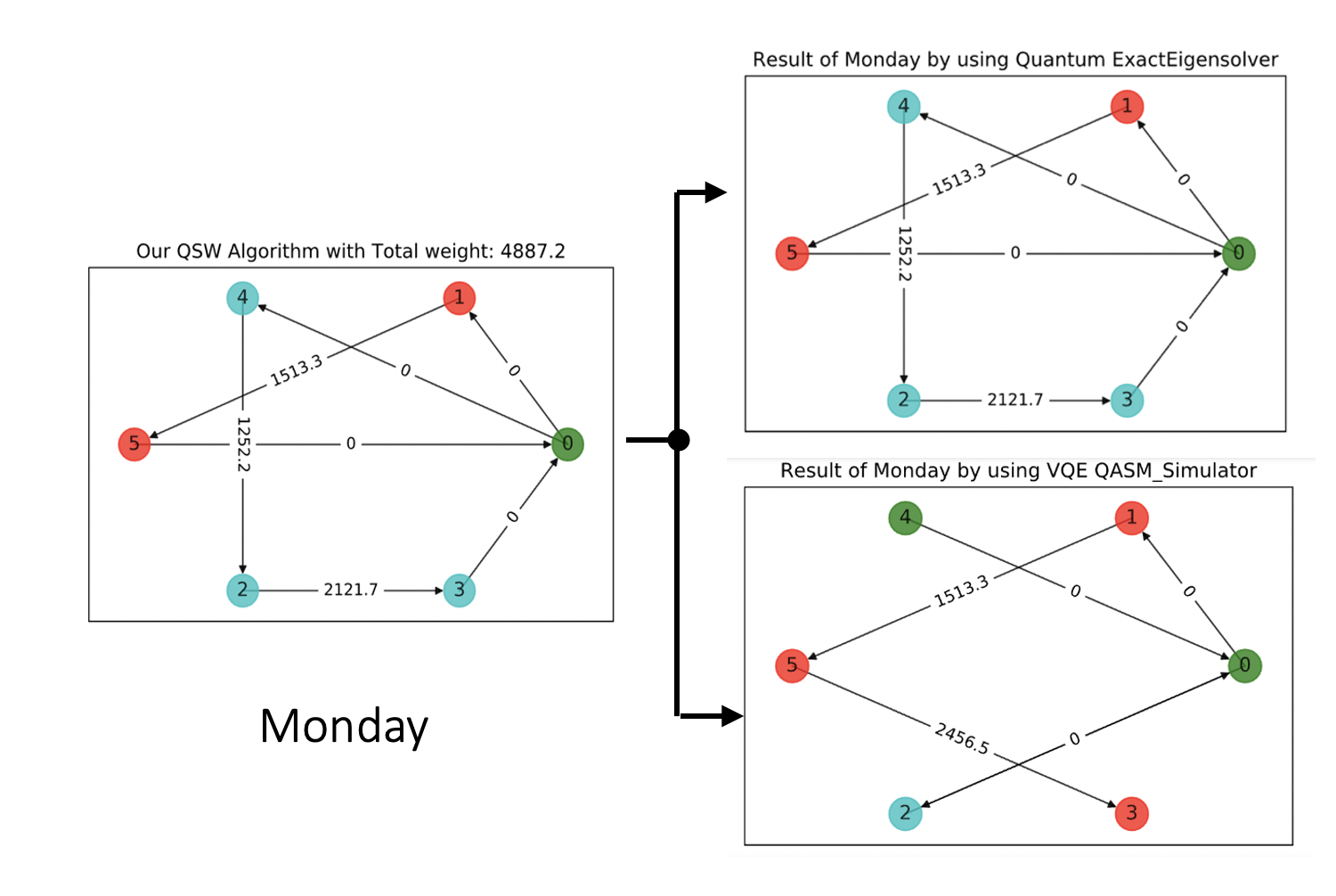}
 \caption{Here we present the results by Monday of our algorithm. We compared the classical solver, the exact quantum solver, and the Variational Quantum Eigensolver. We can observe that the VQE, as an approximated algorithm, didn't have time to find the best solution.}
    \label{fig:comp_Monday}
\end{figure}

\begin{figure}[!h]
    \centering
    \includegraphics[width=0.6\textwidth]{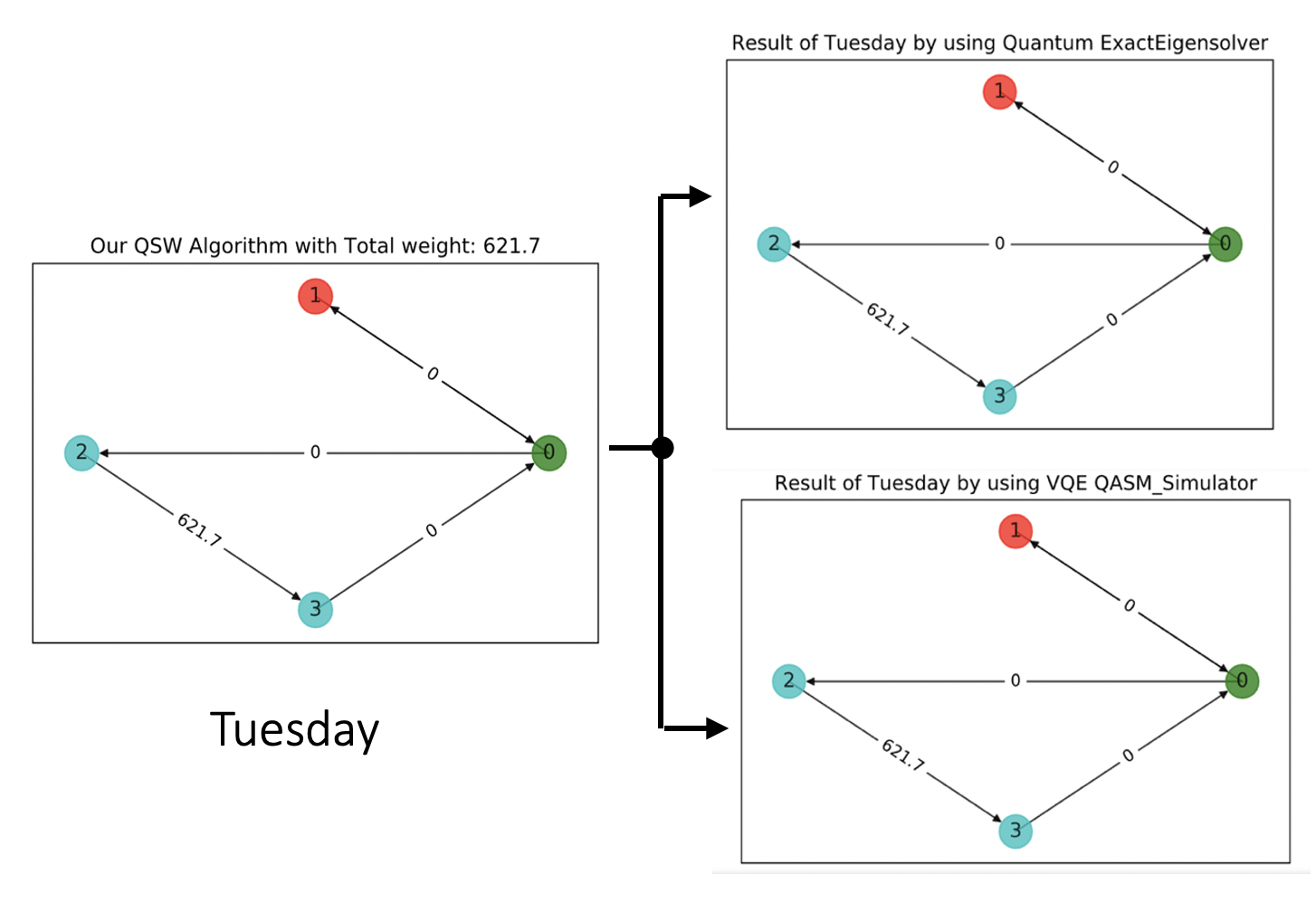}
 \caption{Here we present the results by Tuesday of our algorithm. We are comparing the classical solver and the exact quantum solver, and the Variational Quantum Eigensolver.}
    \label{fig:comp_Tuesday}
\end{figure}

\begin{figure}[!h]
    \centering
    \includegraphics[width=0.6\textwidth]{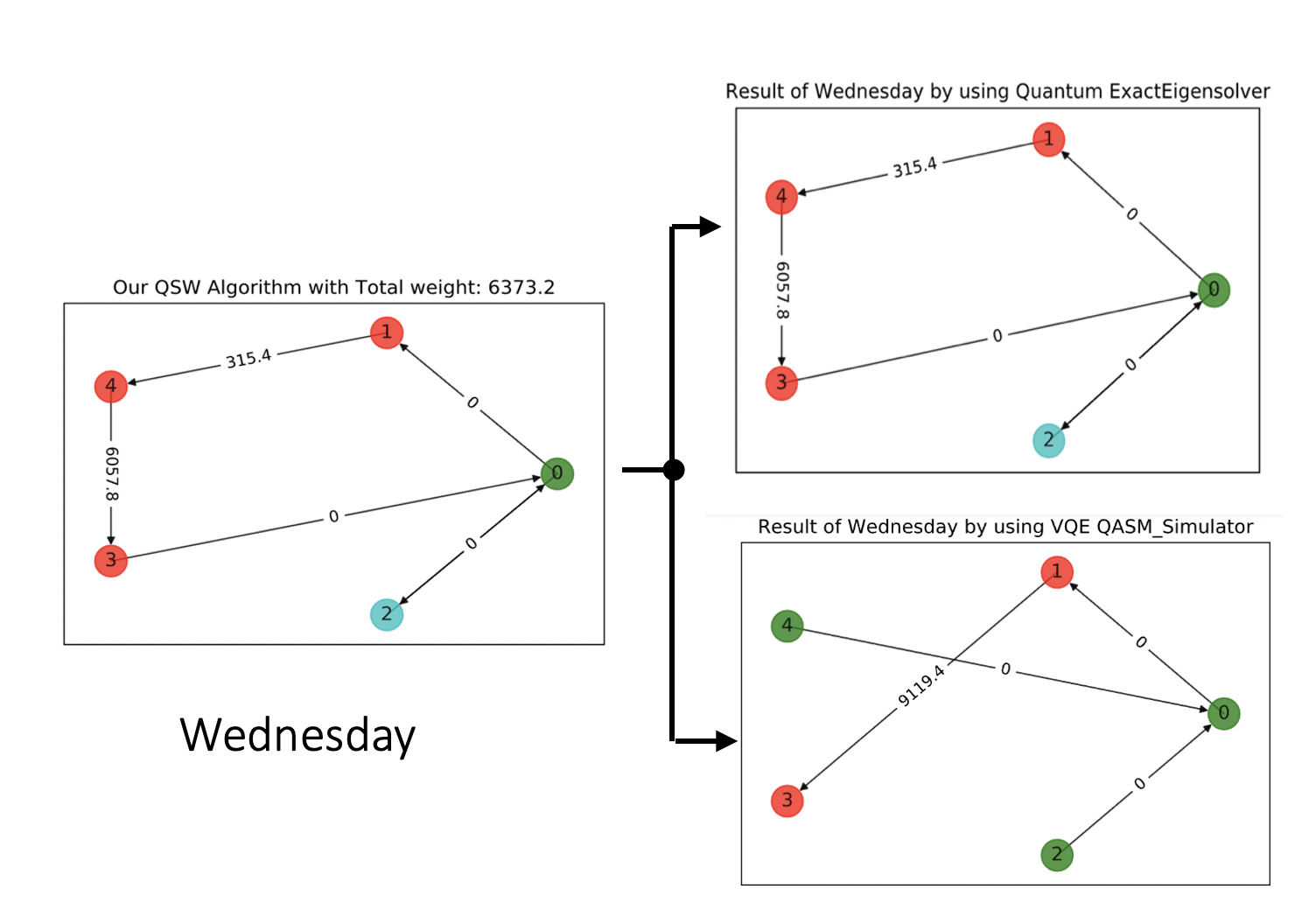}
 \caption{Here we present the results by Wednesday of our algorithm. We are comparing the classical solver and the exact quantum solver, and the Variational Quantum Eigensolver. We can observe that the VQE, as an approximated algorithm, didn't have time to find the best solution.}
    \label{fig:comp_Wednesday}
\end{figure}

\begin{figure}[!h]
    \centering
    \includegraphics[width=0.6\textwidth]{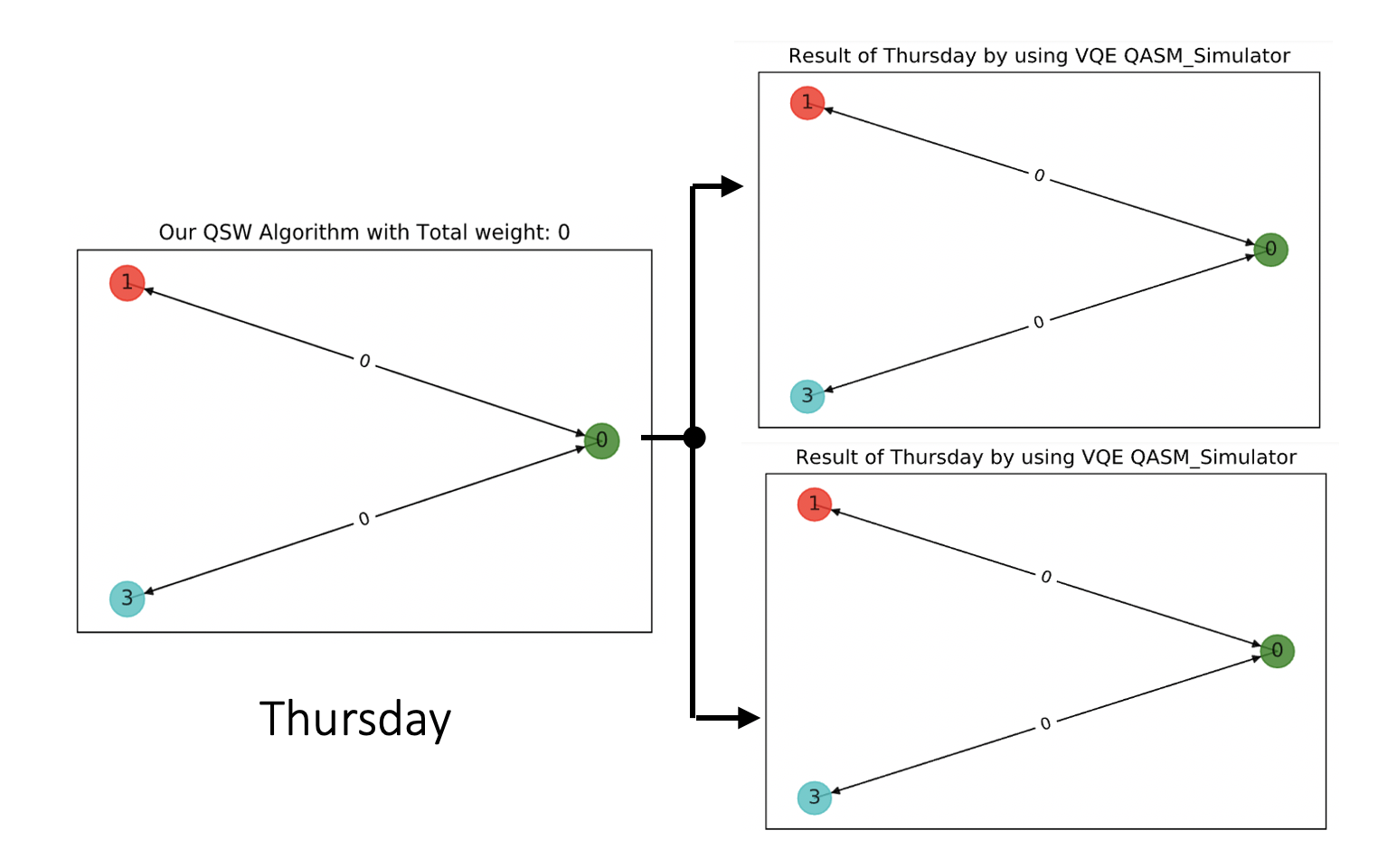}
 \caption{Here we present the results by Thursday of our algorithm. We are comparing the classical solver and the exact quantum solver, and the Variational Quantum Eigensolver.}
    \label{fig:comp_Thursday}
\end{figure}

\begin{figure}[!h]
    \centering
    \includegraphics[width=0.6\textwidth]{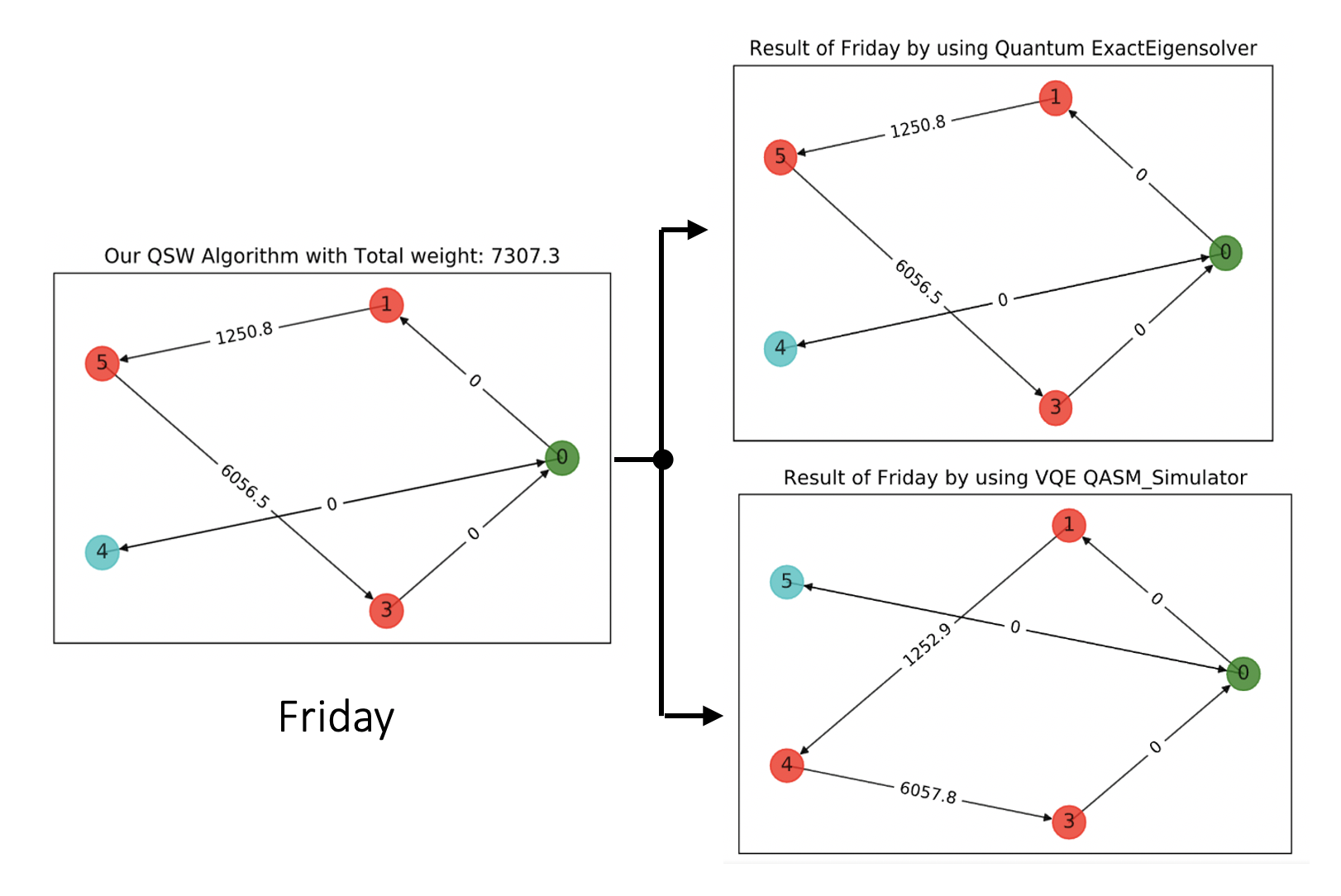}
 \caption{Here we present the results by Friday of our algorithm. We are comparing the classical solver and the exact quantum solver, and the Variational Quantum Eigensolver.}
    \label{fig:comp_Friday}
\end{figure}

\begin{figure}[!h]
    \centering
    \includegraphics[width=0.6\textwidth]{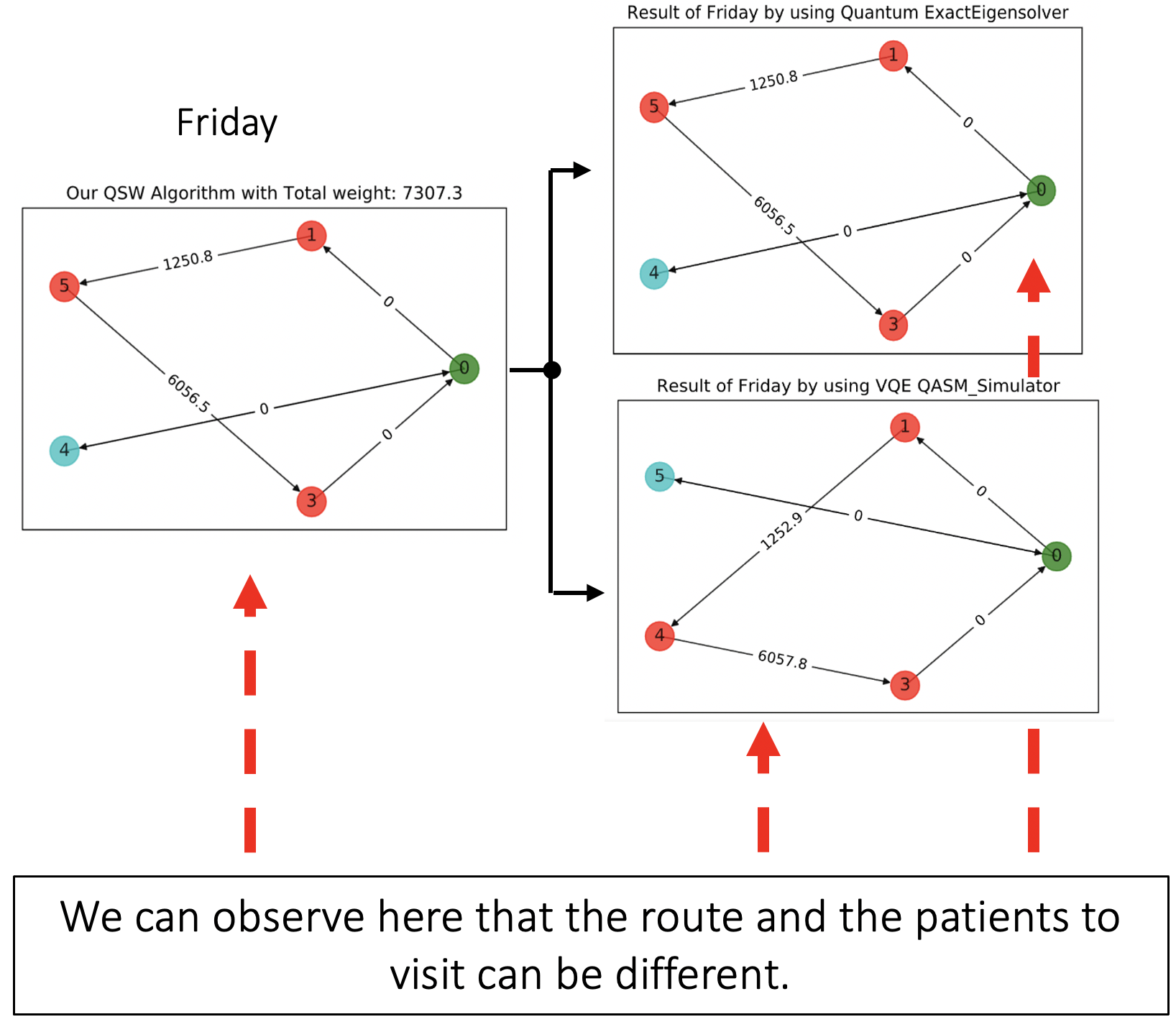}
 \caption{Here we present the results by analysing Friday of our algorithm. We are comparing the classical solver and the exact quantum solver, and the Variational Quantum Eigensolver.}
    \label{fig:Comp_Analysis_Friday}
\end{figure}

\begin{figure}[h!]
    \centering
    \includegraphics[width=0.5\textwidth]{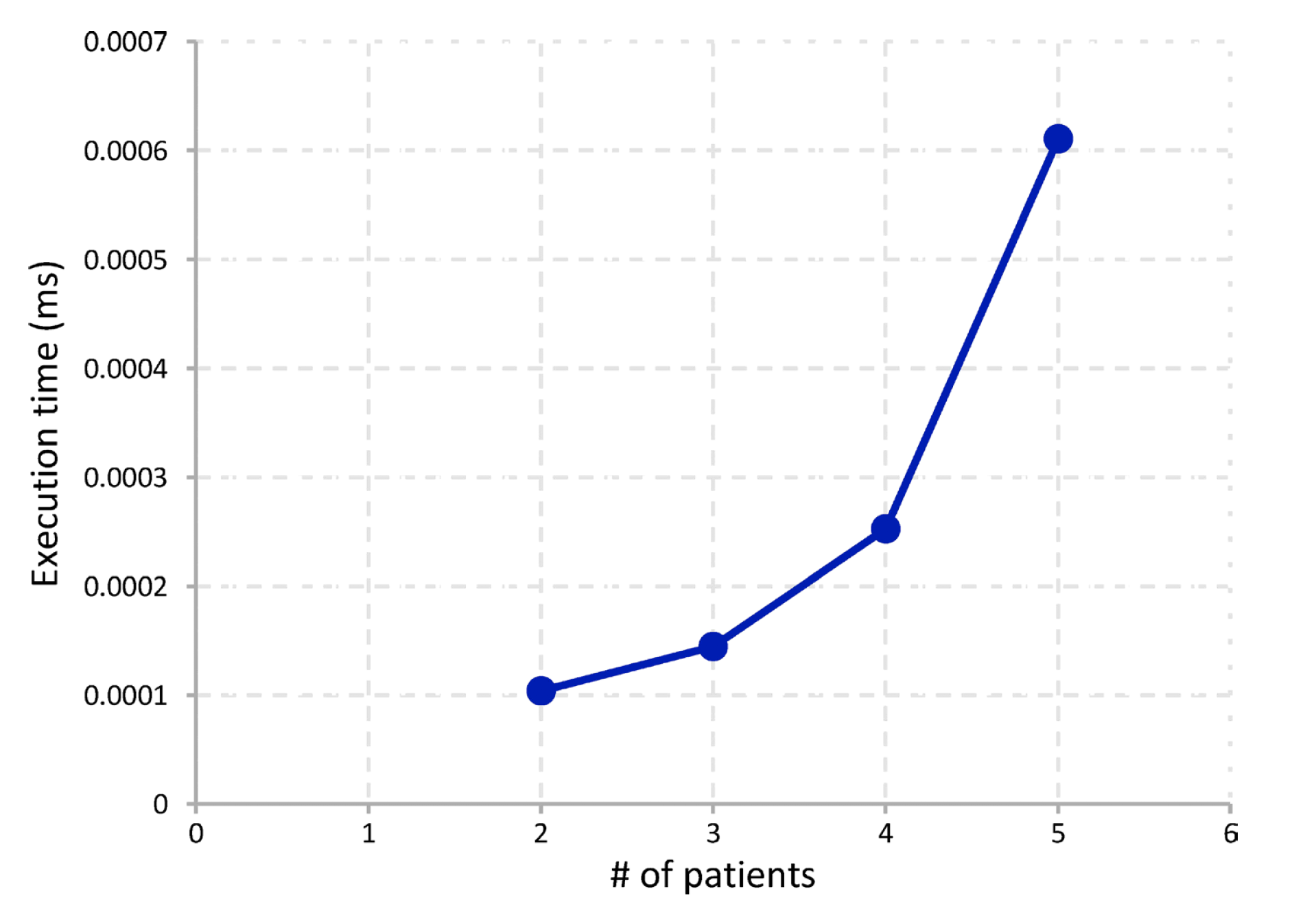}
 \caption{Backtracking’s behaviour depending on the execution time and the number of
patients.}
    \label{fig:Backtracking_Behaviour}
\end{figure}

\begin{figure}[h!]
    \centering
    \includegraphics[width=0.5\textwidth]{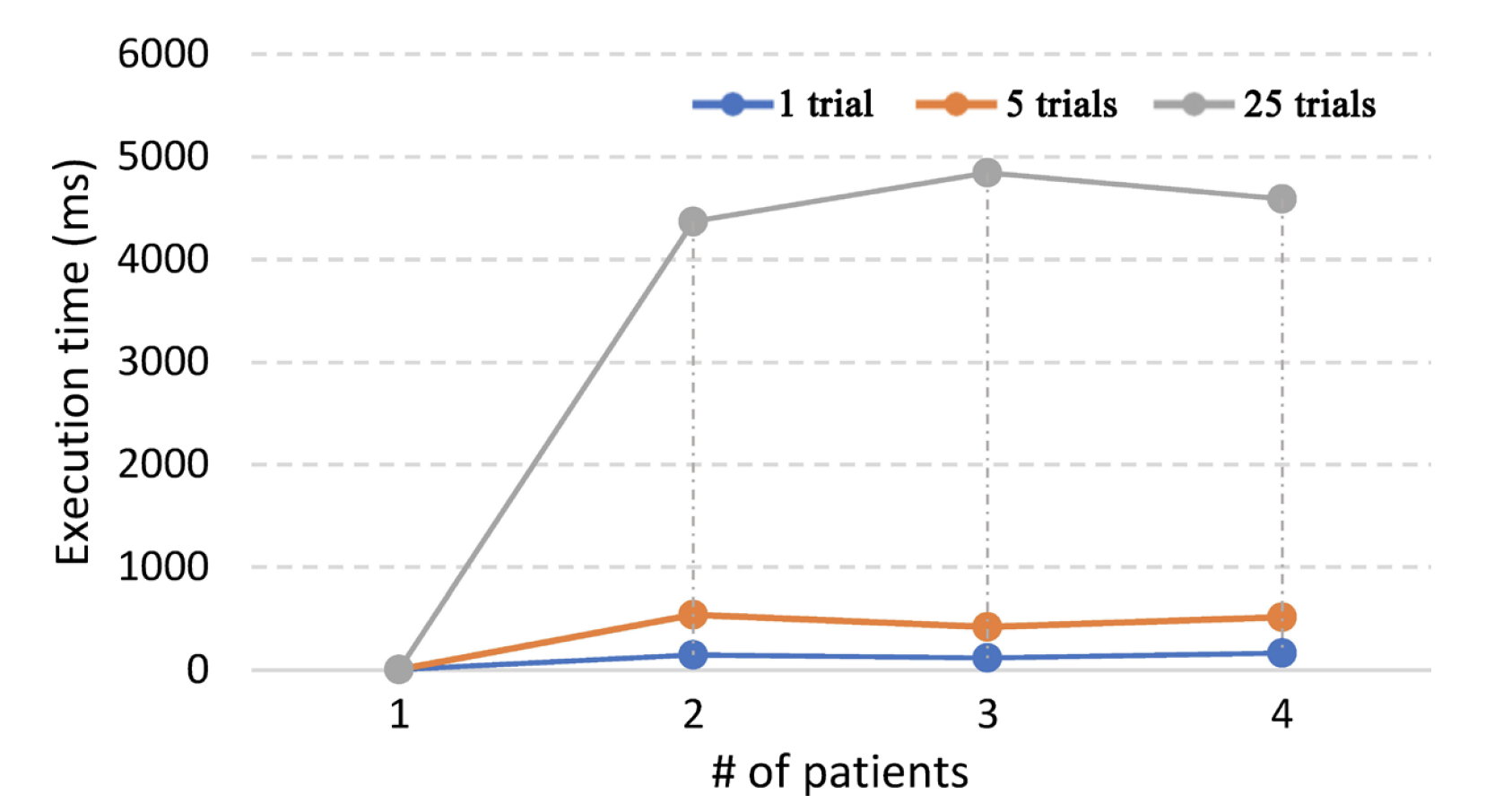}
 \caption{VQE’s behaviour depending on the execution time and the number of patients
and number of trials}
    \label{fig:VQE_Behaviour_}
\end{figure}

The results obtained with the VQE, compared with the classical and exact quantum computing solver, demonstrate that the results obtained are correct. The only significant difference is the time it took for this quantum computers era to compute our algorithm right now. It is worth noting that no such time has anything to do with the complexity class. In contrast, nowadays, what is gained with quantum computers is to observe how well our algorithms work (See Fig. \eqref{fig:SWP_Classical_BackT_Mon-Wed} to \eqref{fig:VQE_Behaviour_}).
\subsection{Backtracking}\label{sec: SWP_Bench_Bactraking}
Backtracking is not the most efficient algorithm to solve this problem, but it is suitable for the type of CSP problem (Constraint Satisfaction Problems). Backtracking depends on a user-given scenario that defines the problem to be solved, the nature of the partial solutions, and how they are scaled into complete solutions. It is, therefore, a meta-heuristic rather than a specific algorithm; it is guaranteed to find all outcomes to a finite (limited) problem in a limited amount of time since backtracking algorithms are generally exponential in both time and space. In our case, we use the Backtracking algorithm to consider all possible issues within the constraints.
We can observe that, while Backtracking presents an exponential behaviour as the number of patients (which would be the nodes of the graph) increases, the VQE algorithm, without considering the cost of evaluation and calibration of the algorithm, has a growth linear. This, as the number of patients grows, will offer more considerable advantages than a classic algorithm, such as Backtracking, since its temporary cost will be much lower for more complex problems.
Finally, we compare the time cost of executing the classic algorithm used with the VQE algorithm on the \textit{ibmq\_16\_melbourne}. To do this, we changed the back-end that allows us to perform the algorithm on a real quantum computer with the highest number of qubits available (15 qubits). 

The \textit{ibmq\_16\_melbourne} returns in the result array a parameter that indicates the total time required to run the algorithm, including the calibrations necessary before running the algorithm in each trial. It is essential to keep in mind since the results’ time will always be very high, but we are interested in the asymptotic trend that the algorithm follows as we increase the total number of patients to visit. Besides, it is necessary to consider that this kind of algorithm shows its potential in the face of highly complex problems. So, the problem we are trying to solve with such a small number of patients does not allow the VQE algorithm to offer an advantage over a classic algorithm. Still, it will enable us to observe its behaviour and validate that it can solve the posed problem. The results are shown in table \eqref{tab:Backtracking_behaviour} and figures \eqref{fig:SWP_Classical_BackT_Mon-Wed} and \eqref{fig:SWP_Classical_BackT_Th-Fri}.

\begin{figure}[!h]
    \centering
    \includegraphics[width=0.5\textwidth]{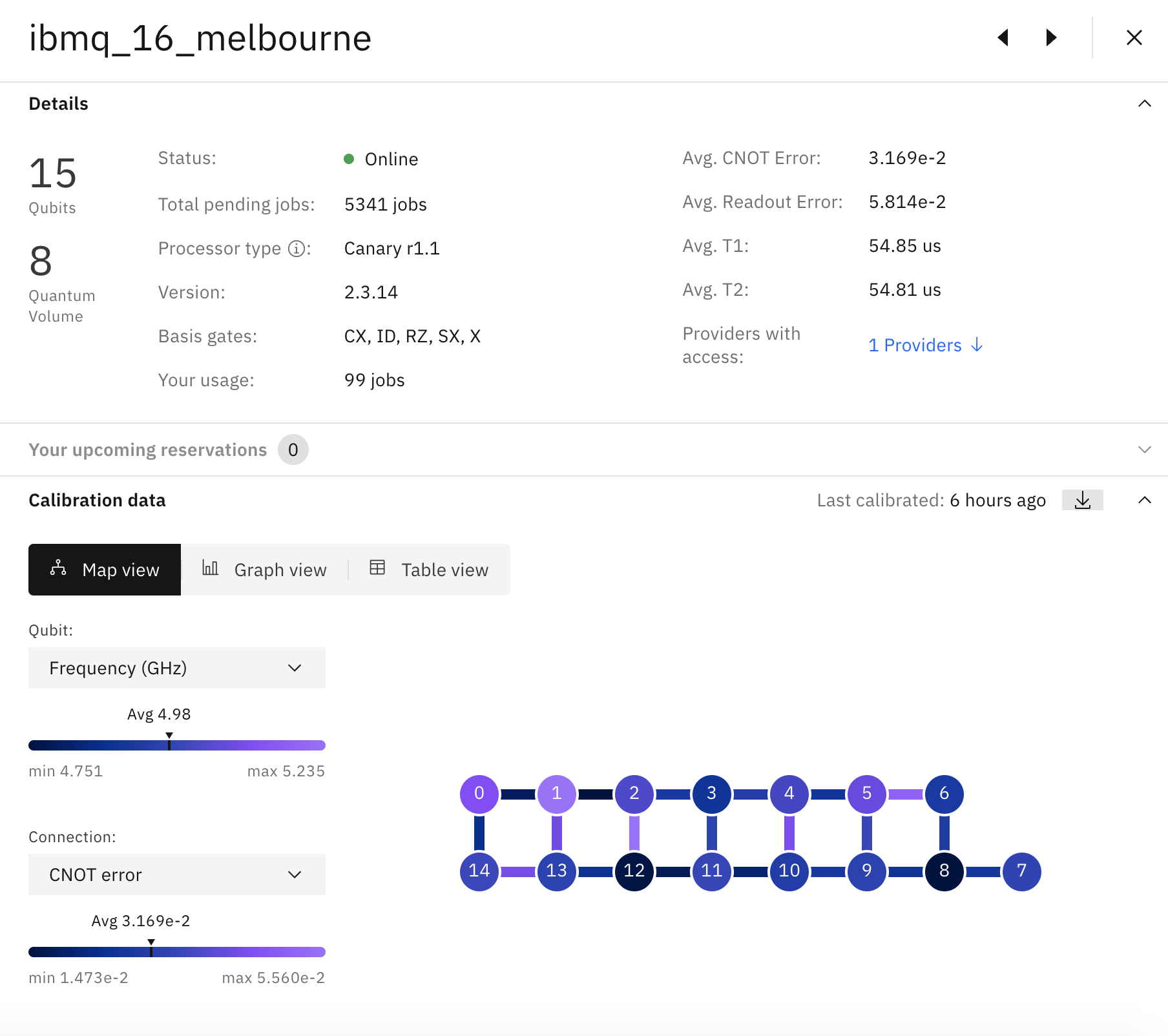}
    \caption{The structure of the $ibmq\_melbourne$ quantum device from IBM. This computer has fifteen superconducting qubits, which are connected by the $C_{NOT}$ gate. The bidirectionality of arrows reflects that each of the qubits can be both a control and a target.}
    \label{fig:Melbourne_node}
\end{figure}

All the tests were done with the calibration from table \eqref{tab:Calibration}—nevertheless, this ref. \cite{Kuzmak2020} could help to dive deeply. Figure \eqref{fig:Melbourne_node} expresses the structure of the $ibmq\_melbourne$ quantum device from IBM. This computer has fifteen superconducting qubits.

\begin{table}[!h]
    \centering
    \label{tab:Calibration}
    \caption{This is $ibmq\_16\_melbourne\_calibrations$}
    \small
    \resizebox{!}{0.8\textheight}{ 
    \begin{turn}{90}
    \begin{tabular}{c|c|c|c|c|c|c|c|}
    \hline
    \multicolumn{8}{|c|}{ $ibmq\_16_melbourne\_calibrations$ } \\
    \hline
       Qubit  &	Frequency (GHz) &	T1 (µs) &	T2 (µs)	 & $\sqrt{x} (sx)$ error &	Single-qubit Pauli-X error &	Readout assignment error &	CNOT error\\
    \hline
	&$5.114621711511489$	&$71.32106756982616$	&$102.41449927678529$
	&$4.183978644302012e-4$	&$4.183978644302012e-4$	&$2.6499999999999968e-2$	& $cx0_14: 2.492e-2 , cx0_1: 1.843e-2$ \\
1   &$5.23503871516284$	&$50.19498723625214$	&$47.728364075207956$	&$1.0042524463122974e-3$	&$1.0042524463122974e-3$	&$3.5700000000000065e-2$	&$cx1_13: 4.305e-2 , cx1_2: 1.473e-2 , cx1_0: 1.843e-2$\\
2	&$5.038356392729712$	&$53.16105211069983$	&$62.779117846203924$	&$6.693469486494128e-4$	&$6.693469486494128e-4$	&$4.1100000000000025e-2$	&$cx2_12: 5.560e-2 , cx2_3: 2.907e-2 , cx2_1: 1.473e-2$ \\
3	&$4.894450631891079$	&$58.021400636079925$	&$17.48180445538179$	&$1.0612091500236409e-3$	&$1.0612091500236409e-3$	&$6.059999999999999e-2$	&$cx3_11: 3.419e-2 , cx3_4: 3.149e-2 , cx3_2: 2.907e-2$ \\
4	&$5.022087248371323$	&$68.25270765048857$	&$73.03990190979944$	&$8.648278230413689e-4$	&$8.648278230413689e-4$	&$4.3399999999999994e-2$	&$cx4_10: 4.481e-2 , cx4_5: 2.751e-2 , cx4_3: 3.149e-2$\\
5	&$5.073224611163873$	&$19.419330885019857$	&$32.372911443214136$	&$2.7881350400977855e-3$	&$2.7881350400977855e-3$	&$5.7800000000000074e-2$	&$cx5_9: 3.543e-2 , cx5_6: 5.156e-2 , cx5_4: 2.751e-2$\\
6	&$4.929465352220578$5	&$67.88136633618807$	&$73.3076496426577$	&$1.6592402096110034e-3$	&$1.6592402096110034e-3$	&$1.866000000000001e-1$	&$cx6_8: 3.061e-2 , cx6_5: 5.156e-2$ \\
7	&$4.983244873960245$	&$42.189182464521444$	&$27.512101284082842$	&$1.6103212138957879e-3$	
&$1.6103212138957879e-3$  &$6.559999999999999e-2$	&$cx7_8: 2.799e-2$ \\
8	&$4.751431321984846$	&$105.48845158260177$	&$78.14201824409058$	&$7.577620344854018e-4$	&$7.577620344854018e-4$	&$3.0000000000000027e-2$	&$cx8_6: 3.061e-2 , cx8_9: 2.650e-2 , cx8_7: 2.799e-2$ \\
9	&$4.973518170784609$	&$36.08378088891272$	&$55.028770007179574$	&$3.210266597002098e-3$	&$3.210266597002098e-3$	&$4.760000000000009e-2$	&$cx9_5: 3.543e-2 , cx9_10: 2.901e-2 , cx9_8: 2.650e-2$\\
10	&$4.944698530540347$	&$64.22423206501318$	&$39.243938767033356$	&$1.1260692077635155e-3$	&$1.1260692077635155e-3$	&$4.1000000000000036e-2$	&$cx10_4: 4.481e-2 , cx10_11: 2.106e-2 , cx10_9: 2.901e-2$ \\
11	&$4.997443578451156$	&$50.11265705810803$	&$85.15292984056299$	&$6.135524302248029e-4$	&$6.135524302248029e-4$	&$4.0100000000000025e-2$	&$cx11_3: 3.419e-2 , cx11_12: 1.786e-2 , cx11_10: 2.106e-2$ \\
12	&$4.763630144177076$	&$73.55506793299364$	&$54.90241515537368$	&$8.832737710164472e-4$	&$8.832737710164472e-4$	&$5.369999999999997e-2$	&$cx12_2:5.560e-2 , cx12_13: 2.609e-2 , cx12_11: 1.786e-2$ \\
13	&$4.973575005661678$	&$25.128388166502205$	&$27.796780492098847$	&$2.7890936362750737e-3$	&$2.7890936362750737e-3$	&$5.940000000000001e-2$	&$cx13_14: 4.387e-2 , cx13_1: 4.305e-2 , cx13_12: 2.609e-2 $\\
14	&$5.007402435608253$	&$37.71533982028009$	&$45.30433211703277$	&$1.5790876435366457e-3$	&$1.5790876435366457e-3$	&$8.299999999999996e-2$	
&$cx14_0:2.492e-2 , cx14_13: 4.387e-2$ \\
\end{tabular}
\end{turn}
}
\end{table}

%%%%%%%%%%%%%%%%%%%% Table No: 10 starts here %%%%%%%%%%%%%%%%%%%%
\begin{table}[!h]
    \centering
    \begin{tabular}{c|c}
    \hline
        Number of patients & Time $(\mu s)$ \\
    \hline
        2 &	0.1039505 \\
        3 &	0.1451969 \\
        4 &	0.2527237 \\
        5 &	0.6108284 \\
    \end{tabular}
    \caption{Experiment scenario using the Backtracking technique with 2, 3, 4 and 5 patients}
    \label{tab:Backtracking_behaviour}
\end{table}
%%%%%%%%%%%%%%%%%%%% Table No: 11 starts here %%%%%%%%%%%%%%%%%%%%
\begin{table}[t!]
\centering
\begin{tabular}{ |c|c|c|c|  }
 \hline
 \multicolumn{4}{|c|}{Temporal cost (ms) regarding the number of patients and trials used} \\
 \hline
 Patients/Trials & 1  & 5  & 25\\
 \hline
2   &	147.253512  &	115.247231  &	164.301820  \\
3   &	536.902598  &	421.142229  &	512.342655 \\
4   &	4371.009830 &	4845.496973 &	4588.167234 \\
 \hline
\end{tabular}
\caption{VQE evolution of trials and number of patients}
\label{tab:VQE_Evolution_trials}
\end{table}
%%%%%%%%%%%%%%%%%%%% Table No: 11 ends here %%%%%%%%%%%%%%%%%%%%
\subsection{Comparison of the algorithms’ complexity}\label{sec:SWP_Complexity_Comparison}
We can observe that, while Backtracking (Fig. \eqref{fig:Backtracking_Behaviour}) and the classical exact solver present an exponential behaviour as the number of patients (which would be the nodes of the graph) increases, the VQE, QAOA algorithm (\eqref{fig:VQE_Behaviour_}), without taking into account the cost of evaluation and calibration of the algorithm, have a logarithmic growth. This, as the number of patients grows, will offer more considerable advantages than a classic algorithm, such as Backtracking, since its temporary cost will be much lower for more complex problems \eqref{fig:Backtracking_Behaviour} and \eqref{fig:VQE_Behaviour_}. Let us insist that the very good scenario of this time benchmark would be with a recursive Grover; nevertheless, it doesn't make a lot of sense compare time in this quantum era.

\subsection{OR Tool SAT}\label{sec:SWP_Bench_OR_Tool}
\begin{figure}[h!]
    \centering
    \includegraphics[width=0.6\textwidth]{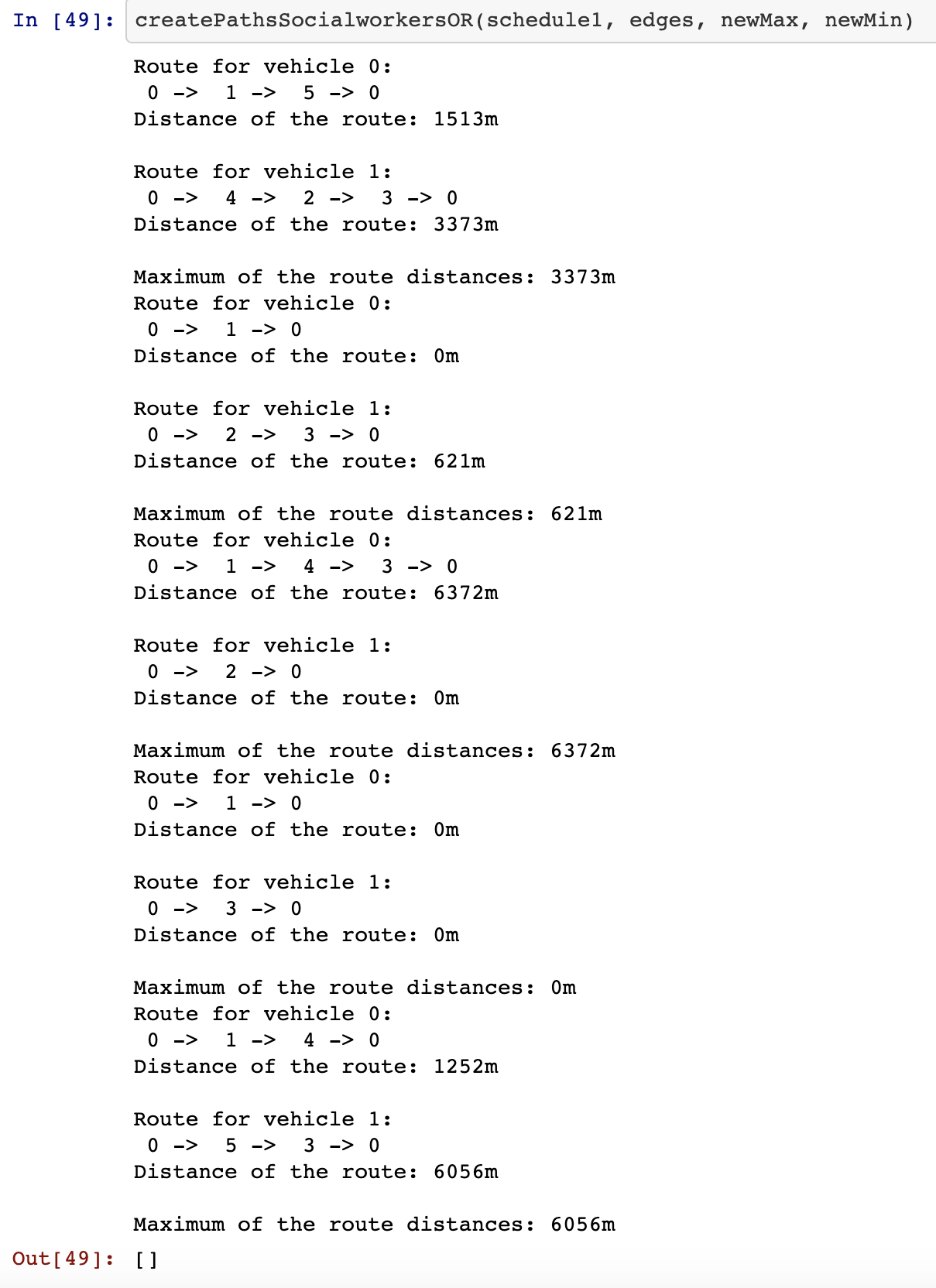}
 \caption{SWP solved by OR-Tool-SAT from google }
    \label{fig:OR_SWP}
\end{figure}

To validate the correct implementation and solution of problems, we use the Google OR-tools API, which is geared towards solving VRP (Google Developers OR Tools, 2019). To use this API, it is required to assign, at least, a distance between nodes. This distance, for this problem, will be the calculation of weights, conditioned by the time limitations between schedules. This means that if a path is not valid, we assign it a weight corresponding to the system maximum (\textit{sys.maxsize}). Thanks to this, the API can recognise that paths are not valid since they can be added to the maximum distance that a vehicle can move in a journey. Therefore, if we also assign this maximum to \textit{sys.maxsize}, we cannot move the paths with the same length, preventing us from considering these paths valid. The result of the algorithm is then as shown in figure \eqref{fig:OR_SWP}.

\subsection{Comparing 10 scenarios of SWP with VQE}\label{sec:SWP_10_VQE}
\begin{figure}[h!]
    \centering
    \includegraphics[width=0.6\textwidth]{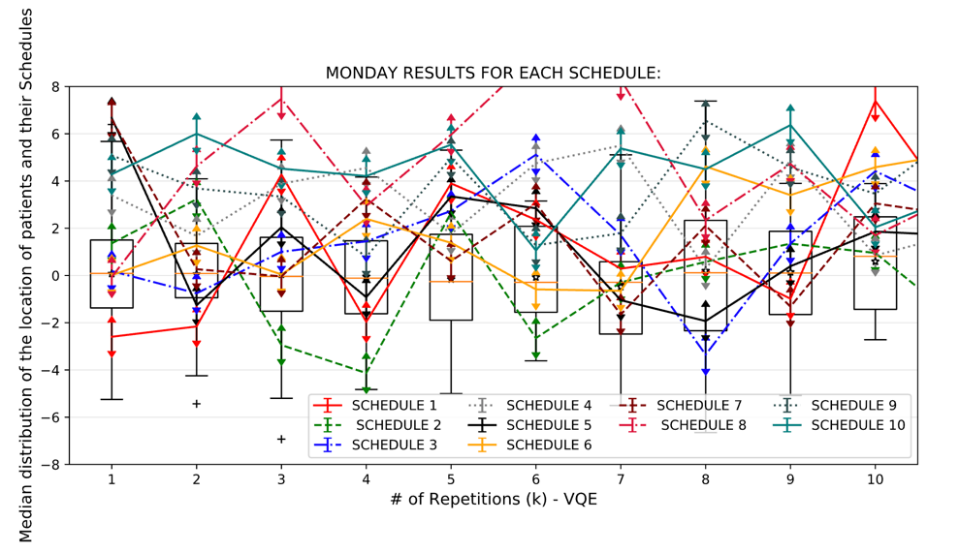}
 \caption{Results of the experimentation by comparing ten scenarios of the combinations of Social Workers' schedules for a fixed patient's location (for Monday). We plot the error bar of the median distribution of the location of each patient and their schedules concerning the number of repetitions of the quantum's instance.}
    \label{fig:Comparing_10_SW_Schedules}
\end{figure}

This section will compare the SWP' schedules in ten scenarios by fixing the patient's location and solving it with the VQE (See Fig. \eqref{fig:Comparing_10_SW_Schedules}). We realise that the QAOA with the same settings as VQE finds the optimal solution with the little sample. We must indeed consider the high values of the shot parameter.

\subsubsection{Comparing 5 scenarios of SWP with VQE and QAOA}
\begin{figure}[h!]
    \centering
    \includegraphics[width=0.6\textwidth]{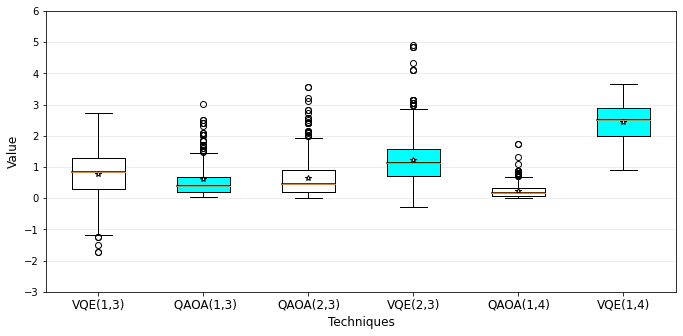}
 \caption{Results of the experimentation by comparing six scenarios of the combinations of Social Workers' Problem on VQE and QAOA. By changing the number of patients, the social worker shot configuration intending to analyse the quantum cost to meet the optimal solution. We realise that the QAOA with the same settings as VQE finds the optimal solution with the little sample. We must indeed consider the high values of the shot parameter. The star is the medium and the red line real the result.}
    \label{fig:Comparing_5_SW_VQE_QAOAS}
\end{figure}

This section will compare the SWP in five scenarios the algorithm's scalability by varying the patient's number and solving it with the VQE and QAOA (See Fig. \eqref{fig:Comparing_5_SW_VQE_QAOAS}). We realised that the QAOA with the same settings as VQE finds the optimal solution with the little sample.  We must indeed consider the high values of the shot parameter.  The star is the medium and the red line real the result.

\subsection{Comparing 10 scenarios of SWP on VQE and QAOA} \label{sec:SWP_10_VQE_QAOA}
\begin{figure}[h!]
    \centering
    \includegraphics[width=0.5\textwidth]{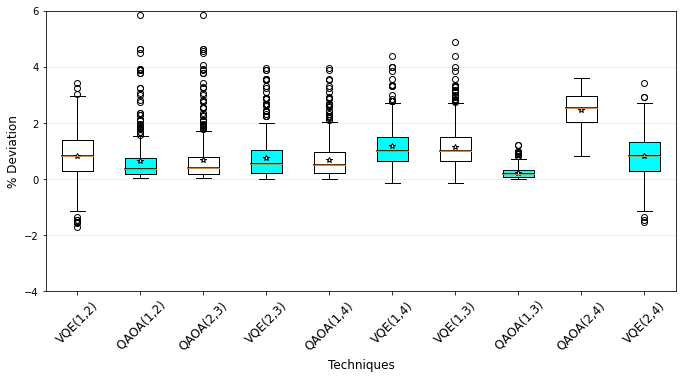}
 \caption{Results of the experimentation by hanging the number of patients, social worker shot configuration intending to analyse the quantum cost to meet the optimal solution. We realise that the QAOA with the same settings as VQE finds the optimal solution with the little sample.}
    \label{fig:Comparing_10_SW_VQE_QAOA}
\end{figure}

This section will compare the SWP in ten scenarios the algorithm's scalability by varying the patient's number and solving it with the VQE and QAOA (See Fig. \eqref{fig:Comparing_10_SW_VQE_QAOA}). We realised that the QAOA with the same settings as VQE finds the optimal solution with the little sample.  We must indeed consider the high values of the shot parameter.  The star is the medium and the red line real the result.
\begin{figure}[h!]
    \centering
    \includegraphics[width=0.5\textwidth]{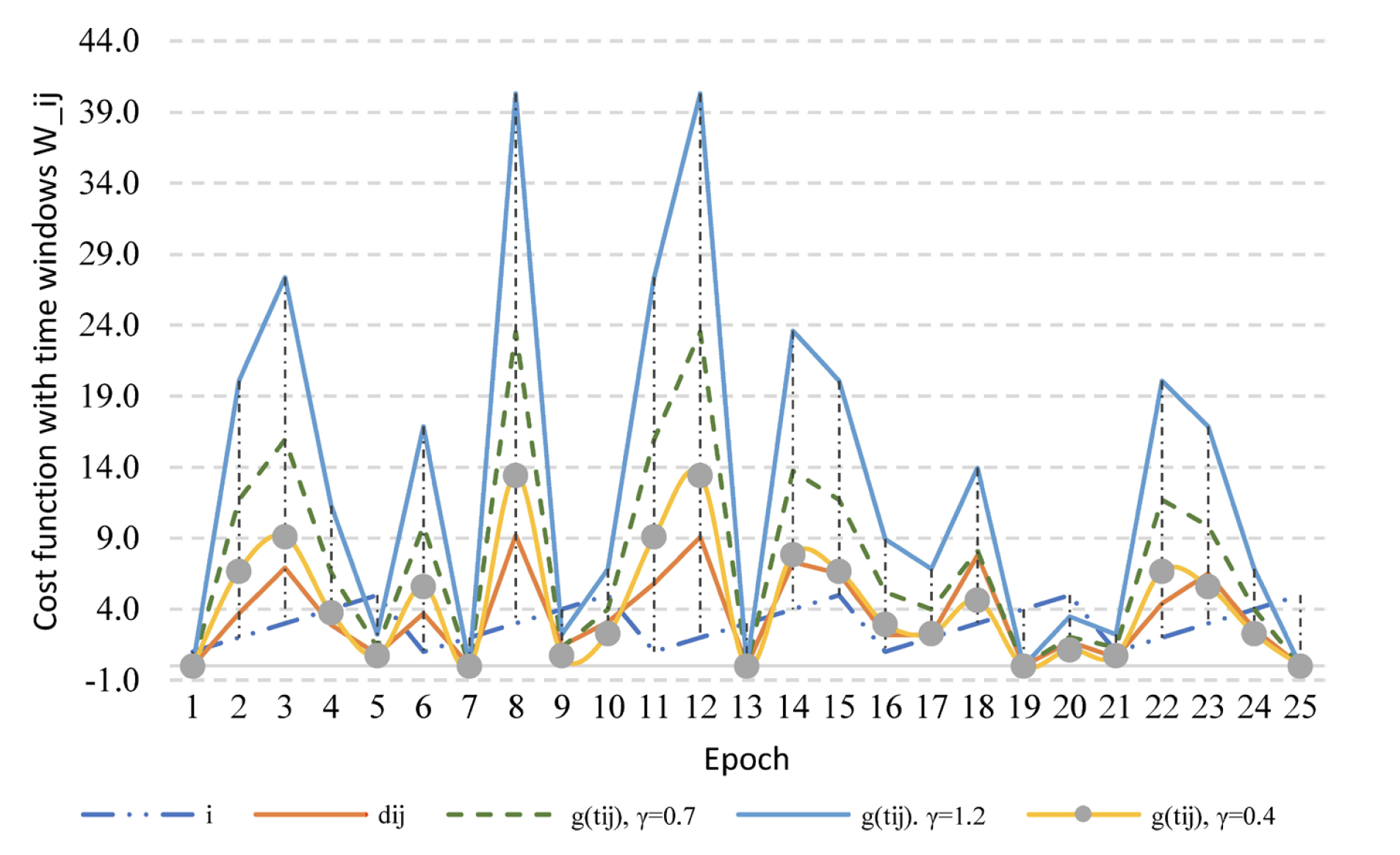}
 \caption{The Standard Deviation expected total anneal time for 98$\%$  per cent success for each mapping value, with the best $\varepsilon$  for each shoot is shown. Our optimal case is for $\varepsilon =0,7$. Our most representative cases are for $\varepsilon =0,4$, $\varepsilon =0,7$ and  $\varepsilon =1,1$.}
    \label{fig:Analysis_epsilon}
\end{figure}

\subsection{Comparing 10 banks of the Social Workers' schedules on QAOA}\label{sec:SWP_Bench_10_SWP_QAOA}
\begin{figure}[h!]
    \centering
    \includegraphics[width=0.5\textwidth]{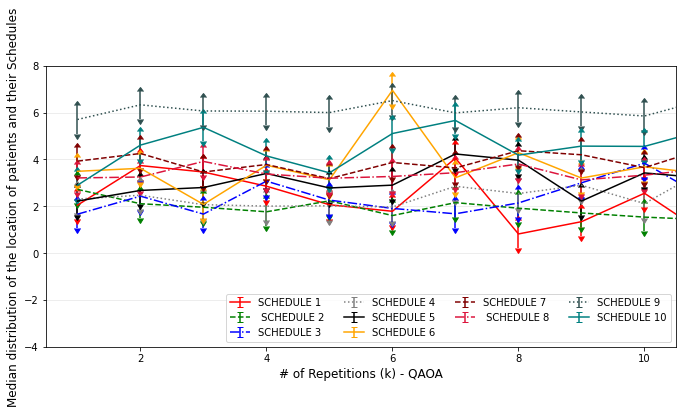}
 \caption{Results of the experimentation by comparing ten scenarios of the combinations of Social Workers' schedules for a fixed patient's location. We plot the error bar of the median distribution of the location of each patient and their schedules concerning the number of repetitions of the quantum's instance.}
    \label{fig:Comparing_10_Median_Err}
\end{figure}

This section will compare the SWP in ten banks schedules by fixing the patient's location and solving it with the QAOA (See Fig. \eqref{fig:Comparing_10_SW_VQE_QAOA} and \eqref{fig:Comparing_10_Median_Err}). We realised that the QAOA with the same settings as VQE finds the optimal solution with the little sample.  We must indeed consider the high values of the shot parameter.  The star is the medium and the red line real the result.

\subsection{Comparing the SWP on all quantum algorithms} \label{sec:SWP_Bench_all}
Since we wanted to make a decent comparison, we mean like many qubit numbers, we had to change the back-end from the \textit{statevector\_simulator} to \textit{qasm\_simulator} and the real quantum computer. We made this change because the superposition calculations of the quantum states or complex amplitudes that the simulator provides to keep track of the algorithm overthrow the computer. It must be said that this has nothing to do with the efficiency of the quantum computer but is a tool that facilitates Qiskit to learn the steps that the algorithm follows.

The results obtained with the \textit{qasm\_simulator} are very similar to the results of the real quantum computer, as seen in the figures \eqref{fig:SWP_Quantum_ADMM} and \eqref{fig:SWP_Quantum_MinimumEigenOptimizer}.

We can observe (Fig. \eqref{fig:Comp_Analysis_Friday}) in this case that the quantum exact solver gave the same result as the backtracking but, in the case of the VQE, the solution provided has one patient changed.
This can be a problem if the patients will not allow a change of social worker daily.
Another improvement is to reduce the computational cost of the top-down algorithm.

 %%%%%%%%%%%%  Starting New Page here %%%%%%%%%%%%%%

\newpage

\section{Conclusion and Further Work}\label{sec:SWP_Conclusion_intro}
This thesis work has studied, implemented, and proved the feasibility of solving the Constraint Search Problem more efficiently with quantum computing. We have researched and implemented, in classic and quantum, several combinatorial optimisation algorithms to solve the problem of social workers visiting patients at their homes. We have also tested the concepts and techniques of combinational optimisation in programming environments, Cirq, Qiskit and Pennylane. Considering this, we analysed the state of maturity of the hardware, the framework, of the scientific community and even the response time of the technical support of each leading company in this quantum computing career. We have opted for IBMQ for all the facilities (framework, hardware, support and libraries) they have given us to be within their community. After the implementation, we have run the algorithms on both quantum simulators (\textit{32q\_ibmq\_qasm\_simulator}) and quantum computers (\textit{ibmq\_16\_melbourne v2.3.0}) from IBM. The results and the discussions are developed in detail in this thesis work.

\subsection{Conclusions}\label{sec:SWP_Conclusion}
We have developed several mathematical formulations and proposed one to solve our problem. First, we reduce the number of the qubits to be proportional to $N^2$ instead of the $N^4$ of the literature \cite{papalitsas2019qubo}.
Our proposed formulation allows us to design a strategy to take advantage of the era we are in (NISQ) (\textbf{Few useful qubits}, limitations in quadratic techniques with inequality restrictions, just to mention these). Moreover, although it seems specific, said formulation is later generalised to encompass more generic applications, and we did a comparative analysis of 5 different implementations made and discussed during this thesis work.

We have analysed the Ising and the QUBO models in-depth to solve quadratic problems. We have also studied the CPLEX optimisation tools such as IBM's docplex under its framework Qiskit.

Qiskit's optimisation tool included the generic Quadratic Programs that help to model any optimisation problem. It is also involved many converters to map the problem to solve to the correct input format. Converters like, \textit{Inequality to Equality} to map inequality constraints into equality constraints with additional slack variables. The converter \textit{Integer To Binary} is useful in the case of the need to convert integer variables into binary variables and corresponding coefficients. \textit{Linear Equality To Penalty} helps to convert equality constraints into additional terms of the object function, and last but not least, the \textit{Quadratic Program To QUBO} converter is used as a wrapper for all the mentioned converters.

We have analysed the Minimum Eigen Optimiser and the optimiser ADMM alternative to VQE and QAOA.

We have also developed some guidelines to make it easier for future PhD or graduate students to take their first firm steps in quantum computing.

\subsection{Future directions}\label{sec:SWP_F_Directions}
As a future line, we will repeat the comparative studies of our formulation with the generalised formulation. But for this comparison, we need to be able to count a quantum computer of at least 49 qubits =  $n \left( n-1 \right)$. Where $n$ is the number of the patients. Another exciting line that we contemplate is the design of an adaptive and specific algorithm that considers any modification or configuration in real-time of the patient or the social worker. 
Another future line is to design the specific VQE for the proposed problem to gain efficiency.
During our experimentation, we realised that the initial configurations are not adequate in some cases; these quantum variational algorithms (VQE, QAOA, ADMM) fail. However, they are today one of the most promising applications of quantum computers in this era (NISQ).
We define this future direction to focus on a unique family of quantum circuits called the Hamiltonian Variational Ansatz (HVA), which the QAOA and adiabatic quantum computing inspire. By studying its entanglement spectrum and energy gradient statistics, scientists are experimenting with how HVA exhibits favourable structural properties and optimisation facilities compared to well-studied hardware ansatz.
A line of future that interests us a lot is Quantum Machine Learning. However, it is worth mentioning that we are already halfway to QML since all the algorithms based on the variational principle lead us towards this goal. Within this objective, one of the algorithms of the Qiskit framework is the Recursive Minimum Eigen Optimiser that takes a Minimum Eigen Optimiser as input and applies the recursive optimisation structure and strategy to reduce the size of the problem one variable at a time.

\section{Summary}

In this section, we have presented our contribution to the scientific community as its implementation. We have proposed, designed, and implemented the algorithm to formulate the problem of social workers who visit patients at their homes. We have tested brute force techniques, Backtracking, Dijkstra, ADMM, Minimum Eigen Optimiser using different Minimum Eigensolver, such as VQE, QAOA or Numpy Minimum Eigensolver (classical), GroverOptimiser, CplexOptimiser and for the classical optimiser Cobyla, SLSQP, SPSA, $ \ldots $ ).

In the Generalisation of the solution section, we generalised the algorithm for generic scheduling and routing problems to be programmed under Quantum Annealing or Gates-Based Quantum Computers.

In the case of Dwave, the formulation already works for the annealing solver. The same goes for Qiskit. In addition, however, we can use the new module from IBMQ (Fig. \eqref{fig:Quantum_Opt_qiskit}) that boosts research, development, and benchmarking of quantum optimisation algorithms for the NISQ era and beyond.

This last formulation (equation \eqref{Hamiltonian_eq} to \eqref{Binary_variable_t_ij}) improves our proposed formulation, in the margin of the social worker waiting time, but requires more binary variables than the first one and therefore more qubits (that right now are lacking and very valuable).

Using the Qiskit library, Quadratic Programming, we will model inequality constraints and much more.

Qiskit, in addition to having converters of inequality to equality constraints, allows modelling from scratch with inequality constraints. So, we have reformulated our system with the library to have inequality restrictions. But as the techniques (slack) of using Qiskit to model, the system applies additional variables that limit us when it comes to having useful qubits. So, both the simulations and the actual tests on the quantum computer are limited since the number of qubits required is higher than the characteristics of the hardware we use.

The number of qubits for integers it is  $\log _{2} \left( ub - lb \right)$ for binaries it is 1:1. Inequality constraints depend on the constraint, i.e. the range of the slack variable to be introduced.

\newpage
\graphicspath{{./media/}}

\chapter{The quantum CBR (qCBR)}\label{sec:12}
\section{Introduction}
From our previous works till now, and following the journey of this thesis \eqref{fig:Phd_overview}, we have seen that the social workers problem (SWP) stands for solving the schedules of social workers visiting patients' homes while fitting both distance and time restrictions~\cite{Atc20} and represents a class of combinatorial optimisation problems, which lie in the NP complexity class.
The standard way to solve this class of problems begins by establishing the cost function. Then, depending on its form, existing linear or quadratic programming methods such as Simplex \cite{Simplex_solver} or Cplex \cite{CPlex_solver} can be applied. More complex cost functions require more sophisticated numerical methods. Depending on the problem's complexity class, the algorithm can be improved by introducing some heuristics or restrictions in the objective function to reduce its computational cost for an approximate solution. When the size of the problem grows, the computational cost may soon become intractable for the current computational paradigms. In addition to the above, solving these problems is more challenging when the input data presents some overlapping issues or when outstanding accuracy is required.

%we observed that many problems can be modelled as the social workers' problem (SWP)\cite{Atc20}, where their resolution may have two straightforward approaches, Top-down solution and how to solve them.
%Top-down resolution is a resolution that can be set as standard: Given a combinatorial optimisation problem, its objective function can be found, and we can use the existing linear o quadratic programming methods of solving as Simplex \cite{Simplex_solver}, CPLEX \cite{CPlex_solver} or numerical to solve it. Depending on the problem's complexity class, the algorithm can be improved by introducing some heuristics or restrictions in the objective function to reduce its computational cost for an approximated solution.

An approach combining adapting Case-Based Reasoning\cite{Aam94} to a quantum computing is proposed to solve this class of problems. This paradigm, denoted Quantum Case-Based Reasoning (qCBR), will address both the overlap in the input data and the accuracy problem. Furthermore, by directly constructing the framework, questions like the actual efficiency of a qCBR implementation at the present level of quantum technology, the tolerance concerning input overlap, the scalability and the applicability to other combinatorial optimisation problems will be discussed.

The proposal seeks to answer the following questions based on experimentation results.
\begin{enumerate}
\item
Can a qCBR be efficiently implemented in this quantum age? 

\item
Could the proposed  qCBR be scalable while maintaining a mean accuracy?
\item.
Can the proposed qCBR be used in other combinational optimisation problems?
\end{enumerate}

\begin{figure}[h!]
\centering
\includegraphics[width=0.7\textwidth]{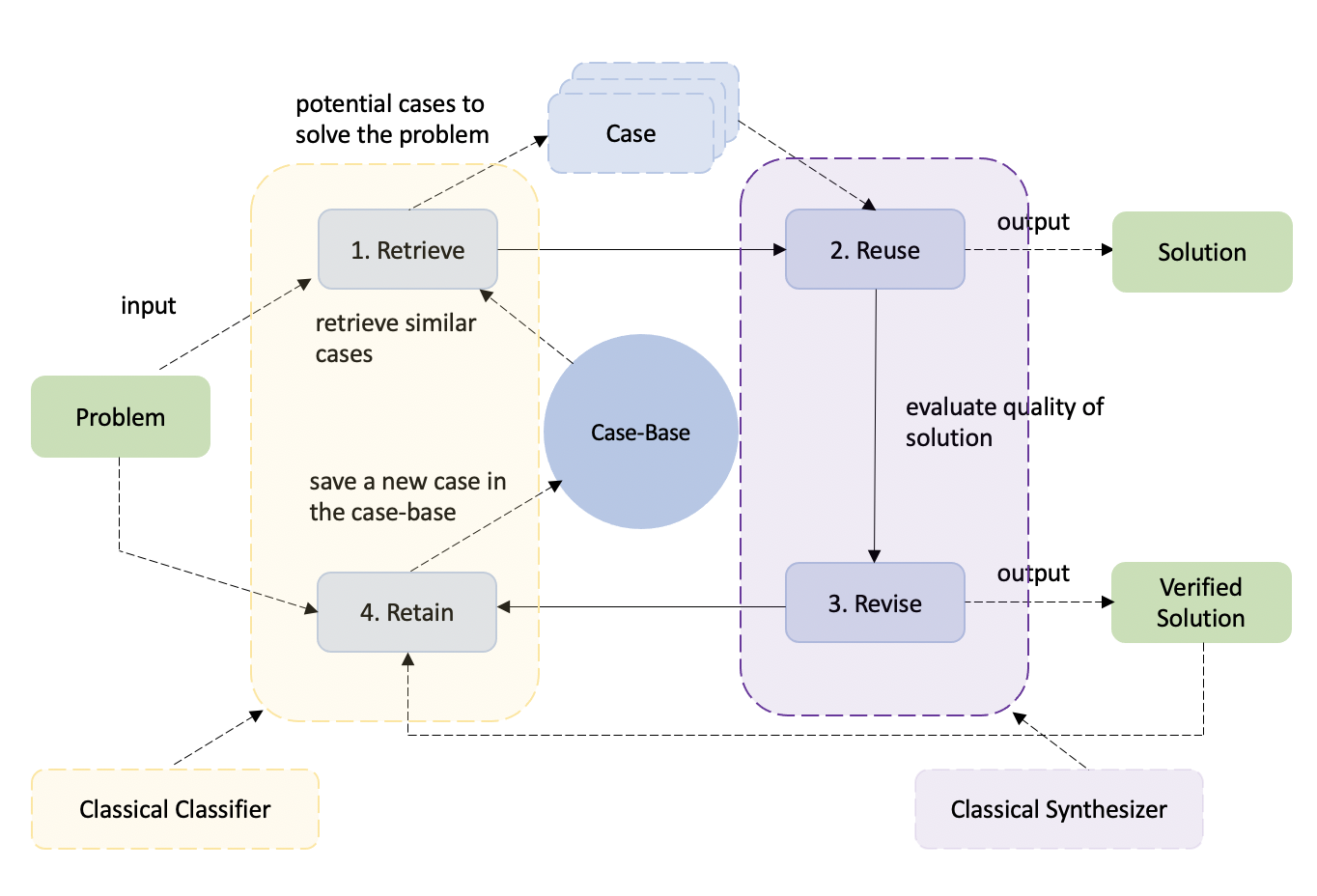}
\caption{Case-Based Reasoning block diagram. For the standard CBR, two essential blocks are distinguished in this work. The classifier and the synthesiser. The classifier is made up of the retrieve and retain blocks and the re-use and revise blocks to make up the synthesis system.}
\label{fig:CBR}
\end{figure}
\begin{table*}[h!]
   \centering
    \begin{tabular}{l|c|c|c|c}
        \textbf{Methods} & \textbf{ Brute force } & \textbf{ k-d tree \cite{Kak05} } & \textbf{ Ball tree \cite{Mun11} } \\
         \hline
        Training time complexity & $O(1)$ & $O(dNlog(N))$ & $O(dNlog(N))$ \\
        Training space complexity & $O(1)$ & $O(dN)$ & $O(dN)$ \\
        Prediction time complexity & $O(KNd)$  & $O(Klog(N))$  & $O(Klog(N))$ \\
        Prediction space complexity & $O(1)$ & $O(1)$ & $O(1)$ \\
    \end{tabular}
    \caption{Table of the NN brute force's, k-d tree's, and Ball tree's complexity method. Where $d$, is the data dimensionality, $N$ is the number of points in the training dataset and $K$ is the algorithm's neighbours' number}
    \label{tab:Complexity_Classifier}
\end{table*}
\subsection{Related Work}\label{sec:Related_work}
This section summarises previous works in CBR, Quantum CBR and Quantum Machine Learning (QML) algorithms.

CBR is a problem solving approach widely considered in the literature with a large record of success. Application examples are a medical reasoning program that improves with experience \cite{koton1989medical}, an individual prognosis of diabetes long-term risks\cite{armengol2001individual}, Case-Based Sequential ordering of songs for playlist recommendation \cite{baccigalupo2006case}, ranking order in financial distress prediction \cite{li2008ranking}, monitoring the elderly at home \cite{koton1989medical}, software control \cite{Aam94}, in the medical field \cite{Aam94}, sequencing problems \cite{Pao01} etc.
In one of our previous works \cite{Atc201,Atc20}, we observed part of the benefits of using the CBR instead of the Top-Down method. And the needs of empowering this problem-solving method based on human learning were seen.

Quantum computing stands as a new computing paradigm based on exploiting the principles of quantum mechanics and establishing the quantum bit (qubit) as the elementary unit of information. It emerged in the early 1990's from algorithms that were able to take advantage of quantum characteristics to show advantages over their classical counterparts, being Shor's algorithm \cite{shor1994algorithms} for integer factorisation and Grover's algorithm \cite{grover1996fast} for searching in an unordered data sequence, the most famous. However, current quantum computing devices suffer from technological limitations, such as the number of qubits available and the noise and decoherence problems, such that they are still no match for their classical counterparts. This situation is known as the Noisy Intermediate-Scale Quantum (NISQ) era \cite{Joh18}. These limitations have forced the scientific community to develop handy tools for hybrid computing, mixing classical and quantum. Taking advantage of the variational principle, it is possible to solve combinatorial optimisation problems and enhance one of this era's most promising fields; quantum machine learning (QML) \cite{Mar14,Adr20}. In this new approach, several techniques and methods already explored in Machine Learning (ML) are being worked on.

In the last two years, the number of algorithms based on QML have increased considerably since the first definition in 2014 \cite{Mar14}. This progress relies on the advances in decoherence  control \cite{decoherence_ctrl,Quantum_decoherence} and error correction systems \cite{Quantum_error_correct} combined with the availability of several quantum server providers in the cloud. Most of these new algorithms take after the variational principle, being the Variational Quantum EigenSolver (VQE) \cite{Peruzzo2014} and the Quantum Approximate Optimisation Algorithm (QAOA) \cite{farhi2014quantum,QAOA_2019,Adapt_QAOA} the most famous. Other promising developments are the Quantum Neural Network (QNN) \cite{QNN_2014,Quest_QNN_2014,Train_QNN_2020}, the Quantum Support Vector Machine (QSVM) \cite{Pat14,SL_Quantum_Feature_Space,SVM_DWAVE} and the data loading system \cite{UAT2021,Mar19}. On the one hand, the following references \cite{a14070194, gonzalezbermejo2021gps} highlight works done in the Top-Down philosophy. On the other hand, references \cite{zhao2021smooth, lamata2020quantum, benfenati2021improved, cerezo2021cost, alonsolinaje2021eva, atchadeadelomou2021quantum} highlight the many contributions in quantum machine learning, from using the properties of quantum computing to finding new drugs as new ways to calculate the expected value, among others.

The literature shows examples of exploiting the possibilities of hybrid (classical-quantum) computing connecting it to CBR. For instance, in reference~\cite{Cognite_CBR_GA} a cognitive engine that uses CBR-QGA to adjust and optimise the radio parameters is presented. An initial quantum bit made up of the matching case parameters is used to avoid blindness of the initial population search and speed up optimisation of the quantum genetic algorithm. References~\cite{Modeling_CBR,CBR_Ontology} propose a new framework that can be adopted in many applications that require Computational Intelligence (CI) solutions. The framework is built under the concepts of Soft Computing (SC), where Fuzzy Logic (FL), Artificial Neural Network (ANN) and Genetic Algorithm (GA) are exploited to perform reasoning tasks based on soft cases. Also studies \cite{review_KNN_CL_QC} focused on some vital blocks of the CBR were reviewed. It has focused on the quantum version of the k-NN algorithm that allows us to understand the fundamentals when transcribing classic machine learning algorithms into their quantum versions. 

Reviewing state of the art, we have seen an interesting field known as Quantum Information Retrieval (QIR) \cite{piwowarski2010can, lebedev2020introductory, kitto2012quantum} that uses the Gleason theorem \cite{GleasonTheorem} on the Measures on the Closed Subspaces of a Hilbert Space for information retrieval geometry \cite{van2004geometry}. It calculates the probability algebraically through the density matrix trace and acts on a quantum projector. The projector can be any concept to recover. However, for the quantum CBR, we are not only interested in a great recovery system, but we also need to provide the qCBR with a synthesiser whose function will be to fine-tune the recovered data in the case of not being the optimal result since the qCBR has the process of "generate" a new outcome based on the retrieved information.

However, no quantum case-based reasoning was found to satisfy the above mentioned requirements.

\section{Case-Based Reasoning}\label{sec:Case_based_reasoning}
CBR \cite{Aam94} is a machine learning technique based on solving new problems using experience, as humans do. The \textit{experience} is represented as a case memory containing previously solved cases. The CBR cycle can be summarised in four steps: (1) Retrieval of the most similar cases, (2) Adaptation to those cases to propose a new solution to the new environment, (3) Validity check of the proposed solution and finally, (4) Storage following a learning policy. In the present work, the proposed qCBR modifies these phases as follows (see Fig.\eqref{fig:CBR} and \eqref{fig:qCBR}).

The CBR technique could be summarised in two large blocks according to their functionality: a classifier and a synthesiser. One of the classical CBR advantages is its classifier's simplicity, being a k-nearest neighbors algorithm (K-NN) \cite{JHF75,Fuk75} classifier a common option. This apparent advantage can lead to collateral problems \cite{Lau11} at the memory level, at the level of slowness when the volume of data grows considerably and at data synthesis. The synthesis block is in charge of adapting the experience and saving the new problem. Such adaptation and classification can be costly (Table \ref{tab:Complexity_Classifier}) for considerably high data volumes \cite{Abd15}. From this follows that a different approach would be required to further empowering this technique.

The proposal of this note is to achieve such empowering in two steps. First by making a CBR with a quantum classifier \cite{Suk191} instead of a classical neural network, KNN \cite{JHF75,Fuk75} or a Support Vector Machine (SVM) \cite{Wil06} since quantum classifiers offer outstanding accuracy and tolerate overlapping problems \cite{8715261}. The second would be changing the classical synthesis technique for the Variational Quantum Eigensolver (VQE) \cite{Alb13,Dao19,Placeholder1} with \textit{Initial\_point} \cite{Qis21}.

\begin{figure}[h!]
\centering
\includegraphics[width=0.4\textwidth]{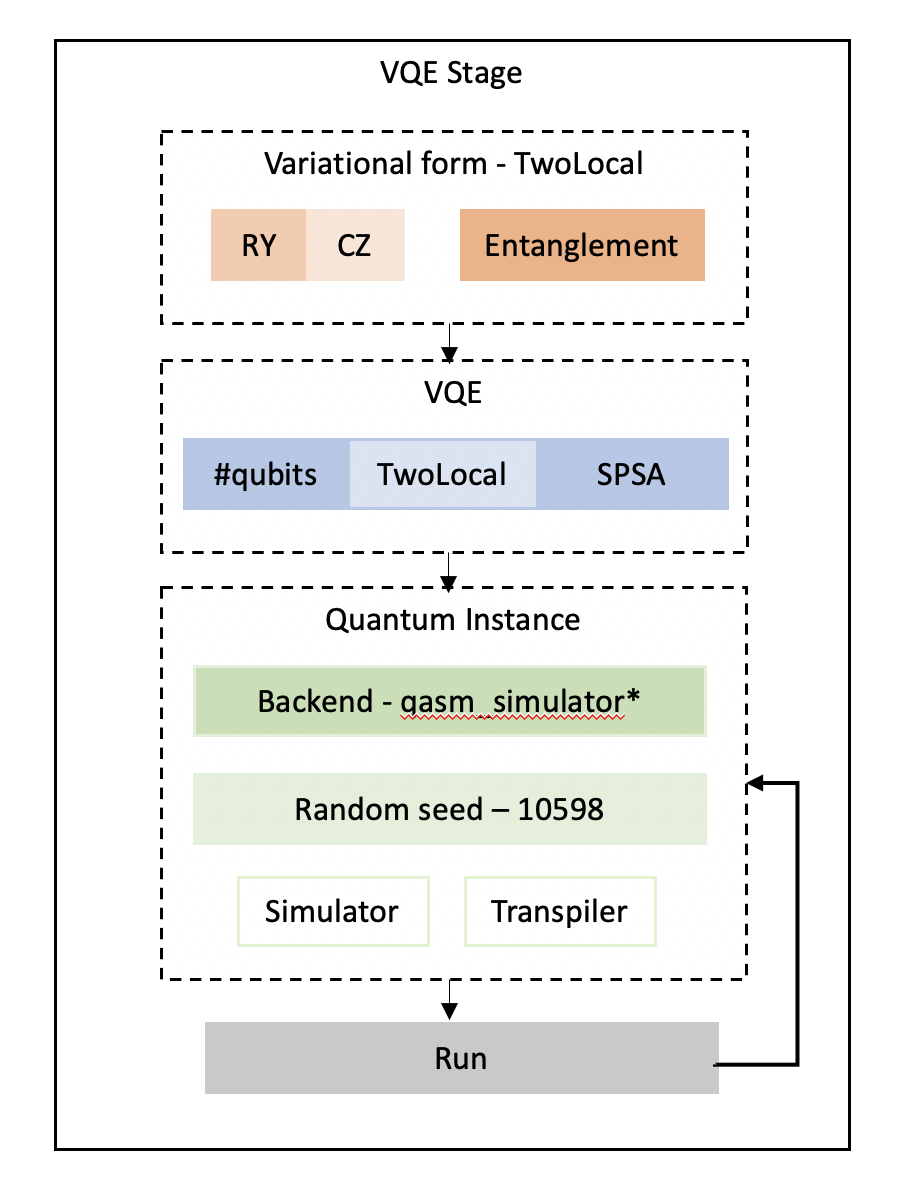}
\caption{VQE working principle based on the quantum variational circuit in Qiskit}
\label{fig:VQE_fig_}
\end{figure}

\section{Implementation}\label{sec:implementation}
Let us recall some concepts above discussed: Quantum circuits are mathematically defined as operations on an initial quantum state. Quantum computing generally makes use of quantum states built from qubits, that is, binary states represented as  $\ket{\psi}=\alpha\ket{0} +\beta\ket{1}$. Their number of qubits $n$  commonly defines the states of a quantum circuit and, in general, the circuit's initial state $\ket\psi_{0}$ is the zero state $\ket{0}$. In general, a quantum circuit implements an internal unit operation $U$  to the initial state $\ket\psi_{0}$ to transform it into the final output state $\ket\psi_{f}$ . This gate $U$  is wholly fixed and known for some algorithms or problems. In contrast, others define its internal functioning through a fixed structure, called Ansatz \cite{Ansatz_best} (Parametrised Quantum Circuit (PQC)), and adjustable parameters $\theta$ \cite{Suk191}. Parameterised circuits are beneficial and have interesting properties in this quantum age since they broadly define the definition of ML and provide flexibility and feasibility of unit operations with arbitrary precision \cite{JBi17,Adr20,Mar14}.

\begin{figure}[!ht]
\centering
\includegraphics[width=0.75\textwidth]{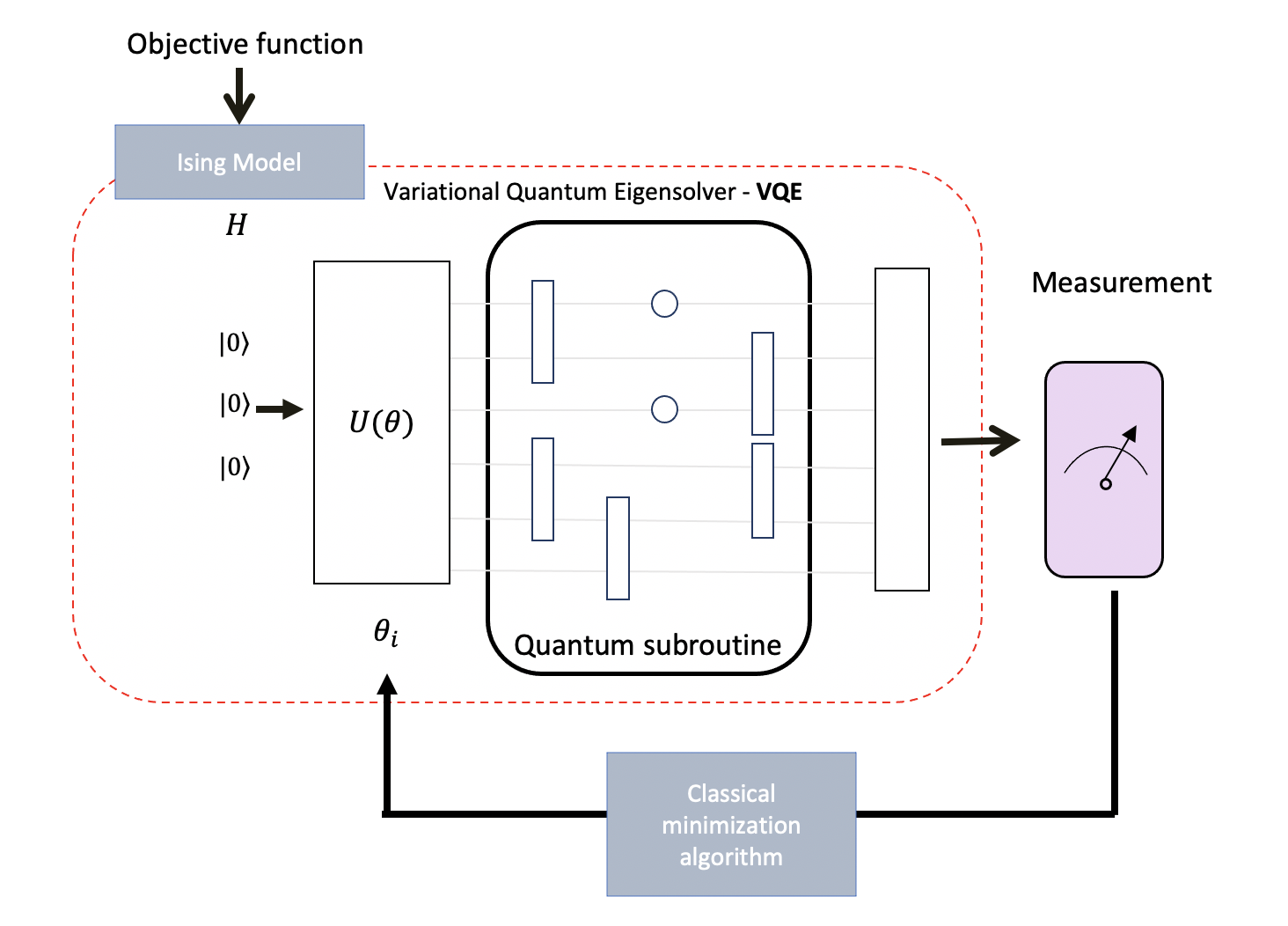}
\caption{VQE working principle based on the quantum variational circuit. Given an objective function that characterises the problem, with the help of the Ising Model block, we pass the objective function from the classical to the quantum domain.
The ansatz is initialised with random values. Then, starting from these initial values (initial position) and depending on the measured value, a classical and external optimiser is used to feedback the new values of the ansatz parameters. So, until reaching the minimum energy value, equivalent to the ground state of our Hamiltonian, defined by the variational principle.}
\label{fig:VQC}
\end{figure}

Figure \eqref{fig:VQC} depicts the concept of hybrid computing (quantum + classical), which defines the NISQ. This takes advantage of quantum computing's capacity to solve complex problems, and the experience of classical optimisation algorithms (COBYLA \cite{The21}, SPSA \cite{Jam01}, BFGS \cite{BFGS_Limted}, etc.) to train variational circuits. Classical algorithms are generally an iterative scheme that searches for better candidates for the parameters $\theta$ at each step.

The value of the hybrid computing idea in the NISQ era is necessary because it allows the scientific community to exploit the powers of both and reap the benefits of the constant acceleration of the oncoming quantum-computer development. With a good optimisation system and a closed-loop system, the non-systematic noises could be automatically corrected during the optimisation process.

Furthermore, with the insertion of information (data) into the variational circuit through the quantum gate $U$, learning techniques can be improved.

The Variational Quantum Circuit (VQC) \cite{Mic00,Mic}, consists of a quantum circuit that defines the base structure similar to neural network architecture (Ansatz) while, the variational procedure can optimise the types of gates (one or two-qubit parametric gates) and their free parameters.
All this is summarised in a few very identifiable steps. 
First, the Ansatz must be designed, using a set of one- and two-qubit parametric gates. The Ansatz of this circuit can follow a particular path by exploiting the problem's characteristics. A critical block is measuring the quantum state resulting from the given Ansatz. Since the VQC is a feedback system, these measurements evaluate a cost function  $CF(\theta)$ that encodes the problem. The classical optimiser has the role of optimising the cost function to find the value of the parameters that minimise it.

\begin{figure}[h!]
\centering
\includegraphics[width=0.9\textwidth]{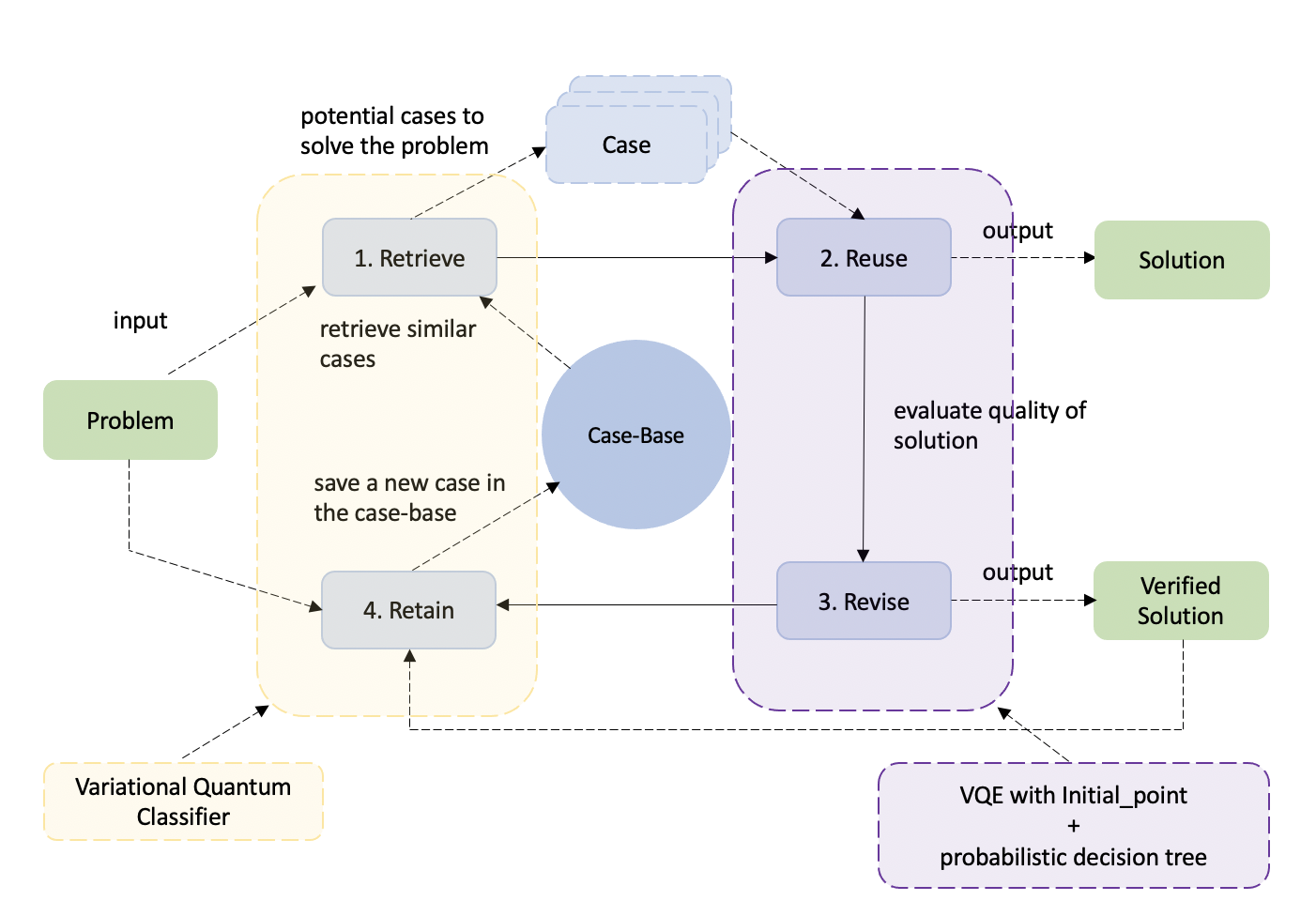}
\caption{Quantum case-based system block diagram. In this scheme, to convert the classic CBR into quantum one, it is proposed to change the retrieve and retain blocks for a quantum variational classifier and the re-use and revise blocks for a synthesis system based on VQE with initial\_point.}
\label{fig:qCBR}
\end{figure}

The work proposed in this article is the implementation of a quantum Case-Based Reasoning (qCBR) based on figure \eqref{fig:CBR}. The strategy to follow is to replace the classical classification: an Artificial Neural Network (ANN) or a Support Vector Machine (SVM) or the KNN with a quantum variational classifier that guarantees the required accuracy. And for the quantum synthesis system, use the VQE with and without \textit{Initial\_point} together with a probabilistic decision tree. Figure \eqref{fig:qCBR} shows the changes that will be introduced to obtain the qCBR, and figure \eqref{fig:Funct_qCBR} shows the detail of the functional blocks implemented with the specific problem of social workers. The two VQE blocks and the Variational Quantum Classifier are presented before detailing them.

\section{Quantum classifiers}
The variational quantum classifier belongs to the variational algorithms like VQE, where classically tunable parameters of a unit circuit are used to minimise the expected value of an observable. The great novelty resides in loading the data in the variational system.

As a universal quantum classifier of $n$ qubits is pursued, a sub-base in the Hilbert vector space of equitably dividing the hyperplane $Z$ is described as follows.
%Let  $ B= \left(\{ i,j,k,l,m,n,o,p  \}\right)$ be a sub-base within the Hilbert vector space, for the space of the classes  $C^{2^{q}}$, the coordinates of the target classes are defined by expression \eqref{measurement_basis} with $q$ as the number of the qubits.

Let $ B= \{ e_{i} : 0 \leq i \leq 2^n-1  \}$ be a sub-base within the Hilbert vector space, for the space of the classes  $C^{2^{n}}$, where $e_{i}$ denotes the vector with a $1$ in the $i_{th}$ coordinate and $0$'s else and where with $n$ is the number of the qubits. \textbf{The $e_i$ are our labels}.

%\begin{equation}
%\label{measurement_basis}
%\begin{aligned}
%&\{i \left( \text{1,0,0,0,0,0,0,0} \right)  ;j \left( \text{0,1,0,0,0,0,0,0} \right) %;k \left( \text{0,0,1,0,0,0,0,0} \right) ; \\
%&l \left( \text{0,0,0,1,0,0,0,0} \right) ; m \left( \text{0,0,0,0,1,0,0,0} \right) ;n %\left( \text{0,0,0,0,0,1,0,0} \right) ; \\
%&o \left( \text{0,0,0,0,0,0,1,0} \right) ;p \left( \text{0,0,0,0,0,0,0,1} \right) \}.
%\end{aligned}
%\end{equation}

The Ansatz design and data loading (variables  ${x_i}$ similar to neural networks)\cite{UAT2021} are given by equation \eqref{Ansatz}, and its analysis is detailed in \ref{sec:App-Ansatz}.
\begin{equation}
\label{Ansatz}
U=\left(\theta ,{x} \right)= R_{x}\left( { \theta _{1}}{x}+{ \theta _{2}} \right) R_{z} \left( { \theta _{3}}\right). 
\end{equation}

\section{Variational Quantum Classifier}\label{sec:App-VQC_Reup}
To date, two dominant categories allow designing quantum classifiers. Although almost all are inspired by the classical classifiers (kernel or neural networks) \cite{Abd15}, there is a new category of classifiers that respond to the current era of quantum computing (NISQ); hybrid and variational classifiers. \\

Let us find the operator that will help us to create our classifier.
\begin{itemize}
 \item  Let $ n $ be the number of qubits.
 \item  Let $ \vec {x} $ be a vector of dimension $ m $. 
 \item  Let $ \vec {\theta} $, a matrix of dimension $ n (m + 1) $.
 \item  Given a row $ i $, we will say that $ \vec{\theta_i} $: = $\vec{\theta_i}^{(w)}$ + $ \theta_i^{(b)}$ 
\begin{itemize}
\item where $\vec {\theta_i}^{(w)}$ has dimension $m$ and $\theta_{i}^{(b)} $ dimension 1.
\end{itemize}
\item  Let us define generically $ \vec{\theta_i} \cdot \vec{x} $: = $ \vec{\theta_i}^{(w)} \cdot \vec{x} $ + $ \theta_ {i}^{(b)} $.
\end{itemize}
Taking into account all, one way to define the model will be:

$$
U (\vec {x}, \vec {\theta}) = \bigotimes_{i = 0} ^ {n - 1} U_{i} (\vec {x}, \vec {\theta_i}),
$$

where: (for example)

$$
U_{i} (\vec {x}, \vec {\theta_i}) = R_y (\vec {\theta_i}^{(b)}) R_z (\vec {\theta_i}\cdot \vec{x}).
$$

\begin{itemize}
 \item Let $\ket{\psi(\vec{x})}$ be a functional quantum state.
 \item Let $f_{i}: \mathbb{R}^{m} \rightarrow \mathbb{C}$ be complex function. 
\end{itemize}

\begin{equation}
\label{qSTATE}
    \ket{\psi(\vec{x})} = \sum_{i=0}^{2^{n}-1}f_i(\vec{x}) \ket i,
\end{equation}

\begin{equation}
\label{qSTATE2}
\sum_{i=0}^{2^{n}-1}  |f_i(\vec{x})|^2 = 1.
\end{equation}

The circuit $\mathcal{U}(\vec{x}, \vec{\theta})$ approximates the state as:
\begin{equation}
\ket{\psi(\vec{x})} \sim \mathcal{U}(\vec{x}, \vec{\theta}) \ket{0}^{\otimes n}, \qquad {\rm with} \qquad
\mathcal{U}(\vec{x}, \vec{\theta}) = \prod_{i=1}^k U(\vec{x},\vec{\theta_i}),
\end{equation}
with better results as the number of layers $k$ increases and $n$ the number of the classes.
\begin{itemize}
 \item $\vec{\theta} = \{ \vec{\theta_i}\}$ is found with classical optimisation techniques.
 \item Cost function $= {\rm Distance}(\ket{\psi(\vec{x})}, \mathcal{U}(\vec{x}, \vec{\theta}) \ket{0}^{\otimes n}).$
\end{itemize}

\begin{figure}[!ht]
\centering
\includegraphics[width=0.7\textwidth]{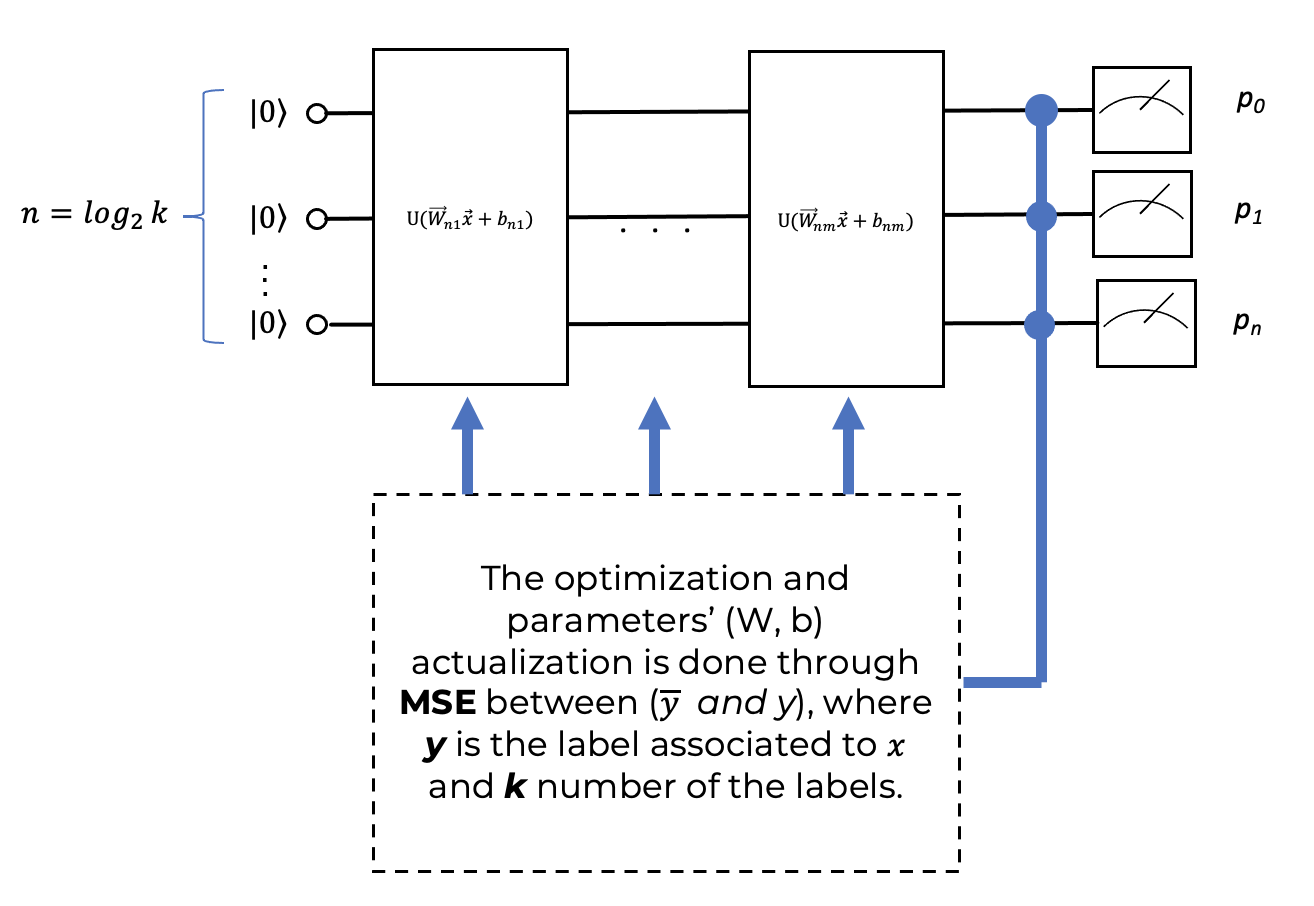}
\caption{This is the variational classifier's diagram block used in the qCBR,  we use the data re-uploading technique to create an n-dimensional classifier as if it were a neural network where the non-linearity of the quantum gates will act as an activation function, and we will use the model $y = Wx + b$.}
\label{fig:class_Gen_}
\end{figure}

We have designed a classifier that emulates neural networks solving the function $Wx + b$, with $W$ and $b$ the parameters and $x$, the sample data to be classified. The non-linearity of the quantum gates is used to implement the activation function $f (Wx + b)$ given $Wx + b$. The figure \eqref{fig:class_Gen_} provides us with the block diagram of the classifier. The optimisation and parameters' $(W, b)$ actualisation are done in the first step, MSE between ($\bar y$ and $y$), where $y$ is the label associated with $x$ and $k$ of the labels.

\begin{figure*}[t!]
\centering
\includegraphics[width=1\textwidth]{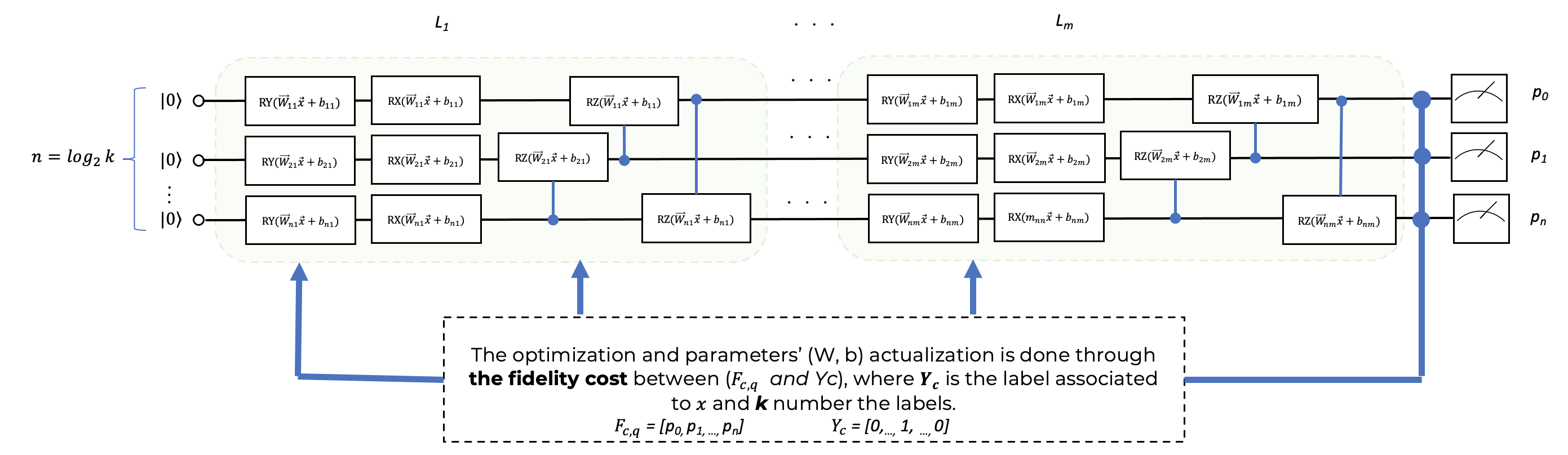}
\caption{This is the design of the classifier implemented in qCBR. Considering that $x$ is the input data of dimension $l$, $Y_{c}$ is the label class of $x$, $k$ is the number of labels, and we use the fidelity cost \eqref{fidelity_cost_function}. It is worth mentioning that the class of the tags $Y_{c}$ and, in this case, coincides with the computational base; thus, we save the target class. In this figure, we got $m$ layers.}
\label{fig:Classi_details}
\end{figure*}

The detailed operations of the classifier are given by the figure \eqref{fig:Classi_details} where the quantum gates, $R_{y}$, $R_{x}$ and $C_{RZ}$ are used to define the block. In our, the optimisation and parameters' ($W$, $b$) actualisation are done through the fidelity cost between $\alpha _{c,q}F_{c,q} \left( \overrightarrow{ \theta },\overrightarrow{ \omega ,}\overrightarrow{x}_{ \mu } \right)$ and $Y_{c} \left( \overrightarrow{x}_{ \mu } \right)$, where $\overrightarrow{x}_{ \mu }$ are the training points and  $\overrightarrow{ \alpha }= \left(  \alpha _{1}, \ldots , \alpha _{C} \right)$  are introduced as class weights to be optimised together with  $\overrightarrow{ \theta } $,  $ \overrightarrow{ \omega ,}$  are the parameters and  $Q$  the numbers of the qubits. Counting on  $Y_{c} \left( \overrightarrow{x}_{ \mu } \right)$  as the fidelity vector for a perfect classification and $F_{c,q} \left( \overrightarrow{ \theta },\overrightarrow{ \omega ,}\overrightarrow{x}_{ \mu } \right) =\langle \psi _{c} \vert   \rho _{q} \left( \overrightarrow{ \theta },\overrightarrow{ \omega ,}\overrightarrow{x} \right)  \vert   \psi _{c} \rangle$. 

In the next section, we will dive deeply into the Ansatz analysis.

\subsection{The Ansatz}\label{sec:App-Ansatz}
\begin{figure}
    \centering
    \includegraphics[width=0.3\textwidth]{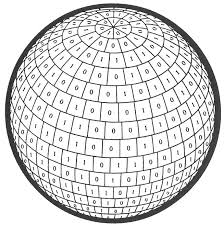}
    \caption{Bloch sphere, where the infinite Hilbert vector space resides. It is worth remembering that Hilbert space is the immense place where the states that describe a quantum system live. Let N be the number of qubits and let $2^N$ be the space dimension. Given the 65-qubit IBM processor, it will be $3.68935 × 10^{19}$ dimensions. Many works of literature lead us to use or, at least, to think about using the quantum computer in Machine Learning (ML).  By definition, the Hilbert vector space has defined these operations (internal and external products and mapping inputs in a large space) that a quantum computer performs natively and very easily \cite{Pet07}.}
    \label{fig:Block_sphere}
\end{figure}

The basic idea that one pursues is to have an ansatz (Parameterised quantum circuits (PQC)) that, formed by basic gates for quantum computing, is the most representative in the Hilbert vector space. In other words, with the control or parameterisation of these parameters, the Ansatz, in particular, maps the maximum number of points within the Bloch sphere (Fig. \eqref{fig:Block_sphere}). Another way to understand the objective of the Ansatz is to find the state vector that best approximates all the points of the Hilbert vector space. We must remember that the Hilbert vector space is the computational space of quantum mechanics, therefore, quantum computing.
The Ansatz design inherited from previous works \cite{Atc201,Atc20} \cite{Suk191}. The way to load the data into the Ansatz is inspired by \cite{UAT2021} where the data (variable  $x$) is entered using the weights and biases scheme. In this case, the single-qubit gate that serves as the building block for all Ansatz is given by equation \eqref{anstaz_1D} similar to neural networks.
\begin{equation}
\label{anstaz_1D}
U\left(  \theta ,x \right) =R_{x} \left(  \theta _{1}x+ \theta _{2} \right) R_{z} \left(  \theta _{3} \right). 
\end{equation}

\begin{figure}[]
\centering
\includegraphics[width=.5\textwidth]{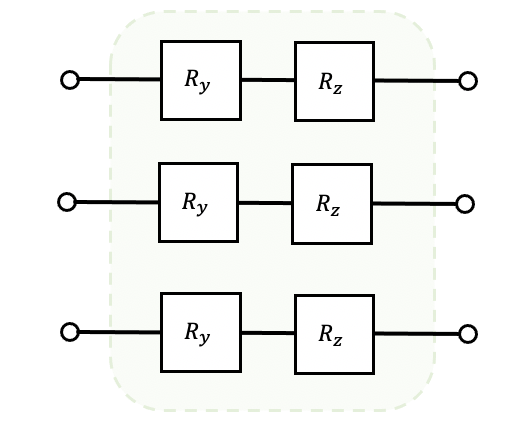}
\caption{$R_y$ and $R_z$ Ansatzes without entanglament used in qCBR experimentation.}
\label{fig:Ansatz_1}
\end{figure}

\begin{figure}	
\centering
\includegraphics[width=.6\textwidth]{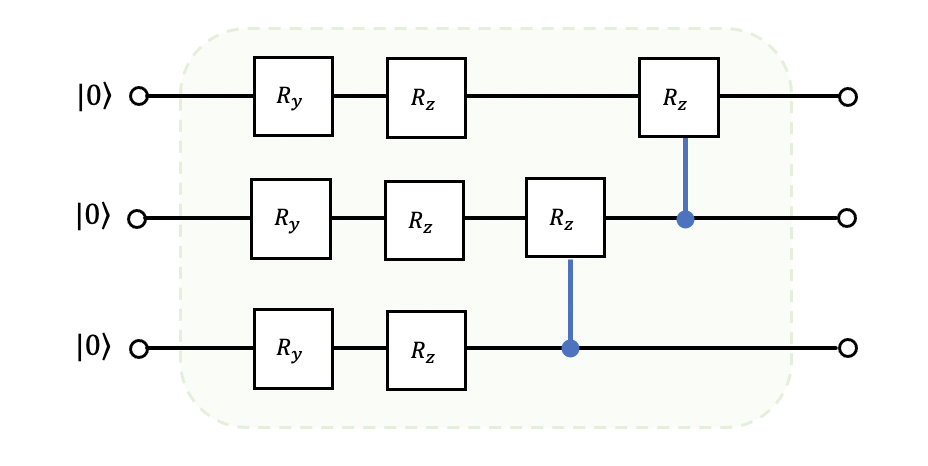}
\caption{$R_y$ and $R_z$ Ansatzes with $CRZ$ entanglement used in qCBR experimentation.}
\label{fig:Ansatz_2}
\end{figure}

\begin{figure}	
\centering
\includegraphics[width=.6\textwidth]{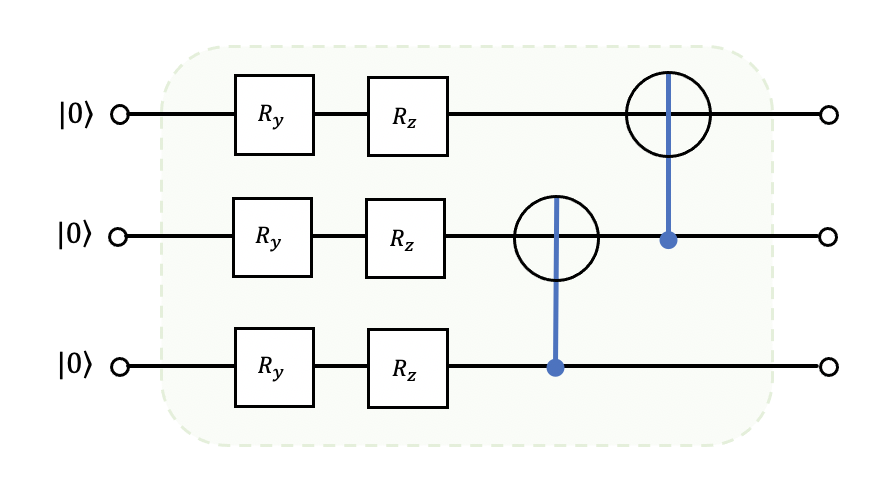}
\caption{$R_y$ and $R_z$ Ansatzes with $CNOT$ entanglement used in qCBR.}
\label{fig:Ansatz_3}
\end{figure}

Being  $\theta$ the vector of the parameters and  $R_{x}$  and  $R_{y}$  the unit gates of qubits used to create the Ansatz. To complement the experimentation scenario, it would be necessary to add the CNOT gate and the CRZ, which are the gates that help to achieve entanglement as seen in figures \eqref{fig:Ansatz_1}\eqref{fig:Ansatz_2} and \eqref{fig:Ansatz_3}.

\begin{figure}[!h]	
    \centering
\includegraphics[width=.7\textwidth]{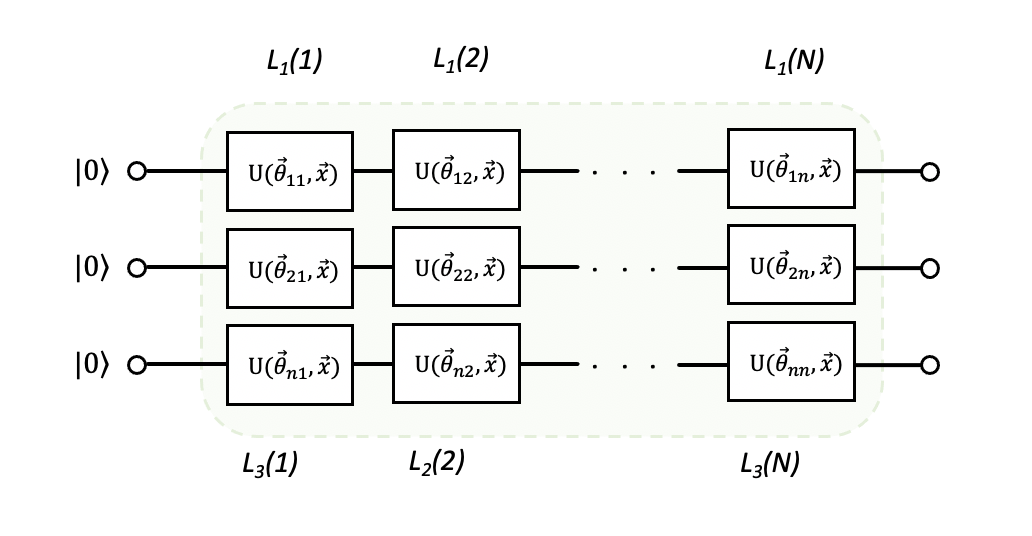}
\caption{Three-qubit quantum classifier circuit without entanglement.}
\label{fig:Ansatz_Comb}
\end{figure}

\begin{figure}[!h]
    \centering
\includegraphics[width=.7\textwidth]{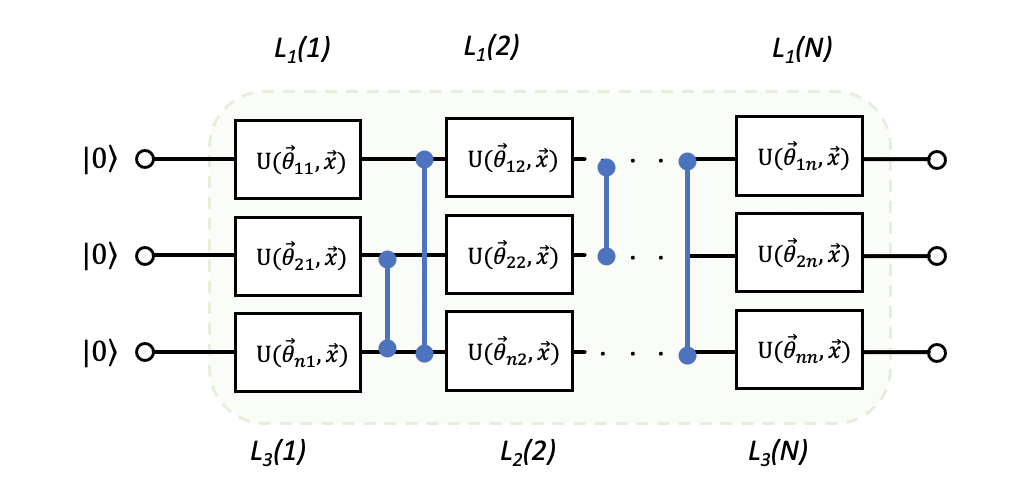}
\caption{Three-qubit quantum classifier circuit with entanglement by using $CZ$ or $CNOT$  gates}
\label{fig:Ansatz_Comb_1}
\end{figure}

The variational quantum classifier structure, figure \eqref{fig:Ansatz_Comb} and \eqref{fig:Ansatz_Comb_1}, is based on layers of trainable circuit blocks  $ L \left( i \right) = \prod_{i,j}^{}U \left( i,j \right)$ and data coding, as shown in the equation \eqref{Ansatz} for 8 dimensional or in \eqref{anstaz_1D} for 2 dimensional data size. Additionally, the entanglement can be achieved using the CRZ or CNOT gates.


The number of parameters to optimise the classifier is given by the equation \eqref{num_parameter}.
\begin{equation}
\label{num_parameter}
  \text{ NumParam}= ( nL\ast 2d \ast L).
\end{equation}
In this case, with  $n$  the number of qubits,  $n = 3$ , $L$ the number of layers (blocks), in the experiment, it is a variable data and  $d$  which is the dimension of the problem. In other words, $d$ varies with the choice of Ansatz and whether or not entanglement is applied. In the case of the entanglements in figure \eqref{fig:Ansatz_2}, the $d$ would be summed 1 ($CRZ$ gate has one parameter), which equates to equation \eqref{num_paramenter_nD}.
\begin{equation}
\label{num_paramenter_nD}
\text{NumParam}= ( nL\ast 2 ( d+1 ) \ast L ). 
\end{equation}
\subsubsection{Fidelity cost function}\label{sec:App-Fidelity}
The similarity function follows the same strategy as the re-uploading and path; the Ansatz is different. It uses the definition of quantum fidelity with several qubits and maximises said average fidelity between the test state and the final state corresponding to its class. Equation \eqref{Cost_Funtion} \cite{Adr20} defines the cost function used.
\begin{equation}
\label{Cost_Funtion}
\small
\begin{split}
    CF \left( \overrightarrow{ \alpha },\overrightarrow{ \theta },\overrightarrow{ \omega } \right) = \frac{1}{2} \sum _{ \mu =1}^{M} \sum _{c=1}^{C}  \left(  \sum _{q=1}^{Q} \left(  \alpha _{c,q}F_{c,q} \left( \overrightarrow{ \theta },\overrightarrow{ \omega ,}\overrightarrow{x}_{ \mu } \right) -Y_{c} \left( \overrightarrow{x}_{ \mu } \right)  \right) ^{2} \right), 
\end{split}
\end{equation}
with
\begin{equation}
\label{fidelity_cost_function}
F_{c,q} \left( \overrightarrow{ \theta },\overrightarrow{ \omega ,}\overrightarrow{x}_{ \mu } \right) =\langle \psi _{c} \vert   \rho _{q} \left( \overrightarrow{ \theta },\overrightarrow{ \omega ,}\overrightarrow{x} \right)  \vert   \psi _{c} \rangle. 
\end{equation}

Where  $\rho _{q}$  is the reduced density matrix of the qubit to be measured,  $M$  is the total number of training point,  $C$  is the total number of the classes,  $\overrightarrow{x}_{ \mu }$ are the training points and  $\overrightarrow{ \alpha }= \left(  \alpha _{1}, \ldots , \alpha _{C} \right)$  are introduced as class weights to be optimised together with  $\overrightarrow{ \theta } $,  $ \overrightarrow{ \omega ,}$  are the parameters and  $Q$  the numbers of the qubits. Counting on  $Y_{c} \left( \overrightarrow{x}_{ \mu } \right)$  as the fidelity vector for a perfect classification. This cost function \eqref{Cost_Funtion} is weighted and averaged over all the qubit that form this classifier.
To complete the hybrid system, it is used for the classical part, the following minimisation methods above cited: L-BFGS-B \cite{ich95}, COBYLA \cite{The21} and SPSA \cite{Jam01}.

\section{Details of qCBR}\label{sec:App-CBR}
 The operation of the \textbf{retrieve} (prediction) block is given by a new case (schedule). In this experimentation, the schedule that best adapts to the latest issue to be solved is recovered with the predict method, which is executed at a time  $O(log(MN))$. It is worth saying that, due to the SWP descriptions, a possible schedule change, a stage of understanding or interpretation is necessary since an adequate resolution of the new schedule cannot be carried out if it is not understood with some completeness. This stage of understanding is a simple decision algorithm with minimal intelligence.
 
\begin{figure*}[t!]
\centering
\includegraphics[width=1\textwidth]{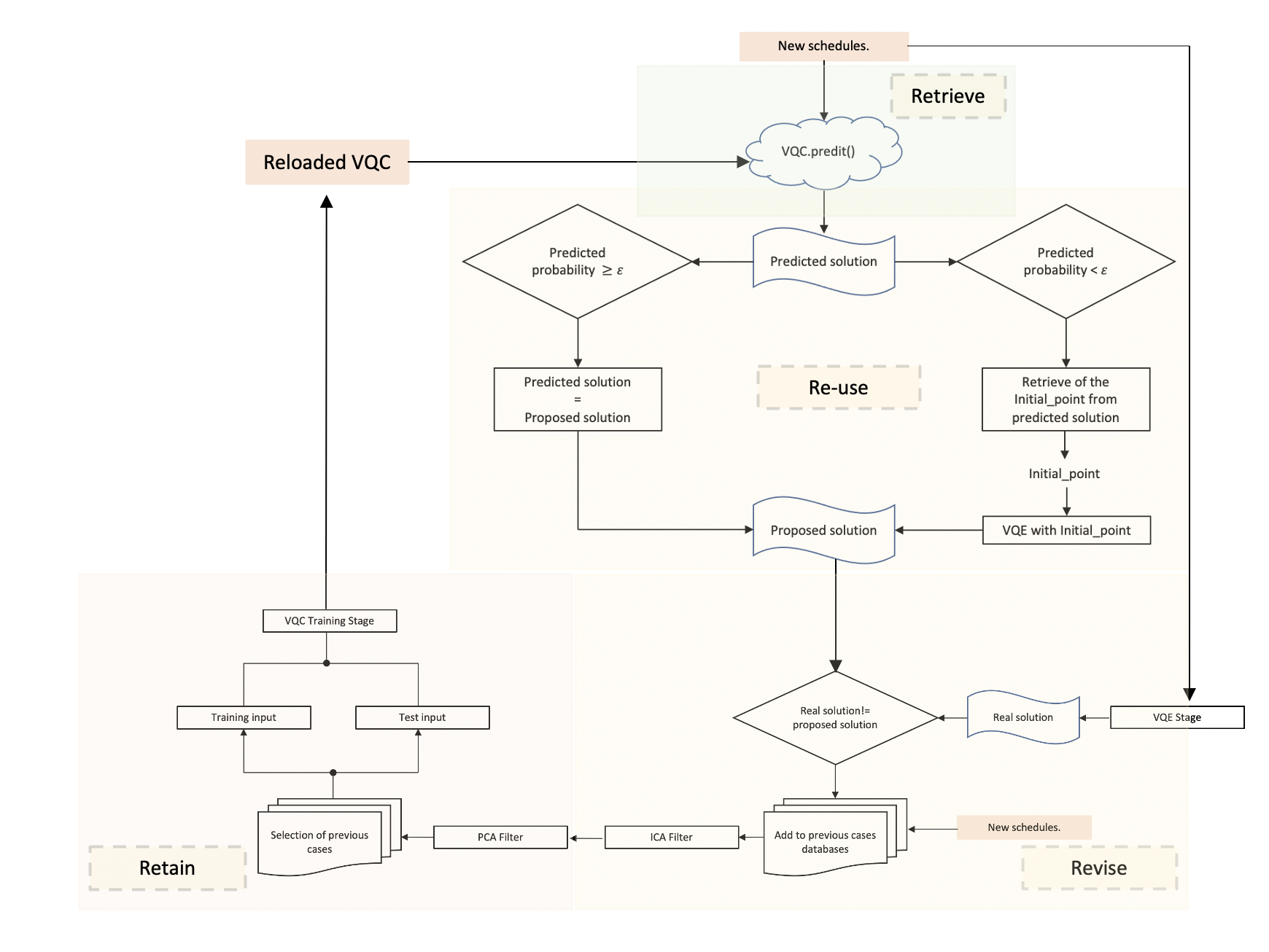}
\caption{Block diagram proposed for the resolution of qCBR, considering a real dataset with an overlap problem between the data components. This block diagram contemplates the treatment of the input data and the use of ICA \cite{Hyv01-interscience} and PCA \cite{Jon14} before training and classifying the data. In this version, a classifier based on the re-uploader has been designed to handle the classification tasks. And for the synthesis tasks, a decision tree passed in the classifier predictions have been used together with the VQE plus the \textit{Initial\_point}}
\label{fig:Funct_qCBR}
\end{figure*}

 Once having the predicted solution, the synthesis block creates a new solution (proposed solution) by combining recovered solutions. To do this, the algorithm is divided into two main lines (figure\eqref{fig:Funct_qCBR}). A line that determines an acceptable degree of error (after a probabilistic study) that the predicted solution can be considered the proposed solution. The second branch is in charge of improving the expected solution towards a better-proposed solution. To do this, the \textit{Initial\_point} associated with the retrieved schedule is retrieved from the case memory, and the Variational Quantum Eigensolver is executed with very few shots (\textit{k shots}). The idea here is to refine the new schedule's similarity with the recovered one. Operating the VQE with \textit{Initial\_point} provides the algorithm with parameter values through the initial point as a starting point for searching for the minimum eigenvalue (similarity between the two times) when the new time's solution point is believed to be close to a matter of the recovered schedule. This is how the \textbf{Re-use} block works. These operations have a complexity of $O(klog(N)+log(NM))$. Where  $N$  is the number of social workers, $M$ is the number of patients, and $k$  is the number of shots.
 
 The algorithm's processes to review the proposed solution are seen below the \textbf{Re-use} block in figure \eqref{fig:Funct_qCBR}. It is essential to classify the best possible solution for the proposed prototype. The best possible solution is calculated with the VQE with the maximum resolution and depth (for the variational part). Once the solution is obtained, it is compared with the proposed solution and said solution with its characteristics is added to the new schedule before storing it (see figures \eqref{fig:qMem} and \eqref{fig:mem1}). The computational complexity of the Revise is determined by $O(log(N)+ICA))$. In this work, access to data (states) is selected by $O(log(MN))$ due to the characteristics of the inner products and superpositions.
 
 One of the most critical blocks in this work is to \textbf{Retain}. This block is the heart of the CBR because it is the classifier and because it is the block that allows us to conclude that it has been learned from the previous cases. Unfortunately, not all instances (schedules) are saved in this job, leading to the excessively slow classifier. Therefore, the best issues (timetables) that summarise all the essential information are retained in this part of the algorithm.

\begin{figure}[h!]
\centering
\includegraphics[width=0.5\textwidth]{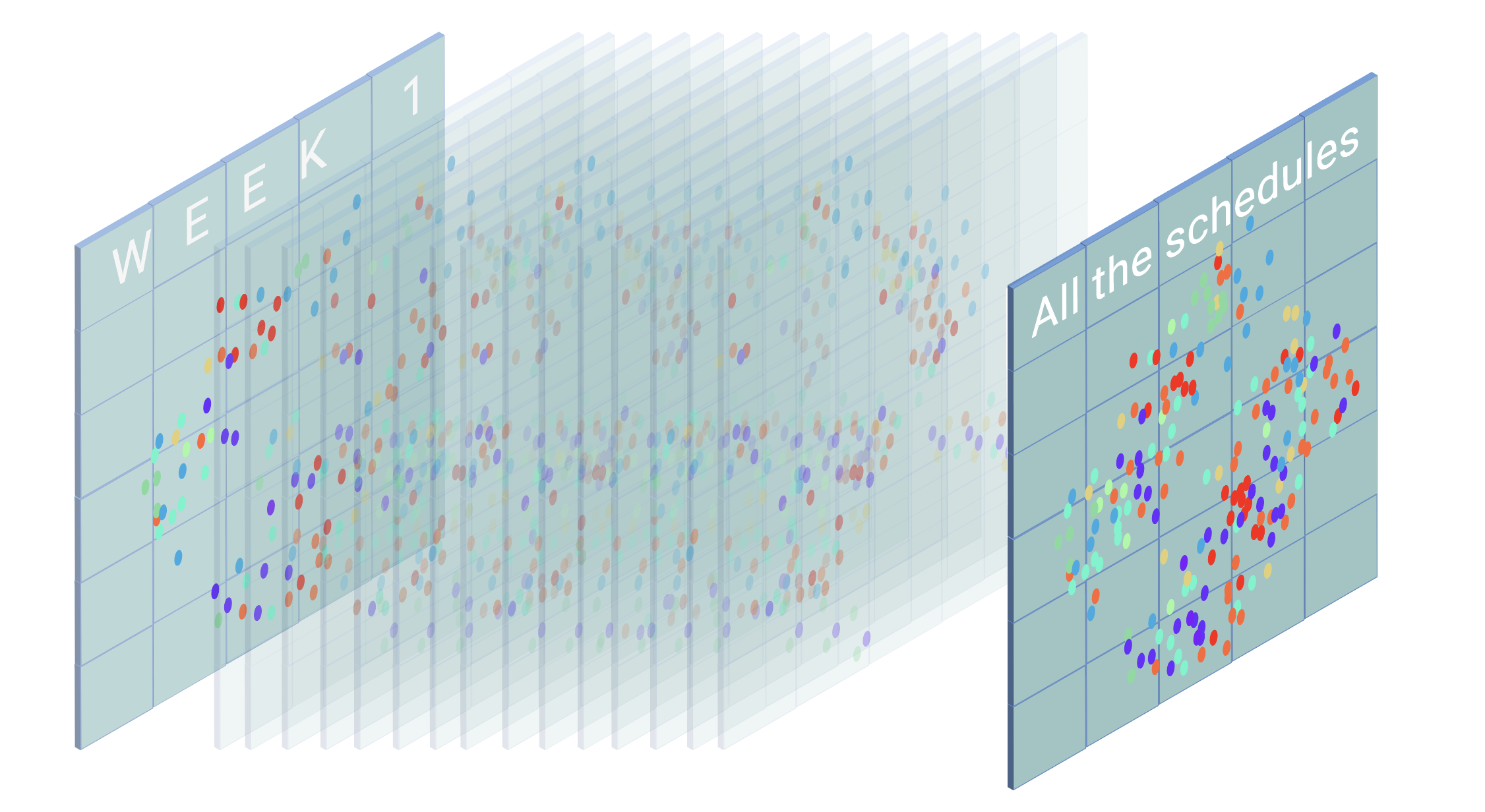}
\caption{Representation of the generation of the  $n$  weekly schedules of the SWP. The last plane, the one to the right of everything, represents all the SWP classes. The overlapping effect generated by the social workers' problem's characteristics and experimentation scenarios can be observed. This experimentation leads us to use the ICA technique to have a resulting dataset regardless of the schedules.}
\label{fig:Representation_SWP}
\end{figure}

\begin{figure}[h!]
\centering
\includegraphics[width=0.7\textwidth]{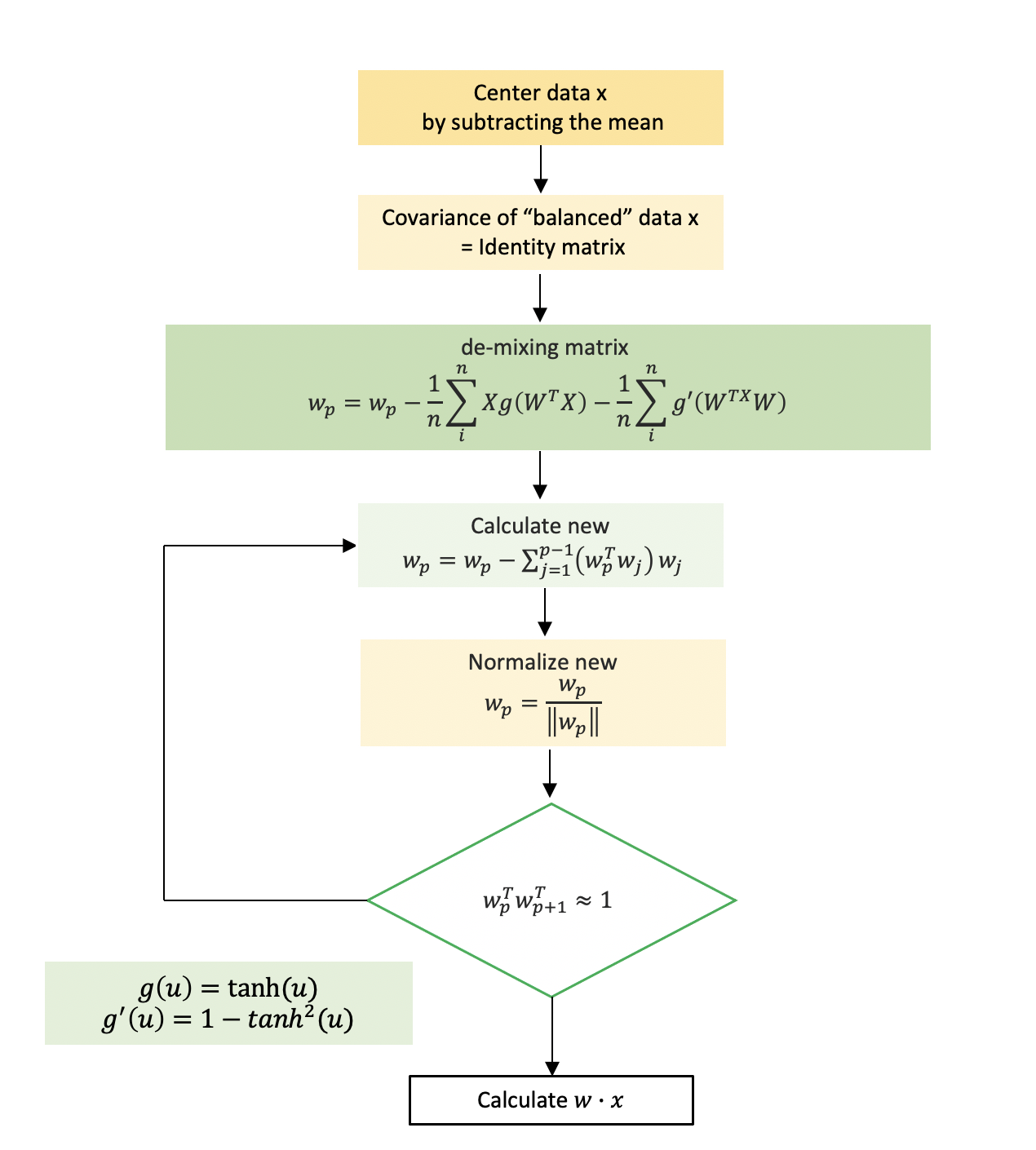}
\caption{Fundamental processes to apply ICA to SWP dataset. The first thing that is done is to centre the data x by subtracting the mean, balancing the data x removing its variance, and calculating the unmixing matrix of W. Then, the new value of w is calculated. Then w is normalised before checking if, with the said value, the algorithm converges or not. If it does not converge, a new w must be recalculated, and if it converges, calculate the scalar product of $\langle x,y \rangle$ to obtain the independent weekly schedules.}
\label{fig:ICA}
\end{figure}

The \textbf{Retain} process begins with the treatment of schedules, searching for the algorithm's best efficiency, which is a challenge to solve in this block.
In the case of SWP, the patient visits hours have a margin range of 30mn. Therefore, if one schedule starts at 9:00, the next could begin at 9:30, leading to a dataset with overlap between plans if many programs have similar time ranges spread over different days of the week. Suppose we add the non-linearity of the data to this issue. In that case, an almost perfect classifier is needed with a mean accuracy greater than 80$\%$  or the treatment and intelligence system classifier can be helped. Then, you can separate the overlapping components of the cases.

In this work, we contemplate both scenarios. First, get an excellent classifier and apply data processing techniques to help a poor classifier.
Using the standard classifier, ICA \cite{Hyv01-interscience} is applied to the original data to reduce the effects of the degree of overlap (figure \eqref{fig:Representation_SWP}) without losing the fundamental characteristics of the data. Figure \eqref{fig:ICA} summarises the processes and operations applied to reduce the overlapping effect observed in the generation of SWP schedules. The complexity of this operation is noted as  $ O(ICA)$ . The PCA is then used to reduce the data dimension from 8 to 2 and apply it to the designed variational classifier with the complexity equal to  $O(PCA)$.
Once the best time is determined, we retain the knowledge acquired at the case's resolution.

\subsection{Memory Structure}
\begin{figure}[h!]
\centering
\includegraphics[width=0.5\textwidth]{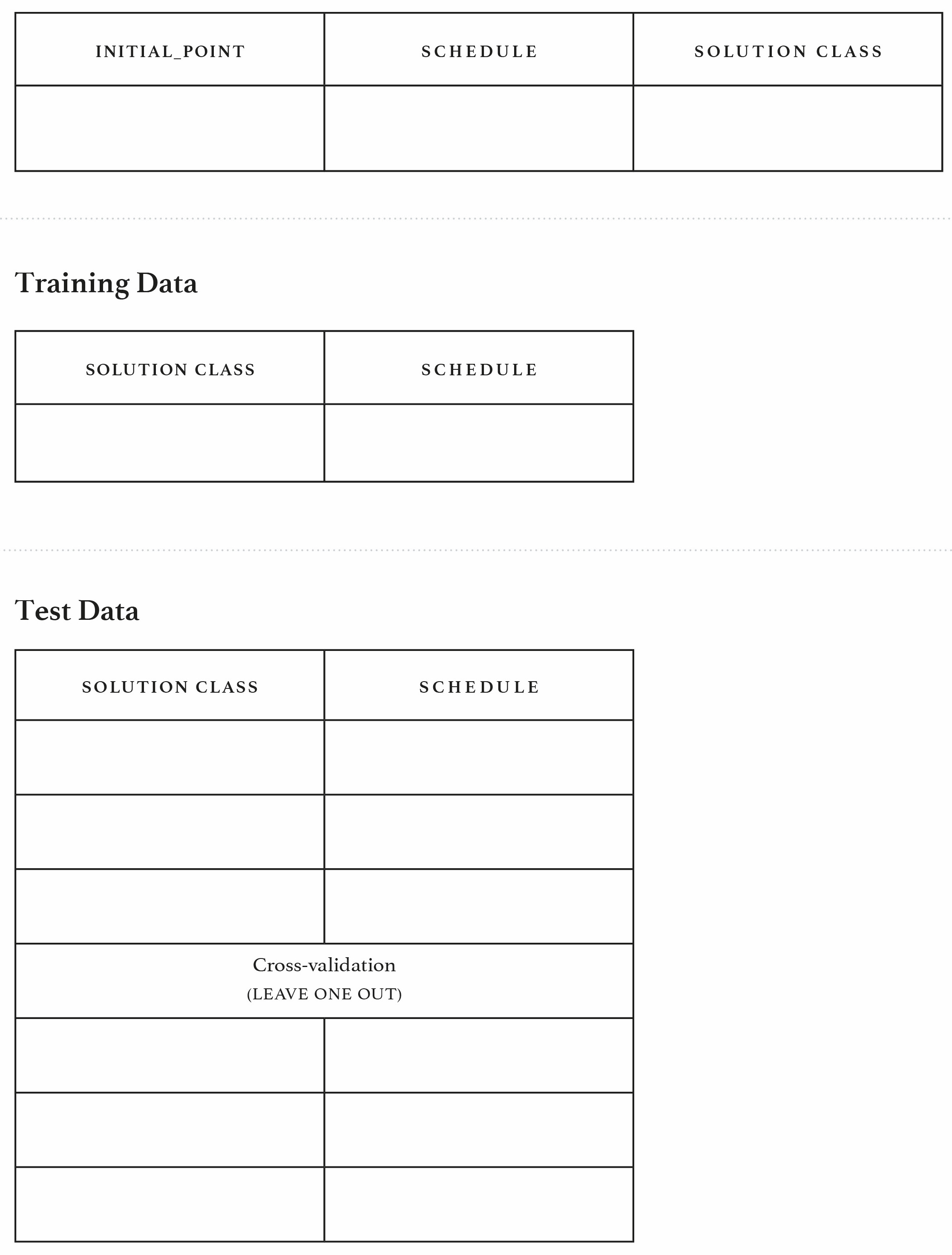}
\caption{qCBR's Memory structure. The use the cross-validation technique helps to improve the quality of the classifier training\cite{Mic}}
\label{fig:qMem}
\end{figure}

\begin{figure}[h!]
\centering
\includegraphics[width=0.5\textwidth]{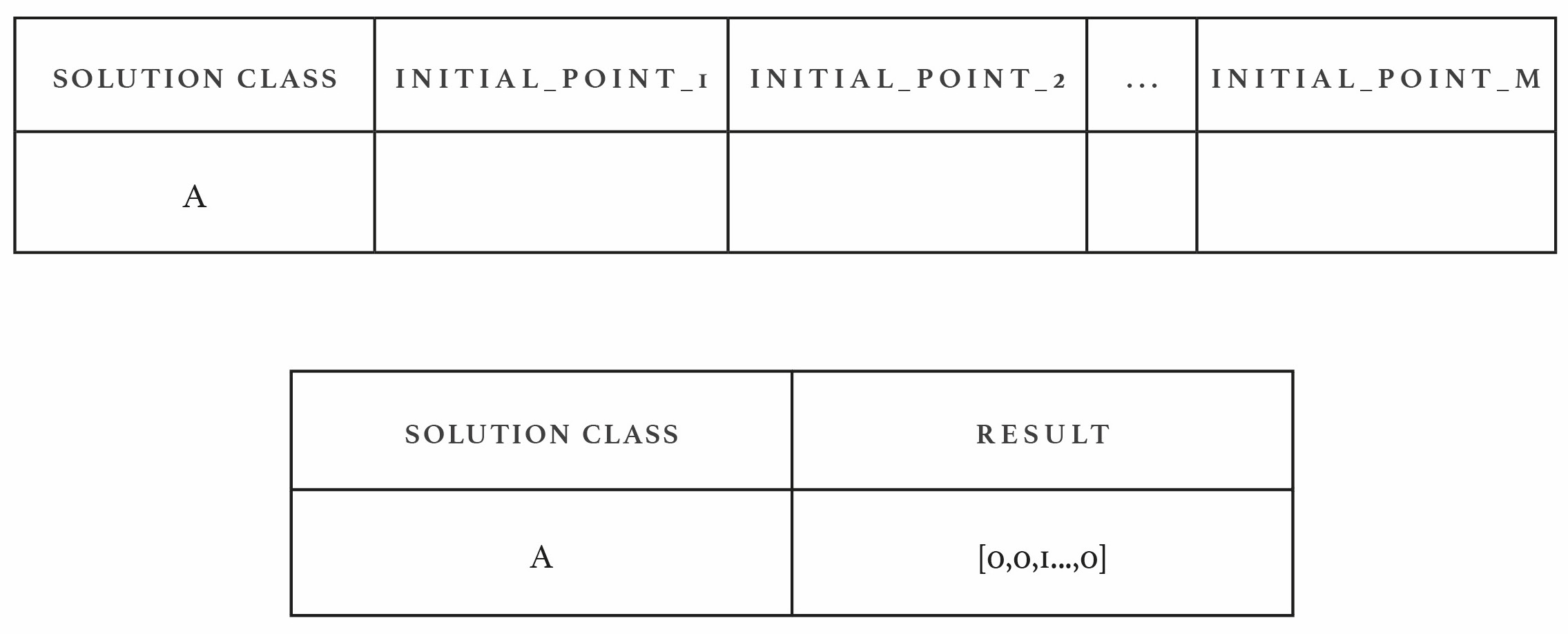}
\caption{Memory structure used for the Training System. A label identifies each class. And in this case, A is one of the labels. For more detail, refer to the code \cite{Ade21}.}
\label{fig:mem1}
\end{figure}

Next, some test benches based on the memory structure described in figure \eqref{fig:qMem} and \eqref{fig:mem1} are defined to train the parameterised quantum circuit, and its performance is analysed in terms of the circuit architecture. Finally, the results and discussions session will emphasise the classifier with or without entanglement and a comparative study with different ansatzes.
\newline

The memory structure of the qCBR' retention system is given in figure \eqref{fig:qMem}. The solution class (target) corresponds to the paths each Social Worker will take between the different patients for a specific schedule, representing these paths as an adjacency matrix, such as:

\begingroup
\renewcommand*{\arraystretch}{0.7}
\begin{equation}
\label{eq:SWAdjacencyMatrixExample}
 SOL_{SWP} = \begin{pmatrix}
    0 & x_{0,1} & x_{0,2}\\
    x_{1,0} & 0 & x_{1,2}\\
    x_{2,0} & x_{2,1} & 0\\
 \end{pmatrix}
\end{equation}
\endgroup

where $x_{i,j}$ is a binary variable, the rows of the matrix represent the origin node and the column the destination node of the path.\\

Each solution class is represented as a label (e.g., 'A') and is related to the different initial points associated with each of the samples that make up the training dataset can be seen. This solution class is also associated with the result of the VQE.

\begin{figure}[h!]
\centering
\includegraphics[width=0.5\textwidth]{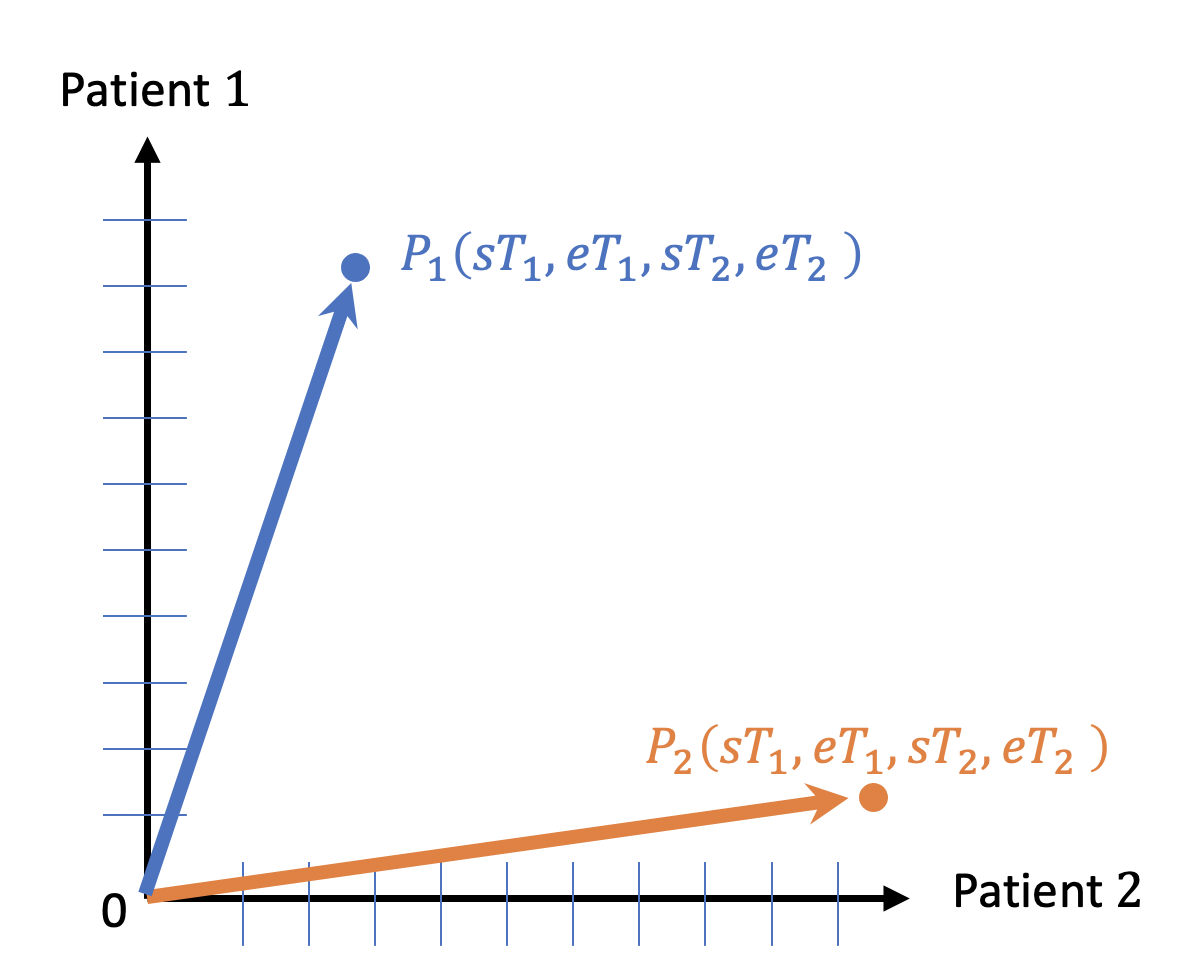}
\caption{The SWP is represented in vector form to take advantage of the Hilbert vector space's characteristics within quantum computing. It is seen that each patient represents a dimension, and the points that make up the dataset are a dimension of $2n$ coordinates, $n$ being the total number of patients. In this figure, to simplify the understanding, we use two patients, therefore, two dimensions}
\label{fig:SWP}
\end{figure}

To fill out the data structure created in the classifier to train and test its predictions, the parameters of the \textit{Initial\_point} obtained by VQE are abstracted from the result. And it is composed of each class's coordinates with the following parameters: start time $sT$ and end time  $eT$  of patient  $1$ to  $n$, where  $n$  is the maximum number of patients in the app. Figure \eqref{fig:SWP} summarises the data's representation and description that make up the training dataset.

Each coordinate's class corresponds to the VQE solution following the memory structure in figure \eqref{fig:qMem} where the class number of the classifier is given by equation \eqref{Num_Sol_SWP} taking into account the conditions that every worker has a patient and that the workers are indistinguishable (that is, it doesn't matter whether the social worker $m_{1}$ takes care of the patient  $n_{1}$  and  $m_{2}$  takes care of $n_{2}$  or vice versa).
\begin{equation}
\label{Num_Sol_SWP}
    N_{SOL_{SWP}}=\frac{1}{m!} \sum _{k=0}^{m-1} \left( -1 \right) ^{k} {m \choose m-k} \left( m-k \right) ^{n}.
\end{equation}

Let  $n$  be the number of patients and  $m$ the number of social workers and knowing that the appearance orders patients in the schedule (from earliest to latest, let  $n_{1}$  be the patient with the earliest plan and  $n_{k}$  be the patient with the latest program).

In this work all the tests done are for  $n = 4$  with the data structure equal to  $( sT_{1},eT_{1},sT_{2},eT_{2},sT_{3} \\,eT_{3},sT_{4},eT_{4})$; an 8-dimensional vector for each social worker visit the patient. In this case, the number of qubits will be defined by  $q=\log _{2} \left( N_{SOL_{SWP}} \right) =3$. These qubits are used to instance the quantum classifier, and it is worth to mention that the classifier must have  $N_{SOL_{SWP}}$ classes.

The detail of the qCBR's implementation and analysis is in section \ref{sec:App-CBR}.

\section{Results}\label{sec:qCBR_result}
When testing the classifier, a section of the sample database, schedules previously solved by the VQE algorithm to obtain its corresponding true solution (\textit{ground truth}), was used as test samples (applying “Leave-one-out” cross-validation). Then, the total accuracy of the classifier predictions was obtained based on the ratio between the number of labels predicted correctly and the total number of labels.

\begin{figure*}[t!]
\centering
\includegraphics[height=6cm]{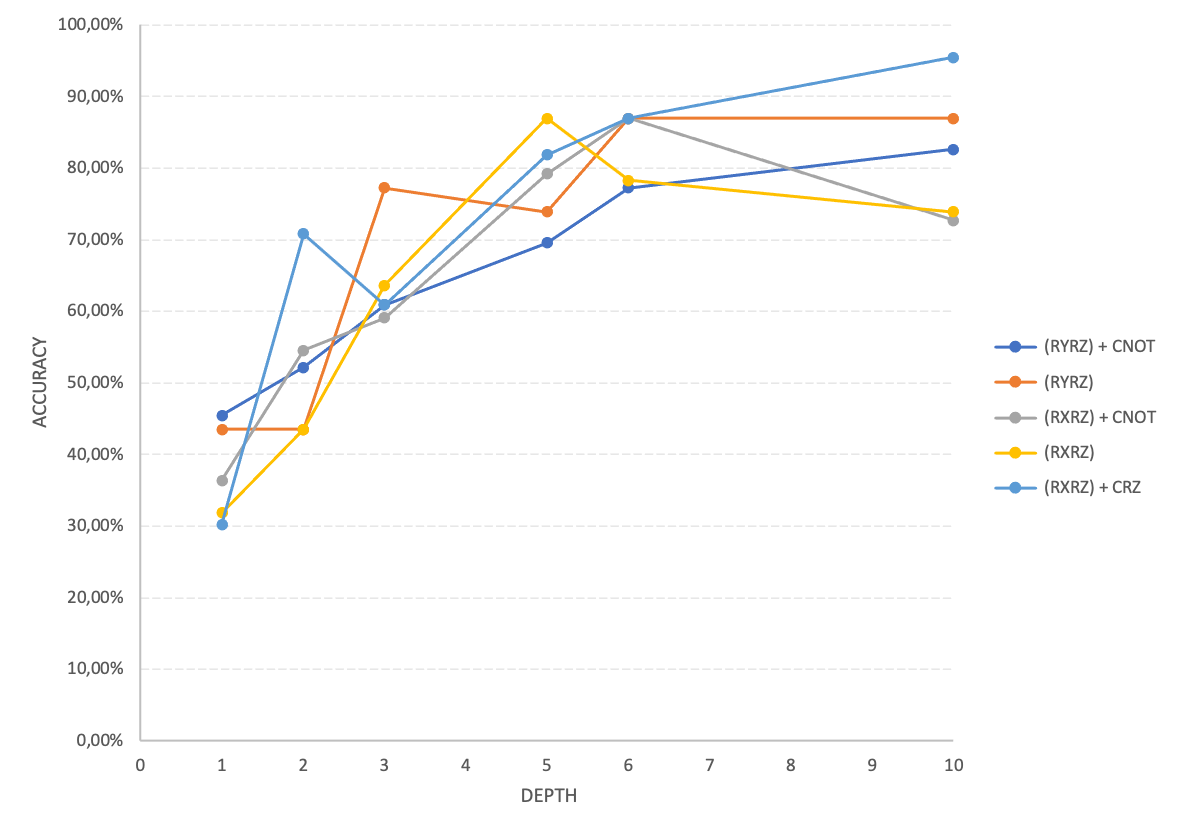}
\includegraphics[height=6cm]{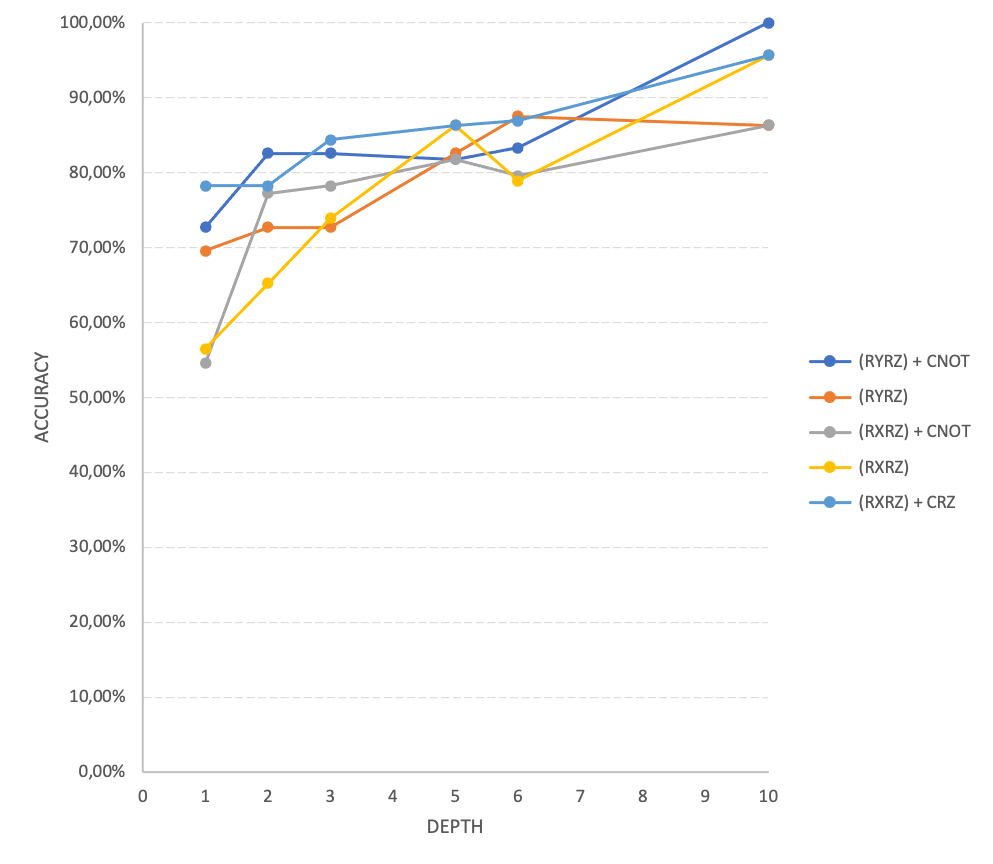}
\caption{Comparative graphs between different ansatzes, taking into account the classifier's accuracy as a function of depth. Represents the evolution of the ansatzes of dimension two and eight.}
\label{fig:Bench_C}
\end{figure*}

\begin{figure}[t!]
\centering
\includegraphics[width=.49\textwidth]{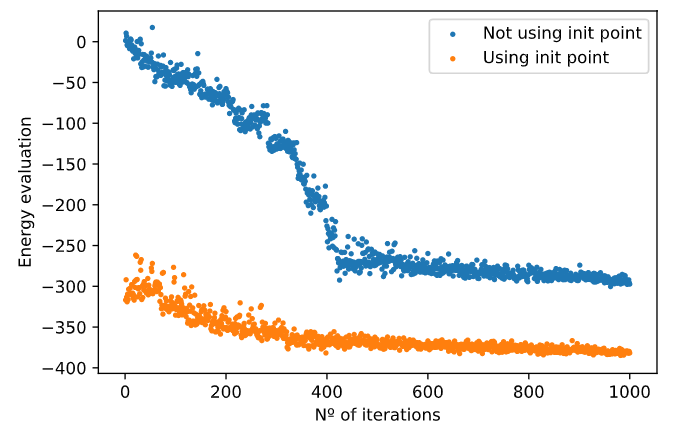}
\caption{Energy comparison between VQE algorithm without using or using the initial point. It is noticeable how the first one tends to stabilise after multiple iterations (approximately 400), starting the search for a minimum from a random starting point (depending on the seed provided). Meanwhile, the second one is capable of stabilising and reach a solution close to the absolute minimum with much less iterations, starting from energy point evaluation close to the real energy solution}
\label{fig:Bench_VQE_Init}
\end{figure}

Figures \eqref{fig:Bench_C} to \eqref{fig:Bench_VQE_Init} show the implementation outcomes performed in \textit{qibo} \cite{qibo} and \textit{qiskit}\cite{mckay2018qiskit,Qis21} to identify the best model architecture and represent functions similar to qCBR.

Tables \eqref{tab:results_qCBR_SW_Full} to \eqref{tab:results_qCBR_SW_4x3SW} show the global results of qCBR solving the SWP. In table \eqref{tab:results_qCBR_SW_Full}, the outcome of the different tested scenarios can be observed. Varying the number of patients, social workers, and the quantum circuit's depth to see the global hit number of the qCBR. In table \eqref{tab:results_qCBR_5x4SW}, we can observe the resolution of the SWP, considering five patients, four social workers and setting the depth of the quantum circuit to eight. Through this scenario, the behaviour of the qCBR can be observed considering the number of cases carried out. It can be seen how the system begins to give more than satisfactory results after exceeding the threshold of the 240 results stored in the case memory.
Table \eqref{tab:results_qCBR_SW_4x3SW} repeats the steps of table \eqref{tab:results_qCBR_5x4SW} with the only change of the input data; the number of patients and social workers.
Tables \eqref{tab:results_CBR_SW_5x4SW} to \eqref{tab:results_CBR_KNN_SW_Full} show the result of the implementation of the classical CBR leveraged on ANN and KNN to solve the SWP.

Tables \eqref{tab:results_qCBR_5x4SW} and \eqref{tab:results_qCBR_SW_4x3SW} represent the outcomes of the qCBR and show better results than the ones obtained with the classical CBR (tables \eqref{tab:results_CBR_SW_5x4SW} and \eqref{tab:results_CBR_SW_4x3SW}). Tables \eqref{tab:results_CBR_KNN_SW_Full}, \eqref{tab:results_CBR_NN_SW_Full} and \eqref{tab:results_qCBR_SW_Full} show the degree of scalability of the qCBR as a function of the variation in the number of patients and social workers. It has also been seen that qCBR is much better shared with overlapping as we wanted to demonstrate.

\begin{table}[t!]
\centering
\begin{tabular}{ |c|c|c|c|c|c|  }
 \hline
 \multicolumn{6}{|c|}{ qCBR solving the Social Workers' Problem } \\
 \hline
 \#Patients & \#SW & \#Qubits &\#Layers  & \#Cases & Accuracy\\
 \hline
 3   & 2 & 6  &  2  & 580 & 82.5\\
 4   & 3 & 12 &  3  & 580 & 82\\
 5   & 2 & 20 &  4  & 580 & 82.5 \\
 5   & 3 & 20 &  5  & 580 & 87\\
 5   & 4 & 20 &  8  & 580 & 92.8\\
 5   & 4 & 20 &  10 & 580 & 100\\
 \hline
\end{tabular}
\caption{The result of the qCBR with a variational classifier and using the VQE and the Initial\_point with some decision trees as a synthesiser \eqref{fig:Funct_qCBR}. This table shows the different studies made as a function of the quantum circuit's depth (layer). number of the patients and the social workers. The accuracy of the classifier is the maximum with the number of layers equal to 10. SW denotes Social Workers.}
\label{tab:results_qCBR_SW_Full}
\end{table}

\begin{table}[t!]
\centering
\begin{tabular}{|c|c|c|  }
 \hline
 \multicolumn{3}{|c|}{ qCBR solving the Social Workers' Problem } \\
 \multicolumn{3}{|c|}{ For 5 patients and 4 socials workers} \\
 \hline
 Layers &\#Cases & Accuracy\\
 \hline

& 20  & -  \\ 
&50  & 12.5\\
&100 & 72.5\\
8 &240 & 92.1\\ 
&340 & 95.5\\
&480 & 97.2\\
&500 & 98.7 \\
&580 & 99.1\\
 \hline
\end{tabular}
\caption{The result of the qCBR for a number of patients and social workers fixed at 5 and 4, respectively. The better behaviour of qCBR can be observed for some cases greater than 240. To have a good functioning of the qCBR, it must be iterated with the social workers' dataset one 239 times. And at the case number of 240, we will have an accuracy of 92\%. The "accuracy" value is the percentage of the number of correct solutions found by the qCBR.
A hyphen (-) denotes that no solution was found within the 20 cases. All these tests were done for the quantum circuit depth (layers) equal to 8.
}
\label{tab:results_qCBR_5x4SW}
\end{table}

\begin{table}[t!]
\centering
\begin{tabular}{|c|c|c|}
 \hline
 \multicolumn{3}{|c|}{ qCBR solving the Social Workers' Problem } \\
 \multicolumn{3}{|c|}{ For 4 patients and 3 socials workers} \\
 \hline
 Layers &\#Cases & Accuracy\\
 \hline

& 20  & -  \\ 
&50  & 11.5\\
&100 & 73.1\\
8 &240 & 91.1\\ 
&340 & 91.9\\
&480 & 96.6\\
&500 & 98.1 \\
&580 & 99.0\\
 \hline
\end{tabular}
\caption{The result of the qCBR for a number of patients and social workers fixed at 4 and 3, respectively. The better behaviour of qCBR can be observed for some cases greater than 240. To have a good functioning of the qCBR, it must be iterated with the social workers' dataset one 239 times. And at the case number of 240, we will have an accuracy of 91\%. The "accuracy" value is the percentage of the number of correct solutions found by the qCBR.
A hyphen (-) denotes that no solution was found within the 20 cases. All these tests were done for the quantum circuit depth (layers) equal to 8.
}
\label{tab:results_qCBR_SW_4x3SW}
\end{table}

\begin{table}[t!]
\centering
\begin{tabular}{|c|c|c|}
 \hline
 \multicolumn{3}{|c|}{ CBR with KNN solving the Social Workers' Problem } \\
 \multicolumn{3}{|c|}{ For 5 patients and 4 socials workers} \\
 \hline
 Layers &\#Cases & Accuracy\\
 \hline
& 20  & -  \\ 
&50  & 42.9\\
&100 & 46.5\\
1 &240 & 52.6\\ 
&340 & 55.3\\
&480 & 56.8\\
&500 & 60.7 \\
&580 & 63.1\\

 \hline
\end{tabular}
\caption{The classical CBR result on KNN classifier for a number of patients and social workers fixed at 5 and 4, respectively. The better behaviour of this CBR can be observed for some cases greater than 240. To have a good functioning of the CBR, it must be iterated with the social workers' dataset one 239 times. And at the case number of 240, we will have an accuracy of 52.6\%. The "accuracy" value is the percentage of the number of correct solutions found by the CBR leveraged on KNN, applying a 10-KFold cross-validation process.
A hyphen (-) denotes that no solution was found within the 20 cases. All these tests were done for the layer equal to 1.}
\label{tab:results_CBR_SW_5x4SW}
\end{table}

\begin{table}[t!]
\centering
\begin{tabular}{|c|c|c|}
 \hline
 \multicolumn{3}{|c|}{ CBR with KNN solving the Social Workers'Problem } \\
 \multicolumn{3}{|c|}{ For 4 patients and 3 socials workers} \\
 \hline
 Layers &\#Cases & Accuracy\\
 \hline
& 20  & -  \\ 
&50  & 55.1\\
&100 & 58.5\\
1 &240 & 70.3\\ 
&340 & 71.1\\
&480 & 73.6\\
&500 & 74.8 \\
&580 & 76.8\\

 \hline
\end{tabular}
\caption{The classical CBR result on KNN classifier for a number of patients and social workers fixed at 4 and 3, respectively. The better behaviour of this CBR can be observed for some cases greater than 100. To have a good functioning of the CBR, it must be iterated with the social workers' dataset one 239 times. And at the case number of 240, we will have an accuracy of 70.3\%. The "accuracy" value is the percentage of the number of correct solutions found by the CBR leveraged on KNN, applying a 10-KFold cross-validation process.
A hyphen (-) denotes that no solution was found within the 20 cases. All these tests were done for the layer equal to 1.}
\label{tab:results_CBR_SW_4x3SW}
\end{table}

\begin{table}[t!]
\centering
\begin{tabular}{ |c|c|c|c|c|  }
 \hline
 \multicolumn{5}{|c|}{  CBR leveraged by CNN solving the Social Workers' Problem } \\
 \hline
 \#Patients & \#SW  &\#Layers  & \#Cases & Accuracy\\
 \hline
 3   & 2 &   2  & 580 & 65.4\\
 4   & 3 &   2  & 580 & 43.3\\
 5   & 2 &  2  & 580 & 37.3 \\
 5   & 3 &  2  & 580 & 26.3\\
 5   & 4 &  2 & 580 & 45.2\\

 \hline
\end{tabular}
\caption{CBR with a neural network classifier and a backtracking algorithm as a synthesiser. SW denotes Social Workers.}
\label{tab:results_CBR_NN_SW_Full}
\end{table}

\begin{table}[t!]
\centering
\begin{tabular}{ |c|c|c|c|c|  }
 \hline
 \multicolumn{5}{|c|}{CBR with KNN solving the Social Workers' Problem } \\
 \hline
 \#Patients & \#SW  &\#Layers  & \#Cases & Accuracy\\
 \hline
 3   & 2 &  1  & 580 & 95.6\\
 4   & 3 &  1  & 580 & 77.8\\
 5   & 2 &  1  & 580 & 47.8 \\
 5   & 3 &  1  & 580 & 44.7\\
 5   & 4 &  1 & 580 & 63.1\\
 \hline
\end{tabular}
\caption{CBR with a KNN classifier and a backtracking algorithm as a synthesiser. SW denotes Social Workers.}
\label{tab:results_CBR_KNN_SW_Full}
\end{table}

Also, we experimented by skipping the Principal Component Analysis (PCA) module \cite{Jon14,Ewi19}, Independent Component Analysis (ICA) \cite{Hyv01_interscience} and creating a classifier of the same dimension as the data (8 dimensions). The results obtained have been very satisfactory at the Ansatz's accuracy and depth level. Still, the need to change the \textit{BFGS} \cite{BFGS_Limted} optimiser to the SPSA \cite{Jam01} has become visible due to its slow convergence for the number of data and high parameters. Figure \eqref{fig:Bench_C} describes the behaviour and compare the two scenarios.

Later the Re-use module was analysed using VQE with \textit{Initial\_point} to synthesise the predicted results. In the graph shown in figure \eqref{fig:Bench_VQE_Init}, it is observed how the algorithm, without initial parameters, tends to use a high energy constant of variation to quickly reach an approximation of the fundamental state. Which makes it have to progressively, after several iterations,  $n$ , reduce said constant to find the local minimum. On the other hand, when using an \textit{Initial\_point}, the algorithm does not need to start with a high variation to reach energy bands close to the ground state since it is much closer to said energy, reducing the number of iterations necessary reach to the local minimum. We can then see how qCBR can afford to run VQE with \textit{Initial\_point} to refine the accuracy of its results since it requires fewer iterations to find the solution closest to the minimum, not assuming such a high computational cost as it would be running VQE without initial parameters.

\begin{table}[t!]
    \centering
    \begin{tabular}{c|c}
         \textbf{Methods}   & \textbf{Complexity} \\
         \hline
         Retrieve & $O(log NM)$  \\
         Re-use &  $ O(Klog(N)+log NM)$ \\
         Revise &  $O(log(N))+ICA)$ \\
         Retain &  $O(log NM)+PCA$ \\
    \end{tabular}
    \caption{Table of the complexity of qCBR counting the PCA cases and the ICA complexity. In this case, K, the number of shots of the VQE, is fixed to 50.}
    \label{tab:qCBR_complexity}
\end{table}

The qCBR complexity (Table \eqref{tab:qCBR_complexity}) is provided below where it can be seen that the \textit{Retain} is the highest cost operation and has an exponential improvement compared to a \textit{Retain} of a classic CBR that is usually of the order of $O( M^{2}(M+N))$ \cite{Pat14}

\section{Discussions}\label{sec:qCBR_Discussions}
Firstly, the proposed qCBR works very well and meets the objectives set using quantum computing to create efficient quantum Case-Based Reasoning. One of the issues to comment on is the improvement observed in figure \eqref{fig:Bench_C}  with respect to the 2 and 8-dimensional classifiers. Due to the small number of depths, but with many more parameters, the 8-dimensional classifiers have an average of about 25$\%$  of improvements over the 2-dimensional ones. With this result, in the case of not wanting an accuracy of around 95$\%$, shallow depth could be used, and computation time saved, depending on the problems.
Despite all these improvements, it is essential to highlight some aspects to refine. In the intelligent system that allows deciding the proposed solution, now, the average of the \textit{Initial\_point} of each solution class samples' \textit{Initial\_point} is used. It could still be seen based on the predicted solution, which \textit{Initial\_point} is the most suitable for the solution to propose. Thus, the cases to be re-used could be better classified.

Also, one of the improvements is to train the classifier with noisy data further so that the qCBR can adapt to real past situations that adjust to the new situation. Because, in practice, there is usually no past case strictly the same as a new one.

The last improvement is to generalise the qCBR to serve various types of problems (betting problem, financial, software maintenance, human reasoning, etc.). To get it, we must focus on designing the memory of the cases so that different data sizes can be indexed and train the classifier with several other data models. 

Secondly, both QIR \cite{lebedev2020introductory} and qCBR work with a data representation model based on a multidimensional vector in Hilbert space.

This offers the possibility for quantum algorithms to perform a clustering or discrimination of the data within this vector space.

The QIR analyses whether a certain entry is related to other types of documents previously studied and how the classic NLP techniques are performed \cite{chowdhury2003natural, liddy2001natural}. To do this, it projects the input vector introduced concerning the bases of the clusters built corresponding to each class with similar patterns.

At the same time, qCBR follows a similar process for predicting whether an input vector corresponds to a previously analysed class and calculates the probability that each type corresponds to the new vector from the proximity of each vector subspaces generated from each category.

The text representation is transformed to a numeric vector from a process called word2vec \cite{goldberg2014word2vec, rong2014word2vec, church2017word2vec} and doc2vec \cite{lau2016empirical, kim2019multi}, and once the vector is obtained, the process to follow is identical to the one to follow by qCBR. In many cases, seeing references \cite{khrennikov2019quantum, bruza2006quantum, lund1996producing, deerwester1990indexing}, QIR and NLP already predefine the classes to be analysed, either Pop, Rock, etc. By predefining that each axis of the Hilbert space corresponds to a type, this process is similar to the qCBR but without the synthesiser's ability.

The clustering process allows the algorithm to create classes and related documents without specifying the categories; therefore, in the case of QIR, it does not move away from an abstraction of the classical problem of "bag-of-words" parsers of spam.

The creation of the SWP vector subspace over the Hilbert vector space is similar in the references \cite{piwowarski2010exploring, Piwowarski2010} where the authors focus on filters, request and document retrieval.

It is worth noting that the qCBR does not present a barren plateau problem due to the low numbers of qubits, shallow quantum circuit and because we have used local cost functions as advocated by the barren plateau theorem \cite{cerezo2021cost}.

\section{Conclusions and further work}\label{sec:qCBR_Conclusions}
We observed the outstanding performance of qCBR compared to its classical counterpart on the average accuracy, scalability and tolerance to an overlapping dataset.
Some of the problems of standard and classical CBR have been mitigated in this work. With the design that has been proposed in this work, it has been possible to measure situations of difficult similarity between cases. Despite the non-linear and overlapping attributes, the classifier has been endowed with characteristics that serve to arrive at two similar topics that may seem quite different by having different values in features, but not very important. In the VQE with \textit{Initial\_point}, we can have different \textit{Initial\_point} associated with each training class sample with the same class. With the technique of the average of the "\textit{Initial\_point}", it is possible to solve this problem by providing the qCBR to distinguish the similarity between cases. Another issue that qCBR mostly solves is the time required to classify a new topic.

With the results of the two implementations (classical and quantum CBR), it is observed that the classical CBR designed with the KNN behaves better for some determined cases (table \eqref{tab:results_CBR_KNN_SW_Full}). It is seen that the system has not finished learning thoroughly (table \eqref{tab:results_CBR_SW_5x4SW} and \eqref{tab:results_CBR_SW_4x3SW}) contrary to the qCBR (table\eqref{tab:results_qCBR_5x4SW} and \eqref{tab:results_qCBR_SW_4x3SW}). This is due to its classifier's accuracy, without forgetting the significant contribution of its synthesis system.

Another improvement that qCBR introduces is when retaining cases, implementing a retention system that maintains model cases and that, together, synthesise the real and most important information.
One of the improvements to consider is the implementation of quantum ICA. In this way, the classical ICA analysis's complexity cost will be significantly reduced. Also counting that the PCA is saved since we have an 8-dimensional classifier, the complexity of the qCBR would be that of the classifier plus some setup constants. 

The other exciting line of the future is to design the memory of cases using the quantum technique of random-access memory (qRAM) \cite {qRAM_} to improve the memory of stored cases.

Now we will generalise the SWP into Batching and Picking problem by defining qRobot.

We will work on qRobot as a quantum computing approach in mobile robot order picking and batching problem solver optimisation in the next session. We will change the social workers by robots and the patient by item to pick and batch.

\section{Summary}
In this section, we have seen how to solve the SWP with a machine learning approach. We have seen how a universal quantum classifier has been created from scratch, developed a formula for calculating the number of solutions, and solved the issue of overfitting and overlapping that the SWP dataset has. In the next chapter, we will see the generalisation of the SWP problem through qRobot.
\newpage
\graphicspath{{./media/}}
\chapter{The qRobot: generalising the SWP}\label{sec:13}
Following the journey of this thesis work, here, we propose and solve the Batching and Picking Problem with the generalisation of the SWP called qRobot.

\section{Introduction}\label{sec:qrobot_introduction}
From DHL, Gartner and others \cite {angeleanu2015new, kuzmicz2015benchmarking, savelsbergh201650th}, we know that the first wave of automation using smart robotics has reached the logistics industry. Driven by rapid technological advancements and increased affordability, robotic solutions (software and hardware) are forcibly entering labour logistics, supporting flawless processes and boosting productivity. Robots, especially mobile, will adopt more roles in the supply chain, helping workers with storage, transportation and little by little, they will expand their service. In fact, in some countries, there are already robotic delivery services \cite{while2021urban}.

We are already living an exponential increment of mail-order shopping, online shopping and supply chain systems, requiring large-scale logistic centres. Almost everyone can order products remotely, and the logistic centre increases its functionalities, including keeping and shipping products. 
While there was a tendency to increase the adoption of automated systems based on robots powered by AI to increase efficiency \cite {van2018robotic, siderska2020robotic, agostinelli2020towards}, COVID-19 introduced the concept of touch-less online shopping that reduces the risk of infections. Smart Warehouses are the epicentre of the cost-efficiency of any e-commerce company \cite{Tompkins2010planning}. 

The emerging field of hybrid (quantum-classical) algorithms joins CPU and QPU \cite {Karalekas2020} to speed up specific calculations within a classical algorithm. This allows for shorter quantum runs that are less susceptible to the cumulative effects of noise and work well in current devices.

Recently the scientific community are researching the real implementation of quantum computing algorithms in mobile platforms because performances are not here yet \cite {Cornet2021}. 

In this work, we demonstrate that we can implement this system in well-known, widely used robotics fields, computer systems like raspberry-pi exploring the performance of a quantum picking and batching model. A hybrid system is proposed to effectively replace the current ones and open the doors to quantum computing in robotics. In addition, we are analysing the results obtained with different public access simulators on the market: IBMQ, Amazon Braket (Dwave), and Pennylane. As far as the authors know, this is the first time this type of implementation has been done.

\section{Work Context}\label{sec:qrobot_Related_work}
According to \cite{chen2016cancer,bustillo2015slaughterhouse,koch2016grouping}, supply chains, warehouses and distribution centres occupy a very important position when storing and serving customer demand. Today, to be competitive within this sector, Logistics 4.0 has been created, known as the set of artificial intelligence technologies and techniques that seek the efficiency of the movements of materials and products in a factory or warehouse. In addition, better time management helps logistics companies find and locate a material, reduce fatigue and possible workplace accidents, and spend less time documenting items.

Many works of literature highlight these factors as the main ones where the loss of time and resources in a process require an urgent solution, and precisely, it is technologies such as Artificial Intelligence and the Internet of Things (IoT), which today allow us to optimise them \cite{albareda2009multi,cergibozan2019order,azadnia2013order}.

Researchers have addressed the multiple order picking planning problems in the last decade. The study of the efficiency of a Warehouse can be addressed based on numerous parameters. According to \cite{vangils2018picking}, there are three key considerations: 1) Performance Measure (time, cost, productivity, and service), 2) How we model the warehouse (Analytical model, Mathematical Model, or Simulation), and the combination of factors (storage location assignment, routing, order batching, or other order picking planning problems).

Based on data from \cite{vangils2018picking}, we can see the percentage of the relevance of the considered order picking planning problems based on the percentage of papers that are related to such challenges: 

\begin{figure}[!ht]
\centering
\includegraphics[width=0.55\textwidth]{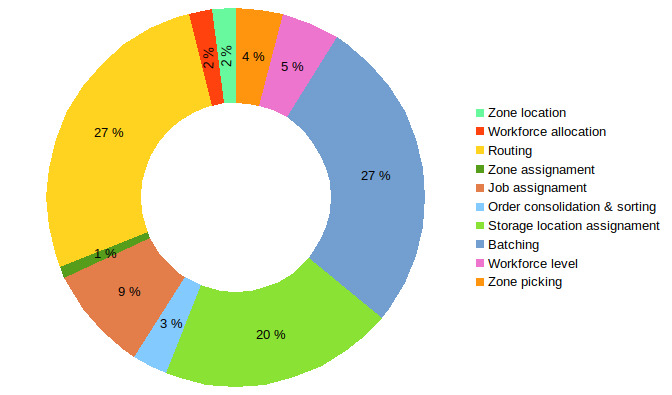}
\caption{Distribution of considered order picking problems based on the percentage of publications.
}
\label{fig:planningprob}
\end{figure}
As we can see in Fig.\ref{fig:planningprob} Picking and Batching are the top priorities based on the research contributions. 

Order preparation (picking) is one of the most frequent and costly operations in labour \cite{chen2016cancer,  bustillo2015slaughterhouse} since it is responsible for recovering the items required by the orders of customer orders (could also be supplied, but in this article, we focus exclusively on sales orders), and to create the batches, grouping several orders of orders in a picking list to collect all the batch demands in a single warehouse tour. In this last part of order preparation, our quantum algorithm comes into action to optimise the routes travelled to achieve efficient picking.

There are many techniques and strategies for solving the picking problem. The most striking are “The selected techniques for evaluation include A *” \cite{duchovn2014path}, “Potential Fields (PF)”, “Rapidly-Exploring Random Trees * (RRT *)” \cite{lavalle2001rapidly, lavalle1998rapidly, cheng2002resolution} and "Variations of the Fast-Marching Method (FMM)"\cite {rawlinson2005fast}.
Other strategies have been explored using the TSP and the VRP as algorithms to solve the picking problem. In this case, if the number of order orders per lot is greater than two \cite {gademannvan}, picking becomes an NP-Hard problem in which the number of possible lots and binary variables increases exponentially with the number of purchase orders \cite{gademannvan}.
From there, several heuristic techniques, methods and algorithms (for example, genetic) were created to relax these difficulties \cite{Cortina2001,azadnia2013order,chen2016cancer,hsu2005batching,koch2016grouping,tsai2008using}. However, and as mentioned above, depending on the volume of data, the computational cost of the algorithm becomes intractable for classical computing.

The latter leads us to explore new approaches to the large-scale picking problem, and one of the approaches to consider to solve this task is quantum computing [6].
Quantum computing could help us change the degree of complexity of the problem, enhanced by its high computing power. Among the great fields where quantum computing is called to stand out is constraint satisfaction problems (CSP) \cite{tsang2014foundations}. One of the useful algorithms in this field is Quadratic Unconstrained Binary Optimisation (QUBO) problems \cite{kochenberger2014unconstrained}.

\begin{figure}[!ht]
\centering
\includegraphics[width=0.75\textwidth]{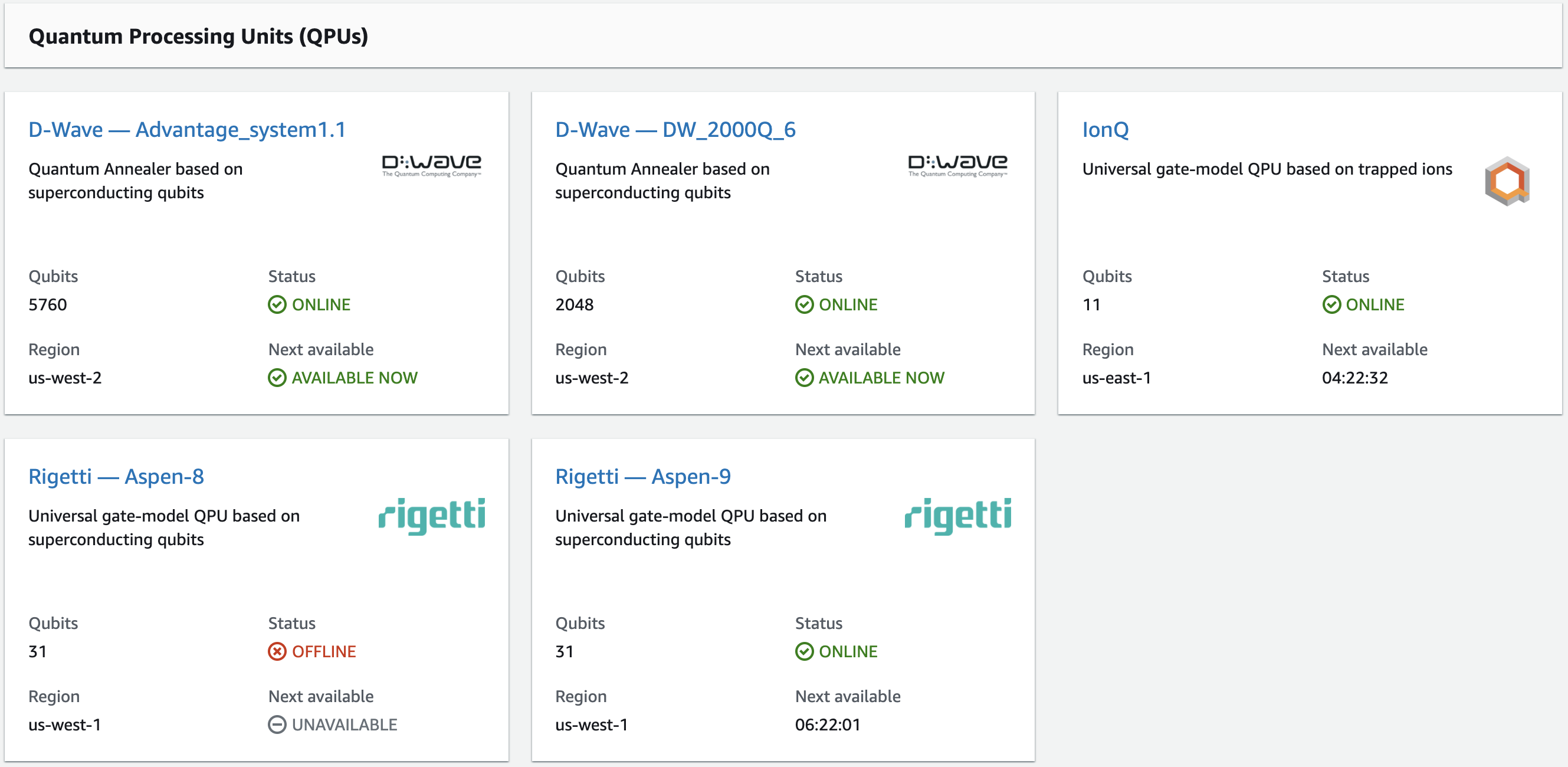}
\caption{Technical Specifications of the Quantum Hardware Technologies (Gate-based superconducting qubits, Gate-based ion traps and Quantum annealing) available in Amazon Braket.}
\label{fig:AWS_QC}
\end{figure}

\begin{figure}[!ht]
\centering
\includegraphics[width=0.75\textwidth]{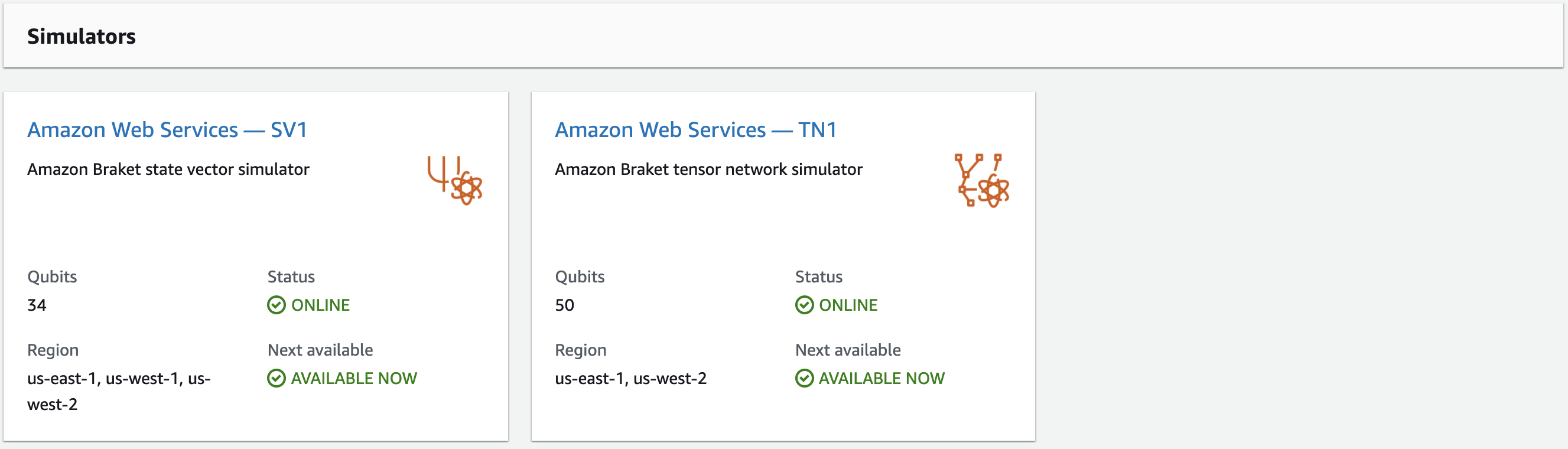}
\caption{Technical Specifications of Quantum Simulators systems where we can see it state vector simulator (34 qubits) and tensor network simulator (50 qubits).}
\label{fig:AWS_Simulador}
\end{figure}

Amazon Braket\cite{AWS_Braket_web} is a cloud-based (Fig.\eqref{fig:AWS_QC} and Fig.\eqref{fig:AWS_Simulador}), fully managed quantum computing service that helps researchers and developers get started into quantum world technology to accelerate research and discovery. Amazon Braket provides a development environment to explore, create, test and run quantum algorithms, quantum circuit simulators, and different quantum hardware technologies. 

We will use all these related works to define an appropriate strategy for our proposal in this NISQ era.

Studying and comparing different optimisation methods of warehouse's challenge, like picking and batching, \cite{vangils2018picking} proposes three options: analytical models, simulation experiments, and mathematical programming. In our approach, we consider the latter. We use a set of mathematical expressions that describe the problem, represented by an objective mathematical function and constraints within the classical context and translate it to the quantum domain.  

While reviewing the state of the art research, this reference \cite {xie2021formulating} was found. The integrated order routing and the batch problem is modelled in such systems as an extended multi-tank vehicle routing problem with network flow formulations of three indices and two commodities. Such a variable neighbourhood search algorithm provides close to optimal solutions within a computational time acceptable for classical but not quantum computing.

This article intends to bring quantum computing to robotics by proposing an approach that combines the experience of classical robotics computing with the computation of complex and high-cost processes by quantum computing. We suggest preparing an environment to execute the quantum algorithms in the mobile and autonomous robot remotely and locally and designing a quantum algorithm that helps the efficiency of the warehouse management.

\begin{figure*}[t!]
\centering
\includegraphics[width=.7\textwidth]{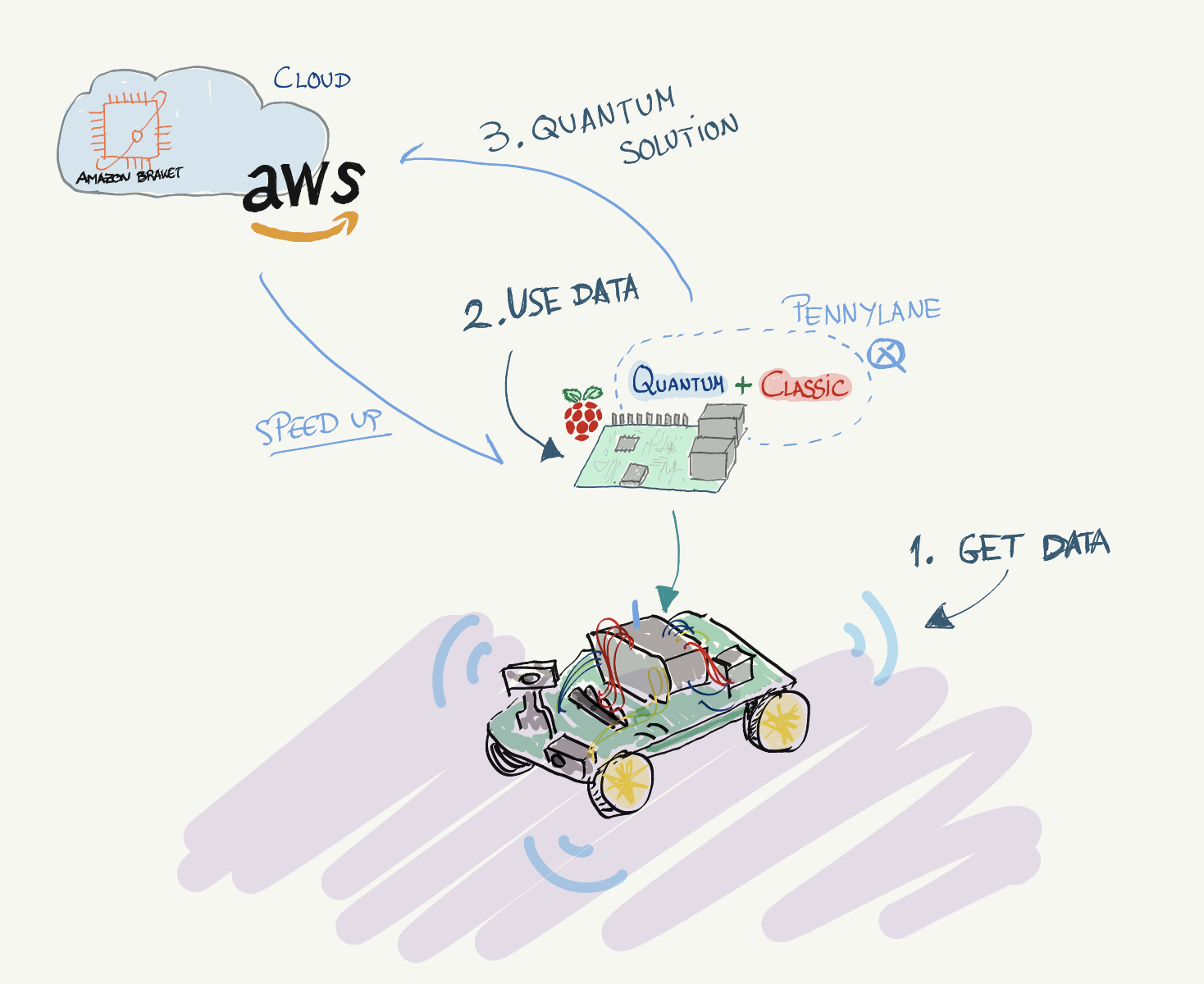}
\caption{We propose a robot that prepares batches and increases the efficiency of picking in a warehouse, taking advantage of the classic Machine Learning experience and leveraging hybrid computing (classical + quantum) in the cloud and distributed. This robot uses the Optimal routing strategy to calculate the shortest route, regardless of the layout or location of the items.}
\label{fig:qRobot}
\end{figure*}

\begin{figure}[!ht]
\centering
\includegraphics[width=0.75\textwidth]{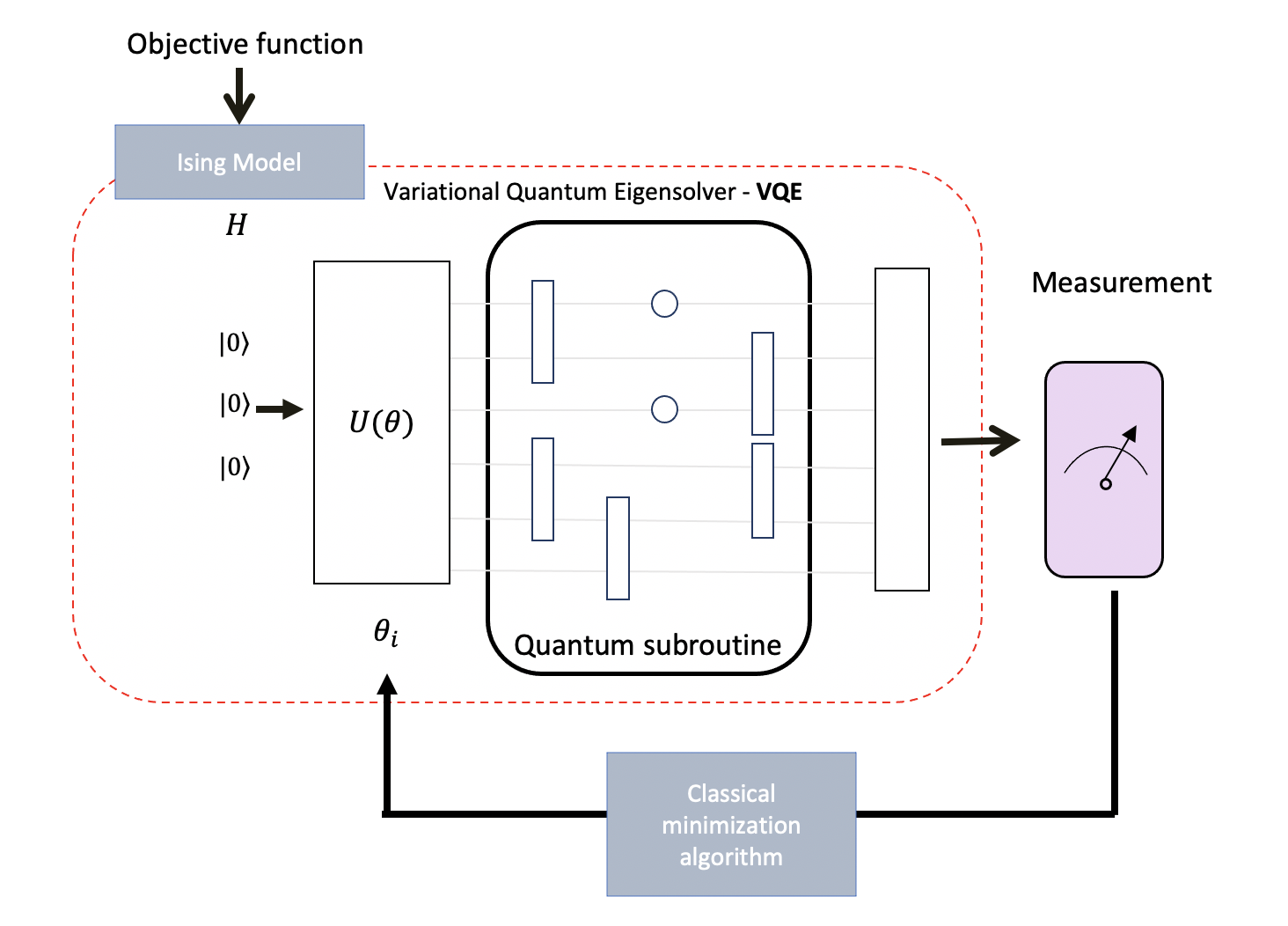}
\caption{VQE working principle based on the quantum variational circuit.}
\label{fig:VQE_f}
\end{figure}

From the variational principle in section\eqref{sec:variational_Cal}, the following equation $\langle H \rangle _{ \psi   \left( \overrightarrow{ \theta } \right) } \geq  \lambda _{i}$ can be reached out. With $\lambda _{i}$  as eigenvector and  $\langle H \rangle _{ \psi \left( \overrightarrow{ \theta } \right)}$  as the expected value. By this way, the VQE finds \eqref{expectative_value} such an optimal choice of parameters $\overrightarrow{\theta }$, that the expected value is minimised and that a lower eigenvalue is located. 
\begin{equation}
\label{expectative_value}
 \langle H \rangle =\langle\psi\left(\theta  \right)\vert H \vert\psi\left(\theta  \right)\rangle. 
\end{equation}
We will use the VQE (Fig. \eqref{fig:VQE_f}) to find the minima of our objective function translated to the Ising model.

\section{Implementation }\label{sec:qRobot_implementation}
\begin{figure}[!ht]
\centering
\includegraphics[width=0.75\textwidth]{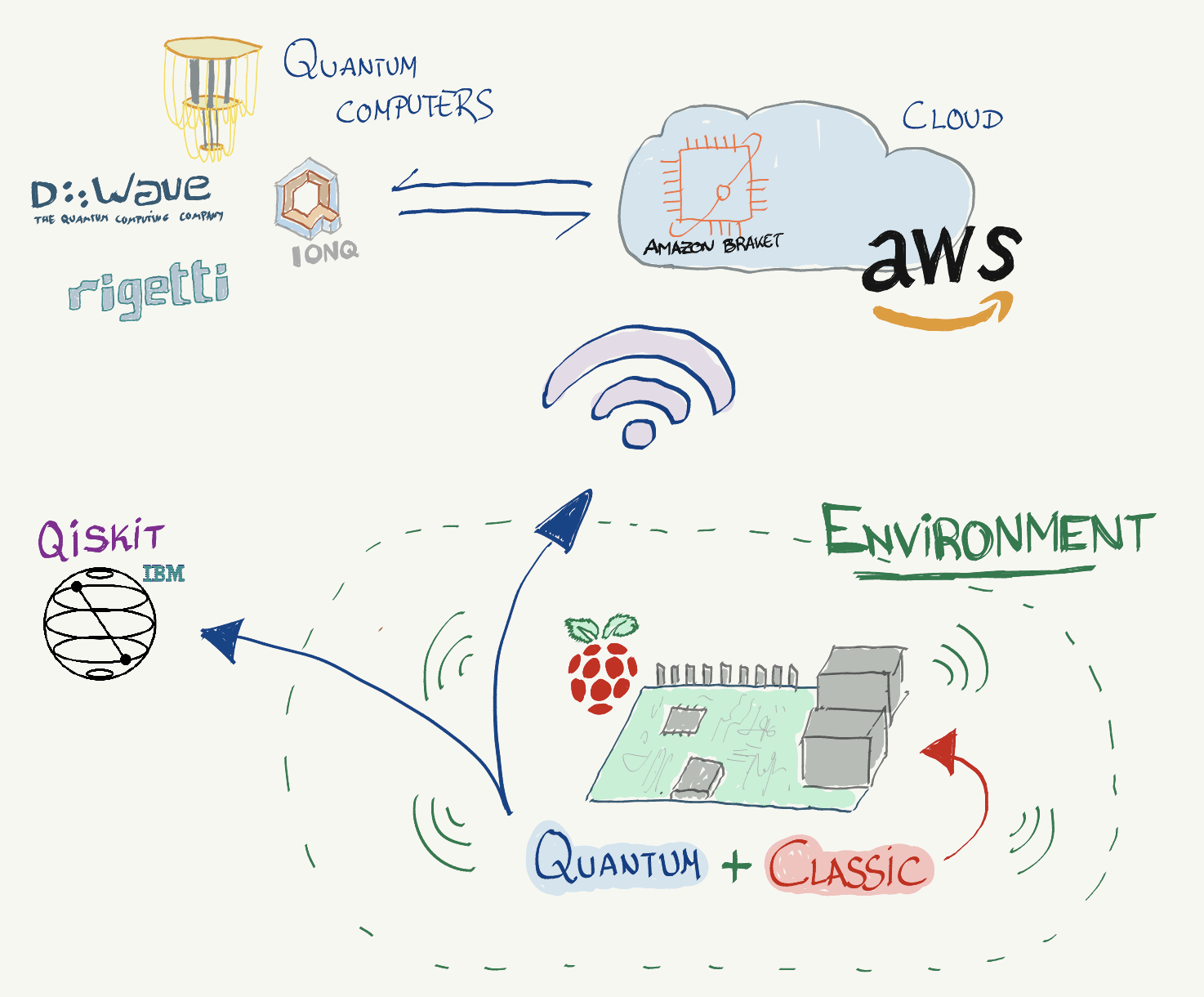}
\caption{We have installed the following frameworks successfully (Qiskit, Pennylane, AWS-Braket) on our Raspberry Pi 4 under the ARM64 operating system. More information about the qRobot platform can be found at Ref.~\cite{qRobotP}}
\label{fig:qRobot_platform}
\end{figure}

To carry out the implementation of our proof of concept (Fig.\eqref{fig:qRobot} and Fig.\eqref{fig:qRobot_platform}), we must first prepare the programming environment.
Considering that the core of our robot will be the Raspberry Pi 4 \cite{Raspberry}, the first thing to do is prepare it so that it can execute quantum algorithms with the guarantees required for the proposed application and especially for future operations on gradients. It is necessary to install an ARM64 operating system \cite{jaggar1997arm,jiang2020power} with all the needed packages to run all the required environments to carry out this project.
We took advantage of the work for Raspberry Pi Os Desktop (32-bit) on which the author describes how to install and run Qiskit - IBM's open-source quantum computing software framework \cite{Qis21}— on a Raspberry Pi to turn it into a quantum computing simulator and use it to access real IBM quantum computers. In our case, we do need ARM64 because we need to execute at least TensorFlow's version 3.2.1. The tasks to convert the Raspberry Pi 4 in our "quantum computer" are described in the following reference \cite{qRobotP}.

After setting up the environment, we will focus on designing and experimenting with the announced proof of concept.

\subsection{The problem's formulation}\label{sec:qRobot_problem_formulation}
In this formulation, we will seek to optimise the collection of the products, and, later, we will make the batches.

\begin{figure}[t!]
\centering
\includegraphics[width=.45\textwidth]{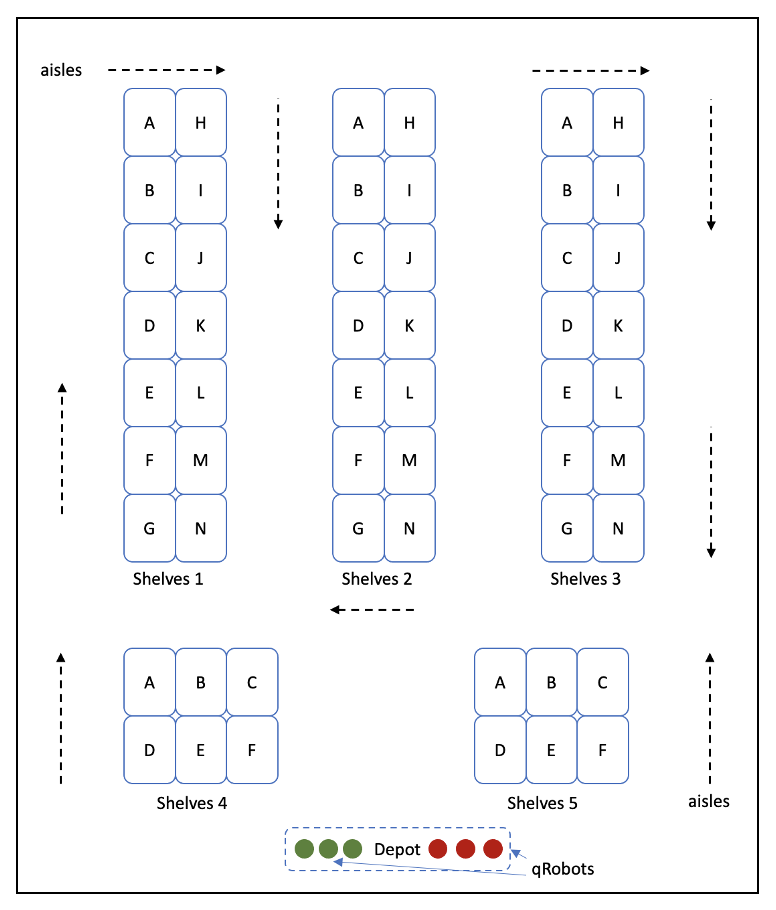}
\caption{Structure of our warehouse with pick locations. The warehouse has a rectangular layout with no unused space. We use all the parallel corridors. This proof of concept contemplates a single warehouse used to take the order and deliver it, and it is also divided into blocks, which contain slots for storing products, and transverse aisles separate them. The cross aisles do not have any products but allow the collector to navigate the warehouse. We base our picking strategy on minimising the route and optimising batch preparation. We do not contemplate shelving of different levels.}
\label{fig:Wharehouse}
\end{figure}

To carry it out, we will consider the following assumptions:
\begin{enumerate}
    \item The strategy we will follow is the picking routing problem to retrieve each lot which the total distance travelled to retrieve all the items in a lot will be calculated.
    \item The warehouse configuration is given in figure\eqref{fig:Wharehouse}.
    \item For the orders of the storage positions, more than one picking robot can be used.
    \item Movements in height are not considered.
    \item Each product is stored in a single storage position and only one product is stored in each storage position.
    \item Each picking route begins and ends at the Depot.
    \item The load capacity for each order will not exceed the load capacity of the picking robot.
    \item At the moment, the division of order orders is not contemplated. That is, only the batches of closed orders can be prepared.
    \item The concept testing will be done on all AWS-Braket, Pennylane, D-Wave and Qiskit environments. And we will stick with the scenario that best benefits our proof of concept.
    \item We will use the docplex \cite{docplex} to model our formulation.
\end{enumerate}

\begin{figure}[!ht]
\centering
\includegraphics[width=0.45\textwidth]{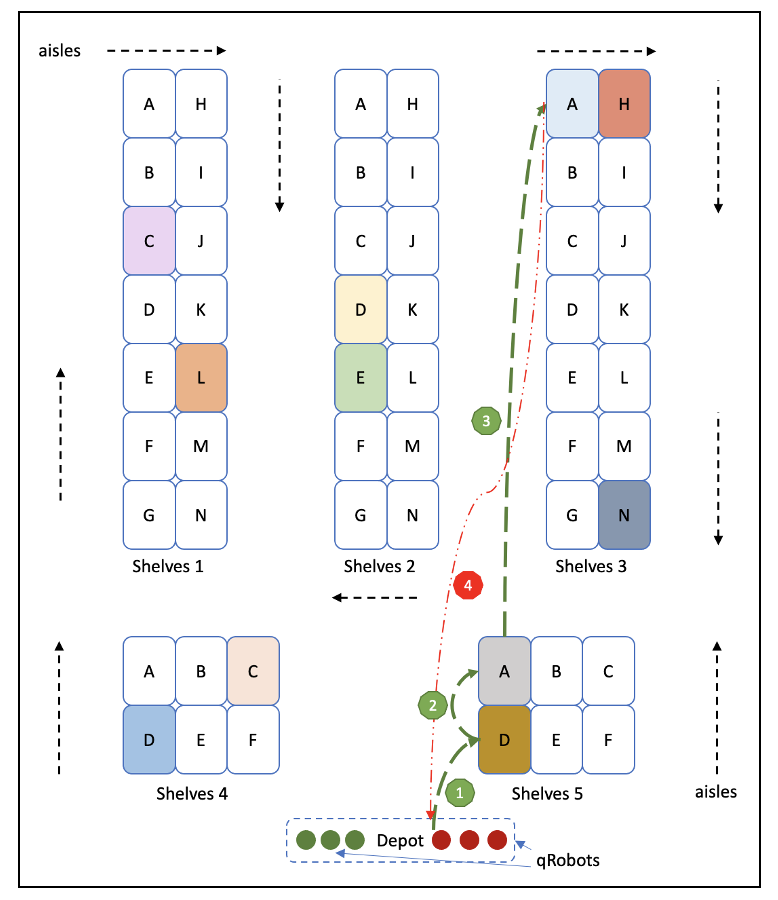}
\caption{Scenario 1, Independent lots. The robot receives the orders and calculates which order is the most optimal according to the coordinates in which each product is found. In this example, lot 4 is the most optimal.}
\label{fig:scenario1}
\end{figure}
\begin{figure}[!ht]
\centering
\includegraphics[width=0.45\textwidth]{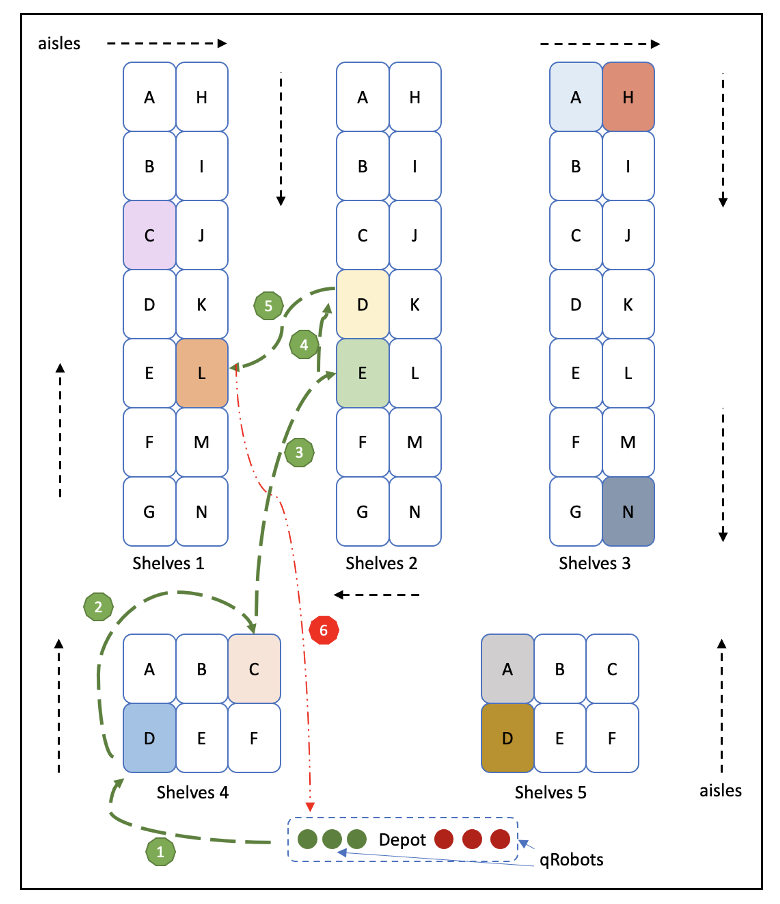}
\caption{Scenario 2, Collecting products in the same route from different batches. The robot will calculate a path that includes all the products to optimise their collection in a single journey. For example, if the product from Lot 2 is next to one from Lot 1, the robot will pick it up and store it in the basket from Lot 2.}
\label{fig:scenario2}
\end{figure}

\subsection{Picking and Batching formulation}\label{sec:q_robot_picking_batching_P}
The formulation is represented as follows. In this scenario, the travel load is described according to the number of robots we have. Let's imagine that we have several robots and that each of them makes a single trip. It would be the same as saying that we have a single robot that makes $n$ trips.

Let $ N_ {0} $ be the origin node, let $ N_ {1} \dots N_{n} $ be the nodes of the products, let $ W_ {1} \dots W_{n} $ be the weights in kg associated with for each product, let $ d_ {i, j} $ be the distance from node $ i $ to $ j $, let $ M $ be the maximum load of the qRobots, let $ K $ be the number of qRobots available, let $t$ be the instant, $i$ the node (product), $p$ the robot and let $ x_ {t , i, p} $ be our binary variable (for example, for $ x_ {2,3,2} = 1 $. It means that at time $ 2 $, the qRobot $ 2 $ is at node 3). In our formulation, time really tells us the order, that is to say $t = 0$ will be the origin $t = 1$ the moment in which it goes for the first batch. At $t = 2$ it will be the moment of the second so on.

\begin{equation}
\label{2BP_Object_Function}
\begin{aligned}
    \min_{x} \quad & \sum_{p=1}^{K}\sum_{t=1}^{n+1}{\sum_{i=0}^{n}\sum_{j=0}^{n} x_{t_{t-1},i,p}x_{t,j,p}d_{i,j}},\\
\end{aligned}
\end{equation}

\begin{equation}
\label{2BP_Object_Function_Res_1}
\begin{aligned}
\textrm{s.t.} \\
    \quad & \sum_{p=1}^{K}x_{0,0,p} = K, \\
\end{aligned}
\end{equation}
\begin{equation}
\label{2BP_Object_Function_Res_2}
\begin{aligned}
    \quad & \sum_{p=1}^{K}x_{n+1,0,p} = K, \\
\end{aligned}
\end{equation}
\begin{equation}
\label{2BP_Object_Function_Res_3}
\begin{aligned}
    \quad & \sum_{t=1}^{n+1}\sum_{i=1}^{n} x_{t,i,p}W{i}\leq M \qquad \forall p \in \{1, ..., K\},\\
\end{aligned}
\end{equation}

\begin{equation}
\label{2BP_Object_Function_Res_4}
\begin{aligned}
    \sum_{i=1}^{n} x_{t,i,p} = 1 \quad \forall t \in \{1, \ldots, n+1\} \\ \quad \forall p \in \{1, ..., K\},\\
\end{aligned}
\end{equation}

\begin{equation}
\label{2BP_Object_Function_Res_5}
\begin{aligned}
    \sum_{t=1}^{n+1}\sum_{p=1}^{K} x_{t,i,p} = 1  \qquad \forall i \in \{1,\ldots, n+1\},\\
\end{aligned}
\end{equation}

\begin{equation}
\label{2BP_Object_Function_Res_6}
\begin{aligned}
    x_{t,i,p} \in \{0,1\} \quad \forall t \in \{0, \ldots, n+1\}\\
    \\ \quad \forall i \in \{0, ..., n\}\\
    \\ \quad \forall p \in \{1, ..., K\}.\\
\end{aligned}
\end{equation}

The equation \eqref{2BP_Object_Function} is our new objective function. Here we minimise the total distance. We add the distance of all the robots travelling simultaneously, and we will check the nodes' distances. Restriction \eqref{2BP_Object_Function_Res_1} establishes that all the qRobots start from Depot. The restriction \eqref{2BP_Object_Function_Res_2}) establishes that all the qRobots end at Depot. The constraint \eqref{2BP_Object_Function_Res_3} establishes any robot $p$ cannot carry more load than allowed. The constraint \eqref{2BP_Object_Function_Res_4} declares that each robot can only be one node at any time. \eqref{2BP_Object_Function_Res_5} establish that throughout the entire route, the robots together pass each node only once and the restriction \eqref{2BP_Object_Function_Res_6} describes that $x_{t, i, p}$ are binary variables.

The number of the qubits to perform this algorithm is equal to $K(n+1)(n+2) + K \lceil log_{2}M\rceil$. At this point, we can only map our objective function in quantum and then solve it with a VQE.

\subsection{Mapping the classical to quantum optimisation} \label{sec:Mapp_Batching_Ising}
A common method for mapping classic optimisation problems to quantum hardware is by coding it into the Hamiltonian\cite{eisberg1985quantum} of an Ising model \cite{lucas2014ising}.
\begin{equation}
\label{Hamiltoninano_ISing}
\begin{aligned}
    H_{Ising} = \sum_{i<j} J_{i,j}\sigma_{i}\sigma_{j} + \sum_{i} h_{i}\sigma_{i}.\\
\end{aligned}
\end{equation}

Where $ \sigma_i $ is the product of $ n $ identity matrices $ I $ except a gate $ Z $ in the i-th position and $ \sigma_i \sigma_j $ product of identities minus gates $ Z $ in positions $ i $ and $ j $.

As we already can build our objective function as a QUBO in the form $\langle x^T\vert Q \vert x\rangle$, now we can map our QUBO to Ising Hamiltonian formulation leads to calculating the values of $J_{ij}$  and $h_{i}$. 

The transformation between QUBO and Ising Hamiltonian is $z_i = 2x_i-1$, where $z_i$ is a new variable that can take the values $-1$ or $1$. This means we will have the algorithm in Ising form by writing an algorithm for QUBO with this single variable change. That is very useful to have the algorithm for various platforms based on quantum gates (IBM Q and Pennylane) or quantum annealing (meanly D-Wave) in case of going from the Hamiltonian form.
we can now solve our Picking and Batching Problem with VQE $\langle \psi(\theta)\vert H \vert \psi(\theta)\rangle$.

\section{qRobot Results}\label{sec:qRobot_result}
\begin{figure*}[t!]
\centering
\includegraphics[height=4cm]{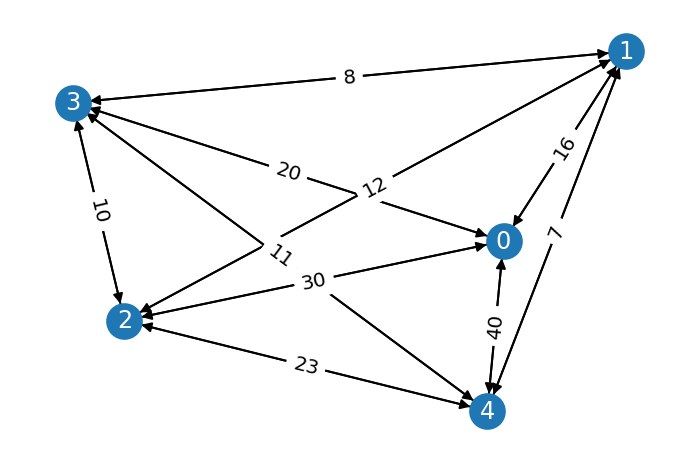}
\includegraphics[height=4cm]{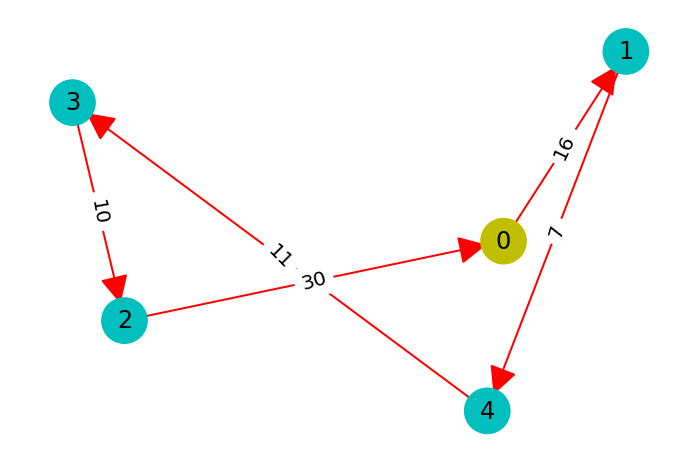}
\includegraphics[height=4cm]{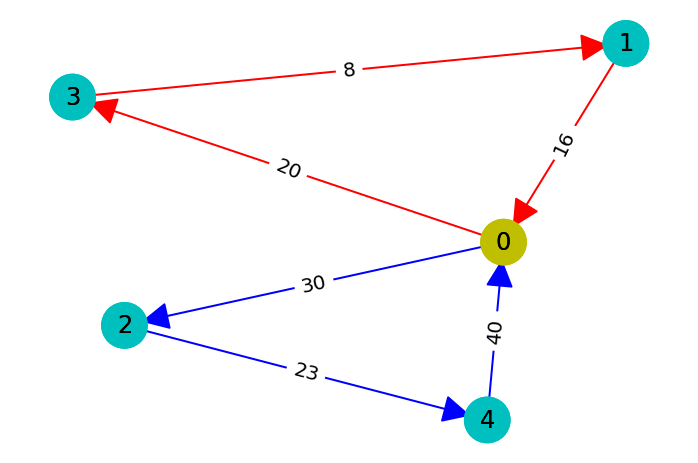}
\includegraphics[height=4cm]{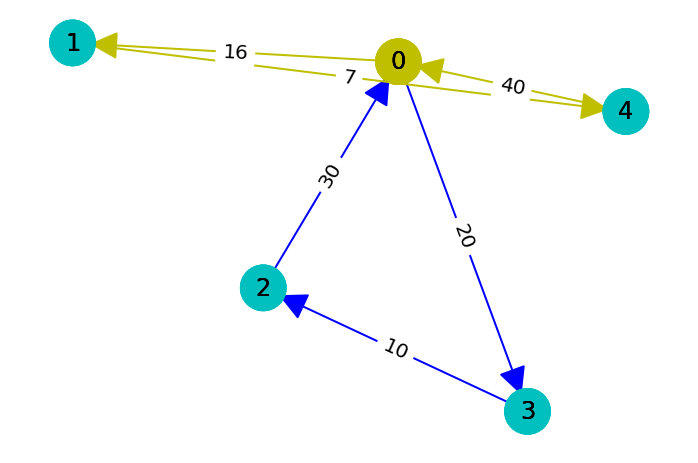}
\includegraphics[height=4cm]{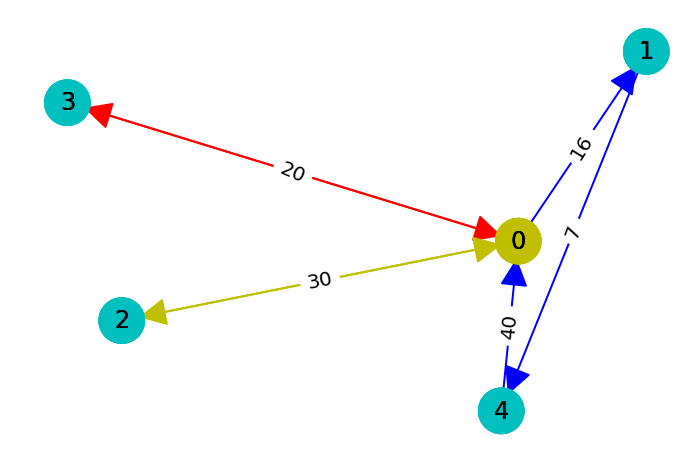}
\includegraphics[height=4cm]{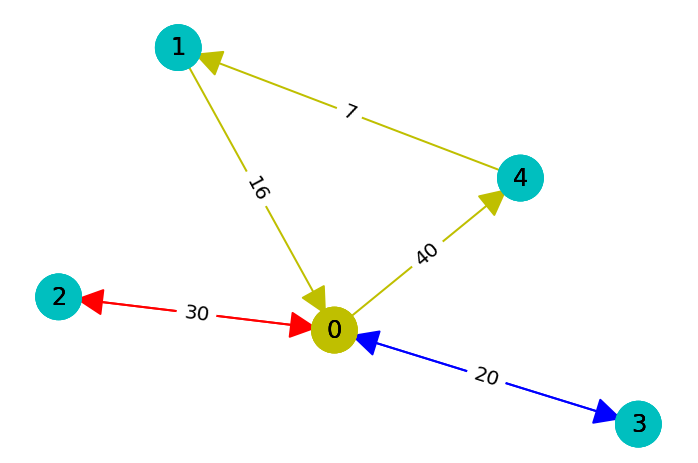}
\caption{In these graphs, we can observe the results of the algorithm in different scenarios. A different colour represents each qRobots; qRobot number $ 1 $ is red, next is blue, and the third is yellow, so on. While the depot is the $ 0 $ node in yellow color and the rest of the nodes are represented in blue. The weights of each item (not normalised) in kg are $ w_{0} = 0, w_{1} = 8, w_{2} = 8, w_{3} = 3$ and $ w_{4} = 3$. The maximum capacity of each qRobots is $ 45 $.
 In this case, we have $ 4 $ items and the possibility of using up to $ 3 $ qRobots.
Reading the images from left to right, we see that the nodes and their respective distances are shown in the first image. The second image shows the result of the algorithm having a qRobot. In the third and fourth images, we can see two different cases solved by two qRobots. And finally, in the fifth and sixth images, we can see two other issues solved by three qRobots.
It is important to highlight that our algorithm in this proof of concept minimises the distance travelled and optimises the number of qRobots necessary to solve the cases presented. If it judges that the task can be performed with a single qRobot, it will not send $ 2 $ qRobots.}
\label{fig:Results_qRobots_4}
\end{figure*}

\begin{figure*}[t!]
\centering
\includegraphics[height=4cm]{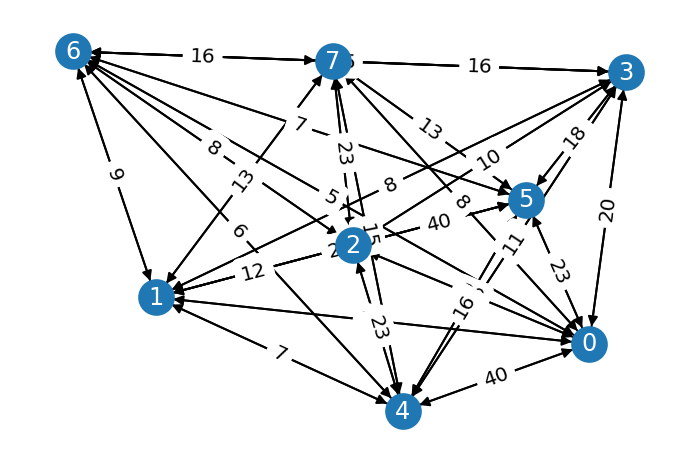}
\includegraphics[height=4cm]{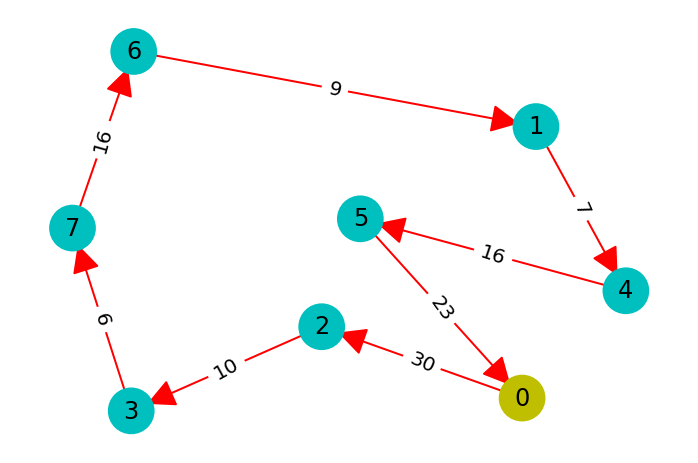}
\includegraphics[height=4cm]{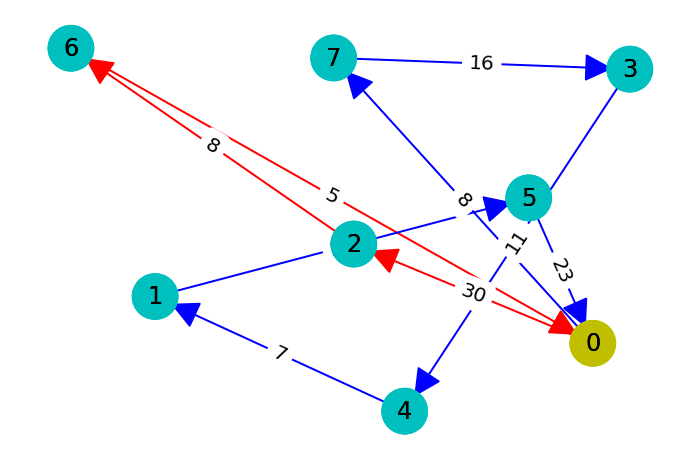}
\includegraphics[height=4cm]{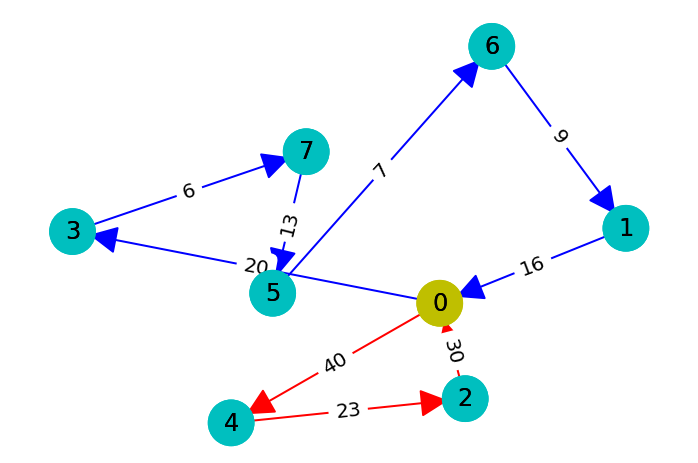}
\includegraphics[height=4cm]{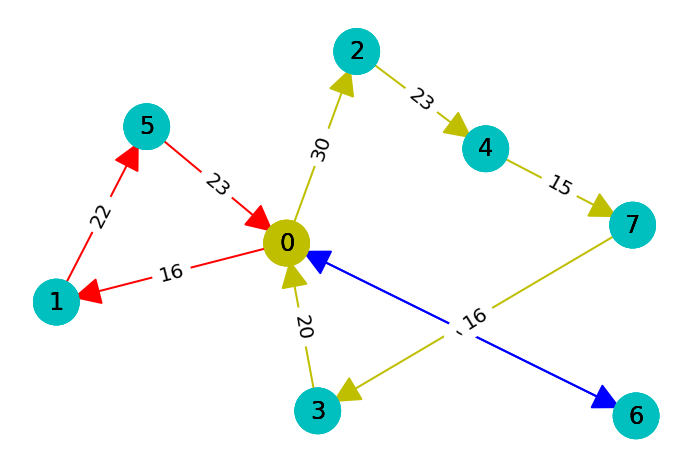}
\includegraphics[height=4cm]{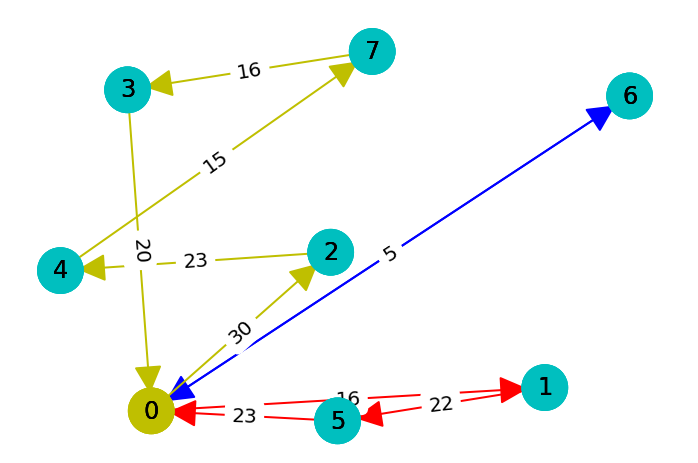}
\caption{In these graphs, we can observe the results of the algorithm in different scenarios. A different color represents each qRobots; qRobot number $ 1 $ is red, the next is blue, the third is yellow, and so on, while the depot is the $ 0 $ node in yellow color, and the rest of the nodes are represented in blue. The weights of each item (not normalised) in kg are $ w_{0} = 0, w_{1} = 8, w_{2} = 8, w_{3} = 3, w_{4} = 3, w_{5} = 1, w_{6} = 2$, and $w_{7} = 4 $. The maximum capacity of each qRobots is $ 45 $.
 In this case, we have $ 7 $ items and the possibility of using up to $ 4 $ qRobots.
Reading the images from left to right, we see that the nodes and their respective distances are shown in the first image. The second image shows the result of the algorithm having a qRobot. In the third and fourth images, we can see two different cases solved by two qRobots. And, finally, in the fifth and sixth images, we can see two other issues solved by three qRobots.
It is important to highlight that our algorithm in this proof of concept minimises the distance travelled and optimises the number of qRobots necessary to solve the cases presented. If it judges that the task can be performed with a single qRobot, it will not send $ 2 $ qRobots.}
\label{fig:Results_qRobots_7}
\end{figure*}

\begin{figure}[!ht]
\centering
\includegraphics[width=0.75\textwidth]{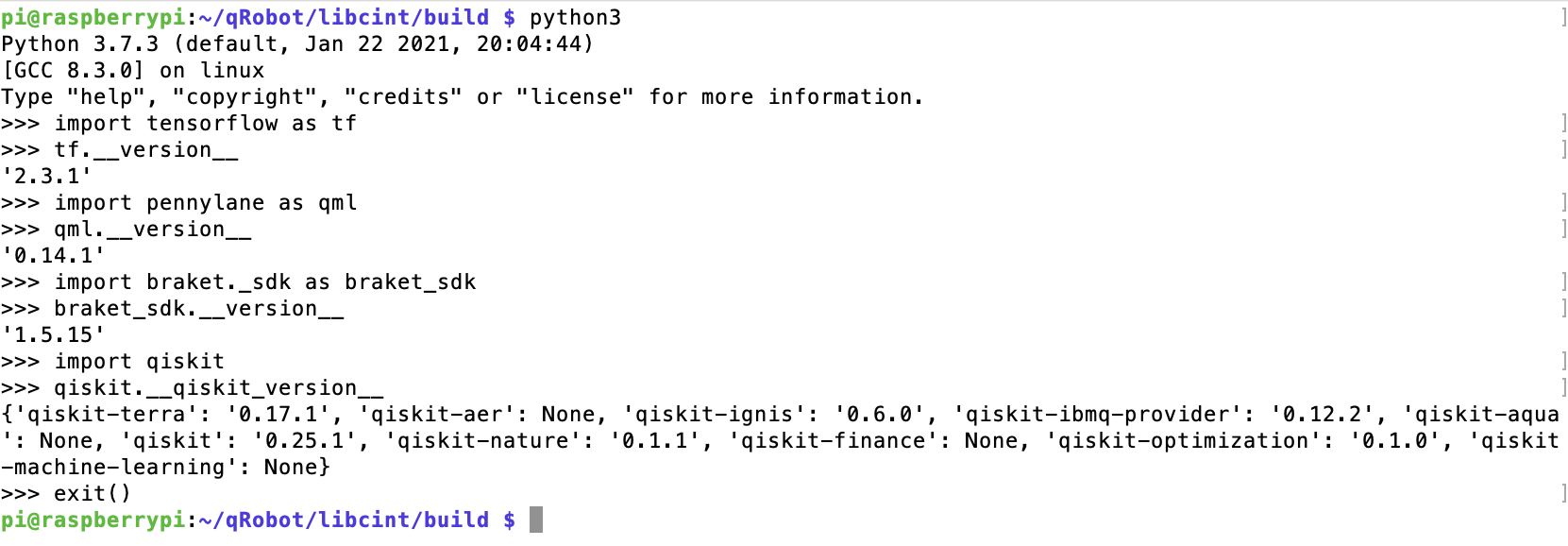}
\caption{This figure shows that we judge important environments to carry out quantum computing to robotics and beyond. We can see the correct installation of TensorFlow 3.2.1 as required for all gradient operations; see the installation of Pennylane version 14.1, the installation of the latest version of Amazon Braket, and all the packages of the newest version of Qiskit 0.25 minus the qiskit-machine-learning package.}
\label{fig:installPackagesTerminal}
\end{figure}

\begin{figure}[!ht]
\centering
\includegraphics[width=0.75\textwidth]{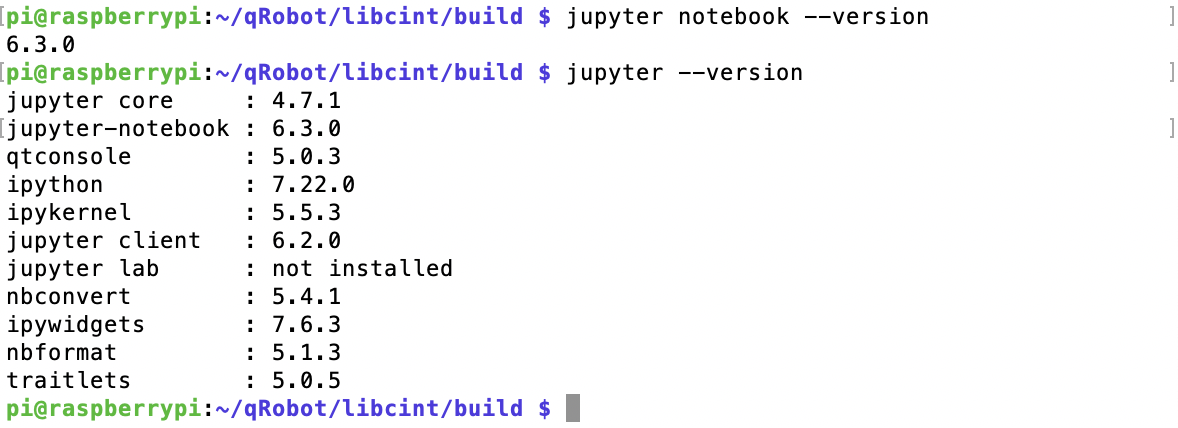}
\caption{In this figure, we can see the correct installation of the Jupyter package and the Jupyter notebook that has been our environment of proof of concept. With this, everything is ready to import or write code in the different frameworks mentioned above (IMBQ, AWS-Braket, Pennylane and D-Wave).}
\label{fig:Jupyter_Package}
\end{figure}

\begin{figure}[!ht]
\centering
\includegraphics[width=0.75\textwidth]{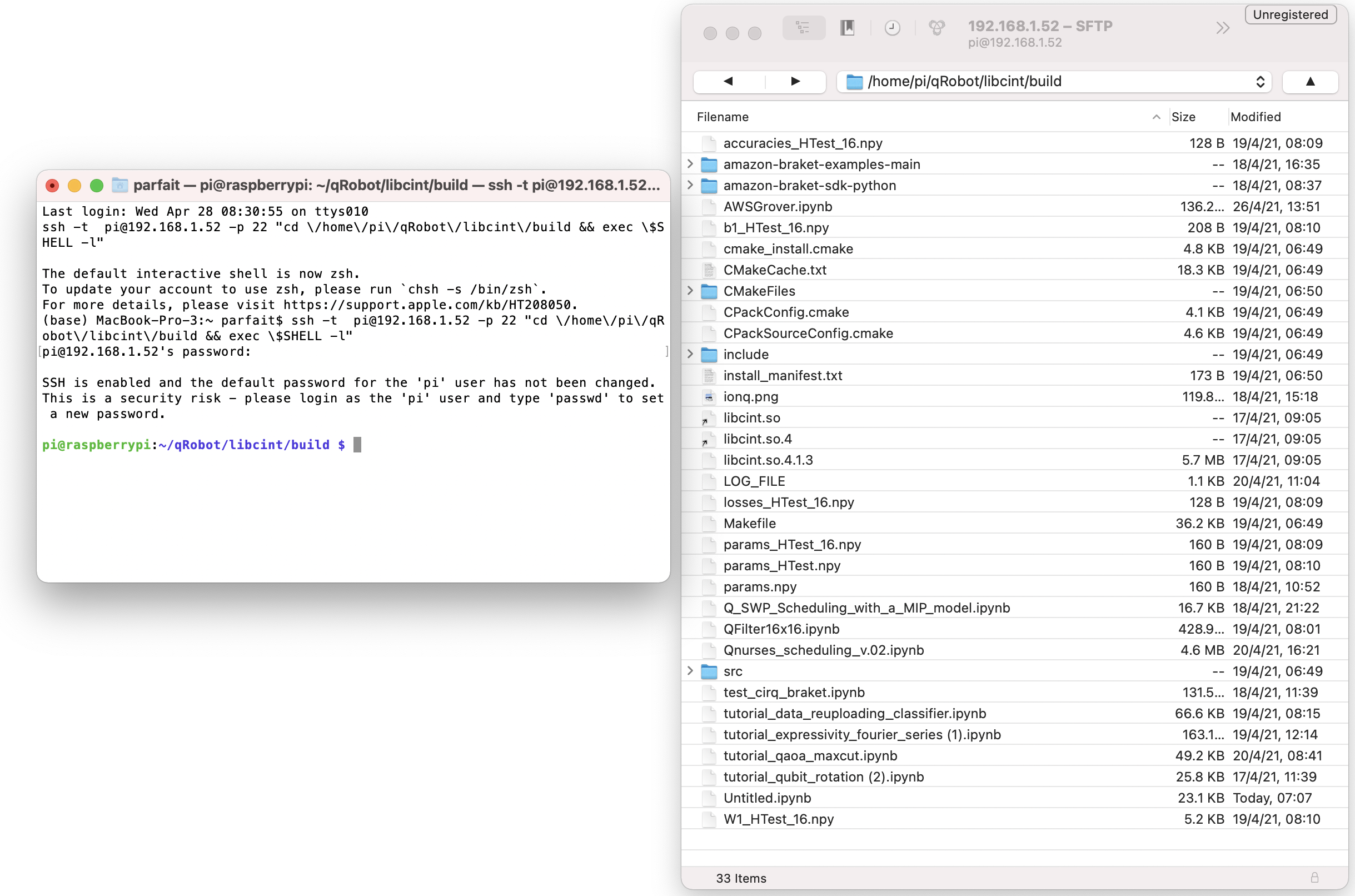}
\caption{This figure shows the files window through the CyberDuck client SSH \cite{T_cyberduck} viewer with the directory and file structure. And on the left, you can see the terminal that gives access to the qRobot. To access the qRobot by SSH, the username and password are required. Everything is configurable \cite{config_Rasp}.}
\label{fig:AccessSSH}
\end{figure}

\begin{figure}[!ht]
\centering
\includegraphics[width=0.45\textwidth]{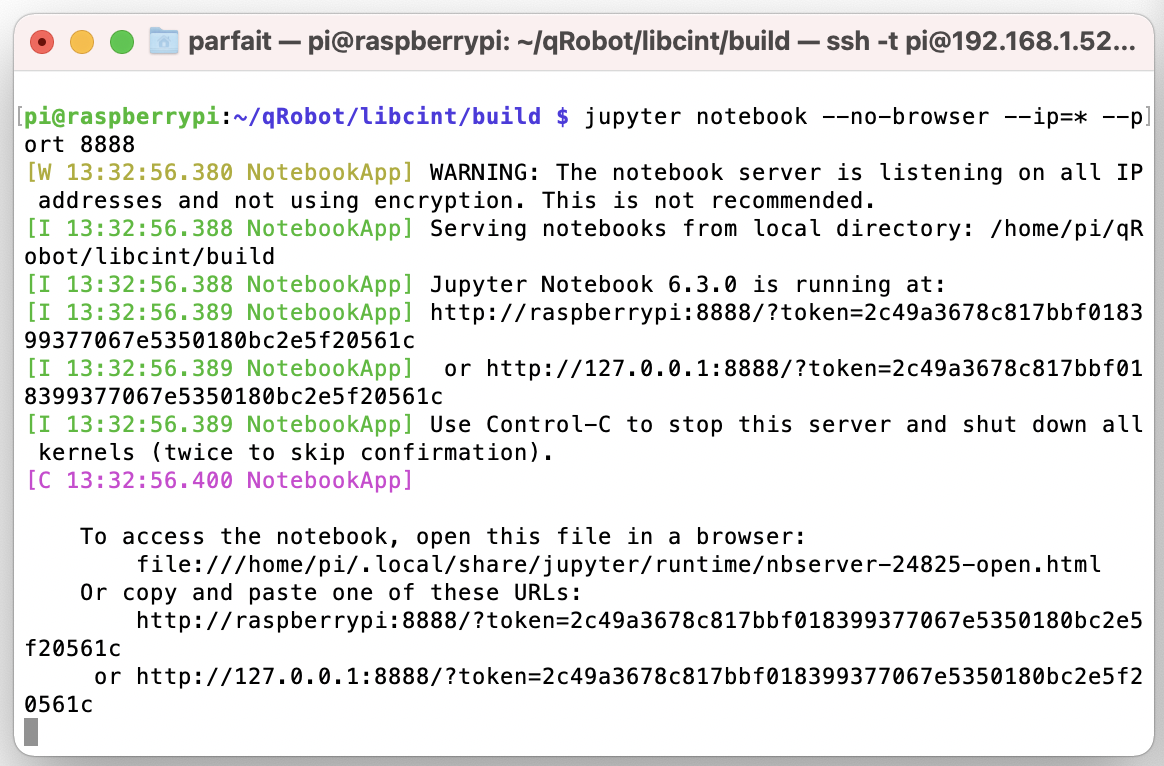}
\includegraphics[width=0.45\textwidth]{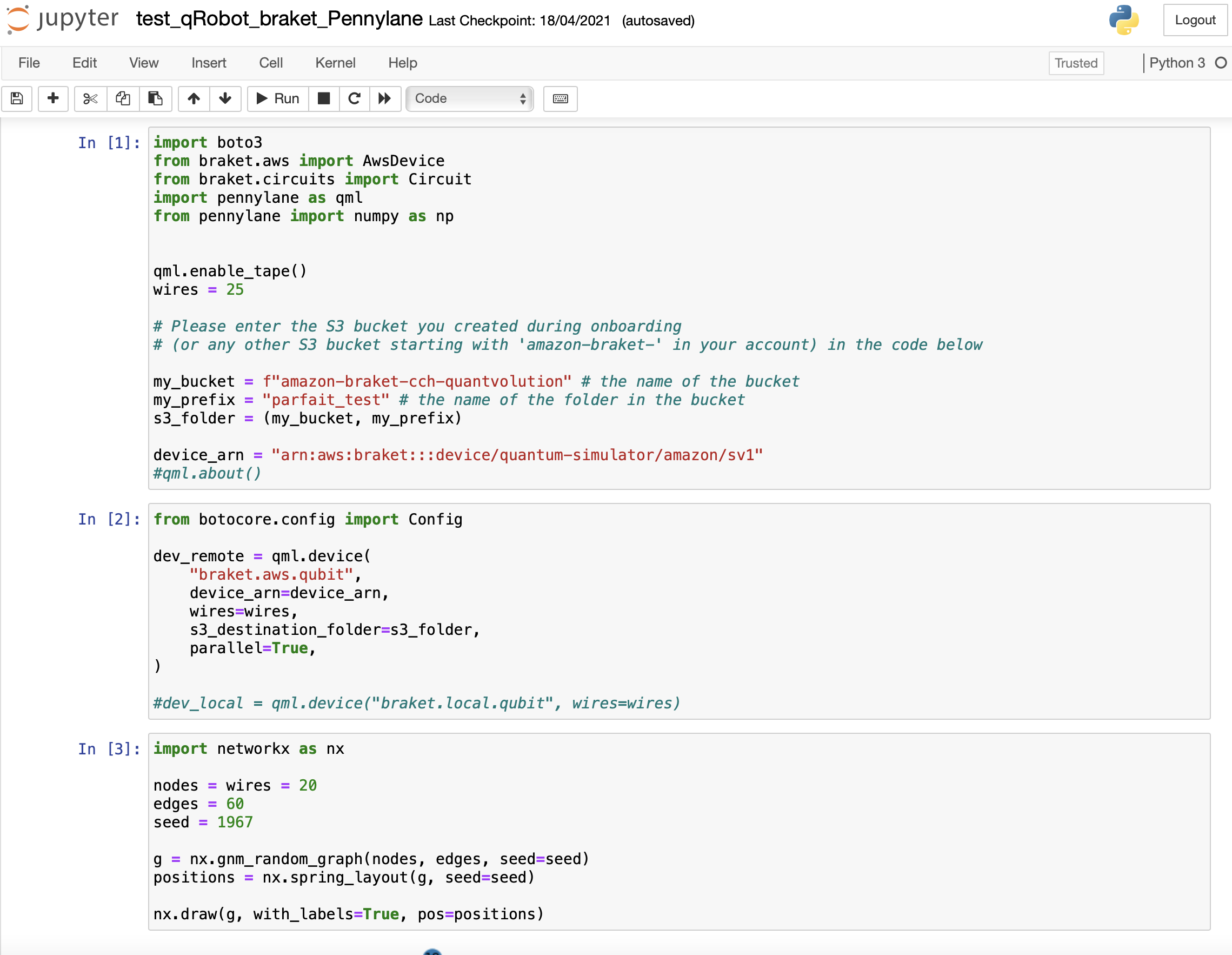}
\includegraphics[width=0.45\textwidth]{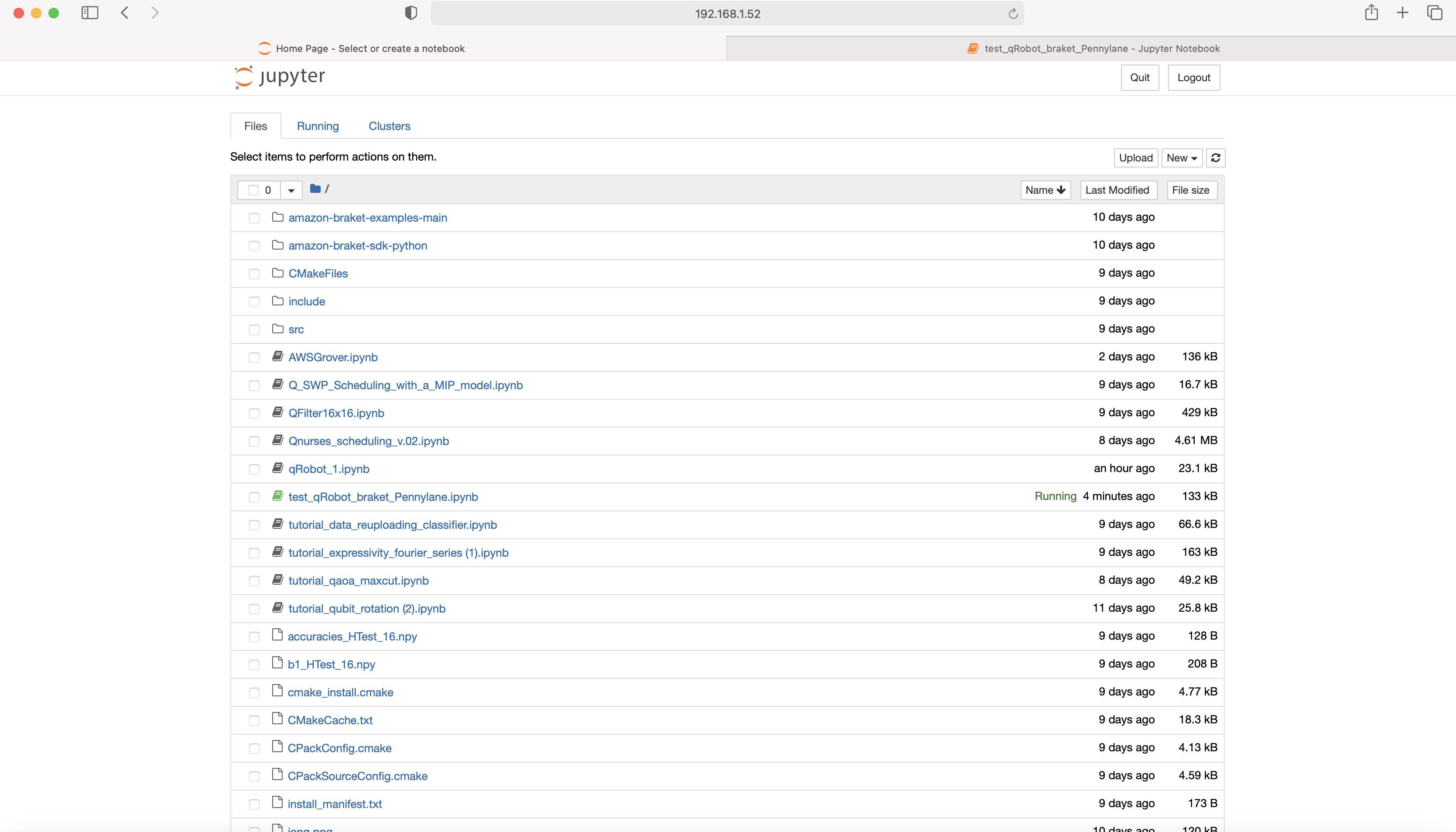}
\includegraphics[width=0.45\textwidth]{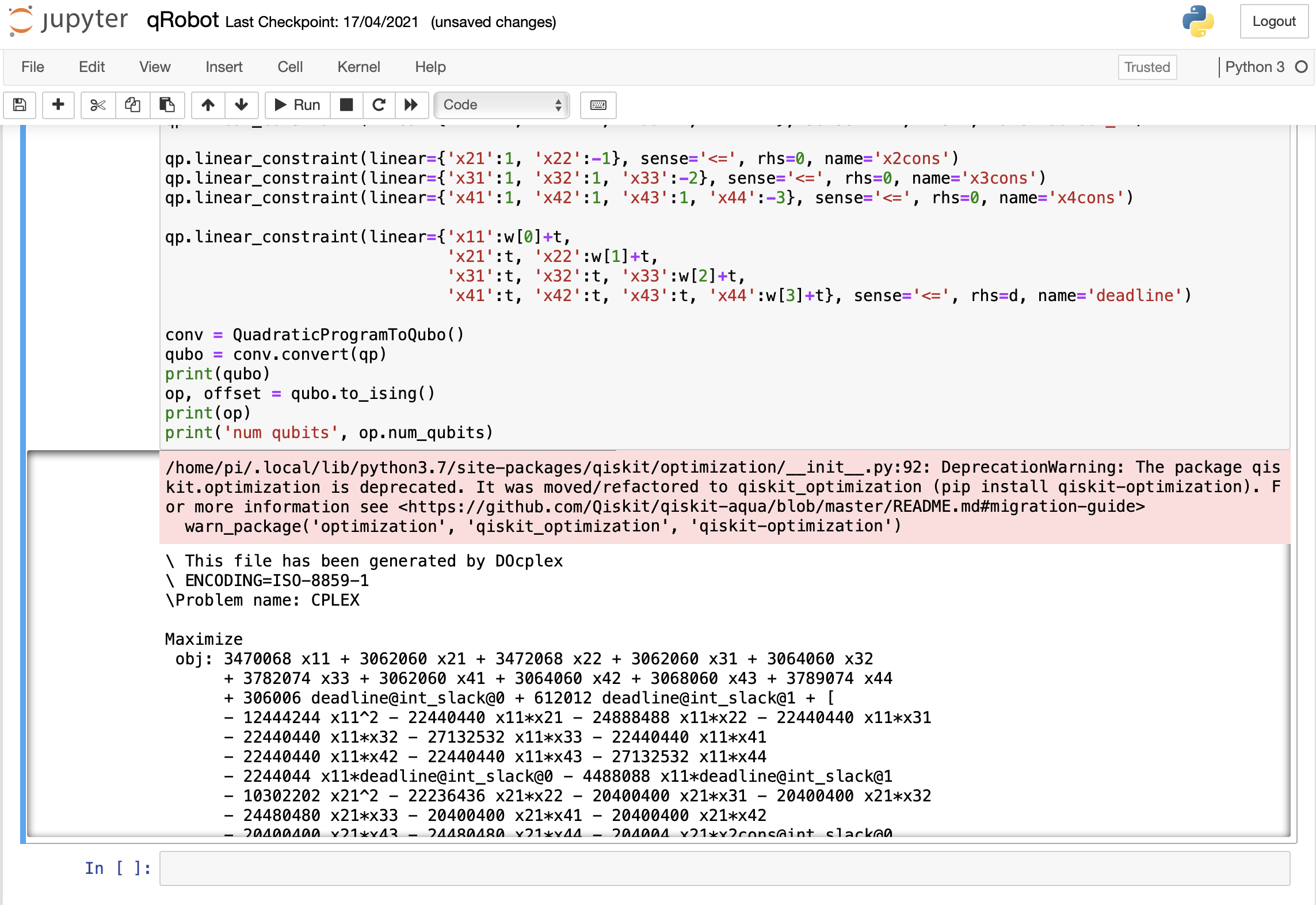}
\caption{}
\label{fig:JupyterWorking}
\end{figure}

Before analysing all the results of our proof of concept in detail, it is of the utmost importance that we validate its operation globally and affirm that qRobot meets our expectations and works as expected.
Let's split the results of this proof of concept in two. 1, the configuration and conversion results of the Raspberry Pi 4 in a quantum computing environment (Fig.\eqref{fig:installPackagesTerminal} to Fig.\eqref{fig:JupyterWorking}) and 2, the picking and batching algorithm results represented by tables \eqref{tab:Benchmark_qrobot_simulators} to \eqref{tab:Table_benchmark_qRobot_With_La} on one side and Fig.\eqref{fig:Results_qRobots_4} and \eqref{fig:Results_qRobots_7} on the other.

\begin{figure}[!ht]
\centering
\includegraphics[width=0.75\textwidth]{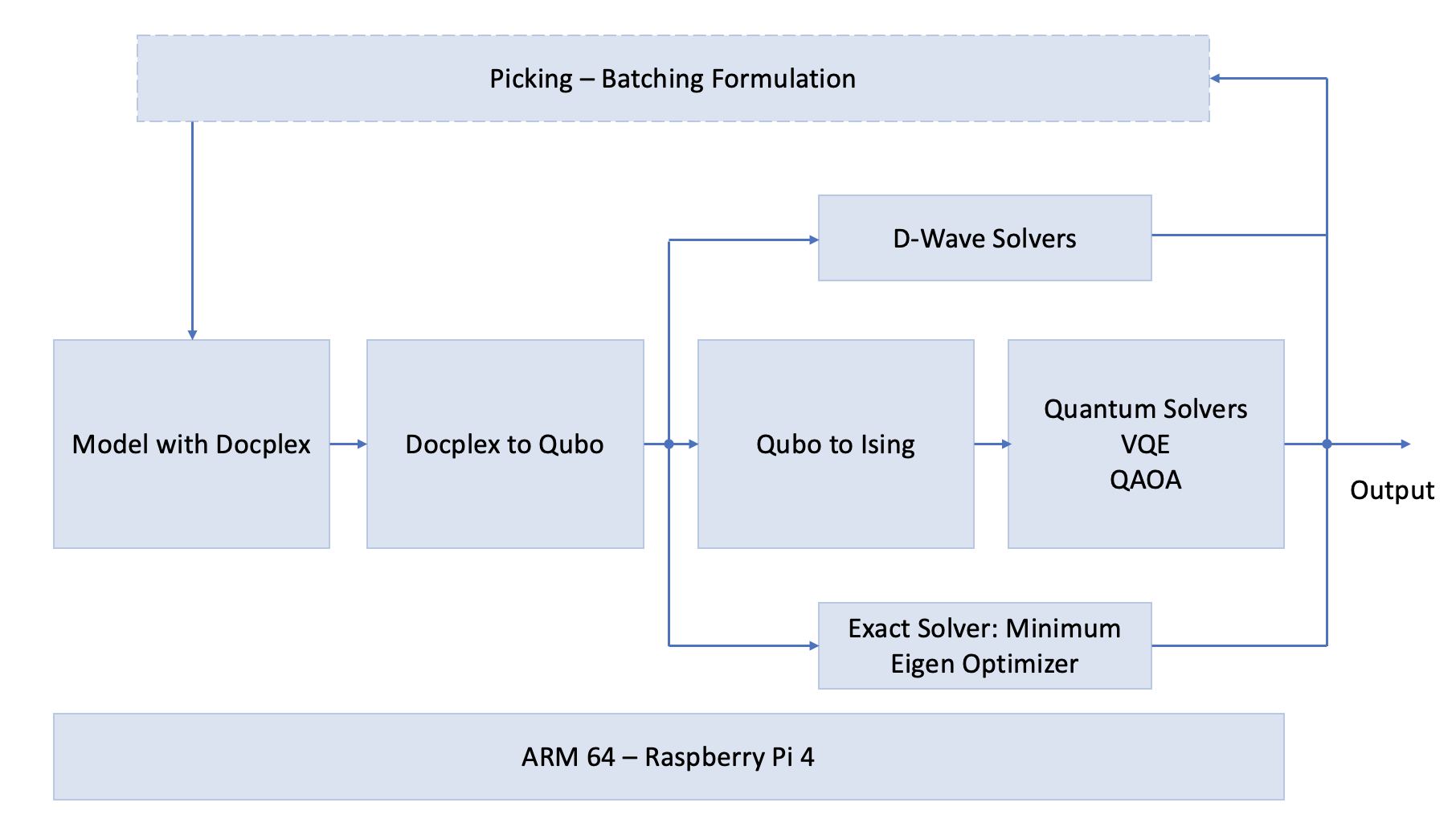}
\caption{qRobot operation diagram. This diagram shows all the necessary blocks and processes that allow modelling the picking-batching problem and its proper functioning on the Raspberry Pi 4.}
\label{fig:qRobot_Blocs}
\end{figure}

The block diagram (Fig.\eqref{fig:qRobot_Blocs}) summarises the result of the implementation of the qRobot. The first thing we did was determine the mathematical model of our problem. We then used the Docplex to model our objective function and its constraints. For our proof of concept, we used the Docplex library packages to move from Docplex to QUBO. We had two possible operations according to our objectives from this point on. First, we modelled the problem for computers based on quantum gate technology like IBMQ, and second, for annealing computers like D-Wave.
Our experiments used the Exact solver and VQE to test the Qiskit framework as samples based on quantum gates. But before using the VQE, we needed to map the QUBO model to the Ising model. Then, when we used the D-Wave computer, we only needed to reform the QUBO output list from Docplex to the QUBO format of the D-Wave computer.

Table \eqref{tab:Benchmark_qrobot_simulators} shows the experimentation results by setting the number of qRobots as their capacities (maximum load) at $1$ and $45$, respectively. We compared the execution time of our algorithm with different public access simulators on the market during this experimentation, solving the problem of picking and batching. We observed that, for issues of this nature, especially due to the number of qubits required in each scenario, the behaviour of the D-Wave is the desired one at the temporal level, comparing it with Gate based Quantum Computing. However, it should be taken into account that, for experiments with numbers of qubits less than $20$, the behaviour of these simulators is equated with the D-Wave. This experimentation helps to have a clear vision about the feasibility of this proof of concept.

Continuing with the analysis of the results, table \eqref{tab:Table_benchmark_qRobot} shows us the computational results of our picking and batching algorithm considering $ 1 $ qRobot through AWS-Braket and on the real quantum computer D-Wave Advantage\_system1\cite{Zaborniak_2021}. The time value is an average and does not count latency time, job creation, and job return time. 

We also analyse the latency time when running the algorithm from the qRobot to the quantum computer. The quantum computer was in Oregon (US) and our qRobot in Barcelona (Spain) and Segovia (Spain) in the tests we had done. Out of all the tests we have run, we had an average latency time of around $ 2 $ seconds plus all job management processes rising to roughly $ 8$ seconds. For the number of qubits greater than $30$, it is very convenient to use AWS-Braket (Advantage\_system1.1) instead of Qiskit or Pennylane for the number of qubits and the execution time; it is differentially better. This scenario makes the use of quantum in robots very viable. For tests with a value of $ M $ less than those in the table, the number of qubits is relaxed, and the execution time is improved. This leads us to normalise the weights of the batches. Since the number of qubits follows the formula $ K (n + 1) (n + 2) + K \lceil log_ {2} M \rceil $, where the $K \lceil log_ {2} M \rceil$ qubits are needed as ancillaries qubits.

We also analyse the quantum real-time execution deeply through table \eqref{tab:Table_benchmark_qRobot_With_La}. We have measured the execution time without counting the latency time, creating jobs, and returning the work.

Fig.\eqref{fig:Results_qRobots_4} offers us the algorithm results in different scenarios where we analyse some important cases, which helped us determine viable strategies within our proof of concept. In addition, it is important to note that our algorithm minimises the distance travelled and optimises the number of qRobots. Finally, the Fig.\eqref{fig:Results_qRobots_7} repeats almost the same scenario but now considering $ 7 $ items with the same number of qRobots.

\begin{table*}[t!]
\centering
\begin{tabular}{ |c|c|c|c|c|  }
 \hline
 \multicolumn{5}{|c|}{The benchmark of the qRobot’s algorithm in different quantum simulators.} \\
 \hline
 \# items & \# qubits & DWave - Time(s)  & IBMq - Time(s) & Pennylane - Time(s) \\
 \hline
 2   & 18 & 1.92  &  1.89  & 1.94 \\
 3   & 26 & 3.2  &  737.46  & 656.93 \\
 4   & 36 & 4.88 &  -  & - \\
 5   & 48 & 7.60 &  -  & - \\
 6   & 62 & 11.16 &  -  & -  \\
 7   & 78 & 15.89 &  - & - \\
 8   & 96 & 21.72 &  -  & -\\
 9   & 116 & 30.18 &  - & -\\
 10  & 138 & 43.29 &  - & - \\
 11  & 162 & 53.28 &  - & - \\
 12  & 188 & 63.45 &  - & - \\
 \hline
\end{tabular}
\caption{ In this experimentation, both the number of qRobots and their capacities (maximum load) are fixed and are worth $1$ and $45$ respectively. We compare the execution time of our algorithm in the different public access simulators in the market, solving the picking and batching problem. We see that for issues of this nature, and especially for the number of qubits required in each scenario, the behaviour of the D-Wave is the desired one at the temporal level, comparing it with technologies based on quantum gates. However, it should be noted that for the experiments on numbers of qubits less than $20$, the behaviour of these simulators is equated with the D-Wave.
This experimentation helps to have a clear vision about the feasibility of this proof of concept.}
\label{tab:Benchmark_qrobot_simulators}
\end{table*}

\begin{table}[!t]
    \centering
    \begin{tabular}{c|c|c|c|c|c|c}
         &  & & AWS-Braket\cite{AWS_Braket} & IBMq\cite{Qis21} &  Pennylane\cite{bergholm2020pennylane} \\
        \#\ items & Capacity & \#\ qubits & Avg Time (s) & Avg Time (s) &  Avg Time (s)  \\
         \hline
        2  & 15 & 10 & $11.23$ & $0.053$ & $0.041$ \\
        3  & 15 & 16 & $22.96$ & $0.40$ & $0.27$ \\
        4  & 15 & 24 &$33.07$ & $537.46$ & $480$ \\
        5  & 15 & 34 &$57.93$ & $-$ & $-$ \\
        6  & 15 & 46 &$118.41$  & $-$  & $-$ \\
        7  & 15 & 60 &$145.83$ & $-$ & $-$ \\
        8  & 15 & 76 &$296.81$ & $-$ & $-$ \\
        9  & 15 & 94 &$335.64$& $-$ & $-$ \\
        10 & 25 & 115 & $427.36$ & $-$ & $-$ \\
        11 & 25 & 137 & $650.25$ & $-$ & $-$ \\
        12 & 25 & 161 & $908.71$ & $-$ & $-$ \\
    \end{tabular}
    \caption{ Table of the computational results of our picking and batching algorithm on only $1$ qRobot. The value of time is an average and includes the waiting time, queue, execution and return of the solution. In the case of K is equal to $2$ for $9$ items with the qRobot capacity equal to $15$, the number of qubits is $188$. The execution time is on AWS Braket and on the D-Wave Advantage\_system1 quantum computer. We can realise that there is a latency time in executing the algorithm from the qRobot to the real quantum computer. In the tests we've done, the quantum computer is in the US West (Oregon). Of all the tests that we have done, we have had an average latency time of about $2$ plus all the work management processes that rises more or less to about $8$ seconds. For the number of qubits exceeding 30, it is very convenient to use AWS-Braket (Advantage\_system1.1)\cite{Zaborniak_2021} instead of Qiskit or Pennylane. By the number of qubits and the execution time, that is differentially better. This scenario makes the use of quantum in robotics very viable. For the tests with a value of $M$ lower than those in the table, the number of qubits is relaxed, and the execution time is improved. This leads us to normalise the weights of the batches. Since the number of qubits follows the formula $K(n+1)(n+2) + K \lceil log_{2}M\rceil$.}
    \label{tab:Table_benchmark_qRobot}
\end{table}

\begin{table}[!t]
    \centering
    \begin{tabular}{c|c|c|c|c|c|c}
         &  & & AWS-Braket\cite{AWS_Braket} & IBMq\cite{Qis21} &  Pennylane\cite{bergholm2020pennylane} \\
        \#\ items & Capacity & \#\ qubits & Avg Time (s) & Avg Time (s) &  Avg Time (s)  \\
         \hline
        2  & 15 & 10 & $0.13$ & $0.053$ & $0.041$ \\
        3  & 15 & 16 & $0.31$ & $0.40$ & $0.27$ \\
        4  & 15 & 24 &$1.69$ & $537.46$ & $480$ \\
        5  & 15 & 34 &$7.93$ & $-$ & $-$ \\
        6  & 15 & 46 &$11.31$  & $-$  & $-$ \\
        7  & 15 & 60 &$22.30$ & $-$ & $-$ \\
        8  & 15 & 76 &$36.11$ & $-$ & $-$ \\
        9  & 15 & 94 &$54.01$& $-$ & $-$ \\
        10 & 25 & 115 & $80.40$ & $-$ & $-$ \\
        11 & 25 & 137 & $139.67$ & $-$ & $-$ \\
        12 & 25 & 161 & $195.60$ & $-$ & $-$ \\
    \end{tabular}
    \caption{ In this table, we only consider the running time of the quantum algorithm on the real quantum computer from the qRobot (Advantage\_system1.1 \cite{Zaborniak_2021}), not counting latency time, job creation, and job return time.}
    \label{tab:Table_benchmark_qRobot_With_La}
\end{table}

\section{Discussions}\label{sec:qRobot_Discussions}
We have achieved that, given a warehouse with a single robot, a list of several products with their respective loads and a list of batches, our system minimises the distance to collect all the products and prepare the batches. Furthermore, this formulation solves the order in which the robot could manage all the products and make the packs pass through the Depot. 
Another important achievement this approach offers is that each robot makes a single trip. However, it is possible to band the code so that if we find ourselves in a situation where there are many batches to create and only a few robots to do the picking, these robots can be made to make the necessary trips if we have $ k $ qRobots that make at most one trip (we will never need more with $n$ batches). In this way, we will obtain all the packages for trips we are interested in doing. A more understandable way of explaining it would be that when the first qRobot has finished its journey, it should only be ordered to do the one that would have made the qRobot $ k + 1 $, which does not exist and so on with all the qRobots $ k + 2 $, $ k + 3 $, $ k + 4 $ etc. until all scheduled batches are finished.

In addition to the processor, quantum computing simulation is closely related to memory. It takes up memory to simulate a quantum computer, but the quantum computer does not need that memory, so it is assumed that it will be better.
In this proof of concept, we got the following results using 8GB of RAM on the Raspberry Pi 4. The algorithm of collection and generation of packages take between $2$ and $450$ seconds to generate the batches and picking. If you want the qRobot to do all these tasks, we must calculate the path before forming the packs.
That said, we must remember that if we want to recalculate new routes when the robot has already left, we must consider a lower latency time but close to said interval. A possible solution would be to choose a Raspberry with more RAM capacity. For example, if we had a 64GB Raspberry Pi, this time would be cut to 2/8, and it would take approximately 56.25 seconds (less than a minute) to create the batches. However, in this era of quantum computing, it is not representative to compare times. The computational differences will be noticed when the problems begin to grow, not on the small scales we are currently dealing with.

Effective viability for today's warehouses would consist of splitting the tasks of the robots and having a qRobot that centralises all the requests and passes them to the fleet of $ n $ qRobots so that they collect the products belonging to each batch.

We also performed tests and developed a system to model the problem and run it on a Dwave.
Despite the optimisation of the algorithm, the number of necessary qubits ($K(n+1)(n+2) + K \lceil log_{2}M\rceil$) and the need for low latency make this code adapted to the Annealing model. For this reason, we have prepared the Raspberry PI to run D'wave directly and under Amazon-braket-ocean-plugin. 
This scenario could have a "reasonable" latency for low data volume. Things that today, computers based on quantum gates cannot offer.

\section{Conclusions and further work}\label{sec:qRobot_Conclusions}

As we have seen, the problem raised throughout this work offers us an efficient way of managing a series of $K$ qRobots to collect a set of orders, optimising the number of robots used. The provided approach applies to a "central computer" capable of carrying out all the calculations and giving each robot orders. However, when we begin to deal with very large problems both in the number of products and the number of robots, the number of qubits required will tend to grow too large. A possible solution is to distribute the calculation of a central computer to each robot so that each one has to calculate its route given a list of products to be collected. In this case, the equations of the problem would not change, just take $ K = 1 $ for each qRobot and apply the technique mentioned at the beginning of the discussion. Although it may not be possible to reach the best solutions, this process of distribution of the calculation would suppose a significant computational cost reduction despite the need to create the batches beforehand. This search for batch creation will be studied in future projects.
On the other hand, it is important to note that the problem dealt with has a QUBO-type formulation, which allows it to be executed in annealing-type quantum computers. This makes a big difference in today's era (NISQ) as we have managed to work with $ ~ $ 200 qubits versus the $ ~ $ 30 qubits that we would have with a gate-based quantum computer. Finally, note that the defined problem minimises the total distance the robots travel, making it worthwhile for not all the robots to come out. For future lines, we will address the same problem. Still, We will continue to try to reduce the total times instead of the distance travelled (as done in this previous work\cite{Atc202}) since this situation is also very important in warehouse logistics.

\section{Summary}
In this section, we have seen how SWP has become widespread through qRobot. This generalisation was carried out through a proof of concept on a Raspberry Pi 4 and used the D-Wave framework (from AWS-Braket platform), Pennylane and Qiskit, to solve the formulation. This is how we wanted to respond to the universality of the algorithm. In the next section, we will see an overview of this thesis work's results, discussions, and conclusion, which has led us to formulate a combinatorial optimisation project with hard constraints, its resolution both Top-down and through QML well as its generalisations through the qRobot. All this to answer the initial hypothesis.

\part{Conclusions}

%\chapter{The results and discussions}

\newpage
\graphicspath{{./media/}}

\chapter{Results} \label{sec:phd_results}
\section{Introduction}
In this section, we will present the results of this thesis.

Figure \eqref{fig:Phd_overview}  helps us to understand the results of this thesis. The results of this thesis work have been broad and positive. In section \eqref{sec:SWP_results}, we can see the detailed results of the SWP in the various scenarios designed.

One of the first results that we would like to highlight is the formulation (\eqref{SWP_FORM_eq} with \eqref{weight_SWP_eq} and \eqref{time_window_SWP_eq}) of the SWP, after its implementation and testing with five different techniques.

We have tested our algorithm on VQE, QAOA, Numpy Minimum Eigensolver classical, GroverOptimiser, CplexOptimiser, Numpy Minimum Eigensolver for QUBOs, Backtracking and CP-SAT Solver from google on the \textit{ibmq-16-melbourne v1.0.0}, \textit{ibmq-qasm-simulator (up to 32 qubits)} with COBYLA and SLSQP as the classical optimiser. We can consider QAOA as a particular case of VQE. This is one of the reasons we can apply the Hamiltonian of the Ising model of our proposed formulation directly to QAOA almost without modification. The Figures \eqref{fig:10_SW_schedules} to \eqref{fig:SWP_Quantum_MinimumEigenOptimizer} show the results of our algorithm once we executed our algorithm under the IBMQ. The optimal visit considers the hours of visits to form the optimal schedule. We have done several experiments with the QML defining different scenes using shot configuration. With our quantum machine, we can configure the number of repetitions of each circuit for the sampling we need. With that, we will be doing machine learning on circuit design for each shot, and when the loop ends, we will get to the ground state energy. Consequently, we solved our problem by creating one quantum circuit for each shot, and the best circuit will be the one that optimises our Social Workers' Problem. We also prepared the solution in QUBO form for the annealing computing.

In section \eqref{sec:qCBR_result}, we can see the detailed results of the qCBR of the different scenarios. These very promising results lead us to consider our approach to this age of quantum computing. Nevertheless, let us summarise the qCBR results here.

Figures \eqref{fig:Bench_C} to \eqref{fig:Bench_VQE_Init} offer the implementation outcomes performed in \textit{Qibo} \cite{qibo} and \textit{Qiskit} \cite{mckay2018qiskit,Qis21} to identify the best model architecture and represent functions similar to qCBR.

Tables \eqref{tab:results_qCBR_SW_Full} to \eqref{tab:results_qCBR_SW_4x3SW} present the global results of qCBR solving the SWP. With table \eqref{tab:results_qCBR_SW_Full}, the result of the different tested scenarios can be observed by varying the number of patients, social workers, and the quantum circuit's depth to see the global hit number of the qCBR. In table \eqref{tab:results_qCBR_5x4SW}, we can observe the resolution of the SWP, taking into account five patients, four social workers and setting the depth of the quantum circuit to eight. With this scenario, it can be observed how the behaviour of the qCBR is a function of the number of cases. It is seen how the system, after a threshold of 240 cases already resolved, begins to give very satisfactory results.
Table \eqref{tab:results_qCBR_SW_4x3SW} repeats the steps of table \eqref{tab:results_qCBR_5x4SW} with the only change being the  input data of the number of patients and social workers.
Tables \eqref{tab:results_CBR_SW_5x4SW} to \eqref{tab:results_CBR_KNN_SW_Full} show the result of the implementation of the classical CBR leveraged on ANN and KNN to solve the SWP.

Tables \eqref{tab:results_qCBR_5x4SW} and \eqref{tab:results_qCBR_SW_4x3SW} show the best performance at the average accuracy level of qCBR for the classic (Table \eqref{tab:results_CBR_SW_5x4SW} and \eqref{tab:results_CBR_SW_4x3SW}). Table \eqref{tab:results_CBR_KNN_SW_Full}, \eqref{tab:results_CBR_NN_SW_Full} and \eqref{tab:results_qCBR_SW_Full} show the degree of scalability of the qCBR as a function of the variation in the number of patients and social workers. It has also been seen that qCBR is much better shared with overlapping just as we had hoped to demonstrate.

One of the greatest achievements of this qCBR is the formula for calculating the number of solutions to know the class number of our quantum classifier. This formula has been key for the optimal functioning of our proposal.

 $$N_{SOL_{SWP}}=\frac{1}{m!} \sum _{k=0}^{m-1} \left( -1 \right) ^{k} {m \choose m-k} \left( m-k \right) ^{n}.$$

With  $n$ the number of patients and  $m$ the number of social workers and knowing that the appearance orders patients in the schedule (from earliest to latest,  $n_{1}$ the patient with the earliest plan and  $n_{k}$ the patient with the newest program).

We have designed qRobot to generalise the SWP and complement our learning of the annealing platform, in this case, D-Wave. Since to date, all our experimentation has been on quantum gate-based technology (IBM and Pennylane).

In section \eqref{sec:qRobot_result}, we can also explore here\cite{qRobotP}, the results of the qRobot as a generalisation of the SWP. In this case, the results to create the platform that has been used to develop the proof of concept.

Let's split the results of this proof of concept in two. 1, the configuration and conversion results of the Raspberry Pi 4 in a quantum computing environment (Fig.\eqref{fig:installPackagesTerminal} to Fig.\eqref{fig:JupyterWorking}) and 2, the picking and batching algorithm results represented by tables \eqref{tab:Benchmark_qrobot_simulators} to \eqref{tab:Table_benchmark_qRobot_With_La} on one side and Fig.\eqref{fig:Results_qRobots_4} and \eqref{fig:Results_qRobots_7} on the other.

The block diagram (Fig.\eqref{fig:qRobot_Blocs}) summarises the result of the implementation of the qRobot. The first thing we did was determine the mathematical model of our problem. We then used the Docplex to model our objective function and its constraints. For our proof of concept, we used the Docplex library packages to move from Docplex to QUBO. We had two possible operations according to our objectives from this point on. First, we modelled the problem for computers based on quantum gate technology like IBMQ, and second, for annealing computers like D-Wave.
Our experiments used the Exact solver and VQE to test the Qiskit framework as samples based on quantum gates. But before using the VQE, we needed to map the QUBO model to the Ising model. Then, when we used the D-Wave computer, we only needed to reform the QUBO output list from Docplex to the QUBO format of the D-Wave computer.

Let us remember that for the qRobot, we created one universal quantum computer on a Raspberry Pi 4. The main idea is to have a mobile robot for the batching and picking problem. The steps to convert the Raspberry Pi 4 into a "quantum computer" are in the \cite{qRobotP}. 

Table \eqref{tab:Benchmark_qrobot_simulators} shows the experimentation results by setting the number of qRobots as their capacities (maximum load) at $1$ and $45$, respectively. We compared the execution time of our algorithm with different public access simulators on the market during this experimentation, solving the problem of picking and batching. We observed that, for issues of this nature, especially due to the number of qubits required in each scenario, the behaviour of the D-Wave is the desired one at the temporal level, comparing it with Gate based Quantum Computing. However, it should be taken into account that, for experiments with numbers of qubits less than $20$, the behaviour of these simulators is equated with the D-Wave. This experimentation helps to have a clear vision about the feasibility of this proof of concept.

Continuing with the analysis of the results, table \eqref{tab:Table_benchmark_qRobot} shows us the computational results of our picking and batching algorithm considering $ 1 $ robot through AWS-Braket and on the real quantum computer D-Wave Advantage\_system1\cite{Zaborniak_2021}. Again, the time value is an average and does not count latency time, job creation, and job return time.

One of the outstanding results of the qRobot is to have achieved a formulation that gives us the minimum (most efficient) number of qubits to carry out the proposed task.
$ K (n + 1) (n + 2) + K \lceil log_ {2} M \rceil $, where the $K \lceil log_ {2} M \rceil$ qubits are needed as ancillaries qubits. With  $ M $  the maximum load of the qRobots and $ K $  the number of qRobots available.

We carried out further experimentation, and more details can be found here \eqref{sec:qRobot_result} should the reader desire. 

\begin{figure}
    \centering
    \includegraphics[width=0.75\textwidth]{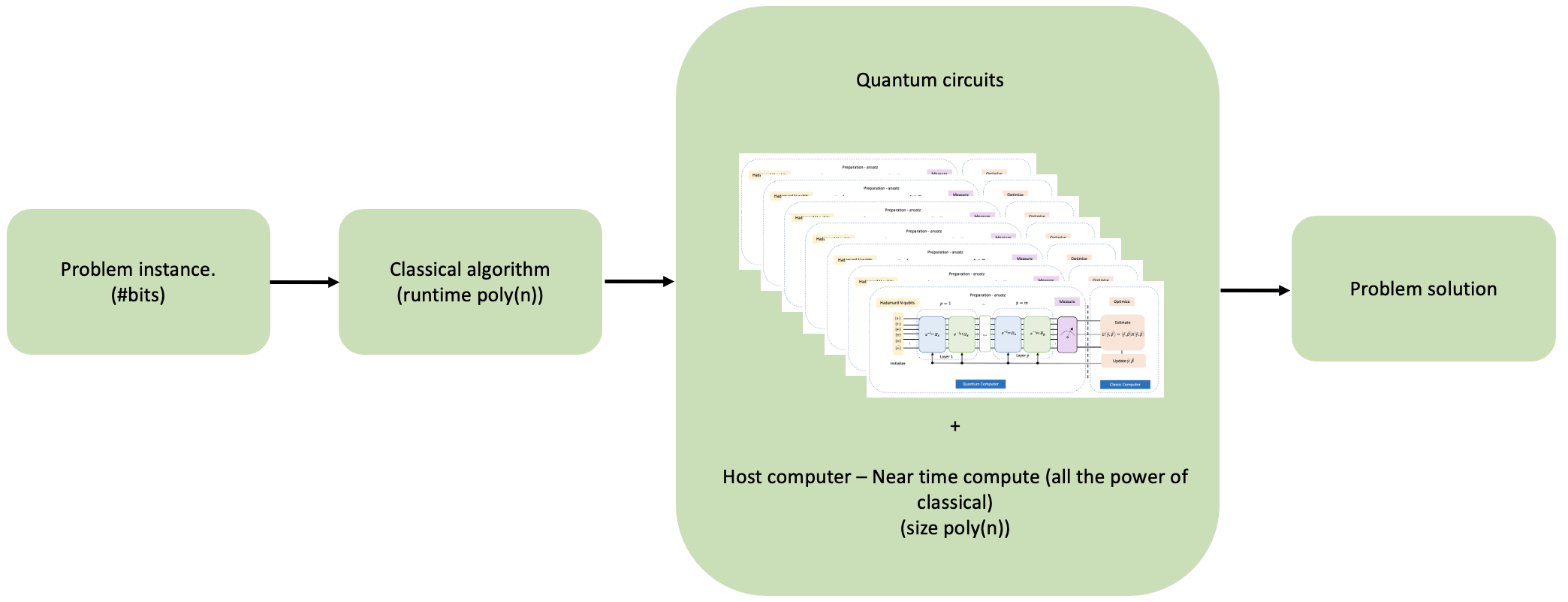}
    \caption{In this figure, we can observe a workflow of a quantum algorithm. First, let us remember that a quantum computer or hardware only executes a quantum circuit. But we can use classical algorithms to generate a description of our quantum circuit efficiently, so it would be smart to use it. Our quantum algorithm is born to execute or answer problems/approaches that are very expensive for classical computing.
A good quantum algorithm would be the polynomial scale with qubits. This figure shows that we can clarify it polynomially given a problem and follow these guidelines.
}
    \label{fig:q-algorith-WF}
\end{figure}

We would like to highlight the importance of figure \eqref{fig:q-algorith-WF} as our algorithm's workflow. We have concluded using classical algorithms to efficiently generate a description of our quantum circuit because our quantum algorithm is designed to execute the very expensive approaches for classical computing. \\

This thesis work has led to two final Master's projects. One of the projects was for the Master of Computer Science and the other, the Master of Mathematics of the University of Valladolid. The result can be seen in this reference \cite{gonzalez2021gps}. The objective of the second project was to capture a real scenario based on the VRP problem, in which we proposed a way to solve it based on quantum models. Subsequently, we analysed the advantages and disadvantages that quantum variational algorithms can offer us compared to the most popular algorithms in classical computing. This Master's was with the University of Ramon Llull.
\newpage
\graphicspath{{./media/}}

\section{Discussion}\label{sec:PhD_Discussions}

In these sections \eqref{sec:SWP_Discussions}, \eqref{sec:qCBR_Discussions} and \eqref{sec:qRobot_Discussions}, we can see the different discussions made throughout this thesis work. These discussions have motivated us to continue investigating the formulation of the SWP, its implementation, experimentation, and the various comparisons to the generalization of the SWP with the qRobot. In addition, our thesis hypothesis has motivated all the following discussions (could quantum computing solve efficiently Hard Constrained Optimization Problems?).

Let us summarise the discussion of this thesis. The first part of this section will be on the validation of the SWP algorithm. 

We can observe that, while Backtracking (Fig. \eqref{fig:Backtracking_Behaviour}) and the CPLEX present an exponential behaviour as the number of patients increases, the VQE, QAOA algorithm (\eqref{fig:VQE_Behaviour_}), without taking into account the cost of evaluation and calibration of the algorithm, have a logarithmic growth. This, as the number of patients grows, will offer more considerable advantages than a classic algorithm, such as Backtracking, since its temporary cost will be much lower for more complex problems \eqref{fig:Backtracking_Behaviour} and \eqref{fig:VQE_Behaviour_}.  

We did several tests and comparisons to validate our formulation and benchmark with other optimisers (quantum and classic) that are Backtracking, QAOA, OR \_Tool from google. Many comparisons were also made to fine-tune the algorithm. The results are shown in table \eqref{tab:Backtracking_behaviour} and figures \eqref{fig:SWP_Classical_BackT_Mon-Wed} and \eqref{fig:SWP_Classical_BackT_Th-Fri}.

Since we wanted to make a strong comparison, such as many qubit numbers, we had to change the back-end from the \textit{statevector\_simulator} to \textit{qasm\_simulator} and the real quantum computer. We made this change because the superposition calculations of the quantum states, or complex amplitudes that the simulator provides to keep track of the algorithm,  overthrow the computer. It must be said that this has nothing to do with the efficiency of the quantum computer but is a tool that facilitates Qiskit to learn the steps that the algorithm follows.

The results obtained with the \textit{qasm\_simulator} are very similar to the results of the real quantum computer, as seen in the figures \eqref{fig:SWP_Quantum_ADMM} and \eqref{fig:SWP_Quantum_MinimumEigenOptimizer}.

We can observe (Fig. \eqref{fig:Comp_Analysis_Friday}) in this case that the Quantum exact solver gives the same result as the Backtracking. So instead, the VQE provides a solution with but with a patient change.
This can be a problem if the patients will not change social workers daily. Another limitation can be the learning process of the top-down approach.
Another improvement would be to reduce the computational cost of the top-down algorithm.

The evaluation of the algorithm on an \textit{ibmq\_16\_melbourne v1.0.0} from IBM was fulfilled. With any change in the input, variables are mapped proportionally to our cost variable within a time window.
We would like to add that the suggested formulation \eqref{descomp_Funct_Objective_eq} and \eqref{T_W_SWP_eq} is not only specific to the proposed problem. It can be used to solve any family planning, scheduling and routing problem related to a list of tasks, restrictions, and allocation of resources on location and time. The test performed and showed in Fig. \eqref{fig:Analysis_epsilon} allows us to see the behaviour of our formulation with the variation of the correction factor  $\varepsilon$. We understand how our time window $T_{ij}= \left(  \tau_{i-} \tau_{j} \right)$ adapts perfectly at the extremes to the cost variable in the distance. This achievement is due to the chosen quadratic function \eqref{T_W_SWP_eq}. We want it to be adapted so that our resultant function weights together the short distances and time, and the long distances and late times.
Other functions can be studied to have a test bench to compare the final results.
QAOA, like VQE, takes a qubit operator from the Hamiltonian of the Ising model. The only mapping that is archived when QAOA builds a variational form based on the qubit operator, and thus, what we understood was that it does not change the original qubit operator input. Figure \eqref{fig:Comparing_5_SW_VQE_QAOAS} and \eqref{fig:Comparing_10_SW_VQE_QAOA} reveal the comparison work between VQE and QAOA algorithms for the same configuration parameters. After several tests, we confirmed that our algorithm takes less execution time with the QAOA than the VQE and requires fewer samples for optimal solutions. But in many cases, we have to increase the seed value to get a reasonably stable result.

If the reader wants to see the graphs in detail and all the discussions, we invite them to section \eqref{sec:SWP_Discussions}.

One of the first things we want to say about the qCBR is that it works very well and is a valid method to solve the SWP through artificial intelligence. However, it must be noted that it is initially required to execute the Top-down algorithm a few times to have a set of initial cases. For more details, we encourage the reader to refer to the \eqref{sec:qCBR_Discussions} section.

One of the issues to comment on is the improvement observed in (figure\eqref{fig:Bench_C}), both two and eight-dimensional classifiers. Due to the small number of depths, but many more parameters, the 8-dimensional classifiers have an average of about 25$\%$ of improvements over the 2-dimensional ones. With this result, in the case of not wanting an accuracy of around 95$\%$, shallow depth could be used, and computation time saved, depending on the problems.
Despite all these improvements, it is essential to highlight some areas of further improvement. For example, in the intelligence system that allows deciding the proposed solution now, the average of the \textit{Initial\_point} of each solution class samples' \textit{Initial\_point} is used. However, it could still be seen based on the predicted solution, which \textit{Initial\_point} is the most suitable for the answer to propose. Thus, the cases to be re-used could be better classified.

Also, one of the improvements is to train the classifier with noisy data further so that the qCBR can adapt to real past situations that adjust to the new situation. Because, in practice, there is usually no past case strictly the same as a new one.

The last improvement is to generalise the qCBR to serve various problems (betting, financial, software maintenance, human reasoning, etc.). For this, it would be a success to focus on the design in memory of cases, so that different data sizes can be indexed and, above all, can train the classifier with several other data models.

In the qRobot's discussions \eqref{sec:qRobot_Discussions}, we can see that this formulation solves the order in which the robot could manage all the products and make the batches passing through the depot.

Another important achievement this approach offers is that each robot makes a single trip. However, it is possible to band the code so that, if we find ourselves in a situation where there are many batches to create and only a few robots to do the picking, these robots can be made to make the necessary trips if we have $ k $ qRobots that make, at most, one trip (we will never need more with $n$ batches). In this way, we will obtain all the packages for trips we are interested in doing. A more understandable way of explaining it would be that when the first qRobot has finished its journey, it should only be ordered to do the one that would have made the qRobot $ k + 1 $, which does not exist and so on with all the qRobots $ k + 2 $, $ k + 3 $, $ k + 4 $ etc. until all scheduled batches are finished.

One of the most important points to comment on here is that our formulation minimises the distance. But, in the case of wanting to reduce the time, we should use a strategy facing the qRobots, but make some small adjustments in the formulation.

We encourage the reader to review each discussion to get a broader idea of the work developed.

One topic of discussion is \textit{Quantum Error Correction} also known as QEC. In this thesis work, we focus on implementing our quantum algorithms that obey some transpiler of each framework; thus, leaving us with little margin in correcting errors. A future approach would be to see how we base our algorithms on the design of all quantum circuits so we can control even more the issue of error controls.

The formulations proposed for both the SWP and qRobot can be seen as an answer to a class of generic problems written in the form of an objective function with some restrictions. This methodology followed during this thesis can be encapsulated as a framework proposal to solve a large part of these problems. But for this, it would be necessary to study the Lagrange multiplier and different forms of penalty function in detail, and inequalities constraints. The latter will be the future lines of this work.

Also, qCBR can be seen as a framework to solve any case-based optimization problem based on a decision-making solver. We want to say that instead of the problem that we are solving here, we can use the qCBR as proposed in this thesis, to solve issues like Max-Cut, Max-Clique, etc. Only the classifier should be well-trained with the new data sample space.
\newpage
\graphicspath{{./media/}}

\section{Conclusions and Further Work}\label{sec:PhD_Conslusions}
\subsection{Conclusions}

Before describing the conclusion of this thesis summarised by Figure \eqref{fig:Phd_overview}, We would like to highlight my satisfaction with the positive results. From the hypothesis question to the SWP\cite{Atc201,Atc20,Atc202,atchadeadelomou2021quantum,adelomou2020formulation,QSWPGithub,AllSWPGithub,Par20,AtchadeAdelomou2020,adelomou2020using,Adelomou2020} and qRobot\cite{a14070194} proofs of concepts. The gold brush has been put in the search for temporary efficiency with the work of EVA \cite{alonsolinaje2021eva}.

With this PhD work, we have responded to the initial hypothesis "could quantum computing solves CSP?" carried this research work in a field where the scientific community is debating, how to join the ML with quantum computing to create more powerful QML tools. We have developed several quantum algorithms for this NISQ era, hybrid algorithms that use the classical part's potential, and quantum mechanics' power to solve combinatorial optimisation problems, such as the Social Workers' Problem.
We have seen that in addition to solving this problem in a top-down way, we can use a methodology based on Artificial Intelligence to solve and offer the resolution of these problems with our qCBR as human beings do. 

A qRobot to generalise an SWP and as a quantum computing approach in mobile robot order picking and batching problem solver optimisation is proposed and tested.

Let us delve deeply into the conclusion of this work. First, we recall that this thesis aims to study the state of the art and design of a series of quantum algorithms on combinatorial optimisation problems. To do this, we define and create a combinatorial optimisation problem with hard restrictions and solve it in two different approaches. In a top-down approach, we formulate the problem mathematically, both classical and quantum. Our proposed formulation allows us to design a strategy to take advantage of this NISQ era (Few useful qubits, decoherence, limitations in quadratic techniques with inequality constraints, etc.). Although it seems specific, said formulation is later generalised to encompass more generic applications. We also make a comparative analysis of five different implementations and discussed them during this thesis work.
Continuing under the top-down approach, we have thoroughly analysed the Ising model and the QUBO model to solve quadratic problems. We have also studied Cplex optimisation tools such as IBM's docplex under their Qiskit framework. The Qiskit optimisation tool included the generic quadratic programs that help to model any optimisation problems. Many converters are also involved in mapping the problem to solve it in the correct input format. Converters like \textit{Inequality to equality} are used to map inequality constraints to equality constraints with additional slack variables. The \textit{Integer To Binary} converter is useful in the case of the need to convert integer variables into binary variables and the corresponding coefficients. \textit{Linear Equality to Penalty} helps to convert equality constraints into additional terms of the SWP's objective function. Finally, the \textit{Quadratic QUBO} converter is used as a container for all the above converters.
We have thoroughly analysed the Minimum Eigen Optimiser and the ADMM optimiser as alternatives to VQE and QAOA.
We have also developed some guidelines to make it easier for prospective PhD or graduate students to take their first firm steps in quantum computing.

On the other hand, we have statistically solved the problem with the qCBR algorithm machine learning technique.

To do this, we have developed the classical and quantum CBR and observed the outstanding performance of qCBR compared to its classical counterpart in precision, scalability and average tolerance to an overlapping dataset. We've mitigated some of the standard and classic CBR issues. With the design that has been proposed, it has been possible to measure situations of difficult similarity between cases. Despite the non-linear and overlapping attributes, the classifier has been endowed with characteristics that arrive at two similar topics that may seem quite different because they have different values in the features, but are not very significant. For example, in the VQE with \textit {Initial\_point}, we can have different \textit {Initial\_point} associated with each training class sample with the same class. With the "\textit {Initial\_point}" averaging technique, it is possible to solve this problem by providing the qCBR to distinguish the similarity between cases. Another problem that qCBR mainly solves is the time required to classify a new topic.

With the results of the two implementations (classical and quantum CBR), it is observed that the classical CBR designed with the KNN behaves better for some determined cases (table \eqref{tab:results_CBR_KNN_SW_Full}). It is seen that the system has not finished learning thoroughly (table \eqref{tab:results_CBR_SW_5x4SW} and \eqref{tab:results_CBR_SW_4x3SW}) contrary to qCBR (table \eqref{tab:results_qCBR_5x4SW} and \eqref{tab:results_qCBR_SW_4x3SW}). This is due to the precision of its classifier, without forgetting the important contribution of its synthesis system.

Another improvement that qCBR introduces is when it comes to retaining cases, implementing a retention system that maintains case models and that, together, synthesise the real and most important information. This would not be possible if a variational quantum classifier were not designed.

Following the thread of this thesis work, we generalise our formulation of the SWP to carry out efficient management of robots by substituting robots for social workers and pick-up requests for patients. We call this generalisation qRobot.

The problem raised throughout this work offers us an efficient way to manage a series of $ K $ qRobots (instead of social workers) to collect a set of orders, optimising the number of robots used. The approach provided is applied to a "central computer" capable of doing all the calculations and then giving the orders to each qRobot. However, when we start to deal with very big problems both in the number of products and the number of robots, the number of required qubits will tend to grow too large. One possible solution is to distribute the calculation from a central computer to each robot so that each has to calculate its route given a list of products to collect. In this case, the equations in the problem would not change, just take $ K = 1 $ for each qRobot and apply the technique mentioned at the beginning of the discussion. Although it may not be possible to arrive at the best solutions, this process of distributing the calculation would mean a significant reduction in computational cost despite the need to create batches in advance. This search for batching will be explored in future projects.
On the other hand, it is important to point out that the problem dealt with has a QUBO-type formulation, which allows its execution in annealing-type quantum computers. This makes a huge difference in the current era (NISQ), as we have managed to work with $ ~ $ 200 qubits versus the $ ~ $ 30 qubits we would have with a gate-based quantum computer. Finally, note that the defined problem minimises the total distance the robots travel, making it worthwhile for not all robots to exit. For future lines of research, we will tackle the same problem. Still, we will continue to try to reduce total times instead of distance travelled (as was done in this previous work \cite{Atc202}) as this situation is also very important in warehouse logistics.

We can also conclude that the scenario designed to answer our thesis question is adequate. It is true that, with quantum computing, we can efficiently solve combinatorial optimisation problems with hard constraints and the qRobot, as a generalisation of the SWP, as the top-down resolution approach, and the qCBR as the resolution approach based on quantum machine learning, has been validated by the quantum scientific community.

To finish this thesis work, I would like to highlight part of my pedagogical contribution to my research group (Data Science 4 Digital Society) and the quantum community as part of the objective of this thesis. Also, how could you monetise this thesis work? Also, I would like to highlight all that I have learned, since one should not end any trip without growing from it. Thank you, Elisabet Golobardes i Ribé! Thank you, Xavier Vilasis Cardona!

\subsection{Further Work}
From a general point of view, we will continue investigating optimisation problems and creating quantum algorithms to benefit society and our research group for the future line of work. Unfortunately, error correction is expensive at the level of the number of qubits, and thus, there is a very large gap between what is needed and what is currently available to build a useful quantum computer. One line of future inquiry that we set ourselves is investigating more about Quantum Error Correction (QEC) and finding more efficient codes.

Before diving into details into each part of our contribution in this work, one line of future reasearch that we find interesting is to develop all our libraries and tutorials for the Pennylane, Qiskit, Qibo and AWS-Braket platforms and frameworks in the form of a license.

One of our future lines of work for the SWP, is to take advantage of the improvements in the number of qubits of quantum computers to validate, based on more tests, the computational complexity that we achieved.
This validation should confirm if we have achieved exponential improvement comparable to the tests done with Backtracking shown from figure \eqref{fig:VQE_Behaviour_} and \eqref{fig:Backtracking_Behaviour}.

One of the qCBR future lines of research, is when retaining cases, implementing a retention system that maintains model cases and that, together, synthesise the real and most important information.
One of the improvements to consider is the implementation of quantum ICA. In this way, the classical ICA analysis's complexity cost will be significantly reduced. Also, counting that the PCA is saved since we have an 8-dimensional classifier, the complexity of the qCBR would be that of the classifier, plus some setup constants. 
The other exciting line of future work, is to design the memory of cases using the quantum technique of random-access memory (qRAM) \cite{qRAM_} to improve the memory of stored instances. It is worth noting that the qCBR does not present a \textbf{barren plateau} problem due to the low numbers of qubits; qCBR is thus a shallow quantum circuit as advocated by the \textbf{barren plateau theorem} \cite{cerezo2021cost}.

One of the future lines of research for the qRobot, is to adapt the formulation of the qRobot for optimisation of time, not distance. This means minimising the time it takes for the robots to collect all the orders or, what is equivalent, minimising the maximum of the lengths that each robot travels.
Another line of future work, is to develop an algorithm \textit{Quantum Annealing}, focusing on exposing the mathematics that helps to understand how to perform good modelling.
Additional work in this would be to try to apply all the improvements achieved in this work to the IBMQ \textit{Docplex} tool \cite{docplex} as a library.

A clear future line of work, is to design a framework to encapsulate the methodology we defined here to solve optimisation problems. But for this, it would be necessary to study the Lagrange multiplier and different forms of penalty function in detail, and inequalities constraints.

\subsection{International contribution and talent pool}

This section will detail our contributions to journals, conferences, and publications of the scientific community and the group of talents we met to develop this project.

In the SWP, we have contributed with the formulation (\eqref{sec:SWPPaperQubo}, \eqref{sec:SWP-Vector-QUBO} and its first intent of generalisation \eqref{sec:SWP-first-generalization}) of the SWP mathematically, both classically and quantum, and made a comparative study on the IBMQ quantum computer. In this paper, we have validated our formulation, both for QUBO and Ising in the scientific community:
\begin{itemize}
    \item \textit{Atchade-Adelomou P., Golobardes Ribé E., Vilasís Cardona X. Formulation of the Social Workers' Problem in quadratic unconstrained binary optimisation form and solve it on a quantum computer. Journal of Computer and Communications. 2020 Nov 5. 
    DOI: 10.4236/jcc.2020.811004}
\end{itemize}

\textbf{Abstract}
The problem of social workers visiting their patients at home is a class of combinatorial optimisation problems and belongs to the type of problems known as NP-Hard. These problems require heuristic techniques to provide an efficient solution in the best of cases. In this article, in addition to providing a detailed resolution of the Social Workers' Problem using the Quadratic Unconstrained Binary Optimization Problems (QUBO) formulation, an approach to mapping the inequality constraints in the QUBO form is given. Finally, we map it in the Hamiltonian of the Ising model to solve it with the Quantum Exact Solver and Variational Quantum Eigensolvers (VQE). The quantum feasibility of the algorithm will be tested on IBMQ computers.

\vspace{3mm}

This paper uses the VQE to create a solution based on the machine learning technique. In addition, we have participated in the international conference on Hybrid Artificial Intelligent Systems.
Our most relevant contributions in this work have been put in the following paper.

\begin{itemize}
    \item \textit{ Atchade-Adelomou P., Golobardes Ribé E., Vilasís Cardona X. (2020) Using the Variational-Quantum-Eigensolver (VQE) to Create an Intelligent Social Workers Schedule Problem Solver. In: de la Cal E.A., Villar Flecha J.R., Quintián H., Corchado E. (eds) Hybrid Artificial Intelligent Systems. HAIS 2020. Lecture Notes in Computer Science, vol 12344. Springer, Cham. https://doi.org/10.1007/978-3-030-61705-9\_21}
\end{itemize}

\textbf{Abstract}
The scheduling problem of social workers is a class of combinatorial optimisation problems that can be solved in exponential time at best. Because it belongs to a class of the issues known as NP-Hard, which have a huge impact on our society, nowadays, the focus on the quantum computer should no longer be just for its enormous computing capacity, but also for the use of its imperfection (Noisy Intermediate-Scale Quantum (NISQ) era) to create a powerful machine learning device that uses the variational principle to solve the optimisation problem by reducing their complexity's class. We propose a formulation of the Vehicle Rooting Problem (VRP) with time windows to efficiently solve the social workers' schedule problem using Variational Quantum Eigensolver (VQE). The quantum feasibility of the algorithm will be modelled with docplex and tested on IBMQ computers.

\vspace{3mm}
In this work, we have collaborated with other techniques based on parameterised circuits to create an Intelligent social workers' schedule problem solver. The following paper has demonstrates our most relevant contributions to this work.
 
\begin{itemize}
    \item \textit{ Atchade-Adelomou P., Golobardes Ribé E., Vilasís Cardona X. Using the Parameterized Quantum Circuit combined with Variational-Quantum-Eigensolver (VQE) to create an Intelligent social workers' schedule problem solver. arXiv preprint arXiv:2010.05863. 2020 Oct 12.}
\end{itemize}

\textbf{Abstract}
The social worker scheduling problem is a class of combinatorial optimisation problems that combines scheduling with routing issues. These types of problems with classical computing can only be solved, in the best of cases, in an approximate way and significantly when the input data does not grow considerably. Today, the focus on the quantum computer should no longer be only on its enormous computing power but also on the use of its imperfection for this era (Noisy Intermediate-Scale Quantum (NISQ)) to create a powerful optimisation and learning device that uses variational techniques. We had already proposed a formulation and solution to this problem using the capacity of the quantum computer. In this article, we present some broad results of the experimentation techniques. Above all, we propose an adaptive and intelligent solution that efficiently recalculates the schedules of social workers, taking into account new restrictions and changes in the initial conditions, using a case-based reasoning system and the variational quantum eigensolver based on a finite-depth quantum circuit that encodes the ground state of the Hamiltonian of social workers.
The quantum feasibility of the algorithm will be modelled with docplex and tested on IBMQ computers.

\vspace{3mm}

In the case of qCBR's contributions, we have created a dataset (from the SWP formulation) with an overfitting and overlapping problem creating a quantum classifier based on the universality theorem. We had to perform a very efficient quantum multiclass classifier and synthesiser to do this. We have also developed a formula for calculating the number of classes of the multiclass classifier. As a sum of our contributions, we have proposed a quantum CBR (qCBR) to solve optimisation problems as we would usually solve them as human beings.
The most relevant contributions have been put in the following paper. Nevertheless, we can find details of the qCBR at reference \eqref{sec:App-CBR}.

\begin{itemize}
    \item \textit{Atchade-Adelomou P, Casado-Fauli D, Golobardes Ribé E, Vilasís Cardona X. quantum Case-Based Reasoning (qCBR). arXiv preprint arXiv:2104.00409. 2021 Apr 1.}

\end{itemize}

\textbf{Abstract}
Case-Based Reasoning (CBR) is an artificial intelligence approach to problem-solving with a good record of success. This article proposes using Quantum Computing to improve some of the key processes of CBR defining a Quantum Case-Based Reasoning (qCBR) paradigm. The focus is on designing and implementing a qCBR based on the variational principle that improves its classical counterpart in terms of average accuracy, scalability and tolerance to overlapping. A comparative study of the proposed qCBR with a classic CBR is performed for the case of the Social Workers' Problem as a sample of a combinatorial optimisation problem with overlapping. The algorithm's quantum feasibility is modelled with docplex, tested on IBMQ computers, and experimented with the Qibo framework.

\vspace{3mm}

In the case of qRobot, we have contributed both to the creation of a "universal" platform to create a quantum computer in a raspberry PI 4 and, on the other hand, to create an efficient formulation for this quantum era that generalises the SWP and solves the problem of batching and picking within from a warehouse. We are pleased to say that the paper, in addition to being published in MDPI magazine \textit{Algorithms}, it was also on the cover. It is worth saying that these improvements could lead to improvements over TSP and VRP.
Our most relevant contributions on the qRobot have been put in the following paper. Nevertheless, details can be found of our formulation at reference \eqref{sec:qRobot_problem_formulation} and we can also find how we have turned the Raspberry PI 4 into a universal computer.

The description step by step about how installing and running Pennylane, AWS-Braket, D-Wave-Ocean, Qiskit, on a Raspberry Pi 4 under the ARM64 \cite{jiang2020power} operating system turn it into a quantum computing simulator and use it to access real quantum computers from IBMQ \cite{Qis21,mckay2018qiskit}, AWS-Braket \cite{AWS_Braket}, D-Wave \cite{dwave_computer}, and Regetti \cite{sete2016functional} can be found here \cite{qRobotP}. These frameworks and packages are required for the proof of concept that we propose \cite{qRobotP}.

\vspace{3mm}

\begin{itemize}
    \item \textit{Atchade-Adelomou, Parfait, Guillermo Alonso-Linaje, Jordi Albo-Canals, and Daniel Casado-Fauli. 2021. "qRobot: A Quantum Computing Approach in Mobile Robot Order Picking and Batching Problem Solver Optimisation" Algorithms 14. https://doi.org/10.3390/a14070194}
\end{itemize}  

\textbf{Abstract}
This article aims to bring quantum computing to robotics. A quantum algorithm is developed to minimise the distance travelled in warehouses and distribution centres where order picking is applied. For this, a proof of concept is proposed through a Raspberry Pi 4, generating a quantum combinatorial optimisation algorithm that saves the distance travelled and the batch of orders to be made. In case of computational need, the robot will be able to parallelise part of the operations in hybrid computing (quantum + classical), accessing CPUs and QPUs distributed in a public or private cloud. Before this, we must develop a stable environment (ARM64) inside the robot (Raspberry) to run gradient operations and other quantum algorithms on IBMQ, Amazon Braket, D'wave and Pennylane locally or remotely. The proof of concept will run in the above quantum environments.
\vspace{3mm}

In the case of QFilter, we have contributed to quantum computing by proposing a method to, on the one hand, bring quantum computing closer to the community of classical computing (machine learning) and, on the other, offer a new approach to solving convolutional networks (CNN) by changing only the classical scalar product by a quantum and keeping the rest of the classical code. This approach and contribution have made us a finalist of about 2000 participants of the Xanadu Hackathon (with the AWS-Braket contest, Google).
The following paper demonstrates our most relevant contributions over the QFilter.

\begin{itemize}
    \item \textit{Atchade-Adelomou P., Alonso-Linaje G. Quantum Enhanced Filter: QFilter. arXiv preprint arXiv:2104.03418. 2021 Apr 7.}
\end{itemize} 

\textbf{Abstract}
Convolutional Neural Networks (CNN) are used mainly to treat problems with many images characteristic of Deep Learning. This work proposes a hybrid image classification model to take advantage of quantum and classical computing. The method will use the potential that convolutional networks have shown in artificial intelligence by replacing classical filters with variational quantum filters. Similarly, this work will compare other classification methods and the system's execution on different servers. Finally, the algorithm's quantum feasibility is modelled and tested on Amazon Braket Notebook instances and experimented on Pennylane's philosophy and framework.

\vspace{3mm}

In the case of EVA, we have contributed to quantum computing by proposing a new method to calculate the quantum expected value based on a single quantum circuit, compared to the VQE technique that divides a given Hamiltonian into several small ones of very shallow depth. The importance of this contribution is enormous in this quantum age, but it will be even more so for the future since we exclusively make one call to the quantum computer. Also, it is worth saying that EVA is intended for chemical applications. With EVA, we could efficiently calculate the energy of some molecules to facilitate the discovery of new drugs.  
Our most relevant contributions regarding EVA can be found in the following paper.

\begin{itemize}
    \item \textit{Alonso-Linaje G, Atchade-Adelomou P. EVA: a quantum Exponential Value Approximation algorithm. arXiv preprint arXiv:2106.08731. 2021 Jun 16.}
\end{itemize} 

\textbf{Abstract}
VQE is currently one of the most widely used algorithms for optimising quantum computers' problems. However, a necessary step in this algorithm is calculating the expectation value given a state, which is calculated by decomposing the Hamiltonian into Pauli operators and obtaining this value for each of them. We have designed an algorithm capable of figuring this value using a single circuit in this work. In addition, a time cost study has been carried out, and it has been found that in certain more complex Hamiltonians, it is possible to obtain a good performance over the current methods.

\subsection{Schedule/Work plan}
In this session, we will observe a series of tasks and time management that summarise this thesis work. Like any project, this needed a work plan, and specific planning, even though part of its tasks could not be limited due to the subject's demand,  and, above all, because of the little information found when we started this journey in the centre of the quantum computing world.
Table \eqref{tab:PhD_Tasks_dev} shows this PhD task development. Also, we can observe below the detail of the tasks that have been carried out.

\begin{table}[!h]
\centering
%\begin{scriptsize}
\caption {PhD Tasks development}
\label{tab:PhD_Tasks_dev}
\begin{tabular}{| c | c | c | c |c | c | c |c | c | c |c | c | c |c | }
\hline 
  month & 0 & 3 & 6 & 9 & 12 & 15 & 18 & 21 & 24 & 27 & 30 & 33 & 36 \\
\hline 
T1  & \cellcolor{green} & \cellcolor{green} & \cellcolor{green} &   &   &   &   &   &   &   &   &   &   \\
\hline
T2  &  &  &  & \cellcolor{green} & \cellcolor{green}  &   &   &    &  &  &  &  &  \\
\hline
T3  &  &  &  &  &  & \cellcolor{green}  &   &  &  &  &  &  & \\
\hline
T4  &  &  &  &  &  &  &  \cellcolor{green} &   &   &   &   &   & \\
\hline
T5  &  &  &  &  &  &  &  \cellcolor{green} &   &  &  &  &  & \\
\hline
T6  &  &  &  &  &  &  &   \cellcolor{green}&   &  &  &  &  & \\
\hline
T7  &   &  &  &  &  &  &  &  \cellcolor{green}  &  \cellcolor{green} &  &  &  & \\
\hline
T8  &   &  &  &  &  &  &  &   &  &  \cellcolor{green} &  \cellcolor{green} &  & \\
\hline
T9  &   &  &  &  &  &  &  &   & \cellcolor{green}  & \cellcolor{green} & \cellcolor{green} &  \cellcolor{green} & \cellcolor{green} \\
\hline
T10  &   &  &  &  &  &  & \cellcolor{green} & \cellcolor{green} & \cellcolor{green} & \cellcolor{green} & \cellcolor{green} &  & \\
\hline
T11  &   &  &  &  &  &  & \cellcolor{green} & \cellcolor{green} & \cellcolor{green} & \cellcolor{green} & \cellcolor{green} & \cellcolor{green}  &  \cellcolor{green} \\
\hline
T12  &   &  &  &  &  &  & \cellcolor{green} & \cellcolor{green} & \cellcolor{green} & \cellcolor{green} & \cellcolor{green} & \cellcolor{green}  &  \cellcolor{green} \\
\hline
\end{tabular}
%\end{scriptsize}
\end{table}

\begin{itemize}
\item \textbf{Task 1:} This task took place during the first six months of the work. Studying the Strengths, Weaknesses, Opportunities, and Threats (SWOT) was crucial for developing this work.
\item \textbf{Task 2:} This task took place during the first six months of the work. This task began with a solid study of the existing CSP algorithms and models. 
\item \textbf{Task 3:} This task took place during the first three months of the work. We learn for better comprehension of the linear, quadratic programming solver's functioning.
\item \textbf{Task 4:} This task took place during the first month of the work. We learned about TSP, JSP and VRP. State of the art research of Heuristic.
\item \textbf{Task 5:} This task took place during the first month of the work. State of the art of Systems and Hamiltonians concepts, Ising Hamiltonian Model, The Hamiltonian of a TSP, The Hamiltonian of a JSP and VRP.
\item \textbf{Task 6:} This task took place during the first month of the work. State of the art of the complexity class and Quantum Complexity class.
\item \textbf{Task 7:} This task took place during the first six months of the work. Design and Model, mathematically the Social Workers' Problem (SWP) as CSP according to state of the art and the ones proposed in the scope of the PhD work. We learned about the players involved in SWP, namely quantum computing frameworks, quantum computers, and quantum research groups. Third, Learn about how to map the SWP in quantum.
\item \textbf{Task 8:} This task took place during the first six months of the work. First, developing and implementing the SWP in this NISQ era. Solve the SWP with QML. This methodology will be based on clustering and classification techniques, resulting in a rule base concerning the SWP. An extensive set of simulation results will be used as the basis for applying the proposed QML method. Third, designing and implementing a quantum Case-Based Reasoning (qCBR) and testing and validating the developed models and approaches.
\item \textbf{Task 9:} Writing at least two papers for top-level conferences and six articles for SCI journals.
\item \textbf{Task 10:} Development and implementation of the QFilter.
\item \textbf{Task 11:} Development and implementation of the qRobot Platform and the picking and batching formulation.
\item \textbf{Task 12:} Development and implementation EVA.
\end{itemize}

\subsection{Pedagogical contribution}
As a PhD work, one of the most important parts besides validating the hypothesis, and  writing the contributions in scientific articles, is self-explanatory and can be reproduced: our job is self-consistent. That is why we have made our contributions academic and catalysts to attract new talent from collaborations in our group and team.

This section will highlight part of my pedagogical contribution to my research group.
\begin{itemize}
    \item Some guidelines and steps to solve combinatorial optimization problems in quantum.
    \item A documentation of the state of the art and compilation and access to the main tutorials of quantum frameworks
    \item a GitHub with the codes and experimentation environments
    \item Complete formulation of a real combinatorial optimization problem (SWP) with restrictions and mapping it in quantum.
    \item Step-by-step development of how to bring a classical to a quantum optimization problem (SWP)
    \item Design, implementation and results of the SWP top-down resolution and quantum Case-Based Reasoning
    \item Design, implementation and comparisons of a variational quantum classifier.
    \item Design, implementation of a universal quantum platform using Raspberry PI.
    \item Design, implementation and comparisons of picking and batching problem.
    \item co-accompany a group of collaborators in quantum computing for DS4DS.
    \item Co-direction of a TFGs in quantum computing.
\end{itemize}
\subsection{Contribution to business vision}
Today, we can already see a lot of changes in quantum computing companies. Several companies are already using this technology to offer consulting tasks at the optimization level or training at the introductory level. Our work goes a little further. However, our group DS4DS could particularise one of our formulations as a Fintech product and base it on the cloud. Developing a subscription/use business model; Pay peruse.

\subsection{Lessons learned (Hard Skills)}
Before ending this session, I would like to leave the evidence of what I learned at a cognitive level and, above all, to record what I have acquired at a skill level. I feel very fortunate to write what I have learned in this section.
\begin{itemize}
\item Quantum mechanics applied to quantum computing
\item Qiskit, Qibo, Pennylane, AWS-Braket, D-Wave, etc. programming environments
\item Mathematical formulations of optimization problems
\item Mapping of classical combinatorial optimization problems in quantum
\item Design and implement classical and quantum classifiers
\item Understand quantum Machine Learning.
\item Write articles
\item Scientific rigour
\item Quantum computing.
\item The era in which quantum computing is found.
\end{itemize}

\subsection{Consolidated skills (Soft Skills)}
Quantum computing is not one of the most intuitive concepts and the easy ones.  In this section I list the soft skills that I did to help me develop and consolidate during this this 3-year journey, to help tackle this complex subject.
\begin{itemize}
\item Self-motivation
\item Overcome the obstacles
\item Find my life
\end{itemize}
%\part{Annexes}
%\input{Annex}

%\bibliographystyle{unsrturl}
%\bibliographystyle{unsrt}
%\bibliography{main}
%\newpage
\printbibliography
\end{document}